\pdfoutput=1

\documentclass[%
    11pt,                   
    a4paper,                
    UKenglish,ngerman,      
    BCOR=5.2mm,               
    cdgeometry=no,          
    DIV=13,                 
    cdfont=false,           
    sfdefaults=false,
    twoside=false,           
    cdmath=false
]{tudscrreprt}

\usepackage[utf8]{inputenc}
\usepackage[T1]{fontenc}
\usepackage{lmodern}

\AfterCalculatingTypearea{%
  \addtolength{\oddsidemargin}{-2.6mm}%
}
\recalctypearea

\usepackage{babel}
\usepackage{isodate}

\usepackage{amsmath,amssymb,mathtools,amsfonts,bbm,mathrsfs}
\usepackage{physics,slashed,tensor}
\usepackage{bbold,simplewick,stackrel}
\numberwithin{equation}{chapter} 
\allowdisplaybreaks

\usepackage{enumitem}

\usepackage{xcolor,graphicx}
\usepackage{makecell}

\usepackage{amsthm}

\newtheoremstyle{definitionwithcolon}
  {\topsep}   
  {\topsep}   
  {\normalfont}  
  {0pt}       
  {\bfseries} 
  {:}         
  {.5em}      
  {}          
\theoremstyle{definitionwithcolon}
\newtheorem{definition}{Definition}[chapter]
\newtheorem{example}{Example}[chapter]
\newtheorem*{definitionsandpropositions}{Definitions and Propositions}

\newtheorem*{solution}{Solution}
\newtheorem*{algorithm}{Algorithm}
\newtheorem*{condition}{Conditions}
\newtheorem*{singlecondition}{Condition}
\newtheorem*{remark}{Remark}
\newtheorem{option}{Option}        

\newenvironment{optionblock}{%
  \setcounter{option}{0}%
}{}
\newcommand{\baseoption}{}%
\newenvironment{suboptions}{%
  \edef\baseoption{\number\numexpr\value{option}+1\relax}
  \setcounter{option}{0}
}{%
  \setcounter{option}{\baseoption}
}
\newenvironment{optionblockroman}{%
  \begingroup
  \setcounter{option}{0}%

}{%
  
  \endgroup
}

\newtheoremstyle{plainwithcolon}
  {\topsep}   
  {\topsep}   
  {\itshape}  
  {0pt}       
  {\bfseries} 
  {:}         
  {.5em}      
  {}          
\theoremstyle{plainwithcolon}
\newtheorem{theorem}{Theorem}[chapter]
\newtheorem{proposition}{Proposition}[chapter]
\newtheorem{lemma}{Lemma}[chapter]
\newtheorem{corollary}{Corollary}[chapter]

\makeatletter
\renewenvironment{proof}[1][\proofname]{%
  \par\pushQED{\qed}%
  \normalfont
  \topsep6\p@\@plus6\p@\relax
  \trivlist
  \item[\hskip\labelsep
        \itshape
    #1:]\ignorespaces
}{%
  \popQED\endtrivlist\@endpefalse
}
\makeatother

\usepackage{chngcntr}
\counterwithout{footnote}{chapter}

\usepackage[backend=biber,style=ieee,citestyle=numeric-comp]{biblatex}
\DeclareBibliographyCategory{ownpubs}
\addtocategory{ownpubs}{Stockinger:2023ndm,Ebert:2024xpy,vonManteuffel:2025swv,Belusca-Maito:2023wah,Aebischer:2023nnv,Kuhler:2024fak}

\usepackage[automark]{scrlayer-scrpage}
\clearpairofpagestyles
\ihead{}                  
\ohead{\headmark}         
\automark[chapter]{chapter}
\KOMAoptions{headsepline=0.5pt}

\usepackage{hyperref} 
\hypersetup{
  pdftitle={Renormalisation of Chiral Gauge Theories with Non-Anticommuting gamma5 at the Multi-Loop Level},
  pdfauthor={Matthias Weißwange}
}
\usepackage[capitalise]{cleveref}
\usepackage{xurl}

\newcommand{\intx}{\int d^4x}
\newcommand{\Dintx}{\int d^Dx}
\newcommand{\projR}{\mathbb{P}_{\mathrm{R}}}
\newcommand{\projL}{\mathbb{P}_{\mathrm{L}}}

\newcommand{\hypL}{\mathcal{Y}_L}
\newcommand{\hypR}{\mathcal{Y}_R}
\newcommand{\hypLR}{\mathcal{Y}_{LR}}
\newcommand{\hypRL}{\mathcal{Y}_{RL}}
\newcommand{\hypS}{\mathcal{Y}_S}

\newcommand{\myTr}{\mathrm{Tr}}
\newcommand{\myTrbig}[1]{\mathrm{Tr}\big(#1\big)}

\newcommand{\trYY}[2]{\mathrm{Tr}\big(Y_{#1}Y_{#2}\big)}
\newcommand{\trYbarYbar}[2]{\mathrm{Tr}\big(Y_{#1}^{\dagger}Y_{#2}^{\dagger}\big)}

\newcommand{\Sfull}{S}
\newcommand{\LaFull}{\mathcal{L}}
\newcommand{\DeltaHat}{\widehat{\Delta}}

\newcommand{\STIopD}{\mathcal{S}_{D}}

\newcommand{\vgb}{\mathcal{V}}
\newcommand{\myGenLe}{T}
\newcommand{\myGenQu}{\mathcal{T}}
\newcommand{\myGenScl}{\Theta}
\newcommand{\myxi}{\xi_{F}}
\newcommand{\myNL}{N_{L}}

\makeatletter
\newcommand{\grouplabel}[1]{%
  \edef\@currentlabel{\baseoption}
  \edef\@currentlabelname{Option~\baseoption}
  \phantomsection
  \label{#1}%
}
\makeatother

\begin{document}

\setkomafont{chapter}{\LARGE\bfseries\sffamily}
\setkomafont{section}{\Large\bfseries\sffamily}
\setkomafont{subsection}{\large\bfseries\sffamily}
\addtokomafont{title}{\Huge\bfseries\sffamily}
\setkomafont{captionlabel}{\bfseries}
\renewcommand*{\captionformat}{: \par}
\setcapindent{0pt}
\setcapwidth{\linewidth}


\pagenumbering{roman}
\setcounter{page}{1}

\selectlanguage{ngerman}

\faculty{Faculty of Physics}
\institute{Institute of Nuclear and Particle Physics}
\chair{Theoretical Particle Physics}
\date{27.11.2025}
\author{Matthias Weißwange}
\title{%
Renormalisation of Chiral Gauge Theories with Non-Anticommuting $\boldsymbol{\gamma_5}$ at the Multi-Loop Level
}
\thesis{diss}
\graduation[Dr. rer. nat.]{Doctor Rerum Naturalium}

\referee{Prof. Dr. Dominik Stöckinger \and Prof. Dr. Adrian Signer}

\maketitle

\clearpage
\thispagestyle{empty}
\null
\clearpage

\selectlanguage{UKenglish}

\TUDoption{abstract}{section,notoc,multiple}

\begin{abstract}[pagestyle=empty.tudheadings,option=nofill]
This thesis presents a comprehensive study of the renormalisation of chiral gauge theories in dimensional regularisation (DReg) at the multi-loop level.
We employ the mathematically consistent Breitenlohner-Maison/`t~Hooft-Veltman (BMHV) scheme with non-anticommuting $\gamma_5$, whose modified algebraic relations induce a spurious violation of gauge and BRST invariance.
A central focus is the systematic restoration of the broken symmetry, for which we provide a transparent and fully algorithmic procedure based on the quantum action principle.
A major achievement of this work is the complete 4-loop renormalisation of an Abelian chiral gauge theory --- the highest-order application of the BMHV scheme to date.
This calculation is made possible by an automated, high-performance computational framework incorporating several optimised algorithms.
Our results demonstrate that a rigorous, self-consistent treatment of $\gamma_5$ is feasible even at very high loop orders.
We further analyse dimensional ambiguities and evanescent details corresponding to different implementations of the regularisation, and identify practically efficient prescriptions for $D$-dimensional fermions and gauge interactions.
Building on these insights, we present the complete 1-loop renormalisation of the full Standard Model (SM) in the BMHV scheme, providing a first step towards a fully self-consistent multi-loop renormalisation of the SM and establishing a solid foundation for future high-precision electroweak phenomenology.

\nextabstract[ngerman]
Diese Dissertation präsentiert eine umfassende Untersuchung der Renormierung chiraler Eichtheorien in dimensionaler Regularisierung (DReg) auf dem Mehrschleifen-Niveau. 
Wir verwenden das mathematisch konsistente Breitenlohner-Maison/`t~Hooft-Veltman (BMHV) Schema mit nicht-antikommutierendem $\gamma_5$, dessen modifizierte algebraische Relationen zu einer Verletzung der Eich- und BRST-Invarianz führen. 
Ein zentraler Schwerpunkt ist die systematische Wiederherstellung der gebrochenen Symmetrie, wofür wir ein transparentes und vollständig algorithmisches Verfahren auf Basis des Quantenwirkungsprinzips vorstellen.
Ein wesentliches Ergebnis dieser Arbeit ist die vollständige 4-Schleifen-Renormierung einer Abelschen chiralen Eichtheorie --- die bislang höchste erreichte Schleifenordnung im BMHV-Schema.
Diese Berechnung wird durch eine automatisierte Hochleistungsrechenumgebung ermöglicht, die mehrere optimierte Algorithmen integriert. 
Unsere Resultate zeigen, dass eine selbstkonsistente Behandlung von $\gamma_5$ auch bei sehr hohen Schleifenordnungen praktisch durchführbar ist.
Darüber hinaus analysieren wir dimensionsbedingte Ambiguitäten und evaneszente Details, die unterschiedlichen Realisierungen der Regularisierung entsprechen, und identifizieren effiziente Vorschriften für $D$-dimensionale Fermionen und Eichwechselwirkungen. 
Aufbauend auf diesen Erkenntnissen präsentieren wir die vollständige 1-Schleifen-Renormierung des gesamten Standardmodells (SM) im BMHV-Schema und bereiten damit den ersten Schritt zu einer vollständig selbstkonsistenten Mehrschleifen-Renormierung des SM sowie eine solide Grundlage für zukünftige Hochpräzisionsphänomenologie elektroschwacher Physik.

\end{abstract}

\clearpage
\thispagestyle{empty}
\null
\clearpage

\markboth{}{} 

\cleardoublepage

\KOMAoptions{headsepline=false} 
\section*{List of Publications}
This thesis is mainly based on the content of the following publications: \phantom{\cite{Stockinger:2023ndm,Ebert:2024xpy,vonManteuffel:2025swv,Belusca-Maito:2023wah,Aebischer:2023nnv,Kuhler:2024fak}}
\begin{refsection}
  \nocite{Stockinger:2023ndm,Ebert:2024xpy,vonManteuffel:2025swv,Belusca-Maito:2023wah,Aebischer:2023nnv,Kuhler:2024fak}
  \printbibliography[category=ownpubs,heading=none]
\end{refsection}

\cleardoublepage

\section*{Acknowledgements}

An dieser Stelle möchte ich all denjenigen danken, die zum Gelingen dieser Arbeit beigetragen und mich während meiner Promotionszeit begleitet haben.

Zunächst gilt mein besonderer Dank meinem Betreuer und Doktorvater Dominik Stöckinger für die hervorragende Betreuung, die große wissenschaftliche Freiheit und die stets offene Tür für Fragen und Diskussionen.
Seine außerordentliche fachliche Kompetenz und sein tiefes Verständnis der Materie haben mich während der gesamten Promotion beeindruckt und meine wissenschaftliche Entwicklung in hohem Maße bereichert.
Für seinen Rat, seine Geduld und seine kontinuierliche Unterstützung bin ich ihm sehr dankbar.

Ein ebenso großer Dank gilt Andreas von Manteuffel, der während meines dreimonatigen Forschungsaufenthalts an der Universität Regensburg wie ein zweiter Betreuer für mich wurde.
Diese Zeit war außerordentlich lehrreich und produktiv.
Es ist alles andere als selbstverständlich, dass er mich darüber hinaus weiterhin intensiv betreut und begleitet hat.
Für die vielen anregenden Diskussionen und die ausgezeichnete Zusammenarbeit bin ich sehr dankbar.

Weiterhin möchte ich mich bei allen aktuellen und ehemaligen Mitgliedern unserer Arbeitsgruppe bedanken --- von Bachelor- und Masterstudenten über Doktoranden bis hin zu Postdocs.
Der stetige wissenschaftliche Austausch, zahlreiche hilfreiche Gespräche und viele heitere Momente haben maßgeblich zu einer angenehmen und motivierenden Arbeitsatmosphäre beigetragen.
Besonders hervorheben möchte ich meine Mitdoktoranden Uladzimir Khasianevich, Paul Kühler, Kilian Möhling und Johannes Wünsche, mit denen ich die Höhen und Tiefen des Promotionsalltags geteilt habe, sowie den Masterstudenten Paul Ebert, den ich während seiner Arbeit betreuen durfte.

Ein besonderer Dank gilt dabei Paul Kühler, der am selben Thema arbeitet und daher viele ähnliche Herausforderungen durchlaufen hat. 
Im Laufe der Zeit ist er zu einem engen Freund geworden, und sowohl unsere zahlreichen fachlichen Diskussionen als auch die gemeinsamen Aktivitäten außerhalb der Universität haben meine Promotionszeit auf vielfältige Weise bereichert.

Darüber hinaus möchte ich mich bei meinen Freunden bedanken, die mich während der gesamten Promotionszeit begleitet haben. 
Ihre Aufmunterung, ihr Verständnis und die vielen gemeinsamen Momente außerhalb des wissenschaftlichen Alltags waren für mich von großem Wert und haben mir geholfen, auch anspruchsvolle Phasen gut zu bewältigen.

Mein tiefster Dank gilt jedoch meiner Familie. 
Meinen Eltern und meinem Bruder danke ich für ihre bedingungslose Unterstützung, ihr Vertrauen und dafür, dass sie mir stets den Rücken freigehalten haben. 
Ohne ihre Unterstützung wäre dieser Weg in seiner heutigen Form nicht möglich gewesen.

Schließlich möchte ich meiner Freundin Leonie Schulp von Herzen danken. 
Sie hat mich während der zweiten Hälfte meiner Promotion begleitet, mir außergewöhnlich viel Verständnis entgegengebracht --- insbesondere in der intensiven Endphase, in der viele lange Arbeitstage und Nächte kaum Zeit für anderes ließen --- und mir mit vielen kleinen Gesten immer wieder neuen Mut gegeben.
Ihre Unterstützung hat mir unglaublich viel bedeutet.

\cleardoublepage

\KOMAoptions{headsepline=0.5pt} 
\tableofcontents
\cleardoublepage

\pagenumbering{arabic}
\setcounter{page}{1}
\chapter{Introduction}

The Standard Model (SM) of particle physics is one of the most profound achievements in modern science.
It provides a mathematically self-consistent and conceptually elegant framework that describes the fundamental building blocks of nature and their interactions --- except for gravity --- with an extraordinary degree of precision.
Its predictions have been confirmed across a vast range of energy scales, from low-energy precision measurements, such as the anomalous magnetic moment $(g-2)$, to high-energy collider experiments, most notably the discovery of the Higgs boson.
As we enter a new era of high-precision experiments, the demand for equally precise theoretical predictions becomes indispensable, both to further strengthen the status of the SM and to uncover potential signs of physics beyond it.
To match experimental accuracy requires the evaluation of higher-order quantum corrections, and thus the computation of multi-loop Feynman diagrams.

The computation of these radiative corrections gives rise to ultraviolet (UV) divergences and therefore requires a consistent framework for handling them.
The presence of UV divergences reflects that quantum fields are operator-valued distributions rather than functions: products of distributions evaluated at coinciding spacetime points are, in general, ill-defined, yet such products appear unavoidably in perturbation theory through time-ordered correlation functions.
Closed loops correspond to such products of distributions.
From a physical perspective, loops can be viewed as the exchange of virtual particles whose momenta are integrated up to arbitrarily large values, which may generate divergent contributions.
The theory of renormalisation provides a systematic framework for extending these distributional products to the full spacetime region, thereby removing all UV divergences and producing well-defined, finite expressions.
In the case of gauge theories, renormalisation must preserve BRST invariance, encoded in the Slavnov-Taylor identity, at each order in perturbation theory to ensure unitarity and obtain a consistent quantum theory.

In practice, a regularisation is usually applied to systematically isolate all divergences before subtracting them by suitable counterterms.
Dimensional regularisation (DReg) has become the standard method for this purpose: it is mathematically consistent, compatible with unitarity and causality, and computationally efficient, while maintaining Lorentz invariance and, in vector-like gauge theories, manifestly preserving gauge invariance.
However, when applied to chiral gauge theories DReg encounters the well-known $\gamma_5$-problem, see Ref.~\cite{Jegerlehner:2000dz}.
Chirality is mathematically encoded by the $\gamma_5$ matrix and strictly tied to the representation theory of the 4-dimensional Lorentz group.
Therefore, $\gamma_5$ is an inherently 4-dimensional object, and its embedding into the $D$-dimensional framework of DReg is nontrivial.
There is no extension of $\gamma_5$ to $D$ dimensions that simultaneously preserves all of its familiar 4-dimensional properties.

A defining feature of the SM is its chiral nature: left- and right-handed fermions $\psi_{L/R}=\frac{1}{2}(\mathbb{1}\mp\gamma_5)\psi$ interact differently with the electroweak gauge bosons.
In other words, the SM and all its extensions for potentially new physics are chiral gauge theories.
Consequently, a fully consistent treatment of $\gamma_5$ is essential not only conceptually but also for phenomenological applications.

The Breitenlohner-Maison/`t~Hooft-Veltman (BMHV) scheme (see Refs.~\cite{tHooft:1972tcz,Breitenlohner:1977hr}) is the only known framework for the treatment of $\gamma_5$ that is proven to be mathematically consistent at all orders of perturbation theory.
In this scheme, $\gamma_5$ is treated manifestly 4-dimensional, thereby abandoning its anticommutativity with the fully $D$-dimensional $\gamma^\mu$-matrices.
However, the mathematical consistency comes at a cost: the modified algebraic relations break gauge and BRST invariance at the regularised level.
This breaking must be compensated by symmetry-restoring counterterms in order to reestablish the validity of the Slavnov-Taylor identity and obtain a consistently renormalised quantum theory.
Since the breaking is spurious --- not a physical anomaly --- the required counterterms are guaranteed to exist by the methods of algebraic renormalisation (see Ref.~\cite{Piguet:1995er}), but their explicit construction poses a considerable computational challenge, especially at the multi-loop level.

The projects underlying this thesis build upon and significantly extend recent progress in the renormalisation of chiral gauge theories within the BMHV framework.
Our research efforts focused on several key developments, with particular emphasis on multi-loop applications.
At the precision frontier, we achieved two important milestones: we published the first complete 3-loop renormalisation (see Ref.~\cite{Stockinger:2023ndm}) and the first complete 4-loop renormalisation (see Ref.~\cite{vonManteuffel:2025swv}) of a chiral gauge theory within the BMHV scheme.
The latter represents the highest-order, and thus most precise, application of the BMHV scheme to date.
These results extend previous state-of-the-art calculations and demonstrate that a rigorous and fully self-consistent treatment of $\gamma_5$ is feasible even at high loop orders.
A major part of this doctoral research work concerned the development of a fully automated, high-performance computational framework tailored specifically to address both the general challenges of higher-order calculations and the additional complications posed by a non-anticommuting $\gamma_5$ in the BMHV scheme.
This computational setup is capable of handling billions of intermediate terms per Feynman diagram and was indispensable for the 4-loop calculation.
Complementing these high-precision computations on the conceptual side, we collaboratively worked on a comprehensive review (see Ref.~\cite{Belusca-Maito:2023wah}), which provides extensive discussion of the theoretical foundations of renormalisation theory, the quantum action principle and the algebraic restoration of symmetries in the present context of DReg with non-anticommuting $\gamma_5$. 
Furthermore, we conducted a detailed investigation of dimensional ambiguities within the BMHV scheme of DReg (see Ref.~\cite{Ebert:2024xpy}), analysing the impact of different $D$-dimensional implementations of fermions and gauge interactions.
These ambiguities arise from the non-uniqueness of the $D$-dimensional extension and essentially correspond to different realisations of the regularisation scheme.
In this way, we established a solid foundation for applications to electroweak physics.
A first concrete step towards this goal is already presented in this thesis: we provide the complete 1-loop renormalisation of the full SM in the BMHV scheme, including an analysis of evanescent gauge interactions in this context.
This constitutes the starting point of a long-term research programme aimed at enabling consistent, high-precision electroweak phenomenology.

For context, we briefly summarise further relevant developments in the literature.
The BMHV scheme has been studied in non-Abelian chiral gauge theories at the 1-loop level, both with and without scalar fields in Refs.~\cite{Martin:1999cc,Belusca-Maito:2020ala,Cornella:2022hkc,OlgosoRuiz:2024dzq}.
The first complete 2-loop renormalisation of a non-Abelian theory in this framework was achieved in Ref.~\cite{Kuhler:2025znv}, with initial discussions in Ref.~\cite{Kuhler:2024fak}.
Detailed 2-loop studies of Abelian chiral gauge theories are provided in Refs.~\cite{Belusca-Maito:2021lnk,Belusca-Maito:2022wem}.
The Abelian Higgs model was analysed in this context in Ref.~\cite{Sanchez-Ruiz:2002pcf}, and recent applications of the BMHV scheme to effective field theories (EFTs) can be found in Refs.~\cite{Naterop:2023dek,DiNoi:2023ygk,Heinrich:2024ufp,Heinrich:2024rtg,Naterop:2024ydo,DiNoi:2025uan}.
Alternative $\gamma_5$ prescriptions have also been proposed --- most notably ``Kreimer's scheme'' (see Refs.~\cite{Kreimer:1989ke,Korner:1991sx,Kreimer:1993bh}) and ``Larin's prescription'' (see Ref.~\cite{Larin:1993tq}) --- but these approaches are not fully self-consistent and free of ambiguities.
At the multi-loop level, their scope of application is limited and not entirely under control, as illustrated by Refs.~\cite{Chetyrkin:2012rz,Zoller:2015tha,Bednyakov:2015ooa,Poole:2019txl,Davies:2019onf,Davies:2021mnc,Herren:2021vdk,Chen:2023lus,Chen:2024zju}.

This thesis is organised as follows.
In chapter~\ref{Chap:Gauge_Theories}, we briefly review the theoretical foundations of gauge theories, chiral fermions, the BRST formalism, and anomalies.
Chapter~\ref{Chap:Renormalisation_and_Symmetries} then discusses the theory of renormalisation and the manifestation of symmetries at the quantum level in terms of the Slavnov-Taylor identity.
Furthermore, we analyse symmetry breakings using the quantum action principle and the methods of algebraic renormalisation, and review the renormalisation of gauge theories.
In chapter~\ref{Chap:DReg}, we develop the framework of dimensional regularisation in detail and present a comprehensive discussion of the $\gamma_5$-problem and the BMHV scheme, including comments on alternative prescriptions together with their drawbacks.
Chapter~\ref{Chap:Multi-Loop_Calculations} introduces the computational framework developed in this work and describes the methods employed for multi-loop calculations.
These include the tadpole decomposition, tensor reduction, the implementation of the Dirac algebra within the BMHV scheme, as well as the integration-by-parts (IBP) reduction of generic Feynman integrals to master integrals.
In chapter~\ref{Chap:Practical_Symmetry_Restoration}, we examine the symmetry breaking induced by the BMHV-regularisation in detail and describe the general procedure of symmetry restoration based on the quantum action principle, which is used to determine the required symmetry-restoring counterterms.
This methodology is then applied to a general Abelian chiral gauge theory at the 1-loop level in chapter~\ref{Chap:General_Abelian_Chiral_Gauge_Theory}, including a detailed investigation of dimensional ambiguities and evanescent aspects of the BMHV implementation.
The main results of the multi-loop calculations are presented in chapter~\ref{Chap:BMHV_at_Multi-Loop_Level}, where we discuss the complete renormalisation of an Abelian chiral gauge theory up to the 4-loop level and analyse the multi-loop properties of the BMHV scheme.
Chapter~\ref{Chap:The_Standard_Model} contains the complete 1-loop renormalisation of the full SM and addresses the implications of evanescent gauge interactions in this context, providing the first step of the SM renormalisation research programme.
Finally, chapter~\ref{Chap:Conclusions} summarises our findings and offers an outlook on future research directions.

\chapter{Gauge Theories}\label{Chap:Gauge_Theories}

The fundamental interactions among elementary particles --- the basic building blocks of nature --- are described by gauge theories.
The underlying gauge principle asserts that the fundamental laws of nature do not depend on the choice of the local reference frame for the internal degrees of freedom.
This requirement leads to gauge invariance, i.e.\ invariance under certain continuous local symmetry transformations known as gauge transformations.
However, gauge symmetry is not a proper physical symmetry, but rather a redundancy in the parametrisation of the internal coordinates used to describe nature; it is therefore more accurately referred to as gauge redundancy.
Maintaining invariance under such local transformations naturally gives rise to gauge bosons --- the mediators of the fundamental physical forces, i.e.\ the force-carrying particles.
For electromagnetism, as well as for the weak and strong interactions, these mediators are spin-1 particles.
Conversely, a Lorentz invariant and self-consistent description of an interacting theory of massless spin-1 particles implies gauge invariance, see e.g.\ Refs.~\cite{Weinberg:1995mt,Weinberg:1996kr,Stoeckinger:2023rqft2,QuevedoStandardModel}.
In particular, locality and manifest Lorentz invariance require a gauge redundancy in the description of the theory.

The Standard Model (SM) of particle physics describes all known elementary particles and their interactions via three of the four fundamental forces, except for gravity.
Once the matter content is specified, the SM follows elegantly from Poincar\'e invariance, the chosen gauge group, and renormalisability.
Poincar\'e invariance encodes spacetime symmetries such as translations and rotations, whereas the gauge group specifies the internal gauge ``symmetry'' of the theory.
In particular, the SM is based on the gauge group
\begin{align}\label{Eq:The-SM-Gauge-Group}
    G_{\mathrm{SM}}=U(1)_Y \times SU(2)_L \times SU(3)_c \, .
\end{align}
Here, $SU(3)_c$ governs the strong interaction mediated by the eight gluons $G^a_\mu$, while $U(1)_Y\times SU(2)_L$ describes the unified electroweak interaction.
The electroweak symmetry is spontaneously broken via the Higgs mechanism, as the Higgs boson has a nonzero vacuum expectation value: $U(1)_Y\times SU(2)_L \to U(1)_Q$.
As a consequence, the gauge bosons $W^{\pm}_\mu$ and $Z_\mu$ associated with the broken generators acquire masses --- consistent with experimental observations --- while the unbroken subgroup $U(1)_Q$ corresponds to the familiar electromagnetic interaction, mediated by the massless photon $A_\mu$.

The Standard Model constitutes a central milestone in our understanding of fundamental physics.
Built upon the principles of quantum mechanics and special relativity, it forms a mathematically consistent and conceptually coherent quantum field theory whose predictions agree with experiment to remarkable precision.
Notable examples include the observation of the massive vector bosons $W^{\pm}_\mu$ and $Z_\mu$, the discovery of the Higgs boson, and the remarkable agreement with high-precision observables such as the anomalous magnetic moment $(g-2)$.

The geometrical anatomy of the gauge principle is closely analogous to the principle of general coordinate invariance in general relativity and shows that geometrical principles are deeply rooted in modern physics.
In the SM, all interactions are mediated by spin-1 gauge bosons associated with $G_{\mathrm{SM}}$, whereas general relativity can be viewed as a gauge theory of diffeomorphism invariance, describing physics independently of the chosen coordinate system, with the spin-2 graviton serving as the mediator.
This analogy highlights a unified perspective in which all four fundamental forces, including gravity, can be understood in terms of underlying geometrical principles, specifically through the language of gauge principles (see Refs.~\cite{Weinberg:1996kr,Boehm:2001Gauge,TongStandardModel,Stoeckinger:2023rqft2}).
Although a proper theory of quantum gravity is still unknown, general relativity can be consistently treated as an effective quantum field theory at energies well below the Planck scale, $E\ll M_P\sim 10^{18} \text{GeV}$.
At higher energies, where quantum effects of gravity become significant, this effective description breaks down and must eventually be replaced by a more fundamental UV-complete theory --- a yet unknown fundamental theory of quantum gravity.

In what follows, we outline the basic conceptual construction of gauge theories in Sec.~\ref{Sec:Gauge_Theories_Basic_Concepts} and introduce chiral gauge theories --- in which gauge bosons couple differently to left- and right-handed fermions --- in Sec.~\ref{Sec:Chiral_Gauge_Theories}.
In Sec.~\ref{Sec:BRST-Symmetry}, we discuss the quantisation of gauge theories and the emergence of BRST invariance, which supersedes gauge invariance at the quantum level.
Finally, Sec.~\ref{Sec:Anomalies} addresses the phenomenon of anomalies, where classical symmetries are broken by quantum effects.
Throughout this discussion, we closely follow Refs.~\cite{Belusca-Maito:2023wah,Weinberg:1995mt,Weinberg:1996kr,Boehm:2001Gauge,Stoeckinger:2019rqft1a,Stoeckinger:2022rqft1b,Stoeckinger:2023rqft2,TongStandardModel,TongGaugeTheory,QuevedoStandardModel}.

\section{Basic Concepts}\label{Sec:Gauge_Theories_Basic_Concepts}

To construct a Lorentz invariant local Lagrangian describing a massless spin-1 particle, the particle is embedded in a vector field $A_\mu(x)$, which makes Lorentz invariance manifest and possesses four internal degrees of freedom. 
However, the irreducible unitary representation of the Poincar\'e group for a massless spin-1 particle has only two degrees of freedom corresponding to its two transverse polarisations.
The remaining two are unphysical and are eliminated by gauge invariance, which identifies the gauge boson with an equivalence class of vector fields.

Following the references cited above, we now briefly outline the construction of a Yang-Mills gauge theory for a non-Abelian Lie group $G$.
We start from a theory of matter fields $\psi(x)=(\psi_1(x),\ldots,\psi_{N_F}(x))$ described by a Lagrangian $\mathcal{L}_\mathrm{matter}(\psi,\partial\psi)$ that is invariant under global transformations of $G$. 
An element of the Lie group $G$ that is continuously connected to the identity can be written in a unitary representation as
\begin{align}\label{Eq:General-Representation-of-Lie-Group}
    U(\theta^a)=e^{-ig\theta^a T^a},
\end{align}
where $\theta^a$ are the group parameters, $g$ is the gauge coupling, and $T^a$ are the generators of the Lie algebra $\mathfrak{g}$ of the group $G$.
The generators $T^a$ are taken to be hermitian to ensure that the representation $U(\theta^a)$ is unitary.
For infinitesimal $\theta^a$, the group action reads 
\begin{align}
    U(\theta^a)=\mathbb{1}-ig\theta^a T^a + \mathcal{O}(\theta^2),   
\end{align}
and a general representation of the Lie group can be obtained by exponentiating this infinitesimal action (cf.\ Eq.~\eqref{Eq:General-Representation-of-Lie-Group}).
The local structure of the Lie group is determined by its Lie algebra $\mathfrak{g}$,
\begin{align}
    [T^a,T^b] = i f^{abc} T^c,
\end{align}
where $f^{abc}$ are the totally antisymmetric structure constants.
The global symmetry of the matter Lagrangian $\mathcal{L}_\mathrm{matter}(\psi,\partial\psi)$ implies that it remains invariant under the transformation of the matter fields in some unitary representation
\begin{align}
    \psi_i(x) \longrightarrow U(\theta^a)_{ij} \psi_j(x).
\end{align}

The central idea of the gauge principle is to promote this global symmetry to a local invariance by allowing the transformation parameters to depend on spacetime $U(\theta^a)=U(\theta^a(x))\equiv U(x)$, with $\theta^a=\theta^a(x)$, $\forall \, x\in\mathbb{M}_4$.
The matter fields then transform locally as
\begin{align}\label{Eq:Local-Gauge-Trafor-of-Matter-Field}
    \psi_i(x)\longrightarrow U(\theta^a(x))_{ij} \psi_j(x).
\end{align}
Physically, this means that the choice of the local reference frame for the internal degrees of freedom is arbitrary and does not affect observable quantities.
However, having a localised coordinate system, i.e.\ spacetime-dependent basis vectors, we can no longer directly compare matter fields $\psi(x)$ and $\psi(x+dx)$ at different spacetime points or define meaningful derivatives in the usual way.
The ordinary derivative $\partial_\mu\psi(x)$ would not transform covariantly anyway and, by itself, cannot yield a gauge invariant Lagrangian due to the spacetime dependence of the local transformation. 
To address this issue, a \emph{parallel transport} is defined that accounts for changes in the local reference frame and allows to compare fields at neighbouring spacetime points in a locally defined internal space.
The parallel-transported field has physically the same field value but may potentially be expressed in a different local coordinate system.
This construction naturally introduces a \emph{connection}, which is identified with the \emph{gauge field} or \emph{gauge potential} $A_\mu(x)$.
This gauge potential is an element of the Lie algebra, and can thus be expressed as a linear combination of the generators as $A_\mu(x)=A^a_\mu(x)T^a$.
For each generator $T^a$, this introduces an associated gauge field $A^a_\mu(x)$.
In this context, a derivative is now defined as 
\begin{align}
    D_\mu = \partial_\mu + i g T^a A^a_\mu(x),
\end{align}
which is the \emph{covariant derivative} and ``replaces'' the ordinary derivative.
It acts on the matter fields as $(D_\mu \psi(x))_i = (\partial_\mu \delta_{ij} + i g T^a_{ij} A^a_\mu(x) ) \psi_j(x)$, and, under local transformations $U(x)$, it transforms covariantly as $D_\mu \psi(x) \longrightarrow U(x) D_\mu \psi(x)$.
The action of the gauge group on the gauge field is 
\begin{align}
    A_\mu(x) \longrightarrow U(x) A_\mu(x) U^{-1}(x) - \frac{1}{ig} (\partial_\mu U(x))U^{-1}(x).
\end{align}
In particular, the gauge field transforms in the adjoint representation of the group.
The curvature of the locally defined internal space is expressed by the \emph{curvature tensor} or \emph{field strength tensor} 
\begin{align}
    F_{\mu\nu} = \frac{1}{ig} [D_\mu,D_\nu] = \partial_\mu A_\nu - \partial_\nu A_\mu + i g [A_\mu,A_\nu],
\end{align}
which transforms covariantly as $F_{\mu\nu}\longrightarrow UF_{\mu\nu}U^{-1}$.
With $F_{\mu\nu} = F^a_{\mu\nu}T^a$, we find 
\begin{align}
    F^a_{\mu\nu} = \partial_\mu A^a_\nu - \partial_\nu A^a_\mu - g f^{abc} A^b_\mu A^c_\nu.
\end{align}
A non-vanishing curvature indicates a genuine effect, implying that one cannot globally transform back to a configuration without connection fields.
As already mentioned, this construction is completely analogous to general relativity and its principle of general coordinate invariance (see Refs.~\cite{Weinberg:1996kr,Boehm:2001Gauge,TongStandardModel,Stoeckinger:2023rqft2}).

For infinitesimal $\theta=T^a\theta^a$, the local gauge transformations are given by
\begin{subequations}\label{Eq:Infinitesimal-Gauge-Transformations}
    \begin{align}
        \psi_i(x) &\longrightarrow \psi_i(x) - i g \theta^a(x) T^a_{ij}\psi_j(x),\\
        A_\mu(x) &\longrightarrow A_\mu(x) + \partial_\mu \theta(x) - ig[\theta(x),A_\mu(x)],\\
        A^a_\mu(x) &\longrightarrow A_\mu^a(x) + \partial_\mu \theta^a(x) - gf^{abc} A_\mu^b(x) \theta^c(x).
    \end{align}
\end{subequations}
The transformation of the gauge boson is universal for all representations and contains the gauge coupling $g$.
For a simple non-Abelian gauge group, this universality implies that a single gauge coupling governs all interactions of the gauge bosons, both among themselves and with matter fields.

Using the field strength tensor $F_{\mu\nu}$, we can construct a kinetic term for the gauge field $A_\mu(x)$, such that it becomes a dynamical and propagating field.
To ensure gauge invariance, ordinary derivatives in the matter Lagrangian $\mathcal{L}_\mathrm{mat}$ are replaced by covariant derivatives: $\partial_\mu\to D_\mu$.
Then, a renormalisable, gauge invariant Lagrangian for this Yang-Mills theory takes the form
\begin{subequations}
    \begin{align}
        \mathcal{L}_\mathrm{inv} &= \mathcal{L}_\mathrm{gauge} + \mathcal{L}_\mathrm{mat},\label{Eq:Def-of-L_inv-for-Gauge-Theory}\\
        \mathcal{L}_\mathrm{gauge} &= -\frac{1}{4} F^{a,\mu\nu}F^a_{\mu\nu},\\
        \mathcal{L}_\mathrm{mat} &= \mathcal{L}_\mathrm{mat}(\psi,D_\mu\psi).
    \end{align}
\end{subequations}
The explicit form of $\mathcal{L}_\mathrm{mat}$ depends on the details of the specific theory under consideration.

\section{Chiral Gauge Theories}\label{Sec:Chiral_Gauge_Theories}

A fundamental insight of particle physics is that electroweak interactions act on chiral fermions, treating left- and right-handed components differently.
In particular, the weak force couples exclusively to left-handed fermions.
As a result, parity is not a fundamental symmetry of nature but is instead violated by the weak interaction --- a fact first conjectured theoretically in Refs.~\cite{Lee:1956qn,Salam:1957st} and experimentally confirmed in Ref.~\cite{PhysRev.105.1413}.
Consequently, the fundamental building blocks of matter are chiral fermions rather than Dirac fermions.
This makes the Standard Model --- and any of its possible extensions --- a chiral gauge theory, in which left- and right-handed fermions transform under different representations of the gauge group (at least for the subgroup $SU(2)_L$) and interact differently with the gauge bosons.
Since particles correspond to unitary representations of the Poincar\'e group, this observation is not surprising but follows naturally from the representation theory of spacetime symmetries, where chirality is intrinsically encoded in the structure of the Lorentz group.

We mainly follow Refs.~\cite{Belusca-Maito:2023wah,Weinberg:1995mt,Weinberg:1996kr,QuevedoStandardModel,TongStandardModel}.
The Poincar\'e group is the isometry group of Minkowski spacetime $(\mathbb{M}_4,\eta)$, consisting of the spacetime translations and Lorentz transformations.
It represents the spacetime symmetry group of special relativity and is realised by a unitary operator $U(\Lambda,a)$ acting on the Hilbert space of states as $\ket{\psi}\to U(\Lambda,a)\ket{\psi}$.

The relevant subgroup is the Lorentz group $O(1,3)$, whose transformations leave the Minkowski metric invariant, i.e.\ $\Lambda^\mathsf{T}\eta\Lambda=\eta$.
It has four disconnected components, but we restrict our attention to the one whose transformations are continuously connected to the identity: the proper orthochronous Lorentz group $SO^+(1,3)$, which is the subgroup of all Lorentz transformations preserving both spatial orientation and direction of time.
The representation theory of this group is best studied by considering its universal covering group $SL(2,\mathbb{C})$, which is simply connected.\footnote{As the covering group, $SL(2,\mathbb{C})$ is obtained by exponentiating the generators of the Lorentz algebra.}

The simplest nontrivial irreducible representations of $SL(2,\mathbb{C})$ (and the Lorentz group) are the fundamental representation $(\frac{1}{2},0)$ and the conjugate fundamental representation $(0,\frac{1}{2})$.
These correspond to left-handed Weyl spinors $\chi_\alpha$ for $(\frac{1}{2},0)$, and right-handed Weyl spinors $\bar\eta^{\dot\alpha}$ for $(0,\frac{1}{2})$ --- complex 2-component spinors characterised by their behaviour under Lorentz transformations.\footnote{Left- and right-handed Weyl spinors have different, independent Lorentz transformation properties.}
They are related by complex conjugation, such that a left-handed spinor can be expressed as a right-handed complex conjugated spinor.
These two representations, embodied by 2-component Weyl spinors, constitute the irreducible building blocks for matter fields.
The 4-component Dirac spinors are reducible representations of the Lorentz group and can be constructed as the direct sum of the two irreducible Weyl representations $(\frac{1}{2},0)$ and $(0,\frac{1}{2})$ as
\begin{align}\label{Eq:Dirac-Spinor-expressed-via-Weyl-Spinors}
  \psi = \begin{pmatrix} \chi_{\alpha} \\ \bar\eta^{\dot\alpha} \end{pmatrix}.
\end{align}
They therefore appear as secondary or composite rather than fundamental objects from a group-theoretical perspective.
All other representations of the Lorentz group, such as $(0,0)$ for scalars and $(\frac{1}{2},\frac{1}{2})$ for vectors, can be built from these Weyl representations.

Reducible structures such as Dirac spinors are required to describe parity transformations, which transform left-handed into right-handed representations and vice versa, $(\frac{1}{2},0) \leftrightarrow (0,\frac{1}{2})$, since they cannot act meaningfully on either representation alone.
Dirac spinors provide the simplest nontrivial representation admitting parity transformations, and theories invariant under parity therefore naturally introduce Dirac spinors.

In the 4-component Dirac spinor formalism, chirality is associated with $\gamma_5$, which is defined in 4 dimensions as
\begin{align}
    \gamma_5 = i \gamma^0\gamma^1\gamma^2\gamma^3=-\frac{i}{4!}\varepsilon_{\mu\nu\rho\sigma} \gamma^\mu\gamma^\nu\gamma^\rho\gamma^\sigma.
\end{align}
From $(\gamma_5)^2=\mathbb{1}$, we can deduce that its eigenvalues are $\pm1$, which define the \emph{chirality} of a state.
Hence, the eigenvalues of $\gamma_5$ are referred to as chirality.
Any Dirac spinor can be decomposed into its left- and right-handed components using the projection operators $\mathbb{P}_\mathrm{L/R}=(\mathbb{1}\mp\gamma_5)/2$, such that
\begin{align}
    \psi = \psi_L + \psi_R,
\end{align}
where $\psi_{L/R}=\mathbb{P}_\mathrm{L/R}\psi$.
Expressing the Dirac spinor in terms of the 2-component Weyl spinors (see Eq.~\eqref{Eq:Dirac-Spinor-expressed-via-Weyl-Spinors}) yields 
\begin{align}
    \psi_L &= 
    \begin{pmatrix} \chi_\alpha \\ 0 \end{pmatrix},
    & \overline{\psi_L} &= (0 \, \ \bar\chi_{\dot\alpha}), &
    \psi_R &= 
    \begin{pmatrix} 0 \\ \bar\eta^{\dot\alpha} \end{pmatrix},
    & \overline{\psi_R} &= (\eta^\alpha \, \ 0).
\end{align}
The spinors $\psi_L$ and $\psi_R$ are eigenspinors of $\gamma_5$ (chirality eigenstates) with eigenvalues $-1$ and $+1$, respectively.
Because $\gamma_5$ and the projectors commute with the Lorentz generators, $\psi_L$ and $\psi_R$ transform independently under Lorentz transformations, forming two invariant subspaces.
Therefore chirality is a Lorentz invariant property, and left- and right-handed fermions can appear independently in a Lorentz invariant theory.
A more detailed discussion can be found in Ref.~\cite{Belusca-Maito:2023wah} and the references therein.

In summary, chiral gauge theories such as the electroweak Standard Model are built upon these fundamental chiral fermions, which emerge naturally from Lorentz group representation theory.
Left- and right-handed fermions can appear in different representations of the gauge group, and hence couple differently to the associated gauge bosons --- as seen in the weak interactions, where left-handed fermions form $SU(2)_L$ doublets while right-handed fermions are $SU(2)_L$ singlets.
The renormalisation of such chiral gauge theories is considerably more intricate than that of vector-like gauge theories, in which all fermions are treated equally.
The reason are difficulties in the regularisation associated with the $\gamma_5$ matrix.
This issue, known as the $\gamma_5$-problem (see Sec.~\ref{Sec:The-g5-Problem}), and the consistent renormalisation of chiral gauge theories are the central topic of this thesis.
Despite its technical complications, $\gamma_5$ is inseparable from the concept of chirality and constitutes the essential mathematical tool for its implementation.
Consequently, $\gamma_5$ with all its intricacies is an unavoidable fact of life (see Ref.~\cite{Jegerlehner:2000dz}).

\section{Quantisation of Gauge Theories and the BRST Formalism}\label{Sec:BRST-Symmetry}

Classical gauge theories provide an elegant geometrical framework for describing the fundamental interactions of nature and play an indispensable role in understanding the Standard Model of particle physics.
However, their quantisation poses challenges absent in theories that only contain scalars or spinors.
The central difficulty arises from gauge invariance: while being the defining principle of the theory, it also introduces a redundancy in the description of the physical degrees of freedom.
Field configurations related by gauge transformations are physically equivalent; physics is thus encoded in equivalence classes.
To construct a consistent, unitary, and Lorentz covariant quantum theory, this redundancy must be handled systematically.
In this section, we briefly outline the difficulties arising in the quantisation of gauge theories, both within canonical quantisation and the path integral formalism.
We then discuss their resolution through the BRST formalism, originally developed in Refs.~\cite{Becchi:1974xu,Becchi:1974md,Becchi:1975nq,Tyutin:1975qk}.
The presentation mainly follows Refs.~\cite{Belusca-Maito:2023wah,Weinberg:1996kr,Stoeckinger:2023rqft2,Boehm:2001Gauge}.

\paragraph{Canonical Quantisation and Faddeev-Popov Approach:}
In canonical quantisation, the conjugate momentum of $A^a_0$ vanishes, i.e.\ $\Pi^a_0=0$, because the Lagrangian does not depend on the time-derivative of $A_0^a$. 
This first-class constraint indicates that $A_0^a$ is non-dynamical and obstructs the canonical commutation relations.
As discussed in Ref.~\cite{Weinberg:1995mt}, canonical quantisation can still be performed by imposing a gauge condition on $A^a_\mu(x)$, such as Coulomb or axial gauge.
This procedure, however, breaks manifest Lorentz invariance and thereby complicates the formulation of a Lorentz invariant and renormalisable $S$-matrix.

The path integral explicitly reveals gauge invariance as the origin of the problem.
A naively defined path integral in a gauge theory is ill-defined and divergent.
In the path integral, one integrates over all possible field configurations, each weighted by the exponential of the action $S_\mathrm{inv}=\int d^4x\,\mathcal{L}_\mathrm{inv}$.
This includes the integration over infinitely many physically equivalent gauge configurations that yield the same value of $S_\mathrm{inv}$, since the action remains constant along a gauge orbit --- a gauge orbit corresponds to the set of field configurations connected by gauge transformations, i.e.\ those that belong to the same equivalence class.
The solution to this problem, proposed in Ref.~\cite{Faddeev:1967fc}, is to integrate only over one representative of each gauge orbit.
Mathematically, this corresponds to ``dividing out'' equivalent gauge configurations from the path integral.
This is achieved by fixing the gauge, i.e.\ by imposing a condition $\mathcal{G}^a(A_\mu;x)=0$, which selects a single representative gauge configuration from each equivalence class (see Ref.~\cite{Boehm:2001Gauge}), ensuring that the integration is performed only once per gauge orbit.
Here, $\mathcal{G}^a(A_\mu;x)$ is a local functional of the gauge fields and specifies the gauge-fixing condition --- for example, $\mathcal{G}^a(A_\mu;x)=\partial^\mu A_\mu^a(x)=0$.\footnote{The solution is not globally unique but admits so-called Gribov ambiguities (see Ref.~\cite{Gribov:1977wm}), which are, however, irrelevant in perturbation theory. For discussions of non-perturbative aspects, see Refs.~\cite{Dudal:2008sp,Sorella:2009vt}.}
Implementing the gauge condition into the path integral introduces the so-called Faddeev-Popov determinant, which can be written as a path integral over Grassmann-valued fields.
This naturally introduces the Faddeev-Popov \emph{ghosts} and \emph{anti-ghosts}, $c^a(x)$ and $\overline{c}^a(x)$, respectively.
With $\mathcal{L}_\mathrm{inv}$ as in Eq.~\eqref{Eq:Def-of-L_inv-for-Gauge-Theory}, one obtains an effective Lagrangian admissible for quantisation
\begin{subequations}\label{Eq:Effective-Lagrangian-Faddeev-Popov-Method}
    \begin{align}
        \mathcal{L}_\mathrm{eff} &= \mathcal{L}_\mathrm{inv} + \mathcal{L}_\mathrm{ghost+fix},\\
        \mathcal{L}_\mathrm{ghost+fix} &= -\frac{1}{2\xi} \mathcal{G}^a(A_\mu;x)\mathcal{G}^a(A_\mu;x) - \int d^4z\, \overline{c}^a(x) \frac{\delta \mathcal{G}^a(A_\mu;x)}{\delta A^c_\rho(z)} D_\rho^{cb} c^b(z),\label{Eq:Ghost-Gauge-Fixing-Term-Faddeev-Popov}
    \end{align}
\end{subequations}
which contains the ghost fields and explicitly depends on the gauge-fixing condition (with gauge parameter $\xi$), and is therefore no longer gauge invariant.
Nevertheless, the $S$-matrix and physical observables remain gauge independent, as discussed below (see Refs.~\cite{Weinberg:1996kr,Boehm:2001Gauge}).

\paragraph{The BRST Formalism:}
Although gauge fixing breaks the original gauge invariance, a new symmetry emerges from the ashes: the BRST symmetry.
The effective Lagrangian in Eq.~\eqref{Eq:Effective-Lagrangian-Faddeev-Popov-Method} is no longer gauge invariant but instead BRST invariant.
Effectively, the classical gauge invariant theory, together with its constraints, is replaced by a physically equivalent theory without constraints but endowed with BRST symmetry.
This new formulation, described by the effective Lagrangian in Eq.~\eqref{Eq:Effective-Lagrangian-Faddeev-Popov-Method}, includes a gauge-fixing term and introduces anticommuting ghost fields.
Hence, the BRST invariance supersedes gauge invariance and can be regarded as its proper substitution.
The ghost fields carry non-vanishing ghost number, $N_\mathrm{gh}(c^a)=+1$ and $N_\mathrm{gh}(\overline{c}^a)=-1$, whereas $\psi$ and $A_\mu$ have zero ghost number.\footnote{Ghost number is conserved.}
The BRST formalism elegantly realises the concept that physics is described by equivalence classes by reformulating it as a \emph{cohomology} problem of the BRST operator, see Ref.~\cite{Kugo:1979gm} as well as Refs.~\cite{Weinberg:1996kr,Boehm:2001Gauge,Stoeckinger:2023rqft2}.
\begin{definition}[BRST Operator]\label{Def:BRST-Operator-s}\ \\
    The fermionic differential operator $s$ acting on the algebra of fields that encodes the gauge invariance of the physical system is called \emph{BRST operator}.
    The BRST operator is nilpotent, $s^2=0$, and carries ghost number $N_\mathrm{gh}(s)=+1$.
\end{definition}
The BRST operator is the exterior derivative acting on the group manifold.
Being a fermionic differential operator, it obeys a graded algebra, so that it anticommutes with fermionic objects:
\begin{subequations}
  \begin{align}
    s(B_1B_2) &= (sB_1)B_2+B_1(sB_2), \\
    s(F_1B_2) &= (sF_1)B_2-F_1(sB_2), \\
    s(F_1F_2) &= (sF_1)F_2-F_1(sF_2),
  \end{align}
\end{subequations}
for bosonic and fermionic expressions $B_i$ and $F_i$, respectively.
The BRST operator always increases the ghost number of the expression it acts on by one, since $N_\mathrm{gh}(s)=+1$.
From its nilpotency, it follows that $\mathrm{Im}(s)=\{X \,|\,X=sY\}\subset \mathrm{Ker}(s)=\{X\,|\,sX=0\}$.
Elements of $\mathrm{Ker}(s)$ are called \emph{BRST closed}, while those of $\mathrm{Im}(s)$ are \emph{BRST exact} and can be written as total BRST variations.

Nilpotency is a crucial property of the BRST operator for the internal consistency of the BRST formalism, which generalises gauge invariance. 
Using its cohomology, field configurations can be classified as follows (see e.g.\ Ref.~\cite{Belusca-Maito:2023wah}):
\begin{itemize}
    \item A field configuration $X$ with $N_\mathrm{gh}(X)=0$ is called \emph{physical} if $sX=0$ but $X\notin\mathrm{Im}(s)$.
    It corresponds to one representative of the associated physical equivalence class.
    Such field configurations correspond to non-trivial elements of the BRST cohomology.
    \item A field configuration $X$ is called ``unphysical'' if $X\in\mathrm{Im}(s)$, i.e.\ $X=sY$.
    Such field configurations correspond to trivial elements of the BRST cohomology, equivalent to the zero-element of the cohomology group.
    \item Two \emph{physical} field configurations $X_1$ and $X_2$, with ghost number zero, are \emph{physically equivalent} if $\exists \,\, Y$, such that $X_2=X_1+sY$.\footnote{Here, $Y$ is an arbitrary field configuration with $N_\mathrm{gh}(Y)=-1$, since $N_\mathrm{gh}(s)=+1$.}
    This means that $X_1$ and $X_2$ belong to the same equivalence class, denoted with $X_1\sim X_2$.
\end{itemize}
Trivial elements of the BRST cohomology are unphysical in the sense that they do not correspond to physical degrees of freedom but rather represent a gauge redundancy --- they are equivalent to the trivial field configuration. 
Following this logic, we distinguish between two types of terms in the Lagrangian:
on the one hand, terms represented by $\mathcal{L}_\mathrm{inv}$ in Eq.~\ref{Eq:Effective-Lagrangian-Faddeev-Popov-Method}, which are BRST invariant because they are gauge invariant but not BRST exact, i.e.\ non-trivial elements of the BRST cohomology.
Hence, $\mathcal{L}_\mathrm{inv}$ constitutes the physical Lagrangian and contains the actual physical content of the theory.
On the other hand, terms represented by $\mathcal{L}_\mathrm{ghost+fix}$ in Eq.~\ref{Eq:Effective-Lagrangian-Faddeev-Popov-Method}, which are BRST exact and thus trivial elements of the BRST cohomology.

The BRST transformations are given by
\begin{subequations}\label{Eq:Definition-of-the-BRST-Transformations}
    \begin{align}
        s\psi_i(x) &= - ig c^a(x)T^a_{ij}\psi_j(x), \label{Eq:BRST-Trafo-of-the-Matter-Field}\\
        s A_\mu^a(x) &= \partial_\mu c^a(x) - g f^{abc} A_\mu^b(x)c^c(x) = (D_\mu c(x))^a,\label{Eq:BRST-Trafo-of-the-Gauge-Field}\\
        s c^a(x) &= \frac{1}{2} g f^{abc} c^b(x) c^c(x),\label{Eq:BRST-Trafo-of-the-Ghost-Field}\\
        s \overline{c}^a(x) &= B^a(x),\label{Eq:BRST-Trafo-of-the-AntiGhost-Field}\\
        s B^a(x) &= 0.\label{Eq:BRST-Trafo-of-the-Nakanishi-Lautrup-Field}
    \end{align}
\end{subequations}
The first two transformations, those of the physical fields in Eqs.~\eqref{Eq:BRST-Trafo-of-the-Matter-Field} and \eqref{Eq:BRST-Trafo-of-the-Gauge-Field}, follow directly from the infinitesimal gauge transformations in Eq.~\eqref{Eq:Infinitesimal-Gauge-Transformations} by replacing $\theta^a(x)\to \vartheta c^a(x)$, with Grassmann-valued parameter $\vartheta$, such that $\delta\phi_i(x) = \vartheta s \phi_i(x)$.
The transformation of the introduced ghost field $c(x)=T^ac^a(x)$ is provided in Eq.~\eqref{Eq:BRST-Trafo-of-the-Ghost-Field}.

In the BRST formalism, the gauge-fixing and ghost terms $\mathcal{L}_\mathrm{ghost+fix}$ can be derived straightforwardly by first introducing a \emph{BRST doublet} $\{\overline{c}^a,B^a\}$, formed by the antighosts $\overline{c}^a$ and the Nakanishi-Lautrup auxiliary fields $B^a$, which transform according to Eqs.~\eqref{Eq:BRST-Trafo-of-the-AntiGhost-Field} and \eqref{Eq:BRST-Trafo-of-the-Nakanishi-Lautrup-Field}.
Introducing such a BRST doublet does not change the cohomology, and therefore leaves the physical content of the theory unchanged, as proven in Ref.~\cite{Piguet:1995er}.
A suitable term of the form $sX=s(\overline{c}^aF^a)$ then generates both the gauge-fixing and ghost terms and can safely be added to the Lagrangian, since it does not modify the physical content of the theory.
Indeed, it represents a BRST exact term and thus a trivial element of the BRST cohomology in the sense discussed above.
Using the gauge-fixing functional $\mathcal{G}^a(A_\mu;x)=\partial^\mu A_\mu^a(x)$ for $F^a=\xi B^a/2 + \mathcal{G}^a(A_\mu;x)$, we obtain
\begin{align}\label{Eq:Gauge-Fixing-Condition-BRST-Formalism}
    \mathcal{L}_\mathrm{ghost+fix} = s \bigg[\overline{c}^a \bigg(\partial^\mu A^a_\mu + \frac{\xi}{2} B^a\bigg)\bigg] = B^a (\partial^\mu A^a_\mu) + \frac{\xi}{2} B^aB^a - \overline{c}^a \partial^\mu(D_\mu c)^a,
\end{align}
which is a standard choice and allows a straightforward quantisation.
The gauge-fixing also ensures that the kinetic operator of the gauge field can be inverted without difficulty, yielding the gauge boson propagator.
Integrating out the auxiliary field gives $B^a(x)=-\partial^\mu A_\mu^a(x)/\xi$, leading to 
\begin{align}\label{Eq:Gauge-Fixing-and-Ghost-Terms-Standard}
    \mathcal{L}_\mathrm{ghost+fix} = - \frac{1}{2\xi} (\partial^\mu A^a_\mu)^2 - \overline{c}^a \partial^\mu(D_\mu c)^a,
\end{align}
which reproduces the expression for $\mathcal{L}_\mathrm{ghost+fix}$, in Eq.~\eqref{Eq:Ghost-Gauge-Fixing-Term-Faddeev-Popov}, obtained in the Faddeev-Popov approach for the choice $\mathcal{G}^a(A_\mu;x)=\partial^\mu A_\mu^a(x)$.

In general, BRST transformations are nonlinear (except for those of the BRST doublet $\{\overline{c}^a,B^a\}$) and therefore correspond to local composite operators (see Sec.~\ref{Sec:GreenFunctions_and_GeneratingFunctionals}).
As such, they receive nontrivial quantum corrections at higher orders in perturbation theory --- in other words, BRST transformations themselves generally renormalise.
To systematically study the renormalisation of gauge theories, it is convenient to couple the BRST transformations $s\phi_i(x)$ to external sources $K_i(x)$ and define an ``external'' Lagrangian as
\begin{align}\label{Eq:Def-of-Lagrangian_ext}
    \mathcal{L}_\mathrm{ext} = K_is\phi_i = \rho^{a,\mu}sA_\mu^a + {\overline{R}}{}^{i} s \psi_i + (s\overline{\psi}_i)R^i + \zeta^a s c^a + \chi^a s\overline{c}^a,
\end{align}
which allows to keep track of the renormalisation of the BRST transformations $s\phi_i(x)$ themselves (cf.\ Batalin-Vilkovisky formalism and see Refs.~\cite{Henneaux:1992ig,Weinberg:1996kr,Boehm:2001Gauge}).
The auxiliary field $B^a$ is not coupled to an external source, since its BRST transformation vanishes.
However, we introduced a source for $\overline{c}^a$ even though this is not strictly necessary, because its BRST transformation is linear and thus does not renormalise.
The external sources are non-dynamical in the sense that they do not propagate and remain classical as they are not integrated over in the path integral.
Their mass dimension is determined by $\mathrm{dim}(K_is\phi_i)\overset{!}{=}4$, while their ghost number follows from $N_\mathrm{gh}(K_is\phi_i)\overset{!}{=}0$, ensuring that the Lagrangian has the correct overall dimension and ghost number.
The properties of the external sources are summarised in Tab.~\ref{Tab:Table-for-External-Sources}.
\begin{table}[t!]
    \centering
    \begin{tabular}{|c||c|c|c|c|} \hline 
        External Source & Mass Dimension & Ghost Number & Statistics & Lorentz Transf. \\\hline\hline
        $\rho_a^\mu$ & $3$ & $-1$ & Fermion & 4-Vector \\ \hline
        $\vcenter{\hbox{$\overline{R}{}^{i}$}}$/$\vcenter{\hbox{$R^i$}}$ & $5/2$ & $-1$ & Boson & Spinor \\ \hline
        $\vcenter{\hbox{$\Upsilon^a$}}$ & $3$ & $-1$ & Fermion & Scalar \\ \hline
        $\zeta^a$ & $4$ & $-2$ & Boson & Scalar \\ \hline
        $\chi^a$ & $2$ & $0$ & Boson & Scalar \\ \hline
    \end{tabular}
    \caption{External sources coupled to the BRST transformations of vector fields $A_\mu^a$, fermions $\psi_i$/$\overline{\psi}_i$, scalars $\phi_a$, ghosts $c^a$, and antighosts $\overline{c}^a$, and their properties.
    Listed are their mass dimension, ghost number, statistics and transformation behaviour under the Lorentz group.}
    \label{Tab:Table-for-External-Sources}
\end{table}

The complete classical Lagrangian of a gauge theory is then given by 
\begin{align}\label{Eq:Full-Lagrangian-for-Quantisation}
    \mathcal{L}_\mathrm{cl} = \mathcal{L}_\mathrm{inv} + \mathcal{L}_\mathrm{ghost+fix} + \mathcal{L}_\mathrm{ext},
\end{align}
which serves as the starting point for quantisation and for the perturbative renormalisation procedure at higher orders.
Since $sK_i=0$ and $s^2=0$, we obtain $s\mathcal{L}_\mathrm{ext}=0$, implying that each term in Eq.~\eqref{Eq:Full-Lagrangian-for-Quantisation} is separately BRST invariant.
Consequently, the full Lagrangian satisfies $s\mathcal{L}_\mathrm{cl}=0$.
The corresponding tree-level action of the gauge theory is
\begin{align}\label{Eq:Full-tree-level-Action-of-a-gauge-theory}
    \Gamma_\mathrm{cl} \equiv S_0 = \int d^4x \, \mathcal{L}_\mathrm{cl}.
\end{align}
The gauge-fixing condition can equally be written as 
\begin{align}\label{Eq:The-Gauge-Fixing-Condition}
    \frac{\delta\Gamma_\mathrm{cl}}{\delta B^a(x)} = \mathcal{G}^a(A_\mu;x) + \xi B^a(x) = \partial^\mu A_\mu^a(x) + \xi B^a(x),
\end{align}
which, for our choice $\mathcal{G}^a(A_\mu;x)=\partial^\mu A_\mu^a(x)$, is linear in the quantum fields.

\paragraph{The physical Hilbert Space and the $S$-Matrix:}
According to Noether's theorem, the BRST symmetry gives rise to a conserved current $j_B^\mu$ and thus a conserved charge $Q_B$.
Upon quantisation, $Q_B$ becomes the BRST operator acting on the space of states.
This operator generates the BRST transformations of all field operators $\phi_i(x)$ as $[Q_B,\phi_i(x)]_\pm = s \phi_i(x)$, where the commutator or anticommutator is chosen according to the statistics of $\phi_i$ (see Ref.~\cite{Boehm:2001Gauge}).
The BRST operator is nilpotent, $Q_B^2=0$, implying $\mathrm{Im}(Q_B)\subset\mathrm{Ker}(Q_B)$.
Importantly, the operator $Q_B$ can be used to reduce the formal Fock space (the full space of states) to the physical Hilbert space, which contains only states of positive norm, by means of the same cohomology arguments introduced above.

The formal Fock space contains both physical and unphysical states. 
The physical states, which have positive norm $\braket{\phi_\mathrm{phys}}>0$, include the fermions and the transversal polarisations of the gauge fields $\ket{\lambda=1,2}$.
In contrast, the unphysical states possess norms of mixed signature and comprise the unphysical polarisations of the gauge fields $\ket{\lambda=L}$ and $\ket{\lambda=S}$, as well as the ghosts $\ket{c}$ and antighosts $\ket{\overline{c}}$.
In fact, since the ghosts and antighosts are spin-0 fermions violating the spin-statistics theorem, their states have negative norm.

The physical Hilbert space is defined through the BRST cohomology of $Q_B$ as
\begin{align}\label{Eq:Def-of-phys-Hilbert-Space}
    \mathcal{H}_\mathrm{phys} = \frac{\mathrm{Ker}(Q_B)}{\mathrm{Im}(Q_B)},
\end{align}
that is, as the space of BRST-closed states modulo BRST-exact states.
Analogous to the discussion above, a state $\ket{\psi}$ is physical --- a representative of a physical equivalence class --- if it lies in the BRST cohomology, i.e.\ if $Q_B\ket{\psi}=0$ but $\ket{\psi}\notin\mathrm{Im}(Q_B)$. 
Two physical states are equivalent (belong to the same equivalence class) if they are related by a total BRST variation $Q_B\ket{\chi}$.
In contrast, BRST-exact states $\ket{\psi}=Q_B\ket{\chi}\in\mathrm{Im}(Q_B)$ are ``unphysical'' in the sense that they are equivalent to the 0-vector.

For the following discussion, we consider a free theory without interactions ($g\to0$), such that the BRST transformations in Eq.~\eqref{Eq:Definition-of-the-BRST-Transformations} simplify and become linear.
Furthermore, we restrict ourselves to a pure gauge theory (without fermions) in order to focus on the essential aspects.
In this setting, the transversal polarisation states of the gauge fields
\begin{align}
    Q_B\ket{\lambda=1,2} = 0
\end{align}
are the physical representatives.
The remaining states are unphysical and can be organised as BRST quartets (see Ref.~\cite{Kugo:1979gm}):
\begin{equation}
    \begin{aligned}
        &\ket{\lambda=L} \xrightarrow{Q_B} \ket{c}\\
        \ket{\overline{c}} \xrightarrow{Q_B} &\ket{\lambda=S},
    \end{aligned}
\end{equation}
which is known as the \emph{quartet mechanism}.
These states are indeed unphysical because they are either $\notin\mathrm{Ker}(Q_B)$ (for $\ket{\lambda=L}$ and $\ket{\overline{c}}$) or $\in\mathrm{Im}(Q_B)$ and therefore equivalent to the 0-vector (for $\ket{c}$ and $\ket{\lambda=S}$).
In this way, physical and unphysical states are cleanly separated within the formal Fock space, yielding a physical Hilbert space of positive-norm states defined by equivalence classes.

In the interacting theory with $g\neq0$, the presence of interactions must not spoil the interpretation of physical equivalence classes.
Therefore, the interaction Hamiltonian or Lagrangian must be defined consistently on the physical Hilbert space, ensuring that interactions neither lead out of this physical subspace nor alter the equivalence of physically equivalent states.
In fact, the BRST construction extends straightforwardly to interacting field theories and applies to asymptotically free states in the LSZ scattering formalism of perturbation theory.
Since BRST symmetry is exact, the BRST charge $Q_B$ commutes with both the Hamiltonian $H$ and the $S$-matrix.
Consequently, if the asymptotic incoming state $\ket{i}$ is annihilated by $Q_B$, the same holds for the asymptotic outgoing state $\ket{f}$.

To analyse the physical $S$-matrix\footnote{The physical $S$-matrix is defined as the projection (or quotient mapping) of the full $S$-matrix onto the physical subspace $\mathcal{H}_\mathrm{phys}$, i.e.\ $S_\mathrm{phys}:\mathcal{H}_\mathrm{phys}\to\mathcal{H}_\mathrm{phys}$ (see Refs.~\cite{Boehm:2001Gauge,Kugo:1979gm}). 
It is thus restricted to $\mathcal{H}_\mathrm{phys}$ and does not map outside this space.
This restriction is well-defined because the $S$-matrix commutes with the BRST charge $Q_B$, guaranteeing that it preserves the physical subspace.} and identify the physical degrees of freedom, one studies the BRST transformation behaviour of the asymptotic fields,\footnote{Recall the correspondence between asymptotic fields and states (see Ref.~\cite{Boehm:2001Gauge}).} which appear in amputated on-shell Green functions used to derive $S$-matrix elements in the LSZ reduction formalism.
Specifically, by taking the external line of $s\phi_i$ on-shell and extracting the pole terms, one obtains the asymptotic transformations $s\phi^\mathrm{as}_i$.
In general, both linear and nonlinear contributions from $s\phi_i$ can occur (cf.\ Eq.~\eqref{Eq:Definition-of-the-BRST-Transformations}).
However, in the LSZ formalism, the nonlinear contributions drop out after amputation and taking the Green functions on-shell.
The reason is that such nonlinear contributions do not enter the LSZ formalism with a simple $1/(p_i^2-m_i^2)$ pole; they therefore vanish upon multiplication by $p_i^2-m_i^2$ and going on-shell.
This leads to a significant simplification and yields
\begin{subequations}
    \begin{align}
        sA_\mu^{\mathrm{as},a}(x) &= \partial_\mu c^{\mathrm{as},a}(x),\\
        s c^{\mathrm{as},a}(x) &= 0,\\
        s \overline{c}^{\mathrm{as},a}(x) &= B^{\mathrm{as},a}(x),\\
        s B^{\mathrm{as},a}(x) &= 0,
    \end{align}
\end{subequations}
which coincide precisely with the BRST transformations of the free fields.\footnote{The equation of motion $B^{\mathrm{as},a}(x)=-\partial^\mu A_\mu^{\mathrm{as},a}(x)/\xi$ relates the auxiliary field with the unphysical component $\ket{\lambda=S}$ of the gauge field.}
Hence, the quartet mechanism applies exactly as in the free theory, with physical Hilbert space $\mathcal{H}_\mathrm{phys}$.
Note that fermions possess only physical degrees of freedom, since the transformations of asymptotic fermion fields vanish (see Eq.~\eqref{Eq:Definition-of-the-BRST-Transformations}).

Gauge independence of the physical $S$-matrix can be established using Ward and Slavnov-Taylor identities, which arise from BRST invariance and express functional relations among Green functions (see Sec.~\ref{Sec:The-Slavnov-Taylor-Identity}).
In particular, the dependence on the gauge parameter $\xi$ enters only through the longitudinal part of the gauge boson propagator (cf.\ Eq.~\eqref{Eq:Gauge-Fixing-and-Ghost-Terms-Standard}).
For instance, in fermionic scattering processes via terms $\propto q_\mu \overline{u}(p')\gamma^\mu u(p)$.
From Ward and Slavnov-Taylor identities it follows that contracting an entire fermion line with the gauge boson momentum yields zero --- the longitudinal momentum contracted with the current vanishes --- reflecting current conservation.
In QED, for example, the $\xi$-independence of matrix elements $\mathcal{M}_{\mu\nu\ldots}$ becomes explicit through the Ward identity $k^\mu\mathcal{M}_{\mu\nu\ldots}=0$, which immediately reveals that corrections to the photon propagator of the form $\xi k_\mu k_\nu$ vanish (provided external fermions are on-shell).
Notably, this Ward identity also guarantees Lorentz invariance.
These arguments generalise straightforwardly to amputated on-shell Green functions and their associated Ward and Slavnov-Taylor identities, ensuring a gauge-independent physical $S$-matrix.

Unitarity of the physical $S$-matrix follows from its commutation with the BRST charge $Q_B$, i.e.\ $[Q_B,S]=0$, a relation that can be established by analysing the action of $Q_B$ on asymptotic states in amputated on-shell Green functions.
Consequently, if $Q_B\ket{\psi}=0$ for a state $\ket{\psi}\in\mathcal{H}_\mathrm{phys}$, then $Q_B S \ket{\psi} = 0$, implying that $S\ket{\psi}\in\mathrm{Ker}(Q_B)$.
Since $Q_B$ is Lorentz invariant and the effective Lagrangian (see Eqs.~\eqref{Eq:Effective-Lagrangian-Faddeev-Popov-Method} and \eqref{Eq:Full-Lagrangian-for-Quantisation}) is hermitian --- ensuring that the full $S$-matrix is unitary --- the unitarity of the physical $S$-matrix follows from the commutation relation $[Q_B,S]=0$ together with the cohomological definition of $\mathcal{H}_\mathrm{phys}$.

In summary, the BRST formalism provides a consistent framework for quantising gauge theories, yielding a well-defined quantum theory with a physical Hilbert space $\mathcal{H}_\mathrm{phys}$ and a physical $S$-matrix acting on it.
The resulting physical $S$-matrix is Lorentz invariant, gauge independent, and unitary, and it preserves the physical subspace $\mathcal{H}_\mathrm{phys}$.
The unphysical ghosts and antighosts must necessarily appear as internal lines in Feynman diagrams to ensure unitarity of the $S$-matrix and the correct imaginary part of its elements, thereby guaranteeing the validity of the optical theorem.
In fact, the unphysical states --- the ghosts and the unphysical degrees of freedom of the gauge bosons --- cancel among themselves, leading to a consistent interacting theory that preserves probability.
At the level of states, the role of the ghosts is to realise the BRST quartet mechanism and the associated structure of states.
Further details can be found in Refs.~\cite{Boehm:2001Gauge,Weinberg:1996kr,Stoeckinger:2023rqft2}.
Finally, the discussion above applies at tree level and does not yet account for renormalisation.
The renormalisation of gauge theories is addressed in Sec.~\ref{Sec:Renormalisation-of-GaugeTheories}, where we show that BRST invariance must be maintained --- through the corresponding Ward and Slavnov-Taylor identities --- order by order in renormalised perturbation theory to ensure the same consistency at higher orders.

\section{Anomalies}\label{Sec:Anomalies}

Symmetries play a central role in our understanding of the fundamental laws of nature.
Their breaking can lead to rich phenomenological consequences, as in the case of spontaneous symmetry breaking, but may also render an entire theory inconsistent, which is the case for the \emph{gauge anomaly}.
In this section, we briefly discuss the implications of anomalies, mainly following Refs.~\cite{Weinberg:1996kr,Boehm:2001Gauge,Stoeckinger:2023rqft2,TongStandardModel,TongGaugeTheory}.

In spontaneous symmetry breaking, a field acquires a non-vanishing vacuum expectation value such that the ground state is no longer invariant under the symmetry, while the Lagrangian itself remains symmetric.
In contrast, an \emph{anomaly} refers to a symmetry breaking induced by quantum effects of order $\mathcal{O}(\hbar)$:
a classical symmetry of the theory ceases to hold at the quantum level.
To avoid misconceptions, it is useful to distinguish between two cases:
\begin{enumerate}[label={(\arabic*)}]
    \item \emph{Anomaly}: The symmetry is broken by quantum effects that cannot be compensated by adjusting counterterms. 
    Such a breaking is a genuine physical effect, independent of the chosen regularisation or renormalisation scheme, and signals that the symmetry is incompatible with the quantum theory.
    \item \emph{Spurious Symmetry Breaking:} The symmetry is broken by the regularisation in such a way that it can be restored by adjusting symmetry-restoring counterterms (see Sec.~\ref{Sec:Algebraic_Renormalisation} and chapter~\ref{Chap:Practical_Symmetry_Restoration}).
    This breaking is merely an artefact of the regularisation and has no physical significance.
\end{enumerate}
Anomalies were first discovered in the context of chiral symmetry breaking in theories with chiral fermions, see Refs.~\cite{Adler:1969gk,Bell:1969ts,Adler:1969er}.
Fundamentally, they arise when the unitarity and causality constraints of renormalisation are incompatible with the symmetry in question.
Their origin can be traced to the path integral measure, as discussed in Refs.~\cite{Fujikawa:1979ay,Fujikawa:1980eg}: 
in a theory with an anomalously broken symmetry, the measure acquires a nontrivial Jacobian and does not stay invariant.
It is important to distinguish between the anomalous breaking of a global symmetry and of gauge invariance --- the latter corresponding to the gauge anomaly.

When a global symmetry is anomalously broken, the corresponding classically conserved Noether current fails to be conserved at the quantum level, i.e.\ $\partial_\mu j^\mu = \mathcal{O}(\hbar) \neq 0$.
While such global anomalies do not threaten the consistency of the quantum theory, they can have observable physical consequences.
An important example is the anomalous breaking of the global $U(1)_A$ symmetry of axial rotations, leading to well-known phenomena discussed below.

First, the decay of the neutral pion $\pi^0$ --- which has a much shorter lifetime than its charged counterparts and predominantly decays as $\pi^0\to \gamma \gamma$ --- provided the historical discovery of anomalies (see Refs.~\cite{Adler:1969gk,Bell:1969ts,Adler:1969er}).
In quantum electrodynamics with massless fermions, the axial symmetry is anomalously broken.
For the corresponding axial Noether current $j_A^\mu=\overline{\psi}_i\gamma^\mu\gamma_5\psi_i$, one finds at the quantum level
\begin{align}\label{Eq:Chiral-Anomaly-or-ABJ-Anomaly}
    \partial_\mu j_A^\mu = \frac{e^2N_f}{16\pi^2} \varepsilon^{\mu\nu\rho\sigma} F_{\mu\nu}F_{\rho\sigma},
\end{align}
which indicates the anomalous breaking of the axial symmetry, also known as \emph{ABJ anomaly}.
Here, $N_f$ denotes the number of fermion flavours. 
Without this anomaly, the matrix element $\langle j_A^\mu j^\nu j^\rho \rangle$ would vanish upon contraction with the pion momentum.
The non-vanishing is caused by the anomalous symmetry breaking of the global chiral symmetry, such that $\partial_\mu j_A^\mu\neq0$, and successfully explains the observed pion decay rate.
In contrast, the current $j_\mu$ associated with the gauged $U(1)$ symmetry remains conserved, i.e.\ $\partial_\mu j^\mu =0$, ensuring the consistency of the gauge theory.

Second, in the context of chiral symmetry breaking in QCD, the global $U(1)_A$ symmetry is anomalously broken.
While the remaining chiral symmetry group is spontaneously broken by the quark condensate, yielding massless Goldstone bosons in the limit of vanishing quark masses, the $\eta'$ meson associated with the $U(1)_A$ symmetry is significantly heavier.
This mass difference originates from the anomalous breaking of the axial symmetry, which prevents the $\eta'$ from being a (pseudo-)Goldstone boson.
Hence, it is neither massless in the limit of vanishing quark masses, nor particularly light in reality.

This situation is entirely different for local gauge invariance.
As discussed above, gauge invariance is not a true symmetry of nature but rather a redundancy of the mathematical formulation.
Nevertheless, it is essential for consistently describing massless spin-1 particles (the gauge bosons). 
As shown in Sec.~\ref{Sec:BRST-Symmetry}, gauge invariance or rather its successor BRST invariance is required to handle unphysical degrees of freedom, separate them from the physical content, and to ensure unitarity of the quantum theory.
Hence, gauge invariance represents a crucial part of the definition of a consistent gauge theory.
If it is anomalously broken by quantum effects, the theory becomes inconsistent and must be abandoned.
Therefore, the absence of gauge anomalies is a necessary condition for any viable quantum field theory, which must be ensured by \emph{anomaly cancellation}.
In chiral gauge theories, anomaly cancellation can be achieved by arranging the matter content such that the individual anomalous contributions cancel.
This is significantly simplified by the fact that gauge anomalies do not renormalise, established by the \emph{Adler-Bardeen theorem} (see Refs.~\cite{Adler:1969er,Piguet:1995er}), which states that gauge anomalies are 1-loop exact.
Ensuring its cancellation at the 1-loop level is sufficient to guarantee its absence to all orders.
Its connection to topology provides another explanation for the non-renormalisation of the gauge anomaly, established via the \emph{Atiyah-Singer index theorem} (see Refs.~\cite{Boehm:2001Gauge,TongGaugeTheory}).
In particular, we obtain
\begin{align}\label{Eq:Atiyah-Singer-Index-Theorem}
    \mathrm{Ind}(i\slashed{D}) = \frac{e^2}{32\pi^2} \int d^4x\, \varepsilon^{\mu\nu\rho\sigma} F_{\mu\nu}F_{\rho\sigma},
\end{align}
where $\mathrm{Ind}(i\slashed{D})=n_+-n_-$ denotes the index of the Dirac operator, given by the difference of the numbers of zero modes of $i\slashed{D}$ with $\gamma_5$ eigenvalues $\pm1$.
This index is a topological invariant, equal to a topological quantum number and therefore an integer.
The RHS of Eq.~\eqref{Eq:Atiyah-Singer-Index-Theorem} is likewise a topological integer that must remain unchanged under renormalisation, therefore excluding higher-order corrections to the anomaly which could otherwise spoil this relation.

Following the notation of Ref.~\cite{TongStandardModel}, the anomaly is encoded in the group-theoretical factor $\mathcal{A}(R)$ for some fermion representation $R$ of the gauge group $G$, for which we find
\begin{align}
    \mathrm{Tr}(T_R^a\{T_R^b,T_R^c\}) = \mathcal{A}(R) \, \mathrm{Tr}(T^a\{T^b,T^c\}) \equiv \mathcal{A}(R) \, d^{abc}
\end{align}
where $T_R^a$ denote the generators in representation $R$, $T^a$ those in the fundamental representation, and $d^{abc}$ is a totally symmetric group invariant.
For the fundamental representation $F$, we find $\mathcal{A}(F)=1$.
Consequently, the \emph{anomaly cancellation condition}
\begin{align}\label{Eq:General-Anomaly-Cancellation-Condition}
    \sum_\mathrm{left} \mathcal{A}(R_L) = \sum_\mathrm{right} \mathcal{A}(R_R),
\end{align}
ensures that the gauge anomaly vanishes to all orders of perturbation theory.
This condition must be satisfied in every consistent gauge theory.
It is automatically fulfilled in vector-like theories, while in the Standard Model its validity is a nontrivial consequence of the specific fermionic content and their charge assignments (see Refs.~\cite{Bouchiat:1972iq,Gross:1972pv,Geng:1989tcu}).\footnote{There are also mixed gauge-gravitational anomalies and nonperturbative effects. However, in the Standard Model gauge-gravitational anomalies also cancel, ensuring perturbative renormalisability.}
In Abelian chiral gauge theories, where left- and right-handed fermions couple through hypercharges $\hypL$ and $\hypR$, respectively, the anomaly cancellation condition reduces to
\begin{align}\label{Eq:Abelian-Anomaly-Cancellation-Condition}
    \mathrm{Tr}(\hypL^3) = \mathrm{Tr}(\hypR^3).
\end{align}
In this thesis, we always assume the absence of gauge anomalies.
For the theories discussed in chapters~\ref{Chap:General_Abelian_Chiral_Gauge_Theory} and \ref{Chap:BMHV_at_Multi-Loop_Level}, this is ensured by imposing the anomaly cancellation condition~\eqref{Eq:Abelian-Anomaly-Cancellation-Condition}, whereas in the Standard Model, discussed in chapter~\ref{Chap:The_Standard_Model}, it follows from its specific fermionic content fulfilling condition~\eqref{Eq:General-Anomaly-Cancellation-Condition}.

\chapter{Renormalisation and Symmetries}\label{Chap:Renormalisation_and_Symmetries}

In perturbative quantum field theory, the computation of quantum corrections --- higher-order contributions in $\hbar$ represented by Feynman diagrams with closed loops --- is of central importance for understanding the fundamental constituents of nature and their interactions.
However, such computations typically give rise to divergences. 
These can be classified into two types: infrared (IR) and ultraviolet (UV) divergences.
The theory of renormalisation addresses the latter, which will be the focus of this chapter.

The origin of UV divergences lies in the fact that loops correspond to the exchange of virtual particles whose momenta are integrated over all values up to infinity, thereby potentially introducing divergences in the momentum-space representation of such Feynman diagrams.
More fundamentally, quantum fields $\phi_i(x)$ are operator-valued distributions rather than operator-valued functions.
The product of distributions evaluated at the same spacetime point is, in general, ill-defined.
Such (time-ordered) products of quantum field operators inevitably appear in the evaluation of scattering processes.
In particular, the free propagator is defined by $\mathcal{P}_{ij}(x-y)=\langle0|T\phi_i(x)\phi_j(y)|0\rangle$, and a closed loop corresponds to the case where the propagator has coinciding end points, i.e.\ a time-ordered product of distributions evaluated at the same spacetime point.

The theory of renormalisation provides a systematic procedure to assign a well-defined meaning to such expressions by extending products of distributions beyond their ``classical'' domain of definition to coinciding spacetime points, thereby eliminating all UV divergences and yielding physically meaningful results.
Practically, this amounts to removing all UV divergences from the momentum-space representation of any Feynman diagram.
In Sec.~\ref{Sec:Renormalisation_Theory}, we review the theory of renormalisation from both perspectives: at the level of time-ordered products of quantum field operators appearing in the scattering operator and at the level of Feynman diagrams.

Another important aspect of renormalisation concerns the fate of symmetries at the quantum level.
As discussed in Sec.~\ref{Sec:Anomalies}, symmetries may be broken by quantum effects, where one distinguishes between genuine anomalies and spurious breakings.
Since some symmetries --- most notably gauge and BRST invariance, used to define the physical Hilbert space and ensure a unitary $S$-matrix, see Sec.~\ref{Sec:BRST-Symmetry} --- are essential for the consistency of the theory, their breaking would render the theory inconsistent.
To study symmetries at the quantum level, we introduce the Slavnov-Taylor identity, which encodes symmetries in form of functional relations between off-shell Green functions, together with the quantum action principle, which relates a potential symmetry breaking to variations of Green functions, in Sec.~\ref{Sec:Symmetries_at_the_Quantum_Level}.
The framework of algebraic renormalisation, introduced in Sec.~\ref{Sec:Algebraic_Renormalisation}, then offers an elegant approach for analysing possible breakings of BRST symmetry based on the cohomology of the BRST operator.
Ultimately, we discuss the renormalisability of gauge theories in general in Sec.~\ref{Sec:Renormalisation-of-GaugeTheories}, and conclude this chapter with a discussion of certain peculiarities of Abelian gauge theories in Sec.~\ref{Sec:Peculiarities_of_Abelian_Gauge_Theories}.

\section{Renormalisation Theory}\label{Sec:Renormalisation_Theory}

As introduced above, renormalisation theory provides the essential framework to construct a well-defined and finite perturbative quantum field theory by addressing the problem of UV divergences that originate from products of quantum field operators evaluated at coinciding spacetime points.

A rigorous and constructive formulation was pioneered and developed in Refs.~\cite{Bogoliubov:1957gp,Hepp:1966eg,Zimmermann:1969jj,Speer:1971fub,Speer:1974cz,Speer:1974hd,Epstein:1973gw}.
In this approach, renormalisation is understood as the systematic construction of the scattering operator $S(\mathbf{g})$, as a formal functional power series of all possible time-ordered products of free quantum field operators.
This construction proceeds inductively and guarantees that the resulting perturbative expansion is UV-finite and consistent with the fundamental physical principles of relativistic covariance, causality and unitarity.
Within this framework, each Feynman diagram is consistently mapped to a finite expression.

From a technical perspective, finite Green functions, $S$-matrix elements and related physical quantities are obtained by the removal of UV divergences.
In practice, this is achieved via local counterterms: divergences are typically first isolated through some regularisation procedure and subsequently subtracted by counterterms added to the Lagrangian. 
These counterterms can be interpreted as reparametrisations of the fields and parameters in the Lagrangian.
In this sense, renormalisation not only provides a subtraction procedure but also a reinterpretation of bare quantities in terms of physical ones.\footnote{Physical quantities are finite, and the renormalisation conditions supply the dictionary for this reinterpretation; for example, defining the physical mass as the pole of the propagator.}

Ultimately, renormalisation offers an effective definition of the perturbative expansion by fixing each contribution to the Gell-Mann-Low formula and thereby providing a perturbative definition of the path integral as the sum of all renormalised Feynman diagrams; see Sec.~\ref{Sec:Regularised-QAP} for the case of dimensional regularisation.

In the following, we adopt this constructive formulation of renormalisation, following the original works cited above, as well as the lectures in Refs.~\cite{Hepp:1971bda,Piguet:1980nr,Stoeckinger:2020mlr}, the textbooks \cite{Bogoliubov1980,Collins_1984,Piguet:1995er,Weinberg:1995mt}, and the reviews in Refs.~\cite{Popineau:2016lhf,Belusca-Maito:2023wah}.
We begin with the formal approach of causal perturbation theory and then turn to the practically more relevant framework at the level of Feynman diagrams.

\paragraph{Causal Perturbation Theory:}
The scattering operator $S(\mathbf{g})$ of the interacting theory governs the transition from an initial state $\ket{i}$ to a final state $\ket{f}$ via $\ket{f}=S(\mathbf{g})\ket{i}$.
In perturbation theory, the building blocks of the inductive construction of the scattering operator $S(\mathbf{g})$ are a finite set of free field operators $\{\phi_i\}$ and the associated Wick monomials $W_i(x)$ --- normal-ordered products of these field operators and their derivatives.
By assigning to each Wick monomial a number-valued test function $g_{i}(x)\in\mathscr{S}(\mathbb{M}_4)$, where $\mathscr{S}(\mathbb{M}_4)$ denotes the Schwartz space of rapidly decreasing smooth functions on Minkowski spacetime, we may define the interaction Lagrangian as
\begin{align}\label{Eq:generalised-int-Lagrangian-BPHZ-EG-PS}
    \mathcal{L}_{\text{int}}(x)=\sum_i g_{i}(x) W_i(x),
\end{align}
where $\{W_i(x)\}_{i\geq1}$ covers all local field monomials of interest, in particular those of the physical Lagrangian and additional composite operators.
The functions $\mathbf{g}(x)=(g_{i}(x))_{i\geq1}$ act as both localised coupling parameters --- which eventually approach the true physical coupling constants in the adiabatic limit (see Refs.~\cite{Epstein:1973gw,Bogoliubov1980}) --- and external sources.
The scattering operator corresponding to this interaction Lagrangian is then defined perturbatively as the formal functional power series 
\begin{align}\label{Eq:ScatteringOperator-FormalPowerSeries}
    S(\mathbf{g})=\mathbb{1} + \sum_{n=1}^{\infty} \frac{i^n}{n!} \int d^4x_1\ldots d^4x_n \sum_{i_1,\ldots,i_n} T[W_{i_1}(x_1)\ldots W_{i_n}(x_n)] \, g_{i_1}(x_1) \ldots g_{i_n}(x_n)
\end{align}
in $\mathbf{g}(x)$, where $T$ denotes time-ordering.
This expansion provides a perturbative definition of the transition amplitudes of the interacting theory.

In the quantum theory, each quantum field operator is defined as an operator-valued distribution on $\mathbb{M}_4$, $\phi:\mathscr{S}(\mathbb{M}_4)\longrightarrow\mathscr{O}(\mathcal{H})$, where 
$\mathscr{S}(\mathbb{M}_4)$ is the Schwartz space of test functions introduced above and 
$\mathscr{O}(\mathcal{H})$ denotes the set of densely defined operators on the Hilbert space $\mathcal{H}$.\footnote{In other words, quantum fields are (tempered) distributions rather than ordinary functions, which constitutes the central difficulty of renormalisation theory.}
Hence, the time-ordered product $T[W_{i_1}(x_1)\ldots W_{i_n}(x_n)]$ of $n$ Wick monomials is a product of operator-valued distributions, which is in general ill-defined if two or more spacetime points $x_i$ coincide ($n\geq2$).
As a consequence, this inevitably leads to the well-known \emph{cursed ultraviolet divergences}, as stated in Ref.~\cite{Popineau:2016lhf}, arising in Feynman integrals.
As announced, renormalisation is a procedure to give such time-ordered products a mathematically well-defined meaning, while preserving the fundamental axioms of quantum field theory.
In this way, the formal power series in Eq.~\eqref{Eq:ScatteringOperator-FormalPowerSeries} is consistently determined order by order in perturbation theory.

These axioms can be formulated as conditions on the scattering operator $S(\mathbf{g})$ enforced by fundamental physical principles as follows:
\begin{condition}[on the physical scattering operator $S(\mathbf{g})$]\
    \begin{enumerate}
        \item[$\mathbf{(0)}$] \textbf{Initial condition:}
        \begin{align}
            S(0) = \mathbb{1}, 
            \qquad\qquad 
            T[W_{i}(x)] = W_i(x).
        \end{align}
        \item[$\mathbf{(a)}$] \textbf{Relativistic covariance:} 
        \begin{align}
            U(\Lambda,a)S(\mathbf{g})U^\dagger(\Lambda,a) = S(\mathbf{g}_a),
        \end{align}
        where $U(\Lambda,a)$ is the usual unitary representation of the Poincaré group on the Hilbert space under consideration, and $\mathbf{g}_a(x)=\mathbf{g}(x-a),\,\,\forall\,a\in\mathbb{M}_4$. 
        Note that translational invariance is already sufficient for a well-defined perturbation theory, see Refs.~\cite{Epstein:1973gw,Popineau:2016lhf} and the remark below. However, in the relativistic setting of particle physics, Lorentz invariance is required as well.
        \item[$\mathbf{(b)}$] \textbf{Unitarity:} 
        \begin{align}
            S^\dagger(\mathbf{g})S(\mathbf{g})=S(\mathbf{g})S^\dagger(\mathbf{g})=\mathbb{1},
        \end{align}
        $\forall$ hermitian interactions $g_i(x)W_i(x)$. 
        The adjoint scattering operator $S^\dagger(\mathbf{g})$ is expressed in terms of anti-time-ordered products $\overline{T}[\bullet]$, see Ref.~\cite{Epstein:1973gw}. 
        \item[$\mathbf{(c)}$] \textbf{Causality:} 
        \begin{align}
            S(\mathbf{g}+\mathbf{h})=S(\mathbf{g})S(\mathbf{h}),
            \qquad\qquad
            \text{if}\quad\text{supp}(\mathbf{g})\gtrsim\text{supp}(\mathbf{h}),
        \end{align}
        where $\text{supp}(\mathbf{g})=\{x\in\mathbb{M}_4\,|\,\mathbf{g}(x)\neq0\}$ is the support of $\mathbf{g}$, and $\text{supp}(\mathbf{g})\gtrsim\text{supp}(\mathbf{h})$ means that all elements in $\text{supp}(\mathbf{h})$ are outside the closed future lightcone of any element in $\text{supp}(\mathbf{g})$: $\text{supp}(\mathbf{h}) \cap (\text{supp}(\mathbf{g})\cup\overline{\mathbb{V}}^+_{\mathbf{g}})=\emptyset$, where $\overline{\mathbb{V}}_\mathbf{g}^+=\bigcup_{x\in\text{supp}(\mathbf{g})}J^+(x)$ is the union of closed future lightcones $J^+$ for all elements in $\text{supp}(\mathbf{g})$.
        Hence, events in $\text{supp}(\mathbf{h})$ cannot be causally influenced by events in $\text{supp}(\mathbf{g})$, such that $S$ factorises as above. 
        Equivalently, the causality condition can be expressed in differential form as
        \begin{align}
            \frac{\delta}{\delta\mathbf{g}(y)}\bigg(\frac{\delta S(\mathbf{g})}{\delta\mathbf{g}(x)} S^\dagger(\mathbf{g})\bigg) =0,
            \qquad\qquad \text{for}\quad y^0<x^0.
        \end{align}
    \end{enumerate}
\end{condition}
These conditions translate directly into constraints on the time-ordered products $T[\bullet]$, see Refs.~\cite{Epstein:1973gw,Belusca-Maito:2023wah,Bogoliubov1980}, which play the central role in renormalisation theory and the perturbative construction of the scatter operator. 
In particular, these time-ordered products are constructed inductively. 
In the induction hypothesis, it is assumed that the construction has already been carried out up to order $n-1$, yielding a set of well-defined operator-valued distributions $T[W_{i_1}(x_1)\ldots W_{i_{n-1}}(x_{n-1})]$, which satisfy the required conditions $\mathbf{(0)}$ to $\mathbf{(c)}$.
In the induction step, it is then shown that the construction can consistently be extended to order $n$.
At this stage, the products of distributions $T[W_{i_1}(x_1)\ldots W_{i_{n}}(x_{n})]$ are initially defined only on $\mathbb{M}_4\setminus\text{diag}(X)$, where $X=(x_1,\ldots,x_n)$ and $\mathrm{diag}(X)=(X|_{x_1=\ldots=x_n})$ denotes the ``main diagonal''.
On this set, where all spacetime points coincide, the product is still ill-defined, leading to UV divergences in the momentum-space representation of corresponding Feynman diagrams with closed loops. 
The task is therefore to extend these time-ordered products of operator-valued distributions to the full space $\mathbb{M}_4$, including $\text{diag}(X)$, in such a way that all conditions $\mathbf{(0)}$ to $\mathbf{(c)}$ are preserved.
In this process, higher-order time-ordered products are expressed in terms of lower-order ones.
This extension, carried out inductively, is the central step of the renormalisation procedure, see Refs.~\cite{Epstein:1973gw,Popineau:2016lhf}, whereby the would-be divergences are absorbed into well-behaved local expressions.

Such a construction indeed exists (see Refs.~\cite{Epstein:1973gw,Bogoliubov1980} for proofs), but the extension is not unique and admits an ambiguity. 
The ambiguity takes to form of an operator-valued distribution supported entirely on the ``main diagonal'' $\mathrm{diag}(X)$, where the original expression was undefined and where the conditions $\mathbf{(0)}$ to $\mathbf{(c)}$ impose no further restrictions.
Concretely, this amounts to a quasilocal\footnote{\label{FN:Polylocal-and-Quasilocal}\emph{Quasilocal} refers to a polylocal operator that vanishes everywhere except at coinciding point $x_1=\ldots=x_n$.\\
Operators $\mathcal{O}(x_1,\ldots,x_n)$ built from $n$ normal-ordered local operators and their derivatives (i.e.\ Wick monomials) are called \emph{polylocal} if they satisfy $[\mathcal{O}_1(x_1,\ldots,x_n),\mathcal{O}_2(y_1,\ldots,y_m)]=0$, whenever all $x_i$ are spacelike separated from all $y_j$, see Ref.~\cite{Bogoliubov1980}.} hermitian operator-valued distribution of the form 
\begin{align}\label{Eq:Ambiguity-as-quasilocal-Operator}
    \mathcal{Q}_n(x_1,\ldots,x_n)=\sum_\alpha P_{\alpha}(x_n)D^\alpha\delta^{(4)}(x_1-x_2)\ldots\delta^{(4)}(x_{n-1}-x_n),
\end{align}
where $P_\alpha(x)$ is some Wick polynomial and $D^\alpha$ a derivative with multi-index $\alpha$.
Such terms vanish everywhere except on $\mathrm{diag}(X)$ and correspond precisely to local counterterms in the Lagrangian.
They reflect the freedom to add finite local counterterms to change the renormalisation scheme.

Hence, this inductive construction not only organises the perturbative series but also exposes the local ambiguities inherent to renormalisation.
Choosing a particular extension is equivalent to fixing a renormalisation scheme.
These insights can be summarised in the following theorem:
\begin{theorem}[Main Theorem of Renormalisation]\label{Thm:Main_Theorem_of_Renormalisation}\ \\
    Let $S_T(\mathbf{g})$ and $S_{T'}(\mathbf{G})$ be two perturbative constructions of the scattering operator as a formal functional power series in $\mathbf{g}=(g_i)_{i\geq1}$ and $\mathbf{G}=(G_i)_{i\geq1}$, see Eq.~\eqref{Eq:ScatteringOperator-FormalPowerSeries}, corresponding to two sequences of time-ordered products of operator-valued distributions $T[\bullet]$ and $T'[\bullet]$, respectively, each with any valid extension to the full space $\mathbb{M}_4$ (renormalisation) as described above.
    Then:
    \begin{enumerate}[label={(\arabic*)}]
        \item Both are valid solutions that yield well-defined, UV-finite results, satisfying relativistic covariance, unitarity and causality at every order.
        \item There exists a local formal power series in $\mathbf{g}$ and its derivatives,
        \begin{align}\label{Eq:Finite-Reparametrisation-formal-Series}
            G_i(x) = g_i(x)+\sum_{n=1}^\infty G_{i,n}(\mathbf{g},D\mathbf{g})(x),
        \end{align}
        such that 
        \begin{align}\label{Eq:Two-Renormalisations-equally-allowed}
            S_T(\mathbf{g}) = S_{T'}(\mathbf{G}).
        \end{align}
    \end{enumerate}
\end{theorem}
A constructive proof of the existence of such a perturbative procedure with valid extension/renormalisation, modulo the ambiguity discussed around Eq.~\eqref{Eq:Ambiguity-as-quasilocal-Operator}, is provided in Ref.~\cite{Epstein:1973gw}.
A general proof that particularly establishes the second part of the theorem based on the causality condition is given in Ref.~\cite{Popineau:2016lhf}.
An equivalent proof of renormalisation at the level of Feynman diagrams is to be found in Ref.~\cite{Hepp:1971bda}. Historically, the first complete proof of these results was given by the BPHZ theorem in Refs.~\cite{Bogoliubov:1957gp,Hepp:1966eg,Zimmermann:1969jj}, formulated in terms of Feynman diagrams and the so-called $\mathcal{R}$-operation, in which UV divergences are subtracted recursively order by order in perturbation theory. This approach will be discussed in the next part of this section. Before that, however, some remarks are in place concerning the physical conditions, the ambiguity of the renormalisation procedure, the asymptotic nature of the formal power series, and a comparison with the approach presented in Ref.~\cite{Bogoliubov1980}:
\begin{remark}[on the physical conditions]\
    \begin{itemize}
        \item Translational invariance is already sufficient for the construction of the scattering operator, see Refs.~\cite{Epstein:1973gw,Popineau:2016lhf}.
        Hence, the methods of perturbative quantum field theory can also be applied to systems that are not Lorentz invariant, such as those encountered in condensed matter physics. 
        However, in the relativistic framework of particle physics, we also require Lorentz invariance --- and thus Poincaré invariance as above. 
        After renormalisation, ensuring Lorentz covariance in a Lorentz invariant theory reduces to a trivial cohomology problem of $SL(2,\mathbb{C})$, the universal covering group of the Lorentz group $SO^+(1,3)$.
        For a finite dimensional representation, this cohomology is trivial according to Ref.~\cite{Popineau:2016lhf}.
        Consequently, a Lorentz covariant renormalisation of a Lorentz invariant theory is always possible.
        \item Unitarity guarantees that the norm is preserved and all probabilities sum to one.
        \item The fundamental requirement of causality ensures that later interactions or events cannot influence earlier ones: an event may affect the future evolution of a system but never its past. 
        In other words, causes always precede their effects.
        \item Local commutativity (microcausality): Causality implies
        \begin{align}
            [S(\mathbf{g}),S(\mathbf{h})]=0,
            \qquad\qquad
            \text{if}\quad\text{supp}(\mathbf{g})\sim\text{supp}(\mathbf{h}),
        \end{align}
        where the supports are spacelike separated, see Refs.~\cite{Bogoliubov1980,Epstein:1973gw}.
        This expresses the fact that no signal can travel faster than the speed of light, i.e.\ information propagates at a finite speed. 
        It translates directly into a condition for the time-ordered products $T[W_{i_1}(x_1)\ldots W_{i_n}(x_n)]$ in Eq.~\eqref{Eq:ScatteringOperator-FormalPowerSeries}, restricting them to be polylocal\footnote{See footnote~\ref{FN:Polylocal-and-Quasilocal} for a definition of polylocal.} operators, see Ref.~\cite{Bogoliubov1980}. 
        Since the individual Wick monomials $W_i(x)$ appearing in the Lagrangian are themselves local operators, this guarantees the fundamentally local nature of interactions: no instantaneous action at a distance.
        \item Additional symmetries, such as gauge and BRST invariance, must often be imposed on the scattering operator --- and thus on the renormalisation procedure --- to ensure a consistent theory describing the underlying physics, thereby further constraining renormalisation, as will be discussed in the following sections.
    \end{itemize}
\end{remark}
\begin{remark}[on the ambiguity in the renormalisation procedure and physical equivalence]\ \\
    As we have seen, the extension of the time-ordered products $T[\bullet]$ to coinciding spacetime points is not unique.
    The resulting ambiguity is local in nature and corresponds to a finite reparametrisation: the divergent part is uniquely determined in any valid extension, since any unsubtracted divergence would render the operator-valued distributions ill-defined, thereby guaranteeing UV finiteness at every order.    
    Crucially, this freedom does not affect the physical meaning of the theory, as it amounts merely to finite local redefinitions of fields and parameters.
    Hence, all admissible renormalisations differ only by finite local counterterms and are physically equivalent.
    This equivalence is made explicit in theorem~\ref{Thm:Main_Theorem_of_Renormalisation}, where $S_T(\mathbf{g})$ and $S_{T'}(\mathbf{G})$ are related through the finite reparametrisation in Eq.~\eqref{Eq:Finite-Reparametrisation-formal-Series}, restricted by power-counting and the physical requirements.
    Choosing a particular extension thus corresponds to selecting a renormalisation scheme, such as MS, $\overline{\mathrm{MS}}$ or on-shell, without altering the underlying physics.
    In other words, any allowed renormalisation describes the same physics.
\end{remark}
\begin{remark}[on the formal power series]\ \\
    The perturbative expansion of the scattering operator in Eq.~\eqref{Eq:ScatteringOperator-FormalPowerSeries} is a formal power series that provides only an asymptotic approximation of the exact quantity.
    While the complete series might be divergent when summed to infinite order, the individual terms at each order --- the primary objects of interest --- are expected to be integrable and well-defined, see Refs.~\cite{Bogoliubov1980,Piguet:1995er,Epstein:1973gw}.
    As discussed, these terms are time-ordered products of operator-valued distributions that are ``smeared'' against sufficiently smooth and rapidly decreasing test functions $\mathbf{g}(x)$, such that, after renormalisation (i.e.\ extension to the full space $\mathbb{M}_4$), the individual integrals converge to well-defined expressions.
    Truncated at a finite order, this asymptotic expansion provides remarkably accurate results for physical observables in weakly coupled theories, in excellent agreement with experiment.
\end{remark}
\begin{remark}[Epstein-Glaser vs.\ Bogoliubov-Shirkov]\ \\
    The approach of Ref.~\cite{Bogoliubov1980} is more heuristic and physically motivated. 
    A smooth measurable map $g:\mathbb{M}_4\longrightarrow[0,1]$ is introduced to ``switch'' interactions on and off: for $g(x)=0$ the interaction is absent, for $0<g(x)<1$ it is partially present, and for $g(x)=1$ it is fully active.
    The interaction Lagrangian is accordingly modified as $\mathcal{L}_{\text{int}}(x)\longrightarrow g(x)\mathcal{L}_{\text{int}}(x)$. 
    Imposing the conditions $\mathbf{(0)}$ to $\mathbf{(c)}$, Ref.~\cite{Bogoliubov1980} shows that a hermitian, relativistically covariant and local Lagrangian $\mathcal{L}_{\text{int}}$ satisfies all requirements and allows for the iterative construction of the scattering operator with $T[\mathcal{L}_{\text{int}}(x_1)\ldots\mathcal{L}_{\text{int}}(x_n)]$, cf.\ Eq.~\eqref{Eq:ScatteringOperator-FormalPowerSeries}, modulo an ambiguity given by quasilocal hermitian operators $\Lambda_n(x_1,\ldots,x_n)$ with coefficient functions involving derivatives and $\delta$-distributions. 
    This ambiguity arises because causality, unitarity, Lorentz covariance, and other symmetries impose recursion relations for the coefficients $S_n(x_1,\ldots,x_n)$, but do not fix them completely for $n \geq 2$. 
    Consequently, the scattering operator is fully specified only once both the interaction Lagrangian $\mathcal{L}_{\text{int}}$ and the infinite sequence $(\Lambda_n)_{n\geq2}$ are fixed, such that the most general solution reads
    \begin{align}\label{Eq:Scattering-Operator-Bogoliubov-Shirkov}
        S(g) = T \Big[e^{i\int d^4x\,\mathcal{L}_{\text{int}}(x;g)}\Big] = \mathbb{1} + \sum_{n=1}^{\infty}\frac{i^n}{n!}\int d^4x_1\ldots d^4x_n \, T[\mathcal{L}_{\text{int}}(x_1;g)\ldots\mathcal{L}_{\text{int}}(x_n;g)],
    \end{align}
    with effective interaction Lagrangian (the most general admissible form)
    \begin{align*}
        \mathcal{L}_{\text{int}}(x;g) = g(x)\mathcal{L}_{\text{int}}(x) + \sum_{n=2}^{\infty} \frac{1}{n!} \int d^4x_1\ldots d^4x_{n-1} \, \Lambda_n(x,x_1,\ldots,x_{n-1})g(x)g(x_1)\ldots g(x_{n-1}).
    \end{align*}
    The physically relevant case is the so-called adiabatic limit, $g(x)=1$, $\forall\,x\in\mathbb{M}_4$, see Ref.~\cite{Bogoliubov1980}.
    The quasilocal nature of $\Lambda_n(x,x_1,\ldots,x_{n-1})$ ensures that all integrals collapse, such that $\mathcal{L}(x;g)$ is indeed local.
    In fact, the $\Lambda_n$-terms play the role of local counterterms:
    on the one hand they cancel the UV divergences, and on the other hand they encode the freedom to add finite local counterterms, i.e.\ the choice of a renormalisation scheme.
    In this way, the connection to the standard counterterm approach becomes manifest and more transparent than in Ref.~\cite{Epstein:1973gw}.

    Thus, the approaches of Refs.~\cite{Epstein:1973gw} and \cite{Bogoliubov1980} are fully equivalent, differing only in their formulation and technical details.
    The Epstein-Glaser framework is mathematically more rigorous and based on distribution theory: renormalisation is reduced to extending distributions from their natural domain of definition to coinciding spacetime points. This extension is always possible (UV finiteness is guaranteed by construction) but not unique, and the resulting ambiguity appears as the freedom to add finite local counterterms, see Eq.~\eqref{Eq:Ambiguity-as-quasilocal-Operator}.
    In the Bogoliubov-Shirkov formulation, which is more heuristic yet physically more motivated, the infinite sequence $(\Lambda_n)_{n\geq2}$ captures the same freedom (finite local counterterms), but simultaneously provides the divergent counterterms --- determined recursively by the $\mathcal{R}$-operation --- making the connection to the standard counterterm method particularly explicit.
\end{remark}

\paragraph{Renormalisation on the level of Feynman Diagrams:} 
In the BPH theorem established in Refs.~\cite{Bogoliubov:1957gp,Hepp:1966eg}, UV divergences are isolated using an intermediate regularisation and subtracted recursively via the $\mathcal{R}$-operation.
Thus, this procedure is directly connected to the practical steps of regularising Feynman diagrams and adding counterterms.
Zimmermann's forest formula, see Ref.~\cite{Zimmermann:1969jj}, provides an explicit solution to the recursion and leads to the BPHZ formulation of renormalisation, in which the subtraction is implemented through repeated Taylor operations applied directly to the integrand.
This produces finite momentum-space integrals without intermediate regularisation, making the method conceptually cleaner, though technically more demanding.

The BPH and BPHZ theorems were the first rigorous proofs that renormalisation yields UV-finite results consistent with covariance, unitarity, and causality.
Their generality extends to all integrals, including those of non-renormalisable theories and composite operators, thereby establishing renormalisation as a consistent and universal procedure.

On the level of Feynman graphs, the main difficulty is the presence of subdivergences, which may be nested or overlapping; these are systematically resolved by the recursive $\mathcal{R}$-operation.
In this section we follow Refs.~\cite{Hepp:1971bda,Collins_1984,Stoeckinger:2020mlr} and review renormalisation in this diagrammatic setting and its relation to the counterterm formalism.
For this purpose, we begin with basic graph-theoretical concepts, for which further details can be found in Refs.~\cite{Hepp:1971bda,Collins_1984,Breitenlohner:1977hr,Stoeckinger:2020mlr}:
\begin{definitionsandpropositions}[Aspects of Graph Theory]\ \\
    Let $G$ be a Feynman graph, then:
    \begin{enumerate}[label={$(\roman*)$}]
        \item A connected graph that remains connected after cutting any line is called \emph{1-particle irreducible} (1PI).
        \item A maximal tree-level subgraph $H_T\subseteq G$ is called a \emph{spanning tree}.\footnote{Here, ``maximal'' means that adding any additional line would necessarily create at least one loop in the graph.}
        There is a 1-to-1 correspondence between all standard momentum routings and spanning trees.
        \item For $H\subseteq G$, the graph $G/H$, obtained by shrinking $H$ to a point, is called \emph{reduced graph}.
        \item The set of non-overlapping (disjoint or nested) 1PI subgraphs of $G$, is called a \emph{forest}: $\mathcal{U}(G)=\{H\subseteq G\,|\,\text{1PI}, \,H_i\cap H_j=\emptyset \,\,\text{or}\,\, H_i\subset H_j \, (i\neq j)\}$, which may also include the empty set.
        A forest that contains the full graph $G$ is called a \emph{full forest}, while one that does not contain $G$ is called a \emph{normal forest}.
        The set of all possible forests of $G$ is denoted by $\mathscr{F}(G)$, and the set of all normal forests by $\overline{\mathscr{F}}(G)$.
        \item A maximal set of non-overlapping 1PI subgraphs of $G$ is called \emph{maximal forest} $\mathcal{C}$. Any maximal forest $\mathcal{C}$ contains $L$ elements, where $L$ is the number of loops of $G$.
        A \emph{labelled maximal forest} $(\mathcal{C},\sigma)$ is obtained by assigning to each $H\in\mathcal{C}$ a characteristic line $l_H \in H$ as a label via the map $\sigma:H \longmapsto l_H$, such that, for $H'\subsetneq H$, one has $\sigma(H)\notin H'$. 
        Such a labelling is always possible but not unique.
        Removing all labelled lines, $G-\sigma(\mathcal{C})$, yields a spanning tree $H_T\subseteq G$.
        For any labelled maximal forest $(\mathcal{C},\sigma)$, we can define a domain
        $\mathcal{D}(\mathcal{C},\sigma)=\{(\alpha_1,\ldots,\alpha_I)\,|\,\alpha_i\geq0,\,\alpha_l\leq\alpha_{\sigma(H)}\,\mathrm{for}\,l\in H\in\mathcal{C}\}$, 
        where $I$ is the number of (internal) lines of $G$.
        This specifies one integration sector in the $\alpha$-space of Schwinger parametrisation.
        The disjoint union of all such integration sectors yields the full $\alpha$-space integration region,\footnote{Here, ``disjoint'' is meant in the sense of integration theory: for two different $(\mathcal{C},\sigma)$ and $(\mathcal{C}',\sigma')$, the intersection $\mathcal{D}(\mathcal{C},\sigma)\cap\mathcal{D}(\mathcal{C}',\sigma')$ is a set of zero measure.} i.e.\ $\bigcup_{(\mathcal{C},\sigma)}\mathcal{D}(\mathcal{C},\sigma)=\{(\alpha_1,\ldots,\alpha_I)\,|\,\alpha_i\geq0,\,\forall\,i\}.$
    \end{enumerate}
\end{definitionsandpropositions}
In renormalisation, an important property of a Feynman graph $G$ is its overall (or superficial) degree of divergence, a basic tool of power-counting.
It provides a first estimate of the UV behaviour of the diagram by indicating whether the graph exhibits an overall divergence (disregarding subdivergences) and, if so, the severity of that divergence. 
This degree directly determines the polynomial structure of the local counterterm required to cancel it.
Specifically, the overall degree of divergence is given as follows:
\begin{definition}[Overall Degree of Divergence]\ \\
    Let $G$ be a Feynman graph whose associated Feynman integral contains (at most) $N_{N}$ loop momenta in the numerator (including the measure) and $N_D$ in the denominator, then:
    \begin{align}
        \omega(G) = N_N-N_D = D L + n_N - N_D
    \end{align}
    is called \emph{overall (or superficial) degree of divergence}, where $n_N=N_N-DL$ denotes the maximal number of loop momenta in the numerator excluding the measure, $D$ is the spacetime dimension, and $L$ is the number of loops of $G$.
    For a quantum field theory with fields of type $A$ and mass dimension $d_A$, the overall degree of divergence of a graph $G$ with $E_A$ external lines of type $A$ and $v_i$ vertices of type $i$ (arising from a local interaction monomial $W_i(x)$) can equivalently be expressed as
    \begin{align}
        \omega(G) = D - \sum_A d_AE_A -\sum_i v_i \big[D-\mathrm{dim}(W_i)\big].
    \end{align}
    Accordingly, we call a graph with
    \begin{enumerate}[label={$(\roman*)$}]
        \item $\omega(G)<0$ power-counting convergent,
        \item $\omega(G)=0$ logarithmically divergent,
        \item $\omega(G)>0$ divergent of degree $\omega(G)$.
    \end{enumerate}
\end{definition}
Based on the overall degree of divergence, we can state the following result, also referred to as Weinberg's theorem:
\begin{theorem}[Convergence Theorem for power-counting finite Integrals]\label{Thm:Weinberg-Theorem}\ \\
    Let $G$ be a Feynman graph with overall degree of divergence $\omega(G)<0$, and $\omega(H)<0$ $\forall$ subgraphs $H\subset G$, then:
    The corresponding integral of $G$ is UV convergent.
\end{theorem}
If all propagators of the graph are massive, IR divergences are absent and the theorem guarantees absolute convergence under the stated conditions.
A detailed discussion and proof of this theorem can be found in Refs.~\cite{Collins_1984,Weinberg:1995mt}.

An important observation of this theorem is the following:
if all lower-loop 1PI subdiagrams of a graph $G$ are convergent, its divergence is determined solely by the overall degree $\omega(G)$.
By taking sufficiently many derivatives w.r.t.\ the external momenta, the degree $\omega(G)$ can be reduced to a negative value;
in particular, after $\omega(G)+1$ derivatives the remaining integral becomes UV finite by theorem~\ref{Thm:Weinberg-Theorem}.
It follows that the overall divergence is necessarily a polynomial in the external momenta (and masses, see Ref.~\cite{Collins_1984}) of degree $\omega(G)\geq0$.
Consequently, the counterterm that cancels this divergence of $G$ is a local polynomial of degree $\omega(G)$, i.e.\ supported at coinciding points in position space.
This observation underpins the locality of counterterms in renormalisation theory, as discussed above.

The main challenge arises from the fact that subgraphs $H\subset G$ can be divergent, giving rise to subdivergences that may be nested or overlapping.
Even after applying derivatives w.r.t.\ the external momenta, multi-loop diagrams with divergent subgraphs remain divergent, and these contributions are generally non-polynomial (non-local).
As mentioned above, handling these subdivergences is the main difficulty of renormalisation:
they must be subtracted recursively for every divergent subgraph, order by order in perturbation theory.
This is achieved by the $\mathcal{R}$-operation, see.~\cite{Bogoliubov:1957gp,Hepp:1966eg,Zimmermann:1969jj,Bogoliubov1980,Collins_1984,Stoeckinger:2020mlr}, which is defined as follows:
\begin{definition}[$\mathcal{R}$-Operation]\label{Def:The_R-Operation}\ \\
    Let $G$ be a Feynman graph and $H_1,\ldots,H_s$ all possible disjoint 1PI subgraphs of $G$, excluding $G$ itself.
    The $\mathcal{R}$-operation is defined by
    \begin{equation}\label{Eq:The-R-Operation}
        \begin{aligned}
            \overline{\mathcal{R}}(G)&=G+\sum_{H_1,\ldots,H_s} {G/_{H_1\cup\ldots\cup H_s}} \,\circ\, C(H_1)\ldots C(H_s),\\
            \mathcal{R}(G) &= \overline{\mathcal{R}}(G)+C(G),
        \end{aligned}
    \end{equation}
    with overall counterterm
    \begin{align}
        C(G)=-\mathcal{K}\cdot\overline{\mathcal{R}}(G).
    \end{align}
    Here, $\overline{\mathcal{R}}(G)$ denotes the subrenormalised graph and $\mathcal{R}(G)$ the fully renormalised graph.
    The operation ``$\circ$'' denotes an insertion-operation: 
    specifically, $G/H\,\circ\,C(H)$ means that the subgraph $H$ is shrunk to a point in $G$, and the associated counterterm $C(H)$ is inserted at that point into the reduced graph $G/H$.  
    The operator $\mathcal{K}$ extracts the divergent part of the expression it acts on, for example by a Taylor expansion in the external momenta, as in BPHZ renormalisation.
\end{definition}
In this context, it is convenient to introduce the shorthand notation
\begin{align}
    \mathcal{K}_H \circ G = G/H \,\circ\,\mathcal{K}\cdot H,
\end{align}
which denotes the replacement of the subgraph $H$ by its divergent part.
When applying multiple such operations, $\mathcal{K}_{H_i}\circ\mathcal{K}_{H_j}$ is always ordered by nesting: the divergence of the innermost subgraph is evaluated first, followed successively by the larger subgraphs, proceeding from innermost to outermost.
With this notation, we can state:
\begin{theorem}[Zimmermann's Forest Formula]\label{Thm:Zimmermanns_Forest_Formula}\ \\
    A solution of the recursive $\mathcal{R}$-operation in Eq.~\eqref{Eq:The-R-Operation} is given by
    \begin{equation}
    \begin{aligned}
        \overline{\mathcal{R}}(G)&=\sum_{\mathcal{U}\in\overline{\mathscr{F}}(G)}\prod_{H\in\,\mathcal{U}}(-\mathcal{K}_H)\circ G, \\
        \mathcal{R}(G)&=\sum_{\mathcal{U}\in\mathscr{F}(G)}\prod_{H\in\,\mathcal{U}}(-\mathcal{K}_H)\circ G = (1-\mathcal{K}_G)\circ \overline{\mathcal{R}}(G).
    \end{aligned}
    \end{equation}
\end{theorem}
The formulation and the inductive proof of the forest formula were first presented in Ref.~\cite{Zimmermann:1969jj}, while more recent discussions can be found in Refs.~\cite{Stoeckinger:2020mlr,Collins_1984}.
Employing Zimmermann's forest formula, renormalisation can be performed directly on any Feynman integrand without introducing an intermediate regularisation, yielding a UV-finite result.
This leads to the framework of BPHZ renormalisation and allows us to formulate the corresponding convergence theorem --- arguably the most important theorem of renormalisation theory:
\begin{theorem}[BPHZ Theorem --- Theorem of Convergence]\label{Thm:BPHZ-Theorem}\ \\
    Let $G$ be a Feynman graph and $\mathscr{F}(G)$ the set of all forests $\mathcal{U}$ of $G$, composed of the subgraphs $H\subseteq G$, then:
    \begin{align}\label{Eq:BPHZ-Subtraction-Formula}
        \mathcal{R}(G)=\sum_{\mathcal{U}\in\mathscr{F}(G)}\prod_{H\in\,\mathcal{U}}(-\mathcal{K}_H)\circ G = (1-\mathcal{K}_G)\circ \overline{\mathcal{R}}(G),
    \end{align}
    defines a fully renormalised graph, whose associated Feynman integral is well-defined and UV convergent.
\end{theorem}
Analogous to theorem~\ref{Thm:Weinberg-Theorem}, the BPHZ theorem guarantees absolute convergence if all propagators of the graph are massive, such that IR divergences are absent.
As already mentioned, the BPHZ theorem was originally established in Refs.~\cite{Bogoliubov:1957gp,Hepp:1966eg,Zimmermann:1969jj}, with modern presentations available in Refs.~\cite{Collins_1984,Weinberg:1995mt,Stoeckinger:2020mlr}.

The BPHZ formulation indeed resolves the problem of subdivergences by recursively subtracting all divergences --- both overall and those of every divergent subdiagram --- by summing over all possible forests of a graph.
In this framework, the operator $\mathcal{K}$ is realised as a Taylor expansion $\mathcal{T}_{H}$ of order $\omega(H)$ in the external momenta of a subgraph $H$, such that $(-\mathcal{K}_H)\circ G\longrightarrow-\mathcal{T}_{H}G$.
In this way, each UV divergent momentum-space integral is rendered finite and well-defined by BPHZ-subtraction of the integrand.
In line with the discussion above, the subtraction procedure is not unique, reflecting the freedom to choose a renormalisation scheme.
BPHZ renormalisation has clear advantages: it acts directly on the integrand without requiring an intermediate regularisation, it is mathematically rigorous and well defined, and it applies systematically to any Feynman graph, independent of the renormalisability of the underlying theory or the choice of regularisation. 
However, the method is also technically demanding. 
The combinatorics of forests makes the procedure cumbersome, practical calculations are difficult, and it is not well suited for theories with complicated symmetries such as gauge theories, where counterterms must satisfy nontrivial relations.
Moreover, since subtractions are made at zero momentum, infrared divergences arise in massless theories, necessitating the BPHZL extension (see the remark on IR divergences below).

A central virtue of the method is that it makes the locality of counterterms manifest: divergences are cancelled by local polynomials in momentum space, which correspond to local operators in position space and hence a reparametrisation of the theory.
This agrees with the discussion below theorem~\ref{Thm:Weinberg-Theorem}, but now it is evident that the statement applies to all divergences, not only to the overall one.
Consequently, the associated counterterms $C(H)=-\mathcal{K}\cdot\overline{\mathcal{R}}(H), \, \forall \,H\subseteq G$, are local polynomials in the external momenta (and masses) of degree $\omega(H)$.
A proof and detailed discussion of this result can be found in Ref.~\cite{Collins_1984}, making use of the fact that the $\mathcal{R}$-operation commutes with derivatives.

The importance of the BPHZ theorem is further underscored by the fact that it provides a mathematically well-defined renormalisation procedure consistent with fundamental principles of quantum field theory, such as unitarity and causality.
Moreover, any other renormalisation scheme may differ from BPHZ only by local counterterms.
Hence, any renormalisation procedure that is equivalent to BPHZ modulo local counterterms is equally admissible and consistent with unitarity and causality. 

These aspects are of great practical relevance, since they permit the use of any self-consistent regularisation and renormalisation scheme --- thus avoiding the technically demanding BPHZ procedure --- provided it is equivalent to BPHZ modulo local counterterms.
This leads us to the following corollary:
\begin{corollary}[Counterterms]\label{Thm:Corollary-on-Counterterm-Method}\ \\
    Let $G$ be a 1PI Feynman graph.
    Applying the $\mathcal{R}$-operation to graphs derived from the tree-level Lagrangian
    \begin{align}\label{Eq:Tree-Level-Lagrangian-Corollary-on-CTs}
        \mathcal{L}_0(x)=\mathcal{L}_\mathrm{free}(x)+\sum_i g_i W_i(x)
    \end{align}
    is equivalent to computing Feynman diagrams with the modified Lagrangian $\mathcal{L}_0+\mathcal{L}_{\mathrm{ct}}$, where 
    \begin{align}\label{Eq:Counterterm-Lagrangian-from-Graphs-G}
        \mathcal{L}_{\mathrm{ct}}(x) = \sum_{G} \mathcal{L}_\mathrm{ct}^{(G)}(x)
    \end{align}
    is a local counterterm Lagrangian obtained from $C(G)=-\mathcal{K}\cdot\overline{\mathcal{R}}(G), \forall \,G$.
    In this formulation, an intermediate regularisation is required, and the renormalisation procedure is performed iteratively order by order.
\end{corollary}
This procedure, the counterterm Lagrangian method, is applied in most practical applications.
After choosing a regularisation scheme --- for example dimensional regularisation, introduced in chapter~\ref{Chap:DReg} --- regularised Feynman diagrams are computed at a given order in perturbation theory.
Tree-level Feynman rules from Eq.~\eqref{Eq:Tree-Level-Lagrangian-Corollary-on-CTs} are of order $\hbar^0$, while each closed loop corresponds to a contribution of order $\hbar$.
Thus, a divergent $L$-loop 1PI diagram $G$ gives rise to a counterterm $C(G)$, and hence a counterterm Feynman rule, of order $\hbar^L$, which allows the sum in Eq.~\eqref{Eq:Counterterm-Lagrangian-from-Graphs-G} to be organised loop by loop.

Summing all Feynman diagrams contributing to a given Green function at fixed order in $\hbar$, with Feynman rules derived from $\mathcal{L}_0+\mathcal{L}_{\mathrm{ct}}$, yields a UV-finite and well-defined result.
In practical terms, this means that all Green functions become finite once the corresponding counterterm diagrams are included.
The modified Lagrangian then gives rise to a finite and well-defined scattering matrix that satisfies the fundamental requirements of unitarity and causality discussed above.

The counterterms are determined iteratively, order by order.
Their finite parts are not uniquely fixed, but are constrained by the renormalisation conditions that define the chosen scheme and symmetry requirements.
In fact, this counterterm method in combination with dimensional regularisation is not only used in most practical calculations but also throughout this thesis, as it is far better suited for gauge theories than a direct application of Eq.~\eqref{Eq:BPHZ-Subtraction-Formula} in the original BPHZ formulation.

Because the structure and number of counterterms determine the consistency and predictive power of the quantum theory, it is natural to classify theories by their degree of renormalisability, following Refs.~\cite{Piguet:1995er,Collins_1984,Weinberg:1995mt}, as follows:
\begin{definition}[Renormalisability]\ \\
    A quantum field theory is called 
    \begin{enumerate}[label={(\arabic*)}]
        \item \emph{superrenormalisable} if only finitely many UV-divergent diagrams exist (divergences occur only up to some finite order in perturbation theory),
        \item \emph{renormalisable} if only finitely many types of UV-divergent diagrams occur, such that all UV divergences can be absorbed by a finite set of counterterms bounded by power-counting and order by order in perturbation theory,
        \item \emph{nonrenormalisable} if new types of UV-divergent diagrams emerge at higher orders, requiring infinitely many counterterms unbounded by power-counting.
    \end{enumerate}
\end{definition}
\begin{proposition}[Renormalisability]\ \\
    Let $\delta_i=D-\mathrm{dim}(W_i)$ denote the dimension of the coupling constant associated with the interaction monomial $W_i$, then:
    \begin{enumerate}[label={(\arabic*)}]
        \item $\delta_i>0, \,\forall\,\,\mathrm{interactions}\,\,i$: the theory is superrenormalisable,
        \item $\delta_i\geq0, \,\forall\,\,\mathrm{interactions}\,\,i$: the theory is renormalisable,
        \item $\delta_i<0$ for at least one interaction $i$: the theory is nonrenormalisable.
    \end{enumerate}
\end{proposition}
In this thesis, we focus exclusively on power-counting renormalisable theories.
In such theories, the overall degree of divergence of a given Green function does not increase with loop order, and all divergences can be cancelled by a finite set of counterterms.
These counterterms correspond to a reparametrisation of the original fields and parameters in the classical Lagrangian. 
Thus, only finitely many physical parameters --- fixed by experiment --- are required, which ensures the predictive power of the theory.

By contrast, nonrenormalisable theories generally require new counterterms at each higher order, leading in principle to infinitely many free parameters.
Nevertheless, such theories --- including general relativity and the 4-Fermi theory --- can be consistently interpreted within the framework of effective field theories (EFTs), where predictive power is recovered order by order in an expansion w.r.t.\ an energy scale below which the EFT provides a valid low-energy approximation of a more fundamental theory. 

Moreover, symmetries can impose strong constraints on the structure of counterterms.
In some cases, they can even reduce the number of independent counterterms to a finite set, such that the theory effectively depends on finitely many physical parameters despite the presence of infinitely many possible counterterms.
Examples include supersymmetric gauge theories and nonlinear sigma models, see Ref.~\cite{Piguet:1995er}.

Before concluding this section, a remark regarding IR divergences is in place:
\begin{remark}[IR divergences]\ \\
IR divergences can arise in theories involving massless particles originating from the low-momentum (long-wavelength) behaviour of loop integrals.
Unlike UV divergences, which stem from large momenta, IR divergences can appear either directly from integration over small loop momenta or indirectly through subtraction terms that misbehave at zero momentum as artifacts of the renormalisation procedure.

A distinction is made between unphysical and physical IR divergences.
The former, which may occur for generic external momenta and correspond to artifacts of renormalisation, are removed by extending the subtraction scheme.
For example, the BPHZ method is generalised to the BPHZL procedure, which subtracts such unphysical IR divergences order by order and ensures the finiteness of Green functions, see Ref.~\cite{Lowenstein:1975ps}.
Physical IR divergences, by contrast, arise only for specific kinematics, such as on-shell external particles, and reflect the long-range nature of interactions mediated by massless particles.
These divergences cancel in physically meaningful observables, such as cross sections, where virtual and real corrections --- i.e.\ loop corrections and the emission of soft or collinear massless particles, respectively --- are combined.
\end{remark}

\paragraph{Concluding Comments on Renormalisation:}
Renormalisation is far more than just a technical procedure for ``sweeping infinities under the carpet''.
It is a fundamental concept that expresses the decoupling of energy scales in nature --- the very reason physics is possible at all. 
Physics at a given energy scale depends on phenomena at much higher energies only through a finite set of renormalised parameters and is otherwise insensitive to the UV details.
At its core, renormalisation is the systematic reparametrisation of the fields and parameters of a theory to absorb UV divergences, while maintaining consistency with fundamental physical principles such as covariance, unitarity and causality.
The divergences that appear in a perturbative quantum field theory are not a pathology, but rather an artifact of extrapolating a theory beyond its natural domain of validity to arbitrarily high energies --- scales where it is at best an effective description of a more fundamental reality.

The BPHZ theorem provides a rigorous mathematical foundation for renormalisation:
UV divergences are local and can systematically be subtracted order by order through local counterterms that preserve causality and unitarity, yielding finite observables.
More generally, the appearance of divergences reflects the ill-defined nature of products of distributions evaluated at coinciding spacetime points; renormalisation corresponds to extending such products beyond their original domain of definition to the full spacetime, a procedure that always exists but is not unique.

Physically, renormalisation corresponds to a reparametrisation of the theory that does not change its physical meaning.
Following the discussion in Ref.~\cite{Boehm:2001Gauge}, this can be illustrated by a simple theory with Lagrangian $\mathcal{L}(\phi_0,\partial\phi_0,g_0)$ and a ``bare'' parameter $g_0$.
Quantum corrections that change the relations between parameters and observables render $g_0$ regularisation-scale-dependent and divergent, making it unphysical.
The free parameter $g_0$ can be determined by relating observables $\sigma_i=\sigma_i(g_0,\Lambda)$, with $\Lambda$ being some regulator, to experimental measurements $E_i$, e.g.\ $E_1=\sigma_1(g_0,\Lambda)$.
Eliminating $g_0$ via $g_0=\sigma_1^{-1}(E_1,\Lambda)$ leads to finite, scheme-independent relations among physical observables, e.g.\ $\sigma_2(g_0,\Lambda)=\sigma_2(\sigma_1^{-1}(E_1,\Lambda),\Lambda)$.
Hence, the remaining observables can be compared with experiment providing testable predictions of the theory.
Introducing a renormalised parameter $g$ via $g_0=g_0(g,\Lambda)$ (and analogously for the fields), with $\sigma_i=\sigma_i(g_0(g,\Lambda),\Lambda)=\sigma'_i(g,\Lambda)$, makes this reparametrisation explicit under which we find equivalence of the physical relations between observables (physics remains unchanged).
The freedom to choose a renormalisation scheme\footnote{Different schemes yield results that differ only by higher-order corrections, which vanish in the all-order limit, see Ref.~\cite{Boehm:2001Gauge}} reflects the insensitivity of low-energy physics to the details of unknown high-energy dynamics, all of which are encoded in the renormalised parameters of the effective Lagrangian.

The scale evolution of the parameters is described by the renormalisation group (RG).
Together with the concept of effective field theories, this leads to an even more profound understanding of renormalisation.
In this framework, also referred to as the Wilsonian picture, every quantum field theory --- including the Standard Model --- is interpreted as an effective field theory valid below some energy scale $\Lambda$, beyond which new physics is expected.
Renormalisation corresponds to integrating out high-energy fluctuations, generating a sequence of effective Lagrangians at successively lower scales.
The RG equations, such as the Callan–Symanzik equation, govern the resulting flow of couplings.
A central insight is that generic RG flows are quickly attracted to trajectories determined by a few ``relevant'' operators, while the vast number of ``irrelevant'' operators, representing details of the UV completion, are suppressed, see Ref.~\cite{Skinner:QFTII}. 
This explains why low-energy physics depends only on finitely many parameters, regardless of the microscopic details of the underlying high-energy theory.

In this sense, renormalisation theory provides a framework that allows us to implement our ignorance and limited knowledge into our theories of nature.
Combined with the concept of effective field theories, it explains why macroscopic phenomena can be described without quantum mechanics, and why elementary particle physics can be understood without a fundamental theory of quantum gravity.
Renormalisation thus serves as the conceptual bridge that ensures predictive power across vastly different energy scales.

\section{Symmetries at the Quantum Level}\label{Sec:Symmetries_at_the_Quantum_Level}

At the quantum level, symmetries manifest themselves as functional relations among Green functions.
Specifically, symmetry relations may generally acquire quantum corrections, i.e.\ they renormalise.
A central question is whether classical symmetries survive quantisation or are broken by quantum effects --- an issue of particular importance for symmetries that are essential for consistency of a theory, such as gauge invariance.
In this section, we first review Green functions and their generating functionals, which encode the full quantum dynamics, see Sec.~\ref{Sec:GreenFunctions_and_GeneratingFunctionals}, with additional remarks on the insertion of composite operators.
We then introduce the quantum action principle, which systematically relates variations of the action, including potential breakings, to variations of Green functions and generating functionals (Sec.~\ref{Sec:TheQuantumActionPrinciple}).
Finally, in Sec.~\ref{Sec:The-Slavnov-Taylor-Identity}, we present the Slavnov–Taylor identity, a functional relation that encodes symmetries at the quantum level and serves as the central tool for ensuring the consistency of gauge theories after renormalisation, cf.\ Sec.~\ref{Sec:Renormalisation-of-GaugeTheories}.
For this purpose, we follow primarily Refs.~\cite{Belusca-Maito:2023wah,Stoeckinger:2020mlr}, supplemented by standard literature in Refs.~\cite{Zinn-Justin:1989rgp,Itzykson:1980rh}.

\subsection{Green Functions and Generating Functionals}\label{Sec:GreenFunctions_and_GeneratingFunctionals}

In quantum field theory, the fundamental objects encoding the complete information of the full, interacting quantum theory are the Green functions (or correlation functions)
\begin{align}\label{Eq:Exact-Green-Functions}
    G_{i_1\ldots i_n}(x_1,\ldots,x_n)\coloneqq\bra{\Omega}T[\Phi^H_{i_1}(x_1)\ldots\Phi^H_{i_n}(x_n)]\ket{\Omega},
\end{align}
which are vacuum expectation values of time-ordered products of quantum field operators.
Here, $\ket{\Omega}$ denotes the exact vacuum of the interacting theory (the ground state of the full Hamiltonian), and $\Phi^H_i(x_i)$ are quantum field operators in the full interacting Heisenberg picture.
Hence, Eq.~\eqref{Eq:Exact-Green-Functions} represents a Green function of the exact quantum theory.

\paragraph{The Generating Functional $Z[J]$:}
A central tool in quantum field theory is the \emph{generating functional} 
\begin{align}\label{Eq:Exact-General-Generating-Functional}
    Z[J]=\mathcal{N}\,\bra{\Omega}Te^{i\int d^4x\,J_i(x)\Phi^{H}_i(x)}\ket{\Omega},
\end{align}
with normalisation constant $\mathcal{N}$ and sources $J_i(x)$, which are number-valued test functions for each quantum field operator.
Derivatives w.r.t.\ the sources $J_i$ generate all Green functions 
\begin{align}\label{Eq:Exact-Green-Functions-from-Generating-Functional}
    \langle\Omega|T[\Phi^H_{i_1}(x_1)\ldots\Phi^H_{i_n}(x_n)]|\Omega\rangle = \frac{1}{Z[0]}\bigg(\frac{\delta}{i\delta J_{i_1}(x_1)}\cdots\frac{\delta}{i\delta J_{i_n}(x_n)}Z[J]\bigg)\Bigg|_{J=0},
\end{align}
of the full, non-perturbative quantum field theory.
Thus, Green functions --- and axiomatic quantum field theory in general --- can be defined without the need of a Lagrangian $\mathcal{L}$, relying solely on operator relations.

In practice, however, essentially all relevant quantum field theories, including the Standard Model, are specified by a Lagrangian.
A Lagrangian provides a constructive definition of the specific theory under consideration, and via canonical or path-integral quantisation establishes relations between the fundamental fields and the resulting Green functions.
This is particularly advantageous because many key aspects, such as symmetries, are most transparently formulated at the Lagrangian level.
With a Lagrangian $\mathcal{L}$, the generating functional admits the path integral representation
\begin{align}\label{Eq:Exact-General-Generating-Functional-Path-Integral}
    Z[J]=\mathcal{N}' \int \mathcal{D}\phi \, e^{i \int d^4x \big[\mathcal{L}(x)+J_i(x)\phi_i(x)\big]},
\end{align}
where $\mathcal{D}\phi$ denotes the path integral measure.
Here the fields $\phi_i$ are number-valued, while quantisation is implemented by the functional integration over all possible field configurations, which then distinguishes the fields $\phi_i$ from the external sources $J_i$.
Although Eq.~\eqref{Eq:Exact-General-Generating-Functional-Path-Integral} is in principle non-perturbative, the path integral itself lacks a fully rigorous mathematical definition and should be understood formally.
It can be made well-defined perturbatively, within a consistent framework of regularisation and renormalisation, as will be discussed in Sec.~\ref{Sec:Regularised-QAP}.

\paragraph{Composite Operators:} 
Before addressing the perturbative formulation of quantum field theory using the Gell-Mann-Low formula, we introduce composite operators and their insertions, which will later be essential for expressing symmetry breakings, see e.g.\ chapter~\ref{Chap:Practical_Symmetry_Restoration}.
\begin{definition}[Composite Operators]\ \\
    A local operator $\mathcal{O}(x)$, defined as a product of elementary quantum field operators and their derivatives evaluated at the same spacetime point, is called a \emph{composite operator}.
\end{definition}
Since quantum fields are operator-valued distributions, such products at coinciding arguments are initially ill-defined, and thus require renormalisation, as discussed in Sec.~\ref{Sec:Renormalisation_Theory}.
Hence, composite operators can only be defined in the quantum theory within a chosen renormalisation procedure.
As discussed, while such a definition is not unique (finite ambiguities correspond to local counterterms), symmetries and renormalisation conditions fix these ambiguities. 
By construction, composite operators are local, and their tree-level Green functions coincide with the ``naive expressions'', while higher-order corrections are renormalisation-scheme dependent.

Composite operators can be seamlessly incorporated into the general framework of Green functions and renormalisation (cf.\ Ref.~\cite{Collins_1984,Stoeckinger:2020mlr,Belusca-Maito:2023wah}). 
As for elementary fields, this is achieved by introducing external sources $K_i(x)$ (number-valued test functions) coupled to the operators, such that
\begin{align}\label{Eq:Lagrangian+SourceTermsOfCompositeOperators}
    \mathcal{L}(x)\longrightarrow\mathcal{L}(x)+\sum_iK_i(x)\mathcal{O}_i(x),
\end{align}
where $K_i$ are classical sources (not quantised) and do not possess internal propagators, analogous to $J_i$.
Green functions involving such local composite operators are then given by
\begin{equation}\label{Eq:Full-Green-Functions-Fields-and-CompositeOperators}
\begin{aligned}
    G_{i_1\ldots i_n}^{k_1\ldots k_m}&(x_1,\ldots,x_n,y_1,\ldots,y_m)=\langle\Omega|T[\Phi^H_{i_1}(x_1)\ldots\Phi^H_{i_n}(x_n)\mathcal{O}^H_{k_1}(y_1)\ldots\mathcal{O}^H_{k_m}(y_m)]|\Omega\rangle\\
    &=\frac{1}{Z[0]}\bigg(\frac{\delta}{i\delta J_{i_1}(x_1)}\cdots\frac{\delta}{i\delta J_{i_n}(x_n)}\frac{\delta}{i\delta K_{k_1}(y_1)}\cdots\frac{\delta}{i\delta K_{k_m}(y_m)}Z[J,K]\bigg)\Bigg|_{J=K=0},
\end{aligned}
\end{equation}
where $\mathcal{O}^H_i(x)$ are composite operators composed of field operators $\Phi^H_j(x)$ of the full, interacting Heisenberg picture.
The corresponding generating functional is given non-perturbatively by
\begin{align}\label{Eq:Exact-General-Generating-Functional-J-and-K}
    Z[J,K]
    =\mathcal{N}\,\langle\Omega|Te^{i\int d^4x\big[J_i(x)\Phi^H_i(x)+K_i(x)\mathcal{O}^H_i(x)\big]}|\Omega\rangle,
\end{align}
or via the path integral
\begin{align}\label{Eq:Exact-General-Generating-Functional-J-and-K-via-PathIntegral}
    Z[J,K]= \mathcal{N}' \int \mathcal{D}\phi \, e^{i \int d^4x \big[\mathcal{L}(x)+J_i(x)\phi_i(x)+K_i(x)\mathcal{O}_i(x)\big]}.
\end{align}

\paragraph{Perturbation Theory and Gell-Mann-Low Formula:}
In perturbation theory, the Gell-Mann-Low formula provides the following construction of the generating functional
\begin{align}\label{Eq:Z[J,K]-via-Gell-Mann-Low}
    Z[J,K]=
    \frac{\langle 0|Te^{i\int d^4x\big[\mathcal{L}_{\mathrm{int}}(x)+J_i(x)\phi_i(x)+K_i(x)\mathcal{O}_i(x)\big]}|0\rangle}{\langle 0|Te^{i\int d^4x\,\mathcal{L}_{\mathrm{int}}(x)}|0\rangle},
\end{align}
where $\ket{0}$ denotes the vacuum of the free theory, $\phi_i(x)$ are the free quantum field operators, and the Lagrangian is split as $\mathcal{L}=\mathcal{L}_\mathrm{free}+\mathcal{L}_\mathrm{int}$, with $\mathcal{L}_\mathrm{free}$ being the Lagrangian of the free theory which is bilinear in the fields. 

General Green functions, including those with insertions of composite operators, are then obtained via the Gell-Mann-Low formula as 
\begin{equation}\label{Eq:Perturbative-Green-Functions-from-Generating-Functional}
    \begin{aligned}
        G_{i_1\ldots i_n}^{k_1\ldots k_m}&(x_1,\ldots,x_n,y_1,\ldots,y_m)\equiv\langle\phi_{i_1}(x_1)\ldots\phi_{i_n}(x_n)\mathcal{O}_{k_1}(y_1)\ldots\mathcal{O}_{k_m}(y_m)\rangle\\
        &= \frac{1}{Z[0]}\bigg(\frac{\delta}{i\delta J_{i_1}(x_1)}\cdots\frac{\delta}{i\delta J_{i_n}(x_n)}\frac{\delta}{i\delta K_{k_1}(y_1)}\ldots\frac{\delta}{i\delta K_{k_m}(y_m)}Z[J,K]\bigg)\Bigg|_{J=K=0}\\
        &=\frac{\bra{0}T[\phi_{i_1}(x_1)\ldots\phi_{i_n}(x_n)\mathcal{O}_{k_1}(y_1)\ldots\mathcal{O}_{k_m}(y_m)\,e^{i\int d^4x\,\mathcal{L}_{\mathrm{int}}(x)}]\ket{0}}{\bra{0}T e^{i\int d^4x\,\mathcal{L}_{\mathrm{int}}(x)}\ket{0}}\\
        &=\frac{\int D\phi \,\phi_{i_1}(x_1)\ldots\phi_{i_n}(x_n)\mathcal{O}_{k_1}(y_1)\ldots\mathcal{O}_{k_m}(y_m)\,e^{i\int d^4x \, \mathcal{L}(x)}}{\int D\phi \,e^{i\int d^4x \, \mathcal{L}(x)}}.
    \end{aligned}
\end{equation}
Thus, Green functions can be constructed perturbatively from free fields, with interactions organised by the exponential of $\mathcal{L}_\mathrm{int}$, and where the equivalence holds order by order in the sense of perturbation theory (up to non-perturbative effects).
In particular, Eq.~\eqref{Eq:Z[J,K]-via-Gell-Mann-Low} and the third line of Eq.~\eqref{Eq:Perturbative-Green-Functions-from-Generating-Functional} show explicitly how the scattering operator --- whose perturbative construction has been discussed in Sec.~\ref{Sec:Renormalisation_Theory}, cf.\ Eqs.~\eqref{Eq:ScatteringOperator-FormalPowerSeries} and \eqref{Eq:Scattering-Operator-Bogoliubov-Shirkov} --- appears in the Gell-Mann-Low formula.

Evaluated through Wick contractions, each term of the expansion gives rise to the familiar Feynman diagrams.
As discussed in Sec.~\ref{Sec:Renormalisation_Theory}, the associated expressions are generally divergent and hence require renormalisation to be well-defined (typically accompanied by a regularisation).
It this sense, the path integral acquires a well-defined meaning perturbatively as the sum of all regularised Feynman diagrams, see Sec.~\ref{Sec:Regularised-QAP}.
Composite operators enter on the same footing as elementary fields, and are thus fully incorporated, including the necessary counterterms for $\mathcal{O}_i$ generated by renormalisation.

Finally, Eq.~\eqref{Eq:Perturbative-Green-Functions-from-Generating-Functional} further illustrates the practical advantage of the Lagrangian formalism: it naturally accommodates the counterterm method by adding $\mathcal{L}_\mathrm{ct}$ and provides a more accessible framework for calculations than a direct application of the $\mathcal{R}$-operation alone.

\paragraph{Generating Functionals for Connected and 1PI Diagrams:}
So far, we have considered full Green functions, which generate all Feynman diagrams, including disconnected components. 
To isolate specific classes of diagrams and condense the information content, it is convenient to introduce generating functionals for connected and 1PI Green functions.

Connected Green functions generate only topologically connected diagrams and are obtained from the generating functional $W[J,K]$ defined via
\begin{align}\label{Eq:Definition-of-W-Functional}
    Z[J,K]=e^{iW[J,K]}.
\end{align}

For renormalisation, and hence for this thesis, the most relevant class of diagrams are the 1PI diagrams (introduced in Sec.~\ref{Sec:Renormalisation_Theory}).
Built directly from the Feynman rules, 1PI diagrams form the elementary building blocks of perturbation theory and contain the complete information about the UV behaviour of the theory, with closed loops naturally appearing only inside a 1PI block.
The corresponding 1PI Green functions are generated by the functional $\Gamma$, also called the \emph{effective (quantum) action} or \emph{vertex functional}, which is defined as the Legendre transformation of $W[J,K]$ w.r.t.\ the sources $J_i$ of the elementary fields (\emph{not} w.r.t.\ $K_i$):
\begin{align}\label{Eq:Definition-of-Gamma-Functional}
    \Gamma[\phi_{\mathrm{cl}},K] = W[J,K] - \int d^4x\, J_i(x) \phi_{i,\mathrm{cl}}(x)\bigg|_{\phi_{\mathrm{cl}}=\frac{\delta W}{\delta J}},
\end{align}
with its inverse given by
\begin{align}
    W[J,K] = \Gamma[\phi_{\mathrm{cl}},K] + \int d^4x\, J_i(x) \phi_{i,\mathrm{cl}}(x)\bigg|_{J=-\frac{\delta \Gamma}{\delta \phi_{\mathrm{cl}}}}.
\end{align}
The arguments of $\Gamma$ are ``classical'' fields defined as
\begin{align}\label{Eq:Definition-of-phi_cl-as-Arguments-of-Gamma}
    \phi_{i,\mathrm{cl}}(x)\coloneqq\langle \phi_i(x) \rangle_{J,K} = \frac{\delta W[J,K]}{\delta J_i(x)},
\end{align}
which are expectation values of the quantum field operators $\phi_i(x)$ in presence of the sources $J_i(x)$ and $K_i(x)$.
These expectation values are number-valued and assumed to vanish for vanishing sources, i.e.\ $J_i=0\longleftrightarrow\phi_{i,\mathrm{cl}}=0$ (for $K=0$):
\begin{align}\label{Eq:Tadpole-Elimination-Condition}
    \frac{\delta W[J,K]}{\delta J_i(x)}\bigg|_{J=K=0} = 0 = \frac{\delta \Gamma[\phi_{\mathrm{cl}},K]}{\delta \phi_{i,\mathrm{cl}}(x)}\bigg|_{\phi_{\mathrm{cl}}=K=0}.
\end{align}
Since the Legendre transform is w.r.t.\ $J_i\longleftrightarrow\phi_i$, the dependence on the sources of composite operators $K_i$ remains unchanged, acting as spectators (see Ref.~\cite{Belusca-Maito:2023wah,Itzykson:1980rh,Zinn-Justin:1989rgp}), such that
\begin{align}\label{Eq:Relation-W-to-Gamma-derivatives-wrt-K}
    \frac{\delta\Gamma[\phi_{\mathrm{cl}},K]}{\delta K_i(x)}=\frac{\delta W[J,K]}{\delta K_i(x)}.
\end{align}

The effective action perturbatively takes the form
\begin{align}
    \Gamma[\phi_{\mathrm{cl}},K] = \Gamma_\mathrm{cl}[\phi_{\mathrm{cl}},K] + \mathcal{O}(\hbar),
\end{align}
with the classical action
\begin{align}
    \Gamma_\mathrm{cl} \equiv S_0 = \int d^4x \, \mathcal{L}_{0}(x).
\end{align}
Thus, $\Gamma$ is the classical action plus all quantum corrections, which justifies the name \emph{effective action}.

For convenience, we henceforth drop the subscript ``$\mathrm{cl}$’’ and write simply $\phi_i$ whenever no confusion can arise.
Furthermore, for the rest of this subsection, we collectively denote the classical fields $\phi_{i,\mathrm{cl}}(x)$ and the sources $K_i(x)$ by $\phi_i(x)$ in order to compactify the notation.

Expanding $\Gamma$ in powers of the fields,
\begin{align}\label{Eq:Expansion-of-Gamma-in-Fields}
    \Gamma[\phi] = \sum_{n=2}^\infty\frac{1}{|n|!}\int d^4x_1\ldots d^4x_n \, \phi_{i_1}(x_1)\ldots\phi_{i_n}(x_n) \, \Gamma_{\phi_{i_1}(x_1)\ldots\phi_{i_n}(x_n)}
\end{align}
with $|n|!=\prod_k n_k!$ (where $n_k$ counts fields of type $k$), the 1PI Green functions are obtained as
\begin{align}
    \Gamma_{\phi_{i_1}(x_1)\ldots\phi_{i_n}(x_n)} = \frac{\delta}{\delta\phi_{i_1}(x_1)}\cdots\frac{\delta}{\delta\phi_{i_n}(x_n)}\Gamma[\phi]\bigg|_{\phi=0} \equiv -i \langle \phi_{i_1}(x_1)\ldots\phi_{i_n}(x_n) \rangle^{\mathrm{1PI}},
\end{align}
where the order of derivatives matters for fermionic fields.
Because tadpoles are eliminated by the condition in Eq.~\eqref{Eq:Tadpole-Elimination-Condition}, the sum in Eq.~\eqref{Eq:Expansion-of-Gamma-in-Fields} starts from $n=2$.

Switching to momentum space via Fourier transformation, we obtain
\begin{align}
    \Gamma[\phi]=\sum_{n=2}^\infty \frac{1}{|n|!}\int \frac{d^4p_1}{(2\pi)^4}\cdots\frac{d^4p_n}{(2\pi)^4} \, \phi_{i_1}(p_1)\ldots\phi_{i_n}(p_n) \, \Gamma_{\phi_{i_1}(p_1)\ldots\phi_{i_n}(p_n)} \, (2\pi)^4 \delta^{(4)}\bigg(\sum_{j=1}^{n}p_j\bigg),
\end{align}
where the tilde to indicate Fourier transformed quantities is omitted for convenience and momentum conservation is enforced by the $\delta$-distribution.
Taking derivatives yields
\begin{align}
    \Gamma_{\phi_{i_1}(p_1)\ldots\phi_{i_n}(p_n)} \, (2\pi)^4 \delta^{(4)}\bigg(\sum_{j=1}^{n}p_j\bigg) = (2\pi)^{4n} \frac{\delta}{\delta\phi_{i_1}(p_1)}\cdots\frac{\delta}{\delta\phi_{i_n}(p_n)}\Gamma[\phi]\bigg|_{\phi=0},
\end{align}
so that the 1PI Green functions in momentum space are 
\begin{align}\label{Eq:1PI-Green-Functions-MomentumSpace}
    i \Gamma_{\phi_{i_1}(p_1)\ldots\phi_{i_n}(p_n)} = \langle \phi_{i_1}(p_1)\ldots\phi_{i_n}(p_n) \rangle^{\mathrm{1PI}},
\end{align}
with all momenta incoming.
These 1PI Green functions are evaluated perturbatively as the sum of all Feynman diagrams with external fields $\phi_{i_1}(p_1),\ldots,\phi_{i_n}(p_n)$ at the given loop order.

\paragraph{Single Operator Insertions:}
In many cases it is convenient to absorb the source terms for composite operators into the Lagrangian, cf.\ Eq.~\eqref{Eq:Lagrangian+SourceTermsOfCompositeOperators}, so that they are fully incorporated into the standard renormalisation procedure.
This applies, for example, to operators associated with nonlinear BRST transformations (see Sec.~\ref{Sec:BRST-Symmetry}).

However, in some situations one is interested only in the insertion of a single special operator. 
In these cases, the operator is treated slightly differently: a source is introduced only temporarily and removed after differentiation.
For a local composite operator $\Delta$ with auxiliary source $Y$, a single insertion into the generating functional is defined by
\begin{align}\label{Eq:SingleOperatorInsertion-Z}
    \Delta\cdot Z[J,K] = \mathcal{N}' \int \mathcal{D}\phi \, \Delta \, e^{i\int d^4x \big[ \mathcal{L}(x) + J_i(x)\phi_i(x) + K_i(x)\mathcal{O}_i(x)\big]} = \frac{\delta Z[J,K,Y]}{i\delta Y}\bigg|_{Y=0},
\end{align}
where in $Z[J,K,Y]$ the source term $Y(x)\Delta(x)$ is introduced into the exponent of $Z[J,K]$ only temporarily for the purpose of differentiation.
In this way, the insertion of $\Delta$ is naturally embedded into the general formalism and renormalisation framework without introducing a permanent $Y$-dependence.

Analogous definitions hold for the other generating functionals,
\begin{equation}\label{Eq:SingleOperatorInsertion-W-Gamma}
    \begin{aligned}
        \Delta\cdot W[J,K] &= \frac{\delta W[J,K,Y]}{\delta Y}\bigg|_{Y=0},\\
        \Delta\cdot \Gamma[\phi,K] &= \frac{\delta \Gamma[\phi,K,Y]}{\delta Y}\bigg|_{Y=0},
    \end{aligned}
\end{equation}
with the relations
\begin{align}\label{Eq:Relation-Z-W-Gamma-for-OperatorInsertion}
    \Delta\cdot\Gamma[\phi,K]=\Delta\cdot W[J,K]\big|_{J=-\frac{\delta\Gamma}{\delta\phi}} = \frac{\big(\Delta\cdot Z[J,K]\big)}{Z[J,K]}\bigg|_{J=-\frac{\delta\Gamma}{\delta\phi}}.
\end{align}

The corresponding 1PI Green functions with a single operator insertion are then given by
\begin{align}
    (\Delta\cdot\Gamma)_{\phi_{i_1}(x_1)\ldots\phi_{i_n}(x_n)}=-i\langle\Delta\phi_{i_1}(x_1)\ldots\phi_{i_n}(x_n)\rangle^{\mathrm{1PI}},
\end{align}
and in momentum space,
\begin{align}\label{Eq:1PI-Green-Functions-MomentumSpace-with-DeltaInsertion}
    i (\Delta\cdot\Gamma)_{\phi_{i_1}(p_1)\ldots\phi_{i_n}(p_n)} = \langle \Delta \phi_{i_1}(p_1)\ldots\phi_{i_n}(p_n) \rangle^{\mathrm{1PI}}.
\end{align}
Diagrammatically, this corresponds to all 1PI diagrams at a given order with the specified external fields and a single additional vertex associated with the composite operator $\Delta(x)$.
In practice, $\Delta$ thus induces an effective interaction vertex with the usual factor of $i$ arising from the exponential in the Gell–Mann–Low formula.
The expressions in Eqs.~\eqref{Eq:1PI-Green-Functions-MomentumSpace} and \eqref{Eq:1PI-Green-Functions-MomentumSpace-with-DeltaInsertion} constitute the essential quantities computed in this thesis, as they capture precisely the UV behaviour relevant for renormalisation.

An important feature of such operator insertions is that the lowest-order term of $\Delta\cdot\Gamma$ coincides with the classical operator,
\begin{align}\label{Eq:Single-Operator-Insertion-lowest-order-Contribution}
    \Delta\cdot\Gamma=\Delta_\mathrm{cl} + \mathcal{O}(\hbar),
\end{align}
where $\Delta_\mathrm{cl}$ is obtained by replacing the quantum field operators in $\Delta$ with their expectation values.
Consequently, if 
\begin{align}\label{Eq:Single-Operator-Insertion-starting-at-L-loop-level}
    \Delta\cdot\Gamma=0+\mathcal{O}(\hbar^L),
\end{align}
then necessarily
\begin{align}\label{Eq:Single-Operator-Insertion-local-at-L-loop-Level}
    \Delta\cdot\Gamma = \hbar^L \times (\text{local terms}) + \mathcal{O}(\hbar^{L+1}).
\end{align}
This property, consistent with the interpretation of $\Gamma$ as the effective action, plays a crucial role in inductive proofs of renormalisability, see sections~\ref{Sec:Algebraic_Renormalisation} and \ref{Sec:Renormalisation-of-GaugeTheories}.

\subsection{The Quantum Action Principle}\label{Sec:TheQuantumActionPrinciple}

As noted in Sec.~\ref{Sec:GreenFunctions_and_GeneratingFunctionals}, a Lagrangian is not strictly required to define a quantum field theory. 
Nevertheless, Green functions defined via the path integral (see Eq.~\eqref{Eq:Exact-General-Generating-Functional-J-and-K-via-PathIntegral}) or the Gell-Mann–Low formula (see Eq.~\eqref{Eq:Z[J,K]-via-Gell-Mann-Low}) inherit structural properties from the underlying Lagrangian, including symmetry properties realised at the quantum level via Ward and Slavnov-Taylor identities. 
The \emph{quantum action principle} provides a general and powerful framework that relates variations of the action --- such as reparametrisations and field transformations --- to variations of Green functions.
It thus offers a systematic method for deriving and analysing symmetry identities and their potential breakings at the quantum level.

The quantum action principle can be derived as a path integral relation that holds formally. 
Since the functional measure is not a priori rigorous and well-defined, its validity must be established within a chosen perturbative setting together with a consistent renormalisation scheme --- typically by representing the measure as the sum of all regularised Feynman diagrams, including required counterterm diagrams.
In dimensional regularisation (DReg), the quantum action principle holds at the regularised and the renormalised level, as originally proven in Ref.~\cite{Breitenlohner:1977hr}.
In Ref.~\cite{Stockinger:2005gx}, this proof has been extended to a mathematically consistent version of dimensional reduction (DRed).
Importantly, a regularisation-independent formulation of the quantum action principle also exists and has been established in the BPHZ framework in Refs.~\cite{Lowenstein:1971vf,Lowenstein:1971jk,Lam:1972mb,Lam:1972fzg,Lam:1973qa,Clark:1976ym}.

In this section, we derive the quantum action principle as a formal path-integral relation, following Refs.~\cite{Belusca-Maito:2023wah,Stoeckinger:2020mlr}.
A rigorous proof within the framework of DReg is provided in Sec.~\ref{Sec:Regularised-QAP}.

\begin{theorem}[Quantum Action Principle]\label{Thm:Formal-QuantumActionPrinciple}\ \\
    Let $\Sfull=\int d^4x\,\LaFull(x)$ be the action, also including source terms $K_i(x)\mathcal{O}_i(x)$. Given a local field transformation $\phi_i(x)\longmapsto\phi_i(x)+\delta\phi_i(x)$ under which the path integral measure is invariant, then:
    The quantum action principle is formally given by
    \begin{enumerate}[label={(\arabic*)}]
        \item its general formulation
        \begin{align}\label{Eq:Formal-QAP-General-Formulation}
            0 = \Big\langle i \int d^4x \, J_i(x)\delta\phi_i(x)+i\int d^4x \frac{\delta\Sfull}{\delta\phi_i(x)}\delta\phi_i(x) \Big\rangle_{J,K},
        \end{align}
        \item its formulation on elementary Green functions
        \begin{align}\label{Eq:Formal-QAP-Formulation-on-explicit-GreenFunctions}
            0 = \delta \big\langle T\phi_{i_1}(x_1)\ldots\phi_{i_n}(x_n) \big\rangle + \big\langle T\phi_{i_1}(x_1)\ldots\phi_{i_n}(x_n) \bigg[ i \int d^4x \frac{\delta\Sfull}{\delta\phi_k(x)}\delta\phi_k(x) \bigg] \big\rangle,
        \end{align}
        with $\delta=\int d^4x \, \delta\phi_{i}(x)\frac{\delta}{\delta\phi_{i}(x)}$,
        \item its formulation via the effective quantum action
        \begin{align}\label{Eq:Formal-QAP-Formulation-wrt-Gamma}
            \int d^4x \frac{\delta\Gamma}{\delta\phi_i(x)}\frac{\delta\Gamma}{\delta K_i(x)} = \bigg[\int d^4x \frac{\delta\Sfull}{\delta\phi_i(x)}\delta\phi_i(x)\bigg]\cdot\Gamma,
        \end{align}
        with a single operator insertion on the RHS.
    \end{enumerate}
\end{theorem}
\begin{proof}
    Suppose a local family of field transformations 
    \begin{align}\label{Eq:Local-Field-Transformation}
        \phi_i(x)\longmapsto\phi'_i(x)=\phi_i(x)+\delta\phi_i(x)=\phi_i(x)+\varepsilon_a F_i^a(\phi(x),\partial\phi(x)),
    \end{align}
    with infinitesimal parameters $\varepsilon_a$ that label the transformations, and local field monomials $F_i^a(\phi(x),\partial\phi(x))$,\footnote{$F_i^a(\phi(x),\partial\phi(x))$ may or may not be a composite operator, depending on whether the transformation is nonlinear or linear.}
    which leave the path integral measure invariant, such that
    \begin{align*}
        \mathcal{D}\phi'=\mathrm{det}\bigg(\frac{\delta\phi'(x)}{\delta\phi(y)}\bigg)\mathcal{D}\phi=\mathcal{D}\phi.
    \end{align*}
    Then, starting from the generating functional (see Eq.~\eqref{Eq:Exact-General-Generating-Functional-J-and-K-via-PathIntegral}),\footnote{Note that the source terms $K_i(x)\mathcal{O}_i(x)$ are absorbed into the action, as stated in theorem~\ref{Thm:Formal-QuantumActionPrinciple}.} we obtain
    \begin{equation*}
        \begin{aligned}
            Z[J,K] &= \mathcal{N}' \int \mathcal{D}\phi \, e^{i \big[\Sfull[\phi] + \int d^4x J_i(x)\phi_i(x)\big]} = \mathcal{N}' \int \mathcal{D}\phi' \, e^{i \big[\Sfull[\phi'] + \int d^4x J_i(x)\phi'_i(x)\big]}\\
            &= \mathcal{N}' \int \mathcal{D}\phi \, e^{i \big[\Sfull[\phi+\delta\phi] + \int d^4x [ J_i(x)\phi_i(x) + J_i(x)\delta\phi_i(x)]\big]}\\
            &= \mathcal{N}' \!\!\int \! \mathcal{D}\phi \Big[ 1 + i \! \int d^4x J_i(x)\delta\phi_i(x)+i \! \int d^4x \frac{\delta\Sfull}{\delta\phi_i(x)}\delta\phi_i(x) \Big] e^{i\big[ \Sfull + \int d^4x J_i(x)\phi_i(x) \big]} + \mathcal{O}(\delta\phi^2),
        \end{aligned}
    \end{equation*}
    where in the first line we performed a change of integration variables, in the second line we identified these new variables $\phi'$ with the transformed fields in Eq.~\eqref{Eq:Local-Field-Transformation}, while using that the measure is invariant, and in the last line we performed an expansion to first order in $\delta\phi$.
    From this result, we can immediately read off the quantum action principle in its general formulation
    \begin{align*}
        0 = \int \mathcal{D}\phi \Big[ i \int d^4x \, J_i(x)\delta\phi_i(x)+i\int d^4x \frac{\delta\Sfull}{\delta\phi_i(x)}\delta\phi_i(x) \Big] e^{i\big[ \Sfull + \int d^4x \, J_i(x)\phi_i(x) \big]},
    \end{align*}
    as stated in Eq.~\eqref{Eq:Formal-QAP-General-Formulation}. 
    The second formulation, expressed in terms of elementary Green functions, follows directly by successively applying functional derivatives w.r.t.\ $iJ_{i_k}(x_k)$ onto Eq.~\eqref{Eq:Formal-QAP-General-Formulation} and subsequently setting $J=0$. This yields
    \begin{equation*}
        \begin{aligned}
            0 &= \big\langle T\delta\phi_{i_1}(x_1)\phi_{i_2}(x_2)\ldots\phi_{i_n}(x_n)\big\rangle + \ldots + \big\langle T\phi_{i_1}(x_1)\ldots\phi_{i_{n-1}}(x_{n-1})\delta\phi_{i_n}(x_n) \big\rangle\\ 
            &+ \big\langle T\phi_{i_1}(x_1)\ldots\phi_{i_n}(x_n)\bigg[ i \int d^4x \frac{\delta\Sfull}{\delta\phi_k(x)}\delta\phi_k(x) \bigg] \big\rangle,
        \end{aligned}
    \end{equation*}
    which is precisely the relation in Eq.~\eqref{Eq:Formal-QAP-Formulation-on-explicit-GreenFunctions}.
    Finally, for the last formulation of the quantum action principle, we start from Eq.~\eqref{Eq:Formal-QAP-General-Formulation}.
    Assuming that the field transformations $\delta\phi_i$ are coupled to sources $K_i$, and using the notion of a single operator insertion according to Eqs.~\eqref{Eq:SingleOperatorInsertion-Z} and \eqref{Eq:SingleOperatorInsertion-W-Gamma}, we may rewrite it as 
    \begin{equation}\label{Eq:IntermediateResult-for-QAP-(1)-to-(3)}
        \begin{aligned}
            0 &= \Big\langle i \int d^4x \, J_i(x)\delta\phi_i(x)+i\int d^4x \frac{\delta\Sfull}{\delta\phi_i(x)}\delta\phi_i(x) \Big\rangle_{J,K}\\
            &= \int d^4x \, J_i(x) \frac{\delta Z[J,K]}{\delta K_i(x)} + i \bigg[\int d^4x \frac{\delta\Sfull}{\delta\phi_i(x)}\delta\phi_i(x)\bigg]\cdot Z[J,K],
        \end{aligned}
    \end{equation}
    where we used that $J_i(x)$ are number-valued test functions that can be pulled out of the path integral. 
    For the first term in Eq.~\eqref{Eq:IntermediateResult-for-QAP-(1)-to-(3)} we obtain 
    \begin{align*}
        \int d^4x \, J_i(x) \frac{\delta Z[J,K]}{\delta K_i(x)} = \int d^4x \, J_i(x) \frac{\delta \Gamma[\phi,K]}{\delta K_i(x)} iZ[J,K] = - \int d^4x \, \frac{\delta\Gamma[\phi,K]}{\delta\phi_i(x)} \frac{\delta \Gamma[\phi,K]}{\delta K_i(x)} iZ[J,K].
    \end{align*}
    In the first step we used Eq.~\eqref{Eq:Definition-of-W-Functional} to express the functional derivative of $Z[J,K]$ through $W[J,K]$, and the fact that $\Gamma$ is the Legendre transformation of $W$ w.r.t.\ the sources $J_i$ (not $K_i$) --- a relation shown in Eq.~\eqref{Eq:Definition-of-Gamma-Functional}, which is ``blind'' w.r.t.\ derivatives of $K_i$, cf.\ Eq.~\eqref{Eq:Relation-W-to-Gamma-derivatives-wrt-K} --- to rewrite it in terms of $\Gamma[\phi,K]$. 
    In the second step, we then substituted $J_i(x)=-\delta\Gamma/\delta\phi_i(x)$.
    For the second term in Eq.~\eqref{Eq:IntermediateResult-for-QAP-(1)-to-(3)} we use the notion of a single operator insertion as introduced in the previous section (see particularly Eq.~\eqref{Eq:Relation-Z-W-Gamma-for-OperatorInsertion}) to obtain 
    \begin{align*}
        i \bigg[\int d^4x \frac{\delta\Sfull}{\delta\phi_i(x)}\delta\phi_i(x)\bigg]\cdot Z[J,K] = i Z[J,K] \, \bigg[\int d^4x \frac{\delta\Sfull}{\delta\phi_i(x)}\delta\phi_i(x)\bigg]\cdot\Gamma[\phi,K].
    \end{align*}
    Combining both results in Eq.~\eqref{Eq:IntermediateResult-for-QAP-(1)-to-(3)}, we find
    \begin{align*}
        0 = i Z[J,K] \bigg( - \int d^4x \, \frac{\delta\Gamma[\phi,K]}{\delta\phi_i(x)} \frac{\delta \Gamma[\phi,K]}{\delta K_i(x)} + \bigg[\int d^4x \frac{\delta\Sfull}{\delta\phi_i(x)}\delta\phi_i(x)\bigg]\cdot\Gamma[\phi,K] \bigg),
    \end{align*}
    which reproduces Eq.~\eqref{Eq:Formal-QAP-Formulation-wrt-Gamma} stated above.
\end{proof}
While the quantum action principle presented above governs the influence of field transformations at the quantum level, similar relations can be obtained for variations of external sources $K_i(x)$ and Lagrangian parameters $\lambda$, see Refs.~\cite{Belusca-Maito:2023wah,Piguet:1995er}.

In particular, for external (non-propagating) sources one finds
\begin{align}
    \frac{\delta Z[J,K]}{\delta K_i(x)} = \int \mathcal{D} \phi \, i\frac{\delta\Sfull}{\delta K_i(x)} \, e^{i\big[ \Sfull + \int d^4x \, J_i(x)\phi_i(x) \big]} = i \delta \phi_i(x)\cdot Z[J,K],
\end{align}
which implies for elementary Green functions
\begin{align}
    -i\frac{\delta}{\delta K_i(x)}\big\langle T\phi_{i_1}(x_1) \ldots\phi_{i_n}(x_n) \big\rangle = \big\langle T\phi_{i_1}(x_1) \ldots\phi_{i_n}(x_n) \frac{\delta\Sfull}{\delta K_i(x)} \big\rangle.
\end{align}

Similarly, for a variation of a parameter $\lambda$ one obtains
\begin{align}
    \frac{\partial Z[J,K]}{\partial\lambda} = \int \mathcal{D} \phi \, i\frac{\partial\Sfull}{\partial \lambda} \, e^{i\big[ \Sfull + \int d^4x \, J_i(x)\phi_i(x) \big]} = i \bigg[\frac{\partial\Sfull}{\partial \lambda}\bigg]\cdot Z[J,K],
\end{align}
leading to
\begin{align}
    -i\frac{\partial}{\partial\lambda}\big\langle T\phi_{i_1}(x_1) \ldots\phi_{i_n}(x_n) \big\rangle = \big\langle T\phi_{i_1}(x_1) \ldots\phi_{i_n}(x_n) \frac{\partial\Sfull}{\partial\lambda} \big\rangle
\end{align}
for Green functions.
At the level of the effective quantum action this becomes
\begin{align}
    \frac{\partial\Gamma}{\partial\lambda} = \bigg[\frac{\partial\Sfull}{\partial \lambda}\bigg]\cdot\Gamma,
\end{align}
whose lowest-order contribution, cf.\ Eq.~\eqref{Eq:Single-Operator-Insertion-lowest-order-Contribution}, is 
\begin{align}
    \bigg[\frac{\partial\Sfull}{\partial \lambda}\bigg]\cdot\Gamma = \frac{\partial\Sfull}{\partial \lambda}+\mathcal{O}(\hbar),
\end{align}
which is the variation of the classical action plus quantum corrections.

An important relation that can be derived from the quantum action principle is the quantum generalisation of equations of motion.
Starting from Eq.~\eqref{Eq:Formal-QAP-General-Formulation} and differentiating w.r.t.\ $\delta/\delta(\delta\phi_k(x))$ yields
\begin{align}\label{Eq:Quantum-Generalised-Equations-of-Motion}
    0 = \Big\langle i J_k(x) + i \frac{\delta\Sfull}{\delta\phi_k(x)} \Big\rangle_{J,K} = i J_k(x) Z[J,K] + i \bigg[\frac{\delta\Sfull}{\delta\phi_k(x)}\bigg]\cdot Z[J,K],
\end{align}
which constitutes a path integral version of the equations of motion.
In terms of Green functions the equations of motion are then given as
\begin{equation*}
    \begin{aligned}
        0 &= \big\langle T \delta_{k i_1}\delta^{(4)}(x-x_{1}) \phi_{i_2}(x_2)\ldots\phi_{i_n}(x_n)\big\rangle + \ldots + \big\langle T \delta_{k i_n} \delta^{(4)}(x-x_n) \phi_{i_1}(x_1)\ldots\phi_{i_{n-1}}(x_{n-1}) \big\rangle\\ 
        &+ \big\langle T\phi_{i_1}(x_1)\ldots\phi_{i_n}(x_n) i \frac{\delta\Sfull}{\delta\phi_k(x)} \big\rangle.
    \end{aligned}
\end{equation*}

The same relation can equivalently be obtained by considering the path integral expression of some functional $\mathcal{F}[\phi]$, shifting the integration variables (with the measure remaining invariant) and differentiating w.r.t.\ this shift, i.e.\ $\delta/\delta(\delta\phi_k(x))$:
\begin{align}\label{Eq:Dyson-Schwinger-Equation}
    0=\int \mathcal{D}\phi \frac{\delta}{\delta\phi_k(x)}\mathcal{F}[\phi]e^{iS},
\end{align}
where we used that the path integral over a total derivative vanishes, just like for standard integrals.
Eq.~\eqref{Eq:Dyson-Schwinger-Equation} is the so-called Dyson-Schwinger equation.
Choosing $\mathcal{F}[\phi]$ to be a string of quantum fields $\phi_{i_1}(x_1)\ldots\phi_{i_n}(x_n)$ reproduces exactly the equations of motion in terms of Green functions given above.
In this sense, the Dyson-Schwinger equation governs the fate of equations of motion in the quantum theory.

Rewriting Eq.~\eqref{Eq:Quantum-Generalised-Equations-of-Motion} in terms of the effective quantum action gives 
\begin{align}\label{Eq:Quantum-Generalised-Equations-of-Motion-for-Gamma}
    \frac{\delta\Gamma[\phi,K]}{\delta\phi_k(x)} = \bigg[\frac{\delta\Sfull}{\delta\phi_k(x)}\bigg] \cdot \Gamma[\phi,K].
\end{align}
Here the quantum nature becomes apparent in a clear way:
the variation of the effective quantum action w.r.t.\ $\phi_k$ is equal to the insertion of the classical variation $\delta\Sfull/\delta\phi_k(x)$ as a composite operator into the full effective quantum action.
At lowest order, cf.\ Eq.~\eqref{Eq:Single-Operator-Insertion-lowest-order-Contribution}, we find
\begin{align}\label{Eq:Quantum-Generalised-Equations-of-Motion-for-Gamma-expansion-in-hbar}
    \bigg[\frac{\delta\Sfull}{\delta\phi_k(x)}\bigg] \cdot \Gamma[\phi,K] = \frac{\delta\Sfull}{\delta\phi_k(x)} + \mathcal{O}(\hbar),
\end{align}
showing that the classical equations of motion reappear as the leading contribution, with quantum corrections systematically added, compatible with the interpretation of Eqs.~\eqref{Eq:Quantum-Generalised-Equations-of-Motion} and \eqref{Eq:Quantum-Generalised-Equations-of-Motion-for-Gamma} as the quantum generalisation of the equations of motion.

\subsection{The Slavnov-Taylor Identity}\label{Sec:The-Slavnov-Taylor-Identity}

Ultimately, this section concerns the symmetry properties of quantum field theories, their formal description, and the fate of classical symmetries at the quantum level. 
In the full quantum theory, the general answer to how symmetries manifest themselves is provided by the \emph{Slavnov-Taylor identities}.
These identities are functional relations among off-shell Green functions, thereby encoding the symmetry at the quantum level.
In other words, the Slavnov-Taylor identities are a quantum manifestation of the classical symmetries.

By relating different Green functions to each other, these identities constrain the structure of possible divergences and consequently the form of admissible counterterms. 
This makes them a powerful tool in the process of renormalisation --- provided, of course, that the regularisation does not violate them in intermediate steps of the calculation (see chapter~\ref{Chap:Practical_Symmetry_Restoration}).
In particular for gauge theories, the Slavnov-Taylor identity associated with BRST symmetry is indispensable: its validity after renormalisation ensures both the internal consistency of the quantum theory and the proof of renormalisability, as will be discussed in Sec.~\ref{Sec:Renormalisation-of-GaugeTheories}.

To derive the Slavnov-Taylor identity, we consider a local symmetry transformation, that is a local transformation as shown in Eq.~\eqref{Eq:Local-Field-Transformation} that leaves the classical action invariant, i.e.\ $\delta S/\delta\phi_i(x)=0$ or equivalently $\Gamma_\mathrm{cl}[\phi,K]=\Gamma_\mathrm{cl}[\phi+\delta\phi,K]+\mathcal{O}(\delta\phi^2)$.
In this case, the Slavnov–Taylor identity emerges directly as the specialisation of the quantum action principle in theorem~\ref{Thm:Formal-QuantumActionPrinciple} to symmetry transformations.
\begin{corollary}[Slavnov-Taylor Identity]\label{Thm:Formal-Slavnov-Taylor-Identity}\ \\
    Let $\Sfull=\int d^4x\,\LaFull(x)$ be the action, also including source terms $K_i(x)\mathcal{O}_i(x)$.
    Given a local symmetry transformation $\phi_i(x)\longmapsto\phi_i(x)+\delta\phi_i(x)$ under which the path integral measure and the action are invariant, then:
    From the quantum action principle the Slavnov-Taylor identity follows as
    \begin{enumerate}[label={(\arabic*)}]
        \item its general formulation
        \begin{align}\label{Eq:Formal-STI-General-Formulation}
            0 = \big\langle i \int d^4x \, J_i(x)\delta\phi_i(x) \big\rangle_{J,K} = i \int d^4x \, J_i(x) \big\langle \delta\phi_i(x) \big\rangle_{J,K},
        \end{align}
        \item its formulation on elementary Green functions
        \begin{align}\label{Eq:Formal-STI-Formulation-on-explicit-GreenFunctions}
            0 = \delta \big\langle T\phi_{i_1}(x_1)\ldots\phi_{i_n}(x_n) \big\rangle,
        \end{align}
        with $\delta=\int d^4x \, \delta\phi_{i}(x)\frac{\delta}{\delta\phi_{i}(x)}$,
        \item its formulation via the effective quantum action
        \begin{align}\label{Eq:Formal-STI-Formulation-wrt-Gamma}
            0=\int d^4x \frac{\delta\Gamma}{\delta\phi_i(x)}\big\langle \delta\phi_i(x) \big\rangle_{J,K}=\int d^4x \frac{\delta\Gamma}{\delta\phi_i(x)}\frac{\delta\Gamma}{\delta K_i(x)}.
        \end{align}
    \end{enumerate}
\end{corollary}
Evidently, the Slavnov-Taylor identity can be understood as a quantum invariance relation for the effective action, given by $\Gamma[\phi,K]=\Gamma[\phi+\langle\delta\phi\rangle_{J,K},K]+\mathcal{O}(\delta\phi^2)$, which is almost the same relation as for the classical action, with the important difference that here the $\phi$'s are expectation values of the quantum field operators.
In particular, the field $\phi$ is shifted by the expectation value of the field variation $\delta\phi$ in the presence of the sources.
For this reason, a crucial distinction arises depending on whether the classical symmetry transformation is linear or nonlinear in the quantum fields.

\paragraph{Linear Symmetry Transformations:}
In the case of a linear classical symmetry
\begin{align}\label{Eq:Linear-Symmetry-Transformation-deltaphi}
    \delta\phi_i(x)=a_i(x)+\int d^4y \, c_{ij}(x,y) \phi_j(y) \equiv F_i(\phi(x)),
\end{align}
with number-valued $a_i$ and $c_{ij}$, the expectation value
\begin{align}\label{Eq:Expectation-Value-of-Linear-Symmetry-Transformation-deltaphi}
    \langle\delta\phi_i(x)\rangle_{J,K} = a_i(x) + \int d^4y \, c_{ij}(x,y) \langle\phi_j(y)\rangle_{J,K} = F_i(\langle\phi(x)\rangle_{J,K}) = \delta \phi^\mathrm{class}_i(x)
\end{align}
is identical to the classical symmetry transformation.
It follows that the full effective quantum action $\Gamma$ is invariant under the same linear symmetry transformations as the classical action $\Gamma_\mathrm{cl}$, as they do not receive nontrivial quantum corrections.\footnote{Recall that the arguments of $\Gamma$ are the classical fields $\phi_{i,\mathrm{cl}}(x)=\langle\phi_i(x)\rangle_{J,K}$, see Eq.~\eqref{Eq:Definition-of-phi_cl-as-Arguments-of-Gamma}, where we have dropped the subscript ``$\mathrm{cl}$'' for notational simplicity.}
The Slavnov-Taylor identity then reduces to 
\begin{align}
    0=\int \delta\phi_i(x)\frac{\delta\Gamma}{\delta\phi_i(x)}, 
\end{align}
having dropped the superscript ``$\mathrm{class}$'', and is often called \emph{Ward identity}.

In quantum field theory, symmetries are typically local.
For a local linear transformation, the kernel takes the form $c_{ij}(x,y)=b_{ij}(x)\delta^{(4)}(x-y)$.

Examples for linear symmetries include baryon and lepton number conservation, global gauge transformations (rigid in spacetime, such that $\vartheta c$ is independent of $x$, and therefore even non-Abelian transformations reduce to linear combinations of the fields), and certain local gauge transformations in Abelian theories, which will be discussed in more detail in Sec.~\ref{Sec:Peculiarities_of_Abelian_Gauge_Theories}.

\paragraph{Nonlinear Symmetry Transformations:}
For nonlinear transformations, where $\delta\phi$ is a product of quantum fields, the symmetry receives nontrivial quantum corrections, i.e.\ it renormalises.
The reason for this is that the expectation value of a product of operators is not equal to the product of their expectation values, i.e.\ it does not factorise: $\langle\phi_j\phi_k\rangle_{J,K}\neq\langle\phi_j\rangle_{J,K}\langle\phi_k\rangle_{J,K}$.
While the RHS corresponds again just to the classical expression, the nonlinear combination on the LHS that we need in the symmetry identity is different and introduces genuine quantum corrections.
To handle this case, one typically introduces external sources $K_i(x)$ that couple to these nonlinear variations $\delta\phi_i(x)$, cf.\ Sec.~\ref{Sec:BRST-Symmetry}, analogous to the treatment of generic composite operators in Sec.~\ref{Sec:GreenFunctions_and_GeneratingFunctionals}.
The Slavnov-Taylor identity then takes its characteristic form shown in Eq.~\eqref{Eq:Formal-STI-Formulation-wrt-Gamma}.
Strictly speaking, Eq.~\eqref{Eq:Formal-STI-Formulation-wrt-Gamma} holds a priori only for $K_i\equiv0$.
However, it is also true for arbitrary $K_i$ provided that the additional terms $\int d^4x\,K_i(x)\delta\phi_i(x)$ in the action are also invariant under the symmetry transformation under consideration.


\paragraph{Slavnov-Taylor Operator:}
In this thesis, we will particularly use the formulation of the Slavnov-Taylor identity in terms of the effective quantum action shown in Eq.~\ref{Eq:Formal-STI-Formulation-wrt-Gamma}.
For this purpose, we introduce the Slavnov-Taylor operator:
\begin{definition}[Slavnov-Taylor Operator]\ \\
    The nonlinear functional operator
    \begin{align}\label{Eq:Slavnov-Taylor-Operator-Formal-Def}
        \mathcal{S}:\mathcal{F}[\phi,K]\longmapsto\int d^4x \frac{\delta\mathcal{F}[\phi,K]}{\delta\phi_i(x)}\frac{\delta\mathcal{F}[\phi,K]}{\delta K_i(x)},
    \end{align}
    acting on any functional $\mathcal{F}$, is called \emph{Slavnov-Taylor operator}.
\end{definition}
With this operator, the Slavnov-Taylor identity in Eq.~\ref{Eq:Formal-STI-Formulation-wrt-Gamma} can compactly be written as
\begin{align}
    \mathcal{S}(\Gamma)=0.
\end{align}
Applying $\mathcal{S}$ on a sum of two functionals $\mathcal{F}$ and $\mathcal{G}$ motivates the following definition:
\begin{definition}[Linearised Slavnov-Taylor Operator]\ \\
    The linear functional operator w.r.t.\ the functional $\mathcal{F}$
    \begin{align}\label{Eq:Linearised-Slavnov-Taylor-Operator-Def}
        \mathcal{S}_\mathcal{F} : \mathcal{G}[\phi,K] \longmapsto \int d^4x \bigg( \frac{\delta\mathcal{F}[\phi,K]}{\delta\phi_i(x)}\frac{\delta\mathcal{G}[\phi,K]}{\delta K_i(x)} + \frac{\delta\mathcal{F}[\phi,K]}{\delta K_i(x)} \frac{\delta\mathcal{G}[\phi,K]}{\delta\phi_i(x)} \bigg),
    \end{align}
    acting on any functional $\mathcal{G}$, is called \emph{linearised Slavnov-Taylor operator}.
\end{definition}
Using this definition, we may write\footnote{Note that $\mathcal{S}_\mathcal{F}\mathcal{G}=\mathcal{S}_\mathcal{G}\mathcal{F}$, provided the differentiated expressions commute.}
\begin{align}
    \mathcal{S}(\mathcal{F}+\mathcal{G}) = \mathcal{S}(\mathcal{F}) + \mathcal{S}_\mathcal{F}\mathcal{G} + \mathcal{S}(\mathcal{G}) =  \mathcal{S}(\mathcal{F}) + \mathcal{S}_\mathcal{F}\mathcal{G} + \mathcal{O}(\mathcal{G}^2),
\end{align}
with the explicit form
\begin{align}
    \mathcal{S}_\mathcal{F} = \int d^4x \bigg( \frac{\delta\mathcal{F}[\phi,K]}{\delta\phi_i(x)}\frac{\delta}{\delta K_i(x)} + \frac{\delta\mathcal{F}[\phi,K]}{\delta K_i(x)} \frac{\delta}{\delta\phi_i(x)} \bigg).
\end{align}

\paragraph{BRST Symmetry:}
In the case of BRST symmetry (see Sec.~\ref{Sec:BRST-Symmetry}), the Slavnov-Taylor framework provides a convenient functional representation of the BRST operator $s$ and naturally leads to the definition of $b$ as linearised Slavnov-Taylor operator w.r.t.\ the classical action.

Let $S_0$ denote the classical action that also includes the source terms $K_i(x)\mathcal{O}_i(x)$, with $\mathcal{O}_i(x)=\delta\phi_i(x)=s\phi_i(x)$ for the BRST transformations.
The complete action is then given by $S_0+S_\mathrm{ct}$.
In this setting, the BRST operator takes the functional form
\begin{align}
    s = \int d^4x \frac{\delta S_0}{\delta K_i(x)} \frac{\delta}{\delta\phi_i(x)}.
\end{align}
The linearised Slavnov-Taylor operator w.r.t.\ the classical action $S_0$ is defined as
\begin{align}\label{Eq:b-The-Linearised-Slavnov-Taylor-Operator-wrt-Classical-Action}
    b \coloneqq \mathcal{S}_{S_{0}} = \int d^4x \bigg( \frac{\delta S_0}{\delta K_i(x)} \frac{\delta}{\delta\phi_i(x)} + \frac{\delta S_0}{\delta\phi_i(x)}\frac{\delta}{\delta K_i(x)} \bigg) = s + \int d^4x \frac{\delta S_0}{\delta\phi_i(x)}\frac{\delta}{\delta K_i(x)}.
\end{align}
With this notation, the Slavnov-Taylor identity for the complete action can be written, to linear order in the counterterms, as
\begin{align}
    \mathcal{S}(S_0+S_\mathrm{ct})=sS_0+bS_{\mathrm{ct}}+\mathcal{O}(S_{\mathrm{ct}}^2),
\end{align}
using that $\mathcal{S}(S_0)=sS_0$.

Nilpotency of the BRST operation, i.e.\ $s^2=0$, is implemented in functional form via 
\begin{subequations}
    \begin{align}
        \mathcal{S}_\mathcal{F}\mathcal{S}(\mathcal{F}) &= 0, \qquad \forall\,\mathcal{F},\label{Eq:First-Nilpotency-Cond-of-lin-STI}\\
        \mathcal{S}_\mathcal{F}\mathcal{S}_\mathcal{F} &= 0, \qquad \mathrm{if} \, \, \mathcal{S}(\mathcal{F})=0,
    \end{align}
\end{subequations}
see Ref.~\cite{Piguet:1995er}.
Evidently, for $\mathcal{F}=S_0$ this implies nilpotency of $b$, i.e.\ $b^2=0$, provided that the classical action $S_0$ satisfies the Slavnov-Taylor identity.

\paragraph{Concluding Comments:}
Taking seriously the path integral necessarily entails a regularisation and renormalisation prescription by which the theory is defined at higher orders: as mentioned above, the measure itself is only well-defined perturbatively within a specific renormalisation scheme.
Similarly, the Lagrangian must in general be supplemented by counterterms, which depend on the chosen scheme.
Consequently, the quantum action principle and the Slavnov–Taylor identity must be established rigorously within a concrete regularisation and renormalisation scheme.
In theorem~\ref{Thm:Formal-QuantumActionPrinciple} and corollary~\ref{Thm:Formal-Slavnov-Taylor-Identity} these relations were introduced as formal path-integral statements, valid for some action $S=\int d^4x \,\mathcal{L}$ without commenting on possible counterterms or the implications of the regularisation.
These issues are addressed in Sec.~\ref{Sec:Regularised-QAP}, where the quantum action principle is rigorously established in dimensional regularisation, both at the regularised and renormalised level (including counterterms). 
The case of the Slavnov-Taylor identity is more nuanced and depends critically on the properties of the regulator.
If the regularisation preserves the symmetry (as in DReg for vector-like gauge theories such as QCD), the Slavnov-Taylor identity holds at both the regularised and renormalised level.
However, the symmetry can, and may in general, be broken by the regularisation, as for example is the case in the BMHV scheme in chiral gauge theories (see Sec.~\ref{Sec:The-BMHV-Scheme}).
In this case, the Slavnov-Taylor identity is violated at the regularised level but, in the absence of a genuine anomaly, can be restored after renormalisation by symmetry-restoring counterterms, as discussed in chapter~\ref{Chap:Practical_Symmetry_Restoration}.
By contrast, in the presence of a genuine anomaly (see Sec.~\ref{Sec:Anomalies}), the Slavnov-Taylor identity cannot be reestablished, neither at the regularised nor renormalised level, reflecting an unavoidable breaking by quantum effects.
In the case of gauge anomalies this invalidates the consistency of the entire theory, see Sec.~\ref{Sec:Renormalisation-of-GaugeTheories}.
The methods of algebraic renormalisation, see Sec.~\ref{Sec:Algebraic_Renormalisation}, provide a rigorous framework to distinguish ``spurious'' breakings, which can be cancelled, from anomalous ones, which cannot.

\section{Algebraic Renormalisation}\label{Sec:Algebraic_Renormalisation}

Algebraic renormalisation provides a powerful and systematic framework for establishing the renormalisability of gauge theories, particularly in the absence of an invariant regularisation procedure.
Employing the quantum action principle (cf.\ Eq.~\eqref{Eq:Formal-QAP-Formulation-wrt-Gamma}), the methods of algebraic renormalisation can be used to establish the validity of the Slavnov-Taylor identity (see Eq.~\eqref{Eq:Formal-STI-Formulation-wrt-Gamma}) associated to BRST invariance, as well as of the gauge condition (see Eq.~\eqref{Eq:The-Gauge-Fixing-Condition}) at higher orders in perturbation theory.
As discussed in Sec.~\ref{Sec:Anomalies}, only non-anomalous symmetries can be consistently maintained at the quantum level.
If a symmetry is broken at higher orders, algebraic renormalisation provides the means to determine whether this breaking is \emph{spurious} or \emph{anomalous} by solving the corresponding cohomology problem.
For BRST invariance, this amounts to studying the BRST cohomology of the associated linearised Slavnov-Taylor operator $b$, introduced in Eq.~\eqref{Eq:b-The-Linearised-Slavnov-Taylor-Operator-wrt-Classical-Action}.
A spurious breaking corresponds to a trivial element of the BRST cohomology and can be removed by adjusting appropriate counterterms, while an anomalous breaking corresponds to a nontrivial element of the BRST cohomology and cannot be compensated --- indicating an incompatibility of the symmetry with the quantum theory.

The framework of algebraic renormalisation has been employed to proof the renormalisability of the Standard Model to all orders, see Ref.~\cite{Kraus:1997bi}.
A similar proof exists in the background field gauge in Ref.~\cite{Grassi:1997mc}, and an extension to the supersymmetric SM can be found in Ref.~\cite{Hollik:2002mv}.
In the following, we focus on the validity of the gauge condition and, in particular, on establishing the Slavnov–Taylor identity, which forms a central aspect of this thesis.
We analyse the BRST cohomology problem of $b$ to distinguish between spurious and anomalous breakings, and illustrate how the former can be removed through symmetry-restoring counterterms.
Our discussion mainly follows Refs.~\cite{Belusca-Maito:2023wah,Piguet:1995er}.
Furthermore, we make use of the lowest-order behaviour of operator insertions as discussed around Eqs.~\eqref{Eq:Single-Operator-Insertion-lowest-order-Contribution}--\eqref{Eq:Single-Operator-Insertion-local-at-L-loop-Level}.

We begin with the gauge-fixing condition:
\begin{proposition}[Gauge-Fixing Condition]\label{Prop:Gauge-Condition-holds-at-all-Orders}\ \\
    The gauge-fixing condition~\eqref{Eq:The-Gauge-Fixing-Condition} holds to all orders in perturbation theory, such that
    \begin{align}\label{Eq:Gauge-Fixing-Condition-at-all-Orders}
        \frac{\delta \Gamma}{\delta B^a} = \frac{\delta \Gamma_\mathrm{cl}}{\delta B^a} = \partial^\mu A_\mu^a + \xi B^a.
    \end{align}
\end{proposition}
\begin{proof}
    The proof proceeds by induction.
    Suppose the gauge-fixing condition holds up to and including order $\mathcal{O}(\hbar^{n-1})$. 
    Using the quantum generalisation of the equations of motion (see Eqs.~\eqref{Eq:Quantum-Generalised-Equations-of-Motion-for-Gamma} and \eqref{Eq:Quantum-Generalised-Equations-of-Motion-for-Gamma-expansion-in-hbar}) together with the behaviour of operator insertions (see Eqs.~\eqref{Eq:Single-Operator-Insertion-lowest-order-Contribution}--\eqref{Eq:Single-Operator-Insertion-local-at-L-loop-Level}), the most general breaking of the gauge-fixing condition reads 
    \begin{align}
        \frac{\delta \Gamma}{\delta B^a} = \partial^\mu A_\mu^a + \xi B^a + \hbar^n \Delta^a + \mathcal{O}(\hbar^{n+1}),
    \end{align}
    where the breaking $\Delta^a$ is a local functional in $A_\mu^a$, $c^a$, $\overline{c}^a$, and $B^a$, of mass dimension 2, as constrained by power counting, and ghost number 0.
    From the commutativity of $B^a$-variations, i.e.\ $[\delta/\delta B^a(x),\delta/\delta B^b(y)]=0$, a consistency condition for $\Delta^a$ can be derived (see Ref.~\cite{Piguet:1995er}), implying that
    \begin{align}
        \Delta^a(x) = \frac{\delta}{\delta B^a(x)} \widetilde{\Delta}.
    \end{align}
    Adding a counterterm $-\widetilde{\Delta}$ at order $\mathcal{O}(\hbar)$ to the effective action $\Gamma$ restores the gauge-fixing condition at this order, thus completing the inductive proof.
\end{proof}
The chosen gauge-fixing condition in Eq.~\eqref{Eq:Gauge-Fixing-Condition-at-all-Orders} is linear in the quantum fields, and therefore cannot receive higher-order corrections from loop diagrams, as no interaction vertices are present.
Although finite counterterms could in principle be added or required by symmetry arguments, such terms can always be removed as shown above.
Hence, the tree-level gauge condition remains valid to all orders.
This argument also applies to linear BRST transformations --- particularly that of the antighost, $\chi^a s\overline{c}^a$ (see Sec.~\ref{Sec:BRST-Symmetry}) --- which therefore do not renormalise.

Before turning to the Slavnov–Taylor identity and possible breakings thereof --- the central issue of this section --- we comment on the ghost equation
\begin{subequations}\label{Eq:Ghost-Eq}
    \begin{align}
        \mathscr{G}\Gamma &= 0, \qquad \mathrm{with} \,\,\,\, \mathscr{G} \coloneqq \frac{\delta}{\delta\overline{c}} + \partial^\mu \frac{\delta}{\delta \rho^\mu},\label{Eq:Ghost-Eq-standard}\\
        \Longleftrightarrow\,\frac{\delta \Gamma|_{\rho\to\hat{\rho}}}{\delta \overline{c}} &= 0, \qquad \mathrm{with} \,\,\,\, \hat{\rho} \coloneqq \rho^\mu + \partial^\mu \overline{c},\label{Eq:Ghost-Eq-with-rho_hat}
    \end{align}
\end{subequations}
which follows from the gauge-fixing condition and the Slavnov-Taylor identity.
Analogously, the ghost equation can be shown to hold to all orders, with a proof equivalent to that of the gauge condition when formulated as in Eq.~\eqref{Eq:Ghost-Eq-with-rho_hat}, see Ref.~\cite{Piguet:1995er}.

We now turn to the core issue of this section --- the implementability of the Slavnov-Taylor identity to all orders in perturbation theory.
This identity, associated with BRST invariance, is essential for the internal consistency and renormalisability of quantum gauge theories (see Sec.~\ref{Sec:Renormalisation-of-GaugeTheories}). 
It ensures that BRST invariance, which is crucial for defining a consistent and unitary quantum theory, persists beyond the classical level.
\begin{theorem}[Renormalisability of the Slavnov-Taylor Identity]\label{Thm:Renormalisability_of_the_Slavnov-Taylor_Identity}\ \\
    In the absence of a gauge anomaly, the Slavnov-Taylor identity can be established to all orders in perturbation theory, such that
    \begin{align}
        \mathcal{S}(\Gamma)=0.
    \end{align}
\end{theorem}
\begin{proof}
    The proof again proceeds by induction.
    At lowest order ($n=0$), the Slavnov-Taylor identity holds by construction of the classical action $\Gamma_\mathrm{cl}=S_0$.
    Suppose now that it holds up to and including order $\mathcal{O}(\hbar^{n-1})$, such that 
    \begin{align}
        \mathcal{S}(\Gamma)=\mathcal{O}(\hbar^n).
    \end{align}
    Here, we omit sub- and superscripts on $\Gamma$ for brevity and to focus on the essential aspects.
    At the $n$-loop level, the effective action
    \begin{align}
        \Gamma_\mathrm{subren}^{(\leq n),\mathrm{fin}} = \Gamma_\mathrm{subren}^{(\leq n)} + S_\mathrm{sct}^{(n)},
    \end{align}
    is subrenormalised up to order $n-1$, satisfies the Slavnov-Taylor identity up to that order, and is finite at order $n$ through the inclusion of the necessary divergent $n$-loop counterterms.
    The only possible remaining issue is a potential breaking of the Slavnov-Taylor identity at order $n$, which will be analysed in the following.
    
    Using the quantum action principle (cf.\ Eq.~\eqref{Eq:Formal-QAP-Formulation-wrt-Gamma}) and the properties of operator insertions (see Eqs.~\eqref{Eq:Single-Operator-Insertion-lowest-order-Contribution}--\eqref{Eq:Single-Operator-Insertion-local-at-L-loop-Level}), we can express the potential breaking as
    \begin{align}\label{Eq:Breaking-of-STI-at-order-n-for-inductive-proof}
        \mathcal{S}(\Gamma) = \hbar^n \Delta\cdot\Gamma = \hbar^n \Delta + \mathcal{O}(\hbar^{n+1}),
    \end{align}
    where $\Delta$ is an integrated local functional in the fields and derivatives of mass dimension 4 (restricted by power-counting), and ghost number 1.
    Recall from Sec.~\ref{Sec:TheQuantumActionPrinciple} that the quantum action principle holding formally has rigorously been established in the BPHZ formalism as well as in DReg (see Sec.~\ref{Sec:Regularised-QAP}).

    Applying the linearised Slavnov-Taylor operator 
    \begin{align}
        \mathcal{S}_\Gamma = b + \mathcal{O}(\hbar),
    \end{align}
    to Eq.~\eqref{Eq:Breaking-of-STI-at-order-n-for-inductive-proof}, using the nilpotency condition in Eq.~\eqref{Eq:First-Nilpotency-Cond-of-lin-STI}, and extracting terms of order $\mathcal{O}(\hbar^n)$ yields the \emph{Wess-Zumino consistency condition}
    \begin{align}\label{Eq:The-Wess-Zumino-Consistency-Condition}
        b\Delta = 0.
    \end{align}
    This purely algebraic condition on the possible breakings, restricts any admissible breaking to local polynomials annihilated by $b$.
    The most general solution of this condition is obtained by solving the cohomology of the operator $b$.
    The unique nontrivial solution of this consistency condition is the gauge anomaly, as proven in Ref.~\cite{Piguet:1995er} (see also the remark below).
    Since, by assumption, the gauge anomaly is absent, all possible solutions of Eq.~\eqref{Eq:The-Wess-Zumino-Consistency-Condition} in the present context are trivial elements of the BRST cohomology, i.e.\ all solutions are $b$-exact.
    Hence, there exists an integrated local functional in the fields and derivatives $\Delta'$ of mass dimension 4 and ghost number 0, such that
    \begin{align}
        \Delta = b \Delta'.
    \end{align}
    Adding a counterterm $-\hbar^n\Delta'$ of $n$-loop order to the action cancels the $\mathcal{O}(\hbar^{n})$-breaking in Eq.~\eqref{Eq:Breaking-of-STI-at-order-n-for-inductive-proof}:
    \begin{align}
        \mathcal{S}(\Gamma-\hbar^n\Delta') = \mathcal{S}(\Gamma) - \hbar^n b\Delta' + \mathcal{O}(\hbar^{n+1})=\mathcal{S}(\Gamma) - \hbar^n \Delta + \mathcal{O}(\hbar^{n+1}) = \mathcal{O}(\hbar^{n+1}).
    \end{align}
    As shown in Ref.~\cite{Piguet:1995er}, this counterterm $-\hbar^n\Delta'$ is compatible with the gauge-fixing condition (see Eq.~\eqref{Eq:Gauge-Fixing-Condition-at-all-Orders}) and the ghost equation (see Eq.~\eqref{Eq:Ghost-Eq}), ensuring that these remain valid.
    This ensures the validity of the Slavnov-Taylor identity at order $\mathcal{O}(\hbar^{n})$, in absence of a gauge anomaly, and hence completes the proof by induction.
\end{proof}
\begin{remark}[BRST Cohomology and the Gauge Anomaly]\ \\
    The Wess-Zumino consistency condition $b\Delta=0$ imposes a condition on all possible breakings $\Delta$, i.e.\ it implies $\Delta\in\mathrm{Ker}(b)$.
    This condition is purely algebraic and finding its most general solution is achieved by studying the cohomology of the nilpotent operator $b$.
    Using the notation from Sec.~\ref{Sec:BRST-Symmetry}, the cohomology classes are determined by the equivalence relation $\Delta_1\sim\Delta_2$, which means that two breakings $\Delta_1$ and $\Delta_2$ are in the same equivalence class if $\exists\,\,\widetilde{\Delta}$, such that $\Delta_2=\Delta_1+b\widetilde{\Delta}$ (see also Ref.~\cite{Piguet:1995er}).
    We distinguish two cases:
    \begin{enumerate}[label={(\arabic*)}]
        \item \emph{Trivial elements of the BRST cohomology} are $b$-exact, i.e.\ they can be written as total $b$-variations $b\widetilde{\Delta}=\Delta\in\mathrm{Im}(b)$, and are thus equivalent to the 0-element of the cohomology, i.e.\ $\Delta\sim0$.
        Such breakings are \emph{spurious} and can be removed by suitable symmetry-restoring counterterms.
        If all solutions of the Wess-Zumino consistency condition are $b$-exact, the Slavnov-Taylor identity can be established to all orders.
        Regularisation-induced breakings in the BMHV scheme (see chapter~\ref{Chap:Practical_Symmetry_Restoration}) are of this type.
        \item \emph{Nontrivial elements of the BRST cohomology} are annihilated by $b$ but are not $b$-exact, i.e.\ $\Delta\in\mathrm{Ker}(b)\land\Delta\notin\mathrm{Im}(b)$. 
        Consequently, these elements cannot be written as total $b$-variations, and therefore such breakings cannot be cancelled by counterterms.
        These breakings represent \emph{anomalies} corresponding to nontrivial, i.e.\ nonvanishing, cohomology classes, i.e.\ $\Delta\sim\mathcal{A}\neq0$.
        In this case, the Slavnov-Taylor identity cannot be maintained at higher orders, which invalidates the theory, rendering it inconsistent and physically meaningless as a quantum theory.
        Fortunately, the unique nontrivial solution (independent of the loop order) is the well-known gauge (ABJ) anomaly (see Ref.~\cite{Piguet:1995er}), which does not renormalise (see Sec.~\ref{Sec:Anomalies}).
        If the anomaly cancellation condition~\eqref{Eq:General-Anomaly-Cancellation-Condition} is satisfied, the anomaly is absent.
    \end{enumerate}
\end{remark}

Finally, we recall an important result proven in Ref.~\cite{Piguet:1995er} concerning the basis of quantum insertions.
Using that the lowest order of an insertion corresponds to the classical operator (see Eq.~\eqref{Eq:Single-Operator-Insertion-lowest-order-Contribution}), we state:
\begin{proposition}[Basis of Insertions]\label{Prop:Basis_of_Insertions}\ \\
    Let $\{\Delta_\mathrm{cl}^p\}_{p\geq1}$ be a basis for classical insertions bounded by power counting, $\mathrm{dim}(\Delta^p_\mathrm{cl})\leq4$.
    Then the set 
    \begin{align}
        \{\Delta^p\cdot\Gamma=\Delta_\mathrm{cl}^p+\mathcal{O}(\hbar)\,\,|\, \, p\in\mathbb{N}, \,\mathrm{dim}(\Delta^p)\leq4\}
    \end{align}
    forms a basis for the quantum insertions of dimension bounded by power counting.
\end{proposition}
This result ensures that any higher-order breaking $\Delta\cdot\Gamma$ can be expressed in terms of a finite basis formed by any quantum extension of the classical insertions.
It constitutes a key ingredient for demonstrating the \emph{stability} (in the sense of Ref.~\cite{Piguet:1995er}) of the renormalisation programme for chiral gauge theories in the BMHV scheme (see chapter~\ref{Chap:Practical_Symmetry_Restoration}, specifically Sec.~\ref{Sec:Symmetry_Restoration_Procedure}).
\begin{definition}[Stability under Renormalisation]\label{Def:Stability_under_Renormalisation}\ \\
    The theory is \emph{stable} under renormalisation, i.e.\ under higher-order corrections, if all invariant counterterms correspond to a reparametrisation of the fields and parameters already present in the classical action, and any potential symmetry breaking can be removed by a finite number of symmetry-restoring counterterms at each order.
\end{definition}

\section{Renormalisation of Gauge Theories}\label{Sec:Renormalisation-of-GaugeTheories}

The renormalisation of gauge theories is considerably more difficult than renormalising non-gauge theories, because systematically subtracting UV divergences order by order is necessary but not sufficient.
The central challenge is to perform the renormalisation in a way that preserves the underlying gauge invariance or more precisely BRST invariance, in order to establish unitarity and Lorentz invariance, and eventually obtain a consistent quantum theory (cf.\ Sec.~\ref{Sec:BRST-Symmetry}).
Because BRST transformations are generally nonlinear (see Eq.~\eqref{Eq:Definition-of-the-BRST-Transformations}) and thus correspond to local composite operators (see Sec.~\ref{Sec:GreenFunctions_and_GeneratingFunctionals}), they acquire nontrivial quantum corrections, i.e.\ they renormalise, which must be taken into account.
To track the renormalisation of the BRST transformations, one can couple them to external sources and include them in the Lagrangian, see Eq.~\eqref{Eq:Def-of-Lagrangian_ext}.
At the quantum level, BRST invariance is encoded by the corresponding Slavnov-Taylor identity (see Sec.~\ref{Sec:The-Slavnov-Taylor-Identity}), whose validity must be established at each order; once in place, it yields a consistent quantum theory.\footnote{Note that here we do not distinguish between Ward identities (for linear symmetries) and Slavnov-Taylor identities but refer to both as Slavnov-Taylor identities (cf.\ Sec.~\ref{Sec:The-Slavnov-Taylor-Identity}).}
In particular, given a valid Slavnov–Taylor identity at some order, the steps from Sec.~\ref{Sec:BRST-Symmetry} can be repeated to establish: 
\begin{itemize}
    \item a physical Hilbert space $\mathcal{H}_\mathrm{phys}$ with positive norm, separated from unphysical states;
    \item unitarity of the physical $S$-matrix restricted on $\mathcal{H}_\mathrm{phys}$ (unphysical external states decouple from physical $S$-matrix elements), such that probability is preserved;
    \item Lorentz invariance in all physical amplitudes;
    \item Gauge independence of the physical $S$-matrix and observables.
\end{itemize}
Thus, the validity of the Slavnov-Taylor identity $\mathcal{S}(\Gamma)=0$ at all orders is essential and can be regarded as the defining equation of a gauge theory.

Altogether, we formulate the following set of requirements for a consistent gauge theory:
\begin{condition}[on Gauge Theories]\ 
\begin{enumerate}[label={(\arabic*)}]
    \item Unitarity, Causality and Poincar\'e invariance
    \item UV-Finiteness
    \item Symmetry requirements: $\mathcal{S}(\Gamma)=0$
    \item Gauge-fixing condition $\delta\Gamma/\delta B^a=\delta\Gamma_\mathrm{cl}/\delta B^a$ and similar relations for non-interacting fields
    \item Renormalisation conditions, such as $\mathrm{MS}$, $\overline{\mathrm{MS}}$ and On-Shell
    \item Power-counting renormalisability: $\mathrm{dim}(W_i)\leq4$ $\forall$ interactions\footnote{Optional: EFTs with $\mathrm{dim}(W_i)>4$ can also consistently be formulated and provide meaningful predictions.}
\end{enumerate}
\end{condition}
If these conditions can be maintained at each order in perturbation theory, the required counterterms are uniquely determined.
These counterterms are both necessary and sufficient to ensure a consistent quantum theory.
In practice, this means that all Feynman diagrams contributing to power-counting divergent 1PI Green functions must be computed within a chosen regularisation scheme, and local counterterms are subsequently adjusted to remove UV divergences and restore any spurious symmetry violations (if present).

In general, symmetries can be broken in higher-order calculations; for example, by the regularisation procedure (see chapter~\ref{Chap:Practical_Symmetry_Restoration}).
We distinguish the following cases:
\begin{enumerate}[label={(\arabic*)}]
    \item $\mathcal{S}(\Gamma_\mathrm{reg})=0$: the symmetry is manifestly preserved at all orders;
    \item $\mathcal{S}(\Gamma_\mathrm{reg})\neq0$: the classical symmetry is broken in one of two ways (cf.\ Sec.~\ref{Sec:Algebraic_Renormalisation}):
    \begin{enumerate}[label={(\alph*)}]
        \item spurious symmetry breaking: the symmetry can be restored, such that $\mathcal{S}(\Gamma_\mathrm{ren})=0$;
        \item anomalous symmetry breaking: the symmetry cannot be restored, i.e.\ $\mathcal{S}(\Gamma_\mathrm{ren})\neq0$.
    \end{enumerate}
\end{enumerate}
Case (1) is the simplest:
the symmetry is manifestly preserved, symmetry-restoring counterterms are not required, and all counterterms are manifestly symmetric, so that multiplicative renormalisation is sufficient.
Moreover, symmetry relations can be used to simplify the computation.
In case (2a), the symmetry is spuriously broken, which complicates the renormalisation:
adjusting appropriate symmetry-restoring counterterms --- whose existence is guaranteed for $b$-exact breakings (see Sec.~\ref{Sec:Algebraic_Renormalisation}) --- restores the Slavnov-Taylor identity and effectively maps the situation back to case (1), albeit with more involved intermediate steps.
Case (2b) violates the renormalisability requirements, the symmetry is lost, and the theory must be discarded.

The renormalisability of non-Abelian gauge theories was established in Refs.~\cite{tHooft:1971qjg,tHooft:1971akt,Lee:1972fj,Lee:1972ocr,Lee:1972yfa,Lee:1973fn} and generalised via BRST and algebraic renormalisation in Refs.~\cite{Becchi:1974xu,Becchi:1974md,Becchi:1975nq,Tyutin:1975qk}; see Refs.~\cite{Piguet:1980nr,Piguet:1995er} for reviews.
The formalism based on the nilpotent BRST charge $Q_B$ to identify $\mathcal{H}_\mathrm{phys}$ and the quartet mechanism (see Sec.~\ref{Sec:BRST-Symmetry}) were formulated in Ref.~\cite{Kugo:1979gm}.
Following Refs.~\cite{Belusca-Maito:2023wah,Weinberg:1996kr,Boehm:2001Gauge,Stoeckinger:2023rqft2,Collins_1984}, and restricting to power-counting renormalisable theories, we formulate:
\begin{theorem}[Renormalisability of Gauge Theories]\ \\
    In the absence of gauge anomalies, gauge theories are renormalisable:
    all UV divergences can be absorbed through a reparametrisation of the fields and parameters of the theory, consistent with causality and translational invariance.
    The Slavnov-Taylor identity can be established at all orders, ensuring unitarity, Lorentz invariance and gauge independence, thereby yielding a consistent quantum theory.
\end{theorem}
\begin{proof}
    The proof again proceeds by induction.
    At the classical level ($n=0$), the theory is UV finite and satisfies the Slavnov-Taylor identity by construction (see Sec.~\ref{Sec:BRST-Symmetry}).
    Now, suppose that the theory is renormalised up to order $\mathcal{O}(\hbar^{n-1})$, with renormalised effective action $\Gamma_\mathrm{ren}^{(\leq n-1)}$ that is finite, obeys all defining equations --- particularly the Slavnov-Taylor identity --- and yields a consistent theory.
        
    In case (1), the Slavnov-Taylor identity is manifestly preserved and multiplicative renormalisation is sufficient, since the divergences are restricted by BRST invariance.
    For example, in a non-Abelian gauge theory as discussed in Sec.~\ref{Sec:BRST-Symmetry}, multiplicative renormalisation is given by
    \begin{align}\label{Eq:Multiplicative-Renormalisation}
        g &\longrightarrow g_\mathrm{bare}=Z_g \, g, \nonumber\\
        \psi &\longrightarrow \psi_\mathrm{bare}=\sqrt{Z_\psi} \, \psi, \nonumber\\
        A^{a,\mu} &\longrightarrow A^{a,\mu}_\mathrm{bare} = \sqrt{Z_A} \, A^{a,\mu}, \nonumber\\
        c^a &\longrightarrow  c^a_\mathrm{bare} = \sqrt{Z_c} \, c^a, \nonumber\\
        B^a,\overline{c}^a, \xi &\longrightarrow \frac{1}{\sqrt{Z_A}} B^a, \frac{1}{\sqrt{Z_A}} \overline{c}^a, Z_A \xi, \nonumber\\
        \Longrightarrow \mathcal{L}_\mathrm{cl} &\longrightarrow \mathcal{L}_\mathrm{bare}=\mathcal{L}_\mathrm{cl}+\mathcal{L}_\mathrm{ct}.
    \end{align}
    The renormalisation transformations of $\overline{c}^a$ and $B^a$ already incorporate their BRST relation (see Eq.~\eqref{Eq:Definition-of-the-BRST-Transformations}); they therefore renormalise in the same way.
    The gauge-fixing term does not renormalise (bilinear in the fields, no interactions), which is reflected in the transformation behaviour of $\xi$.
    
    Case (2b) is excluded by assuming the absence of gauge anomalies; thus, only case (2a) remains.
    According to theorem~\ref{Thm:Renormalisability_of_the_Slavnov-Taylor_Identity} (see Sec.~\ref{Sec:Algebraic_Renormalisation}), symmetry-restoring counterterms reestablish the broken Slavnov-Taylor identity and thereby reduce this situation to case (1).
    
    Therefore, at order $\mathcal{O}(\hbar^{n})$, our starting point is the regularised effective action
    \begin{align}
        \Gamma_\mathrm{reg}^{(\leq n)} = \Gamma_\mathrm{reg,fin}^{(\leq n)} + \Gamma_\mathrm{reg,div}^{(n)},
    \end{align}
    where $\Gamma_\mathrm{reg,fin}^{(\leq n)}$ is finite, fully renormalised up to order $\mathcal{O}(\hbar^{n-1})$, contains all finite contributions up to order $\mathcal{O}(\hbar^{n})$, and satisfies the Slavnov-Taylor identity through order $\mathcal{O}(\hbar^{n})$.
    Hence, $\Gamma_\mathrm{reg,fin}^{(\leq n)}$ may already include symmetry-restoring counterterms up to $\mathcal{O}(\hbar^{n})$, if required.
    The remaining, unrenormalised divergences at order $\mathcal{O}(\hbar^{n})$ are represented by $\Gamma_\mathrm{reg,div}^{(n)}$.
    According to the general theorem of renormalisation (see Sec.~\ref{Sec:Renormalisation_Theory}), these divergences are restricted to local polynomials in external momenta and masses, with homogeneous degree determined by the overall degree of divergence of the corresponding 1PI Green functions.    
    Consequently, $\Gamma_\mathrm{reg,div}^{(n)}$ is a local functional, so that 
    \begin{align}\label{Eq:General-Ansatz-For-CTs-in-Renormalisability-Proof}
        S_\mathrm{ct}^{(n)}=\int \mathcal{L}_\mathrm{ct}^{(n)} = - \Gamma_\mathrm{reg,div}^{(n)},
    \end{align}
    provides the general Ansatz for the local counterterms --- constraint by power counting --- that cancel all divergences.
    
    The divergences and their corresponding counterterms are constrained by BRST invariance, as required for the consistency of the quantum theory.
    Determining their most general form necessitates taking into account the renormalisation of the BRST transformations themselves.
    This can be achieved by considering renormalised Green functions of the form
    \begin{align*}
        s_\mathrm{ren}\langle\ldots\rangle_\mathrm{ren}=\langle\ldots (s_\mathrm{ren}\phi_i)\ldots\rangle_\mathrm{ren}+\ldots,    
    \end{align*} 
    where $s_\mathrm{ren}$ is the renormalised BRST operator, such that $s_\mathrm{ren}(\mathcal{L}_\mathrm{cl}+\mathcal{L}_\mathrm{ct})=0$ (see Refs.~\cite{Collins_1984,Stoeckinger:2023rqft2}).
    
    However, in this work we employ the methods of algebraic renormalisation, using the Slavnov-Taylor identity (see e.g.\ Refs.~\cite{Belusca-Maito:2023wah,Weinberg:1996kr,Boehm:2001Gauge,Stoeckinger:2023rqft2}).
    In particular, after coupling the BRST transformations to external sources (see Eq.~\eqref{Eq:Def-of-Lagrangian_ext}), the Slavnov-Taylor identity, established up to order $\mathcal{O}(\hbar^{n})$, imposes the following condition:
    \begin{align}
        0 &= \mathcal{S}(\Gamma_\mathrm{reg}^{(\leq n)}) = \int \frac{\delta (\Gamma_\mathrm{reg,fin}^{(\leq n)} + \Gamma_\mathrm{reg,div}^{(n)}) }{\delta \phi_i(x)} \frac{\delta (\Gamma_\mathrm{reg,fin}^{(\leq n)} + \Gamma_\mathrm{reg,div}^{(n)}) }{\delta K_i(x)} 
        \nonumber \\
        &= \int \frac{\delta (\Gamma_\mathrm{reg,fin}^{(\leq n)})}{\delta \phi_i(x)} \frac{\delta (\Gamma_\mathrm{reg,div}^{(n)})}{\delta K_i(x)} + \int \frac{\delta (\Gamma_\mathrm{reg,div}^{(n)})}{\delta \phi_i(x)} \frac{\delta (\Gamma_\mathrm{reg,fin}^{(\leq n)})}{\delta K_i(x)} + \int \frac{\delta (\Gamma_\mathrm{reg,div}^{(n)})}{\delta \phi_i(x)} \frac{\delta (\Gamma_\mathrm{reg,div}^{(n)})}{\delta K_i(x)} + \mathrm{fin}
        \nonumber \\
        &= \int \frac{\delta \Gamma_\mathrm{cl}}{\delta \phi_i(x)} \frac{\delta \Gamma_\mathrm{reg,div}^{(n)}}{\delta K_i(x)} + \int \frac{\delta \Gamma_\mathrm{reg,div}^{(n)}}{\delta \phi_i(x)} \frac{\delta \Gamma_\mathrm{cl}}{\delta K_i(x)} + \mathrm{fin} + \mathcal{O}(\hbar^{n+1})
        \nonumber \\
        &= \mathcal{S}(\Gamma_\mathrm{cl}+\Gamma_\mathrm{reg,div}^{(n)}) + \mathrm{fin} + \mathcal{O}(\hbar^{n+1}). \nonumber
    \end{align}
    Hence, the $\mathcal{O}(\hbar^{n})$-divergences are constrained by the Slavnov-Taylor identity as 
    \begin{align}
        \mathcal{S}(\Gamma_\mathrm{cl}+\Gamma_\mathrm{reg,div}^{(n)}) = 0.
    \end{align}
    Using the invariance of the classical action $\Gamma_\mathrm{cl}$, we obtain the following consistency condition for the divergences --- and consequently also for the counterterms (see Eq.~\eqref{Eq:General-Ansatz-For-CTs-in-Renormalisability-Proof}):
    \begin{align}
        b\Gamma_\mathrm{reg,div}^{(n)}=0 \quad \Longrightarrow \quad b S_\mathrm{ct}^{(n)} = 0.
    \end{align}
    This Wess-Zumino-type consistency condition coincides with that obtained for possible breakings in Eq.~\eqref{Eq:The-Wess-Zumino-Consistency-Condition}, here establishing an analogous consistency requirement for the divergences.
    In particular, the condition $b S_\mathrm{ct}^{(n)}$ constrains both divergent and possible finite parts of the counterterms.
    The most general solution to this equation (further restricted by power counting to terms with $\mathrm{dim}\leq4$) determines the admissible counterterm structure, which is necessary and sufficient to cancel all divergences and, if required, to restore the Slavnov-Taylor identity.
    Hence, upon including the local counterterm action $S_\mathrm{ct}^{(n)}$, all divergences can be cancelled, yielding a finite and symmetric theory at order $\mathcal{O}(\hbar^{n})$, which completes the proof by induction.

    Finally, with finiteness and BRST invariance, $\mathcal{S}(\Gamma_\mathrm{ren})=0$, established to all orders, the physical properties follow analogously to Sec.~\ref{Sec:BRST-Symmetry}.
    In particular, one must consider Slavnov-Taylor identities for $S$-matrix elements, which requires studying the finite parts of renormalised, amputated, on-shell Green functions $\langle\ldots\rangle_{\mathrm{ren,OS}}^\mathrm{amp}$ within the LSZ reduction formalism.
    Applying the Slavnov-Taylor identity to such Green functions (cf.\ Eq.~\eqref{Eq:Formal-STI-Formulation-on-explicit-GreenFunctions}) leads to relations of the form
    \begin{align*}
        0 = \delta \langle \phi_1\ldots\phi_n\rangle_{\mathrm{ren,OS}}^\mathrm{amp} = \text{``}s_\mathrm{ren}\langle\phi_1\ldots\phi_n\rangle_{\mathrm{ren,OS}}^\mathrm{amp}\text{''},
    \end{align*}
    which must be analysed together with the transformation behaviour of the asymptotic fields (cf.\ Sec.~\ref{Sec:BRST-Symmetry}).
    In this way, all required properties can indeed be established, yielding a unitary, Lorentz invariant, and gauge-independent physical $S$-matrix on $\mathcal{H}_\mathrm{phys}$.
    Further details can be found in Refs.~\cite{Belusca-Maito:2023wah,Weinberg:1996kr,Boehm:2001Gauge,Stoeckinger:2023rqft2,Collins_1984}.
\end{proof}

\begin{remark}[on spontaneously broken Gauge Theories]\ \\
    Importantly, all of the above considerations are formal, since the $S$-matrix in gauge theories with massless gauge bosons is not, strictly speaking, well-defined due to IR divergences.
    Nevertheless, such theories remain valid quantum theories of nature with remarkable predictive power, as observables are both UV and IR finite.
    As discussed in Sec.~\ref{Sec:Renormalisation_Theory}, IR divergences cancel at the level of cross sections once virtual and real corrections are combined.

    Formal arguments, however, can be made rigorously for theories containing only massive vector bosons, obtained by spontaneous breaking of the gauge symmetry via the Higgs mechanism.
    In such theories, corresponding scalar fields exhibit inhomogeneous BRST transformations involving a linear ghost term, which complicates the analysis of the transformation behaviour of the asymptotic fields and introduces additional unphysical degrees of freedom (cf.\ Sec.~\ref{Sec:BRST-Symmetry}).
    Nevertheless, all preceding arguments can be consistently translated to the case of spontaneously broken gauge theories.

    In the above arguments, the implication was one-sided: from a valid Slavnov-Taylor identity, a consistent and unitary quantum theory follows.
    This naturally raises the question of whether the converse holds --- namely, whether the existence of a consistent quantum theory automatically implies the validity of the Slavnov-Taylor identity.

    In Refs.~\cite{Cornwall:1973tb,Cornwall:1974km}, the authors present a tree-level argument showing that theories with massive vector bosons must necessarily be spontaneously broken gauge theories in order to maintain unitarity of the $S$-matrix.
    In other words, Higgs boson contributions are essential for obtaining unitary results in vector-boson scattering processes.
    Hence, spontaneously broken gauge theories are the unique solution to theories of massive vector bosons.
    Analogous reasoning can be extended to general gauge theories.
    Suppose one begins with a Lagrangian that satisfies the Slavnov-Taylor identity and fulfils unitarity, causality, and relativistic covariance at tree level.
    At higher orders, the remaining freedom lies in finite local counterterms, which are themselves constrained by unitarity, causality, and relativistic covariance (see Sec.~\ref{Sec:Renormalisation_Theory}).
    Given the close connection between unitarity and Lorentz invariance on one hand, and the Slavnov–Taylor identity associated with BRST invariance on the other, the inclusion of any counterterm that violates the Slavnov–Taylor identity would presumably also violate unitarity and/or Lorentz invariance.
    This argument, however, does not constitute a rigorous proof.
    As it is neither required for nor central to the objectives of this thesis, it will not be pursued further here.
\end{remark}

\section{Peculiarities of Abelian Gauge Theories}\label{Sec:Peculiarities_of_Abelian_Gauge_Theories}

While Abelian and non-Abelian gauge theories can both be quantised within the BRST framework (see Sec.~\ref{Sec:BRST-Symmetry}) and are governed by the corresponding Slavnov-Taylor identity, the simpler group structure of Abelian theories entails distinctive features.
For instance, Abelian gauge theories contain less interactions --- in particular no gauge boson self-interactions --- than non-Abelian theories, with corresponding implications for higher-order corrections.
These characteristics lead to certain simplifications in the quantum theory but also necessitate additional symmetry constraints to ensure a consistent renormalisation.
In particular, whereas non-Abelian gauge theories acquire their rigidity and the universality of their coupling constant directly from the nontrivial structure of their Lie algebra, Abelian theories are less constrained by their gauge group and therefore require supplementary conditions to achieve the same physical consistency.
Specifically, a non-renormalisation of the charge $Q$ associated with the fields must be established to all orders.
Comprehensive discussions of Abelian gauge theories in this context can be found in Refs.~\cite{Becchi:1974md,Kraus:1995jk,Haussling:1996rq,Haussling:1998pp,Grassi:1997mc}.
We published a detailed presentation of the peculiarities of Abelian gauge theories in Ref.~\cite{Belusca-Maito:2023wah}, which we follow in this thesis.

Similar to chapter~\ref{Chap:Gauge_Theories}, we start from the gauge invariant classical Lagrangian of an Abelian theory with gauge group $U(1)_Q$ and associated charge $Q_{ij}=Q_i\delta_{ij}$ (no summation implied):
\begin{align}
    \mathcal{L}_\mathrm{inv} = \overline{\psi}_i i \slashed{D}_{ij} \psi_j - \frac{1}{4} F^{\mu\nu} F_{\mu\nu},
\end{align}
with covariant derivative $D_{ij}^\mu=\partial^\mu \delta_{ij} + i e Q_i \delta_{ij} A^\mu$ and field strength tensor $F_{\mu\nu}=\partial_\mu A_\nu-\partial_\nu A_\mu$ of the gauge boson $A_\mu$.
For clarity, we consider a vector-like gauge theory; the arguments extend to Abelian chiral gauge theories.
The gauge-fixing and ghost terms are
\begin{align}\label{Eq:Abelian-Gauge-Fixing-Term}
    \mathcal{L}_{\mathrm{ghost+fix}} & =  s \bigg[\overline{c} \bigg(\partial^\mu A_\mu + \frac{\xi}{2} B\bigg)\bigg] = B (\partial^\mu A_\mu) + \frac{\xi}{2} B^2 - \overline{c} \partial^\mu\partial_\mu c \,,
\end{align}
where we already used the Abelian BRST transformations in Eq.~\eqref{Eq:AbelianBRSTTrafos}.
The equation of motion of the Nakanishi-Lautrup auxiliary field gives $B = - (\partial_{\mu} A^{\mu}) /\xi$.
Coupling the nontrivial BRST transformations to external sources yields
\begin{align}
  \mathcal{L}_{\mathrm{ext}} = \rho^{\mu} sA_{\mu} + {\overline{R}}{}^{i} s\psi_{i} + R^{i} s\overline{\psi}_{i}\,.
\end{align}
The classical action, serving as the foundation for quantisation, is then given as in Eq.~\eqref{Eq:Full-Lagrangian-for-Quantisation}.

A crucial distinction compared to non-Abelian theories arises in the structure of the BRST transformations, dictated by the gauge group.
In a non-Abelian gauge theory, the transformations (cf.\ Eq.~\eqref{Eq:Definition-of-the-BRST-Transformations}) for the gauge and ghost fields are nonlinear, involving the structure constants $f^{abc}$.
Consequently, the composite operators corresponding to these transformations, such as $(sA^a_\mu)$, are themselves subject to renormalisation.
In stark contrast, for an Abelian gauge theory, the BRST transformation for the gauge field is linear in the fields, while the transformation for the ghost field vanishes identically.
In particular, the Abelian BRST transformations are given by 
\begin{subequations}\label{Eq:AbelianBRSTTrafos}
    \begin{align}
        sA_{\mu}(x) & = \partial_{\mu} c(x)\,,\\
        s\psi_{i}(x) & = - i e Q_{i} c(x) \psi_{i}(x)\,, \\
        s\overline{\psi}_{i}(x) &= i e Q_{i} c(x) \overline{\psi}_{i}(x)\,,\\
        sc(x) & = 0\,, \\
        s\overline{c}(x) & =  B(x)\,,\\
        sB(x) &= 0\,.
    \end{align}
\end{subequations}
A significant consequence of this linearity is that these transformations do not receive nontrivial quantum corrections, as discussed in Sec.~\ref{Sec:The-Slavnov-Taylor-Identity} around Eqs.~\eqref{Eq:Linear-Symmetry-Transformation-deltaphi} and \eqref{Eq:Expectation-Value-of-Linear-Symmetry-Transformation-deltaphi}.
Specifically, for the Abelian gauge boson this means that the expectation value $\langle sA_\mu \rangle_{J,K}$ is identical with the classical expression $(sA_\mu)^\mathrm{class}$.
Evidently, the BRST transformations of the fermions $\psi_i$ and $\overline{\psi}_i$ remain nonlinear in the fields, while all other transformations are indeed either linear or vanish identically (see Eq.~\eqref{Eq:AbelianBRSTTrafos}).
However, another crucial aspect of Abelian gauge theories is that the ghost $c$ and antighost $\overline{c}$ completely decouple from the rest of the theory, meaning that they do not appear in any interactions and cannot run in loops (contrary to non-Abelian theories).
The external sources $R^i$ and ${\overline{R}}{}^{i}$ decouple similarly, as they are non-dynamical, and thus non-propagating, number-valued fields that can at most appear as external legs but not as internal lines in loop diagrams.
As a result, none of the Abelian BRST transformations in Eq.~\eqref{Eq:AbelianBRSTTrafos} acquire nontrivial quantum corrections; in other words, they do not renormalise.
An important feature of linear classical symmetries of the classical action $\Gamma_\mathrm{cl}$, and such that do not undergo renormalisation, is that they automatically remain symmetries of the full effective quantum action $\Gamma$.

This simplification, however, comes at a cost and reveals a deeper issue.
In a non-Abelian theory with a simple gauge group $G$ and gauge coupling $g$, the commutation relations of the associated Lie algebra $\mathfrak{g}$ fix the form of the generators $T^a$ up to a choice of representation.
This powerfully constrains all interactions to be governed by a single, universal gauge coupling $g$.
In particular, the couplings of all matter fields to the gauge fields are $\propto g T^a$, while couplings governing gauge boson self-interactions are $\propto g f^{abc}$, both uniquely determined up to the universal gauge coupling $g$.
In contrast, the group structure of an Abelian theory is far less restrictive.
The generator is simply the charge operator $Q$, and any diagonal matrix constitutes a valid representation.
Without additional constraints, the individual charges $Q_i$ of the different fermions could, in principle, be arbitrary real numbers and renormalise independently at each order in perturbation theory --- their values could even vary from order to order.
Hence, group theory alone does not prevent the charges $Q_i$ to obtain quantum corrections.
Such a scenario, however, would violate the fundamental principle of charge universality, which is experimentally well-established.

To ensure a physically consistent renormalisation with fixed charges and a single, universally renormalising gauge coupling $e$, the theory must be defined with additional all-order linear identities.
In particular, this is achieved either by the local Ward identity or by the antighost equation.
These linear identities correspond to functional derivatives of the classical action $\Gamma_\mathrm{cl}$ and are postulated as defining relations of the theory.
Being linear, they do not renormalise --- loop corrections that could potentially violate them require interactions and thus at least three dynamical fields --- and therefore hold identically for the full quantum action $\Gamma$ (like the linear BRST transformations).

We begin with the antighost equation, the variation of $\Gamma_{\mathrm{cl}}$ w.r.t.\ the ghost:
\begin{align}\label{Eq:AntiGhostEquationAbelianCase}
  \frac{\delta \Gamma_{\mathrm{cl}}}{\delta c(x)} = \Box \overline{c}(x) + \partial_{\mu} \rho^{\mu}(x) - i e Q_{i} {\overline{R}}{}^{i}(x) \psi_{i}(x) + i e Q_{i} \overline{\psi}_{i}(x) R^{i}(x)\,.
\end{align}
Additionally, varying $\Gamma_{\mathrm{cl}}$ w.r.t.\ the antighost and
the external source of the photon yields
\begin{align}
  \frac{\delta \Gamma_{\mathrm{cl}}}{\delta \overline{c}(x)} & = - \Box c(x)\,,
  &
  \frac{\delta \Gamma_{\mathrm{cl}}}{\delta \rho_{\mu}(x)} &= sA^{\mu}(x) = \partial^{\mu}c(x)\,,
\end{align}
which combine into the ghost equation (cf.\ Eq.~\eqref{Eq:Ghost-Eq})
\begin{align}\label{Eq:GhostEquationAbelianCase}
  \mathscr{G}\Gamma_\mathrm{cl} \equiv \left( \frac{\delta}{\delta \overline{c}} + \partial_{\mu} \frac{\delta}{\delta \rho_{\mu}}\right) \Gamma_{\mathrm{cl}} = 0\,.
\end{align}
The gauge fixing condition follows from the variation of $\Gamma_{\mathrm{cl}}$ w.r.t.\ the auxiliary field $B$:
\begin{align}\label{Eq:GaugeFixingAbelianCase}
  \frac{\delta \Gamma_{\mathrm{cl}}}{\delta B(x)} = \xi B(x) + \partial_{\mu} A^{\mu}(x)\,.
\end{align}
All identities in Eqs.~\eqref{Eq:AntiGhostEquationAbelianCase}--\eqref{Eq:GaugeFixingAbelianCase} are linear in dynamical fields and therefore receive no loop corrections.
Moreover, as shown in Sec.~\ref{Sec:Algebraic_Renormalisation}, the gauge-fixing condition and the ghost equation can be established to all orders, even in non-Abelian gauge theories.
As a consequence, their all-order validity can indeed be imposed as part of the definition of the theory, i.e.\footnote{%
In Abelian gauge theories with spontaneous symmetry breaking, some of these identities fail to hold.
However, the introduction of appropriate background fields, makes it possible to derive a consistent local Ward identity and/or an Abelian antighost equation (see Refs.~\cite{Haussling:1996rq,Grassi:1997mc}).}
\begin{align}\label{Eq:RquireAllOrdersRelations}
    \frac{\delta \Gamma}{\delta c(x)} & \stackrel{!}{=} \frac{\delta \Gamma_{\mathrm{cl}}}{\delta c(x)},
    &
    \frac{\delta \Gamma}{\delta \overline{c}(x)} &\stackrel{!}{=} \frac{\delta \Gamma_{\mathrm{cl}}}{\delta \overline{c}(x)},
    &
    \frac{\delta \Gamma}{\delta \rho_{\mu}(x)} & \stackrel{!}{=} \frac{\delta \Gamma_{\mathrm{cl}}}{\delta \rho_{\mu}(x)},
    &
    \frac{\delta \Gamma}{\delta B(x)}  &\stackrel{!}{=} \frac{\delta \Gamma_{\mathrm{cl}}}{\delta B(x)}.
\end{align}
The antighost equation (see Eq.~\eqref{Eq:AntiGhostEquationAbelianCase}) is particularly important, as it directly involves the fermion fields $\psi_i$ and their charges $Q_i$.
Imposing its validity on the full quantum action $\Gamma$ enforces a rigid relation that prevents the charges from renormalising independently, thereby fixing them to all orders.

In contrast, the other functional derivatives of the classical action are nonlinear in dynamical fields, and are presented here for completeness as well as for the subsequent derivation of the local Ward identity:
\begin{subequations}\label{Eq:nonlinearfunctionalderivatives}
\begin{align}
  \frac{\delta \Gamma_{\mathrm{cl}}}{\delta R^{i}(x)} & = s\overline{\psi}_{i}(x) = i e Q_{i} c(x) \overline{\psi}_{i}(x)\,,\\
  \frac{\delta \Gamma_{\mathrm{cl}}}{\delta \psi_{i}(x)} & = i
  \partial_\mu\overline{\psi}_{i}(x) \gamma^\mu + e Q_{i} \overline{\psi}_{i}(x) \slashed{A}(x) + i e Q_{i} {\overline{R}}{}^{i}(x) c(x)\,,\\
  \frac{\delta \Gamma_{\mathrm{cl}}}{\delta {\overline{R}}{}^{i}(x)} & = s\psi_{i}(x) = - i e Q_{i} c(x) \psi_{i}(x)\,,\\
  \frac{\delta \Gamma_{\mathrm{cl}}}{\delta \overline{\psi}_{i}(x)} & = i \slashed{\partial} \psi_{i}(x) - e Q_{i} \slashed{A}(x) \psi_{i}(x) - i e Q_{i} R^{i}(x) c(x)\,.
\end{align}
\end{subequations}

The more familiar local Ward identity follows by acting with $\delta/\delta c(x)$ on the Slavnov-Taylor identity
\begin{align} \label{Eq:STIAbelianCase}
  0 = \mathcal{S}({\Gamma}) = \int d^4x \left( \frac{\delta \Gamma}{\delta {\overline{R}}{}^{i}} \, \frac{\delta \Gamma}{\delta \psi_{i}} + \frac{\delta \Gamma}{\delta R^{i}} \, \frac{\delta \Gamma}{\delta \overline{\psi}_{i}} + \frac{\delta \Gamma}{\delta \rho^{\mu}} \, \frac{\delta \Gamma}{\delta A_{\mu}} + B \, \frac{\delta \Gamma}{\delta \overline{c}} \right),
\end{align}
using the antighost equation $\delta \Gamma/\delta c(x)$, which is valid at all orders, and carefully accounting for the fermionic nature of the functional derivative $\delta/\delta c(x)$.
In particular, we obtain
\begin{align}\label{Eq:STIabelianWIderivation} 
    0  = \frac{\delta \mathcal{S}({\Gamma})}{\delta c(x)}
    & = \int d^{4}y \Bigg[ \left(\frac{\delta}{\delta c(x)} \frac{\delta \Gamma}{\delta {\overline{R}}{}^{i}(y)}\right) \frac{\delta \Gamma}{\delta \psi_{i}(y)} + \frac{\delta \Gamma}{\delta {\overline{R}}{}^{i}(y)} \left(\frac{\delta}{\delta c(x)}\frac{\delta \Gamma}{\delta \psi_{i}(y)}\right)
    \nonumber\\
    &\hspace{1.4cm} + \left(\frac{\delta}{\delta c(x)} \frac{\delta \Gamma}{\delta R^{i}(y)}\right) \frac{\delta \Gamma}{\delta \overline{\psi}_{i}(y)} + \frac{\delta \Gamma}{\delta R^{i}(y)} \left(\frac{\delta}{\delta c(x)} \frac{\delta \Gamma}{\delta \overline{\psi}_{i}(y)}\right)
    \nonumber\\
    &\hspace{1.4cm} + \left(\frac{\delta}{\delta c(x)}\frac{\delta \Gamma}{\delta \rho^{\mu}(y)}\right)\frac{\delta \Gamma}{\delta A_{\mu}(y)} - \frac{\delta \Gamma}{\delta \rho^{\mu}(y)} \left(\frac{\delta}{\delta c(x)} \frac{\delta \Gamma}{\delta A_{\mu}(y)}\right)
    \nonumber\\
    &\hspace{1.4cm} + B \left(\frac{\delta}{\delta c(x)} \frac{\delta \Gamma}{\delta \overline{c}(y)}\right) \Bigg]
    \nonumber\\
    & = \int d^{4}y \Bigg[ \left(\frac{\delta}{\delta {\overline{R}}{}^{i}(y)} \frac{\delta \Gamma}{\delta c(x)}\right) \frac{\delta \Gamma}{\delta \psi_{i}(y)} - \frac{\delta \Gamma}{\delta {\overline{R}}{}^{i}(y)} \left(\frac{\delta}{\delta \psi_{i}(y)} \frac{\delta \Gamma}{\delta c(x)} \right)
    \nonumber\\
    &\hspace{1.4cm} + \left(\frac{\delta}{\delta R^{i}(y)} \frac{\delta \Gamma}{\delta c(x)} \right) \frac{\delta \Gamma}{\delta \overline{\psi}_{i}(y)} - \frac{\delta \Gamma}{\delta R^{i}(y)} \left(\frac{\delta}{\delta \overline{\psi}_{i}(y)} \frac{\delta \Gamma}{\delta c(x)} \right)
    \nonumber\\
    &\hspace{1.4cm} - \left(\frac{\delta}{\delta \rho^{\mu}(y)} \frac{\delta \Gamma}{\delta c(x)} \right)\frac{\delta \Gamma}{\delta A_{\mu}(y)} - B \left(\frac{\delta}{\delta \overline{c}(y)} \frac{\delta \Gamma}{\delta c(x)} \right) \Bigg]
    \nonumber\\
    & = - i e Q_{i} \psi_{i}(x) \frac{\delta \Gamma}{\delta \psi_{i}(x)} + i e Q_{i} {\overline{R}}{}^{i}(x) \frac{\delta \Gamma}{\delta {\overline{R}}{}^{i}(x)} + i e Q_{i} \overline{\psi}_{i}(x) \frac{\delta \Gamma}{\delta \overline{\psi}_{i}(x)}
    \nonumber\\
    &\hspace{0.44cm} - i e Q_{i} R^{i}(x) \frac{\delta \Gamma}{\delta R^{i}(x)} - \partial_{\mu} \frac{\delta \Gamma}{\delta A_{\mu}(x)} - \Box B(x)\,.
\end{align}
Rearranging the last line yields the functional form of the local Abelian Ward identity,
\begin{align}\label{Eq:FunctionalAbelianWardId}
  \left( \partial_{\mu} \frac{\delta}{\delta A_{\mu}(x)} + i e Q_{i} \sum_{\Psi} (-1)^{n_{\Psi}} \, \Psi(x) \frac{\delta}{\delta \Psi(x)} \right) \Gamma = - \Box B(x) \,,
\end{align}
where the sum runs over all charged fields and their associated sources, i.e.\ $\Psi\in\{\psi_{i},\overline{\psi}_{i},R^{i},{\overline{R}}{}^{i}\}$ and $n_{\Psi}\in\{0,1,0,1\}$.
This identity generates the well-known Ward identities, such as the transversality condition of the gauge boson self energy and the relation between the fermion self energy and the fermion--gauge boson vertex.
The functional Ward identity can be established to all orders, and thus again fixes the charges $Q_i$, ensuring a consistent and universal renormalisation of the Abelian gauge coupling $e$.
In particular, a single counterterm is sufficient for its renormalisation.

In summary, the peculiarities of Abelian gauge theories stem from their simple group structure.
While this leads to the non-renormalisation of BRST transformations and the decoupling of ghosts, it does not by itself ensure charge universality.
The theory must therefore be supplemented with additional linear constraints --- the antighost equation and the Ward identity --- which are part of its defining relations and constrain the renormalisation procedure.
These identities guarantee a consistent perturbative expansion with a single, universal gauge coupling.
In the presence of spurious symmetry breakings, the Ward identities derived from Eq.~\eqref{Eq:FunctionalAbelianWardId} can be used to determine symmetry-restoring counterterms and interpret the breaking and subsequent restoration of the Slavnov-Taylor identity.

\chapter{Dimensional Regularisation and Renormalisation}\label{Chap:DReg}

As discussed in the previous chapter, and particularly in Sec.~\ref{Sec:Renormalisation_Theory}, perturbative quantum field theory requires a renormalisation procedure to cancel UV divergences and obtain finite results.
While a regularisation is not always necessary --- e.g.\ in the BPHZ formalism (see theorem~\ref{Thm:BPHZ-Theorem}) --- the standard counterterm method, most common in practical applications, relies on an intermediate regularisation to isolate divergences before subtracting them by counterterms (cf.\ corollary~\ref{Thm:Corollary-on-Counterterm-Method} and the text below).

Various schemes exist, including momentum cut-off, Pauli-Villars, analytic regularisation and dimensional regularisation.
Any admissible scheme must be mathematically consistent and compatible with fundamental principles such as unitarity and causality.
Further desirable properties include a valid and well-defined quantum action principle, symmetry preservation, and computational efficiency.
The latter properties are, however, not necessarily required; for example broken symmetries that are not anomalous can be restored during renormalisation.
For an overview and comparison of schemes we refer to Ref.~\cite{Gnendiger:2017pys}.

In this thesis we exclusively employ dimensional regularisation (DReg).
DReg and its variants are the most widely used regularisation schemes due to their practical utility in explicit computations:
they preserve Lorentz invariance and gauge invariance of vector-like gauge theories, leave the structure of loop integrals essentially unchanged (allowing efficient integration techniques), and transparently isolate divergences as $1/(D-4)$ poles (see Refs.~\cite{Belusca-Maito:2023wah,Gnendiger:2017pys}).
Moreover, they have been proven to be mathematically consistent to all orders, equivalent to the BPHZ formalism, and compatible with unitarity and causality.

The central idea of DReg is to extend the dimension of spacetime from 4 to formally $D\in\mathbb{C}$ dimensions, typically $D=4-2\epsilon$ (a convention also adopted in this thesis).
Loop integrals then become $D$-dimensional and are regulated such that divergences appear as poles in $1/\epsilon$; for $\epsilon\neq0$ all integrals are convergent.
These foundations are reviewed in Sec.~\ref{Sec:Spacetime_and_Integrals_in_DReg}, following Refs.~\cite{Collins_1984,Belusca-Maito:2023wah}.
Originally, DReg was pioneered by the authors of Refs.~\cite{tHooft:1972tcz,Bollini:1972ui,Ashmore:1972uj}, with important developments in Refs.~\cite{Wilson:1972cf,Speer:1974cz,Breitenlohner:1975hg,Breitenlohner:1976te,Breitenlohner:1977hr,Collins_1984}.
The embedding of algebraic objects such as the metric, Dirac matrices and other Lorentz covariants into $D$ dimensions is discussed in Sec.~\ref{Sec:Elements_of_D-dim_Spacetime}. 

The only drawback is the so-called $\gamma_5$-problem (see Sec.~\ref{Sec:The-g5-Problem}):
$\gamma_5$ and the Levi-Civita tensor $\varepsilon^{\mu\nu\rho\sigma}$ are manifestly 4-dimensional, and their embedding into $D$ dimensions is intricate.
The only known scheme that is mathematically consistent to all orders is the Breitenlohner-Maison/ `t~Hooft-Veltman (BMHV) scheme (see Sec.~\ref{Sec:The-BMHV-Scheme}), which however breaks gauge and BRST invariance, requiring symmetry-restoring counterterms and resulting in a considerably more complicated renormalisation procedure (see chapter~\ref{Chap:Practical_Symmetry_Restoration}).
Thus, dimensional schemes lack a gauge-invariant treatment of chiral symmetries.

Another important feature of DReg, not shared by all schemes, is that the regularisation can be formulated at the Lagrangian level: a formally $D$-dimensional Lagrangian generates Feynman rules and directly produces fully regularised diagrams (see Sec.~\ref{Sec:D-dim_Lagrangian}).
This allows to study symmetries and other properties at the Lagrangian level.
Our notation and the basic organisation of the renormalisation procedure in DReg are presented in Sec.~\ref{Sec:Notation_and_Organisation_of_Dimensional_Renormalisation}.

Finally, Sec.~\ref{Sec:Renormalisation_in_DReg} presents the fundamental theorem of renormalisation in DReg, establishing its mathematical consistency and equivalence with the BPHZ formalism, while Sec.~\ref{Sec:Regularised-QAP} provides a rigorous proof of the quantum action principle within DReg.
Both results were first obtained and proven in Ref.~\cite{Breitenlohner:1977hr}.

\section{Spacetime and Integrals in $D$ Dimensions}\label{Sec:Spacetime_and_Integrals_in_DReg}

As indicated above, the basic idea of DReg is to regularise momentum integrals by extending the dimensionality of spacetime from its physical value to $D\in\mathbb{C}$.
Essentially, two approaches have been developed in the literature for the implementation of DReg.

On the one hand by defining $D$-dimensional loop momentum integrals in terms of Schwinger parametrisation that satisfy the required properties, as pioneered by the authors of Refs.~\cite{Speer:1974cz,Breitenlohner:1977hr}.
Here, the dependence on the dimensionality $D$ only enters via the Symanzik polynomial $\mathcal{U}$.
Other $D$-dimensional objects, such as the metric or $\gamma$-matrices, are just defined formally with desired algebraic properties, which, however, does not amount to a fully mathematically self-consistent definition.

On the other hand, the approach of Refs.~\cite{Wilson:1972cf,Collins_1984} provides an explicit mathematical definition of the $D$-dimensional space and the objects therein.
Here the associated integral operations are defined constructively together with proofs of existence and uniqueness.

Both approaches are ultimately equivalent and for a recent review we refer the reader to Ref.~\cite{Belusca-Maito:2023wah}.
In the following, we adopt the axiomatic approach in the spirit of Refs.~\cite{Wilson:1972cf,Collins_1984}.

\begin{definition}[Dimensional Regularisation] \label{Def:DReg}\ \\
    The 4-dimensional spacetime $(\mathbb{M}_4,\eta_{[4]})$ is extended to formally $D\in\mathbb{C}$ dimensions:
    \begin{align}
        (\mathbb{M}_4,\eta_{[4]}) \longrightarrow (\mathbb{M}_{D},\eta_{[D]}),
    \end{align} 
    with decomposition
    \begin{align}\label{Eq:Decomposition-of-D-dim-Spacetime}
        \mathbb{M}_{D} = \mathbb{M}_{4} \oplus \mathbb{M}_{D-4}, \qquad\quad
        \eta_{[D]}^{\mu\nu} = \eta_{[4]}^{\mu\nu} + \eta_{[D-4]}^{\mu\nu}, 
    \end{align}
    where $\mathbb{M}_{D}\supset\mathbb{M}_{4}$ is a quasi $D$-dimensional space, that is, an $\infty$-dimensional vector space in which the objects and operations contained therein 
    formally exhibit $D$-dimensional algebraic behaviour, such that particularly $\eta_{[D]}^{\mu\nu}\eta_{[D],\mu\nu}=D$.  
    Along these lines all spacetime integrals, and thus all momentum integrals, become formally $D$-dimensional,
    \begin{align}\label{Eq:DReg-Integrals-4-to-D}
        \int \frac{d^4k}{(2\pi)^4} \longrightarrow \mu^{4-D} \int \frac{d^Dk}{(2\pi)^D},
    \end{align}
    where $\mu$ denotes the renormalisation scale (an artificial mass scale).
\end{definition}

Note that in the following, the subscripts are dropped for convenience, and instead 4- and $(D-4)$-dimensional quantities are denoted with overbars and hats, respectively: 
$\eta_{[D]}^{\mu\nu}\equiv\eta^{\mu\nu}$, $\eta_{[4]}^{\mu\nu}\equiv\overline{\eta}^{\mu\nu}$, $\eta_{[D-4]}^{\mu\nu}\equiv\widehat{\eta}^{\mu\nu}$, and similarly for momenta $p^{\mu}=\overline{p}^{\mu}+\widehat{p}^{\mu}$ and other Lorentz covariants such as $X^{\mu}=\overline{X}^{\mu}+\widehat{X}^{\mu}$.

\begin{remark}[$\infty$-Dimensionality]\
    \begin{itemize}
        \item As a consequence of $\mathbb{M}_D$ being an $\infty$-dimensional vector space, $D$-dimensional momenta $p$ have infinitely many components, and arbitrary sets of momenta may need to be treated as linearly independent (there is no finite basis spanning the entire space), see Ref.~\cite{Collins_1984,Wilson:1972cf}.
        \item The first four components of the quasi $D$-dimensional momenta $(p^{\mu}) = (p^0,p^1,p^2,p^3,\ldots)$ correspond to physical momenta, while all other components vanish in the limit $D=4$.
        \item The embedding $\mathbb{M}_4 \subset \mathbb{M}_D$ is essential for mathematical consistency. Assuming the opposite leads to inconsistencies, as discussed in the context of DRed in Ref.~\cite{Belusca-Maito:2023wah,Gnendiger:2017pys}.
    \end{itemize}   
\end{remark}

A central object in DReg is the $D$-dimensional integration, as indicated in Eq.\ \eqref{Eq:DReg-Integrals-4-to-D}.
Hence, the integration operation over functions of elements of the $\infty$-dimensional vector space $\mathbb{M}_D$ must be axiomatically defined, such that it resembles the desired $D$-dimensional properties and includes a smooth limit $D\to4$ for integrals that are convergent in 4 dimensions.
In particular, existence and uniqueness of the $D$-dimensional integration needs to be established.
Following Refs.~\cite{Belusca-Maito:2023wah,Wilson:1972cf,Collins_1984}, we define:
\begin{definition}[$D$-dimensional Integration] \label{Def:D-dim-Integration}\ \\
    The operation of $D$-dimensional integration over the $\infty$-dimensional vector space $\mathbb{M}_D$, 
    \begin{align}\label{Eq:quasi-D-dim-Integration-Operation}
        \int d^Dk \, f(k),
    \end{align}
    for any function $f(k)$ of a quasi $D$-dimensional vector $k\in\mathbb{M}_D$,
    is defined as a functional of $f$ that satisfies the following axioms:
    \begin{enumerate}[label={\textbf{(\alph*)}}]
    \item Linearity: $\forall$ $a,b\in\mathbb{C}$, and functions $f$, $g$, we have
    \begin{align}
        \int d^Dk \, ( a f(k) + b g(k) ) = a \int d^Dk \, f(k) + b \int d^Dk \, g(k), 
    \end{align}
    \item \label{itm:Translation-Invariance} Translation invariance:
    \begin{align}
        \int d^Dk \, f(k+p) = \int d^Dk \, f(k), \qquad \forall \, p \in \mathbb{M}_D,
    \end{align}
    (rotational covariance is required as well).
    \item Scaling: 
    \begin{align}
        \int d^Dk \, f(sk) = s^{-D}\int d^Dk \, f(k), \qquad \forall \, s \in \mathbb{R},
    \end{align}
    defining the dimensionality to be $D$.
    \item Normalisation via the $D$-dimensional Gaussian integral:
    \begin{align}\label{Eq:D-dim-Gaussian-Integral}
        \int d^Dk \, e^{-k^2} = \pi^{D/2},
    \end{align}
    as naively expected (here calculated with $D$-dimensional Euclidean metric).
\end{enumerate}
\end{definition}

For the current purpose, we are working in quasi $D$-dimensional Euclidean spacetime, i.e.\ with $D$-dimensional Euclidean metric $\delta^{\mu\nu}$, as this is generally sufficient, see Refs.\ \cite{Belusca-Maito:2023wah,Collins_1984}, and Minkowski spacetime can be obtained via Wick rotation.
Applying $D$-dimensional spherical coordinates to evaluate the rotationally symmetric Gaussian integral in Eq.\ \eqref{Eq:D-dim-Gaussian-Integral}, we find
\begin{align}
    \pi^{D/2} = \int d^Dk \, e^{-k^2} = \int d^{D-1}\Omega \int_0^{\infty}dk \, k^{D-1} \, e^{-k^2},
\end{align}
determining the surface of a quasi $D$-dimensional unit sphere to be 
\begin{align}\label{Eq:D-dim-Sphere-Surface}
    \Omega_D \equiv \int d^{D-1}\Omega = \frac{2\pi^{D/2}}{\Gamma(D/2)}.
\end{align}

The axioms in Def.~\ref{Def:D-dim-Integration} establish uniqueness of the integration operation.
In order to further ensure existence, and thus self-consistency, the authors of Refs.~\cite{Wilson:1972cf,Collins_1984} provide a constructive definition of the quasi $D$-dimensional integration operation in Eq.\ \eqref{Eq:quasi-D-dim-Integration-Operation}:
introducing a parallel subspace of finite, integer dimension $N\in\mathbb{N}_0$, such that the space $\mathbb{M}_D$ decomposes into a parallel and an orthogonal subspace, the vector $k\in\mathbb{M}_D$ also decomposes as $k = k_{||} + k_{\perp}$.
Then the integration can be expressed as a sequence of ordinary integrals as 
\begin{align}\label{Eq:Constructive-Definition-of-D-dim-Integration}
    \int d^Dk \, f(k) = \Omega_{D-N} \int d^N k_{||} \int_0^{\infty} dk \, k^{D-N-1} f(k_{||},k^2),
\end{align}
where spherical coordinates and Eq.\ \eqref{Eq:D-dim-Sphere-Surface} are applied to the $(D-N)$-dimensional integration over the orthogonal subspace.
This result is independent of the choice of the finite, integer-dimensional parallel subspace of $k_{||}$, see Ref.~\cite{Collins_1984}.
Moreover, this definition holds not only for scalar functions $\{f(k)\}$ but also for tensor functions $\{f^{\mu\nu\cdots}(k)\}$, by applying it component-wise.

\paragraph{Additional properties} of the quasi $D$-dimensional integration that follow from the above axioms and constructive definition are:
\begin{enumerate}[label={\textbf{(\alph*)}},start=5]
    \item Scaleless integrals vanish:
    \begin{align}
        \int d^Dk \, \big(k^2\big)^{\alpha} = 0.
    \end{align}
    \item \label{itm:Commutation-of-Differentiation} Commutation with $D$-dimensional differentiation:
    \begin{align}
        \frac{\partial}{\partial p}\int d^Dk \, f(k,p,\ldots) = \int d^Dk \, \frac{\partial}{\partial p} f(k,p,\ldots).
    \end{align}
    \item Integration by parts: 
    \begin{align}
        \int d^Dk \frac{\partial f(k)}{\partial k} = 0,
    \end{align}
    as implied by properties \ref{itm:Translation-Invariance} and \ref{itm:Commutation-of-Differentiation}.
    \item\label{itm:independence_of_order_or_integration_in_D} Independence of the order of integration:
    \begin{align}
        \int d^Dp \int d^Dk \, f(p,k) = \int d^Dk \int d^Dp \, f(p,k).
    \end{align}
    \item\label{itm:Smooth-Limit-to-4-dim} Provided the integral is finite in $4$ dimensions and the integrand is analytic, its extension to formally $D$ dimensions is analytic in both $D$ and the external momenta $p_i$ in a neighbourhood of $D=4$.
    Furthermore, it recovers the original result for $D=4$.
\end{enumerate}

\begin{remark}[Consistency, handling divergences and physical implications]\
    \begin{itemize}
        \item Existence and uniqueness of the $D$-dimensional integration ensure mathematical consistency of DReg. This guarantees that final results are unique and independent of the particular organisation of the calculation, which is essential for the internal consistency of the regularisation scheme.
        \item The dependence on $D$ in the exponent, see Eq.\ \eqref{Eq:Constructive-Definition-of-D-dim-Integration}, controls the asymptotic behaviour of the integrand for both the ultraviolet (large $k$) and the infrared (small $k$) limit, and hence makes the regularisation effect explicit.
        For functions $f$ that exhibit at most power-like divergences at these limits, there exists a range in $D$ for which the $k$-integral converges, and is thus well-defined.
        For arbitrary $D\in\mathbb{C}$, the value of such an integral is then obtained via analytical continuation.
        Additionally, Eq.\ \eqref{Eq:D-dim-Sphere-Surface} illustrates that DReg, with $D\in\mathbb{C}$, relies on the analytic properties of the $\Gamma$-function, cf.\ Ref.~\cite{Weinzierl:2022eaz}.
        \item As implied in the previous remark, DReg regularises not only UV divergences, but also IR divergences (arising from massless particles).
        \item Divergences manifest as simple or higher-order poles in $\epsilon=(4-D)/2$, i.e.\ as terms $\sim1/\epsilon^k$ with $k\in\mathbb{N}$.
        \item Property~\ref{itm:Smooth-Limit-to-4-dim}, together with the embedding $\mathbb{M}_4\subset\mathbb{M}_D$, which allows for the decomposition of quasi $D$-dimensional spacetime (see Eq.\ \eqref{Eq:Decomposition-of-D-dim-Spacetime}) and momenta $p^{\mu}=\overline{p}^{\mu}+\widehat{p}^{\mu}$, ensure a smooth and consistent limit to the physical case of 4 dimensions.
        \item The renormalisation scale $\mu$ has mass dimension $[\mu]=1$, such that integrals retain their physical mass dimension in $D$ dimensions, see Eq.\ \eqref{Eq:DReg-Integrals-4-to-D}.
        Importantly, physical observables are completely independent of this unphysical scale $\mu$.
    \end{itemize}
\end{remark}

\section{Elements of $D$-dimensional Spacetime}\label{Sec:Elements_of_D-dim_Spacetime}

As discussed in Sec.~\ref{Sec:Spacetime_and_Integrals_in_DReg}, the vector space $\mathbb{M}_D$ is $\infty$-dimensional and so the $D$-dimensional metric $\eta^{\mu\nu}$, Dirac matrices $\gamma^{\mu}$, momenta $p^{\mu}$ and other Lorentz covariants have infinitely many components.
The embedding $\mathbb{M}_4\subset\mathbb{M}_D$ ensures the presence of manifestly 4-dimensional objects alongside quasi $D$-dimensional ones, which is essential for a smooth limit $D\to4$.

A self-consistent scheme requires not just stating the properties of quasi $D$-dimensional objects, but an explicit construction of such objects that resemble the desired $D$-dimensional algebraic behaviour, despite their infinite number of components.
Such a construction was given in Ref.~\cite{Collins_1984}.
Following Refs.~\cite{Belusca-Maito:2023wah,Collins_1984}, we provide a constructive definition of the metric --- particularly the lower-index metric $\eta_{\mu\nu}$ --- and index contractions via integrals, as well as an inductive construction of the $\gamma^{\mu}$-matrices.

For the constructive definition of the metric in quasi $D$-dimensional spacetime, we again work in Euclidean signature, as the Minkowski case can be obtained by Wick rotation.
As a contravariant tensor, the metric is defined by its components
    \begin{align}
        \delta^{\mu\nu} =
        \begin{cases}
            1, \quad &\text{if} \,\,\, \mu = \nu,\\
            0, \quad &\text{else},
        \end{cases}
    \end{align}
see Ref.~\cite{Collins_1984}.
However, naively summing over all components is not meaningful in $\mathbb{M}_D$ due to the infinite number of components, leading to
\begin{align}
    \sum_{\mu,\nu=1}^{\infty} \delta^{\mu\nu}\delta_{\mu\nu}=\infty,
\end{align}
which clearly does not reproduce the correct $D$-dimensional algebraic behaviour.
For this reason, we proceed with the following definition:
\begin{definition}[The metric and index contractions in quasi $D$ dimensions]\label{Def:metric_and_contractions_in_D-dim} \ \\
    The metric is a multilinear map $\delta$ of type $(0,2)$ --- that is, a covariant tensor of rank $(0,2)$ or equivalently a 2-form --- that acts on contravariant tensors $T$ (with components $T^{\mu\nu}$) as 
    \begin{align}\label{Def:Contraction_in_D-Dimensions_via_Integral}
         \delta(T) = \frac{D\Gamma(D/2)}{\pi^{D/2}} \int d^Dk \, T^{\mu\nu} k_{\mu}k_{\nu} \, \delta(k^2 - 1).
    \end{align}
    Accordingly, index contraction $\delta_{\mu\nu}T^{\mu\nu}$ is defined by Eq.~\eqref{Def:Contraction_in_D-Dimensions_via_Integral}, i.e.\ $\delta_{\mu\nu}T^{\mu\nu}\equiv\delta(T)$.
    The individual components $\delta_{\mu\nu}$ of the metric are thus given by
    \begin{align}
    \delta_{\mu\nu} = \frac{D\Gamma(D/2)}{\pi^{D/2}} \int d^Dk \, k_{\mu} k_{\nu} \, \delta(k^2-1).
\end{align}
\end{definition}
Consequently, the metric and index/tensor contractions are defined via quasi $D$-dimensional integration, where the integrals are evaluated as discussed in Sec.~\ref{Sec:Spacetime_and_Integrals_in_DReg}.
\begin{remark}[Metric and index contractions in quasi $D$ dimensions]\
    \begin{itemize}
        \item In Euclidean spacetime, we have $k^2=\delta^{\mu\nu}k_{\mu}k_{\nu}$ and $k_{\mu}=k^{\mu}$.
        \item Importantly, contraction with $\delta_{\mu\nu}$ is defined via $D$-dimensional integration, see Eq.~\eqref{Def:Contraction_in_D-Dimensions_via_Integral}, rather than by explicit summation over components. 
        Here, index contraction is performed prior to integration, and in general, these two operations do not commute.
        \item Acting on itself, this definition yields the desired algebraic property,
            \begin{align}\label{Eq:Effective-D-dim-Behaviour-of-Metric}
                \delta_{\mu\nu}\delta^{\mu\nu} &= \delta(\delta) = \frac{D\Gamma(D/2)}{\pi^{D/2}} \int d^Dk \, k^2 \, \delta(k^2-1) = D,
            \end{align}
        which shows that the metric defined in this way exhibits the correct $D$-dimensional behaviour.
        Furthermore, component-wise, we see that
            \begin{align}
                \delta_{\mu\nu} = \delta^{\mu\nu},
            \end{align}
        which are the expected component values of $\delta_{\mu\nu}$.
        \item Because $D$-dimensional integrations commute (see property~\ref{itm:independence_of_order_or_integration_in_D}), the metric can be moved inside or outside the integral: 
            \begin{align}
                \delta_{\mu\nu} \int d^Dk \, k^{\mu}k^{\nu} \, f(k) = \int d^Dk \, \delta_{\mu\nu} k^{\mu}k^{\nu} \, f(k) = \int d^Dk \, k^2 f(k).
            \end{align}
        Thus, contraction with $\delta_{\mu\nu}$ commutes with $D$-dimensional integration. 
        \item The quasi $D$-dimensional Minkowski metric $\eta_{\mu\nu}$ is obtained via Wick rotation.
        \item The metric defined in Def.~\ref{Def:metric_and_contractions_in_D-dim} establishes a natural isomorphism between the vector space $\mathbb{M}_D$ and its dual, enabling the raising and lowering of indices. 
        \item Def.~\ref{Def:metric_and_contractions_in_D-dim} extends to other covariant tensors, and hence defines general tensor contractions.
        \item For tensors with only finitely many nonvanishing components, this contraction prescription reproduces the standard operation of explicit summation over components.
    \end{itemize}
\end{remark}
\noindent In this way, all desired properties of the $D$-dimensional metric are established by explicit construction.

Similarly, we provide an explicit construction of the quasi $D$-dimensional $\gamma^{\mu}$-matrices that realise the $D$-dimensional Clifford algebra,
\begin{align}\label{Eq:D-dim-Clifford-Algebra}
    \{\gamma^{\mu},\gamma^{\nu}\} = 2 \eta^{\mu\nu} \mathbb{1},
\end{align}
and, upon contraction, yield
\begin{align}\label{Eq:D-dim-Contraction-of-2-gammas}
    \gamma^{\mu}\gamma_{\mu} = D \mathbb{1},
\end{align}
even though they have infinitely many components.
Here, we state our construction from Ref.~\cite{Belusca-Maito:2023wah}, which follows the general approach of Ref.~\cite{Collins_1984}, but differs w.r.t.\ hermiticity, reality, and charge conjugation properties.
In particular, it is defined to preserve 4-dimensional properties and relations (except those involving $\gamma_5$; see Sec.~\ref{Sec:The-g5-Problem}) in quasi $D$ dimensions.
This is made explicit in the following definition:
\begin{definition}[Construction of $\gamma^{\mu}$ in quasi $D$ dimensions]\label{Def:Dirac_matrices_in_D-dim}\ \\
    The formally $D$-dimensional $\gamma^{\mu}$-matrices are $\infty$-dimensional block matrices, given explicitly by
    \begin{align}\label{Eq:Def-of-quasi-D-dim-gammas}
        \gamma^{\mu} =
        \begin{cases}
            \begin{pmatrix}
                \gamma_{[4]}^\mu & 0 & 0 & \cdots \\
                0 & \gamma_{[4]}^\mu & 0 & \cdots \\
                0 & 0 & \gamma_{[4]}^\mu & \cdots \\
                \vdots & \vdots & \vdots & \ddots
            \end{pmatrix}, \qquad & \text{for} \quad \mu \in \{0,1,2,3\},
            \vspace{1.5ex}\\
            \begin{pmatrix}
                \gamma^\mu_{\langle2^{2\mu+1}\rangle} & 0 & \cdots \\
                0 & \gamma^\mu_{\langle2^{2\mu+1}\rangle} & \cdots \\
                \vdots & \vdots & \ddots
                \end{pmatrix}, \qquad & \text{for} \quad \mu \geq 4.
        \end{cases} 
    \end{align}
    For $\mu\in\{0,1,2,3\}$, the blocks $\gamma^{\mu}_{[4]}$ are the standard $(4\times4)$ $\gamma$-matrices in 4 dimensions.
    For $\mu\geq4$, $\gamma^\mu_{\langle2^{2\mu+1}\rangle}$ is a real, anti-hermitian matrix of dimension $2^{(2\mu+1)}$, defined by
    \begin{align}
        \gamma^\mu_{\langle2^{2\mu+1}\rangle} = 
            \begin{pmatrix}
                0 & \widehat{\gamma}_{\langle 4^{\mu} \rangle} \\
                - \widehat{\gamma}_{\langle 4^{\mu} \rangle} & 0 
            \end{pmatrix},
            \qquad \text{for} \quad \mu \geq 4,
    \end{align}
    where $\widehat{\gamma}_{\langle 4^{k} \rangle}$ is a real, hermitian $4^k$-dimensional matrix composed of $\pm(\gamma_5)_{[4]}$-blocks on its diagonal,
    \begin{align}
        \widehat{\gamma}_{\langle 4 \rangle} = (\gamma_5)_{[4]}, 
        \qquad \quad 
        \widehat{\gamma}_{\langle 4^{k+1} \rangle} =
            \begin{pmatrix}
                \widehat{\gamma}_{\langle 4^{k} \rangle} & 0 & 0 & 0 \\
                0 & - \widehat{\gamma}_{\langle 4^{k} \rangle} & 0 & 0 \\
                0 & 0 & - \widehat{\gamma}_{\langle 4^{k} \rangle} & 0 \\
                0 & 0 & 0 & \widehat{\gamma}_{\langle 4^{k} \rangle}
            \end{pmatrix} \quad \text{for} \quad k\geq1,
    \end{align}
    that satisfies $(\widehat{\gamma}_{\langle 4^{k} \rangle})^2=1$.
    Here, $(\gamma_5)_{[4]}=i\gamma_{[4]}^0\gamma_{[4]}^1\gamma_{[4]}^2\gamma_{[4]}^3$ denotes the standard 4-dimensional $\gamma_5$-matrix (see Sec.~\ref{Sec:g5-in-D-dim} for more information on $\gamma_5$ in $D$ dimensions).
    Further, we assume some representation, for example the chiral representation
    \begin{align}
        \gamma_{[4]}^{\mu} = 
        \begin{pmatrix}
            0 & \sigma^{\mu}\\
            \overline{\sigma}^{\mu} & 0
        \end{pmatrix},
        \qquad
        (\gamma_5)_{[4]} = 
        \begin{pmatrix}
            -\mathbb{1}_{2\times2} & 0 \\
            0 & \mathbb{1}_{2\times2}
        \end{pmatrix},
    \end{align}
    with $(\sigma^{\mu}) = (\mathbb{1}_{2\times2},\sigma^k)$, $(\overline{\sigma}^{\mu}) = (\mathbb{1}_{2\times2},-\sigma^k)$, and $\sigma^k$ the Pauli matrices,
    such that
    \begin{align}\label{Eq:Hermiticity-Reality-Charge-Conjugation-in-4-dim}
        (\gamma_{[4]}^{\mu})^\dagger = \gamma^{0}_{[4]}\gamma_{[4]}^{\mu}\gamma^{0}_{[4]},
        \quad
        (\gamma_{[4]}^{\mu})^* = \gamma^{2}_{[4]}\gamma_{[4]}^{\mu}\gamma^{2}_{[4]},
        \quad
        (\gamma_{[4]}^{\mu})^\mathsf{T} = - C_{[4]}^{-1}\gamma_{[4]}^{\mu}C_{[4]},
        \quad 
        C_{[4]} = i \gamma_{[4]}^{0} \gamma_{[4]}^{2},
    \end{align}
    holds and all $\gamma$-matrices are real except for $\gamma_{[4]}^2$, which is imaginary.
\end{definition}
\begin{remark}[$\gamma^{\mu}$ in quasi $D$ dimensions]\
    \begin{itemize}
        \item The $\gamma$-matrices defined in Eq.~\eqref{Eq:Def-of-quasi-D-dim-gammas} indeed possess all the desired properties in quasi $D$ dimensions. In particular, they satisfy Eqs.~\eqref{Eq:D-dim-Clifford-Algebra} and \eqref{Eq:D-dim-Contraction-of-2-gammas}.
        \item With this construction, the first four $\gamma$-matrices, i.e.\ those with $\mu\in\{0,1,2,3\}$, are essentially the same as in 4 dimensions, representing direct extensions of the 4-dimensional Dirac matrices to $D$-dimensions.
        \item As in 4 dimensions, all $\gamma$-matrices are real except for $\gamma^2$, which is imaginary. 
        \item The $D$-dimensional $\gamma$-matrices obey the same relations for hermitian conjugation, complex conjugation and charge conjugation as in 4 dimensions (see Eq.~\eqref{Eq:Hermiticity-Reality-Charge-Conjugation-in-4-dim}), thereby enabling the definition of spinors and their adjoints and charge conjugates on $\mathbb{M}_D$.
    \end{itemize}
\end{remark}

Having established the explicit construction of the metric and $\gamma$-matrices in $D$ dimensions, we turn to a discussion of key properties and relations of these and other $D$-dimensional objects.

\paragraph{Properties of $D$-dimensional Covariants:}
With Def.~\ref{Def:metric_and_contractions_in_D-dim} for the $D$-dimensional metric in mind, we may write
\begin{align}
    \eta_{\mu\nu} = \eta^{\mu\nu} =
    \begin{cases}
        +1, \quad & \text{for} \, \mu=\nu=0\\
        -1, \quad & \text{for} \, \mu=\nu=1,2,\ldots\\
        0,  \quad & \text{for} \, \mu\neq\nu
    \end{cases}
\end{align}
satisfying
\begin{align}
    \eta^{\mu\nu}\eta_{\mu\nu} = D,
\end{align}
where, as discussed above, contraction is defined via integration.
The metric admits the decomposition
\begin{align}\label{Eq:Metric-split}
    \eta_{\mu\nu} = \overline{\eta}_{\mu\nu} + \widehat{\eta}_{\mu\nu},
\end{align}
as shown in Eq.~\eqref{Eq:Decomposition-of-D-dim-Spacetime}, where the 4-dimensional part $\overline{\eta}^{\mu\nu}$ and the $(D-4)$-dimensional part $\widehat{\eta}^{\mu\nu}$ satisfy
\begin{align}
    \overline{\eta}^{\mu\nu}\overline{\eta}_{\mu\nu} = 4, 
    \qquad\qquad
    \widehat{\eta}^{\mu\nu}\widehat{\eta}_{\mu\nu} = D-4.
\end{align}
The metric tensors act as projection operators on the respective spaces:
\begin{equation}
    \begin{aligned}
        \eta^{\mu\nu}\overline{\eta}_{\nu\rho} &= \overline{\eta}^{\mu\nu}\eta_{\nu\rho} = \overline{\eta}^{\mu\nu}\overline{\eta}_{\nu\rho} = \overline{\eta}^{\mu}_{\phantom{\mu}\rho},\\
        \eta^{\mu\nu}\widehat{\eta}_{\nu\rho} &= \widehat{\eta}^{\mu\nu}\eta_{\nu\rho} = \widehat{\eta}^{\mu\nu}\widehat{\eta}_{\nu\rho} = \widehat{\eta}^{\mu}_{\phantom{\mu}\rho},\\
        \overline{\eta}^{\mu\nu}\widehat{\eta}_{\nu\rho} &= \widehat{\eta}^{\mu\nu}\overline{\eta}_{\nu\rho} = 0.
    \end{aligned}
\end{equation}
Accordingly, any Lorentz vector $X^{\mu}$ can generally be decomposed as
\begin{align}\label{Eq:Split-of-generic-Lorentz-Covariant-X}
    X^{\mu} = \overline{X}^{\mu} + \widehat{X}^{\mu}, 
    \qquad 
    \overline{X}^{\mu} = \overline{\eta}^{\mu}_{\phantom{\mu}\nu} X^{\nu}, 
    \qquad 
    \widehat{X}^{\mu} = \widehat{\eta}^{\mu}_{\phantom{\mu}\nu} X^{\nu},
\end{align}
such that for scalar products we obtain
\begin{align}
    X^2 = \overline{X}^2 + \widehat{X}^2, 
    \qquad 
    X^{\mu}Y_{\mu} = \overline{X}^{\mu}\overline{Y}_{\mu} + \widehat{X}^{\mu}\widehat{Y}_{\mu},
    \qquad
    \overline{X}^{\mu} \widehat{Y}_{\mu} = 0.
\end{align}

By Def.~\ref{Def:Dirac_matrices_in_D-dim}, the $D$-dimensional $\gamma$-matrices on $\mathbb{M}_D$ are $\infty$-dimensional block matrices that satisfy
\begin{align}
    \{\gamma^{\mu},\gamma^{\nu}\} = 2 \eta^{\mu\nu} \mathbb{1},
    \qquad\qquad
    \gamma^{\mu}\gamma_{\mu} = D \mathbb{1}.
\end{align}
Like other $D$-dimensional objects, the $\gamma$-matrices may be decomposed into 4-dimensional and $(D-4)$-dimensional parts, $\overline{\gamma}^{\mu}$ and $\widehat{\gamma}^{\mu}$, respectively, as 
\begin{align} \label{Eq:gamma-split}
    \gamma^{\mu} = \overline{\gamma}^{\mu} + \widehat{\gamma}^{\mu},
\end{align}
satisfying
\begin{equation}\label{Eq:Gamma-Split-Algebra}
    \begin{aligned}
        \{\gamma^{\mu},\overline{\gamma}^{\nu}\} = \{\overline{\gamma}^{\mu},\overline{\gamma}^{\nu}\} &= 2 \overline{\eta}^{\mu\nu} \mathbb{1},
        &
        \qquad\qquad
        \gamma_{\mu} \overline{\gamma}^{\mu} = \overline{\gamma}_{\mu} \overline{\gamma}^{\mu} &= 4 \mathbb{1},\\
        \{\gamma^{\mu},\widehat{\gamma}^{\nu}\} = \{\widehat{\gamma}^{\mu},\widehat{\gamma}^{\nu}\} &= 2 \widehat{\eta}^{\mu\nu} \mathbb{1},
        &
        \qquad\qquad
        \gamma_{\mu} \widehat{\gamma}^{\mu} = \widehat{\gamma}_{\mu} \widehat{\gamma}^{\mu} &= (D-4) \mathbb{1},\\
        \{\overline{\gamma}^{\mu},\widehat{\gamma}^{\nu}\} &= 0,
        &
        \qquad\qquad
        \overline{\gamma}_{\mu} \widehat{\gamma}^{\mu} &= 0.
    \end{aligned}
\end{equation}
As discussed above, in quasi $D$ dimensions there exists a representation of the $\gamma$-matrices (see the construction in Def.~\ref{Def:Dirac_matrices_in_D-dim}) such that
\begin{align}
        (\gamma^{\mu})^\dagger = \gamma^{0}\gamma^{\mu}\gamma^{0},
        \qquad
        (\gamma^{\mu})^* = \gamma^{2}\gamma^{\mu}\gamma^{2},
        \qquad
        (\gamma^{\mu})^\mathsf{T} = - C^{-1}\gamma^{\mu}C,
        \qquad 
        C = i \gamma^{0} \gamma^{2},
\end{align}
with $C^{-1}=C^\dagger=C^\mathsf{T}$ and $C^{\mathsf{T}}=-C$, precisely as in 4 dimensions.
This allows for the definition of spinors $\psi$ on $\mathbb{M}_D$ with
\begin{align}
    \overline{\psi} = \psi^\dagger\gamma^0, \qquad\qquad \psi^C = C \, \overline{\psi}^\mathsf{T},
\end{align}
where further
\begin{align}
    (\psi^C)^C=\psi , \qquad\qquad (\overline{\psi})^C=-\psi^\mathsf{T}C^{-1}=\overline{\psi^C},
\end{align}
in quasi $D$ dimensions.
Consequently, 
\begin{align}
    \overline{\psi_1} \Gamma \psi_2 &= \overline{\psi_2^C} \Gamma^C \psi_1^C,
    &\text{with}& 
    &\Gamma^C &= - C \Gamma^\mathsf{T} C,\\
    \big( \overline{\psi_1} \Gamma \psi_2 \big)^{\dagger} &= \overline{\psi_2} \, \overline{\Gamma} \, \psi_1,
    &\text{with}& 
    &\overline{\Gamma} &= \gamma^0 \Gamma^\dagger \gamma^0,
\end{align}
for some anticommuting spinors $\psi_1$, $\psi_2$, and 
\begin{align}
    \{ \mathbb{1}, \gamma_5, \gamma^{\mu}, \gamma^{\mu} \gamma_5 \}^C 
    &= \{\mathbb{1}, \gamma_5, - \gamma^{\mu}, -\gamma_5 \gamma^{\mu}\},\\
    \overline{\{ \mathbb{1}, \gamma_5, \gamma^{\mu}, \gamma^{\mu} \gamma_5 \}} 
    &= \{\mathbb{1}, - \gamma_5, \gamma^{\mu}, -\gamma_5 \gamma^{\mu}\},
\end{align}
hold in quasi $D$ dimensions.

Furthermore, Ref.~\cite{Collins_1984} defines the trace of a matrix $A$ in quasi $D$ dimensions (requiring linearity, cyclicity and commutation with contraction by $\eta_{\mu\nu}$) as
\begin{align}
    \myTr(A) \equiv \myTr(\mathbb{1}) \mathop{\text{lim}}_{N \, \to \, \infty} \, \frac{1}{N} \sum_{k=1}^N A_{kk},
\end{align}
using that, on $\mathbb{M}_D$, the $\gamma$-matrices are $\infty$-dimensional block matrices with finite dimensional blocks.
Here, we have $\myTr(\mathbb{1})=f(D)$, with $f(4)=4$ to reproduce the 4-dimensional case.
Since the difference $f(D)-f(4)$ only corresponds to a renormalisation group transformation, we set $f(D)=f(4)\,\,\forall\,D$, such that
\begin{equation}\label{Eq:GammaTraceDefinitions}
    \begin{aligned}
        \mathrm{Tr}(\mathbb{1}) = 4,
        \qquad\qquad
        \mathrm{Tr}(\gamma^{\mu}) = 0.
    \end{aligned}
\end{equation}
This fully determines all traces that contain any number $\gamma^\mu$, as such traces can always be reduced to expressions of the form ``$\myTr(\mathbb{1})\times\eta\text{'s}$''.

To conclude this subsection, we comment on evanescent objects, which have already appeared above and will play an important role throughout this thesis:
\begin{definition}[Evanescent Objects]\label{Def:Evanescent-Objects}\ \\
    Objects $\widehat{X}$ that vanish in $D=4$ dimensions are called \emph{evanescent}.
\end{definition}
\begin{remark}[Evanescent Objects]\
    \begin{itemize}
        \item Specifically, these are objects $\widehat{X}\in\mathbb{M}_{D-4}\subset\mathbb{M}_D$, such as $\widehat{\eta}_{\mu\nu}$, $\widehat{\gamma}^{\mu}$, and $\widehat{p}^{\mu}$.
        \item While purely evanescent objects vanish in $D=4$, those accompanied by $\epsilon$-poles, i.e.\ terms of the form $\widehat{X}/\epsilon^k$, with $k\in\mathbb{N}$, do not vanish and may appear as counterterms in $\mathcal{L}_{\mathrm{ct}}$. Hence, we distinguish:
        \begin{itemize}
            \item[(a)] \emph{Finite evanescent terms:} Terms built from evanescent objects whose coefficients are free of $\epsilon$-poles, as well as any finite term with coefficient $\sim\epsilon^k$ ($k\in\mathbb{N}$) as in Ref.~\cite{Weisswange:2022jqx}.
            \item[(b)] \emph{Divergent evanescent terms:} Terms built from evanescent objects whose coefficients contain poles in $\epsilon$, i.e.\ $\widehat{X}/\epsilon^k$ as described above.
        \end{itemize}
    \end{itemize}
\end{remark}


\section{The $\gamma_5$-Problem}\label{Sec:The-g5-Problem}

A central challenge in the renormalisation of chiral gauge theories is the so-called $\gamma_5$-problem. Setting up a consistent regularisation and renormalisation of such theories requires a coherent treatment of $\gamma_5$, which has proven to be difficult and has attracted considerable attention over the past decades.
In particular, within dimensional regularisation schemes, the $\gamma_5$-problem arises from the inherently 4-dimensional nature of $\gamma_5$ and the difficulty of embedding it consistently into the $D$-dimensional framework. 
A mathematically rigorous approach to this problem is provided by the Breitenlohner-Maison/`t~Hooft-Veltman (BMHV) scheme, which treats $\gamma_5$ as a manifestly 4-dimensional object and abandons its anticommutation properties with fully $D$-dimensional $\gamma^{\mu}$-matrices.
The BMHV scheme is the only known framework that is fully self-consistent at all orders in perturbation theory. 

In this section, we first introduce $\gamma_5$ within the framework of DReg and explicitly demonstrate the associated complications.
We then present the BMHV scheme, discussing its treatment of $\gamma_5$ and the modified algebraic relations that arise.
Finally, we briefly review alternative approaches for the treatment of $\gamma_5$ and their limitations, and comment on the persistence of the $\gamma_5$-problem even in non-dimensional regularisation schemes.

\subsection{$\gamma_5$ in $D$ Dimensions}\label{Sec:g5-in-D-dim}

In 4 dimensions, $\gamma_5$ is simply defined as the product of all four $\gamma$-matrices,\footnote{Neither $\gamma_5$ nor $\varepsilon^{\mu\nu\rho\sigma}$ are equipped with overbars, as they are manifestly 4-dimensional objects and will be treated as such throughout this thesis.}
\begin{align}
    \gamma_5 \coloneqq i \overline{\gamma}^0 \overline{\gamma}^1 \overline{\gamma}^2 \overline{\gamma}^3 = - \frac{i}{4!} \varepsilon_{\mu\nu\rho\sigma}\overline{\gamma}^{\mu}\overline{\gamma}^{\nu}\overline{\gamma}^{\rho}\overline{\gamma}^{\sigma},
\end{align}
where $\varepsilon^{\mu\nu\rho\sigma}$ is the totally antisymmetric Levi-Civita tensor (density) with sign convention
\begin{align}
    \varepsilon^{0123} = - \varepsilon_{0123} = + 1.
\end{align}
There are three essential relations that hold for $\gamma_5$ and traces of $\gamma$-matrices in 4 dimensions:
\begin{subequations}\label{Eq:The_3_Properties_of_g5_and_Tr}
    \begin{align}
        \{\gamma_5,\overline{\gamma}^{\mu}\} &= 0, \label{Eq:Anticommutation_of_g5_in_4D}\\
        \myTr(\gamma_5\overline{\gamma}^{\mu}\overline{\gamma}^{\nu}\overline{\gamma}^{\rho}\overline{\gamma}^{\sigma}) &= - 4 i \varepsilon^{\mu\nu\rho\sigma}, \label{Eq:Trace_of_g5_and_4_gammas_in_4D}\\
        \myTr(\Gamma_1\Gamma_2) &= \myTr(\Gamma_2\Gamma_1), \label{Eq:Cyclicity_of_gamma-Traces}
    \end{align}
\end{subequations}
where the last equation represents cyclicity of $\gamma$-traces.
In $D\neq4$ dimensions, however, these three relations cannot simultaneously be retained consistently, leading to the central difficulty:
\begin{proposition}[The $\gamma_5$-Problem]\ \\
    The set of relations 
    \begin{enumerate}[label=({\roman*})]
        \item anticommutation, Eq.~\eqref{Eq:Anticommutation_of_g5_in_4D},
        \item the trace identity, Eq.~\eqref{Eq:Trace_of_g5_and_4_gammas_in_4D},
        \item cyclicity of $\gamma$-traces, Eq.~\eqref{Eq:Cyclicity_of_gamma-Traces},
    \end{enumerate}
    are mutually incompatible in $D\neq4$ dimensions. 
    That is, requiring all three relations to hold simultaneously in $D$ dimensions necessarily leads to a contradiction.
\end{proposition}
\begin{proof}
Let $t^{\mu_1\mu_2\mu_3\mu_4}=\myTr(\gamma^{\mu_1}\gamma^{\mu_2}\gamma^{\mu_3}\gamma^{\mu_4}\gamma_5)$ denote the trace over four $D$-dimensional $\gamma$-matrices and $\gamma_5$.
We then find:
\begin{align}
    D t^{\mu_1\mu_2\mu_3\mu_4} &= \myTr(\gamma_{\alpha}\gamma^{\alpha}\gamma^{\mu_1}\gamma^{\mu_2}\gamma^{\mu_3}\gamma^{\mu_4}\gamma_5)\nonumber\\
    &= \myTr((2\gamma_\alpha\eta^{\alpha\mu_1}-\gamma_\alpha\gamma^{\mu_1}\gamma^{\alpha})\gamma^{\mu_2}\gamma^{\mu_3}\gamma^{\mu_4}\gamma_5)
    \nonumber\\
    &= \ldots
    \nonumber\\
    &= 8 t^{\mu_1\mu_2\mu_3\mu_4} + \myTr(\gamma_{\alpha}\gamma^{\mu_1}\gamma^{\mu_2}\gamma^{\mu_3}\gamma^{\mu_4}\gamma^{\alpha}\gamma_5)
    \nonumber\\
    &= (8-D)t^{\mu_1\mu_2\mu_3\mu_4},
\end{align}
where we have used $D\mathbb{1}=\gamma_{\alpha}\gamma^{\alpha}$, and repeatedly applied the $D$-dimensional Clifford algebra, see Eq.~\eqref{Eq:D-dim-Clifford-Algebra}, to anticommute $\gamma^{\alpha}$ through the $\gamma$-string $\gamma^{\mu_1}\gamma^{\mu_2}\gamma^{\mu_3}\gamma^{\mu_4}$.
In the final step, we have made use of the anticommutation property of $\gamma_5$ in $D$-dimensions, cf.\ Eq.~\eqref{Eq:Anticommutation_of_g5_in_4D}, together with cyclicity of $\gamma$-traces, see Eq.~\eqref{Eq:Cyclicity_of_gamma-Traces}.
Ultimately, this results in
\begin{align}\label{Eq:Inconsistency-due-to-the-gamma5-Problem}
    (4-D) t^{\mu_1\mu_2\mu_3\mu_4} = 0,
\end{align}
which can only be satisfied in $D=4$ dimensions, or if the trace itself vanishes for $D\neq4$.
Thus, imposing Eq.~\eqref{Eq:Anticommutation_of_g5_in_4D} and Eq.~\eqref{Eq:Cyclicity_of_gamma-Traces} together with Eq.~\eqref{Eq:Trace_of_g5_and_4_gammas_in_4D} leads to a contradiction in $D$ dimensions.
In other words, it is impossible for all three relations in Eqs.~\eqref{Eq:The_3_Properties_of_g5_and_Tr} to hold simultaneously in $D\neq4$ dimensions (without inconsistencies).
\end{proof}

The origin of this problem is that $\gamma_5$ is strictly tied to 4 dimensions (chirality in the 4-dimensional Lorentz group, cf.\ Sec.~\ref{Sec:Chiral_Gauge_Theories}) and the $\varepsilon$-tensor relies on ``index counting''.
Index counting refers to expressing tensors or products of $\gamma$-matrices in terms of a finite set of basis elements.
In 4 dimensions, for example, products of $\gamma$'s such as $\overline{\gamma}^{\mu}\overline{\gamma}^{\nu}\overline{\gamma}^{\rho}\overline{\gamma}^{\sigma}\ldots$, can be rewritten as a linear combination of 16 basis matrices (cf.\ Fierz identities).
Likewise, for any object carrying five or more different Lorentz indices in 4 dimensions, we can deduce that at least two of these indices must be equal.
In quasi $D$ dimensions, however, this is not possible anymore: $\mathbb{M}_D$ is an $\infty$-dimensional vector space, and thus no finite basis exists.\footnote{It is possible to construct more linearly independent products of Lorentz covariants by introducing further indices, as there is no maximal number of linearly independent elements.}
As a result, $\gamma_5$ and the $\varepsilon$-symbol can only uniquely be well defined in integer dimensions and lose their unambiguous meaning in the $\infty$-dimensional case.

Hence, along these lines, the Levi-Civita tensor $\varepsilon^{\mu\nu\rho\sigma}$ is also defined in strictly 4 dimensions, with
\begin{align}\label{Eq:Manifestly-4-dim-epsilon-tensor-id}
    \varepsilon^{\mu_1\mu_2\mu_3\mu_4}\varepsilon_{\nu_1\nu_2\nu_3\nu_4} = - \sum_{\pi\in S_4} \text{sgn}(\pi) \prod_{i=1}^{4} \overline{\eta}^{\mu_i}_{\phantom{\mu_i}\nu_{\pi(i)}},
\end{align}
holding only with purely 4-dimensional metric tensors $\overline{\eta}_{\mu\nu}$.
Further properties include
\begin{align}
    \varepsilon^{\mu_1\mu_2\mu_3\mu_4} &= \text{sgn}(\pi) \,  \varepsilon^{\mu_{\pi(1)}\mu_{\pi(2)}\mu_{\pi(3)}\mu_{\pi(4)}},\\
    0 &= \sum_{\pi\in S_5} \text{sgn}(\pi) \, \varepsilon^{\mu_{\pi(1)}\mu_{\pi(2)}\mu_{\pi(3)}\mu_{\pi(4)}} \, \overline{\eta}_{\mu_{\pi(5)}\nu},
\end{align}
where $S_n$ is the permutation group of $n$ elements and $\pi$ is a permutation w.r.t.\ it.

Because a consistent scheme must admit a smooth limit to 4 dimensions, it is not possible to simply set the trace $t^{\mu_1\mu_2\mu_3\mu_4}$ in Eq.~\eqref{Eq:Inconsistency-due-to-the-gamma5-Problem} to zero.
Such a choice would obstruct a continuous limit $D\to4$, as it would discard physical contributions from certain Feynman diagrams involving these traces --- even terms required for unitarity and causality.
To obtain a consistent regularisation scheme, we must therefore abandon the simultaneous validity of all identities in Eqs.~\eqref{Eq:The_3_Properties_of_g5_and_Tr} in $D\neq4$ dimensions, without setting the trace $t^{\mu_1\mu_2\mu_3\mu_4}$ itself to zero.

This issue is of central importance, since the Standard Model and possible extensions thereof are chiral gauge theories and thus directly affected by the $\gamma_5$-problem.
Accordingly, there are many proposals in the literature for handling $\gamma_5$, see Refs.~\cite{Belusca-Maito:2023wah,Jegerlehner:2000dz} for reviews.
In principle, there are the following approaches to resolve the $\gamma_5$-problem:
\begin{solution}[of the $\gamma_5$-Problem]\
    \begin{enumerate}[label=(\arabic*)]
        \item The Breitenlohner-Maison/`t~Hooft-Veltman scheme (Refs.~\cite{tHooft:1972tcz,Breitenlohner:1977hr}):
        $\gamma_5$ is defined strictly 4-dimensionally and its anticommutation property with fully $D$-dimensional $\gamma^\mu$-matrices is abandoned in $D\neq4$ (cf.\ Eq.~\eqref{Eq:Anticommutation_of_g5_in_4D}).
        The first 4 dimensions are thus singled out and treated differently from the remaining $(D-4)$, resulting in modified algebraic relations.
        This scheme is the only one known to be mathematically well-defined and completely self-consistent at all orders.
        However, it breaks gauge and BRST invariance at the regularised level, due to the modified algebra, requiring symmetry restoration. 
        A discussion of this scheme is provided in Sec.~\ref{Sec:The-BMHV-Scheme} below, and it is used throughout this thesis.
        \item Kreimer's scheme (Refs.~\cite{Kreimer:1989ke,Korner:1991sx,Kreimer:1993bh}): An alternative proposal which is based on giving up cyclicity of $\gamma$-traces, cf.\ Eq.~\eqref{Eq:Cyclicity_of_gamma-Traces}. 
        In particular, the trace is redefined as a linear functional that loses cyclicity in $D\neq4$ and coincides with the standard Dirac trace only in $D=4$ dimensions.
        Its value depends on the so-called ``reading point'', i.e.\ where one starts reading the trace. 
        By keeping full anticommutativity, the scheme promises to preserve gauge and BRST symmetry at all orders. 
        However, this scheme is not completely self-consistent and its properties at the multi-loop level are not fully under control. 
        The literature shows that there are ambiguities depending on the choice of the specific reading point. 
        Additionally, its symmetry-conserving properties are in question as well. 
        Applications and drawbacks are discussed in Sec.~\ref{Sec:Alternative-g5-Schemes} below.
        \item Ad hoc prescriptions and using ``tricks'' to circumvent or resolve the problem in a given application at a given order, e.g.\ Larin's prescription (see Ref.~\cite{Larin:1993tq} and Sec.~\ref{Sec:Alternative-g5-Schemes}), and/or ``ignoring the problem'' as far as possible at lower orders. Clearly, such approaches do not amount to a consistent regularisation scheme and cannot be used universally.
    \end{enumerate}
\end{solution}

\paragraph{Local differences and consequences of an inconsistent treatment of $\gamma_5$:}
As discussed above, the BMHV scheme is the only framework that has been proven to be self-consistent to all orders in perturbation theory.
Nevertheless, as long as the difference between an alternative prescription and the BMHV scheme amounts to \emph{local hermitian counterterms}, both renormalisation procedures yield physically equivalent results consistent with unitarity and causality, since such a difference corresponds merely to a local reparametrisation in the sense of theorem~\ref{Thm:Main_Theorem_of_Renormalisation}.
In particular, a scheme employing a naively anticommuting $\gamma_5$, where the Ward and Slavnov-Taylor identities are manifestly preserved, would lead to the same physical predictions as the self-consistent BMHV scheme \emph{if and only if} the deviation between them is local.

At the 1-loop level, this condition is typically satisfied: the discrepancy between a naive $\gamma_5$ treatment and the BMHV prescription reduces to local counterterms.
Consequently, the naive scheme remains compatible with unitarity and causality, preserves gauge invariance through manifestly valid Ward and Slavnov–Taylor identities, and produces correct physical results.
The only remaining issue concerns the gauge anomaly, which is non-local, and a fully naive treatment would incorrectly suggest that all currents are conserved.
However, the absence of gauge anomalies is guaranteed by an appropriate choice of matter content (see Sec.~\ref{Sec:Anomalies}).

Beyond 1-loop ($L\geq2$), the situation becomes more intricate due to the appearance of subdivergences.
In such cases, the difference between a naive $\gamma_5$ scheme and the consistent BMHV prescription can encounter a non-local subdivergence.
When this occurs, the deviation is no longer local, and no counterterm can relate the two schemes.
As a result, the naive prescription becomes inconsistent and fails to reproduce the correct physical predictions.
In certain special cases, the difference has been shown to remain local even at higher orders --- for example, in the two-loop studies of Refs.~\cite{Heinemeyer:2004yq,Stockinger:2004vf}.
However, this is an exceptional situation rather than the general rule.
In most cases, the difference becomes non-local at higher orders, and naive $\gamma_5$ schemes therefore yield incorrect results.

Inconsistencies arising from an improper or naive treatment of $\gamma_5$ may manifest in several ways, including:
\begin{itemize}
    \item differing final results depending on the organisation of the calculation;
    \item incorrect predictions for physical observables;
    \item improper cancellation of IR divergences between virtual and real corrections due to mismatched $\gamma_5$ contributions;
    \item residual non-local divergences persisting even after subrenormalisation;
    \item violation of unitarity, leading to a breakdown of the optical theorem.
\end{itemize}
To date, the BMHV scheme remains the only $\gamma_5$-prescription that has been rigorously proven to be consistent to all orders in perturbation theory, and will be used throughout this thesis.

\subsection{The Breitenlohner-Maison/`t~Hooft-Veltman Scheme}\label{Sec:The-BMHV-Scheme}

The Breitenlohner-Maison/`t~Hooft-Veltman (BMHV) scheme (Refs.~\cite{tHooft:1972tcz,Breitenlohner:1977hr}) is defined by keeping $\gamma_5$ manifestly 4-dimensional even in $D\neq4$ dimensions, thereby abandoning full anticommutativity (cf.\ Eq.~\eqref{Eq:Anticommutation_of_g5_in_4D}) with $\gamma^\mu$. 
In detail, it is specified as follows:
\begin{definition}[BMHV Scheme]\label{Def:The_BMHV_Scheme}\ \\
    In $D\neq4$ dimensions, the spacetime $(\mathbb{M}_{D},\eta_{[D]})$ admits the decomposition $\mathbb{M}_{D} = \mathbb{M}_{4} \oplus \mathbb{M}_{D-4}$ as introduced in Def.~\ref{Def:DReg}, together with the corresponding decompositions of the metric $\eta^{\mu\nu}$ (Eq.~\eqref{Eq:Metric-split}), the $\gamma^\mu$-matrices (Eq.~\eqref{Eq:gamma-split}), and other Lorentz covariants (see Sec.~\ref{Sec:Elements_of_D-dim_Spacetime}). 
    The matrix $\gamma_5$ is kept strictly 4-dimensional and defined as
    \begin{align}\label{Eq:BMHV-Definition-of-g5}
        \gamma_5 \coloneqq i \overline{\gamma}^0 \overline{\gamma}^1 \overline{\gamma}^2 \overline{\gamma}^3 = - \frac{i}{4!} \varepsilon_{\mu\nu\rho\sigma}\overline{\gamma}^{\mu}\overline{\gamma}^{\nu}\overline{\gamma}^{\rho}\overline{\gamma}^{\sigma},
    \end{align}
    embedded in the 4-dimensional subspace $\mathbb{M}_{4}\subset\mathbb{M}_{D}$.
    Likewise, the Levi-Civita symbol remains manifestly 4-dimensional, $\varepsilon_{\mu\nu\rho\sigma}\equiv\overline{\varepsilon}_{\mu\nu\rho\sigma}$.
    Consequently, the (anti-)commutation relations of $\gamma_5$ with the $\gamma^\mu$-matrices are modified in $D$ dimensions, reflecting its inherently 4-dimensional nature and giving rise to the \emph{BMHV algebra}:
    \begin{equation}\label{Eq:BMHV-Algebra}
    \begin{aligned}
        \{\gamma_5,\overline{\gamma}^{\mu}\} &= 0,
        &
        \qquad
        [\gamma_5,\overline{\gamma}^{\mu}] &= 2\gamma_5\overline{\gamma}^{\mu},\\
        \{\gamma_5,\widehat{\gamma}^{\mu}\} &= 2\gamma_5\widehat{\gamma}^{\mu}
        &
        \qquad
        [\gamma_5,\widehat{\gamma}^{\mu}] &= 0,\\
        \{\gamma_5,\gamma^{\mu}\} &= 2 \gamma_5\widehat{\gamma}^{\mu},
        &
        \qquad
        [\gamma_5,\gamma^{\mu}] &= 2\gamma_5\overline{\gamma}^{\mu},
    \end{aligned}
    \end{equation}
    where the last line follows from the first two lines.
\end{definition}

From this definition it is evident that, in $D$-dimensions, a manifestly 4-dimensional $\gamma_5$ no longer satisfies the same anticommutation relations with the $D$-dimensional $\gamma^\mu$-matrices as in 4 dimensions, since it only ``lives'' in the 4-dimensional subspace $\mathbb{M}_{4}\subset\mathbb{M}_{D}$.
In this sense, the BMHV scheme singles out the first four dimensions and treats them differently from the remaining $(D-4)$.
As a result, $\gamma_5$ anticommutes solely with the 4-dimensional components $\overline{\gamma}^{\mu}$ but commutes with the $(D-4)$-dimensional components $\widehat{\gamma}^{\mu}$, as expressed in Eq.~\eqref{Eq:BMHV-Algebra}.
This modification vanishes smoothly in the limit $D\to4$.

The modified algebra necessitates a consistent treatment of the decomposition into 4- and $(D-4)$-dimensional components (see Eqs.~\eqref{Eq:Decomposition-of-D-dim-Spacetime}, \eqref{Eq:Metric-split} and \eqref{Eq:gamma-split}).
In ``naive'' DReg computations, e.g.\ in vector-like gauge theories without $\gamma_5$, one typically does not need to distinguish between these subspaces: 
all objects are extended formally to $D$ dimensions, the calculation is carried out there, and the limit $D\to4$ is taken at the end.
In contrast, the BMHV scheme requires explicit separation into 4- an $(D-4)$-dimensional parts, as both may generally appear in intermediate steps due to the modified algebra in Eq.~\eqref{Eq:BMHV-Algebra}.
Thus, keeping $\gamma_5$ manifestly 4-dimensional --- consistent with its role in defining chirality in the 4-dimensional Lorentz group (cf.\ sections~\ref{Sec:Chiral_Gauge_Theories} and \ref{Sec:g5-in-D-dim}) --- together with the corresponding algebraic modifications and the explicit decomposition of the quasi $D$-dimensional vector space, constitutes the essence of the BMHV scheme.

Unlike in some alternative prescriptions, cyclicity of the trace is preserved in $D$ dimensions.
Furthermore, the 4-dimensional identity
\begin{equation}
\mathrm{Tr}(\gamma_5\overline{\gamma}^{\mu}\overline{\gamma}^{\nu}\overline{\gamma}^{\rho}\overline{\gamma}^{\sigma}) = -4i\varepsilon^{\mu\nu\rho\sigma}    
\end{equation}
remains valid for the $4$-dimensional components of the $\gamma$-matrices, which is essential for a smooth limit to 4 dimensions and even for maintaining unitarity and causality (see Sec.~\ref{Sec:g5-in-D-dim}).

The BMHV scheme is mathematically well-defined and self-consistent: final results do not depend on the organisation of the computation (no ambiguities), and the framework is compatible with unitarity and causality.
It has been proven consistent to all orders, see Refs.~\cite{Breitenlohner:1977hr}.
However, due to the modified algebraic relations, gauge and BRST invariance are violated at intermediate steps of the calculation, i.e.\ the regularisation does not preserve gauge invariance in chiral theories.
This constitutes the main drawback of the scheme.
The source of the breaking lies in the BMHV algebra, which leads to non-vanishing evanescent components of the $D$-dimensional fermion kinetic term that necessarily mix chiralities.
A comprehensive analysis of this symmetry breaking is presented in Sec.~\ref{Sec:Regularisation-Induced_Symmetry_Breaking}.
The resulting violation of the Slavnov-Taylor identity must be compensated by appropriate symmetry-restoring counterterms (cf.\ Sec.~\ref{Sec:Renormalisation-of-GaugeTheories}).
Since the breaking is a regularisation-induced artefact rather than a genuine anomaly, it can always be removed in anomaly-free theories, making full symmetry restoration possible (see Sec.~\ref{Sec:Algebraic_Renormalisation}).
Nevertheless, this considerably complicates the renormalisation procedure, as discussed in chapter~\ref{Chap:Practical_Symmetry_Restoration}, in particular Sec.~\ref{Sec:BMHV-Specific_Challenges}.

\subsection{Alternative $\gamma_5$-Schemes in $D$ Dimensions}\label{Sec:Alternative-g5-Schemes}

We briefly comment on alternative prescriptions for the treatment of $\gamma_5$ in DReg --- in particular Larin's prescription and Kreimer's scheme --- that aim to preserve BRST invariance by allowing a fully anticommuting $\gamma_5$ and thereby simplify the renormalisation of chiral gauge theories.
However, their scope of application at the multi-loop level is limited and not entirely under control.
Typically, such prescriptions are restricted to specific perturbative orders or particular classes of diagrams and often require additional external arguments to resolve ambiguities.
In practical applications, the genuine integer-dimensional identity in Eq.~\eqref{Eq:Manifestly-4-dim-epsilon-tensor-id} is often, but incorrectly, elevated to quasi $D$ dimensions, leading to internal inconsistencies.
In what follows, we first demonstrate that such an improper extension of Eq.~\eqref{Eq:Manifestly-4-dim-epsilon-tensor-id} to quasi $D$ dimensions --- which is often done in alternative $\gamma_5$ prescriptions --- indeed leads to inconsistencies, and subsequently review Larin's prescription and Kreimer's scheme.
We then comment on practical applications and inherent limitations.

\paragraph{Genuine Integer-Dimensional Identities:}
The identity in Eq.~\eqref{Eq:Manifestly-4-dim-epsilon-tensor-id} is a genuine integer-dimensional identity (see Ref.~\cite{Bruque:2018bmy}), meaning it is valid only in a vector space of finite, integer dimension --- specifically, in 4 dimensions in the present case.
As discussed in Sec.~\ref{Sec:Spacetime_and_Integrals_in_DReg}, the quasi $D$-dimensional vector space $\mathbb{M}_D$ of DReg is actually $\infty$-dimensional, so Eq.~\eqref{Eq:Manifestly-4-dim-epsilon-tensor-id} does not hold in $D$ dimensions but only within the 4-dimensional subspace $\mathbb{M}_4$, with strictly 4-dimensional metric tensors $\overline{\eta}^{\mu\nu}$.
Promoting this identity to quasi $D$ dimensions by replacing the 4-dimensional metrics on the RHS of Eq.~\eqref{Eq:Manifestly-4-dim-epsilon-tensor-id} with $D$-dimensional ones necessarily leads to inconsistencies.
The resulting discrepancy is an evanescent ambiguity, i.e.\ of order $\mathcal{O}(\epsilon)$.
To illustrate this, we consider the RHS of Eq.~\eqref{Eq:Manifestly-4-dim-epsilon-tensor-id} with the 4-dimensional metric tensors replaced by $D$-dimensional ones,
\begin{align}
    E^{\mu_1\mu_2\mu_3\mu_4}_{\nu_1\nu_2\nu_3\nu_4} \coloneqq - \sum_{\pi\in S_4} \mathrm{sgn}(\pi) \prod_{i=1}^{4} \eta^{\mu_i}_{\phantom{\mu_i}\nu_{\pi(i)}}.
\end{align}
Now we consider the fully contracted product of four $\varepsilon$-symbols,
\begin{align}
    \varepsilon^{\mu\nu\rho\sigma} \varepsilon_{\alpha\beta\lambda\kappa} \varepsilon_{\mu\nu\rho\sigma} \varepsilon^{\alpha\beta\lambda\kappa}.
\end{align}
In terms of $E^{\mu_1\mu_2\mu_3\mu_4}_{\nu_1\nu_2\nu_3\nu_4}$ this can be evaluated in two ways
\begin{align}
    r_1 \coloneqq E^{\mu\nu\rho\sigma}_{\alpha\beta\lambda\kappa} E_{\mu\nu\rho\sigma}^{\alpha\beta\lambda\kappa}\,, 
    \qquad 
    r_2 \coloneqq E^{\mu\nu\rho\sigma}_{\mu\nu\rho\sigma} E^{\alpha\beta\lambda\kappa}_{\alpha\beta\lambda\kappa} \,.
\end{align}
In 4 dimensions both expressions yield the same result,
\begin{align}
    \overline{r_1} = 576 = \overline{r_2},
\end{align}
whereas in $D\neq4$ dimensions they differ:
\begin{align}
    r_1 = 24 D (D-1) (D-2) (D-3), \qquad r_2 = \big[D(D-1)(D-2)(D-3)\big]^2.
\end{align}
For $D=4-2\epsilon$, the difference becomes
\begin{align}\label{Eq:Ambiguity-or-order-epsilon-LCTensor-Contraction-in-D-dim}
    \Delta r \coloneqq r_1 - r_2 = 16 \epsilon \big( 150 + \mathcal{O}(\epsilon) \big) = \mathcal{O}(\epsilon),
\end{align}
depending on how exactly the contraction is arranged.
As expected, the resulting ambiguity is of order $\mathcal{O}(\epsilon)$, confirming that Feynman diagrams involving such contractions are ambiguous and yield results that depend on the specific organisation of the calculation.
Consequently, Eq.~\eqref{Eq:Manifestly-4-dim-epsilon-tensor-id} is mathematically consistent only in $4$ dimensions.

\paragraph{Larin's Prescription:}
Larin's method is not a self-contained scheme but rather a calculational prescription developed for specific applications, designed to ensure that the axial anomaly retains its 1-loop character in DReg (see Ref.~\cite{Larin:1993tq}).
The prescription employs a symmetric extension of the axial current to $D$ dimensions,
\begin{align}\label{Eq:Symmetric-Extension-In-Larins-Prescription}
    [J_{5}^{\mu,a}]_{[4]} = \overline{\psi} \overline{\gamma}^{\mu} \gamma_5 T^a \psi \longrightarrow [J_{5}^{\mu,a}]_{[D]} = \overline{\psi} \frac{1}{2} \big( \gamma^{\mu} \gamma_5 - \gamma_5 \gamma^{\mu} \big) T^a \psi = - \frac{i}{3!} \varepsilon^{\mu\nu\rho\sigma} \overline{\psi} \gamma_{\nu}\gamma_{\rho}\gamma_{\sigma} T^a \psi,
\end{align}
where $( \gamma^{\mu} \gamma_5 - \gamma_5 \gamma^{\mu} )/2 = - i \varepsilon^{\mu\nu\rho\sigma} \gamma_{\nu}\gamma_{\rho}\gamma_{\sigma}/(3!)$.
This leads to the practical replacement rule
\begin{align}\label{Eq:Larin's-Prescription}
        \gamma^{\mu}\gamma_5 \longrightarrow - \frac{i}{3!} \varepsilon^{\mu\nu\rho\sigma} \gamma_{\nu}\gamma_{\rho}\gamma_{\sigma},
\end{align}
which constitutes the core of Larin's prescription.
Up to this point, the approach is neither inconsistent with nor contradictory to the BMHV scheme; in fact, it can be interpreted as the symmetric $D$-dimensional extension of $\overline{\gamma}^\mu\gamma_5$ to $( \gamma^{\mu} \gamma_5 - \gamma_5 \gamma^{\mu} )/2$ (cf.\ Sec.~\ref{Sec:Regularisation-Induced_Symmetry_Breaking}).
The difficulties arise in two respects.
First, in the presence of more than one $\gamma_5$, Larin's prescription effectively assumes naive anticommutation, allowing $\gamma_5$ matrices to be rearranged and combined using $\gamma_5^2 = \mathbb{1}$.
Second, the prescription relies on elevating the genuine 4-dimensional identity in Eq.~\eqref{Eq:Manifestly-4-dim-epsilon-tensor-id} to quasi $D$ dimensions.
As shown above, such an extension inevitably introduces ambiguities of order $\mathcal{O}(\epsilon)$ (see Eq.~\eqref{Eq:Ambiguity-or-order-epsilon-LCTensor-Contraction-in-D-dim}).
Consequently, the prescription is only valid in applications where these evanescent terms do not affect renormalisation constants, such as the cases considered in Ref.~\cite{Larin:1993tq}.
Notably, Ref.~\cite{Larin:1993tq} also required an additional finite renormalisation constant to restore the Ward identities, which is fixed by demanding that the renormalised axial vertex coincide with the renormalised vector vertex multiplied by $\gamma_5$.

\paragraph{Kreimer's Scheme:}
Kreimer's scheme, introduced in Refs.~\cite{Kreimer:1989ke,Korner:1991sx,Kreimer:1993bh}, was proposed as a self-contained framework for the treatment of $\gamma_5$, in which the cyclicity of Dirac traces is abandoned in $D$ dimensions in order to maintain a fully anticommuting $\gamma_5$.
The trace over Dirac matrices is thereby defined as a linear functional that loses cyclicity for $D\neq4$, and coincides with the standard Dirac trace only in $D=4$ dimensions.
Losing cyclicity implies that the value of a trace depends on the position where the reading of the trace starts, referred to as ``reading point''. 
For this reason, the method is often called the reading-point prescription.
As an example, the symmetric extension of $\overline{\gamma}^\mu\gamma_5$ to $( \gamma^{\mu} \gamma_5 - \gamma_5 \gamma^{\mu} )/2$ (cf.\ Eq.~\eqref{Eq:Symmetric-Extension-In-Larins-Prescription}) in Larin's prescription can be interpreted as an average of two adjacent reading-points (see Ref.~\cite{Bednyakov:2015ooa})
\begin{align}
    \mathrm{Tr}(\ldots \gamma^{\mu}\gamma_5) \longrightarrow \frac{1}{2} \Big[ \mathrm{Tr}(\ldots \gamma^{\mu}\gamma_5) - \mathrm{Tr}(\gamma^{\mu} \ldots \gamma_5) \Big].
\end{align}
By construction, this approach retains full anticommutativity of $\gamma_5$ and thereby promises to preserve BRST invariance manifestly, suggesting a potential practical advantage over the BMHV scheme.
However, despite this appealing feature, Kreimer's prescription is not fully self-consistent and suffers from intrinsic ambiguities, particularly at higher orders, as demonstrated in Refs.~\cite{Zoller:2015tha,Bednyakov:2015ooa,Poole:2019txl,Davies:2019onf,Davies:2021mnc,Herren:2021vdk,Chen:2023lus,Chen:2024zju}.
The central issue is the reading point ambiguity: final results depend on the specific choice of the reading point.
This problem becomes particularly apparent in diagrams with two or more internal fermion loops containing $\gamma_5$, as shown in Refs.~\cite{Zoller:2015tha,Bednyakov:2015ooa}.
Moreover, Ref.~\cite{Bednyakov:2015ooa} indicated that placing the reading points at external vertices can violate gauge invariance (see also Ref.~\cite{Korner:1991sx}), thus casting doubt on the symmetry-preserving properties of the scheme.
The ambiguity cannot be resolved within the framework itself and requires additional, external arguments to fix the result (see particularly Refs.~\cite{Poole:2019txl,Davies:2019onf,Davies:2021mnc,Herren:2021vdk}), revealing a fundamental conceptual inconsistency.
A further inconsistency arises when the genuine 4-dimensional identity in Eq.~\eqref{Eq:Manifestly-4-dim-epsilon-tensor-id} is improperly applied with fully $D$-dimensional metric tensors, as often done in schemes with naively anticommuting $\gamma_5$, such as reading-point implementations.
This introduces ambiguities of order $\mathcal{O}(\epsilon)$, precisely as explained above for Larin's prescription.

\paragraph{Practical Applications:}
Although none of the alternative prescriptions for handling $\gamma_5$ is fully self-consistent or applicable at arbitrarily high orders, they have nonetheless been employed successfully in several important multi-loop computations.
This success is largely due to their restricted use in specific contexts, often accompanied by additional external arguments to resolve inherent ambiguities.
In Ref.~\cite{Chetyrkin:2012rz}, Eq.~\eqref{Eq:Larin's-Prescription} was applied together with the use of Eq.~\eqref{Eq:Manifestly-4-dim-epsilon-tensor-id} involving fully $D$-dimensional metric tensors to compute 3-loop $\beta$-functions.
The authors explicitly parametrised the ambiguity (cf.\ Eq.~\eqref{Eq:Ambiguity-or-order-epsilon-LCTensor-Contraction-in-D-dim}) and demonstrated that it does not affect the divergent contributions, and hence none of the relevant renormalisation constants.
This calculation was later extended to the 4-loop level in Refs.~\cite{Zoller:2015tha,Bednyakov:2015ooa}, where the authors encountered the reading-point ambiguity.
The ambiguity arose in diagrams with two internal fermion loops containing $\gamma_5$, depending on the position of the reading points and whether $\gamma_5$ was naively anticommuted within the trace.
Specifically, the 4-loop contribution to the strong gauge coupling contained an ambiguous term parametrised using the prefactor $R\in\{1,2,3\}$, corresponding respectively to the choices where both reading points are located at external vertices ($R=1$), one external and one internal vertex ($R=2$), or both internal vertices ($R=3$).
In Ref.~\cite{Bednyakov:2015ooa}, external arguments were invoked to identify $R=3$ as the correct value.
Subsequently, the ambiguity was resolved more systematically in Refs.~\cite{Poole:2019txl,Davies:2019onf,Davies:2021mnc,Herren:2021vdk} using the so-called \emph{Weyl consistency conditions}, which relate certain coefficients of $\beta$-functions across different loop orders (e.g.\ between the 4-loop gauge, 3-loop Yukawa, and 2-loop scalar $\beta$-functions).
Enforcing these relations confirmed $R=3$ as the consistent solution.

Despite these successful applications, the dependence on external consistency arguments and the restriction to specific perturbative orders highlight the intrinsic limitations of these alternative $\gamma_5$ prescriptions.
They do not provide fully self-consistent or unambiguous frameworks, their scope of application at the multi-loop level is limited, and their use requires careful case-by-case justification.

\subsection{$\gamma_5$ in non-dimensional Regularisation Schemes}

In Sec.~\ref{Sec:g5-in-D-dim}, the $\gamma_5$-problem was introduced as a fundamental issue of dimensional regularisation, arising from the impossibility of maintaining all relations in Eq.~\eqref{Eq:The_3_Properties_of_g5_and_Tr} simultaneously in $D\neq4$ dimensions.
This naturally raises the question of whether non-dimensional, or fixed-dimensional, regularisation schemes offer improved properties regarding the treatment of $\gamma_5$.
However, Ref.~\cite{Bruque:2018bmy} demonstrated that such schemes encounter essentially the same difficulties as dimensional regularisation: they exhibit analogous consistency problems when dealing with $\gamma_5$.
The underlying reason is that formal manipulations of Lorentz indices and the Dirac algebra are subject to the same ambiguities, irrespective of whether the scheme is dimensional or fixed-dimensional.
In particular, contractions of Lorentz indices do not commute with renormalisation in any scheme that preserves both numerator–denominator consistency and shift invariance of loop momenta.
This non-commutativity invalidates the use of genuine integer-dimensional identities (cf.\ Eq.~\eqref{Eq:Manifestly-4-dim-epsilon-tensor-id}) prior to renormalisation.
Therefore, in fixed-dimensional schemes, a quasi $n$-dimensional space must be introduced --- analogous to that in dimensional reduction --- where ``index-counting'' is not possible. 
Ref.~\cite{Bruque:2018bmy} showed that the $\gamma_5$-problem persists in any regularisation framework satisfying the following three properties:
\begin{itemize}
    \item linearity,
    \item numerator-denominator consistency,
    \item shift invariance of loop momenta (equivalently, vanishing surface integrals).
\end{itemize}
These properties form the conceptual core of the problem.
Nevertheless, each of them is physically and mathematically well motivated.
Linearity follows directly from the linear structure of renormalisation, while numerator–denominator consistency guarantees the proper cancellation of terms between numerators and denominators (see Refs.~\cite{Bruque:2018bmy,Breitenlohner:1977hr} for more information).
Shift invariance, in turn, is directly related to translational invariance, ensures momentum routing independence and the vanishing of surface integrals. 
Together with numerator-denominator consistency, it is considered to be essential for establishing a valid quantum action principle (as it is typically required in the proof, see Refs.~\cite{Bruque:2018bmy,Breitenlohner:1977hr}), and thus for gauge invariance after renormalisation.
All these properties are satisfied within dimensional regularisation and are generally desirable.
However, they unavoidably lead to the $\gamma_5$-problem: in particular, preserving manifest shift invariance requires a modification of the algebra --- also in non-dimensional schemes --- so that the $\gamma_5$-problem persists.

\section{$D$-dimensional Lagrangian and Green Functions}\label{Sec:D-dim_Lagrangian}

An important feature of DReg is that the regularisation can be formulated directly in terms of a regularised Lagrangian, see Ref.~\cite{Belusca-Maito:2023wah} for a review.
This connection to a regularised Lagrangian is not shared by all regularisation schemes; for instance, momentum cut-off or analytic regularisation do not admit such a formulation at the Lagrangian level.
As a consequence, regularised Feynman diagrams in DReg can be obtained directly from a dimensionally extended Gell-Mann-Low formula (cf.\ Eqs.~\eqref{Eq:Z[J,K]-via-Gell-Mann-Low} and \eqref{Eq:Perturbative-Green-Functions-from-Generating-Functional}), which enables a systematic analysis of the symmetries and other structural properties of the regularised Green functions and, consequently, of the regularised theory itself.
An important example is the regularised quantum action principle, which can rigorously be established to all orders in the framework of DReg based on the properties of the $D$-dimensional Lagrangian (see Sec.~\ref{Sec:Regularised-QAP}).

The extension of a 4-dimensional Lagrangian to D-dimensions is achieved by interpreting all its building blocks --- fields $\phi_i(x)$, derivatives $\partial_\mu$, metric tensors $\eta^{\mu\nu}$, and Dirac matrices $\gamma^\mu$ --- as objects within the quasi $D$-dimensional space $\mathbb{M}_D$, as discussed in Sec.~\ref{Sec:Elements_of_D-dim_Spacetime}.
In particular, the construction of $D$-dimensional $\gamma^\mu$ matrices (see Def.~\ref{Def:Dirac_matrices_in_D-dim}) naturally implies a $D$-dimensional extension of the 4-component Dirac spinors $\psi(x)$.
In quasi $D$ dimensions, embedded in $\mathbb{M}_D$, these Dirac spinors then have infinitely many components, analogous to $\gamma^\mu$, see Eq.~\eqref{Eq:Def-of-quasi-D-dim-gammas}. 
Importantly, the 2-component Weyl spinors introduced in Sec.~\ref{Sec:Chiral_Gauge_Theories} are not known to admit an extension to $D$ dimensions, since they are explicitly tied to the representation theory of the Lorentz group and the Clifford algebra in 4 dimensions.
In particular, Weyl spinors are irreducible representations of the Lorentz group distinguished by their chirality, which relies crucially on the algebraic property that, in 4 dimensions, $\gamma_5$ anticommutes with $\overline{\gamma}^\mu$, ensuring that left- and right-handed spinors transform independently under Lorentz transformations and do not mix.
Because of the inherent 4-dimensional nature of $\gamma_5$ --- and thus of chirality --- Weyl spinors must be rewritten in terms of Dirac spinors (see Eq.~\eqref{Eq:Dirac-Spinor-expressed-via-Weyl-Spinors}) to allow for an extension to formally $D$ dimensions within the framework of DReg.
Hence, the extension of chiral gauge theories involving $\gamma_5$ to $D$ dimensions remains possible.
However, due to the modified algebraic relations (see Eq.~\eqref{Eq:BMHV-Algebra}), the resulting $D$-dimensional Lagrangian may no longer be invariant under formally $D$-dimensional Lorentz transformations.
This is not problematic, since even in such cases physically relevant 4-dimensional Lorentz invariance remains manifestly preserved.

The extension of the Lagrangian to $D$ dimensions is not unique.
Ambiguities arise, particularly in theories with chiral interactions involving $\gamma_5$. 
Because of modified algebraic properties of $\gamma_5$ in $D \neq 4$ dimensions, expressions equivalent in 4 dimensions --- such as $\overline{\psi}\gamma^{\mu}\projL\psi$ and $\overline{\psi}\projR\gamma^{\mu}\psi$ --- differ in $D$ dimensions.
This difference corresponds to an evanescent term, and additional evanescent terms, which vanish for $D=4$ dimensions, may always be introduced.
This issue will be discussed in detail in Sec.~\ref{Sec:Regularisation-Induced_Symmetry_Breaking}.
In fact, such ambiguities can appear even in expressions without $\gamma_5$, and thus also in vector-like gauge theories, since interaction terms can be chosen to remain purely 4-dimensional --- for instance, $\overline{\psi}\overline{\slashed{A}}\psi$ in QED or $\phi^\dagger \overline{A}_\mu {\overline{\partial}}{}^\mu\phi$ for scalar gauge interactions.
In vector-like theories, however, this is not advisable, as it unnecessarily complicates calculations.
While the extension of the Lagrangian to $D$ dimensions in DReg is not unique, the freedom associated with evanescent terms does not compromise the consistency of the procedure.
Any admissible and self-consistent extension yields physically equivalent results after renormalisation (see Ref.~\cite{Breitenlohner:1977hr} and Sec.~\ref{Sec:Renormalisation_in_DReg}).
In particular, the proof of the fundamental theorem of renormalisation in DReg is independent of such evanescent details.

Despite this non-uniqueness, a critical constraint must be imposed on the kinetic part of the Lagrangian. 
As usual, the $D$-dimensional Lagrangian can be decomposed into a free and an interaction part, i.e.\
\begin{align}
    \mathcal{L}^{(D)} = \mathcal{L}_\mathrm{free}^{(D)} + \mathcal{L}_\mathrm{int}^{(D)},
\end{align}
where $\mathcal{L}_\mathrm{free}^{(D)}$ is bilinear in the fields and defines the free theory, while $\mathcal{L}_\mathrm{int}^{(D)}$ contains the interactions.
To ensure that loop integrals are properly regularised, the kinetic operators in $\mathcal{L}_{\mathrm{free}}^{(D)}$, must be fully $D$-dimensional.
This requirement essentially fixes the free Lagrangian.
For example, the fermion kinetic term must be of the form $\overline{\psi}(i\gamma^\mu \partial_\mu - m)\psi\equiv\overline{\psi}\mathcal{D}^{(D)}\psi$, where both $\gamma^\mu$ and $\partial_\mu$ are strictly $D$-dimensional. 
The corresponding propagator in momentum space,
\begin{align}
    \widetilde{\mathcal{P}}^{(D)}=\langle0|T\psi\overline{\psi}|0\rangle^{\mathrm{F.T.}} = \frac{i}{\slashed{p}-m} = \frac{i(\slashed{p}+m)}{p^2-m^2},
\end{align}
(with $+i\varepsilon$-prescription suppressed and satisfying $\widetilde{\mathcal{D}}^{(D)}\widetilde{\mathcal{P}}^{(D)}=i$) then depends on the fully $D$-dimensional momentum $p$ in the denominator, which is essential for proper regularisation. 
A kinetic operator restricted to 4 dimensions, i.e.\ with 4-dimensional derivative $\overline{\partial}_\mu$, would produce an unregularised propagator with purely 4-dimensional momentum $\overline{p}$ in the denominator, and thus unregularised loop integrals. 
Such a choice is therefore not admissible.
Formally, one could also construct a regularised fermion propagator as $\widetilde{\mathcal{P}}^{(D)}=i(\overline{\slashed{p}}+m)/[p^2-m^2]$, with 4-dimensional numerator but fully $D$-dimensional denominator.
However, such a propagator cannot be derived from any $D$-dimensional Lagrangian, thereby breaking the direct correspondence between regularised Feynman diagrams and the regularised Lagrangian.
Consequently, essential theoretical features --- such as the validity of the regularised quantum action principle (see Sec.~\ref{Sec:Regularised-QAP}) --- are no longer guaranteed in this framework.
For this reason, such a choice will not be employed here.

With a well-defined $\mathcal{L}_{\mathrm{free}}^{(D)}$, consistent with the arguments above, and a chosen interaction part $\mathcal{L}_{\mathrm{int}}^{(D)}$, the regularised Feynman diagrams are generated by the $D$-dimensional version of the Gell-Mann-Low formula (cf.\ Eq.~\eqref{Eq:Z[J,K]-via-Gell-Mann-Low}): 
\begin{align}
    Z[J,K]=
    \frac{\langle 0|Te^{i\mu^{4-D}\int d^Dx\big[\mathcal{L}^{(D)}_{\mathrm{int}}(x)+J_i(x)\phi_i(x)+K_i(x)\mathcal{O}_i(x)\big]}|0\rangle}{\langle 0|Te^{i\mu^{4-D}\int d^Dx\,\mathcal{L}^{(D)}_{\mathrm{int}}(x)}|0\rangle},
\end{align}
where $\mu$ denotes the renormalisation scale introduced to maintain the correct mass dimension of the Lagrangian, see Eq.~\eqref{Eq:DReg-Integrals-4-to-D}.
The explicit correspondence between the $D$-dimensional Lagrangian and the regularised Green functions constitutes one of the most powerful features of DReg.
For brevity, we will henceforth omit the superscript $(D)$ on $D$-dimensional expressions such as the Lagrangian whenever no ambiguity can arise.

\section{Notation and Organisation of Dimensional Renormalisation}\label{Sec:Notation_and_Organisation_of_Dimensional_Renormalisation}

As discussed in the introduction of this chapter, in DReg divergences are isolated as poles in $1/(D-4)$, or equivalently as poles in $1/\epsilon$ for $D=4-2\epsilon$.
These divergences are removed iteratively, order by order in perturbation theory (i.e.\ loop by loop), by adding appropriate local counterterms to the Lagrangian (see Sec.~\ref{Sec:Renormalisation_Theory}).
This section outlines the organisation of the renormalisation procedure and establishes the notation used throughout this thesis.
We largely follow the conventions that we introduced in Ref.~\cite{Belusca-Maito:2023wah}, with a minor modification in the notation distinguishing fixed-order expressions from those containing all terms up to and including a given order.

The perturbative renormalisation procedure is organised as an expansion in powers of $\hbar$, which is equivalent to an expansion in the number of loops.
The starting point is the classical action, which is of zeroth order in $\hbar$, defines the theory and is denoted by $S_0\equiv\Gamma_\mathrm{cl}$.
The counterterm action $S_\mathrm{ct}$ is constructed as a power series in $\hbar$, where each contribution $S_\mathrm{ct}^{(n)}$ is designed to cancel all divergences and, if necessary, to restore the symmetries and enforce the renormalisation conditions at the corresponding loop order:
\begin{align}\label{Eq:Counterterm-Action-Perturbative-Expansion}
    S_\mathrm{ct} = \sum_{n=1}^\infty S_\mathrm{ct}^{(n)}.
\end{align}
The sum of the classical action and the counterterm action defines the bare action,
\begin{align}
    S_\mathrm{bare} = S_0 + S_\mathrm{ct}.
\end{align}
The bare action thus contains all the terms required to render the theory finite and to ensure that it satisfies the essential symmetries at each order in perturbation theory.

Throughout this thesis, we use a superscript $(n)$ to label quantities of precisely order $n$, and a superscript $[n]\equiv(\leq n)$ to label quantities up to and including order $n$, i.e.\
\begin{align}
    S_\mathrm{ct}^{[L]} \equiv S_\mathrm{ct}^{(\leq L)} = \sum_{n=1}^L S_\mathrm{ct}^{(n)},
\end{align}
which represents the counterterm action containing all counterterms up to and including order $\mathcal{O}(\hbar^L)$.\footnote{This notation differs slightly from that in Ref.~\cite{Belusca-Maito:2023wah}, where we used a superscript $n$ to label quantities of precisely order $n$, and $(n)$ to label quantities up to and including order $n$.}

The subrenormalised effective quantum action at order $n$ is denoted by $\Gamma_\mathrm{subren}^{(n)}$, and is obtained by using the Feynman rules from the tree-level action $S_0$ together with the counterterm action $S_\mathrm{ct}^{[n-1]}$, which includes all counterterms up to order $(n-1)$.
The next step in the iterative procedure is to construct and include the counterterms of order $n$.
At each order, the counterterm action can be decomposed into a divergent/singular part, $S_\mathrm{sct}^{(n)}$, and a finite part, $S_\mathrm{fct}^{(n)}$, such that
\begin{align}
    S_\mathrm{ct}^{(n)} = S_\mathrm{sct}^{(n)} + S_\mathrm{fct}^{(n)}.
\end{align}
When the regularisation does not preserve the symmetry --- such as in the BMHV scheme (see sections~\ref{Sec:The-BMHV-Scheme} and \ref{Sec:Regularisation-Induced_Symmetry_Breaking}) --- the divergent counterterms can further be decomposed into a symmetric part, $S_\mathrm{sct,inv}^{(n)}$, and a breaking part, $S_\mathrm{sct,break}^{(n)}$, i.e.\ $S_\mathrm{sct}^{(n)}=S_\mathrm{sct,inv}^{(n)}+S_\mathrm{sct,break}^{(n)}$.
Adding the divergent $n$-loop counterterms to the subrenormalised effective quantum action at the same order yields
\begin{align}\label{Eq:Determination-of-Singular-Counterterms}
    \mathop{\mathrm{lim}}_{D \, \to \, 4} \Big( \Gamma_\mathrm{subren}^{(n)} + S_\mathrm{sct}^{(n)} \Big) = \mathrm{finite},
\end{align}
which determines the singular counterterms $S_\mathrm{sct}^{(n)}$ unambiguously.

Finite counterterms $S_\mathrm{fct}^{(n)}$ serve two purposes.
First, they can be used to satisfy symmetry requirements if necessary (and if possible).
In particular, if a symmetry is spuriously broken --- as in the BMHV scheme --- the breaking can be removed by appropriately adjusting finite symmetry-restoring counterterms (see Sec.~\ref{Sec:Algebraic_Renormalisation}).
Second, they encode the remaining freedom in the renormalisation procedure --- beyond the cancellation of UV divergences and symmetry restoration --- which corresponds to the choice of a renormalisation scheme. 
Finite symmetric counterterms can then be fixed by imposing renormalisation conditions, for instance through on-shell conditions.
In minimal subtraction schemes, however, finite counterterms are introduced solely to restore symmetries when required and are otherwise absent.
Consequently, in such schemes the role of all counterterms reduces to the essential aspects that ensure the physical consistency of the theory: the cancellation of UV divergences and the restoration of symmetries.

Including all required counterterms up to order $n$, the dimensionally renormalised effective quantum action is given by 
\begin{align}\label{Eq:Dimensionally-Renormalised-Effective-Action-in-D-dim}
    \Gamma_\mathrm{DRen}^{(n)} \equiv \Gamma_\mathrm{subren}^{(n)} + S_\mathrm{sct}^{(n)} + S_\mathrm{fct}^{(n)}.
\end{align}
The effective action $\Gamma_\mathrm{DRen}^{(n)}$ is finite and essentially renormalised (also satisfying the symmetry requirements), but may still contain $\epsilon=(4-D)/2$ and finite evanescent terms, as introduced in Def.~\ref{Def:Evanescent-Objects} and elaborated at the end of Sec.~\ref{Sec:Elements_of_D-dim_Spacetime}.
Taking the limit $D\to 4$ and setting all evanescent quantities to zero --- these are finite since all divergences up to the considered order have been removed --- yields the completely renormalised effective quantum action,
\begin{align}\label{Eq:Fully-Renormalised-Effective-Action-in-4D}
    \Gamma^{(n)} \equiv \Gamma^{(n)}_\mathrm{ren} \equiv \mathop{\mathrm{LIM}}_{D \, \to \, 4} \, \Gamma_\mathrm{DRen}^{(n)},
\end{align}
where $\mathop{\mathrm{LIM}}_{D \, \to \, 4}$ denotes both the limit $D \to 4$ and dropping finite evanescent terms.
This iterative process ensures that the theory is finite and well-defined at every order in perturbation theory.

\section{Renormalisation in Dimensional Regularisation}\label{Sec:Renormalisation_in_DReg}

In the preceding sections, we established dimensional regularisation as a framework for isolating divergences during the renormalisation procedure, which are subsequently removed through appropriate counterterms.
Beyond its computational convenience and its manifest preservation of Lorentz invariance and, in vector-like gauge theories, of gauge invariance, it is essential to ensure that DReg constitutes a mathematically consistent regularisation scheme.
In particular, dimensional renormalisation must be proven to yield correct renormalised results and to be physically equivalent to BPHZ renormalisation, thereby ensuring compatibility with unitarity and causality.
Moreover, all required subtractions must be implementable through a local counterterm Lagrangian.
These properties were rigorously proven in Ref.~\cite{Breitenlohner:1977hr}, which established the convergence theorem for DReg and verified its mathematical consistency.
In the following, we mainly follow the more recent and accessible presentations of Refs.~\cite{Belusca-Maito:2023wah,Stoeckinger:2020mlr}.
\begin{theorem}[Fundamental Theorem of Renormalisation in Dimensional Regularisation]\label{Thm:Fundamental_Theorem_of_Renormalisation_in_DReg}\ \\
    Let $G$ be a 1PI Feynman graph with $L$ loops, overall degree of divergence $\omega(G)$, and subgraphs $H\subset G$.
    The associated regularised Feynman integral (or amplitude) in the dimensionally extended spacetime $(\mathbb{M}_D,\eta)$, defined through $D$-dimensional integrals and covariants, constitutes a meromorphic function of the complex spacetime dimension $D$ (or, equivalently, of $\epsilon=(4-D)/2$).
    All propagators in the amplitude are defined using the $+i\varepsilon$-prescription with $\varepsilon>0$.
    The amplitude of the subrenormalised graph $\overline{\mathcal{R}}(G)$ is obtained by applying the subtraction procedure --- implemented via Zimmermann's forest formula or, equivalently, through the recursive $\mathcal{R}$-operation (see Sec.~\ref{Sec:Renormalisation_Theory}) --- for all proper 1PI subgraphs $H$ of $G$.
    Its singularities in the limit $D\to4$ manifest as $1/\epsilon^n$ poles.
    More precisely, the singular part of its Laurent expansion in $\epsilon$ reads
    \begin{align}\label{Eq:The_Singular_Part_in_DReg}
        \overline{\mathcal{R}}(G)\big|_{\mathrm{div}} = \sum_{n=1}^L \frac{P_G^{(n)}(\mathbf{p},\mathbf{m})}{\epsilon^n} = \frac{1}{\epsilon} P_G^{(1)}(\mathbf{p},\mathbf{m})+\ldots+\frac{1}{\epsilon^L} P_G^{(L)}(\mathbf{p},\mathbf{m}),
    \end{align}
    where each coefficient $P_G^{(n)}(\mathbf{p},\mathbf{m})$ is a homogeneous polynomial of degree $\omega(G)$ in the external momenta $\mathbf{p}=(p_1,\ldots,p_E)$ and masses $\mathbf{m}=(m_1,\ldots,m_s)$ of $G$.
    After completion of the renormalisation procedure, i.e.\ after subtraction of the overall divergence of $\overline{\mathcal{R}}(G)$, the amplitude of the renormalised graph $\mathcal{R}(G)$ is analytic in $\epsilon$ in a neighbourhood of $\epsilon=0$.
    Consequently, the amplitude is finite, and the limit $\epsilon\to0$ exists.
    For $\varepsilon>0$, the renormalised amplitude is a $C^\infty$-function of external momenta $\mathbf{p}$ and masses $\mathbf{m}$, whereas in the limit $\varepsilon\to0^+$, its dependence on $\mathbf{p}$ and $\mathbf{m}$ assumes the character of a tempered distribution.
\end{theorem}
This theorem establishes the validity of the renormalisation procedure within the framework DReg.
Specifically, for any Feynman graph $G$, it ensures that the subtraction of divergent subdiagrams can be performed recursively, using lower-order results obtained in previous steps.
Once all subdivergences have been removed --- equivalently, once all corresponding counterterm graphs have been added --- an overall divergence remains.
This remaining divergence is local and can therefore be cancelled by introducing an appropriate counterterm in the counterterm Lagrangian with the usual properties.
After this final subtraction, the result is finite: the renormalised amplitude is a $C^\infty$-function of the physical variables $(\mathbf{p},\mathbf{m})$ and analytic in $\epsilon$, implying the limit $\epsilon\to0$ exists.

The proof of the theorem is rather technical and lengthy, and it is not necessary for this thesis.
For details, we refer to the original work in Ref.~\cite{Breitenlohner:1977hr}, which provides the proof for theories without massless particles, and to Refs.~\cite{Breitenlohner:1975hg,Breitenlohner:1976te}, where the case of massless particles is discussed.
More recent and pedagogically accessible presentations of the theorem and its proof can be found in Refs.~\cite{Belusca-Maito:2023wah,Stoeckinger:2020mlr}.
We conclude this section with a few additional remarks on the theorem and its implications.
\begin{remark}[on the fundamental Theorem of Renormalisation in DReg]\
    \begin{itemize}
        \item The nontrivial aspect of this result lies in the fact that DReg correctly reproduces the pole structure in the form of local polynomials (cf.\ Sec.~\ref{Sec:Renormalisation_Theory}) and ensures the analyticity of the finite remainder.
        \item At all intermediate stages, the calculation must be carried out in the quasi $D$-dimensional space with $\epsilon\neq0$ and non-vanishing evanescent components.
        The coefficients of the $1/\epsilon^n$ poles can themselves be evanescent (see Def.~\ref{Def:Evanescent-Objects} in Sec.~\ref{Sec:Elements_of_D-dim_Spacetime}), and the 4-dimensional limit is therefore not admissible within the coefficients $P_G^{(n)}$.
        The subtraction procedure requires the removal of all divergences, including those with evanescent coefficients.
        Consequently, the counterterm Lagrangian may contain evanescent operators --- without any 4-dimensional analogue --- to ensure complete cancellation of divergences.
        \item In the amplitude of the fully renormalised graph $\mathcal{R}(G)$, the limit $\epsilon\to0$ can be taken, and $D$-dimensional quantities may be replaced by their 4-dimensional counterparts, i.e.\ evanescent components and finite evanescent terms can be set to zero.
        Consequently, the limit $\mathop{\mathrm{LIM}}_{D \, \to \, 4}$ defined in Eq.~\eqref{Eq:Fully-Renormalised-Effective-Action-in-4D} exists.
        Renormalised amplitudes obtained in DReg thus yield a finite quantum field theory consistent with unitarity and causality, as discussed in Sec.~\ref{Sec:Renormalisation_Theory} and the references therein.
        \item The limit $\varepsilon\to0^+$ associated with the $+i\varepsilon$-prescription in the propagators has been analysed in detail in Ref.~\cite{Hepp:1966eg}.
        It was shown that this limit, taken simultaneously with the removal of the regulator, exists and yields a tempered distribution that provides a proper Lorentz-covariant continuation of the original expression with the correct analytic properties (as in the BPHZ formalism).
        \item Importantly, renormalisation in DReg differs from BPHZ renormalisation only by finite local and hermitian counterterms at each order, i.e.\ a reparametrisation in the sense of theorem~\ref{Thm:Main_Theorem_of_Renormalisation}.
        Since unitarity and causality are established for BPHZ renormalisation (see Sec.~\ref{Sec:Renormalisation_Theory}), and local hermitian counterterms can be added without affecting these properties, DReg is physically equivalent to BPHZ and therefore fully compatible with unitarity and causality.
        This equivalence was already demonstrated in Ref.~\cite{Speer:1974cz} and also discussed in the original proof of theorem~\ref{Thm:Fundamental_Theorem_of_Renormalisation_in_DReg} in Ref.~\cite{Breitenlohner:1977hr}.
        \item Subintegration consistency: the evaluation of a subgraph $H\subset G$ --- in particular, its subrenormalised amplitude $\overline{\mathcal{R}}(H)$ and the corresponding counterterm $C(H)=-\mathcal{K}\cdot\overline{\mathcal{R}}(H)$ --- depends only on the graph $H$ itself and its internal structure, and not on its embedding within a larger graph.
        In other words, the result is independent of the external environment in which it is inserted, and the associated counterterm is universal.
        This independence is of fundamental importance for the mathematical consistency of the renormalisation procedure.
        Since the counterterm Lagrangian $\mathcal{L}_\mathrm{ct}$ can be defined only once, the counterterms associated with a given divergent subgraph $H$ must be uniquely determined by evaluating that subgraph in isolation.
        Once fixed, the corresponding counterterm contribution must be identical whenever the subgraph is inserted into a larger diagram.
        Subintegration consistency is therefore a nontrivial but essential property for self-consistency.
        \item Subtracting all UV divergences appearing in Eq.~\eqref{Eq:The_Singular_Part_in_DReg} uniquely determines the divergent counterterms and thereby defines the minimal subtraction scheme. 
        Any other renormalisation scheme can be obtained by adding finite hermitian local counterterms compatible with the defining symmetries of the theory.
        \item $\mu$-dependence: the divergences of the subrenormalised graph $\overline{\mathcal{R}}(G)$ are independent of the renormalisation scale $\mu$, while the $\mu$-depending parts are completely finite.
        Consequently, differentiation w.r.t.\ $\mu$ does not generate any $1/\epsilon$-divergences, such that $\partial/\partial\mu\,\overline{\mathcal{R}}(G)=\mathrm{finite}$, which is important in formulating the theory of the renormalisation group.
        \item In the minimal subtraction scheme, $\mathcal{L}_\mathrm{ct}$ is independent of the renormalisation scale $\mu$.
        The theorem thus establishes the existence of a renormalisation scheme, namely the minimal subtraction scheme, in which the counterterm Lagrangian $\mathcal{L}_\mathrm{ct}$ is $\mu$-independent and depends polynomially on the masses.
        This property is nontrivial, as the requirement of locality alone only enforces polynomial dependence on the momenta.
        \item As long as the momenta appearing in the denominators of the propagators in loop diagrams are fully $D$-dimensional to guarantee a proper regularisation, the proof of theorem~\ref{Thm:Fundamental_Theorem_of_Renormalisation_in_DReg} is ``blind'' to the detailed evanescent structure of the numerators and interaction vertices.
        In principle, purely 4-dimensional gauge interactions are possible; however, this would violate gauge invariance, as occurs in the BMHV scheme (see Sec.~\ref{Sec:Regularisation-Induced_Symmetry_Breaking}).
        Adopting a purely 4-dimensional numerator in the fermion propagator, i.e.\ $\overline{\slashed{p}}$, would break the connection to the regularised Lagrangian (see Sec.~\ref{Sec:D-dim_Lagrangian}), and is therefore not employed (although formally possible).
        Thus, while some freedom regarding evanescent details exists, such modifications typically deteriorate the properties of the regularisation scheme. 
        \item The quantum action principle holds in DReg, as proven in Ref.~\cite{Breitenlohner:1977hr} (see also Sec.~\ref{Sec:Regularised-QAP}).
        It provides an essential foundation for establishing gauge invariance at the level of the fully renormalised Green functions (see sections~\ref{Sec:Algebraic_Renormalisation} and \ref{Sec:Symmetry_Restoration_Procedure}).
    \end{itemize}
\end{remark}

\begin{remark}[Equivalence to the Counterterm Approach]\ \\
    Notably, the subtraction procedure used in the proof of theorem~\ref{Thm:Fundamental_Theorem_of_Renormalisation_in_DReg} (see Ref.~\cite{Breitenlohner:1977hr}) is based on the recursive $\mathcal{R}$-operation (see Eq.~\eqref{Eq:The-R-Operation}), analogous to the BPHZ formalism.
    This provides an explicit construction of the counterterms through the $\mathcal{R}$-operation.
    In this framework, the reduced graphs correspond to counterterm Feynman diagrams, with counterterm insertions determined by the divergences of the subrenormalised subgraphs that have been shrunk to points.
    Since the overall divergences of the subrenormalised graphs are local polynomials (see Eq.~\eqref{Eq:The_Singular_Part_in_DReg}), the resulting counterterms can be expressed in terms of a counterterm Lagrangian $\mathcal{L}_\mathrm{ct}$.
    The complete counterterm Lagrangian is obtained by summing over all power-counting divergent 1PI graphs, and is constructed iteratively order by order in perturbation theory.
    This establishes the equivalence with the counterterm approach (cf.\ corollary~\ref{Thm:Corollary-on-Counterterm-Method} in Sec.~\ref{Sec:Renormalisation_Theory}).
    The counterterm formulation within DReg serves as the basis for all computations presented in this thesis.
\end{remark}

\begin{remark}[on Renormalisation]\ \\
    In Sec.~\ref{Sec:Renormalisation_Theory}, we discussed the inherent freedom of renormalisation, which corresponds to a finite reparametrisation (adding finite local counterterms), without referring to any particular regularisation scheme.
    In particular, the imaginary parts are determined by unitarity (cf.\ optical theorem), while the real parts are fixed by causality (cf.\ analyticity) up to local counterterms.
    Here, we revisit this point in the context of different regularisation schemes.
    As long as the fundamental requirements --- most notably unitarity and causality, and, in the case of gauge theories, the Slavnov–Taylor identity governing BRST invariance --- are satisfied after renormalisation, any consistent regularisation scheme yields a correctly renormalised theory.
    This may be illustrated schematically as follows:
    \begin{center}
        \includegraphics[width=1\linewidth]{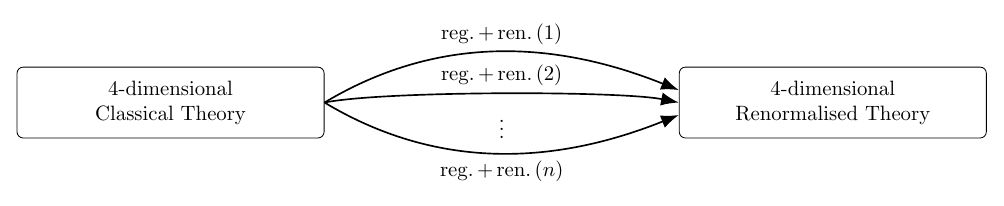}
    \end{center}
    In particular, any valid regularisation and renormalisation procedure yields a renormalised theory that is physically equivalent to BPHZ renormalisation (see Sec.~\ref{Sec:Renormalisation_Theory}) in the sense that they differ only by finite local counterterms, are consistent with the fundamental requirements, and lead to identical predictions for physical observables.
    For analytic regularisation, this equivalence has been proven in Refs.~\cite{Hepp:1971bda,Speer:1971fub}, and for DReg in Refs.~\cite{Speer:1974cz,Breitenlohner:1977hr}.
    In both cases the required finite renormalisation is also shown to be bounded by power counting.
    
    Within DReg, this freedom also manifests internally through the non-uniqueness of the $D$-dimensional extension and the presence of evanescent ambiguities (see above and Sec.~\ref{Sec:D-dim_Lagrangian}), provided the regularising effect is not compromised.
    This feature will be analysed in detail for chiral gauge theories in Sec.~\ref{Sec:Dimensional_Ambiguities_and_Evanescent_Shadows}, with concrete results presented in chapter~\ref{Chap:General_Abelian_Chiral_Gauge_Theory}.

    For gauge theories, it was shown in Sec.~\ref{Sec:Algebraic_Renormalisation} that, provided the classical action satisfies the Slavnov-Taylor identity and gauge anomalies are absent, a renormalised theory can always be constructed that preserves the Slavnov-Taylor identity --- potentially after including symmetry-restoring counterterms.
    
    The whole procedure can be summarised as follows:
    starting from a theory with divergent Green functions as discussed in Sec.~\ref{Sec:Renormalisation_Theory}, we assign any valid regularisation scheme, yielding a family of theories labelled by a regulator (e.g.\ $\Lambda$, $\epsilon$, \ldots). 
    Renormalisation then systematically removes all divergences through a consistent subtraction scheme, resulting in finite Green functions and allowing the regulator to be safely removed (e.g.\ $\epsilon\to0$).
    The resulting renormalised theory still forms a family of theories labelled by free parameters.
    Fixing these free parameters by specifying a renormalisation scheme (different schemes are different parametrisations) and relating them to physical observables eventually yields the fully renormalised theory with finite Green functions, fixed parameters, and no remaining freedom.
    Although individual results for observables may depend on the renormalisation scale (e.g.\ $\mu$), relations between observables do not, and are uniquely determined by the renormalised theory.
    The physics is therefore described unambiguously by this fully renormalised formulation.
\end{remark}

\section{The Regularised Quantum Action Principle}\label{Sec:Regularised-QAP}

The relations formulated within the path integral formalism in Sec.~\ref{Sec:Symmetries_at_the_Quantum_Level} --- in particular, the quantum action principle in Sec.~\ref{Sec:TheQuantumActionPrinciple} and the Slavnov-Taylor identity in Sec.~\ref{Sec:The-Slavnov-Taylor-Identity} --- have so far been of a formal nature.
In DReg, these statements remain valid and can be established rigorously: 
the regularised quantum action principle holds to all orders.
Concretely, all relations are to be interpreted as identities between regularised Feynman diagrams in $D$ dimensions.
Including the corresponding counterterm diagrams, these expressions are properly renormalised and therefore finite.
The all-order proof for DReg was first given in Ref.~\cite{Breitenlohner:1977hr} and extended to DRed in Ref.~\cite{Stockinger:2005gx}.
This is crucial, since the quantum action principle underpins the renormalisation of gauge theories (see Sec.~\ref{Sec:Algebraic_Renormalisation}) and is the central tool for chiral gauge theories in the BMHV scheme (see Sec.~\ref{Sec:Symmetry_Restoration_Procedure}).
In what follows, we establish the regularised quantum action principle in DReg, following mainly Refs.~\cite{Belusca-Maito:2023wah,Stoeckinger:2020mlr,Stoeckinger:2023rqft2}.

\paragraph{Preliminaries:}
The path integral is defined through a limiting procedure, and the properties of the integration measure depend on the chosen regularisation and renormalisation.
In DReg, the path integral is effectively defined perturbatively in terms of dimensionally regularised Feynman diagrams, together with their properties, i.e.\
\begin{align}
    \int \mathcal{D}\phi \, \mathcal{F}[\phi] e^{i\int d^Dx\,\mathcal{L}} \coloneqq \sum (\mathrm{all\,regularised\,Feynman\,diagrams\,in\,} D\neq4),
\end{align}
for any functional of fields $\mathcal{F}[\phi]$.
Since DReg is mathematically self-consistent (as established in Sec.~\ref{Sec:Renormalisation_in_DReg}), this yields a consistent perturbative definition of the path integral.
Moreover, in DReg the measure is invariant under any field variation $\phi_i(x)\longmapsto\phi_i(x)+\delta\phi_i(x)$, so that 
\begin{align}
    0 = \int \mathcal{D}\phi \, \Big[ \delta \mathcal{F}[\phi]+i\mathcal{F}[\phi]\int d^Dx \, \delta\mathcal{L} \Big] e^{i\int d^Dx\,\mathcal{L}},
\end{align}
where $\delta \mathcal{F}[\phi]$ are the terms linear in $\delta \phi_i$.

A key ingredient in the diagrammatic proof of the regularised quantum action principle is that the $D$-dimensional propagators are inverses of the $D$-dimensional kinetic operators appearing in regularised Lagrangian.
In other words, it relies on a direct correspondence between the regularised Feynman diagrams and the regularised Lagrangian (see Sec.~\ref{Sec:D-dim_Lagrangian}), which we impose as a condition:
\begin{singlecondition}[on Dimensional Regularisation]\ \\
    There exists a one-to-one correspondence between regularised Feynman diagrams and the regularised Lagrangian in $D$ dimensions.
    In particular, this correspondence must hold for the free Lagrangian $\mathcal{L}_\mathrm{free}$ and the propagators in $D$ dimensions:
    \begin{align}\label{Eq:Condition-on-DReg-1to1-Correspondence}
        \mathcal{L}_\mathrm{free} = \frac{1}{2} \phi_i \mathcal{D}_{ij} \phi_j \quad \longleftrightarrow \quad \widetilde{\mathcal{P}}_{ij} = \langle0|T\phi_i\phi_j|0\rangle^\mathrm{F.T.},
    \end{align}
    such that $\widetilde{\mathcal{D}}_{ij} \widetilde{\mathcal{P}}_{jk} = i \delta_{ik}$, where $\widetilde{\mathcal{P}}_{ij}$ provides the Feynman rule for the regularised propagator.
\end{singlecondition}
For example, for fermions this explicitly reads
\begin{align}
    \mathcal{L}_\mathrm{free} = \overline{\psi} (i \slashed{\partial} - m) \psi \quad \longleftrightarrow \quad \widetilde{\mathcal{P}} = \frac{i}{\slashed{p} - m} = \frac{i(\slashed{p}+m)}{p^2 - m^2}.
\end{align}

\paragraph{The Regularised Quantum Action Principle:}
In what follows, we consider the complete and regularised Lagrangian $\mathcal{L}$, corresponding to the $D$-dimensional bare Lagrangian $\mathcal{L}_\mathrm{bare}^{(D)}$, which defines the ``full'' action
\begin{align}\label{Eq:Full-Action-For-Reg-QAP}
    S = \int d^Dx \, \mathcal{L}(x) = S_0 + S_\mathrm{ct} = S_\mathrm{free} + S_\mathrm{int} + S_\mathrm{ext} + S_\mathrm{ct}.
\end{align}
Here, $S_\mathrm{free}$ denotes the free action, $S_\mathrm{int}$ the interaction part, $S_\mathrm{ext}$ the term in which the symmetry transformations $\delta\phi_i(x)$ are coupled to external sources $K_i(x)$ (and, when working with generating functionals, may additionally include the sources $J_i(x)$ coupled to the elementary fields $\phi_i(x)$), and $S_\mathrm{ct}$ represents the counterterm action.
For convenience, we define
\begin{align}
    S_\mathrm{INT} = S_\mathrm{int} + S_\mathrm{ext} + S_\mathrm{ct},
\end{align}
which collects all terms beyond the bilinear free action.
Thus, the complete regularised action in DReg can be written compactly as $S=S_\mathrm{free}+S_\mathrm{INT}$. 
With these definitions in place, we now state the theorem:
\begin{theorem}[Regularised Quantum Action Principle]\label{Thm:Regularised-QuantumActionPrinciple}\ \\
    In dimensional regularisation, let $\Sfull=\int d^Dx\,\LaFull(x)$ be the complete, regularised action as defined in Eq.~\eqref{Eq:Full-Action-For-Reg-QAP}, which admits a one-to-one correspondence to regularised Feynman rules as introduced above. 
    Given a local field transformation $\phi_i(x)\longmapsto\phi_i(x)+\delta\phi_i(x)$, the regularised quantum action principle is given by
    \begin{enumerate}[label={(\arabic*)}]
        \item its general formulation
        \begin{align}\label{Eq:Regularised-QAP-General-Formulation}
            0 = \Big\langle i \int d^Dx \, J_i(x)\delta\phi_i(x)+i\int d^Dx \frac{\delta\Sfull}{\delta\phi_i(x)}\delta\phi_i(x) \Big\rangle_{J,K},
        \end{align}
        \item its formulation on elementary Green functions
        \begin{align}\label{Eq:Regularised-QAP-Formulation-on-explicit-GreenFunctions}
            0 = \delta \big\langle T\phi_{i_1}(x_1)\ldots\phi_{i_n}(x_n) \big\rangle + \big\langle T\phi_{i_1}(x_1)\ldots\phi_{i_n}(x_n) \bigg[ i \int d^Dx \frac{\delta\Sfull}{\delta\phi_k(x)}\delta\phi_k(x) \bigg] \big\rangle,
        \end{align}
        with $\delta=\int d^Dx \, \delta\phi_{i}(x)\frac{\delta}{\delta\phi_{i}(x)}$,
        \item its formulation via the effective quantum action
        \begin{align}\label{Eq:Regularised-QAP-Formulation-wrt-Gamma}
            \int d^Dx \frac{\delta\Gamma}{\delta\phi_i(x)}\frac{\delta\Gamma}{\delta K_i(x)} = \bigg[\int d^Dx \frac{\delta\Sfull}{\delta\phi_i(x)}\delta\phi_i(x)\bigg]\cdot\Gamma,
        \end{align}
        with a single operator insertion on the RHS.\footnote{Note that sub- or superscripts on the effective action $\Gamma$ are omitted here (and below), as the quantum action principle applies universally, i.e.\ without counterterms (unrenormalised), with a subset of counterterms (partially renormalised) or with all counterterms (the bare action).\label{ftn:on-Gamma-in-regularised-QAP}}
    \end{enumerate}
\end{theorem}
\begin{proof}
    We proof the regularised quantum action principle in its formulation on elementary Green functions (see Eq.~\eqref{Eq:Regularised-QAP-Formulation-on-explicit-GreenFunctions}); all other formulations follow by equivalence.
    We begin by decomposing the regularised action as $S=S_\mathrm{free}+S_\mathrm{INT}$, which is a unique split, yielding the starting expression
    \begin{align}\label{Eq:Starting-Expression-for-proof-of-regularised-QAP}
        0 &= \sum_{k=1}^n \big\langle T\phi_{i_1}(x_1)\ldots (\delta\phi_k) \ldots\phi_{i_n}(x_n) \big\rangle_{J,K} + \big\langle T\phi_{i_1}(x_1)\ldots\phi_{i_n}(x_n) \bigg[ i \int d^Dx \frac{\delta S_\mathrm{free}}{\delta\phi_j(x)}\delta\phi_j(x) \bigg] \big\rangle_{J,K} \nonumber \\ 
        &\hspace{0.65cm} + \big\langle T\phi_{i_1}(x_1)\ldots\phi_{i_n}(x_n) \bigg[ i \int d^Dx \frac{\delta S_\mathrm{INT}}{\delta\phi_j(x)}\delta\phi_j(x) \bigg] \big\rangle_{J,K}\\
        &\eqqcolon (A) + (B_1) + (B_2), \nonumber
    \end{align}
    where the last line is introduced to give each of the three expressions a label.
    All terms in this expression are to be understood in terms of regularised Feynman diagrams in DReg, obtained perturbatively from the Gell-Mann-Low formula in $D$ dimensions, evaluated via Wick contractions and with Feynman rules derived from the ``full'' action $S$.
    Explicitly, the Gell-Mann-Low formula yields $\langle\phi_1\ldots\phi_n\rangle_{J,K}=\langle0|T\phi_1\ldots\phi_n\mathrm{exp}(i S_\mathrm{INT})|0\rangle/\langle0|T\mathrm{exp}(i S_\mathrm{INT})|0\rangle$.
    Since the denominator will always be the same expression, we focus on the numerator.
    
    The free action can generally be written as
    \begin{align}
        S_\mathrm{free} = \int d^Dx \, \frac{1}{2} \phi_i \mathcal{D}_{ij} \phi_j,
    \end{align}
    which, due to the one-to-one correspondence with regularised Feynman diagrams, generates the properly regularised propagators $\widetilde{\mathcal{P}}_{ij}$ satisfying $\widetilde{\mathcal{D}}_{ij} \widetilde{\mathcal{P}}_{jk} = i \delta_{ik}$ (cf.\ Eq.~\eqref{Eq:Condition-on-DReg-1to1-Correspondence}).
    Its variation, required in Eq.~\eqref{Eq:Starting-Expression-for-proof-of-regularised-QAP}, is given by
    \begin{align}\label{Eq:The-Variation-of-the-Free-Action}
        \frac{\delta S_\mathrm{free}}{\delta\phi_i}\delta\phi_i = \delta \phi_i \mathcal{D}_{ij} \phi_j.
    \end{align}
    
    Each of the three expressions in Eq.~\eqref{Eq:Starting-Expression-for-proof-of-regularised-QAP} is evaluated at some fixed order $\mathcal{O}(S_\mathrm{INT}^N)$.
    For the $k$-th summand of the first term, we obtain
    \begin{align}\label{Eq:A_Contribution-QAP}
        (A)_k \Big|_{\mathcal{O}(S_\mathrm{INT}^N)} \propto\frac{1}{N!}\,\big\langle 0 \big| T\phi_1\ldots(\delta\phi_k)\ldots\phi_n \, \underbrace{(i S_\mathrm{INT})\ldots(i S_\mathrm{INT})}_{\text{$N$ factors}} \big|0\big\rangle,
    \end{align}
    while the third term, involving $\delta S_\mathrm{INT}$, yields
    \begin{align}\label{Eq:B_2-Contribution-QAP}
        (B_2) \Big|_{\mathcal{O}(S_\mathrm{INT}^N)} \propto \frac{1}{(N-1)!}  \big\langle 0 \big| T\phi_1\ldots\phi_n \bigg[ i \int d^Dx \frac{\delta S_\mathrm{INT}}{\delta\phi_i} \delta \phi_i \bigg] \underbrace{(i S_\mathrm{INT})\ldots(i S_\mathrm{INT})}_{\text{$(N-1)$ factors}} \big| 0 \big\rangle.
    \end{align}
    For the second term, involving the free action $S_\mathrm{free}$, we use Eq.~\eqref{Eq:The-Variation-of-the-Free-Action} to obtain
    \begin{align}\label{Eq:B_1-Contribution-QAP}
        (B_1) \Big|_{\mathcal{O}(S_\mathrm{INT}^N)} \propto \frac{1}{N!}\,
        \contraction[1.5ex]{\big\langle 0 \big| T \phi_1 \ldots \phi_k \ldots \phi_n \bigg[ i \int d^Dx \,}{\delta \phi_i}{\mathcal{D}_{ij}}{\!\!\!\! \phi_j}
        \contraction[2.25ex]{\big\langle 0 \big| T \phi_1 \ldots}{\!\!\phi_k}{\ldots \phi_n \bigg[ i \int d^Dx \, \delta \phi_i \mathcal{D}_{ij}}{\, \, \phi_j}
        \contraction[3ex]{\big\langle 0 \big| T \phi_1 \ldots \phi_k \ldots \phi_n \bigg[ i \int d^Dx \, \delta \phi_i \mathcal{D}_{ij}}{\,\phi_j}{\bigg] \, (i S_\mathrm{INT})}{\: \ldots}
        \big\langle 0 \big| T \phi_1 \ldots \phi_k \ldots \phi_n \bigg[ i \int d^Dx \, \delta \phi_i \mathcal{D}_{ij} \phi_j \bigg] \, \underbrace{(i S_\mathrm{INT})\ldots(i S_\mathrm{INT})}_{\text{$N$ factors}} \big| 0 \big\rangle.
    \end{align}
    The next step is to evaluate the relevant types of Wick contractions of the field operator $\phi_j$ appearing in the inserted ``special'' vertex operator $\delta \phi_i \mathcal{D}_{ij} \phi_j$, which are indicated in Eq.~\eqref{Eq:B_1-Contribution-QAP}.
    Specifically, $\phi_j$ can be contracted with $\delta\phi_j$ inside the same vertex operator (contraction~$(1)$), with one of the external field operators $\phi_k$ (contraction~$(2)$), or with a field operator inside one of the $S_\mathrm{INT}$ factors (contraction~$(3)$).
    These three types of contractions give rise to distinct classes of Feynman diagrams.

    Contraction~$(1)$ gives rise to Feynman diagrams that contain a closed loop starting and ending at the ``special'' vertex, with a propagator $\widetilde{\mathcal{P}}_{jl}$ connecting the field $\phi_j$ to some field $\phi_l$ inside the composite operator $\delta\phi_j$.
    Hence, we obtain 
    \begin{align}
        (1) = \ldots \int d^Dk \, \widetilde{\mathcal{D}}_{ij} \widetilde{\mathcal{P}}_{jl} = \ldots \int d^Dk \, i \delta_{il} = 0, 
    \end{align}
    where we used that the propagator cancels the kinetic operator as required and that the loop integral over a constant, i.e.\ a scaleless integral, vanishes identically in DReg.

    Contraction~$(2)$ gives rise to Feynman diagrams where the ``special'' vertex is contracted with an external field $\phi_k$.
    Evaluating the Wick contraction in momentum space gives
    \begin{equation}
        \begin{aligned}
            (2) &= \bigg[ \ldots 
            \contraction[1.75ex]{}{\!\!\phi_k}{\ldots i \int d^Dx \,\delta\phi_i\mathcal{D}_{ij}}{\,\,\phi_j}
            \phi_k \ldots i \int d^Dx \,\delta\phi_i\mathcal{D}_{ij}\phi_j \ldots \bigg]^\mathrm{F.T.}
            = \ldots \, i \delta\phi_i \widetilde{\mathcal{D}}_{ij}\langle0|T\phi_j\phi_k|0\rangle^\mathrm{F.T.}\\
            &= \ldots \, i \delta\phi_i \widetilde{\mathcal{D}}_{ij} \widetilde{\mathcal{P}}_{jk} = \ldots ( -\delta \phi_k ).
        \end{aligned}
    \end{equation}
    In particular, the $\int d^Dx$ integral from the ``special'' vertex operator effectively cancels and the $k$-th external field operator $\phi_k$ is replaced by $-\delta\phi_k$.
    Hence, contraction~$(2)$ in Eq.~\eqref{Eq:B_1-Contribution-QAP} yields precisely the negative contribution of Eq.~\eqref{Eq:A_Contribution-QAP}.
    This type of contraction must be carried out for each of the $n$ external field operators, yielding $(B_1)|_{(2)}=-(A)$.
    Consequently, this contribution of $(B_1)$ exactly cancels the entire contribution of $(A)$.

    Finally, contraction~$(3)$ produces the class of Feynman diagrams in which the ``special'' vertex is connected to an internal vertex from $S_\mathrm{INT}$ located somewhere inside the diagram.
    Specifically, the field $\phi_j$ is contracted with a field $\phi_l$ inside one of the $S_\mathrm{INT}$ factors.
    Evaluating one of these Wick contractions in momentum space gives
    \begin{equation}
        \begin{aligned}
            (3) &= \bigg[ \ldots \Big( i \int d^Dx \,\delta\phi_i\mathcal{D}_{ij}
            \contraction[1.75ex]{}{\!\!\phi_j}{\Big) (i}{\!\!\!S_\mathrm{INT}}
            \phi_j \Big) (i S_\mathrm{INT}) \ldots \bigg]^\mathrm{F.T.}
            = \ldots \int i \delta\phi_i \widetilde{\mathcal{D}}_{ij}\langle0|T\phi_j\phi_l|0\rangle^\mathrm{F.T.} \, i \frac{\delta S_\mathrm{INT}}{\delta \phi_l}\\
            &= \ldots \int i \delta\phi_i \widetilde{\mathcal{D}}_{ij} \widetilde{\mathcal{P}}_{jl} \, i \frac{\delta S_\mathrm{INT}}{\delta \phi_l} = \ldots \bigg( - i \int \delta \phi_i \frac{\delta S_\mathrm{INT}}{\delta \phi_i} \bigg),
        \end{aligned}
    \end{equation}
    and since there are $N$ identical factors of $S_\mathrm{INT}$ in Eq.~\eqref{Eq:B_1-Contribution-QAP}, this term acquires a factor $N$, yielding a total contribution of 
    \begin{align}
        - \frac{N}{N!}  \big\langle 0 \big| T\phi_1\ldots\phi_n \bigg[ i \int d^Dx \frac{\delta S_\mathrm{INT}}{\delta\phi_i} \delta \phi_i \bigg] \underbrace{(i S_\mathrm{INT})\ldots(i S_\mathrm{INT})}_{\text{$(N-1)$ factors}} \big| 0 \big\rangle.
    \end{align}
    Hence, contraction~$(3)$ in Eq.~\eqref{Eq:B_1-Contribution-QAP} yields precisely the negative of Eq.~\eqref{Eq:B_2-Contribution-QAP}.
    Thus, $(B_1)|_{(3)}=-(B_2)$, meaning that this contribution of $(B_1)$ cancels the complete contribution of $(B_2)$.

    Collecting all results, we find that $(B_1)=(B_1)|_{(1)}+(B_1)|_{(2)}+(B_1)|_{(3)}=-(A)-(B_2)$, which holds for all $N$, i.e.\ all orders in $S_\mathrm{INT}$.
    Therefore,
    \begin{align}
        (A) + (B_1) + (B_2) = 0,
    \end{align}
    which completes the proof.
\end{proof}

\begin{remark}[on the regularised Quantum Action Principle in DReg]\ \\
    The proof hinges on three DReg-specific ingredients:
    \begin{itemize}
        \item the diagrammatic realisation via the $D$-dimensional Gell-Mann-Low formula;
        \item the exact relation $\widetilde{\mathcal{D}}_{ij} \widetilde{\mathcal{P}}_{jk} = i \delta_{ik}$ following from the one-to-one correspondence between regularised propagators and the regularised free Lagrangian;
        \item the vanishing of scaleless integrals.
    \end{itemize}
    These steps are not guaranteed in other regularisation schemes; consequently, the quantum action principle must be established for each regularisation scheme individually.
    A distinctive advantage of DReg is that the quantum action principle already holds at the regularised level.
\end{remark}

Having established the regularised quantum action principle, the remaining identities discussed in Sec.~\ref{Sec:Symmetries_at_the_Quantum_Level} --- the Ward and Slavnov-Taylor identities, and the generalised equations of motion --- follow as special cases.
Analogous to the quantum action principle, they are to be understood as identities between regularised Feynman diagrams in $D$ dimensions.
In other words, the relations in Sec.~\ref{Sec:Symmetries_at_the_Quantum_Level} extend straightforwardly to $D\neq4$ dimensions and are rigorously established as special cases of the regularised quantum action principle. 
In particular, the $D$-dimensional Slavnov-Taylor operator (cf.\ Eq.~\eqref{Eq:Slavnov-Taylor-Operator-Formal-Def}) acts as
\begin{align}\label{Eq:D-dim-ST-Operator}
    \mathcal{S}_D:\mathcal{F}[\phi,K]\longmapsto\int d^Dx \frac{\delta\mathcal{F}[\phi,K]}{\delta\phi_i(x)}\frac{\delta\mathcal{F}[\phi,K]}{\delta K_i(x)},
\end{align}
for any functional $\mathcal{F}$.
Analogous $D$-dimensional extensions exist for the linearised Slavnov-Taylor operator, and, in particular, for the BRST operator $s_D$ and the linearised Slavnov-Taylor operator w.r.t.\ the $D$-dimensional classical action $b_D$ (cf.\ Sec.~\ref{Sec:Symmetries_at_the_Quantum_Level}).
With this notation, the regularised quantum action principle, in its formulation w.r.t.\ the $D$-dimensional effective quantum action given in Eq.~\eqref{Eq:Regularised-QAP-Formulation-wrt-Gamma}, takes the form
\begin{align}\label{Eq:Regularised-QAP-wrt-Gamma-in-terms-of-STI-Operator}
    \mathcal{S}_D(\Gamma) = \bigg[\int d^Dx \frac{\delta\Sfull}{\delta\phi_i(x)}\delta\phi_i(x)\bigg]\cdot\Gamma,
\end{align}
that is, $\mathcal{S}_D(\Gamma)$ is related to a single operator insertion of the variation of the complete, regularised action, i.e.\ the $D$-dimensional bare action.
This relation is the starting point for symmetry restoration in chiral gauge theories (see Sec.~\ref{Sec:Symmetry_Restoration_Procedure}).

\paragraph{Gauge invariance in Dimensional Regularisation:}
Consider a symmetry transformation $\phi_i(x)\longmapsto\phi_i(x)+\delta\phi_i(x)$ that leaves the classical action invariant.
The $D$-dimensional Slavnov-Taylor identity w.r.t.\ to the $D$-dimensional effective quantum action (cf.\ Eq.~\eqref{Eq:Formal-STI-Formulation-wrt-Gamma}) reads
\begin{align}\label{Eq:D-dim-STI-wrt-Gamma}
    \mathcal{S}_D(\Gamma) = \int d^Dx \frac{\delta\Gamma}{\delta\phi_i(x)}\frac{\delta\Gamma}{\delta K_i(x)} \overset{?}{=} 0.
\end{align}
When the variations $\delta\phi_i$'s correspond to BRST transformations, Eq.~\eqref{Eq:D-dim-STI-wrt-Gamma} expresses BRST invariance at the quantum level for dimensionally regularised Green functions.
However, this identity does not automatically hold at the regularised level in $D$ dimensions; hence the ``$\overset{?}{=}$'' in Eq.~\eqref{Eq:D-dim-STI-wrt-Gamma}. 
Its validity must, in general, be established explicitly --- which is possible by using the regularised quantum action principle.
Although the path integral measure in DReg is invariant, Eq.~\eqref{Eq:D-dim-STI-wrt-Gamma} holds to all orders if and only if one can construct a $D$-dimensional bare action $S_\mathrm{bare}^{(D)}$ --- including counterterms --- that is invariant under the symmetry in question.
If the 4-dimensional classical action $S_\mathrm{0}^{4D}$ cannot be extended to $D$-dimensions without violating the symmetry, Eq.~\eqref{Eq:D-dim-STI-wrt-Gamma} is broken (at the regularised level), and the corresponding breaking must be analysed, as done in Sec.~\ref{Sec:Algebraic_Renormalisation}.
For spurious breakings, the symmetry can be restored by adding appropriate finite counterterms, thereby reconstructing a symmetric bare action.\footnote{Even if $S_\mathrm{bare}^{(D)}$ is not invariant, expectation values may accidentally satisfy the Ward and Slavnov–Taylor identities at low orders (i.e.\ some Green functions vanish so the breaking is not yet visible). In generic situations, however, higher-order Green functions become nonzero and the identities fail, so they do not hold to all orders (see e.g.\ supersymmetry, cf.\ Refs.~\cite{Belusca-Maito:2023wah,Stoeckinger:2020mlr}).}
In contrast, if the breaking is anomalous, no such restoration is possible (see Sec.~\ref{Sec:Algebraic_Renormalisation}).

It should be emphasised that, even for spurious breakings, the Slavnov–Taylor identity does not need to hold in $D$ dimensions, but only after taking the limit $\mathop{\mathrm{LIM}}_{D \, \to \, 4}$ (as defined in Eq.~\eqref{Eq:Fully-Renormalised-Effective-Action-in-4D}), i.e.\footnote{Here we specifically refer to the dimensionally renormalised effective quantum action $\Gamma_\mathrm{DRen}$ as introduced in Eq.~\eqref{Eq:Dimensionally-Renormalised-Effective-Action-in-D-dim}, since we are considering the symmetry requirement at the renormalised level (cf.\ footnote~\ref{ftn:on-Gamma-in-regularised-QAP}).}
\begin{align}\label{Eq:Ulitmate_Symmetry_Requirement_QAP-Chapter}
    \mathop{\mathrm{LIM}}_{D \, \to \, 4} \, \mathcal{S}_D(\Gamma_\mathrm{DRen}) = 0,
\end{align}
which represents the ultimate symmetry requirement.

\begin{remark}[on Anomalies in DReg]\ \\
    In DReg, the functional measure is invariant under field variations $\phi_i(x)\longmapsto\phi_i(x)+\delta\phi_i(x)$.
    Symmetry breaking therefore manifests not through a non-invariant measure, but through the impossibility of extending an invariant 4-dimensional Lagrangian to a $D$-dimensional regularised Lagrangian that preserves this symmetry.\footnote{In such cases, the path-integral measure can be said to be \emph{effectively} non-invariant.}
    If the breaking is spurious (e.g.\ regularisation-induced), it can be removed by symmetry-restoring counterterms, in accordance with Sec.~\ref{Sec:Algebraic_Renormalisation}.
    If the breaking is anomalous, it cannot be removed: the $D$-dimensional Lagrangian necessarily violates the symmetry, reflecting the symmetry breaking introduced by the process of regularisation and renormalisation (here defined through regularised Feynman diagrams).
    Different schemes may display the anomaly in distinct technical forms, but its presence is independent of the regularisation procedure and represents a genuine physical effect.
\end{remark}

\chapter{Multi-Loop Techniques}\label{Chap:Multi-Loop_Calculations}

In this chapter, we present the computational framework employed for all calculations performed in this thesis and discuss the methods and techniques underlying the implementation.
These methods and their implementation are applicable to a wide class of quantum field theories; thus, the discussion is formulated in a largely model-independent way.
Where appropriate, however, specific examples are provided to illustrate key aspects.

We begin in Sec.~\ref{Sec:Computational_Setup} with a concise overview of the overall workflow and the utilised software tools.
Section~\ref{Sec:Tadpole_Decomposition} then establishes the procedure used to extract UV divergences via a tadpole decomposition.
In Sec.~\ref{Sec:Tensor_Reduction}, we describe the tensor reduction procedure applied in the computation of Feynman integrals, together with the underlying group-theoretical principles and their concrete implementation.
Section~\ref{Sec:Implementation_of_the_BMHV_Algebra} discusses the implementation of the BMHV algebra, focusing in particular on the treatment of Dirac traces involving $\gamma_5$.
The practical implementation of counterterms in our computational setup is briefly illustrated in Sec.~\ref{Sec:Practical_Implementation_of_Counterterms}.
Finally, in Sec.~\ref{Sec:IBP-Reduction}, we comment on the use of integration-by-parts (IBP) reductions for expressing multi-loop integrals in terms of a minimal set of master integrals.
Together, these components constitute a fully automated and internally consistent computational framework capable of performing multi-loop calculations within the BMHV scheme.

\section{Computational Setup}\label{Sec:Computational_Setup}

In the course of this work, two complementary computational frameworks have been developed to carry out the renormalisation of various chiral gauge theories within the BMHV scheme and to compute the corresponding Green functions.
The first setup is implemented in \texttt{Mathematica}~\cite{mathematica}, while the second framework --- with considerably higher performance --- is based primarily on \texttt{FORM}~\cite{Vermaseren:2000nd,Ruijl:2017dtg,FORM:Manual}.
In the following, a brief overview of the \texttt{Mathematica}-based setup is provided, which was employed for the computations presented in Ref.~\cite{Stockinger:2023ndm}.
Subsequently, the more performant \texttt{FORM}-based setup is described in detail, as it served as the foundation for the first 4-loop computation within the BMHV framework (see Ref.~\cite{vonManteuffel:2025swv} and chapter~\ref{Chap:BMHV_at_Multi-Loop_Level}) as well as for the Standard Model applications (see chapter~\ref{Chap:The_Standard_Model}) and will also provide the basis for future projects.

\paragraph{Comment on the \texttt{Mathematica}-based Setup:}
In this setup, the majority of calculations are carried out in \texttt{Mathematica}.
Feynman diagrams --- including those with an insertion of the operator $\Delta$ (see chapter~\ref{Chap:Practical_Symmetry_Restoration}) --- are generated using \texttt{FeynArts}~\cite{Hahn:2000kx}, and most symbolic manipulations, in particular those involving Dirac algebra operations, are handled with \texttt{FeynCalc}~\cite{Mertig:1990an,Shtabovenko:2016sxi,Shtabovenko:2020gxv,Shtabovenko:2021hjx}.
Integral reductions are performed with the \texttt{C++} version of \texttt{FIRE}~\cite{Smirnov:2019qkx,Smirnov:2023yhb}, which is interfaced to \texttt{Mathematica} via \texttt{FeynHelpers}~\cite{Shtabovenko:2016whf}.

This setup was successfully tested and applied to Abelian chiral gauge theories at the 3-loop level, as reported in Ref.~\cite{Stockinger:2023ndm}.
However, already in this context, performance limitations became apparent, with \texttt{Mathematica} forming a significant bottleneck --- particularly during the evaluation of Dirac traces involving many $\gamma$-matrices, high-rank tensor reductions, and the manipulation of large algebraic expressions.
To overcome these limitations and to extend the applicability of the calculations both to higher loop orders (such as 4-loop computations) and to more complex models (such as the Standard Model), a second, substantially more powerful setup based on \texttt{FORM} was developed.
The \texttt{Mathematica}-based framework will therefore not be discussed further in this thesis.
Instead, the subsequent section presents the structure and workflow of the \texttt{FORM}-based computational framework.

\paragraph{\texttt{FORM}-Setup Workflow:}
To illustrate the workflow of the automated \texttt{FORM}-based setup, we largely follow our original description presented in Ref.~\cite{vonManteuffel:2025swv}.
In this framework, Green functions are computed by processing the amplitudes of Feynman diagrams generated by \texttt{QGRAF}~\cite{Nogueira:1991ex} through a sequence of \texttt{FORM} routines.
In the initial step, the relevant Feynman rules are inserted to construct algebraic expressions for multi-leg Green functions.
Since the focus of the computation lies on the extraction of UV divergences required for renormalisation, we employ a tadpole decomposition that systematically maps all Feynman integrals to fully massive, single-scale vacuum bubbles.
The details of this procedure are discussed in Sec.~\ref{Sec:Tadpole_Decomposition}.

Our objective is to express the resulting vacuum bubble diagrams in terms of scalar integrals of the form
\begin{equation}\label{Eq:IntegralFamily}
    \begin{aligned}
        I(\nu_1,\ldots,\nu_N) = 
        \big(\mu^{4-D}\big)^L 
        \int \Bigg( \prod_{i=1}^{L} \frac{d^Dk_i}{(2\pi)^D} \Bigg)  
        \frac{1}{\mathcal{D}_1^{\nu_1} \cdots \mathcal{D}_N^{\nu_N}},
    \end{aligned}
\end{equation}
where $L$ denotes the loop-order, $D=4-2\epsilon$ the spacetime dimension, and $N=N(L)=L(L+1)/2$ the number of distinct propagators.
In the case of vacuum bubble topologies, the denominators are independent of external momenta, and each propagator can therefore be written as
\begin{equation}\label{Eq:General-Propagator-Denominator-for-Vacuum-Bubbles}
    \begin{aligned}
        \mathcal{D}_i = q_i^2 - M^2,
    \end{aligned}
\end{equation}
where $q_i$ represents a linear combination of the loop momenta.
For each loop order, a single set of linearly independent propagator momenta,
\begin{equation}
    \begin{aligned}
        B_L = \Big\{ q_i \, \big| \, i \in \{1,\ldots,N(L)\}; \, \{q_i^2\} \,\, \text{linearly independent} \Big\},
    \end{aligned}
\end{equation}
is sufficient to cover all relevant Feynman diagrams.
We follow the conventions of Ref.~\cite{Luthe:2015ngq}, with the explicit momentum configurations for the propagators summarised in Tab.~\ref{Tab:PropagatorMomentumConfigurations}.
\begin{table}[t]
    \centering
    \begin{tabular}{|c|c|c|c|c|}
        \hline
                 &$B_1$&$B_2$    &$B_3$    &$B_4$        \\ \hline
         $q_1$   &$k_1$&$k_1$    &$k_1$    &$k_1$        \\
         $q_2$   &     &$k_2$    &$k_2$    &$k_2$        \\
         $q_3$   &     &$k_1-k_2$&$k_3$    &$k_3$        \\
         $q_4$   &     &         &$k_1-k_2$&$k_4$        \\
         $q_5$   &     &         &$k_1-k_3$&$k_1-k_4$    \\
         $q_6$   &     &         &$k_2-k_3$&$k_2-k_4$    \\
         $q_7$   &     &         &         &$k_3-k_4$    \\
         $q_8$   &     &         &         &$k_1-k_2$    \\
         $q_9$   &     &         &         &$k_1-k_3$    \\
         $q_{10}$&     &         &         &$k_1-k_2-k_3$\\ \hline
    \end{tabular}
    \caption{Propagator momenta $q_i$ for fully massive single scale vacuum bubbles up to $L=4$ loops, see Ref.\ \cite{Luthe:2015ngq}.}
    \label{Tab:PropagatorMomentumConfigurations}
\end{table}

The assignment of loop momenta is not unique, which may lead to mismatches between the momenta in the denominators of our diagrams and those of the target integral families.
To resolve this, we employ the program \texttt{Feynson}~\cite{Maheria:2022dsq}, which systematically performs loop-momentum shifts to enforce a uniform momentum configuration.
In this way, all propagator momenta coincide with those specified in Tab.~\ref{Tab:PropagatorMomentumConfigurations}.
Additional remarks on this step can be found in Sec.~\ref{Sec:IBP-Reduction}.

At this stage of the computation, tensor integrals with open Lorentz indices on loop momenta in their numerators are typically encountered.
These tensor integrals are reduced to scalar integrals $I(\nu_1,\ldots,\nu_N)$, defined in Eq.~\eqref{Eq:IntegralFamily}, by applying a tensor reduction procedure.
Since this step is among the most computationally expensive parts of the calculation, we have developed an efficient implementation in \texttt{FORM}, which is described in Sec.~\ref{Sec:Tensor_Reduction}.

Once the integrals are scalarised, the Dirac algebra is evaluated.
Our computations employ a non-anticommuting $\gamma_5$ within the framework of the BMHV scheme, which is efficiently handled through a dedicated \texttt{FORM} routine.
This procedure has been optimised for performance by exploiting as many built-in \texttt{FORM} operations as possible, facilitated by separating traces over 4-dimensional and evanescent Dirac matrices.
Further details of this implementation are discussed in Sec.~\ref{Sec:Implementation_of_the_BMHV_Algebra}.

In the final stage of the computation, we reduce the general scalar integrals $I(\nu_1,\ldots,\nu_N)$ to a finite set of master integrals using integration-by-parts (IBP) identities.
Our primary tool for this task is the \texttt{C++} version of \texttt{FIRE}~\cite{Smirnov:2019qkx,Smirnov:2023yhb}.
To ensure a minimal basis of preferred master integrals, particularly at the 4-loop level, we use \texttt{Reduze2}~\cite{vonManteuffel:2012np} to identify and implement sector symmetry relations.
We performed cross-checks and basis changes with \texttt{Kira}~\cite{Maierhofer:2017gsa,Klappert:2020nbg,Klappert:2019emp,Klappert:2020aqs} and the private code \texttt{Finred}~\cite{vonManteuffel:2014ixa,vonManteuffel:2016xki,Peraro:2016wsq}, which is based on finite field arithmetic.
The solutions for the master integrals are taken from the literature:
in particular, Refs.~\cite{Schroder:2005va,Martin:2016bgz} for the 2- and 3-loop cases, and Ref.~\cite{Czakon:2004bu} for the 4-loop case, while the numerical 4-loop results in Ref.~\cite{Schroder:PrivateComm} have additionally been used for consistency checks.
A concise overview of this procedure is provided in Sec.~\ref{Sec:IBP-Reduction}.

In our computational framework, \texttt{Mathematica} is used only as a high-level interface for managing the different software tools and generating \texttt{FORM} code.
All performance-critical operations are delegated to carefully optimised \texttt{FORM} programs and dedicated IBP solvers.
This strategy eliminates the major computational bottleneck encountered in the earlier \texttt{Mathematica}-based setup and in previous calculations, such as those reported in Refs.~\cite{Stockinger:2023ndm,Kuhler:2024fak}.

The computational framework described above has been thoroughly tested and successfully applied up to the 4-loop level.
In particular, we carried out a complete 4-loop renormalisation of vector-like quantum electrodynamics (QED) and verified our results against the literature (see App.~\ref{App:Renormalisation_of_QED} for an overview of these results).
Moreover, the 4-loop results presented in Ref.~\cite{vonManteuffel:2025swv} were found to be local and internally consistent, since they passed the nontrivial consistency check comparing the symmetry-breaking divergent contributions from ordinary 1PI Green functions and from $\Delta$-operator inserted 1PI Green functions (see chapters~\ref{Chap:Practical_Symmetry_Restoration} and \ref{Chap:BMHV_at_Multi-Loop_Level}).
Furthermore, we successfully reproduced our previous 1-, 2- and 3-loop results for the Abelian chiral gauge theory discussed in chapter~\ref{Chap:BMHV_at_Multi-Loop_Level}, in agreement with Refs.~\cite{Belusca-Maito:2021lnk,Stockinger:2023ndm}.

\section{Tadpole Decomposition}\label{Sec:Tadpole_Decomposition}

For a complete renormalisation of chiral gauge theories in the BMHV scheme, it is necessary to compute not only two-point but also three-, four- and even five-point Green functions (see chapters \ref{Chap:General_Abelian_Chiral_Gauge_Theory} to \ref{Chap:The_Standard_Model}).
At the multi-loop level, such calculations are highly demanding: they require the reduction of the associated Feynman integrals to master integrals via integration-by-parts (IBP) relations in the presence of multiple scales, and the subsequent evaluation of the master integrals themselves (at higher orders, e.g.\ 4 loops, solutions for several master integrals with multiple scales/legs are still unavailable in the literature).
A further complication arises from IR divergences in the computation of counterterms, cf.\ remark on IR divergences in Sec.~\ref{Sec:Renormalisation_Theory}.

To address these challenges, we aim for methods that reduce the computational complexity while preserving the correct UV behaviour.
There are essentially two approaches that achieve this by effectively reducing all Feynman integrals to single-scale integrals --- making the computation of master integrals at higher orders feasible --- without changing their (overall) UV contributions.
Both rely on \emph{infrared rearrangement} and find their justification by the fact that counterterms are local polynomials in external momenta and (for mass-independent schemes) internal masses, see Sec.~\ref{Sec:Renormalisation_Theory}.

For completeness, we briefly mention the approach based on the $\mathcal{R}^*$-operation, originally proposed in Refs.~\cite{Chetyrkin:1982nn,Chetyrkin:1984xa,Smirnov:1985yck} as a generalisation of the $\mathcal{R}$-operation (see Def.~\ref{Def:The_R-Operation}) that can handle both UV and IR divergences.
In this method, an infrared rearrangement (first suggested in Ref.~\cite{Vladimirov:1979zm}) is achieved by setting all propagator masses to zero and rerouting the external momenta in such a way that the problem reduces to massless propagator-type integrals depending on a single external momentum as the only scale.
This procedure generally leads to (unphysical) IR divergences, which are then removed recursively by the $\mathcal{R}^*$-operation.
For a recursive solution in the spirit of Zimmermann's forest formula (theorem~\ref{Thm:Zimmermanns_Forest_Formula}), together with detailed discussions and proofs of several theorems see Ref.~\cite{Chetyrkin:2017ppe}, for an extension to arbitrary numerators see Ref.~\cite{Herzog:2017bjx}, and for a pedagogical introduction see Ref.~\cite{Kleinert:2001}.
Although powerful, this approach is not employed in the present work and will not be discussed further.

Instead, we adopt the \emph{tadpole decomposition}, first introduced in Refs.~\cite{Misiak:1994zw,Chetyrkin:1997fm}, to extract UV divergences.
Here, an infrared rearrangement is realised by introducing an auxiliary mass scale $M^2$ into all propagators and recursively decomposing the integrand into leading and subleading UV contributions through the so-called tadpole expansion/decomposition.
Once the expansion is carried out to sufficient order, those subleading terms that have become power-counting finite can be truncated and discarded, leaving only single-scale, fully massive vacuum bubble integrals (tadpoles) depending on $M^2$, with all momentum dependence shifted completely to the numerator.
In this way, UV divergences can be extracted efficiently, while potential IR divergences are avoided by construction through the presence of $M^2$ in every propagator.

\paragraph{``Exact'' Tadpole Decomposition:}
Specifically, a generic denominator of the form
\begin{align}\label{Eq:Generic-Propagator-Denominator}
    \mathcal{D}_e(l_e(k_1,\ldots,k_L),q_e(p_1,\ldots,p_E),m_e)=(l_e(k_1,\ldots,k_L)+q_e(p_1,\ldots,p_E))^2-m_e^2,
\end{align}
depending on the loop momenta $\mathbf{k}=(k_1,\ldots,k_L)$ and external momenta $\mathbf{p}=(p_1,\ldots,p_E)$ through linear combinations
\begin{equation}
    \begin{aligned}
        l_e(k_1,\ldots,k_L)=\sum_{j=1}^L a_{ej} \, k_j, \qquad
        q_e(p_1,\ldots,p_E)=\sum_{j=1}^E b_{ej} \, p_j,
    \end{aligned}
\end{equation}
with $a_{ej},b_{ej}\in\{-1,0,1\}$, and the physical/internal mass $m_e$, can be rewritten as
\begin{align}\label{Eq:Denominator-Decomposition}
    \mathcal{D}_e(l_e(\mathbf{k}),q_e(\mathbf{p}),m_e) = [l^2_e(\mathbf{k})-M^2] + [ 2 \, l_e(\mathbf{k}) \cdot q_e(\mathbf{p}) + q^2_e(\mathbf{p}) - m_e^2 + M^2 ],
\end{align}
where the aforementioned auxiliary mass scale $M^2$ has been introduced.
Upon inversion and partial fractioning, the propagator becomes
\begin{align}\label{Eq:Exact-Tadpole-Decomposition}
    \frac{1}{\mathcal{D}_e(l_e,q_e,m_e)} = \frac{1}{l^2_e-M^2} - \frac{2 \, l_e \cdot q_e + q^2_e - m_e^2 + M^2}{l^2_e-M^2} \frac{1}{\mathcal{D}_e(l_e,q_e,m_e)},
\end{align}
henceforth suppressing the explicit dependence on $\mathbf{k}$ and $\mathbf{p}$ for brevity and readability.
In this way, the propagator decomposes into a leading UV contribution of order $\mathcal{O}(1/l_e^2)$ of tadpole type and a subleading remainder of order $\mathcal{O}(1/l_e^3)$, i.e.\ suppressed by an additional power of $1/l_e$.
This decomposition is an exact relation and can be applied recursively, thereby lowering the degree of divergence of the remnant successively.
In practice, the expansion is carried out up to the order required to capture all UV-divergent contributions, which is essentially determined by the corresponding degree of divergence. 
Power-counting finite terms can then be discarded; after \emph{proper subtraction of subdivergences}, this does not affect the overall UV behaviour.
As a result, all Feynman integrals are mapped to fully massive single-scale vacuum bubble integrals depending on $M^2$, providing an efficient way to extract UV divergences.\footnote{Note that this method does not correctly reproduce the finite parts of the integrals.}

Although Eq.~\eqref{Eq:Exact-Tadpole-Decomposition} is exact, its direct application proves to be inconvenient in automated multi-loop computations.
While the concept is simple, higher-order applications involve nontrivial complications due to subdivergences.
These issues have been analysed in detail in Ref.~\cite{Lang:2020nnl} (see also Refs.~\cite{Stockinger:2023ndm,vonManteuffel:2025swv}).

In particular, for multi-loop diagrams with subdivergences, it is not sufficient to consider only the overall degree of divergence of the whole diagram alone.
Instead, the degrees of divergence of the subdiagrams must also be taken into account.
As explained in Ref.~\cite{Lang:2020nnl}, a multi-loop application requires to organise the decomposition in terms of so-called ``loop chains'' and applying Eq.~\eqref{Eq:Exact-Tadpole-Decomposition} for each chain.
The respective expansion order is then determined by the maximum of the overall divergence degree and those of the relevant subdiagrams (those that contain the corresponding loop chain).

Another technical complication arises from the fact that the decomposition can still contain several unnecessary subleading terms with higher UV suppression than required.
This occurs because, in the second term on the RHS of Eq.~\eqref{Eq:Denominator-Decomposition}, contributions of order $\mathcal{O}(l_e^1)$ are treated on the same footing as those of order $\mathcal{O}(l_e^0)$.
Such unnecessary terms can systematically be eliminated by strict power counting and the introduction of an additional truncation operator, as discussed in Ref.~\cite{Lang:2020nnl}, though this further complicates the implementation. 

The most crucial issue of the exact decomposition method is its momentum-routing dependence in intermediate steps of the procedure.
For consistency, a one-to-one correspondence is required between the expansions of the genuine $L$-loop Feynman diagrams and their associated counterterm-inserted diagrams of lower loop order.
Individually, these contributions lose momentum-routing invariance after truncation of the tadpole expansion; only their sum remains momentum-routing independent.
In fact, a loop momentum shift $l_e\rightarrow l_e+\Delta q_e$ (with some linear combination of external momenta $\Delta q_e$) can transform a term of order $\mathcal{O}((1/l_e)^N)$ into terms of order $\mathcal{O}((1/l_e)^{\geq N})$ upon tadpole decomposition.
Some of these higher-order terms may subsequently be truncated, inducing an intermediate routing dependence, see Ref.~\cite{Lang:2020nnl}.
Consequently, independent choices of momentum routing in genuine $L$-loop diagrams and their corresponding $(<L)$-loop counterterm diagrams can spoil the cancellation of subdivergences.
In summary, a consistent application of the tadpole decomposition in Eq.~\eqref{Eq:Exact-Tadpole-Decomposition} at higher orders requires careful control of a coherent momentum routing across all diagrams.

\paragraph{The Principle of the Tadpole Decomposition via Taylor Expansion:}
Since the ``exact'' decomposition is not convenient for computer implementations, we employ an improved tadpole decomposition (see Refs.~\cite{Stockinger:2023ndm,vonManteuffel:2025swv}), as already implied in Refs.~\cite{Misiak:1994zw,Chetyrkin:1997fm} and systematically discussed in Ref.~\cite{Lang:2020nnl}, to circumvent these difficulties.
We first provide the conceptual idea based on Taylor expansion and then illustrate a proof of its validity using the $\mathcal{R}$-operation.

As before, an auxiliary mass scale $M^2$ is introduced uniformly into all propagators.
The integrand is then Taylor-expanded in the external momenta (and in the physical masses, if present), with the expansion truncated at an order determined by the overall degree of divergence $\omega$ of the considered Feynman diagram (or equivalently of the associated Green function).
For example, starting from the denominator in Eq.~\eqref{Eq:Generic-Propagator-Denominator}, we insert the auxiliary mass $\mathcal{D}_e(l_e,q_e,m_e)\rightarrow\mathcal{D}_e(l_e,q_e,m_e,M^2)=(l_e+q_e)^2-m_e^2-M^2$, and subsequently Taylor-expand the propagator in the external momenta $\mathbf{p}$ and the physical mass $m_e$, obtaining
\begin{equation}\label{Eq:TadpoleDecomposition_Taylor-Expansion}
    \begin{aligned}
        \frac{1}{\mathcal{D}_e(l_e,q_e,m_e,M^2)}
        &= \frac{1}{l_e^2 - M^2} 
        - \frac{2 \, l_e \cdot q_e}{[l_e^2 - M^2]^2} 
        - \frac{q_e^2-m_e^2}{[l_e^2 - M^2]^2} + \frac{4 (l_e \cdot q_e)^2}{[l_e^2 - M^2]^3}\\  
        &+ \frac{4 (q_e^2-m_e^2) \,l_e \cdot q_e}{[l_e^2 - M^2]^3} 
        - \frac{8 (l_e \cdot q_e)^3}{[l_e^2 - M^2]^4}\\
        &+ \frac{(q_e^2-m_e^2)^2}{[l_e^2 - M^2]^3}
        - \frac{12 (q_e^2-m_e^2) \,( l_e \cdot q_e)^2}{[l_e^2 - M^2]^4}
        + \frac{16 (l_e \cdot q_e)^4}{[l_e^2 - M^2]^5}
        + \mathcal{O}\big(\mathrm{scales}^5\big).
    \end{aligned}
\end{equation}
Since UV divergences are \emph{homogeneous} polynomials of degree $\omega$ (see Sec.~\ref{Sec:Renormalisation_in_DReg}), we expect keeping only terms of exactly this degree, rather than all terms up to and including order $\omega$, to be sufficient.
This reduction in the number of generated terms is crucial for efficiency and performance in multi-loop calculations.
However, discarding terms of order $\neq\omega$ must be handled with care, as explained below.

Rearranging the terms on the RHS of Eq.~\eqref{Eq:TadpoleDecomposition_Taylor-Expansion} gives
\begin{align}\label{Eq:Taylor-Expansion-of-Denominator-without-M}
    \frac{1}{l_e^2 - M^2} - \frac{2 \, l_e \cdot q_e + q_e^2 - m_e^2}{[l_e^2 - M^2]^2} + \frac{(2 \, l_e \cdot q_e + q_e^2 - m_e^2)^2}{[l_e^2 - M^2]^3} + \ldots,
\end{align}
which reproduces the recursive application of the exact tadpole decomposition in Eq.~\eqref{Eq:Exact-Tadpole-Decomposition}, except for numerator terms proportional to $M^2$.
Omitting such terms $\propto M^2$ requires a systematic compensation at each order, particularly in multi-loop diagrams with subdivergences (and when terms of order $\neq\omega$ are discarded).
This compensation is implemented through auxiliary counterterms proportional to $M^2$, which consistently cancel the effect of the missing numerator terms. 

\paragraph{Preliminaries:}
Before formulating and proving a theorem on the tadpole decomposition, we first need to clarify some conceptual aspects.
To define a convenient Taylor expansion operator, we rescale the external momenta and physical masses by a parameter $\lambda$, i.e.\ $\mathbf{p}\rightarrow\lambda\mathbf{p}$ (so that $q_e\rightarrow\lambda q_e$) and $m_e\rightarrow\lambda m_e$.
In addition, the auxiliary mass is rescaled according to $M^2 \rightarrow (1+\varrho^2)M^2$, since we will also need to perform Taylor expansions w.r.t.\ $M^2$.
At this point the question arises why we adopt the particular rescaling $M^2 \rightarrow (1+\varrho^2)M^2$.
The reason lies in the construction of the tadpole decomposition via Taylor expansion, see Eq.~\eqref{Eq:TadpoleDecomposition_Taylor-Expansion}. 
For a propagator denominator with this rescaling,
\begin{equation}
    \begin{aligned}
    \frac{1}{\mathcal{D}^{\nu_e}_e(l_e,\lambda q_e,\lambda m_e,(1+\varrho^2)M^2)}
    &= \frac{1}{[(l_e+\lambda q_e)^2-\lambda^2 m_e^2 - (1+\varrho^2)M^2]^{\nu_e}} \\
    &= \frac{1}{[(l_e+\lambda q_e)^2-\lambda^2 m_e^2 - \varrho^2 M^2 - M^2]^{\nu_e}},
\end{aligned}
\end{equation}
we can expand in $q_e$, $m_e$ and $M$ while still maintaining the IR-regulating dependence on $M^2$ after setting the expansion parameters $(\lambda,\varrho)$ to zero in the Taylor expansion.
With this rescaling at hand, the integrand becomes
\begin{align}
    \mathcal{I}(\mathbf{k},\mathbf{p},\mathbf{m}) \longrightarrow \mathcal{I}(\mathbf{k},\lambda\mathbf{p},\lambda\mathbf{m},(1+\varrho^2)M^2) = \frac{\mathcal{N}(\mathbf{k},\lambda\mathbf{p},\lambda\mathbf{m})}{\displaystyle \prod_{e=1}^N \mathcal{D}_e^{\nu_e}(l_e(\mathbf{k}),\lambda q_e(\mathbf{p}),\lambda m_e,(1+\varrho^2)M^2)},
\end{align}
with numerator $\mathcal{N}$.
For $\lambda=1$ and $\varrho^2=-1$, this reduces to the physical integrand $\mathcal{I}(\mathbf{k},\mathbf{p},\mathbf{m})$.

By construction, the functional dependence on the auxiliary mass $M$ is always quadratic, and thus also on the associated rescaling parameter $\varrho$.
Hence, only even numbers of derivatives w.r.t.\ $\varrho$ survive in the Taylor expansion, while odd numbers always vanish.
This allows us to restrict to derivatives w.r.t.\ $\varrho^2$ and to use the relation
\begin{align}\label{Eq:Quadratic-Functional-Dependence}
    \frac{1}{n!} \frac{\partial^n}{\partial x^n} f(x^2) \bigg|_{x=0} = \frac{1}{(n/2)!} \bigg( \frac{\partial}{\partial x^2} \bigg)^{n/2} f(x^2) \bigg|_{x^2=0}, \qquad n \ \text{even}.
\end{align}

Next, we compare the contribution of a fixed-order Taylor term when the derivatives are taken w.r.t.\ $\varrho$ or to $M$:
\begin{equation}\label{Eq:n-th_Taylor_Term_Difference}
    \begin{aligned}
        \frac{(M^2)^n}{n!} \bigg[\bigg(\frac{\partial}{\partial M^{2}}\bigg)^n f(M^2)\bigg|_{M^2=0}\bigg] &= \frac{1}{n!} (M^2)^n f^{(n)}(0),\\
        \frac{1}{n!} \bigg(\frac{\partial}{\partial \varrho^{2}}\bigg)^n f((1+\varrho^2)M^2)\bigg|_{\varrho^2=0} &= \frac{1}{n!} (M^2)^n f^{(n)}(M^2).
    \end{aligned}
\end{equation}
The two results differ in general.
However, if $f(x^2)$ is a polynomial of \emph{homogeneous} degree $n$, the RHS of both equations in \eqref{Eq:n-th_Taylor_Term_Difference} yield the same result, since the $n$ derivatives strip off the entire dependence on the argument.
This observation motivates:
\begin{lemma}\label{Lemma:n-th_Taylor_Term_for_homogeneous_polynomials}\ \\
    Let $P$ and $Q$ be two homogeneous polynomials of degrees $\omega$ and $d_Q$, respectively, related by
    \begin{align}\label{Eq:homogeneous_polynomial_in_p_m_M}
        P(\mathbf{p},\mathbf{m},M^2) = P_0 \, Q(\mathbf{p},\mathbf{m}) (M^2)^{\frac{\omega-d_Q}{2}},
    \end{align}
    with $P_0=\mathrm{const}$ and $\omega,d_Q,(\omega-d_Q)/2\in \mathbb{N}_0$, then:
    \begin{align}
        \Bigg(\frac{\partial^{\frac{\omega-d_Q}{2}}P}{\partial (M^2)^{\frac{\omega-d_Q}{2}}}\Bigg)(p,m,0) = \Bigg(\frac{\partial^{\frac{\omega-d_Q}{2}}P}{\partial (M^2)^{\frac{\omega-d_Q}{2}}}\Bigg)(p,m,M^2). 
    \end{align}
\end{lemma}
The proof follows by explicit evaluation, noting that the $(\omega-d_Q)/2$ derivatives remove the entire dependence on $M^2$.\footnote{Note that $d_Q$ may be zero, in which case $Q$ reduces to a constant.}

The Taylor operator in the external momenta and physical masses (cf.\ Eq.~\eqref{Eq:TadpoleDecomposition_Taylor-Expansion}), which extracts the term of \emph{exactly} order $\omega$, is defined as
\begin{align}\label{Eq:Taylor-Operator-wrt-m-p}
    \mathcal{T}^{(\omega)}_{\{\mathbf{p},\mathbf{m}\}} f(\mathbf{p},\mathbf{m},M) \longrightarrow \mathcal{T}^{(\omega)}_{\{\lambda\}} f(\lambda,\varrho) = \frac{1}{\omega!} \frac{\partial^\omega}{\partial\lambda^\omega} f(\lambda,\varrho) \bigg|_{\lambda=\varrho=0},
\end{align}
and similarly for the auxiliary mass (parametrised via $\varrho$).
The combined operator then becomes
\begin{align}\label{Eq:Taylor-Operator-wrt-m-p-M}
    \mathcal{T}^{(\omega)}_{\{\mathbf{p},\mathbf{m},M\}} f(\mathbf{p},\mathbf{m},M) \longrightarrow \mathcal{T}^{(\omega)}_{\{\lambda,\varrho\}} f(\lambda,\varrho) = \sum_{n=0}^\omega \frac{1}{n!(\omega-n)!} \frac{\partial^n}{\partial\lambda^n}\frac{\partial^{\omega-n}}{\partial\varrho^{\omega-n}} f(\lambda,\varrho) \bigg|_{\lambda=\varrho=0}.
\end{align}
As noted above, if $f$ depends only quadratically on $M$ (and hence on $\varrho$), odd numbers of derivatives w.r.t.\ $\varrho$ vanish, while even numbers reduce according to Eq.~\eqref{Eq:Quadratic-Functional-Dependence}.
It is therefore convenient to introduce, for even $n\in\mathbb{N}_0$, the operator 
\begin{align}\label{Eq:Taylor-Operator-lambda-rho-separately}
    \mathcal{T}^{(\omega-n,n)}_{\{\lambda,\varrho^2\}} f(\lambda,\varrho^2) = \frac{1}{(\omega-n)!(n/2)!} \frac{\partial^{\omega-n}}{\partial\lambda^{\omega-n}}\frac{\partial^{n/2}}{\partial(\varrho^2)^{n/2}} f(\lambda,\varrho^2) \bigg|_{\lambda=\varrho^2=0}, \quad \omega\geq n, \,\,\, n \ \text{even},
\end{align}
which corresponds to a single term in the sum of Eq.~\eqref{Eq:Taylor-Operator-wrt-m-p-M}, already incorporating the $\varrho^2$-dependence in the sense of Eq.~\eqref{Eq:Quadratic-Functional-Dependence}.
For later use, we also introduce the Heaviside step function
\begin{align}
    \Theta(x)=
    \begin{cases}
        0, \quad &x < 0\\
        1, \quad &x \geq0.
    \end{cases}
\end{align}

Finally, recall from Sec.~\ref{Sec:Renormalisation_Theory} the operator $\mathcal{K}$, which extracts the UV divergent part of a Feynman integral.
For an $L$-loop function 
\begin{align}
    F^{(L)}(\epsilon) = \sum_{n=1}^L \frac{C^{(L)}_n}{\epsilon^n} + C^{(L)}_0 + \mathcal{O}(\epsilon),
\end{align}
its action in a minimal subtraction scheme (assumed here) is given by
\begin{align}
    \mathcal{K}\cdot F^{(L)}(\epsilon) = \sum_{n=1}^L \frac{C^{(L)}_n}{\epsilon^n}.
\end{align}
Moreover, we will also need the interplay between derivatives and the $\mathcal{R}$-operation:
\begin{lemma}\label{Lemma:R-operation_commutes_with_derivatives}\ \\
    Derivatives w.r.t.\ external momenta $p_i$ (and masses $m_i$) commute with the $\overline{\mathcal{R}}$-operation,
    \begin{align}
        [\partial,\overline{\mathcal{R}}] = 0,
    \end{align}
    and likewise with the full $\mathcal{R}$-operation, provided that no IR divergences are generated.
\end{lemma}
A proof of this lemma can be found in Refs.~\cite{Caswell:1981ek,Kleinert:2001}.

\paragraph{Tadpole Decomposition via Taylor Expansion:}
With all preliminaries in place and using the Taylor operators in Eqs.~\eqref{Eq:Taylor-Operator-wrt-m-p}--\eqref{Eq:Taylor-Operator-lambda-rho-separately}, we can now state the central result of this section:
\begin{theorem}[Tadpole Decomposition]\label{Thm:Tadpole-Decomposition}\ \\
    Let $G$ be a 1PI Feynman graph of $L$-loop order with overall degree of divergence $\omega$, external momenta $\mathbf{p}$ and physical masses $\mathbf{m}$.
    Further, let $G_M$ denote the corresponding graph in which the auxiliary mass scale $M^2$ is introduced into all propagators, together with the rescaling $(\mathbf{p},\mathbf{m},M^2)\rightarrow(\lambda\mathbf{p},\lambda\mathbf{m},(1+\varrho^2)M^2)$.
    
    Then the overall counterterm of the graph $G$ is given by
    \begin{align}\label{Eq:Counterterm-from-Tadpole-Decomposition-Method}
        C(G) = C\big(\mathcal{T}^{(\omega)}_{\{\lambda,\varrho\}}G_M\big) - C(M_G^2) = \mathop{\mathrm{lim}}_{M^2 \, \to \, 0} C\big(\mathcal{T}^{(\omega)}_{\{\lambda,\varrho\}}G_M\big) = C\big(\mathcal{T}^{(\omega)}_{\{\lambda\}}G_M\big),
    \end{align}
    where the counterterm of a graph is defined, following Sec.~\ref{Sec:Renormalisation_Theory}, via the $\mathcal{R}$-operation as
    \begin{align}
        C(G)=-\mathcal{K}\cdot\overline{\mathcal{R}}(G).
    \end{align}
    Here $C(M_G^2)$ denotes the auxiliary counterterms proportional to $M^2$, obtained as
    \begin{align}\label{Eq:Auxiliary-Counterterms-in-Tadpole-Decomposition-Method}
        C(M_G^2)=\Theta(\omega-2) \Bigg[ - \mathcal{K}\cdot\overline{\mathcal{R}}\Bigg(\sum_{n=1}^{\lfloor\omega/2\rfloor}\mathcal{T}_{\{\lambda,\varrho^2\}}^{(\omega-2n,2n)}G_M\Bigg) \Bigg].
    \end{align}
    The auxiliary counterterms $C(M_H^2)$ of all divergent subgraphs $H\subset G$ must be included in the renormalisation procedure of Eq.~\eqref{Eq:Counterterm-from-Tadpole-Decomposition-Method} to ensure a consistent subtraction of subdivergences.
    Accordingly, the limit in Eq.~\eqref{Eq:Counterterm-from-Tadpole-Decomposition-Method} must only be taken after proper subrenormalisation.
\end{theorem}
\begin{proof}\ \\
    First, we recall from Sec.~\ref{Sec:Renormalisation_in_DReg} that the UV divergent part $\mathcal{K}\cdot\overline{\mathcal{R}}(G)$ of a Feynman graph $G$ is a local polynomial in the physical variables $(\mathbf{p},\mathbf{m})$ of homogeneous degree $\omega$, with $\epsilon$-poles bounded by the number of loops $L$.
    It has the form
    \begin{align}\label{Eq:CT-Result_original_Graph_G}
        C(G)=-\mathcal{K}\cdot\overline{\mathcal{R}}(G)=-\sum_{n=1}^L \frac{P_G^{(n)}(\mathbf{p},\mathbf{m})}{\epsilon^n}=-\mathcal{P}_G(\mathbf{p},\mathbf{m},\epsilon).
    \end{align}
    For the modified graph $G_M$, obtained by inserting the auxiliary mass $M^2$ into all propagators, we similarly obtain
    \begin{align}\label{Eq:CT-Result_modified_Graph_G_M}
        C(G_M)=-\mathcal{K}\cdot\overline{\mathcal{R}}(G_M)=-\sum_{n=1}^L \frac{P_{G_M}^{(n)}(\mathbf{p},\mathbf{m},M^2)}{\epsilon^n}=-\mathcal{P}_{G_M}(\mathbf{p},\mathbf{m},M^2,\epsilon),
    \end{align}
    with purely quadratic dependence on $M$.
    
    In the tadpole decomposition, however, we do not evaluate $G_M$ directly but replace it by the fixed-order term $\mathcal{T}^{(\omega)}_{\{\lambda,\varrho\}}G_M$ of its Taylor expansion, whose order equals the graph's degree of divergence $\omega$.
    The associated counterterm satisfies
    \begin{equation}\label{Eq:CT-G_M-Invariant_under_Taylor_Expansion}
        \begin{aligned}
            C\big(\mathcal{T}^{(\omega)}_{\{\lambda,\varrho\}}G_M\big) &= - \mathcal{K}\cdot\overline{\mathcal{R}}\big(\mathcal{T}^{(\omega)}_{\{\lambda,\varrho\}}G_M\big)\\
            &= - \mathcal{T}^{(\omega)}_{\{\lambda,\varrho\}} \, \mathcal{K}\cdot\overline{\mathcal{R}}(G_M) = \mathcal{T}^{(\omega)}_{\{\lambda,\varrho\}} \, C(G_M)\\
            &= - \mathcal{T}^{(\omega)}_{\{\lambda,\varrho\}} \sum_{n=1}^L \frac{P_{G_M}^{(n)}(\lambda\mathbf{p},\lambda\mathbf{m},(1+\varrho^2)M^2)}{\epsilon^n}
            = - \sum_{n=1}^L \frac{P_{G_M}^{(n)}(\mathbf{p},\mathbf{m},M^2)}{\epsilon^n}\\
            &= C(G_M).
        \end{aligned}        
    \end{equation}
    Here, we first apply the definition of the counterterm via the recursive $\overline{\mathcal{R}}$-operation.
    In the second line, we commute the Taylor operator $\mathcal{T}^{(\omega)}_{\{\lambda,\varrho\}}$ with the $\overline{\mathcal{R}}$-operation using that it is a combination of derivatives (see Eq.~\eqref{Eq:Taylor-Operator-wrt-m-p-M}) and utilising lemma~\ref{Lemma:R-operation_commutes_with_derivatives}.
    The third line exploits that a Taylor operator extracting exactly the order-$\omega$ term reproduces the homogeneous polynomial of order $\omega$ it is applied to, with lemma~\ref{Lemma:n-th_Taylor_Term_for_homogeneous_polynomials} ensuring the correct $M^2$-dependence.
    Hence, the Tadpole-expanded graph $G_M$ yields the same counterterm as $G_M$ itself.

    Since $G_M$ reproduces the original graph $G$ as $G_M \to G$ in the limit $M^2\to 0$ (equivalently $\lambda=1$, $\varrho^2=-1$ after rescaling), and since $G_M$ depends quadratically on $M^2$, locality implies that the difference between Eqs.~\eqref{Eq:CT-Result_original_Graph_G} and \eqref{Eq:CT-Result_modified_Graph_G_M} must be a homogeneous polynomial of degree $\omega$ in $(\mathbf{p},\mathbf{m})$ \emph{and} $M^2$ (if non-vanishing).
    In particular, we find
    \begin{equation}\label{Eq:Difference_CT_G_and_CT_GM}
        \begin{aligned}
            C(G_M) - C(G) &= -\sum_{n=1}^L \frac{P_{G_M}^{(n)}(\mathbf{p},\mathbf{m},M^2)}{\epsilon^n} + \sum_{n=1}^L \frac{P_G^{(n)}(\mathbf{p},\mathbf{m})}{\epsilon^n}\\
            &= - \mathcal{T}^{(\omega)}_{\{\lambda,\varrho\}} \Bigg( \sum_{n=1}^L \frac{P_{G_M}^{(n)}(\lambda\mathbf{p},\lambda\mathbf{m},(1+\varrho^2)M^2)}{\epsilon^n} - \sum_{n=1}^L \frac{P_G^{(n)}(\lambda\mathbf{p},\lambda\mathbf{m})}{\epsilon^n} \Bigg)\\
            &= - \bigg( \mathcal{T}^{(\omega)}_{\{\lambda\}} + \Theta(\omega-2)\sum_{k=1}^{\lfloor\omega/2\rfloor}\mathcal{T}_{\{\lambda,\varrho^2\}}^{(\omega-2k,2k)} \bigg) \\
            &\hspace{0.8cm}\times\Bigg( \sum_{n=1}^L \frac{P_{G_M}^{(n)}(\lambda\mathbf{p},\lambda\mathbf{m},(1+\varrho^2)M^2)}{\epsilon^n} - \sum_{n=1}^L \frac{P_G^{(n)}(\lambda\mathbf{p},\lambda\mathbf{m})}{\epsilon^n} \Bigg)\\
            &= - \sum_{n=1}^L \Bigg[ \mathcal{T}^{(\omega)}_{\{\lambda\}} \Bigg( \frac{P_{G_M}^{(n)}(\lambda\mathbf{p},\lambda\mathbf{m},(1+\varrho^2)M^2)}{\epsilon^n} - \frac{P_G^{(n)}(\lambda\mathbf{p},\lambda\mathbf{m})}{\epsilon^n} \Bigg) \\
            &\hspace{0.8cm} + \Theta(\omega-2)\sum_{k=1}^{\lfloor\omega/2\rfloor}\mathcal{T}_{\{\lambda,\varrho^2\}}^{(\omega-2k,2k)} \frac{P_{G_M}^{(n)}(\lambda\mathbf{p},\lambda\mathbf{m},(1+\varrho^2)M^2)}{\epsilon^n} \Bigg]\\
            &= - \Theta(\omega-2) \sum_{k=1}^{\lfloor\omega/2\rfloor}\mathcal{T}_{\{\lambda,\varrho^2\}}^{(\omega-2k,2k)} \sum_{n=1}^L \frac{P_{G_M}^{(n)}(\lambda\mathbf{p},\lambda\mathbf{m},(1+\varrho^2)M^2)}{\epsilon^n}\\
            &= - \Theta(\omega-2) \sum_{n=1}^L \frac{Q^{(n)}_{G,0} Q^{(n)}_G(\mathbf{p},\mathbf{m}) (M^2)^{\frac{\omega-d_Q}{2}}}{\epsilon^n}
        \end{aligned}        
    \end{equation}
    Here, we used again that the Taylor operator $\mathcal{T}^{(\omega)}_{\{\lambda,\varrho\}}$ exactly reproduces the homogeneous polynomial of degree $\omega$ it acts on.
    Then we decomposed the Taylor operator into a part with only $\lambda$-derivatives and a part involving derivatives w.r.t.\ $\varrho^2$ (see Eqs.~\eqref{Eq:Taylor-Operator-wrt-m-p}--\eqref{Eq:Taylor-Operator-lambda-rho-separately}), using that the dependence on $\varrho$ is quadratic.
    We note that only the UV-divergent contribution of $G_M$ carries $\varrho$-dependence.
    The Heaviside function enforces that the difference vanishes for $\omega\leq1$ and, if present, contains at least one power of $M^2$ (corresponding to at least one $\varrho^2$-derivative).
    The penultimate line uses that the Taylor operator $\mathcal{T}^{(\omega)}_{\{\lambda\}}$ simply extracts the $(\mathbf{p},\mathbf{m})$-parts of both polynomials, which must be identical for $G$ and $G_M$ since their dependence on $(\mathbf{p},\mathbf{m})$ is unchanged and the limit $G_M \to G$ as $M^2\to 0$ exists.
    In the last line, the difference is parametrised in analogy with lemma~\ref{Lemma:n-th_Taylor_Term_for_homogeneous_polynomials}; $Q^{(n)}_{G,0}$ is a constant, $Q^{(n)}_G(\mathbf{p},\mathbf{m})$ is a homogeneous polynomial in $(\mathbf{p},\mathbf{m})$ of degree $0\leq d_Q < \omega$, the exponent $(\omega-d_Q)/2\in \mathbb{N}$ is a strictly positive integer, and $\Theta(\omega-2)$ ensures a $M^2$-depending or otherwise vanishing difference.
    Thus, we find 
    \begin{align}
        P_{G_M}^{(n)}(\mathbf{p},\mathbf{m},M^2) = P_{G}^{(n)}(\mathbf{p},\mathbf{m})+\Theta(\omega-2) Q^{(n)}_{G,0} Q^{(n)}_G(\mathbf{p},\mathbf{m}) (M^2)^{\frac{\omega-d_Q}{2}}, \qquad \forall \,\, n,
    \end{align}
    and
    \begin{align}
        \mathop{\mathrm{lim}}_{M^2 \, \to \, 0} P_{G_M}^{(n)}(\mathbf{p},\mathbf{m},M^2) \equiv P_{G_M}^{(n)}(\mathbf{p},\mathbf{m},0) = P_G^{(n)}(\mathbf{p},\mathbf{m}), \qquad \forall \,\, n,
    \end{align}
    due to the local nature of subrenormalised UV-divergent contributions.

    From Eqs.~\eqref{Eq:CT-G_M-Invariant_under_Taylor_Expansion} and \eqref{Eq:Difference_CT_G_and_CT_GM} we deduce that the difference $C(G_M)-C(G)$ is indeed equal to the auxiliary counterterm $C(M_G^2)$ of Eq.~\eqref{Eq:Auxiliary-Counterterms-in-Tadpole-Decomposition-Method}.
    Furthermore, since the difference, if present, must depend on $M^2$, and $\mathcal{T}^{(\omega)}_{\{\lambda\}}$ extracts only the $(\mathbf{p},\mathbf{m})$-parts, Eq.~\eqref{Eq:Counterterm-from-Tadpole-Decomposition-Method} holds true and provides the desired expression for $C(G)$.
    
    Due to the difference $C(M_G^2)=C(G_M)-C(G)$ at each order, consistency requires proper subrenormalisation, including all non-vanishing auxiliary counterterms $C(M_H^2)$ associated with the divergent subgraphs $H\subset G$.
\end{proof}
In this thesis, we only work with power-counting renormalisable theories, where the degree of divergence of a Feynman graph is equal to the mass dimension of the corresponding Green function and bounded by the spacetime dimension. This motivates the following corollary:
\begin{corollary}[Application of the Tadpole Decomposition]\label{Corollary:Practical_Application_of_the_Tadpole_Decomposition}\ \\
    Let $G$ and $G_M$ be 1PI Feynman graphs as in theorem~\ref{Thm:Tadpole-Decomposition}, in a power-counting renormalisable theory with $\omega<4$.
    Then the overall counterterm of $G$ is given by 
    \begin{align}\label{Eq:Application-of-Tadpole-Decomposition-CT-of-G}
        C(G) = C\big(\mathcal{T}^{(\omega)}_{\{\lambda\}}G_M\big) = - \mathcal{K}\cdot\overline{\mathcal{R}}\big(\mathcal{T}_{\{\lambda\}}^{(\omega)}G_M\big),
    \end{align}
    and the auxiliary counterterm is obtained as
    \begin{align}\label{Eq:Application-of-Tadpole-Decomposition-Auxiliary-Counterterm}
        C(M_G^2)=\Theta(\omega-2) \Big[ - \mathcal{K}\cdot\overline{\mathcal{R}}\big(\mathcal{T}_{\{\lambda\}}^{(\omega-2)}G_M\big) \Big],
    \end{align}
    which is required for consistent subrenormalisation.
\end{corollary}
\begin{proof}
    The determination of the counterterm $C(G)$ in Eq.~\eqref{Eq:Application-of-Tadpole-Decomposition-CT-of-G} follows directly from theorem~\ref{Thm:Tadpole-Decomposition} (cf.\ Eq.~\eqref{Eq:Counterterm-from-Tadpole-Decomposition-Method}).
    The evaluation of the auxiliary counterterm in Eq.~\eqref{Eq:Application-of-Tadpole-Decomposition-Auxiliary-Counterterm} also proceeds along the same lines as the proof of theorem~\ref{Thm:Tadpole-Decomposition}.
    Starting from Eq.~\eqref{Eq:Difference_CT_G_and_CT_GM} and using that $\omega<4$, we obtain
    \begin{equation}
        \begin{aligned}
            C(M_G^2)&=C(G_M)-G(G)\\
            &= - \Theta(\omega-2) \sum_{k=1}^{\lfloor\omega/2\rfloor}\mathcal{T}_{\{\lambda,\varrho^2\}}^{(\omega-2k,2k)} \sum_{n=1}^L \frac{P_{G_M}^{(n)}(\lambda\mathbf{p},\lambda\mathbf{m},(1+\varrho^2)M^2)}{\epsilon^n}\\
            &= - \Theta(\omega-2) \mathcal{T}_{\{\lambda,\varrho^2\}}^{(\omega-2,2)} \sum_{n=1}^L \frac{P_{G_M}^{(n)}(\lambda\mathbf{p},\lambda\mathbf{m},(1+\varrho^2)M^2)}{\epsilon^n}\\
            &= - \frac{\Theta(\omega-2)}{(\omega-2)!} \frac{\partial^{\omega-2}}{\partial\lambda^{\omega-2}}\frac{\partial}{\partial\varrho^2} \sum_{n=1}^L \frac{P_{G_M}^{(n)}(\lambda\mathbf{p},\lambda\mathbf{m},(1+\varrho^2)M^2)}{\epsilon^n} \bigg|_{\lambda=\varrho^2=0}\\
            &= - \Theta(\omega-2) \frac{1}{(\omega-2)!} \frac{\partial^{\omega-2}}{\partial\lambda^{\omega-2}} \sum_{n=1}^L \frac{P_{G_M}^{(n)}(\lambda\mathbf{p},\lambda\mathbf{m},(1+\varrho^2)M^2)}{\epsilon^n} \bigg|_{\lambda=\varrho^2=0}\\
            &= - \Theta(\omega-2) \, \mathcal{T}_{\{\lambda\}}^{(\omega-2)} \, \mathcal{K}\cdot\overline{\mathcal{R}}(G_M)\\
            &= - \Theta(\omega-2) \, \mathcal{K}\cdot\overline{\mathcal{R}}\big(\mathcal{T}_{\{\lambda\}}^{(\omega-2)}G_M\big).
        \end{aligned}
    \end{equation}
    Here we first used that the sum over $k$ in the second line reduces to a single term, since $\omega<4$ in power-counting renormalisable theories.
    The corresponding Taylor operator is then written explicitly using its definition in Eq.~\eqref{Eq:Taylor-Operator-lambda-rho-separately}.
    Because the $P_{G_M}^{(n)}$ are homogeneous polynomials whose $\varrho$-part, if present, is exactly proportional to $\varrho^2$ (no higher powers in $(1+\varrho^2)M^2$ occur by power counting), lemma~\ref{Lemma:n-th_Taylor_Term_for_homogeneous_polynomials} ensures that the derivative w.r.t.\ $\varrho^2$ reproduces the polynomial itself once $\varrho^2=0$ is imposed.
    With the chosen rescaling $(1+\varrho^2)M^2$, simply setting $\varrho^2=0$ suffices to obtain the same result, such that the $\varrho^2$-derivative can be dropped.
    The remaining $\lambda$-derivatives are collected into the Taylor operator $\mathcal{T}_{\{\lambda\}}^{(\omega-2)}$ (see Eq.~\eqref{Eq:Taylor-Operator-wrt-m-p}).
    Finally, lemma~\ref{Lemma:R-operation_commutes_with_derivatives} allows to commute the derivatives --- and hence 
    the Taylor operator --- with the $\overline{\mathcal{R}}$-operation, yielding the stated result.
\end{proof}
\begin{remark}[on the Tadpole Decomposition]\ \\
    The crucial difference between theorem~\ref{Thm:Tadpole-Decomposition} and corollary~\ref{Corollary:Practical_Application_of_the_Tadpole_Decomposition} is that the corollary involves derivatives only w.r.t.\ $\lambda$, and not w.r.t.\ $\varrho$.
    Consequently, the rescaling $M^2\to (1+\varrho^2)M^2$, originally introduced to preserve the IR-regulating effect of $M^2$ during Taylor expansion in this parameter, becomes unnecessary.
    The expansion may therefore be restricted to $\lambda$, and can in fact be carried out directly in the physical parameters $(\mathbf{p},\mathbf{m})$, omitting the rescaling entirely and without the subtleties concerning the IR behaviour discussed in the preliminaries.

    In principle, this simplification is possible for arbitrary $\omega$.
    However, in power-counting renormalisable theories, where $\omega<4$, the auxiliary counterterm, if present, reduces to a homogeneous polynomial strictly proportional to $M^2$ and can be extracted by a single Taylor operator, see Eq.~\eqref{Eq:Application-of-Tadpole-Decomposition-Auxiliary-Counterterm}.
    Thus, only two cases occur:
    \begin{enumerate}[label={$(\arabic*)$}]
        \item $\omega\leq1$: No auxiliary counterterm is generated, i.e.\ $C(M_G^2)=0$.
        \item $2\leq\omega<4$: The auxiliary counterterm is strictly proportional to $M^2$, with general form
        \begin{enumerate}[label={$(\alph*)$}]
            \item $\omega=2$: $C(M_G^2)= \alpha(\epsilon) M^2$,
            \item $\omega=3$: $C(M_G^2)= \sum_i \alpha_i(\epsilon) \, p_i M^2 + \sum_j \beta_j(\epsilon) \, m_j M^2$.
        \end{enumerate}
    \end{enumerate}
    
    Both the auxiliary mass parameter $M^2$ and the auxiliary counterterms serve purely technical purposes in the evaluation of Feynman diagrams.
    They are not intrinsic to the theory and can therefore be regarded as a mathematical trick.
    In particular, the auxiliary counterterm associated with the gauge boson does not affect gauge invariance, as it never enters the action.

    A further subtlety is that the Tadpole decomposition does not commute with numerator-denominator cancellations.
    For example, inserting a counterterm $\propto\slashed{l}_e$ into a massless fermion line naively suggests the cancellation of $l_e^2$ between numerator and denominator.
    However, such cancellation must not be performed before introducing $M^2$ into all propagators.
    With fully massive propagators the above cancellation is obstructed, and in fact consistency requires that the auxiliary mass is inserted into all propagators of the integrand prior to any such simplification.
    In summary,
    $$\frac{\slashed{l}_e\slashed{l}_e\slashed{l}_e}{[l_e^2]^2}=\frac{\slashed{l}_e l_e^2}{[l_e^2]^2}=\frac{\slashed{l}_e}{l_e^2} \longrightarrow \frac{\slashed{l}_e\slashed{l}_e\slashed{l}_e}{[l_e^2-M^2]^2} = \frac{\slashed{l}_e l_e^2}{[l_e^2-M^2]^2} \neq \frac{\slashed{l}_e}{l_e^2-M^2}.$$

    The improved Tadpole decomposition based on Taylor expansion overcomes the shortcomings of the ``exact'' tadpole decomposition introduced above.
    It manifestly preserves momentum-routing invariance at every step of the procedure and requires only the overall degree of divergence (not that of the subdiagrams), making it well suited for automated multi-loop computations.
    Its main drawback, however, is the proliferation of terms generated by the decomposition.
    This issue is alleviated by the fact that only fixed-order terms of degree $\omega$, and, for $2\leq\omega<4$, terms of order $\omega-2$ to account for auxiliary counterterms $\propto M^2$, are needed.\footnote{These extra terms of order $\omega-2$ are the reason why discarding terms $\neq\omega$ must be handled with care, as mentioned above.}
    Nevertheless, for $\omega\geq1$ the growth in number of terms can still be substantial, especially for Green functions with higher degrees of divergence.
    This becomes particularly relevant for $\Delta$-operator inserted Green functions (see Eq.~\eqref{Eq:1PI-Green-Functions-MomentumSpace-with-DeltaInsertion}), since the ghost has mass dimension $0$.
    For instance, the ghost-gauge boson Green function $\langle \Delta \mathcal{V}^{B}_\mu c^A \rangle^{\mathrm{1PI}}$ has degree of divergence $\omega=3$, while the 4-point function $\langle \Delta \mathcal{V}^{D}_\rho \mathcal{V}^{C}_\nu \mathcal{V}^{B}_\mu c^A \rangle^{\mathrm{1PI}}$ has degree of divergence $\omega=1$, making them particularly expensive in terms of memory and runtime.
\end{remark}

Based on corollary~\ref{Corollary:Practical_Application_of_the_Tadpole_Decomposition} and the discussion above, we propose the following streamlined algorithm for the practical application of the tadpole decomposition method, which was also employed in the computations of this thesis:
\begin{algorithm}[Practical Application of the Tadpole Decomposition]\ \\
    In a power-counting renormalisable theory with $\omega<4$, apply the following steps to every Feynman diagram, including counterterm diagrams:
    \begin{enumerate}[label={$(\arabic*)$}]
        \item Insert an auxiliary mass scale $M^2$ into all propagators: $G\to G_M$.
        \item Perform a Taylor expansion in the external momenta and physical masses $(\mathbf{p},\mathbf{m})$:
            \begin{enumerate}[label={$(\alph*)$}]
                \item $\omega\leq1$: extract the order-$\omega$ term using the Taylor operator $\mathcal{T}^{(\omega)}_{\{\mathbf{p},\mathbf{m}\}}$.
                \item $2\leq\omega<4$: extract the order-$\omega$ \emph{and} order-$(\omega-2)$ terms using $\mathcal{T}^{(\omega)}_{\{\mathbf{p},\mathbf{m}\}}$ and $\mathcal{T}^{(\omega-2)}_{\{\mathbf{p},\mathbf{m}\}}$.
            \end{enumerate}
        \item Reduce the resulting fully massive single-scale vacuum bubble integrals via IBP relations and evaluate the corresponding master integrals.
        \item Determine the counterterm $C(G)$ form the order-$\omega$ terms, and, if present, the auxiliary counterterm $C(M_G^2)$ from the order-$(\omega-2)$ terms.
        For consistent subrenormalisation, all lower-order auxiliary counterterms $C(M_H^2)$ of divergent subgraphs $H\subset G$ must be included in the counterterm diagrams.
    \end{enumerate}
\end{algorithm}

\begin{remark}[Auxiliary Counterterms in the context of this Thesis]\ \\
    In line with the original work of Refs.~\cite{Misiak:1994zw,Chetyrkin:1997fm}, we also encountered auxiliary counterterms for the gauge fields in the computations of this thesis.
    Within the BMHV scheme, such terms can appear as mass counterterms for either the 4-dimensional or the evanescent component of the gauge boson.
    They also arise in the renormalisation of the insertion operator $\Delta$ (see chapter~\ref{Chap:Practical_Symmetry_Restoration}).
    For example, the $\Delta$-inserted Green function $\langle \Delta \mathcal{V}^{B}_\mu c^A \rangle^{\mathrm{1PI}}$, which has divergence degree~$3$ and encodes the breaking of gauge boson transversality, produces an auxiliary counterterm~$\propto M^2 \overline{p}^\mu$.
    Due to evanescent contributions, which can cancel $1/\epsilon$-poles, this counterterm even acquires a finite part.
    In the Standard Model (see chapter~\ref{Chap:The_Standard_Model}), the ghost-scalar 3-point functions $\langle \Delta \phi_b\phi_a^\dagger c^A \rangle^{\mathrm{1PI}}$ and $\langle \Delta \phi_b\phi_a c^A \rangle^{\mathrm{1PI}}$ provide examples of mass dimension~$2$ Green functions that generate auxiliary counterterms~$\propto M^2$, despite their 3-point structure.
    
    Analogous to the gauge boson, the scalar self energy produces a mass counterterm~$\propto M^2$.
    By contrast, fermion self energies do not yield such counterterms, as they have mass dimension~$1$.
    Likewise, although the ghost self energy has mass dimension~$2$, it does not generate an auxiliary counterterm.
    The reason is that ghost fields always couple through a ghost-gauge boson vertex proportional to the momentum of the outgoing ghost, making the corresponding integrand proportional to the external momentum.
    As a result, the counterterm is a quadratic polynomial only in the external momentum, without any $M^2$ contribution.
\end{remark}

\begin{remark}[``Exact'' vs.\ Improved Tadpole Decomposition]\ \\
    In this section, we introduced the rescaling $M^2\to (1+\varrho^2)M^2$ for the auxiliary mass parameter.
    By contrast, Ref.~\cite{Lang:2020nnl} employed $M^2\to (1-\varrho^2)M^2$, i.e.\ with opposite sign.
    The difference amounts to a global minus sign in all terms involving an odd number of $\varrho^2$-derivatives, such as those generated by the Taylor operator in Eq.~\eqref{Eq:Taylor-Operator-lambda-rho-separately}.
    
    On the one hand, the advantage of our convention is that, for $\omega<4$, the operators $\mathcal{T}^{(\omega-2,0)}_{\{\lambda,\varrho^2\}}$ (acting with $(\omega-2)$ $\lambda$-derivatives, followed by setting $\lambda=\varrho^2=0$) and $\mathcal{T}^{(\omega-2,2)}_{\{\lambda,\varrho^2\}}$ (a single $\varrho^2$-derivative in addition to the $(\omega-2)$ $\lambda$-derivatives and then setting $\lambda=\varrho^2=0$) yield identical results when applied to the modified graph $G_M$.
    This property was explicitly used in corollary~\ref{Corollary:Practical_Application_of_the_Tadpole_Decomposition} to eliminate $\varrho^2$-derivatives altogether.
    With the opposite convention, the two contributions would instead differ by a global sign.
    This is not incorrect, but it must be accounted for when determining auxiliary counterterms.

    On the other hand, the convention of Ref.~\cite{Lang:2020nnl} offers a conceptual advantage when comparing the improved tadpole decomposition via Taylor expansion with the ``exact'' decomposition of Eq.~\eqref{Eq:Exact-Tadpole-Decomposition}.
    To illustrate this, consider the Taylor expansion of the denominator $\mathcal{D}_e(l_e,\lambda q_e,\lambda m_e,(1+\varrho^2)M^2)$ in $(\lambda,\varrho^2)$ around $\lambda=\varrho^2=0$, which produces
    \begin{align}\label{Eq:Taylor-Expansion-of-Denominator-with-plus-convention}
        \frac{1}{l_e^2 - M^2} - \frac{2 \, l_e \cdot q_e + q_e^2 - m_e^2 - M^2}{[l_e^2 - M^2]^2} + \frac{(2 \, l_e \cdot q_e + q_e^2 - m_e^2 - M^2)^2}{[l_e^2 - M^2]^3} + \ldots,
    \end{align}
    whereas the other sign convention, i.e.\ $\mathcal{D}_e(l_e,\lambda q_e,\lambda m_e,(1-\varrho^2)M^2)$, yields
    \begin{align}\label{Eq:Taylor-Expansion-of-Denominator-with-minus-convention}
        \frac{1}{l_e^2 - M^2} - \frac{2 \, l_e \cdot q_e + q_e^2 - m_e^2 + M^2}{[l_e^2 - M^2]^2} + \frac{(2 \, l_e \cdot q_e + q_e^2 - m_e^2 + M^2)^2}{[l_e^2 - M^2]^3} + \ldots,
    \end{align}
    so that both expansions differ only by the sign of the numerator $M^2$-term (existent contrary to the expansion in Eq.~\eqref{Eq:Taylor-Expansion-of-Denominator-without-M}).
    Now, Eq.~\eqref{Eq:Taylor-Expansion-of-Denominator-with-minus-convention}, corresponding to the sign convention of Ref.~\cite{Lang:2020nnl}, exactly resembles the recursive application of the exact decomposition in Eq.~\eqref{Eq:Exact-Tadpole-Decomposition}, including all numerator terms~$\propto M^2$.
    Our convention (see Eq.~\eqref{Eq:Taylor-Expansion-of-Denominator-with-plus-convention}), by contrast, flips their sign.

    If one applies the full Taylor series up to and including order $\omega$, the sign choice in Ref.~\cite{Lang:2020nnl} has the appealing feature that contributions from $(\omega-2)$ $\lambda$-derivatives cancel exactly against those from $(\omega-2)$ $\lambda$-derivatives plus one $\varrho^2$-derivative, and so forth at higher orders.
    This cancellation eliminates the need for auxiliary counterterms, since the $M^2$ numerator terms are preserved rather than omitted.
    In this sense, the Taylor expansion then fully reproduces the ``exact'' tadpole decomposition.
    In our convention, these contributions add instead of cancelling, as they are equivalent.
    A fact that we exploited in corollary-\ref{Corollary:Practical_Application_of_the_Tadpole_Decomposition} to avoid $\varrho^2$-derivatives entirely.

    From a practical standpoint, however, this difference is not a drawback.
    We never intended to keep the entire Taylor series up to and including order $\omega$, since this leads to an immense proliferation in the number of terms.
    Restricting the expansion to fixed-order operators that extract only the required contributions, as done in our algorithm, is far more efficient.
    Moreover, the fact that $(\omega-2)$ $\lambda$-derivative terms equal those obtained from $(\omega-2)$ $\lambda$-derivatives plus one $\varrho^2$-derivative makes our convention especially convenient in the proofs, as all derivatives are treated on the same footing without having to deal with minus signs.

    In summary, all variants of the tadpole decomposition method yield the same physical counterterms, provided they are applied consistently:
    \begin{itemize}
        \item In the ``exact'' tadpole decomposition, momentum routing and recursion order must be carefully controlled.
        \item In the ``full'' Taylor expansion, the sign convention $(1-\varrho^2)M^2$ form Ref.~\cite{Lang:2020nnl} resembles the complete recursive application of the exact decomposition and thus avoids auxiliary counterterms. 
        \item In the improved tadpole decomposition with fixed-order Taylor expansion adopted in this thesis, both sign conventions are admissible, but must be applied consistently when constructing auxiliary counterterms to ensure correct signs.
    \end{itemize}
\end{remark}
To conclude this section, we present a representative example illustrating the concepts discussed above.
Specifically, we consider the $\Delta$-operator inserted ghost-gauge boson Green function $\langle \Delta B^\mu c \rangle^{\mathrm{1PI}}$ in an Abelian chiral gauge theory within the BMHV scheme (see chapters~\ref{Chap:General_Abelian_Chiral_Gauge_Theory} and \ref{Chap:BMHV_at_Multi-Loop_Level}). 
This example is chosen for three reasons:
first, it involves an operator-inserted Green function specific to the BMHV framework employed in this thesis and is absent in vector-like gauge theories;
second, it possesses an overall degree of divergence $\omega=3$, making it a nontrivial case;
and third, in the right-handed model discussed in chapter~\ref{Chap:BMHV_at_Multi-Loop_Level}, the only other Green function that gives rise to a nonvanishing auxiliary counterterm is the gauge boson self energy $\langle B^\nu B^\mu\rangle^{\mathrm{1PI}}$, which is a familiar example in the context of the tadpole decomposition.
The present example serves as a schematic illustration with generic coefficients, while explicit results for the auxiliary counterterms of both $\langle B^\nu B^\mu\rangle^{\mathrm{1PI}}$ and $\langle \Delta B^\mu c \rangle^{\mathrm{1PI}}$, computed up to the 4-loop level, are provided in Sec.~\ref{Sec:AuxiliaryCTs-RHTheory}.
\begin{example}[Ghost-Gauge Boson $\langle \Delta B^\mu c \rangle^{\mathrm{1PI}}$]\ \\
    The degree of divergence of this Green function is $\omega = 3$.
    The overall counterterm for the complete Green function $\mathcal{G}^\mu$ is given by
    \begin{align}
        C(\mathcal{G}^\mu)=-\mathcal{A}^\mu_{\phantom{\mu}\nu\rho\sigma}(\epsilon) p^\nu p^\rho p^\sigma.
    \end{align}
    The Green function with $M^2$-inserted propagators yields the same counterterm plus an additional auxiliary term proportional to $M^2$,
    \begin{align}
        C(\mathcal{G}^\mu_M)=-\mathcal{A}^\mu_{\phantom{\mu}\nu\rho\sigma}(\epsilon) p^\nu p^\rho p^\sigma -\mathcal{B}^\mu_{\phantom{\mu}\nu}(\epsilon) M^2 p^\nu.
    \end{align}
    For the rescaled diagrams (covering both sign conventions) the counterterm reads
    \begin{align}
        C(\widetilde{\mathcal{G}}^{\mu}_{M})=-\lambda^3\mathcal{A}^\mu_{\phantom{\mu}\nu\rho\sigma}(\epsilon) p^\nu p^\rho p^\sigma - \lambda (1\pm\varrho^2) \mathcal{B}^\mu_{\phantom{\mu}\nu}(\epsilon) M^2 p^\nu.
    \end{align}
    Applying the algorithm introduced above with the sign convention $(1+\varrho^2)M^2$ gives
    \begin{align}\label{Eq:Example-Tadpole-Algorithm}
        C\Big[\big(\mathcal{T}_{\{\lambda\}}^{(3)} + \mathcal{T}_{\{\lambda\}}^{(1)}\big)\widetilde{\mathcal{G}}^{\mu}_{M}\Big] = -\mathcal{A}^\mu_{\phantom{\mu}\nu\rho\sigma}(\epsilon) p^\nu p^\rho p^\sigma -\mathcal{B}^\mu_{\phantom{\mu}\nu}(\epsilon) M^2 p^\nu = C(\mathcal{G}^\mu) + C(M_{\mathcal{G}^\mu}^2),
    \end{align}
    with Taylor operator as in Eq.~\eqref{Eq:Taylor-Operator-wrt-m-p}. 
    Using instead the full Taylor expansion with the convention $(1-\varrho^2)M^2$ yields
    \begin{equation}\label{Eq:Example-Exact-Tadpole-Decomposition}
        \begin{aligned}
            C\Big[\sum_{n=0}^3 &\,\mathcal{T}_{\{\lambda,\varrho\}}^{(n)} \widetilde{\mathcal{G}}^{\mu}_{M}\Big] = C\Big[\big( \mathcal{T}_{\{\lambda\}}^{(1)} + \mathcal{T}_{\{\lambda\}}^{(3)} + \mathcal{T}_{\{\lambda,\varrho^2\}}^{(1,2)} \big) \widetilde{\mathcal{G}}^{\mu}_{M}\Big]\\
            &= -\mathcal{B}^\mu_{\phantom{\mu}\nu}(\epsilon) M^2 p^\nu - \mathcal{A}^\mu_{\phantom{\mu}\nu\rho\sigma}(\epsilon) p^\nu p^\rho p^\sigma +\mathcal{B}^\mu_{\phantom{\mu}\nu}(\epsilon) M^2 p^\nu
            = \mathcal{A}^\mu_{\phantom{\mu}\nu\rho\sigma}(\epsilon) p^\nu p^\rho p^\sigma = C(\mathcal{G}^\mu),
        \end{aligned}
    \end{equation}
    where we used all three Taylor operators from Eqs.~\eqref{Eq:Taylor-Operator-wrt-m-p}--\eqref{Eq:Taylor-Operator-lambda-rho-separately}.
    Evidently, all approaches yield the same physical counterterm $C(\mathcal{G}^\mu)$.
    In particular, the proposed algorithm reproduces both the correct physical counterterm $C(\mathcal{G}^\mu)$ and the auxiliary counterterm $C(M_{\mathcal{G}^\mu}^2)$ (see Eq.~\ref{Eq:Example-Tadpole-Algorithm}),
    whereas Eq.~\eqref{Eq:Example-Exact-Tadpole-Decomposition} reproduces only the physical counterterm, as the $M^2$-dependent terms cancel, in agreement with the discussion above.
\end{example}

\section{Tensor Reduction}\label{Sec:Tensor_Reduction}

Computations in gauge theories inevitably give rise to tensor integrals $I^{\mu_1\cdots\mu_r}$, whose rank is denoted by $r$.
The standard procedure for evaluating these integrals consists of reducing them to scalar integrals via tensor reduction.
In this step, a tensor integral is expressed as a linear combination of all Lorentz-covariant tensors $T_a^{\mu_1\cdots\mu_r}$ compatible with the symmetries:
\begin{equation}\label{Eq:GeneralTensorReduction}
    \begin{aligned}
        I^{\mu_1\cdots\mu_r}(p_1,\ldots,p_E) = 
        \sum_{a=1}^{n} C_a(p_1,\ldots,p_E) \, T_{a}^{\mu_1\cdots\mu_r}(p_1,\ldots,p_E).
    \end{aligned}
\end{equation}
The basis tensors $T_a^{\mu_1\cdots\mu_r}$ are independent of the perturbative order, whereas the coefficients $C_a$ are composed of scalar loop integrals.

The tadpole decomposition introduced in Sec.~\ref{Sec:Tadpole_Decomposition} simplifies the problem by mapping all integrals to vacuum bubble tensor integrals which are independent of external momenta $p_i$.
Consequently, both the basis tensors $T_a^{\mu_1\cdots\mu_r}$ and the coefficients $C_a$ do not depend on external momenta.
In this case, the basis tensors can be constructed solely from products of metric tensors $\eta^{\mu\nu}$.
Since these metric tensors are taken to be $D$-dimensional, the resulting projectors are also strictly $D$-dimensional, so that the coefficients $C_a$ which are to be determined hence consist of scalar integrals with purely $D$-dimensional integrands that can be handled with standard IBP-techniques.

For the efficient tensor reduction of such vacuum bubble tensor integrals, we implemented the method proposed in Refs.~\cite{Herzog:2017ohr,Ruijl:2018poj}.
During the course of the projects presented in this thesis, two other groups independently published related work on tensor reduction, including external momenta and providing additional mathematical insights, see Refs.~\cite{Anastasiou:2023koq} and \cite{Goode:2024mci}.
In particular, Ref.~\cite{Goode:2024mci} contains a detailed discussion of the method used here --- naming it \textit{orbit partition approach} --- and published the tensor reduction software tool \texttt{OPITeR}~\cite{Goode:2024cfy}. 
Nevertheless, we developed and documented our own implementation of this method for the reduction of vacuum bubble tensor integrals, see Ref.~\cite{vonManteuffel:2025swv}.
Since all computations in the projects of this thesis have been performed with our own \texttt{FORM} programs, we provide a self-contained discussion of the method here.
The following presentation is largely based on our publication~\cite{vonManteuffel:2025swv}, complemented by the discussions in Refs.~\cite{Goode:2024mci,gallian2016contemporary,dummit2003abstract,fraleigh2002first}.

In essence, the tensor reduction consists of two main tasks.
The first task is the construction of a basis $\mathcal{B}=\{T_a\}_{a=1}^n$ of all independent tensor structures $T_a^{\mu_1\cdots\mu_r}$ relevant to the given class of tensor integrals (cf.\ Eq.~\eqref{Eq:GeneralTensorReduction}), and the determination of the corresponding projectors $P_a^{\mu_1\cdots\mu_r}$.
These projectors form the dual basis $\mathcal{B}^*$ and can be constructed from the basis tensors $T_a^{\mu_1\cdots\mu_r}$ themselves.
They are defined by canonical duality, which requires the orthogonality relation
\begin{equation}\label{Eq:OrthogonalityRelation}
    \begin{aligned}
        P_a^{\mu_1\cdots\mu_r} \, T_{b,\mu_1\cdots\mu_r} = \delta_{ab}\,.
    \end{aligned}
\end{equation}
The second task is the determination of the coefficients $C_a$ (see Eq.~\eqref{Eq:GeneralTensorReduction}), which are obtained upon projection using condition~\eqref{Eq:OrthogonalityRelation}:
\begin{equation}\label{Eq:ProjectionOfCoefficients}
    \begin{aligned}
        C_a = P_a^{\mu_1\cdots\mu_r} \, I_{\mu_1\cdots\mu_r}.
    \end{aligned}
\end{equation}

The first task can be performed once and prior to the actual computation of Green functions.
The resulting projectors are collected in so-called tensor reduction tables.
By contrast, the second task must be carried out for each tensor integral during the explicit calculation.
Before discussing the construction of the tensor reduction tables (task one) and their application to the tensor integrals (task two), we first introduce the necessary group-theoretical concepts underlying the method, primarily following Refs.~\cite{Goode:2024mci,gallian2016contemporary,dummit2003abstract,fraleigh2002first}.

\paragraph{Group-theoretical Concepts:} 
The orbit partition approach provides a systematic method based on group theory for constructing the required projectors efficiently.
The key idea is to exploit the symmetry properties of tensors under index permutations.
To this end, we consider the action of the permutation group $S_r$ on rank-$r$ tensors, with group action denoted by ``$\circ$'' for some permutation $\sigma\in S_r$.
This action amounts to permuting the tensor indices.
For example, the permutation 
\begin{align}\label{Eq:Example-for-S_8-Permutation}
    \sigma = 
    \begin{pmatrix}
        1 & 2 & 3 & 4 & 5 & 6 & 7 & 8\\
        3 & 8 & 5 & 1 & 4 & 2 & 7 & 6
    \end{pmatrix} \in S_8
\end{align}
acts on a rank-$8$ tensor as $\sigma\circ T^{\mu_1\mu_2\mu_3\mu_4\mu_5\mu_6\mu_7\mu_8} = T^{\mu_3\mu_8\mu_5\mu_1\mu_4\mu_2\mu_7\mu_6}$.

This group action is compatible with index contraction in the following sense:
for two rank-$r$ tensors $A$ and $B$, we have
\begin{align}
    A \cdot B = (\sigma \circ A)\cdot(\sigma \circ B), \qquad \forall \, \sigma \in S_r,
\end{align}
where $A\cdot B \equiv A^{\mu_1\ldots\mu_r}B_{\mu_1\ldots\mu_r}$.
Defining $C\coloneqq \sigma\circ B$, so that $A \cdot B = (\sigma \circ A)\cdot C$ and $B=\sigma^{-1}\circ C$, this identity can be rewritten as 
\begin{align}\label{Eq:GroupAction_in_context_of_Tensor_Contraction}
    A \cdot (\sigma^{-1}\circ C) = (\sigma \circ A) \cdot C, \qquad \forall \, \sigma \in S_r.
\end{align}

The symmetries of a tensor $T_a$ are precisely the permutations that leave it invariant (possibly up to a sign in the case of antisymmetric indices), i.e.\ $\sigma\circ T_a=T_a$.
The set of such permutations forms a subgroup of $S_r$, called the stabiliser subgroup:
\begin{definition}[Stabiliser]\label{Def:Stabliliser_Subgroup}\ \\
    For a tensor $T_a$, the subgroup 
    \begin{align}
        H(T_a) = \{h \in S_r \,|\, h \circ T_a = s_a(h) T_a\} \subset S_r,
    \end{align}
    composed of all permutations that leave the tensor $T_a$ invariant is called \emph{stabiliser} of $T_a$ in $S_r$.
    Here, $s_a(h) = \pm 1$ encodes a possible sign change if $h$ swaps antisymmetric indices.
\end{definition}
The stabiliser subgroup $H(T_a)$ thus encodes the index permutation symmetries of $T_a$:
for any $h\in H(T_a)$, we say that $h$ ``fixes'' $T_a$ through $h \circ T_a = s_a(h) T_a$, with the sign function $s_a$ satisfying
\begin{enumerate}[label={$(\roman*)$}]
    \item $s_a(h)=s_a(h^{-1})$,
    \item $s_a(hg)=s_a(h)s_a(g), \quad \forall \, g\in H(T_a)$,
    \item $s_a(\mathrm{id})=+1$,
\end{enumerate}
and $s_a(h)=-1$ whenever $h$ exchanges antisymmetric indices.

Elements of $H(T_a)$ are not necessarily symmetries of other tensors.
However, the space of tensors must be closed under the action of $H(T_a)$, which naturally introduces the notion of \emph{orbits}. 
Following Refs.~\cite{gallian2016contemporary,fraleigh2002first}:
\begin{definition}[Orbit]\label{Def:Orbit_in_General}\ \\
    Let $G$ be a group of permutations acting on a set $X$.
    For any $x\in X$, the subset 
    \begin{align}
        \mathcal{O}_G(x)=\{\sigma\circ x \, | \, \sigma\in G\}\subset X
    \end{align}
    is called the \emph{orbit} of $x$ under $G$.
    Its cardinality is denoted by $|\mathcal{O}_G(x)|$.
\end{definition}
For example, for the permutation in Eq.~\eqref{Eq:Example-for-S_8-Permutation}, the complete set of orbits is  $\{1,3,4,5\}$, $\{2,6,8\}$ and $\{7\}$.
In the present context of tensor reduction we are specifically interested in orbits under stabilisers:
\begin{definition}[Orbit under Stabilisers]\label{Def:Orbit_under_the_Stabiliser}\ \\
    Let $H(T_a)$ be the stabiliser of the tensor $T_a\in\mathcal{B}$ in $S_r$.
    For any $T_c \in \mathcal{B}$, the equivalence class
    \begin{align}
        \tensor*[_r]{\Theta}{_c^{(a)}} = \{ h \circ T_c \,|\, h \in H(T_a) \} \subset \mathcal{B},
    \end{align}
    is called the \emph{orbit} of $T_c$ under the stabiliser $H(T_a)\subset S_r$.
    That is, $\tensor*[_r]{\Theta}{_c^{(a)}}$ consists of all tensors in $\mathcal{B}$ that can be reached from $T_c$ by an element of $H(T_a)$ (including $T_c$ itself).
\end{definition}
Thus, the orbit of $T_c$ under $H(T_a)$ is generated by the repeated action of all $h\in H(T_a)$.
By group properties, the action of $H(T_a)$ partitions $\mathcal{B}$ into disjoint subsets corresponding to these orbits.
Two tensors $T_c$ and $T_d$ belong to the same orbit (under $H(T_a)$) if and only if $\exists \, \, h \in H(T_a)$, such that $T_d = h \circ T_c$, i.e.\ they are related by a symmetry of $T_a$. 
Orbits are therefore equivalence classes of tensors under the relation
\begin{align}\label{Eq:Equivalence-Relation-for-2-Tensors-wrt-Orbits}
    T_c \sim T_d \quad \Longleftrightarrow \quad \exists \, h \in H(T_a) \,\,\, \mathrm{with} \,\,\, h \circ T_c = T_d.
\end{align}

The subset of permutations in $H(T_a)$ that also stabilise $T_c$ is
\begin{align}
    H(T_a,T_c)=H(T_a)\cap H(T_c)=\{h\in H(T_a)\,|\,h\circ T_c = s_c(h)T_c\}.
\end{align}
For finite dimensional $H(T_a)$, the orbit-stabiliser theorem together with Lagrange's theorem relate the cardinality of an orbit and the order of the stabiliser via $|\tensor*[_r]{\Theta}{_c^{(a)}}|=|H(T_a)|/|H(T_a,T_c)|$, see Refs.~\cite{gallian2016contemporary,fraleigh2002first}.

By construction, all elements of an orbit share characteristic symmetry properties.
This motivates the following (see Ref.~\cite{Goode:2024mci}):
\begin{proposition}[Orbit Sum]\label{Prop:Invariant_Sum}\ \\
    The sum over the orbit $\tensor*[_r]{\Theta}{_c^{(a)}}$ 
    \begin{align}\label{Eq:Invariant_Sum_General}
        \mathcal{J}^{(a)}_c 
        = \frac{1}{|H(T_a,T_c)|} \sum_{h\in H(T_a)} \!\! s_a(h) \, h \circ T_c,
    \end{align}
    possesses the same symmetry properties as $T_a$, i.e.\ it is invariant under the action of $H(T_a)$.
    In other words, both $T_a$ and $\mathcal{J}^{(a)}_c$ are stabilised under the action of $H(T_a)$.
\end{proposition}
\begin{proof}
    For any $h'\in H(T_a)$ we find
    \begin{equation*}
        \begin{aligned}
            h' \circ \mathcal{J}_c^{(a)} &= h' \circ \frac{1}{|H(T_a,T_c)|} \sum_{h\in H(T_a)} \!\! s_a(h) \, h \circ T_c\\
            &= \frac{1}{|H(T_a,T_c)|} \sum_{h\in H(T_a)} \!\! s_a(h) \, h' \circ h \circ T_c\\
            &= \frac{1}{|H(T_a,T_c)|} \sum_{h''\in H(T_a)} \!\! s_a((h')^{-1}h'') \, h'' \circ T_c\\
            &= s_a(h') \frac{1}{|H(T_a,T_c)|} \sum_{h''\in H(T_a)} \!\! s_a(h'') \, h'' \circ T_c\\
            &= s_a(h') \mathcal{J}_{c}^{(a)},
        \end{aligned}
    \end{equation*}
    where we defined $h''=h'\circ h\in H(T_a)$, relabelled the summation variable in the third line and used the properties $s_i(hg)=s_i(h)s_i(g)$ as well as $s_a((h')^{-1})=s_a(h')$, to obtain $s_a(h)=s_a((h')^{-1}h'h)=s_a((h')^{-1}h'')=s_a((h')^{-1}s_a(h'')=s_a(h')s_a(h'')$.
\end{proof}
\begin{remark}[on orbit sums]\
    \begin{itemize}
        \item The prefactor $1/|H(T_a,T_c)|$ in Eq.~\eqref{Eq:Invariant_Sum_General} accounts for overcounting when $|\tensor*[_r]{\Theta}{_c^{(a)}}|<|H(T_a)|$, because in this case the sum generates the same term more than once.
        \item Defining $\mathcal{G}_c(T_a)$ as a set of permutations that generate each tensor in the orbit $\tensor*[_r]{\Theta}{_c^{(a)}}$ exactly once, the sum may equivalently be written as
        $$\mathcal{J}^{(a)}_c = \!\!\sum_{h\in\mathcal{G}_c(T_a)} \!\! s_a(h) \, h \circ T_c.$$
        \item These orbit sums $\mathcal{J}^{(a)}_c$ serve as the natural building blocks for constructing the projectors $P_a$, since they inherit the same symmetries as $T_a$.
    \end{itemize}    
\end{remark}
For a tensor basis of $n=|\mathcal{B}|$ elements, one obtains $n$ orbits $\tensor*[_r]{\Theta}{_c^{(a)}}$.
However, distinct tensors $T_c$ and $T_d$ may belong to the same orbit, i.e.\ $\tensor*[_r]{\Theta}{_c^{(a)}} = \tensor*[_r]{\Theta}{_d^{(a)}}$ (cf.\ Eq.~\eqref{Eq:Equivalence-Relation-for-2-Tensors-wrt-Orbits}).
To avoid redundancy, we restrict to disjoint orbits by selecting orbit representatives:
\begin{definition}[Orbit Representatives]\label{Def:Orbit_Representative}\ \\
    Let $\Omega^{(a)}$ denote the set of disjoint orbits under $H(T_a)$.
    An \emph{orbit representative} is any tensor $T_c$ chosen to uniquely label its orbit $\tensor*[_r]{\Theta}{_c^{(a)}}\in\Omega^{(a)}$.
    The set $\mathscr{R}^{(a)}$ of orbit representatives contains exactly one element from each orbit in $\Omega^{(a)}$.
\end{definition}
\begin{remark}[on disjoint orbits and orbit representatives]\
    \begin{itemize}
        \item With $m$ disjoint orbits, the tensor basis with $n$ elements partitions as
        $$\mathcal{B}=\bigsqcup_{b=1}^n \tensor*[_r]{\Theta}{_b^{(a)}}=\bigcup_{c=1}^m \tensor*[_r]{\Theta}{_c^{(a)}},$$
        where $\bigsqcup$ denotes disjoint union.
        \item The choice of representatives in $\mathscr{R}^{(a)}$ is arbitrary.
        \item The tensor $T_a$ defining the stabiliser $H(T_a)$ is always included, as its orbit is trivial: $\tensor*[_r]{\Theta}{_a^{(a)}}=\{T_a\}$.
    \end{itemize}     
\end{remark}
Finally, the number $m$ of disjoint orbits (or equivalently of orbit representatives) equals the number of disjoint \emph{cycles}.
Following Refs.~\cite{gallian2016contemporary,fraleigh2002first}:
\begin{definition}[Cycle]\label{Def:Cycle}\ \\
    A permutation $\sigma\in S_r$ that has at most one orbit with more than one element is called a \emph{cycle}.
\end{definition}
As an example, the permutation in Eq.~\eqref{Eq:Example-for-S_8-Permutation} decomposes into three disjoint cycles: $(1,3,5,4)$, $(2,8,6)$ and $(7)$, corresponding to $1 \to 3 \to 5 \to 4 \to 1$, $2 \to 8 \to 6 \to 2$ and $7 \to 7$. 
Every permutation of a finite set can be written uniquely as a product of such disjoint cycles.

Note that in the context of tensor reduction, we identify the specific orbit structure through a ``graphical representation'' of the disjoint cycle decomposition of a permutation.
Here, cycles are visualised as closed contraction loops formed by contracted indices of metric tensors.
This graphical viewpoint provides a direct and practical way of determining which tensors belong to which orbits (see below).

\paragraph{Constructing the Tensor Reduction Tables:}
For vacuum bubble tensor integrals, non-vanishing contributions only arise for even tensor rank $r$, since the only possible Lorentz covariants for such integrals are metric tensors $\eta^{\mu\nu}$.
Accordingly, the tensor basis $\mathcal{B}$ is formed by all distinct products of $r/2$ metric tensors.
It is therefore not sufficient to simply consider all permutations in $S_r$, since this would generate the same basis tensors multiple times due to the inherit symmetry of expressions such as $\eta^{\mu_1\mu_2}\ldots\eta^{\mu_{r-1}\mu_r}$, where each $\eta^{\mu\nu}$ is symmetric.
Instead, one must identify and factor out equivalent tensors that arise from exchanging pairs of $\eta$'s (described by $S_{r/2}$) or swapping indices within a given $\eta$ (described by $(S_2)^{r/2}$).
These symmetries are collected in the stabiliser subgroup $S_{r/2} \times (S_2)^{r/2}$ in $S_r$.
After factoring them out, i.e.\ selecting only one element of each coset of $S_{r/2} \times (S_2)^{r/2}$ in $S_r$, the relevant set of permutations is $S_2^r$, which generates all distinct products of metric tensors of rank $r$.
The tensor basis in present case is therefore given by $\mathcal{B}=\{\sigma\circ T^{\mu_1\cdots\mu_r}\,|\,\sigma\in S^r_2\}$.
Following Eq.~\eqref{Eq:GeneralTensorReduction}, the vacuum bubble tensor integrals can thus be decomposed into a linear combination of $n=|S_{2}^{r}|=(r-1)!!=r!/(2^{r/2}(r/2)!)$ independent tensors of the form 
\begin{equation}\label{Eq:GeneralTensorStructureForVacuumBubbles}
    \begin{aligned}
        T_{\sigma}^{\mu_1\cdots\mu_r} = \eta^{\mu_{\sigma(1)}\mu_{\sigma(2)}} \cdots \eta^{\mu_{\sigma(r-1)}\mu_{\sigma(r)}},
    \end{aligned}
\end{equation}
each composed of $r/2$ metric tensors.

The projectors themselves must also be constructed from the available tensor structures, i.e.\ they are linear combinations of products of metric tensors.
Thus, any projector $P_a$ can be expressed as a linear combination of the basis tensors as
\begin{equation}\label{Eq:GeneralProjectors}
    \begin{aligned}
        P_a^{\mu_1\cdots\mu_r} = \sum_{b=1}^{n} A^{(a)}_{b} \, T_b^{\mu_1\cdots\mu_r},
    \end{aligned}
\end{equation}
with coefficients $A^{(a)}_b$ to be determined.

Contracting a fixed projector $P^{\mu_1\cdots\mu_r}_a$ with all tensors of the basis $\mathcal{B}=\{T_{b}^{\mu_1\cdots\mu_r}\}_{b=1}^{n}$ and using the orthogonality condition~\eqref{Eq:OrthogonalityRelation} yields a system of $n$ equations for the $n$ unknown coefficients $A^{(a)}_b$.
Solving this system determines the projector $P_a^{\mu_1\cdots\mu_r}$.
In fact, all projectors $P_a^{\mu_1\cdots\mu_r}$, $\forall\,a\in\{1,\ldots,n\}$, are obtained in this way, as they are related by index permutations:
\begin{proposition}\ \\
    Let $\sigma\in S_2^r$ be a permutation such that $\sigma \circ T_a^{\mu_1\cdots\mu_r} = T_{b}^{\mu_1\cdots\mu_r}$, then $P_b^{\mu_1\cdots\mu_r} = \sigma \circ P_a^{\mu_1\cdots\mu_r}$.
\end{proposition}
\begin{proof}
    Let $P_a$ be the projector dual to the tensor $T_a$, and suppose $\sigma\in S_2^r$ such that $T_{b}=\sigma \circ T_a$ with $b\neq a$. 
    Then
    \begin{equation*}
        \begin{aligned}
            (\sigma\circ P_a) \cdot T_c &= P_a \cdot (\sigma^{-1} \circ T_c)\\
            &= \delta_{bc}\\
            &= P_b \cdot T_c.
        \end{aligned}
    \end{equation*}
    In the first line, Eq.~\eqref{Eq:GroupAction_in_context_of_Tensor_Contraction} was used to shift $\sigma$ from $P_a$ to $T_c$.
    The second line follows from orthogonality (see Eq.~\eqref{Eq:OrthogonalityRelation}), which gives a non-vanishing result only if $\sigma^{-1} \circ T_c = T_a$, i.e.\ if $T_c=T_b$ by assumption.
    Hence, $\sigma\circ P_a = P_b$.
\end{proof}

In practice, solving this system of equations amounts to inverting an $n \times n$ matrix.
This quickly becomes computationally challenging, since the number $n$ of independent tensors grows very rapidly with the rank $r$, as illustrated in Tab.~\ref{Tab:TensorSize}.
\begin{table}[t]
    \centering
    \begin{tabular}{|c||c|c|c|c|c|c|c|c|} \hline 
        $r$ & 2 & 4 & 6 & 8 & 10 & 12 & 14 & 16\\ \hline 
        $n=|S_{2}^{r}|$ & 1 & 3 & 15 & 105 & 945 & 10395 & 135135 & 2027025\\ \hline 
        $m=p(r/2)$ & 1 & 2 & 3 & 5 & 7 & 11 & 15 & 22\\ \hline
    \end{tabular}
    \caption{Number $n$ of independent tensor structures of rank $r$ vacuum tensor integrals and the number $m$ of integer partitions of $r/2$, for $r\leq16$.}
    \label{Tab:TensorSize}
\end{table}

This difficulty can be overcome by exploiting symmetry considerations, as discussed in Refs.~\cite{Herzog:2017ohr,Ruijl:2018poj,Anastasiou:2023koq,Goode:2024mci,vonManteuffel:2025swv}, which substantially reduce the size of the linear system to be solved.
Specifically, for a rank-$r$ tensor integral the projector $P_a^{\mu_1\cdots\mu_r}$ must share the same index-permutation symmetries as the associated basis tensor $T_{a}^{\mu_1\cdots\mu_r}$.
As outlined above, these symmetries are encoded in the corresponding \emph{stabiliser} $H(T_a)$.

By the group-theoretical methods introduced above, we recall that for fixed $a\in\{1,\ldots,n\}$, the entire set of basis tensors $\mathcal{B}=\{T_b^{\mu_1\cdots\mu_r}\}_{b=1}^n$ can be partitioned into $m$ disjoint \emph{orbits} $\{\tensor*[_r]{\Theta}{_c^{(a)}}\}_{c=1}^{m}$, each consisting of tensors related by the action of $H(T_a)$.
As established in proposition~\ref{Prop:Invariant_Sum}, the sum over all tensors in an orbit $\tensor*[_r]{\Theta}{_c^{(a)}}$, $\forall \,c\in\{1,\ldots,m\}$, yields an object (see Eq.~\eqref{Eq:Invariant_Sum_General}) that is invariant under $H(T_a)$, and thus exhibits exactly the same symmetry properties as $T_a^{\mu_1\cdots\mu_r}$.
Consequently, symmetries enforce that all tensors belonging to the same orbit contribute with a common coefficient $A_c^{(a)}$, so that Eq.~\eqref{Eq:GeneralProjectors} can be rewritten as
\begin{equation}\label{Eq:MinimalProjectors}
    \begin{aligned}
        P_a^{\mu_1\cdots\mu_r} = \sum_{c=1}^{m} A^{(a)}_{c} \Bigg( \sum_{ T \in \tensor*[_r]{\Theta}{_c^{(a)}} } T^{\mu_1\cdots\mu_r} \Bigg),
    \end{aligned}
\end{equation}
with only $m$ independent coefficients $A^{(a)}_{c}$.
Since each orbit sum on the RHS of Eq.~\eqref{Eq:MinimalProjectors} is, by construction, stabilised under $H(T_a)$, this Ansatz automatically ensures the required symmetry properties of the projectors.

The number of orbits $m$ depends on the tensor rank $r$.
Both $m$ and the specific tensors belonging to a given orbit can be determined with the help of \emph{cycles}.
In this context, a \emph{cycle} corresponds to a closed contraction loop formed between the Lorentz indices of the metric tensors in the ``reference tensor'' $T_a^{\mu_1\cdots\mu_r}$ (which defines the stabiliser under consideration) and those in another tensor $T_{b,\mu_1\cdots\mu_r}$.
Orbits are then uniquely classified by the cycle structure defined in this way.
Hence, two tensors $T_{b,\mu_1\cdots\mu_r}$ and $T_{d,\mu_1\cdots\mu_r}$ belong to the same orbit $\tensor*[_r]{\Theta}{_c^{(a)}}$ precisely when they are characterised by the same cycle structure upon contraction with $T_a^{\mu_1\cdots\mu_r}$.
In other words, we exploit distinct patterns of index contractions.
The maximal number of different types of such contraction cycles corresponds to the number of integer partitions of $r/2$.
Accordingly, the growth of $m$ with $r$ is much slower than that of $n$, as displayed in Tab.~\ref{Tab:TensorSize}.
This observation allows for a significant simplification:
instead of contracting the projector $P_a^{\mu_1\cdots\mu_r}$ (for fixed $a\in\{1,\ldots,n\}$) with all $n$ tensors of the basis, it is sufficient to contract with the orbit representatives, i.e.\ with a single representative from each of the $m$ orbits $\tensor*[_r]{\Theta}{_c^{(a)}}$.
Applying the orthogonality relation~\eqref{Eq:OrthogonalityRelation} then yields a system of only $m$ equations for $m$ coefficients $A^{(a)}_{c}$, in contrast to the original $n$-dimensional system (cf.\ Tab.~\ref{Tab:TensorSize}).

For clarity, we label the orbits $\{\tensor*[_r]{\Theta}{_c^{(a)}}\}_{c=1}^{m}$ in descending order of the number of contraction cycles:
the orbit with $c=1$ contains tensors with the maximal number of cycles with $T_{a}^{\mu_1\cdots\mu_r}$, while $c=m$ corresponds to tensors with the minimal number of cycles, i.e.\ a single contraction loop.
In particular, the $c=1$ orbit contains only the tensor $T_{a}^{\mu_1\cdots\mu_r}$ itself.
The construction of orbits in terms of contraction cycles is most transparent in an explicit example.

\begin{example}[Tensor Reduction Tables for Rank $r=6$]\label{Example:Tensor-Reduction-Rank-6}\ \\
    To illustrate the orbit partition approach for constructing the tensor reduction tables described above, we consider the case of rank $r=6$.
    After identifying the $n=15$ independent tensors $\mathcal{B}=\{\sigma \circ T^{\mu_1\cdots\mu_6} \, | \, \sigma \in S_{2}^{6}\}$, we fix, w.l.o.g., $a=1$ and construct $P_1^{\mu_1\cdots\mu_6}$ according to Eq.~\eqref{Eq:MinimalProjectors}.
    To this end, we analyse the Lorentz-index contractions of the reference tensor
    \begin{align}
        T_1^{\mu_1\cdots\mu_6}=\eta^{\mu_{1}\mu_{2}}\eta^{\mu_{3}\mu_{4}}\eta^{\mu_{5}\mu_{6}}
    \end{align}
    with $\{T_{b,\mu_1\cdots\mu_6}\}_{b=1}^{15}$ in order to determine the orbits $\{\tensor*[_6]{\Theta}{_c^{(1)}}\}_{c=1}^{m}$ (with $m$ a priori unknown).
    
    For $r=6$ one can have at most $m=3$ different cycle structures, precisely matching the number of integer partitions of $r/2$, i.e.\ $p(3)=3$.
    According to our labelling convention, the first orbit $\tensor*[_6]{\Theta}{_1^{(1)}}$ contains only $T_{1,\mu_1\cdots\mu_6}$ itself: 
    it is the unique tensor fixed by $H(T_1)$ and inherently exhibits the full permutation symmetry of $T_1^{\mu_1\cdots\mu_6}$.
    Furthermore, it is the only tensor that realises the maximal number of contraction cycles with $T_1^{\mu_1\cdots\mu_6}$, provided by
    \begin{equation}\label{Eq:ContractionCycles-3}
        \begin{aligned}
            \contraction[1.25ex]{ \!\! }{ \eta^{\mu_{1}\mu_{2}} }{ \eta^{\mu_{3}\mu_{4}} \eta^{\mu_{5}\mu_{6}} \, }{ \eta_{\mu_{1}\mu_{2}} }
            \bcontraction[1.25ex]{ \,\,\, }{ \eta^{\mu_{1}\mu_{2}} }{ \eta^{\mu_{3}\mu_{4}} \eta^{\mu_{5}\mu_{6}} \, }{ \eta_{\mu_{1}\mu_{2}} }
            \contraction[2ex]{ \eta^{\mu_{1}\mu_{2}} \!\! }{ \eta^{\mu_{3}\mu_{4}} }{ \eta^{\mu_{5}\mu_{6}} \, \eta_{\mu_{1}\mu_{2}} }{ \eta_{\mu_{3}\mu_{4}} }
            \bcontraction[2ex]{ \eta^{\mu_{1}\mu_{2}} \,\,\, }{ \eta^{\mu_{3}\mu_{4}} }{ \eta^{\mu_{5}\mu_{6}} \, \eta_{\mu_{1}\mu_{2}} }{ \eta_{\mu_{3}\mu_{4}} }
            \contraction[2.75ex]{ \eta^{\mu_{1}\mu_{2}} \eta^{\mu_{3}\mu_{4}} \!\! }{ \eta^{\mu_{5}\mu_{6}} }{ \, \eta_{\mu_{1}\mu_{2}} \eta_{\mu_{3}\mu_{4}} }{ \eta_{\mu_{5}\mu_{6}} }
            \bcontraction[2.75ex]{ \eta^{\mu_{1}\mu_{2}} \eta^{\mu_{3}\mu_{4}} \,\,\, }{ \eta^{\mu_{5}\mu_{6}} }{ \, \eta_{\mu_{1}\mu_{2}} \eta_{\mu_{3}\mu_{4}} }{ \eta_{\mu_{5}\mu_{6}} }
            \eta^{\mu_{1}\mu_{2}} \eta^{\mu_{3}\mu_{4}} \eta^{\mu_{5}\mu_{6}} \, \eta_{\mu_{1}\mu_{2}} \eta_{\mu_{3}\mu_{4}} \eta_{\mu_{5}\mu_{6}}.
        \end{aligned}
    \end{equation}
    The cycles are specifically given by $1\rightarrow2\rightarrow1$, $3\rightarrow4\rightarrow3$ and $5\rightarrow6\rightarrow5$, fixing $m=3$ for $r=6$.
    Hence, the first orbit contains only the reference tensor itself as proposed:
    \begin{equation}\label{Eq:Rank6-SymmetrySet-1}
        \begin{aligned}
            \tensor*[_6]{\Theta}{_1^{(1)}} \equiv \{T_{1,\mu_1\cdots\mu_6} = \eta_{\mu_{1}\mu_{2}} \eta_{\mu_{3}\mu_{4}} \eta_{\mu_{5}\mu_{6}}\}.
        \end{aligned}
    \end{equation}
    The second orbit $\tensor*[_6]{\Theta}{_2^{(1)}}$ consists of all tensors admitting exactly two contraction cycles with $T_1^{\mu_1\cdots\mu_6}$.
    For example, for the tensor $T_{4,\mu_1\cdots\mu_6}=\eta_{\mu_{1}\mu_{3}}\eta_{\mu_{2}\mu_{4}}\eta_{\mu_{5}\mu_{6}}$ we find
    \begin{equation}\label{Eq:ContractionCycles-2}
        \begin{aligned}
            \contraction[1.25ex]{ \!\! }{ \eta^{\mu_{1}\mu_{2}} }{ \eta^{\mu_{3}\mu_{4}} \eta^{\mu_{5}\mu_{6}} \, }{ \eta_{\mu_{1}\mu_{3}} }
            \bcontraction[2ex]{ \,\,\, }{ \eta^{\mu_{1}\mu_{2}} }{ \eta^{\mu_{3}\mu_{4}} \eta^{\mu_{5}\mu_{6}} \, \eta_{\mu_{1}\mu_{3}} }{ \!\!\!\!\!\!\!\!\!\! \eta_{\mu_{2}\mu_{4}} }
            \bcontraction[1.25ex]{ \eta^{\mu_{1}\mu_{2}} \!\! }{ \eta^{\mu_{3}\mu_{4}} }{ \eta^{\mu_{5}\mu_{6}} \, }{ \quad\,\,\, \eta_{\mu_{1}\mu_{3}} }
            \contraction[2ex]{ \eta^{\mu_{1}\mu_{2}} \,\,\, }{ \eta^{\mu_{3}\mu_{4}} }{ \eta^{\mu_{5}\mu_{6}} \, \eta_{\mu_{1}\mu_{3}} }{ \eta_{\mu_{2}\mu_{4}} }
            \contraction[2.75ex]{ \eta^{\mu_{1}\mu_{2}} \eta^{\mu_{3}\mu_{4}} \!\! }{ \eta^{\mu_{5}\mu_{6}} }{ \, \eta_{\mu_{1}\mu_{2}} \eta_{\mu_{3}\mu_{4}} }{ \eta_{\mu_{5}\mu_{6}} }
            \bcontraction[2.75ex]{ \eta^{\mu_{1}\mu_{2}} \eta^{\mu_{3}\mu_{4}} \,\,\, }{ \eta^{\mu_{5}\mu_{6}} }{ \, \eta_{\mu_{1}\mu_{2}} \eta_{\mu_{3}\mu_{4}} }{ \eta_{\mu_{5}\mu_{6}} }
            \eta^{\mu_{1}\mu_{2}} \eta^{\mu_{3}\mu_{4}} \eta^{\mu_{5}\mu_{6}} \, \eta_{\mu_{1}\mu_{3}} \eta_{\mu_{2}\mu_{4}} \eta_{\mu_{5}\mu_{6}},
        \end{aligned}
    \end{equation}
    and thus yields the two cycles $1\rightarrow3\rightarrow4\rightarrow2\rightarrow1$ and $5\rightarrow6\rightarrow5$.
    Similarly, we find exactly two such cycles for all other $T_{b,\mu_1\cdots\mu_6}\in\tensor*[_6]{\Theta}{_2^{(1)}}$.
    Collecting them gives
    \begin{equation}\label{Eq:Rank6-SymmetrySet-2}
        \begin{aligned}
            \tensor*[_6]{\Theta}{_2^{(1)}}\equiv
            \{
            &\eta_{\mu_{1}\mu_{2}} \eta_{\mu_{3}\mu_{5}} \eta_{\mu_{4}\mu_{6}},
            \eta_{\mu_{1}\mu_{2}} \eta_{\mu_{3}\mu_{6}} \eta_{\mu_{4}\mu_{5}},
            \eta_{\mu_{1}\mu_{3}} \eta_{\mu_{2}\mu_{4}} \eta_{\mu_{5}\mu_{6}}\\
            &\eta_{\mu_{1}\mu_{4}} \eta_{\mu_{2}\mu_{3}} \eta_{\mu_{5}\mu_{6}},
            \eta_{\mu_{1}\mu_{5}} \eta_{\mu_{2}\mu_{6}} \eta_{\mu_{3}\mu_{4}},
            \eta_{\mu_{1}\mu_{6}} \eta_{\mu_{2}\mu_{5}} \eta_{\mu_{3}\mu_{4}}
            \}.
        \end{aligned}
    \end{equation}
    The remaining tensors form the third and last orbit $\tensor*[_6]{\Theta}{_3^{(1)}}$, characterised by a single contraction cycle with $T_1^{\mu_1\cdots\mu_6}$. 
    An example is $T_{10,\mu_1\cdots\mu_6}=\eta_{\mu_{1}\mu_{5}}\eta_{\mu_{2}\mu_{3}}\eta_{\mu_{4}\mu_{6}}$, for which we find
    \begin{equation}\label{Eq:ContractionCycles-1}
        \begin{aligned}
            \contraction[1.25ex]{ \!\! }{ \eta^{\mu_{1}\mu_{2}} }{ \eta^{\mu_{3}\mu_{4}} \eta^{\mu_{5}\mu_{6}} \, }{ \eta_{\mu_{1}\mu_{5}} }
            \bcontraction[2.75ex]{ \,\,\, }{ \eta^{\mu_{1}\mu_{2}} }{ \eta^{\mu_{3}\mu_{4}} \eta^{\mu_{5}\mu_{6}} \, \eta_{\mu_{1}\mu_{5}} }{ \!\!\!\!\!\!\!\!\!\! \eta_{\mu_{2}\mu_{3}} }
            \contraction[2.75ex]{ \eta^{\mu_{1}\mu_{2}} \!\! }{ \eta^{\mu_{3}\mu_{4}} }{ \eta^{\mu_{5}\mu_{6}} \, \eta_{\mu_{1}\mu_{5}} }{ \quad\,\,\, \eta_{\mu_{2}\mu_{3}} }
            \bcontraction[2ex]{ \eta^{\mu_{1}\mu_{2}} \,\,\, }{ \eta^{\mu_{3}\mu_{4}} }{ \eta^{\mu_{5}\mu_{6}} \, \eta_{\mu_{1}\mu_{5}} \eta_{\mu_{2}\mu_{3}} }{ \!\!\!\!\!\!\!\!\!\! \eta_{\mu_{4}\mu_{6}} }
            \bcontraction[1.25ex]{ \eta^{\mu_{1}\mu_{2}} \eta^{\mu_{3}\mu_{4}} \!\! }{ \eta^{\mu_{5}\mu_{6}} }{ \, }{ \quad\,\,\, \eta_{\mu_{1}\mu_{5}} }
            \contraction[2ex]{ \eta^{\mu_{1}\mu_{2}} \eta^{\mu_{3}\mu_{4}} \,\,\, }{ \eta^{\mu_{5}\mu_{6}} }{ \, \eta_{\mu_{1}\mu_{5}} \eta_{\mu_{2}\mu_{3}} }{ \eta_{\mu_{5}\mu_{6}} }
            \eta^{\mu_{1}\mu_{2}} \eta^{\mu_{3}\mu_{4}} \eta^{\mu_{5}\mu_{6}} \, \eta_{\mu_{1}\mu_{5}} \eta_{\mu_{2}\mu_{3}} \eta_{\mu_{4}\mu_{6}},
        \end{aligned}
    \end{equation}
    producing only the single cycle $1\rightarrow5\rightarrow6\rightarrow4\rightarrow3\rightarrow2\rightarrow1$.
    The complete third orbit is
    \begin{equation}\label{Eq:Rank6-SymmetrySet-3}
        \begin{aligned}
            \tensor*[_6]{\Theta}{_3^{(1)}}\equiv
            \{
            &\eta_{\mu_{1}\mu_{3}} \eta_{\mu_{2}\mu_{5}} \eta_{\mu_{4}\mu_{6}},
            \eta_{\mu_{1}\mu_{3}} \eta_{\mu_{2}\mu_{6}} \eta_{\mu_{4}\mu_{5}},
            \eta_{\mu_{1}\mu_{4}} \eta_{\mu_{2}\mu_{5}} \eta_{\mu_{3}\mu_{6}},
            \eta_{\mu_{1}\mu_{4}} \eta_{\mu_{2}\mu_{6}} \eta_{\mu_{3}\mu_{5}},\\
            &\eta_{\mu_{1}\mu_{5}} \eta_{\mu_{2}\mu_{3}} \eta_{\mu_{4}\mu_{6}},
            \eta_{\mu_{1}\mu_{5}} \eta_{\mu_{2}\mu_{4}} \eta_{\mu_{3}\mu_{6}},
            \eta_{\mu_{1}\mu_{6}} \eta_{\mu_{2}\mu_{3}} \eta_{\mu_{4}\mu_{5}},
            \eta_{\mu_{1}\mu_{6}} \eta_{\mu_{2}\mu_{4}} \eta_{\mu_{3}\mu_{5}}
            \}.
        \end{aligned}
    \end{equation}

    Accordingly, the projector $P_1^{\mu_1\cdots\mu_6}$ then takes the form of Eq.~\eqref{Eq:MinimalProjectors} with $m=3$ coefficients $\{A^{(1)}_c\}_{c=1}^3$ and orbit sums built from the tensors in Eqs.~\eqref{Eq:Rank6-SymmetrySet-1}, \eqref{Eq:Rank6-SymmetrySet-2} and \eqref{Eq:Rank6-SymmetrySet-3}.
    Contracting $P_1^{\mu_1\cdots\mu_6}$ with one representative of each orbit (e.g.\ the tensors used in the demonstrations in Eqs.~\eqref{Eq:ContractionCycles-3}, \eqref{Eq:ContractionCycles-2} and \eqref{Eq:ContractionCycles-1}: 
    $T_1^{\mu_1\cdots\mu_6}$, $T_4^{\mu_1\cdots\mu_6}$ and $T_{10}^{\mu_1\cdots\mu_6}$, respectively) and invoking orthogonality (see Eq.~\eqref{Eq:OrthogonalityRelation}) yields
    \begin{equation}
        \begin{aligned}
            1 &= P_1^{\mu_1\cdots\mu_6} T_{1,\mu_1\cdots\mu_6}
                = D^3 A_1^{(1)} + 6 D^2 A_2^{(1)} + 8 D A_3^{(1)},\\
            0 &= P_1^{\mu_1\cdots\mu_6} T_{4,\mu_1\cdots\mu_6} 
                = D^2 A_1^{(1)} + D \big( D^2 + D + 4 \big) A_2^{(1)} + 4D \big( D + 1 \big) A_3^{(1)},\\
            0 &= P_1^{\mu_1\cdots\mu_6} T_{10,\mu_1\cdots\mu_6} 
                = D A_1^{(1)} + 3 D \big( D + 1 \big) A_2^{(1)} + D \big( D^2 + 3 D + 4 \big) A_3^{(1)}.
        \end{aligned}
    \end{equation}
    Solving these three equations gives
    \begin{equation}\label{Eq:ProjectorCoeffs-Rank6}
        \begin{aligned}
            A^{(1)}_1 &= \frac{D^2+3D-2}{(D^2+2D-8)(D^2+D-2)D},\\
            A^{(1)}_2 &= - \frac{1}{(D^2+2D-8)(D-1)D},\\
            A^{(1)}_3 &= \frac{2}{(D^2+2D-8)(D^2+D-2)D},
        \end{aligned}
    \end{equation}
    which determine the projector $P_1^{\mu_1\cdots\mu_6}$.
    All remaining projectors $P_{b\neq1}^{\mu_1\cdots\mu_6}$ follow by permuting Lorentz indices according to proposition~\ref{Prop:Invariant_Sum}.
    This is conveniently implemented in \texttt{FORM} using so-called wildcards.
\end{example}

\begin{remark}[Tensor Reduction at higher Rank and the Structure of Contraction Cycles]\ \\
    Tracing the metric tensor yields $\eta^{\mu\nu}\eta_{\mu\nu}=D$.
    In example~\ref{Example:Tensor-Reduction-Rank-6} discussed above, each tensor is a product of three metric tensors.
    One might therefore naively try to determine the orbits by counting the power of $D$ obtained upon contraction of the basis tensors with the chosen reference tensor.
    For $r=6$, see example~\ref{Example:Tensor-Reduction-Rank-6}, this gives
    \begin{equation*}
        \begin{aligned}
            T_1^{\mu_1\cdots\mu_6} \, T_{1,\mu_1\cdots\mu_6} &= \eta^{\mu_{1}\mu_{2}}\eta^{\mu_{3}\mu_{4}}\eta^{\mu_{5}\mu_{6}} \, \eta_{\mu_{1}\mu_{2}} \eta_{\mu_{3}\mu_{4}} \eta_{\mu_{5}\mu_{6}} = D^3,\\
            T_1^{\mu_1\cdots\mu_6} \, T_{4,\mu_1\cdots\mu_6} &= \eta^{\mu_{1}\mu_{2}}\eta^{\mu_{3}\mu_{4}}\eta^{\mu_{5}\mu_{6}} \, \eta_{\mu_{1}\mu_{3}}\eta_{\mu_{2}\mu_{4}}\eta_{\mu_{5}\mu_{6}} = D^2,\\
            T_1^{\mu_1\cdots\mu_6} \, T_{10,\mu_1\cdots\mu_6} &= \eta^{\mu_{1}\mu_{2}}\eta^{\mu_{3}\mu_{4}}\eta^{\mu_{5}\mu_{6}} \, \eta_{\mu_{1}\mu_{5}}\eta_{\mu_{2}\mu_{3}}\eta_{\mu_{4}\mu_{6}} = D,
        \end{aligned}
    \end{equation*}
    which indeed separates the three orbits for $r=6$.
    
    However, this criterion fails for higher tensor rank.
    Starting at $r=8$, the number of different types of cycles (and hence orbits) exceeds the number of possible powers of $D$, cf.\ Tab.~\ref{Tab:TensorSize}.
    The reason is that each closed contraction loop --- regardless of its length (number of indices involved) --- contributes a single factor of $D$.
    Counting powers of $D$ only detects the number of cycles, not their length partition, and thus does not resolve the complete structure.
    The full cycle structure --- defined by index contraction with the reference tensor and uniquely classifying all disjoint orbits --- includes the lengths of the individual cycles.
    Consequently, the number of orbits equals the number of integer partitions of $r/2$ (see Tab.~\ref{Tab:TensorSize}), which can be larger than the maximal $D$-power.
        
    As an explicit illustration, consider $r=10$, with $T_1^{\mu_1\cdots\mu_{10}}=\eta^{\mu_{1}\mu_{2}}\eta^{\mu_{3}\mu_{4}}\eta^{\mu_{5}\mu_{6}}\eta^{\mu_{7}\mu_{8}}\eta^{\mu_{9}\mu_{10}}$ as the reference tensor.
    There are seven disjoint orbits, corresponding to the seven integer partitions of $r/2=5$.
    They are characterised by different cycle structures or equivalently by \emph{cycle-length partition} (number of disjoint cycles and their lengths measured in contracted indices; the associated partition of $r/2$ is obtained by halving each length). 
    Orbit representatives and their contraction cycles with the reference tensor $T_1^{\mu_1\cdots\mu_{10}}$ are (cf.\ example~\ref{Example:Tensor-Reduction-Rank-6}):
    \begin{itemize}
        \item 5-cycle orbit (integer partition of $r/2$: $1+1+1+1+1$): this is the orbit that only contains the reference tensor itself and admits the maximal number of five cycles with two elements each, i.e.\ $1\rightarrow2\rightarrow1$, $3\rightarrow4\rightarrow3$, $5\rightarrow6\rightarrow5$, $7\rightarrow8\rightarrow7$ and $9\rightarrow10\rightarrow9$.
        \item 4-cycle orbit (integer partition of $r/2$: $2+1+1+1$): this orbit admits one cycle with four and three cycles with two elements, e.g.\ $T_2^{\mu_1\cdots\mu_{10}}=\eta^{\mu_{1}\mu_{2}}\eta^{\mu_{3}\mu_{4}}\eta^{\mu_{5}\mu_{6}}\eta^{\mu_{7}\mu_{9}}\eta^{\mu_{8}\mu_{10}}$ with cycles $1\rightarrow2\rightarrow1$, $3\rightarrow4\rightarrow3$, $5\rightarrow6\rightarrow5$, $7\rightarrow9\rightarrow10\rightarrow8\rightarrow7$.
        \item 3-cycle orbit (integer partition of $r/2$: $2+2+1$): two cycles with four and one cycle with two elements, e.g.\ $T_{22}^{\mu_1\cdots\mu_{10}}=\eta^{\mu_{1}\mu_{2}}\eta^{\mu_{3}\mu_{5}}\eta^{\mu_{4}\mu_{6}}\eta^{\mu_{7}\mu_{9}}\eta^{\mu_{8}\mu_{10}}$ with cycles $1\rightarrow2\rightarrow1$, $3\rightarrow5\rightarrow6\rightarrow4\rightarrow3$, $7\rightarrow9\rightarrow10\rightarrow8\rightarrow7$.
        \item 3-cycle orbit (integer partition of $r/2$: $3+1+1$): one cycle with six and two with two elements, e.g.\ $T_{82}^{\mu_1\cdots\mu_{10}}=\eta^{\mu_{1}\mu_{2}}\eta^{\mu_{3}\mu_{4}}\eta^{\mu_{5}\mu_{7}}\eta^{\mu_{6}\mu_{9}}\eta^{\mu_{8}\mu_{10}}$ with cycles $1\rightarrow2\rightarrow1$, $3\rightarrow4\rightarrow3$, $5\rightarrow7\rightarrow8\rightarrow10\rightarrow9\rightarrow6\rightarrow5$.
        \item 2-cycle orbit (integer partition of $r/2$: $3+2$): one cycle with six and one with four elements, e.g.\ $T_{162}^{\mu_1\cdots\mu_{10}}=\eta^{\mu_{1}\mu_{3}}\eta^{\mu_{2}\mu_{4}}\eta^{\mu_{5}\mu_{7}}\eta^{\mu_{6}\mu_{9}}\eta^{\mu_{8}\mu_{10}}$ with cycles $1\rightarrow3\rightarrow4\rightarrow2\rightarrow1$, $5\rightarrow7\rightarrow8\rightarrow10\rightarrow9\rightarrow6\rightarrow5$.
        \item 2-cycle orbit (integer partition of $r/2$: $4+1$): one cycle with eight and one with two elements, e.g.\ $T_{322}^{\mu_1\cdots\mu_{10}}=\eta^{\mu_{1}\mu_{2}}\eta^{\mu_{3}\mu_{5}}\eta^{\mu_{4}\mu_{7}}\eta^{\mu_{6}\mu_{9}}\eta^{\mu_{8}\mu_{10}}$ with cycles $1\rightarrow2\rightarrow1$, $3\rightarrow5\rightarrow6\rightarrow9\rightarrow10\rightarrow8\rightarrow7\rightarrow4\rightarrow3$.
        \item 1-cycle orbit (integer partition of $r/2$: $5$): this orbit admits only a single cycle with all ten elements, e.g.\ $T_{562}^{\mu_1\cdots\mu_{10}}=\eta^{\mu_{1}\mu_{3}}\eta^{\mu_{2}\mu_{5}}\eta^{\mu_{4}\mu_{7}}\eta^{\mu_{6}\mu_{9}}\eta^{\mu_{8}\mu_{10}}$ with cycles $1\rightarrow3\rightarrow4\rightarrow7\rightarrow8\rightarrow10\rightarrow9\rightarrow6\rightarrow5\rightarrow2\rightarrow1$.
    \end{itemize}
    Note that the cycle length (in indices) is always twice the corresponding part in the partition of $r/2$, as the number of indices is given by $r$.
    This example demonstrates that different orbits can share the same number of cycles --- and hence yield the same power of $D$ upon contraction with the reference tensor --- while differing in their cycle-length partition (e.g.\ for three cycles: $2+2+1$ vs.\ $3+1+1$).
    Therefore, the correct and complete classification of orbits is given by the integer partitions of $r/2$, not by $D$–power counting.
\end{remark}

\begin{remark}[on Contraction Cycles]\ \\
    As a contextual aside, we explain the correspondence between our ``contraction-cycle'' picture and the standard textbook cycle notation for permutations.
    The crucial link is that tensor contraction defines a permutation on the set of indices, i.e.\ contracting the reference tensor $T_1^{\mu_1\cdots\mu_r}$ with another basis tensor $T_{b,\mu_1\cdots\mu_r}$ induces the corresponding permutation on the index labels: 
    start from a label on $T_1$, follow its contracted partner to $T_b$, return to $T_1$, and iterate.
    The resulting closed walk is a cycle; collecting all such walks yields the disjoint cycle decomposition of the permutation.

    For instance, in example~\ref{Example:Tensor-Reduction-Rank-6} for $r=6$, the permutation needed to go from $T_1$ to $T_4$ is
    \begin{align*}
        \sigma_{1\to4}=
        \begin{pmatrix}
            1 & 2 & 3 & 4 & 5 & 6\\
            3 & 1 & 4 & 2 & 6 & 5
        \end{pmatrix} \in S_2^6.
    \end{align*}
    In cycle notation, this can be written as the two disjoint cycles $(1,3,4,2)$ and $(5,6)$, i.e.\ $\sigma_{1\to4}=(1,3,4,2)(5,6)$, which matches exactly the two contraction cycles identified above (one of length $4$ and one of length $2$).
    Thus, contraction cycles are simply a ``graphical representation'' of the disjoint cycle decomposition of the induced permutation. 
    Counting index-contraction loops gives the number of cycles, and recording their lengths gives the cycle-length partition; together, these data label the $H(T_a)$-orbits and explain why the graphical method correctly classifies tensors into orbits.
\end{remark}

Applying this procedure to all relevant tensor ranks, we precomputed the coefficients and projectors up to rank $r\leq16$.
They are tabulated and called at run time by the tensor reduction routine in our \texttt{FORM} program.

\paragraph{Applying the Tensor Reduction to Tensor Integrals:}
Unlike the projector tables, which are precomputed once, extracting the coefficients must be performed for each tensor integral during the calculation. 
Concretely, the coefficients $C_a$ of a tensor integral are determined by applying the precomputed projectors via Eq.~\eqref{Eq:ProjectionOfCoefficients}.
A naive approach would involve projecting the integral with each of the $n=(r-1)!!$ projectors and solve for every coefficient separately, which quickly affects the computational performance adversely.
This step can be optimised by exploiting the symmetries of the integrand's numerator, which must be preserved under tensor reduction.

The numerator of a tensor integral --- being a product of loop momenta --- often carries permutation symmetries of its Lorentz indices.
For example, a factor $k_1^{\mu_1}k_1^{\mu_2}$ in the numerator is symmetric under the exchange of indices $\mu_1\leftrightarrow\mu_2$.
The set of all such permutations that leave the entire integrand invariant forms the stabiliser $H(I)=\{h\in S_r\,|\,h\circ I^{\mu_1\cdots\mu_r} = I^{\mu_1\cdots\mu_r}\}$ of the full tensor integral in $S_r$.
Applying this invariance to the basis decomposition (see Eq.~\eqref{Eq:GeneralTensorReduction}) yields a powerful constraint.
For any $h\in H(I)$, we find
\begin{equation}\label{Eq:Exploit-Invariance-of-Tensor-Integrand}
    \begin{aligned}
        \sum_{a=1}^{n} C_a \, T_{a}^{\mu_1\cdots\mu_r} = I^{\mu_1\cdots\mu_r} = h \circ I^{\mu_1\cdots\mu_r} = h\circ \bigg(\sum_{a=1}^{n} C_a \, T_{a}^{\mu_1\cdots\mu_r}\bigg) = \sum_{a=1}^{n} C_a \big(h\circ T_{a}^{\mu_1\cdots\mu_r}\big).
    \end{aligned}
\end{equation}
Since the basis tensors $\{T_a\}_{a=1}^n$ are linearly independent, equality of the expansions implies that whenever $h\circ T_a=T_b$ one must have $C_a=C_b$.
Thus, all basis tensors related by $H(I)$ --- i.e.\ that can be transformed into one another by the symmetries of the integrand --- form an orbit under $H(I)$ and share the same scalar coefficient.

In the spirit of the orbit partition approach used above for constructing projectors, the decomposition in Eq.~\eqref{Eq:GeneralTensorReduction} can therefore be regrouped into a sum over $H(I)$-invariant orbit sums, each multiplied by one independent coefficient.
Practically, this reduces the number of coefficients to compute --- one representative coefficient for each orbit under integrand's symmetry group $H(I)$ instead of originally $n$.

As an illustration, consider the rank $r=6$ tensor integral
\begin{equation}\label{Eq:Rank6-TensorIntegral}
    \begin{aligned}
        I^{\mu_1\cdots\mu_6} 
        &=
        \big(\mu^{4-D}\big)^L 
        \int \Bigg( \prod_{i=1}^{L} \frac{d^Dk_i}{(2\pi)^D} \Bigg) 
        \frac{k_1^{\mu_1}k_1^{\mu_2}k_2^{\mu_3}k_2^{\mu_4}k_2^{\mu_5}k_2^{\mu_6}}{\mathcal{D}_1^{\nu_1} \cdots \mathcal{D}_N^{\nu_N}}.
    \end{aligned}
\end{equation}
The numerator $\mathcal{N}^{\mu_1\cdots\mu_6}=k_1^{\mu_1}k_1^{\mu_2}k_2^{\mu_3}k_2^{\mu_4}k_2^{\mu_5}k_2^{\mu_6}$ is invariant under permutations of the first two Lorentz indices $\{\mu_1,\mu_2\}$, and independently of the last four Lorentz indices $\{\mu_3,\mu_4,\mu_5,\mu_6\}$.
The symmetry group is thus given by $H(I) \cong S_2 \times S_4$.
The action of this symmetry group on the $15$ rank-$6$ basis tensors partitions $\mathcal{B}=\{T_a\}_{a=1}^{15}$ into only two orbits.
Consequently, the decomposition of the tensor integral (cf.\ Eq.~\eqref{Eq:GeneralTensorReduction}) reduces to
\begin{equation}\label{Eq:TensorReduced-Rank6-TensorIntegral}
    \begin{aligned}
        I^{\mu_1\cdots\mu_6} &= C_1 \, \big( T_1 + T_2 + T_3 \big)^{\mu_1\cdots\mu_6}
        + C_4 \, \big( T_4 + T_5 + T_6 + \ldots + T_{15} \big)^{\mu_1\cdots\mu_6},
    \end{aligned}
\end{equation}
with only two (instead of $n=15$) independent coefficients.
One orbit is formed by $\{T_1,T_2,T_3\}$ and the other by $\{T_4,\ldots,T_{15}\}$, such that each of the two terms in Eq.~\eqref{Eq:TensorReduced-Rank6-TensorIntegral} is individually invariant under the symmetry group $H(I)$ of the integrand in Eq.~\eqref{Eq:Rank6-TensorIntegral}.
Projecting with one representative projector for each coefficient (see Eq.~\eqref{Eq:ProjectionOfCoefficients}) yields\footnote{Using the rank-$6$ projector coefficients $A^{(1)}_c$ from Eq.~\eqref{Eq:ProjectorCoeffs-Rank6}.}
\begin{equation}
    \begin{aligned}
        C_1 &= P_1^{\mu_1\cdots\mu_6} \, I_{\mu_1\cdots\mu_6}
        =
        \int \Bigg( \prod_{i=1}^{L} \frac{d^Dk_i}{(2\pi)^D} \Bigg) 
        \frac{\big(\mu^{4-D}\big)^L}{\mathcal{D}_1^{\nu_1} \cdots \mathcal{D}_N^{\nu_N}} \frac{(D+3)k_1^2k_2^4-4(k_1\cdot k_2)^2k_2^2}{(D-1)D(D+2)(D+4)},\\
        C_4 &= P_4^{\mu_1\cdots\mu_6} \, I_{\mu_1\cdots\mu_6}
        =  
        \int \Bigg( \prod_{i=1}^{L} \frac{d^Dk_i}{(2\pi)^D} \Bigg)
        \frac{\big(\mu^{4-D}\big)^L}{\mathcal{D}_1^{\nu_1} \cdots \mathcal{D}_N^{\nu_N}} \frac{-k_1^2k_2^4+D(k_1\cdot k_2)^2k_2^2}{(D-1)D(D+2)(D+4)}.
    \end{aligned}
\end{equation}
Hence, it suffices to contract with one projector per $H(I)$-orbit; the chosen projector must share the index-permutation symmetries of the corresponding orbit sum.
This symmetry exploitation substantially reduces the computational effort required to obtain the scalar integrals.

A convenient way to identify the terms forming the $H(I)$-invariant orbit sums is to contract the integrand with a totally symmetric tensor built from Kronecker deltas,
\begin{align*}
\sum_{\sigma\in S_2^r} \sigma \circ (\delta^{\mu_1\mu_2}\ldots\delta^{\mu_{r-1}\mu_r}),    
\end{align*}
which projects onto the required invariant combinations, as mentioned in Ref.~\cite{Goode:2024mci}.
In \texttt{FORM}, we implemented this using the built-in functions \texttt{dd\_} and \texttt{distrib\_} to generate the invariant combinations while avoiding duplicates, as suggested in Ref.~\cite{Ruijl:2018poj}.

\section{Implementation of the BMHV Algebra}\label{Sec:Implementation_of_the_BMHV_Algebra}

The $D$-dimensional Dirac algebra --- its traces and (anti-)commutation relations --- has been introduced in chapter~\ref{Chap:DReg}; see in particular Sec.~\ref{Sec:Elements_of_D-dim_Spacetime} for the general $D$-dimensional framework and Sec.~\ref{Sec:The-BMHV-Scheme} for the BMHV scheme with its modified algebraic relations.
Recall that in the BMHV prescription $\gamma_5$ does not anticommute with the fully $D$-dimensional $\gamma^\mu$-matrices, leading to altered algebraic relations summarised in Eq.~\eqref{Eq:BMHV-Algebra}, known as BMHV algebra.

\texttt{FORM} does not provide a native implementation of the BMHV algebra, nor of the explicit decomposition of $D$-dimensional objects (see Sec.~\ref{Sec:Elements_of_D-dim_Spacetime}, in Eqs.~\eqref{Eq:Metric-split}, \eqref{Eq:Split-of-generic-Lorentz-Covariant-X}, \eqref{Eq:gamma-split} and \eqref{Eq:Gamma-Split-Algebra}); it can handle only strictly $4$- or strictly $n$-dimensional $\gamma$-matrices.
We therefore supplied the missing relations by hand and organised the implementation so that cancellations occur as early as possible, improving computational performance at higher orders.
For example, any contraction between 4-dimensional and evanescent structures vanishes, such as $\overline{\eta}_{\mu\nu}\mathrm{Tr}(\ldots\widehat{\gamma}^{\nu}\ldots)=0$.

Evaluating traces of $\gamma$-matrices requires particular care.
Although \texttt{FORM} is equipped with very efficient trace routines, issues arise for $D$-dimensional traces:
they cannot handle $\gamma_5$, and mixed traces involving $\gamma$-matrices of different dimensionality --- as encountered in the BMHV setting --- are not supported, since the corresponding BMHV algebra is not implemented.
However, strictly $4$-dimensional $\gamma$-traces are heavily optimised using dedicated rules and identities, such as the so-called \emph{Chisholm identity}
\begin{align}\label{Eq:Chisholm_Identity}
    \overline{\gamma}_\mu \mathrm{Tr}(\overline{\gamma}^\mu S) = 2 (S + S^R),
\end{align}
which is valid only in $4$ dimensions.
Here $S$ is a string of an odd number of 4-dimensional $\gamma$-matrices, i.e.\ $S=\overline{\gamma}^{\alpha_1}\ldots\overline{\gamma}^{\alpha_{2n+1}} \mathbb{\Lambda}$ with $\mathbb{\Lambda}\in\{\mathbb{1},\gamma_5,\projL,\projR\}$, and $S^R$ denotes the same string in reversed order.
Clearly, this identity generalises to $\overline{\gamma}^\rho\overline{\gamma}_\mu\overline{\gamma}^\sigma\mathrm{Tr}(\overline{\gamma}^\mu S) = 2\overline{\gamma}^\rho(S + S^R)\overline{\gamma}^\sigma$.
Sometimes the relation
\begin{align}\label{Eq:Chisholm-2}
    \overline{\gamma}^{\mu}\overline{\gamma}^{\nu}\overline{\gamma}^{\rho} 
        = \overline{\eta}^{\mu\nu}\overline{\gamma}^{\rho}
        - \overline{\eta}^{\mu\rho}\overline{\gamma}^{\nu}
        + \overline{\eta}^{\nu\rho}\overline{\gamma}^{\mu}
        + i\varepsilon^{\mu\nu\rho\sigma}\overline{\gamma}_{\sigma}\gamma_5
\end{align}
is referred to as Chisholm identity; here, we adopt the conventions of Ref.~\cite{FORM:Manual}.
The identity in Eq.~\eqref{Eq:Chisholm-2} is also strictly $4$-dimensional.
These algorithms are highly efficient, well-established and thoroughly tested.
Furthermore, they keep term proliferation under control, see Ref.~\cite{FORM:Manual}.
In order to benefit from such optimisations, we factorise every Dirac trace into a $D$-dimensional piece without $\gamma_5$ and a strictly $4$-dimensional piece in which $\gamma_5$ may possibly appear.

As a first step, each $\gamma^{\mu}$-matrix is expressed in terms of $\overline{\gamma}^{\mu}$ and $\widehat{\gamma}^{\mu}$. 
For $D$-dimensional matrices, and depending on the encountered situation, this is done either by $\gamma^{\mu}=\overline{\gamma}^{\mu}+\widehat{\gamma}^{\mu}$ or
\begin{equation}\label{Eq:Projector-Action}
    \begin{aligned}
        \mathrm{Tr}(\ldots\mathbb{P}_{\alpha}\gamma^{\mu}\mathbb{P}_{\beta}\ldots)
        = 
        \begin{cases}
            \mathrm{Tr}(\ldots\overline{\gamma}^{\mu}\projR\ldots), &\alpha=\mathrm{L},\,\beta=\mathrm{R},\\
            \mathrm{Tr}(\ldots\overline{\gamma}^{\mu}\projL\ldots), &\alpha=\mathrm{R},\,\beta=\mathrm{L},\\
            \mathrm{Tr}(\ldots\widehat{\gamma}^{\mu}\projR\ldots), &\alpha=\mathrm{R},\,\beta=\mathrm{R},\\
            \mathrm{Tr}(\ldots\widehat{\gamma}^{\mu}\projL\ldots), &\alpha=\mathrm{L},\,\beta=\mathrm{L},\\
        \end{cases}
    \end{aligned}
\end{equation}
where $\mathbb{P}_\mathrm{L/R}=(\mathbb{1}\mp\gamma_5)/2$.
While the latter variant is computationally more efficient, as it avoids an increase in the number of terms, the first approach is always possible even without projectors.
We then bring every $\gamma$-trace into a \emph{normal form}, with all evanescent matrices to the left and all $4$-dimensional matrices to the right:
\begin{definition}[Normal Form of Dirac Traces]\label{Def:Dirac-Trace-Normal-Form}\ \\
    A trace is in \emph{normal form} if it reads
    \begin{equation}\label{Eq:TraceNormalForm}
        \begin{aligned}
            \mathrm{Tr}(\widehat{\gamma}^{\nu_1}\ldots\widehat{\gamma}^{\nu_m}\overline{\gamma}^{\mu_1}\ldots\overline{\gamma}^{\mu_n}\mathbb{\Lambda}),
        \end{aligned}
    \end{equation}
    with $\mathbb{\Lambda}\in\{\mathbb{1},\gamma_5,\projL,\projR\}$.
\end{definition}
\begin{proposition}[Normal Form]\label{Prop:on-Normal-Form-for-Dirac-Traces}\ \\
    Every Dirac trace can be brought into normal form~\eqref{Eq:TraceNormalForm}.
\end{proposition}
\begin{proof}
    As mentioned above, every $D$-dimensional $\gamma$-matrix can be expressed in terms of its 4- and $(D-4)$-dimensional components, either explicitly as $\gamma^\mu=\overline{\gamma}^\mu+\widehat{\gamma}^\mu$ or according to Eq.~\eqref{Eq:Projector-Action}.
    In this way, each Dirac trace contains only 4- and $(D-4)$-dimensional objects.
    Subsequently, the (anti-)commutation relations $\overline{\gamma}^{\mu}\widehat{\gamma}^{\nu}=-\widehat{\gamma}^{\nu}\overline{\gamma}^{\mu}$ (see Eq.~\eqref{Eq:Gamma-Split-Algebra}) together with those of $\gamma_5$ (see Eq.~\eqref{Eq:BMHV-Algebra}) are applied to move all evanescent matrices to the left and all $4$-dimensional matrices to the right within the trace.
\end{proof}
Clearly, such traces are nonzero only for even $n$ and $m$ (see Sec.~\ref{Sec:Elements_of_D-dim_Spacetime}, Eq.~\eqref{Eq:GammaTraceDefinitions}).
Then, a trace of this form factorises as follows.
\begin{proposition}[Dimensional Factorisation of Dirac Traces]\label{Prop:Dimensional_Factorisation_of_Dirac_Traces}\ \\
    Any $D$-dimensional Dirac trace in normal form~\eqref{Eq:TraceNormalForm} factorises into the product of one trace over strictly 4-dimensional and one trace over strictly $(D-4)$-dimensional matrices as
    \begin{equation}\label{Eq:TraceFactorisationTheorem}
        \begin{aligned}
            \mathrm{Tr}(\widehat{\gamma}^{\nu_1}\ldots\widehat{\gamma}^{\nu_m}\overline{\gamma}^{\mu_1}\ldots\overline{\gamma}^{\mu_n}\mathbb{\Lambda})
            = \frac{1}{4} \mathrm{Tr}(\widehat{\gamma}^{\nu_1}\ldots\widehat{\gamma}^{\nu_m}) \mathrm{Tr}(\overline{\gamma}^{\mu_1}\ldots\overline{\gamma}^{\mu_n}\mathbb{\Lambda}).
        \end{aligned}
    \end{equation}
\end{proposition}
\begin{proof}
    We first take $\mathbb{\Lambda}=\mathbb{1}$ and discuss the generalisation to $\mathbb{\Lambda}\in\{\mathbb{1},\gamma_5,\projL,\projR\}$ below.
    
    For the trace over strictly $4$-dimensional matrices, i.e.\ $\mathrm{Tr}(\overline{\gamma}^{\mu_1}\ldots\overline{\gamma}^{\mu_n}\mathbb{1})$, we obtain
    \begin{equation}\label{Eq:GammaTrace-4-Dim}
        \begin{aligned}
        \mathrm{Tr}(\overline{\gamma}^{\mu_1}\ldots\overline{\gamma}^{\mu_n})
        &= \mathrm{Tr}(\overline{\gamma}^{\mu_2}\ldots\overline{\gamma}^{\mu_n}\overline{\gamma}^{\mu_1})
        = \mathrm{Tr}\Big(\frac{1}{2}\{\overline{\gamma}^{\mu_1},\overline{\gamma}^{\mu_2}\ldots\overline{\gamma}^{\mu_n}\}\Big)\\
        &= \sum_{k=2}^{n} (-1)^k \, \overline{\eta}^{\mu_1\mu_k} \, \mathrm{Tr}(\overline{\gamma}^{\mu_2}\ldots\overline{\gamma}^{\mu_{k-1}}\overline{\gamma}^{\mu_{k+1}}\ldots\overline{\gamma}^{\mu_{n}})\\
        &= \sum_{k=1}^{(n-1)!!} (-1)^{\sigma_k(1,\ldots,n)} \, \overline{\eta}^{\mu_{\sigma_k(1)}\mu_{\sigma_k(2)}}\ldots\overline{\eta}^{\mu_{\sigma_k(n-1)}\mu_{\sigma_k(n)}} \, \mathrm{Tr}(\mathbb{1})\\
        &= 4 \sum_{k=1}^{(n-1)!!} (-1)^{\sigma_k(1,\ldots,n)} \, \overline{\eta}^{\mu_{\sigma_k(1)}\mu_{\sigma_k(2)}}\ldots\overline{\eta}^{\mu_{\sigma_k(n-1)}\mu_{\sigma_k(n)}},
        \end{aligned}
    \end{equation}
    which is a well-known formula.
    Here, cyclicity of the trace was used in the first line, the anticommutator was evaluated in the second line (see Eq.~\eqref{Eq:Gamma-Split-Algebra}), the third line was obtained by recursion, and Eq.~\eqref{Eq:GammaTraceDefinitions} was applied in the final step.
    Furthermore, $(n-1)!!=n!/(2^{n/2}(n/2)!)$, and
    \begin{equation}
        \begin{aligned}
            (-1)^{\sigma_k(1,\ldots,n)} = 
            \begin{cases}
                1, &\text{for even permutations of $\{1,\ldots,n\}$},\\
                -1, &\text{for odd permutations of $\{1,\ldots,n\}$}.\\
            \end{cases}
        \end{aligned}
    \end{equation}
    An analogous recursion yields
    \begin{equation}\label{Eq:GammaTrace-Evanescent}
        \begin{aligned}
            \mathrm{Tr}(\widehat{\gamma}^{\nu_1}\ldots\widehat{\gamma}^{\nu_m})
            &= \sum_{k=2}^{m} (-1)^k \, \widehat{\eta}^{\nu_1\nu_k} \, \mathrm{Tr}(\widehat{\gamma}^{\nu_2}\ldots\widehat{\gamma}^{\nu_{k-1}}\widehat{\gamma}^{\nu_{k+1}}\ldots\widehat{\gamma}^{\nu_{m}})\\
            &= \sum_{k=1}^{(m-1)!!} (-1)^{\sigma_k(1,\ldots,m)} \, \widehat{\eta}^{\nu_{\sigma_k(1)}\nu_{\sigma_k(2)}}\ldots\widehat{\eta}^{\nu_{\sigma_k(m-1)}\nu_{\sigma_k(m)}} \, \mathrm{Tr}(\mathbb{1})\\
            &= 4 \sum_{k=1}^{(m-1)!!} (-1)^{\sigma_k(1,\ldots,m)} \, \widehat{\eta}^{\nu_{\sigma_k(1)}\nu_{\sigma_k(2)}}\ldots\widehat{\eta}^{\nu_{\sigma_k(m-1)}\nu_{\sigma_k(m)}},
        \end{aligned}
    \end{equation}
    for traces over evanescent $\widehat{\gamma}$'s.
    Finally, evaluating the LHS of Eq.~\eqref{Eq:TraceFactorisationTheorem} gives
    \begin{align}\label{Eq:GammaTrace-General}
        &\mathrm{Tr}(\widehat{\gamma}^{\nu_1}\ldots\widehat{\gamma}^{\nu_m}\overline{\gamma}^{\mu_1}\ldots\overline{\gamma}^{\mu_n}) \nonumber\\
        &= (-1)^{n+m} \, \mathrm{Tr}(\widehat{\gamma}^{\nu_2}\ldots\widehat{\gamma}^{\nu_m}\widehat{\gamma}^{\nu_1}\overline{\gamma}^{\mu_2}\ldots\overline{\gamma}^{\mu_n}\overline{\gamma}^{\mu_1})
        = \mathrm{Tr}(\widehat{\gamma}^{\nu_2}\ldots\widehat{\gamma}^{\nu_m}\widehat{\gamma}^{\nu_1}\overline{\gamma}^{\mu_2}\ldots\overline{\gamma}^{\mu_n}\overline{\gamma}^{\mu_1}) \nonumber\\
        &= \mathrm{Tr}\Big(\frac{1}{2}\{\widehat{\gamma}^{\nu_1},\widehat{\gamma}^{\nu_2}\ldots\widehat{\gamma}^{\nu_m}\}\frac{1}{2}\{\overline{\gamma}^{\mu_1},\overline{\gamma}^{\mu_2}\ldots\overline{\gamma}^{\mu_n}\}\Big) \nonumber\\
        &= \sum_{k=2}^{m} \sum_{l=2}^{n} (-1)^k (-1)^l \, \widehat{\eta}^{\nu_1\nu_k} \, \overline{\eta}^{\mu_1\mu_l} \, \mathrm{Tr}(\widehat{\gamma}^{\nu_2}\ldots\widehat{\gamma}^{\nu_{l-1}}\widehat{\gamma}^{\nu_{l+1}}\ldots\widehat{\gamma}^{\nu_{m}}\overline{\gamma}^{\mu_2}\ldots\overline{\gamma}^{\mu_{k-1}}\overline{\gamma}^{\mu_{k+1}}\ldots\overline{\gamma}^{\mu_{n}}) \nonumber\\
        &= \sum_{k=1}^{(m-1)!!} (-1)^{\sigma_k(1,\ldots,m)} \, \widehat{\eta}^{\nu_{\sigma_k(1)}\nu_{\sigma_k(2)}}\ldots\widehat{\eta}^{\nu_{\sigma_k(m-1)}\nu_{\sigma_k(m)}} \\
        &\qquad \times \sum_{l=1}^{(n-1)!!} (-1)^{\sigma_l(1,\ldots,n)} \, \overline{\eta}^{\mu_{\sigma_l(1)}\mu_{\sigma_l(2)}}\ldots\overline{\eta}^{\mu_{\sigma_l(n-1)}\mu_{\sigma_l(n)}} \, \mathrm{Tr}(\mathbb{1}) \nonumber\\
        &= 4 \sum_{k=1}^{(m-1)!!} (-1)^{\sigma_k(1,\ldots,m)} \, \widehat{\eta}^{\nu_{\sigma_k(1)}\nu_{\sigma_k(2)}}\ldots\widehat{\eta}^{\nu_{\sigma_k(m-1)}\nu_{\sigma_k(m)}} \nonumber\\
        &\qquad \times \sum_{l=1}^{(n-1)!!} (-1)^{\sigma_l(1,\ldots,n)} \, \overline{\eta}^{\mu_{\sigma_l(1)}\mu_{\sigma_l(2)}}\ldots\overline{\eta}^{\mu_{\sigma_l(n-1)}\mu_{\sigma_l(n)}}, \nonumber
    \end{align}
    where cyclicity of the trace and $\{\overline{\gamma}^{\mu},\widehat{\gamma}^{\nu}\}=0$ were used, and only traces with even $n$ and $m$ contribute non-vanishing results in the second line.
    Rewriting the expression using anticommutators gives the third line, while evaluating these anticommutators yields the fourth.
    Recursion leads to the penultimate step, and the final equality follows from Eq.~\eqref{Eq:GammaTraceDefinitions}.
    Comparing Eqs.~\eqref{Eq:GammaTrace-4-Dim} and \eqref{Eq:GammaTrace-Evanescent} with Eq.~\eqref{Eq:GammaTrace-General} then reproduces the factorisation formula~\eqref{Eq:TraceFactorisationTheorem} for $\mathbb{\Lambda}=\mathbb{1}$.
    
    The additional factor of $1/4$ on the RHS of Eq.~\eqref{Eq:TraceFactorisationTheorem} arises because the product of two traces of the identity appears there, while only a single trace occurs on the LHS.
    This difference must be taken into account given that $\mathrm{Tr}(\mathbb{1})=f(D)=4$ (see Sec.~\ref{Sec:Elements_of_D-dim_Spacetime}, Eq.~\eqref{Eq:GammaTraceDefinitions}).

    For $\mathbb{\Lambda}\in\{\mathbb{1},\gamma_5,\projL,\projR\}$ the same steps apply, invoking the (anti-)commutation relations of $\gamma_5$ from Eq.~\eqref{Eq:BMHV-Algebra} in the first two steps of the evaluation of the LHS of Eq.~\eqref{Eq:TraceFactorisationTheorem} for general $\mathbb{\Lambda}$ (cf.\ Eq.~\eqref{Eq:GammaTrace-General} for the $\mathbb{\Lambda}=\mathbb{1}$ case).
    In addition, the following trace identities are used:
    \begin{equation}
        \begin{aligned}
            \mathrm{Tr}(\gamma_5) &= 0,\\
            \mathrm{Tr}(\overline{\gamma}^{\mu}\overline{\gamma}^{\nu}\gamma_5) &= 0,\\
            \mathrm{Tr}(\overline{\gamma}^{\mu_1}\ldots\overline{\gamma}^{\mu_{2n+1}}\gamma_5) &= 0,
        \end{aligned}
    \end{equation}
    and, for $n=4$,
    \begin{equation}
        \begin{aligned}
            \mathrm{Tr}(\gamma_5\overline{\gamma}^{\mu}\overline{\gamma}^{\nu}\overline{\gamma}^{\rho}\overline{\gamma}^{\sigma}) = -4i\varepsilon^{\mu\nu\rho\sigma}.
        \end{aligned}
    \end{equation}
    Even $n\geq6$ can be treated recursively using
    \begin{equation}
        \begin{aligned}
            \gamma_5\overline{\gamma}^{\mu}\overline{\gamma}^{\nu}\overline{\gamma}^{\rho} 
            = \overline{\eta}^{\mu\nu}\gamma_5\overline{\gamma}^{\rho}
            - \overline{\eta}^{\mu\rho}\gamma_5\overline{\gamma}^{\nu}
            + \overline{\eta}^{\nu\rho}\gamma_5\overline{\gamma}^{\mu}
            - i\varepsilon^{\mu\nu\rho\sigma}\overline{\gamma}_{\sigma},
        \end{aligned}
    \end{equation}
    which directly follows from Eq.~\eqref{Eq:Chisholm-2}.
\end{proof}

In \texttt{FORM}, both types of traces can now be evaluated using the built-in $\gamma$-trace algorithms, thereby utilising all algorithmic benefits and optimisations.
In particular, traces over evanescent matrices, $\mathrm{Tr}(\widehat{\gamma}^{\nu_1}\ldots\widehat{\gamma}^{\nu_m})$, can be handled by the $n$-dimensional routines, whereas traces of the form $\mathrm{Tr}(\overline{\gamma}^{\mu_1}\ldots\overline{\gamma}^{\mu_n}\mathbb{\Lambda})$ can be processed by the program's dedicated $4$-dimensional algorithms.
In the latter case, the occurrence of $\gamma_5$ does not cause any complications, since these traces are evaluated strictly within $4$ dimensions.
All algebraic manipulations leading to Eq.~\eqref{Eq:TraceFactorisationTheorem} must, of course, consistently respect the BMHV-algebra given in Eq.~\eqref{Eq:BMHV-Algebra}, which has been implemented explicitly in our \texttt{FORM} routines.

\section{Practical Implementation of Counterterms}\label{Sec:Practical_Implementation_of_Counterterms}

All renormalised Green functions in this thesis are computed using the counterterm method introduced in Sec.~\ref{Sec:Renormalisation_Theory} (see corollary~\ref{Thm:Corollary-on-Counterterm-Method}).
At $L$-loop order, the evaluation of a given Green function thus requires not only the set of all genuine $L$-loop Feynman diagrams contributing to it, but also all diagrams of lower loop order containing counterterm insertions.
In this section, we present the implementation of these counterterm insertions by means of explicit examples.
For propagators, we illustrate the method using the gauge boson and fermion propagators, and for interaction vertices we consider the gauge interaction as a representative case.
Throughout this section, the counterterm coefficients, denoted by the various $\delta Z$'s, serve as placeholders; their explicit values are presented in chapters~\ref{Chap:General_Abelian_Chiral_Gauge_Theory}--\ref{Chap:The_Standard_Model}.

\paragraph{Propagator Counterterms:}
Counterterm insertions in propagators are implemented in \texttt{FORM} through modified propagator expressions that are applied recursively, depending on the desired insertion order.
For fermions, the recursive expression reads
\begin{equation}\label{Eq:RecursiveFermionPropagator}
    \begin{aligned}
        \mathcal{D}_{f,ij}(k) &= \frac{i \slashed{k}}{k^2} \bigg( \delta_{ij} 
        + i
        \!\!\!\!\!\!\sum_{\alpha,\beta\in\{\mathrm{L},\mathrm{R}\}}\!\!\!\!\!\!\delta Z^{f}_{\overline{\alpha}\beta,ik} \,
        \mathbb{P}_{\alpha} \slashed{k} \mathbb{P}_{\beta} \,
        \mathcal{D}_{f,kj}(k) \bigg),
    \end{aligned}
\end{equation}
where an index with a bar, e.g.\ $\overline{\alpha}$, indicates opposite chirality, i.e.\ $\overline{\alpha}=\mathrm{R}$ for $\alpha=\mathrm{L}$ and vice versa.
For the gauge boson, the corresponding propagator is given by
\begin{equation}\label{Eq:RecursiveGaugeBosonPropagator}
    \begin{aligned}
        \mathcal{D}_g^{\mu\nu}(k) = \frac{-i}{k^2} \Big(\eta^{\mu\rho} &- \big(1 - \xi\big) \frac{k^{\mu}k^{\rho}}{k^2}\Big) 
        \Big( \delta_{\rho}^{\phantom{\rho}\nu} - i \Big[ 
        - \delta\overline{Z}^{g}_{M} M^2 \overline{\eta}_{\rho\sigma} 
        - \delta\widehat{Z}^{g}_{M} M^2 \widehat{\eta}_{\rho\sigma}\\
        &+ \delta Z^g_D \big( \eta_{\rho\sigma}k^2 - k_{\rho}k_{\sigma} \big) 
        + \delta \overline{Z}^g_{4D} \big( \overline{\eta}_{\rho\sigma}\overline{k}^2 - \overline{k}_{\rho}\overline{k}_{\sigma} \big)\\
        &+ \delta \overline{Z}^g_{11} \, \overline{\eta}_{\rho\sigma}\overline{k}^2
        + \delta \widehat{Z}^g_{12} \, \overline{\eta}_{\rho\sigma}\widehat{k}^2
        + \delta \widehat{Z}^g_{21} \, \widehat{\eta}_{\rho\sigma}\overline{k}^2
        + \delta \widehat{Z}^g_{22} \, \widehat{\eta}_{\rho\sigma}\widehat{k}^2\\
        &- \delta \overline{Z}^g_{33} \, \overline{k}_{\rho}\overline{k}_{\sigma}
        - \delta \widehat{Z}^g_{34} \, \overline{k}_{\rho}\widehat{k}_{\sigma}
        - \delta \widehat{Z}^g_{43} \, \widehat{k}_{\rho}\overline{k}_{\sigma}
        - \delta \widehat{Z}^g_{44} \, \widehat{k}_{\rho}\widehat{k}_{\sigma}
        \Big] \mathcal{D}_g^{\sigma\nu}(k) \Big).
    \end{aligned}
\end{equation}
This recursive construction follows the strategy of Ref.~\cite{Luthe:2017ttg}, but has been adapted to the specific requirements of the BMHV scheme.
The propagators include counterterm contributions built from all possible Lorentz covariants, ensuring a complete and consistent basis.\footnote{The $\delta Z$'s do not in general arise from a purely multiplicative renormalisation, since they also account for symmetry-restoring counterterms.}
This also includes the counterterms~$\propto M^2$ that incorporate the auxiliary counterterms required by the tadpole decomposition method introduced in Sec.~\ref{Sec:Tadpole_Decomposition}.
For each renormalisation constant $\delta Z$, a perturbative expansion is implied.
In the gauge boson propagator, the structures in the first and last two lines of Eq.~\eqref{Eq:RecursiveGaugeBosonPropagator} already form a complete basis.
For convenience, additional $D$- and $4$-dimensional transversal combinations, shown in the second line, are included as they frequently occur in practical computations, albeit at the cost of mild redundancy.
In non-Abelian gauge theories or models with scalar fields, analogous recursive definitions are required for the ghost and scalar propagators, respectively.

\paragraph{Vertex Counterterms:}
Counterterm insertions for interaction vertices are implemented straightforwardly.
For example, the gauge interaction takes the form
\begin{equation}\label{Eq:GaugeInteraction-Implementation}
    \begin{aligned}
        - i g \!\!\!\!\!\!\sum_{\alpha,\beta\in\{\mathrm{L},\mathrm{R}\}}\!\!\!\!\! \big( \mathcal{Y}_{\overline{\alpha}\beta,ij} 
        + \delta \mathcal{Y}_{\overline{\alpha}\beta,ij} \big)
        \mathbb{P}_{\alpha} \gamma^{\mu} \mathbb{P}_{\beta},
    \end{aligned}
\end{equation}
which covers both left- and right-handed fermions and naturally accommodates evanescent gauge interactions.

\paragraph{The Right-Handed Model as an Example:}
As an illustration --- and in anticipation of the discussion in chapter~\ref{Chap:BMHV_at_Multi-Loop_Level} --- we briefly outline the structure of the counterterms in the right-handed model considered there. 
Unlike the more general theory discussed in chapter~\ref{Chap:General_Abelian_Chiral_Gauge_Theory}, the right-handed model features only a nonzero right-handed hypercharge, $\mathcal{Y}_{RR}\equiv\mathcal{Y}_{R}$, while all other hypercharge matrices vanish. 
Consequently, left-handed and evanescent gauge interactions are absent, which reduces the number of required counterterms.
From the set of counterterms introduced above, only $\delta Z^{f}_{RR}$, $\delta\overline{Z}^{g}_{M}$, $\delta\overline{Z}^{g}_{4D}$, $\delta\overline{Z}^{g}_{11}$, $\delta\widehat{Z}^{g}_{12}$ and $\delta \mathcal{Y}_{RR}$ remain relevant in this case.
In chapter~\ref{Chap:BMHV_at_Multi-Loop_Level}, these renormalisation constants will be decomposed into symmetric contributions $\delta Z$, non-symmetric divergent parts $\delta X$ containing $1/\epsilon$-poles, and finite symmetry-restoring counterterms $\delta F$.
With this convention, the corresponding $L$-loop renormalisation constants in the right-handed model take the form
\begin{equation}
    \begin{aligned}
        \delta Z^{f,(L)}_{RR,ij} &= \frac{g^{2L}}{(16 \pi^2)^L} \Big( \delta Z^{(L)}_{\psi,ij} + \delta \overline{X}^{(L)}_{\overline{\psi}\psi,ij} + \delta F_{\psi\overline{\psi}, ij}^{(L)} \Big),\\
        \delta \overline{Z}^{g,(L)}_{4D} &= \frac{g^{2L}}{(16 \pi^2)^L} \, \delta Z^{(L)}_{B},\\ 
        \delta \overline{Z}^{g,(L)}_{11} &= \frac{g^{2L}}{(16 \pi^2)^L} \Big( \delta \overline{X}^{(L)}_{BB} + \delta F_{BB}^{(L)} \Big),\\ 
        \delta \widehat{Z}^{g,(L)}_{12} &= \frac{g^{2L}}{(16 \pi^2)^L} \, \delta \widehat{X}^{(L)}_{BB},\\ 
        \delta \mathcal{Y}^{(L)}_{RR,ij} &= \frac{g^{2L}}{(16 \pi^2)^L} \, \big(\mathcal{Y}_R\big)_{ik} \delta Z^{(L)}_{\psi,kj},
    \end{aligned}
\end{equation}
with explicit expressions provided in chapter~\ref{Chap:BMHV_at_Multi-Loop_Level}.
In addition, although not shown explicitly here, the complete renormalisation procedure also requires symmetry-restoring quartic gauge boson counterterms $\delta \overline{Z}_{4g}$ and quantum corrections to the $\Delta$-operator (see chapter~\ref{Chap:Practical_Symmetry_Restoration}).
Both of which are presented in chapter~\ref{Chap:BMHV_at_Multi-Loop_Level}, where the complete set of counterterms required up to and including the 4-loop level are presented.

\section{Integration by Parts Reduction and Master Integrals}\label{Sec:IBP-Reduction}

Multi-loop computations in quantum field theory generate a vast number of (scalar) Feynman integrals --- recall that tensor integrals can be reduced to scalar integrals via tensor reduction (see Sec.~\ref{Sec:Tensor_Reduction}).
To render these calculations tractable, all integrals are systematically expressed in terms of a minimal set of independent master integrals (MIs), which are then solved.
This reduction is accomplished through integration-by-parts (IBP) identities.\footnote{The integral reduction based on IBP identities is not the only approach to evaluating multi-loop diagrams; however, it is highly efficient and has become the standard method in practical computations.}
In this section, we briefly outline the basic concepts of IBP reduction and then describe the practical reduction strategy adopted for the computations in this thesis, which makes use of the software tools introduced in Sec.~\ref{Sec:Computational_Setup}.

Feynman integrals can be organised into integral families (or topologies), where all integrals within a family share the same set of denominators but differ in the integer powers of these denominators, called indices.
Integrals belonging to the same family are not independent but related by a system of linear relations, which are the aforementioned IBP identities.\footnote{There are additional linear relations, such as those arising from discrete symmetries and shift relations, which are not IBP identities.}
These identities follow from the fundamental property of dimensional regularisation that the integral of a total derivative over the $D$-dimensional momentum space vanishes.

For any $L$-loop integral, the IBP identities are obtained from the vanishing of
\begin{align} 
    \int \left(\prod_{i=1}^{L} \frac{d^D k_i}{(2\pi)^D}\right) \frac{\partial}{\partial k_j^\mu} \left( \frac{v^\mu}{\mathcal{D}_1^{\nu_1} \cdots \mathcal{D}_N^{\nu_N}} \right) = 0\,, 
\end{align}
where the $k_i$ are the loop momenta, $v^\mu$ is an arbitrary vector constructed from loop and external momenta, and $\mathcal{D}_b$ are the propagator denominators with indices $\nu_b$.
Applying the chain rule generates a large set of linear relations among integrals of the same family but with shifted indices.
These relations establish that all integrals in a given family span a finite-dimensional vector space over the field of rational functions in the spacetime dimension $D$ and other kinematic variables. 
Consequently, every integral in the family can be expressed as a linear combination of a finite set of basis integrals, the master integrals $M_a$:
\begin{align}\label{Eq:IBP-Integral-as-lin-comb-of-MIs}
    I = \sum_a c_a M_a,
\end{align}
with coefficients $c_a$ that are rational functions of $D$ and the kinematic variables.
The standard method for solving this system is the Laporta algorithm~\cite{Laporta:2000dsw}, which systematically applies Gaussian elimination based on a predefined ordering of integrals.
For comprehensive discussions of the IBP method and its modern implementations, see Refs.~\cite{Weinzierl:2022eaz,Smirnov:2019qkx,Smirnov:2023yhb,vonManteuffel:2012np,Maierhofer:2017gsa,Klappert:2020aqs,Chetyrkin:1981qh,Laporta:2000dsw} and the references therein.

\paragraph{Uniform Momentum Configuration:}
As discussed in Sec.~\ref{Sec:Computational_Setup}, the assignment of loop momenta in Feynman diagrams is not unique.
This arbitrariness complicates the classification of integrals into distinct integral families, since diagrams corresponding to the same topology may differ by their specific momentum routing.
To resolve this ambiguity, we employ the program \texttt{Feynson}, which systematically maps the arbitrary momentum configuration of each diagram to a predefined canonical basis for its integral family.
The algorithm achieves this by representing the integral's propagator structure as a graph and determining a canonical spanning tree that uniquely defines the loop-momentum basis (see Ref.~\cite{Maheria:2022dsq}).
Subsequently, \texttt{Feynson} solves a linear system to identify the required momentum shifts that transform the original routing into the canonical one.
This procedure constitutes a crucial preprocessing step in our workflow, ensuring that all integrals share a uniform momentum configuration (see Tab.~\ref{Tab:PropagatorMomentumConfigurations}) before being passed to the IBP reduction stage.

\paragraph{Practical Reduction Strategy:}
In this work, a multi-stage workflow is employed to perform the IBP reduction, combining the complementary capabilities of the different programs introduced in Sec.~\ref{Sec:Computational_Setup}.
This approach ensures an efficient reduction to a minimal and user-defined set of preferred master integrals (MIs).
For fully massive, single-scale vacuum integrals, the minimal basis of MIs contains one, two, five, and nineteen elements at the 1-, 2-, 3-, and 4-loop levels, respectively.
We adopt the same master integrals as in Refs.~\cite{Schroder:2005va,Martin:2016bgz,Czakon:2004bu} and use the momentum conventions of Ref.~\cite{Luthe:2015ngq} for the propagators (see Tab.~\ref{Tab:PropagatorMomentumConfigurations}).\footnote{At 4 loops, the master-integral basis used in Ref.~\cite{Czakon:2004bu} differs slightly from that of Ref.~\cite{Luthe:2015ngq}; the two can be related via IBP identities.}
The final minimal basis of preferred master integrals is denoted by $\mathcal{M}=\{M_a\}_{a\geq1}$.
The IBP reduction proceeds through the following sequence of steps:
\begin{enumerate}[label={(\arabic*)}]
    \item Initial Reduction with \texttt{FIRE}: An initial reduction of all required scalar integrals is carried out using the \texttt{C++} version of \texttt{FIRE}.
    While highly efficient, a single IBP solver run often produces a non-minimal set of master integrals, since certain linear relations --- especially those connecting different sectors --- may remain undetected.
    This stage therefore yields an intermediate, non-minimal basis $\mathcal{M}'$.
    \item Symmetry Analysis with \texttt{Reduze2}:
    The non-minimal basis $\mathcal{M}'$ is subsequently processed using \texttt{Reduze2}, which identifies additional relations among the master integrals arising from integral and graph symmetries.  
    These symmetries correspond to transformations of the loop momenta (e.g.\ permutations or shifts) that leave the set of propagator denominators invariant.  
    \texttt{Reduze2} analyses a graph representation of the integral family, determines its automorphisms, and systematically generates the associated sector relations, mapping equivalent integrals across sectors.  
    Solving the resulting extended system reduces the initial basis to the minimal, user-defined set $\mathcal{M}$, and produces replacement rules expressing all redundant MIs in terms of $\mathcal{M}$.
    \item Final Reduction with \texttt{FIRE}:
    The relations generated by \texttt{Reduze2} are translated into a format compatible with \texttt{FIRE}.  
    A second, final reduction is then performed using these rules, ensuring that every scalar integral is expressed as a linear combination of the elements in the minimal basis $\mathcal{M}$, as in Eq.~\eqref{Eq:IBP-Integral-as-lin-comb-of-MIs}.  
    This two-stage procedure guarantees that the final reduction tables contain only the preferred MIs.
    \item Cross-Checks: 
    The consistency and correctness of the resulting reduction tables are verified through independent cross-checks using \texttt{Kira} and \texttt{Finred}.
    \item Integration into the Computational Framework:
    Finally, the resulting reduction tables are converted into a \texttt{FORM}-readable format using \texttt{Mathematica}, allowing all scalar integrals in the main workflow to be automatically substituted by their master-integral decompositions.
\end{enumerate}

\paragraph{Master Integrals:}
The complexity of the IBP reduction grows rapidly with the number of physical scales in the problem, such as distinct masses or external momenta.
In the present case of fully massive, single-scale vacuum bubble integrals, only two variables appear: the spacetime dimension $D$ and the auxiliary mass $M$.
The dependence on $M$, however, can be removed by rescaling the loop momenta according to 
\begin{align}\label{Eq:Momentum-Rescaling-for-Integral-Evaluation}
    k_i\longrightarrow l_i\coloneqq \frac{k_i}{M},
\end{align}
where the rescaled momenta $l_i$ are dimensionless.
After this transformation, the integrals entering the IBP reduction depend solely on the single variable $D$, and the coefficients $c_a$ in Eq.~\eqref{Eq:IBP-Integral-as-lin-comb-of-MIs} become rational functions of $D$ only.
This considerably simplifies the reduction and substantially improves computational efficiency.

Starting from the general $L$-loop fully massive single-scale vacuum bubble integral of Eq.~\eqref{Eq:IntegralFamily}, with propagator denominators defined in Eq.~\eqref{Eq:General-Propagator-Denominator-for-Vacuum-Bubbles}, the rescaling yields
\begin{equation}
\begin{aligned}\label{Eq:General-Tadpole-Integral-rescaled}
    I(\nu_1,\ldots,\nu_N) &= \big(\mu^{4-D}\big)^L \int \Bigg( \prod_{i=1}^{L} \frac{d^Dk_i}{(2\pi)^D} \Bigg) \frac{1}{\mathcal{D}_1^{\nu_1} \cdots \mathcal{D}_N^{\nu_N}} \\
    &= \bigg(\frac{1}{(4\pi)^2}\bigg)^L \bigg(\frac{e^{\epsilon \gamma_E}}{\pi^{D/2}}\bigg)^L \bigg(\frac{\overline{\mu}^2}{M^2}\bigg)^{L\,\frac{4-D}{2}} \big(M^2\big)^{2L-\boldsymbol{\nu}} \int \Bigg( \prod_{i=1}^{L}d^Dl_i \Bigg) \frac{1}{\widetilde{\mathcal{D}}_1^{\nu_1} \cdots \widetilde{\mathcal{D}}_N^{\nu_N}},
\end{aligned}
\end{equation}
with rescaled denominators
\begin{align}
    \widetilde{\mathcal{D}}_i = \widetilde{q}_i^2 - 1, \qquad \text{with} \quad \widetilde{q}_i \coloneqq \frac{q_i}{M}.
\end{align}
Here, the $\widetilde{q}_i$ are linear combinations of the rescaled loop momenta $l_i$, analogous to how the $q_i$ are defined in terms of the original momenta $k_i$.
We also define $\boldsymbol{\nu} = \sum_i \nu_i$ and introduce the $\overline{\mathrm{MS}}$ scale $\overline{\mu}^2 \coloneqq 4\pi e^{-\gamma_E}\mu^2$.

Relating Eq.~\eqref{Eq:General-Tadpole-Integral-rescaled} to the convention used in Ref.~\cite{Czakon:2004bu}, where the integrals are denoted by $\mathrm{PR}(\nu_1,\ldots,\nu_N)$, we find
\begin{align}
    I(\nu_1,\ldots,\nu_N) = \bigg(\frac{i}{(4\pi)^2}\bigg)^L \bigg(\frac{\overline{\mu}^2}{M^2}\bigg)^{L\,\frac{4-D}{2}} \big(M^2\big)^{2L-\boldsymbol{\nu}} \, \mathrm{PR}(\nu_1,\ldots,\nu_N).
\end{align}
In Refs.~\cite{Luthe:2015ngq,Schroder:PrivateComm}, where we denote their integrals by $J(\nu_1,\ldots,\nu_N)$, the evaluation is performed in Euclidean spacetime, and the 1-loop tadpole of unit mass to the power of the loop order $L$ is factored out, such that
\begin{align}
    I(\nu_1,\ldots,\nu_N) = (-1)^{\boldsymbol{\nu}} \bigg(\frac{i}{(4\pi)^2}\bigg)^L \bigg(\frac{\overline{\mu}^2}{M^2}\bigg)^{L\,\frac{4-D}{2}} \big(M^2\big)^{2L-\boldsymbol{\nu}} \big(e^{\epsilon\gamma_E} \,\Gamma(1-D/2)\big)^L \,J(\nu_1,\ldots,\nu_N).
\end{align}
Both representations of the vacuum bubble master integrals, $\mathrm{PR}(\nu_1,\ldots,\nu_N)$ and $J(\nu_1,\ldots,\nu_N)$, have the advantage that they are free of mass scales and depend only on $\epsilon = (4 - D)/2$, making them particularly convenient for $\epsilon$-expansions.

In all calculations presented in this thesis, we employ the rescaled form of the integrals, cf.\ Eq.~\eqref{Eq:General-Tadpole-Integral-rescaled}.
In particular, all momenta are rescaled according to Eq.~\eqref{Eq:Momentum-Rescaling-for-Integral-Evaluation}, such that the integrals are expressed in dimensionless variables.
The IBP reduction is then performed in this dimensionless representation as outlined above, and the solutions for the master integrals are inserted in a form depending only on $\epsilon = (4 - D)/2$.
In this way, the rational arithmetic can be carried out in \texttt{FORM} with high efficiency, operating exclusively on rational functions of $\epsilon$.
Finally, the results are re-rescaled to restore the correct physical mass dimension in the divergent terms entering the counterterms.

\chapter{Symmetry Restoration}\label{Chap:Practical_Symmetry_Restoration}

The renormalisation of vector-like gauge theories within a regularisation scheme that manifestly preserves BRST invariance at all orders can straightforwardly be achieved (cf.\ Sec.~\ref{Sec:Renormalisation-of-GaugeTheories}).
In contrast, the renormalisation of chiral gauge theories presents a significant challenge, since no gauge invariant regularisation scheme is known that simultaneously maintains chiral symmetry.
This incompatibility gives rise to the well-known $\gamma_5$-problem (see Sec.~\ref{Sec:The-g5-Problem}).
Since electroweak interactions act on chiral fermions, the consistent renormalisation of chiral gauge theories is not only of academic interest but of substantial phenomenological and practical importance.
In Sec.~\ref{Sec:The-BMHV-Scheme}, we have introduced the BMHV scheme, which provides a consistent treatment of $\gamma_5$ in chiral gauge theories within dimensional regularisation schemes.
This scheme is the only known framework that has been proven to be self-consistent at all orders in perturbation theory (cf.\ Sec.~\ref{Sec:Alternative-g5-Schemes}).
However, as discussed, the BMHV scheme violates gauge and BRST invariance at intermediate steps of the calculation, which requires the subsequent restoration via symmetry-restoring counterterms.

Here, we first analyse the regularisation-induced symmetry breaking in the BMHV scheme in Sec.~\ref{Sec:Regularisation-Induced_Symmetry_Breaking}, focusing on its origin from the modified algebra (Sec.~\ref{Sec:Origin_of_Symmetry_Breaking_in_BMHV_Scheme}), the ambiguities arising from the non-uniqueness of the $D$-dimensional extension and their implications (Sec.~\ref{Sec:Dimensional_Ambiguities_and_Evanescent_Shadows}), and the general structure of the tree-level breaking (Sec.~\ref{Sec:General_Structure_of_the_Symmetry_Breaking}).
Subsequently, we discuss in detail the symmetry restoration procedure in Sec.~\ref{Sec:Symmetry_Restoration_Procedure}, and finally we conclude with remarks on computational challenges specific to the application of the BMHV scheme to chiral gauge theories in Sec.~\ref{Sec:BMHV-Specific_Challenges}.

\section{Regularisation-Induced Symmetry Breaking}\label{Sec:Regularisation-Induced_Symmetry_Breaking}

A classical symmetry can, and may in general, be broken by the regularisation --- as is the case in the BMHV scheme --- so that the associated Slavnov-Taylor identity is violated,
\begin{align}
    \mathcal{S}(\Gamma_\mathrm{reg})\neq0.
\end{align}
The Slavnov-Taylor identity is the quantum manifestation of classical symmetries, expressing them as functional relations among off-shell Green functions (see Sec.~\ref{Sec:The-Slavnov-Taylor-Identity}).
As established in Sec.~\ref{Sec:Renormalisation-of-GaugeTheories}, the Slavnov-Taylor identity corresponding to gauge and BRST invariance must be satisfied after renormalisation in order to obtain a consistent quantum theory that admits a unitary and gauge independent $S$-matrix, acting within a physical Hilbert space of positive norm states.
Consequently, any symmetry breaking induced by the regularisation must be compensated in the course of renormalisation by adding suitable symmetry-restoring counterterms, which satisfy 
\begin{align}
    \mathcal{S}(\Gamma^{(\leq n)}_\mathrm{subren}) = -b S^{(n)}_\mathrm{ct},
\end{align}
ensuring that the renormalised theory fulfils
\begin{align}
    \mathcal{S}(\Gamma_\mathrm{ren}) = 0
\end{align}
at every order in perturbation theory.
In the absence of genuine anomalies, the existence of such counterterms is guaranteed by the methods of algebraic renormalisation (see Sec.~\ref{Sec:Algebraic_Renormalisation}).
Hence, the mathematical consistency of the BMHV scheme comes at the cost of a more complicated renormalisation procedure.

\subsection{Symmetry Breaking in the BMHV Scheme}\label{Sec:Origin_of_Symmetry_Breaking_in_BMHV_Scheme}

The origin of symmetry breaking in the BMHV scheme can be traced to the fact that the evanescent part of the fermion kinetic term necessarily mixes fermions of different chiralities, which in chiral gauge theories transform differently under gauge transformations, and a mismatch between the manifestly $D$-dimensional fermion kinetic term and the gauge interaction term.
In gauge theories, the fermion kinetic term and the gauge interaction terms are not individually gauge invariant; gauge invariance holds only for an appropriate combination of both, which can be expressed through a covariant derivative.
In the BMHV scheme, the $4$-dimensional part of the fermionic Lagrangian can --- by construction to ensure a smooth limit to the original gauge invariant $4$-dimensional theory --- still be written in terms of a covariant derivative and thus remains gauge invariant.
However, no such covariant formulation can be found for the evanescent components.
As a result, the evanescent part cannot be expressed in terms of a covariant derivative that transforms correctly under gauge and BRST transformations, and therefore explicitly breaks BRST symmetry.
In other words, the breaking of gauge and BRST invariance in the BMHV schemes reflects our inability to formulate a fully invariant regularised Lagrangian in $D$ dimensions.

For an illustration, we consider an Abelian chiral gauge theory in which the gauge boson couples exclusively to the right-handed fermion component.
The physical field content is $\{B^\mu,{\psi_{R}}_i\}$.
The renormalisation of this model up to the 4-loop level will be presented in chapter~\ref{Chap:BMHV_at_Multi-Loop_Level}.
In Sec.~\ref{Sec:Dimensional_Ambiguities_and_Evanescent_Shadows}, we provide a generalisation that also includes left-handed and evanescent gauge interactions.

Within DReg, the kinetic terms must be strictly $D$-dimensional in order to obtain propagators with fully $D$-dimensional momenta in their denominators and thereby ensure a proper regularisation of loop integrals, since $4$-dimensional propagator denominators would not provide sufficient regularisation (see Sec.~\ref{Sec:D-dim_Lagrangian}).
Hence, the fermion kinetic term in $D$ dimensions reads
\begin{align}\label{Eq:D-dim-Fermion-kinetic-Term}
    \overline{\psi}_i i \slashed{\partial} \psi_i 
    = {\overline{\psi_{L}}}_i i \overline{\slashed{\partial}} {\psi_{L}}_i
                  +
                  {\overline{\psi_{R}}}_i i \overline{\slashed{\partial}} {\psi_{R}}_i
                  +
                  {\overline{\psi_{L}}}_i i \widehat{\slashed{\partial}} {\psi_{R}}_i
                  +
                  {\overline{\psi_{R}}}_i i \widehat{\slashed{\partial}} {\psi_{L}}_i,
\end{align}
where the fermion is split into left- and right-handed components as $\psi=\projL\psi+\projR\psi=\psi_L+\psi_R$.
The first two terms involve $4$-dimensional derivatives and do not mix chiralities, while the last two contain evanescent derivatives and mix left- and right-handed components.
The reason that the evanescent terms do not vanish are the modified algebraic relations of the BMHV scheme --- the BMHV algebra (see Sec.~\ref{Sec:The-BMHV-Scheme}) --- according to which the evanescent $\widehat{\gamma}$'s commute rather than anticommute with $\gamma_5$.
In chiral gauge theories, $\psi_L$ and $\psi_R$ generally have different gauge quantum numbers and transformation properties, so that the evanescent terms necessarily violate gauge invariance.

It is important to note that, in order to construct a $D$-dimensional fermion kinetic term in a model that contains only right-handed fermions ${\psi_R}_i$ (as in our example), it is necessary to introduce a fictitious \emph{sterile} left-handed partner field ${\psi_L}_i$, which appears only in the kinetic term.
\begin{definition}[Sterile Quantum Fields]\label{Def:Sterile_Quantum_Field}\ \\
    A quantum field $\psi^{\mathrm{st}}$ that exclusively appears in $\mathcal{L}_\mathrm{free}$ and does not participate in any interaction is called \emph{sterile}.
    Such a field is characterised by the absence of higher-order corrections, i.e.\
    \begin{align}\label{Eq:Linearity-for-Sterile-Quantum-Fields}
        \frac{\delta \Gamma_\mathrm{cl}}{\delta \psi^{\mathrm{st}}} = \frac{\delta \Gamma}{\delta \psi^{\mathrm{st}}} = \text{linear in quantum fields},
    \end{align}
    at all orders of perturbation theory. 
\end{definition}
Such sterile fields are auxiliary constructs introduced solely for the $D$-dimensional formulation (if required).
They are gauge singlets with vanishing BRST transformations but may be assigned the same global quantum numbers as their interacting counterparts of opposite chirality.

The extension of the interaction Lagrangian --- especially the chiral gauge interactions --- to $D$ dimensions is not unique (see sections~\ref{Sec:D-dim_Lagrangian} and \ref{Sec:Renormalisation_in_DReg}) and allows for many \emph{inequivalent} but \emph{equally correct} formulations.
In particular, the right-handed current ${\overline{\psi_{R}}}_i\gamma^\mu{\psi_{R}}_j$ can be extended in the following three independent ways:
\begin{align}
    \overline{\psi}_i \gamma^\mu \projR \psi_j, 
    \qquad
    \overline{\psi}_i \projL \gamma^\mu \psi_j, 
    \qquad
    \overline{\psi}_i \projL \gamma^\mu \projR \psi_j.
\end{align}
These expressions differ because $\projL\gamma^\mu \neq \gamma^\mu \projR$ in $D$ dimensions; their difference, $\gamma^\mu \projR - \projL\gamma^\mu = \widehat{\gamma}^\mu \gamma_5$, is an evanescent term that vanishes in $D=4$ dimensions.
Each choice yields a valid $D$-dimensional extension and produces the same physical results.
However, intermediate calculations and the resulting counterterms will differ by means of a local reparametrisation of the theory (cf.\ Sec.~\ref{Sec:Renormalisation_Theory}).
Following Res.~\cite{Stockinger:2023ndm,vonManteuffel:2025swv,Belusca-Maito:2020ala,Belusca-Maito:2021lnk,Belusca-Maito:2023wah,Jegerlehner:2000dz,Martin:1999cc,
Sanchez-Ruiz:2002pcf,Cornella:2022hkc}, we adopt the most symmetric and natural choice, $\overline{\psi}_i\mathbb{P}_{\mathrm{L}}\gamma^{\mu}\mathbb{P}_{\mathrm{R}}\psi_j = \overline{\psi}_i\mathbb{P}_{\mathrm{L}}\overline{\gamma}^{\mu}\mathbb{P}_{\mathrm{R}}\psi_j = {\overline{\psi_{R}}}_i\overline{\gamma}^{\mu}{\psi_{R}}_j$, which preserves the information that right-handed fermions occur on both sides of the interaction and typically leads to the simplest intermediate expressions (see particularly Refs.~\cite{Belusca-Maito:2023wah,Jegerlehner:2000dz}).

With this extension, the $D$-dimensional fermionic Lagrangian may be written as
\begin{align}
    \mathcal{L}_\mathrm{fermion}=\mathcal{L}_\mathrm{fermion,inv} + \mathcal{L}_\mathrm{fermion,evan} = \overline{\psi}_i i \slashed{\partial} \psi_i - g {\mathcal{Y}_R}_{ij} {\overline{\psi_{R}}}_i\overline{\slashed{B}}{\psi_{R}}_j,
\end{align}
where
\begin{subequations}\label{Eq:L_fermion_as_subequation}
    \begin{align}
        \mathcal{L}_\mathrm{fermion,inv} &= \overline{\psi}_i i \overline{\slashed{\partial}} \psi_i - g {\mathcal{Y}_R}_{ij} {\overline{\psi_{R}}}_i\overline{\slashed{B}}{\psi_{R}}_j, \label{Eq:L_fermion-inv}\\
        \mathcal{L}_\mathrm{fermion,evan} &= \overline{\psi}_i i \widehat{\slashed{\partial}} \psi_i. \label{Eq:L_fermion-evan}
    \end{align}
\end{subequations}
The first part, Eq.~\eqref{Eq:L_fermion-inv}, contains only the 4-dimensional component of the derivative and the gauge field, and can be rewritten as
\begin{align}\label{Eq:L_fermion_inv_with_cov-div}
    \mathcal{L}_\mathrm{fermion,inv} = {\overline{\psi_{L}}}_i i \overline{\slashed{\partial}} {\psi_{L}}_i + {\overline{\psi_{R}}}_i i \overline{\slashed{\partial}} {\psi_{R}}_i - g {\mathcal{Y}_R}_{ij} {\overline{\psi_{R}}}_i\overline{\slashed{B}}{\psi_{R}}_j = {\overline{\psi_{L}}}_i i \overline{\slashed{\partial}} {\psi_{L}}_i + {\overline{\psi_{R}}}_i i \overline{\slashed{D}}_{ij} {\psi_{R}}_j,
\end{align}
with 4-dimensional covariant derivative $\overline{\slashed{D}}_{ij} = \overline{\slashed{\partial}} \delta_{ij} + i g {\mathcal{Y}_R}_{ij} \overline{\slashed{B}}$, and sterile ${\psi_{L}}_i$.
Evidently, this part of the fermionic Lagrangian --- which resembles the 4-dimensional fermionic contribution --- is gauge and BRST invariant, 
\begin{align}
    s_D \mathcal{L}_\mathrm{fermion,inv} = 0,
\end{align}
as expected.
The second contribution, Eq.~\eqref{Eq:L_fermion-evan}, is the purely evanescent part of the fermion kinetic term 
\begin{align}\label{Eq:L_fermion-evan_2}
    \mathcal{L}_\mathrm{fermion,evan} = \overline{\psi}_i i \widehat{\slashed{\partial}} \psi_i = {\overline{\psi_{L}}}_i i \widehat{\slashed{\partial}} {\psi_{R}}_i + {\overline{\psi_{R}}}_i i \widehat{\slashed{\partial}} {\psi_{L}}_i,
\end{align}
which mixes left- and right-handed chiral components with different gauge transformation properties, and indeed explicitly breaks gauge invariance:
\begin{align}
    s_D \mathcal{L}_\mathrm{fermion,evan} \neq 0.
\end{align}

The corresponding $D$-dimensional action can analogously be written as 
\begin{align}\label{Eq:Split-of-S_0-into-inv-and-evan-Part}
    S_0 = S_{0,\mathrm{inv}} + S_{0,\mathrm{evan}},
\end{align}
with evanescent part
\begin{align}
    S_{0,\mathrm{evan}} = \int d^Dx \, \mathcal{L}_\mathrm{fermion,evan} = \int d^Dx \, \overline{\psi}_i i \widehat{\slashed{\partial}} \psi_i.
\end{align}
Here, $S_{0,\mathrm{inv}}$ contains $\mathcal{L}_\mathrm{fermion,inv}$ together with the gauge, gauge-fixing, ghost, and external source terms, all of which remain invariant.
Preparing for higher-order considerations, the $D$-dimensional tree-level BRST operator $s_D$ is replaced by the corresponding $D$-dimensional Slavnov-Taylor operator $\mathcal{S}_D$ (see sections~\ref{Sec:BRST-Symmetry}, \ref{Sec:The-Slavnov-Taylor-Identity} and \ref{Sec:Regularised-QAP}).
Acting with $\mathcal{S}_D$ on the $D$-dimensional tree-level action yields
\begin{align}\label{Eq:Tree-Level-BRST-Breaking-Introductory_Example}
    \mathcal{S}_D(S_0) = \mathcal{S}_D(S_{0,\mathrm{evan}}) \equiv \widehat{\Delta},
\end{align}
where we introduced the breaking operator 
\begin{equation}\label{Eq:Delta_Hat_Introductory_Example}
    \begin{aligned}
        \widehat{\Delta} = - \Dintx \, g \,  {\mathcal{Y}_R}_{ij} \, c \, 
                    \bigg\{
                    \overline{\psi}_i \bigg(
                    \overset{\leftarrow}{\widehat{\slashed{\partial}}} \mathbb{P}_{\mathrm{R}} 
                    + \overset{\rightarrow}{\widehat{\slashed{\partial}}} \mathbb{P}_{\mathrm{L}}
                    \bigg) \psi_j 
                    \bigg\}
                = \Dintx \, \widehat{\Delta}(x).
    \end{aligned}
\end{equation}
The nonvanishing of $\widehat{\Delta}$ confirms that BRST invariance is broken already at tree-level in the BMHV scheme. 
Since the evanescent component of the fermion kinetic term (see Eq.~\eqref{Eq:L_fermion-evan_2}) is required to construct a $D$-dimensional fermion propagator and obtain properly regularised loop integrals, this breaking is an inevitable consequence of the BMHV algebra (see Eq.~\eqref{Eq:BMHV-Algebra}).

The BRST breaking (see Eq.~\eqref{Eq:Tree-Level-BRST-Breaking-Introductory_Example}) can be expressed as a composite operator (see Eq.~\eqref{Eq:Delta_Hat_Introductory_Example}), whose insertions into Green functions (see Sec.~\ref{Sec:GreenFunctions_and_GeneratingFunctionals}) serve as the central tool for analysing higher-order symmetry breakings and deriving the corresponding symmetry-restoring counterterms, as will be discussed in Sec.~\ref{Sec:Symmetry_Restoration_Procedure}.
In particular, an operator insertion of $\widehat{\Delta}$ corresponds to an insertion of an associated interaction vertex with Feynman rule derived from Eq.~\eqref{Eq:Delta_Hat_Introductory_Example},
\begin{equation}\label{Eq:Delta-Tree-Level-Interaction-Vertex}
    \begin{tabular}{rl}
        \raisebox{-42pt}{\includegraphics[scale=0.65]{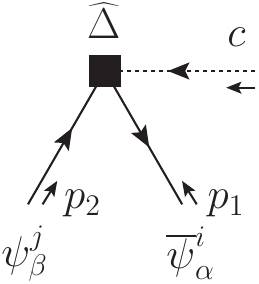}}&
        $\begin{aligned}
            &= - g \, {\mathcal{Y}_R}_{ij} \left(\widehat{\slashed{p}}_1 \mathbb{P}_{\mathrm{R}} + \widehat{\slashed{p}}_2 \mathbb{P}_{\mathrm{L}} \right)_{\alpha\beta}.
        \end{aligned}$
    \end{tabular}
\end{equation}
Due to its evanescent nature, the tree-level breaking $\widehat{\Delta}$ vanishes in $D=4$ dimensions --- a purely regularisation-induced effect.
At higher orders, however, quantum corrections generate $1/\epsilon$ poles, leading to nonvanishing contributions in power-counting divergent Green functions with insertions of $\widehat{\Delta}$.

\subsection{Dimensional Ambiguities and Evanescent Shadows}\label{Sec:Dimensional_Ambiguities_and_Evanescent_Shadows}
As discussed in sections~\ref{Sec:D-dim_Lagrangian} and \ref{Sec:Origin_of_Symmetry_Breaking_in_BMHV_Scheme}, the extension of a 4-dimensional action to $D$ dimensions is not unique, and consequently, the implementation of the BMHV scheme itself is not uniquely defined.
Different consistent choices for the regularised action are possible, each leading --- after proper renormalisation --- to a valid theory that yields identical predictions for physical observables, albeit with different intermediate expressions and counterterms.
In the previous section, we adopted the most symmetric extension of the right-handed interaction current, corresponding to the choice used in e.g.\ Refs.~\cite{Stockinger:2023ndm,vonManteuffel:2025swv,Belusca-Maito:2020ala,Belusca-Maito:2021lnk,Belusca-Maito:2023wah}.
This naturally raises the question of how alternative extensions influence the symmetry breaking and whether such freedom can be exploited to simplify the symmetry restoration procedure.

To address this question, we examine in the present subsection a general class of possible implementations and modifications within the BMHV framework.
Specifically, we investigate different possibilities for the dimensional continuation of fermions, which encompass both alternative implementations of gauge interactions and different options for constructing a $D$-dimensional kinetic term (cf.\ Eq.~\eqref{Eq:D-dim-Fermion-kinetic-Term}).
Although the kinetic term is uniquely fixed in the sense that is must be fully $D$-dimensional (as discussed in sections~\ref{Sec:D-dim_Lagrangian} and \ref{Sec:Origin_of_Symmetry_Breaking_in_BMHV_Scheme}), some freedom remains when fermions possess both left- and right-handed components, depending on whether or not sterile partner fields are introduced.
These ambiguities correspond to evanescent details of a particular dimensional realisation of the BMHV scheme.
Their implications for the regularisation-induced symmetry breaking are analysed in general terms in Sec.~\ref{Sec:General_Structure_of_the_Symmetry_Breaking}, and in specific applications in chapters~\ref{Chap:General_Abelian_Chiral_Gauge_Theory} (see particularly Sec.~\ref{Sec:Results-Shedding_Light_on_Evanescent_Shadows}) and \ref{Chap:The_Standard_Model}.

The discussion presented here follows the analysis that we published in Ref.~\cite{Ebert:2024xpy}.
In line with that work, we omit non-Abelian complications and restrict our attention to different realisations of dimensionally regularised fermions.
Concretely, we focus on an Abelian chiral gauge theory that accommodates left- and right-handed as well as evanescent gauge interactions.
A comprehensive discussion of this model and its complete renormalisation is presented in chapter~\ref{Chap:General_Abelian_Chiral_Gauge_Theory}.
For illustrative purposes, we use the electron and a neutrino as representative examples in the following discussion and generalise to generic fermions at the end of this section.

\paragraph{Fermion Kinetic Term:} 
We begin with a neutrino as a model case for fermions that come with only one chirality --- recall that, in the electroweak Standard Model, neutrinos are massless and appear only as left-handed neutrinos and right-handed antineutrinos.
This situation is analogous to the right-handed model used as an example in Sec.~\ref{Sec:Origin_of_Symmetry_Breaking_in_BMHV_Scheme}.
We express each (anti)neutrino in terms of 4-component Dirac spinors $\nu(x)$, which are eigenstates of $\gamma_5$, such that $\nu = \mathbb{P}_{\mathrm{L}}\nu\equiv \nu{}_{L}$, since the 2-component notation is not known to be extendable to $D$ dimensions (see Sec.~\ref{Sec:D-dim_Lagrangian} and Ref.~\cite{Belusca-Maito:2023wah}).
In 4 dimensions the kinetic term reads
\begin{align}
    \mathcal{L}^{4D}_{\mathrm{kin},\nu} =
    {\overline{\nu{}_L}} i
    \overline{\slashed{\partial}} {\nu{}_L} .
\end{align}
As discussed above, fully $D$-dimensional fermion kinetic terms are required for properly regularised propagators in loop diagrams.
Hence, DReg implicitly requires the introduction of a fictitious sterile right-handed neutrino $\nu^{\mathrm{st}}_R$, such that
\begin{align}
  \nu=\nu_L+\nu^{\mathrm{st}}_R
  =\mathbb{P}_{\mathrm{L}}\nu+\mathbb{P}_{\mathrm{R}}\nu,
\end{align}
which permits the construction of the $D$-dimensional neutrino kinetic term
\begin{align}
          \mathcal{L}_{\mathrm{kin},\nu} = {\overline{\nu}} i {\slashed{\partial}}{\nu} 
                  = {\overline{\nu_L}} i
                  \overline{\slashed{\partial}} {\nu_L}
                  +
                  {\overline{\nu^{\mathrm{st}}_R}} i
                  \overline{\slashed{\partial}} {\nu^{\mathrm{st}}_R}
                  +
                  {\overline{\nu_L}} i
                  \widehat{\slashed{\partial}} {\nu^{\mathrm{st}}_R}
                  +
                  {\overline{\nu^{\mathrm{st}}_R}} i
                  \widehat{\slashed{\partial}} {\nu_L}.
\end{align}
In this way, ${\slashed{\partial}}$ is fully $D$-dimensional and gives rise to an appropriate $D$-dimensional propagator.
For fermions that come with only one chirality, there is no alternative: 
a sterile partner field of opposite chirality is required to obtain a fully $D$-dimensional kinetic term.

As before, the evanescent pieces of the kinetic term mix left- and right-handed components.
While this is inevitable in the BMHV framework, the evanescent terms would be absent in a scheme with naively anticommuting $\gamma_5$.
Because $\nu^{\mathrm{st}}_R$ is non-interacting and specifically has no gauge interactions, its local gauge and transformations differ from those of $\nu_L$.
As a result, the evanescent part of $\mathcal{L}_{\mathrm{kin},\nu}$ breaks BRST invariance precisely in the manner discussed in Sec.~\ref{Sec:Origin_of_Symmetry_Breaking_in_BMHV_Scheme}; see Eq.~\eqref{Eq:Tree-Level-BRST-Breaking-Introductory_Example}.
Assigning to $\nu^{\mathrm{st}}_R$ the same global quantum numbers as $\nu_L$ --- so that both transform identically under global symmetries --- ensures that the kinetic term at least remains invariant w.r.t.\ the global symmetries.

Now we turn to the electron/positron, described by the 4-component Dirac spinor field $e(x)$. 
Decomposing into chiral components yields
\begin{align}
  e=
  \mathbb{P}_{\mathrm{L}}e + \mathbb{P}_{\mathrm{R}}e
  \equiv e_L + e_R ,
\end{align}
with $e_L$ and $e_R$ carrying different gauge quantum numbers.
In 4 dimensions the kinetic term is
\begin{align}\label{Eq:4D-Electron-Kinetic-Term}
          \mathcal{L}^{4D}_{\mathrm{kin},e} & =
                  {\overline{e}} i
                  \overline{\slashed{\partial}} {e}
                  =
                  {\overline{e_L}} i
                  \overline{\slashed{\partial}} {e_L}
                  +
                  {\overline{e_R}} i
                  \overline{\slashed{\partial}} {e_R}.
\end{align}
In contrast to the neutrino case, both chiralities are physical here; a sterile partner field is therefore not necessarily required.
Nevertheless, introducing sterile partners allows for alternative $D$-dimensional realisations.
In particular, for the extension to $D$ dimensions there are now several options:
\begin{optionblock}
\begin{option}[$\psi_L + \psi_R \longrightarrow \psi$]\label{Opt:Option1}\
    The most direct choice is to work only with the physical components, giving rise to the natural Dirac spinor combination
    \begin{align}
        e= e_L + e_R,
    \end{align}
    and thus to the $D$-dimensional kinetic term
    \begin{align}
        \mathcal{L}_{\mathrm{kin},e} = {\overline{e}} i {\slashed{\partial}} {e}.
    \end{align}
    In this case, no additional sterile fields are introduced, and the kinetic term decomposes as
    \begin{align}\label{Eq:Lkin-Electron-D-dim-Decomposed}
          \mathcal{L}_{\mathrm{kin},e} = 
                  {\overline{e_L}} i
                  \overline{\slashed{\partial}} {e_L}
                  +
                  {\overline{e_R}} i
                  \overline{\slashed{\partial}} {e_R}
                  +
                  {\overline{e_L}} i
                  \widehat{\slashed{\partial}} {e_R}
                  +
                  {\overline{e_R}} i
                  \widehat{\slashed{\partial}} {e_L}.
    \end{align}
    While this is the most natural approach, the evanescent $\widehat{\slashed{\partial}}$-terms mix physical fields with different gauge quantum numbers (e.g.\ different hypercharges).
    Consequently, they break not only local gauge and BRST invariance but also global gauge invariance.
\end{option}
\begin{suboptions}
\begin{option}[$\psi_L + \psi_R^{\mathrm{st}} \longrightarrow \psi_1$ and $\psi_L^\mathrm{st} + \psi_R \longrightarrow \psi_2$]\label{Opt:Option2}\grouplabel{Opt:Opt2-GroupRef}\
    As an alternative, the two gauge multiplets $e_L$ and $e_R$ can be separated in order to retain at least global gauge invariance.
    Analogous to the neutrino construction, this approach introduces two fictitious sterile partner fields $e^{\mathrm{st}}_L$ and $e^{\mathrm{st}}_R$, in order to define two Dirac spinors
    \begin{align}\label{Eq:Electron-2-Spinors-Option2}
        e_1 &= e_L + e^{\mathrm{st}}_R,
        &
        e_2 &= e^{\mathrm{st}}_L + e_R.
    \end{align}
    The kinetic term then takes the form
    \begin{align}
          \mathcal{L}_{\mathrm{kin},e} =
                  {\overline{e_1}} i
                  {\slashed{\partial}} {e_1}
                  +
                  {\overline{e_2}} i
                  {\slashed{\partial}} {e_2} ,
    \end{align}
    and its decomposition reads
    \begin{equation}\label{Eq:Lkin-Electron-Option2-Decomposition}
        \begin{aligned}
          \mathcal{L}_{\mathrm{kin},e} & =
                  {\overline{e_L}} i
                  \overline{\slashed{\partial}} {e_L}
                  +
                  {\overline{e^{\mathrm{st}}_R}} i
                  \overline{\slashed{\partial}} {e^{\mathrm{st}}_R}
                  +
                  {\overline{e_L}} i
                  \widehat{\slashed{\partial}} {e^{\mathrm{st}}_R}
                  +
                  {\overline{e^{\mathrm{st}}_R}} i
                  \widehat{\slashed{\partial}} {e_L}\\
                  &+ {\overline{e_R}} i
                  \overline{\slashed{\partial}} {e_R}
                  +
                  {\overline{e^{\mathrm{st}}_L}} i
                  \overline{\slashed{\partial}} {e^{\mathrm{st}}_L}
                  +
                  {\overline{e_R}} i
                  \widehat{\slashed{\partial}} {e^{\mathrm{st}}_L}
                  +
                  {\overline{e^{\mathrm{st}}_L}} i
                  \widehat{\slashed{\partial}} {e_R}.
        \end{aligned}
    \end{equation}
    As a result, the global symmetry can be preserved at the cost of a proliferation of terms.
    However, local gauge and BRST invariance remain violated by the evanescent mixing, as in all other realisations.
\end{option}
\begin{option}[$\psi_L + \psi_R^{\mathrm{st}} \longrightarrow \psi_1$ and $\psi_L^C + {\psi^C_R}^{\mathrm{st}} \longrightarrow \psi_2$]\label{Opt:Option2b}\
    In this variant, the right-handed spinor $e_R$ is replaced by an equivalent left-handed charge-conjugated $e^C_L$ (with opposite charges).
    Accordingly, the second term in Eq.~\eqref{Eq:4D-Electron-Kinetic-Term} can be rewritten as ${\overline{e^C_L}} i \overline{\slashed{\partial}} {e^C_L}$.
    Proceeding as in Option~\ref{Opt:Option2}, but using $e^C_L$ in place of $e_R$, one defines $e_2=e^C_L+{e^C_R}^{\mathrm{st}}$.
    A potential advantage is that all physical fields are left-handed, while all sterile fields are right-handed.
    Apart from this bookkeeping benefit, the properties are the same as in Option~\ref{Opt:Option2}.
    This approach is often convenient in settings such as supersymmetry or grand unification.
\end{option}
\end{suboptions}
\end{optionblock}

Although we illustrated the constructions with a neutrino and the electron, the discussion carries over to any fermion $\psi$.
In dimensional regularisation and the BMHV framework one works with Dirac spinors $\psi=\psi_L+\psi_R$, where either both chiral components are physical (potentially with different gauge quantum numbers) or one of them is a fictitious sterile partner.
The chosen $D$-dimensional realisation determines the evanescent mixing terms and thereby the precise form of the symmetry breaking encountered at the regularised level.

\paragraph{Propagators and Drawback of Option~\ref{Opt:Opt2-GroupRef}:} 
Introducing sterile partner fields in Options~\ref{Opt:Option2} and \ref{Opt:Option2b} leads to mixed (off-diagonal) kinetic terms in Eq.~\eqref{Eq:Lkin-Electron-Option2-Decomposition}.
This leads to an intricate issue for massive fermions, where the resulting propagator admits off-diagonal terms and requires a diagonalisation.
This problem has been discussed in Ref.~\cite{Ebert:2024xpy}, where the propagator has been derived for the massive case.
    
Although the discussion so far has been restricted to massless fermions, let us temporarily consider
physical fermion masses --- recall that in chiral gauge theories such masses originate only from spontaneous symmetry breaking.
The mentioned issue arises only for Option~\ref{Opt:Opt2-GroupRef} and is unavoidable in theories like the electroweak Standard Model (EWSM), where fermions are massive in the broken phase.
Again working with the electron for illustrative purposes, the kinetic term in the massive case decomposes as
    \begin{equation}\label{Eq:Lkin-Electron-MassiveCase}
        \begin{aligned}
          \mathcal{L}_{\mathrm{kin},e} =
                  \Big[
                  {\overline{e_L}} i
                  \overline{\slashed{\partial}} {e_L}
                  +
                  {\overline{e^{\mathrm{st}}_R}} i
                  \overline{\slashed{\partial}} {e^{\mathrm{st}}_R}
                  +
                  {\overline{e_L}} i
                  \widehat{\slashed{\partial}} {e^{\mathrm{st}}_R}
                  +
                  {\overline{e^{\mathrm{st}}_R}} i
                  \widehat{\slashed{\partial}} {e_L}
                  +(L\leftrightarrow R)
                  \Big]
                  - m(\overline{e_L}e_R+\overline{e_R}e_L),
        \end{aligned}
    \end{equation}
to be compared with Eq.~\eqref{Eq:Lkin-Electron-Option2-Decomposition}.
Combining chiral fermions into full Dirac spinors as in Eq.~\eqref{Eq:Electron-2-Spinors-Option2}, the Lagrangian can be rewritten as
    \begin{equation}\label{Eq:Lkin-Electron-MassiveCase-MatrixForm}
        \begin{aligned}
            \mathcal{L}_{\mathrm{kin},e} & =
                \overline{E} \mathcal{D} E =
                \begin{pmatrix}
                    \overline{e_1} & \overline{e_2}\\
                \end{pmatrix}
                \begin{pmatrix}
                    i \slashed{\partial} & -m\projR\\
                    -m\projL & i \slashed{\partial} \\
                \end{pmatrix}
                \begin{pmatrix}
                    e_1\\
                    e_2
                \end{pmatrix}.
        \end{aligned}
    \end{equation}
The inversion of the such defined matrix $\mathcal{D}$ yields the propagator matrix $\mathcal{P}$, i.e.\ 
$\mathcal{P}\mathcal{D}=\mathcal{D}\mathcal{P}=\mathbb{1}$. 
In momentum space, with incoming momentum $p$, one obtains (see Ref.~\cite{Ebert:2024xpy})
\begin{equation}\label{Eq:MassivePropagatorMatrix}
    \widetilde{\mathcal{P}} = 
    \begin{pmatrix}
            \frac{p^2\slashed{p}-m^2\overline{\slashed{p}}\projL}{p^4-m^2\overline{p}^2}& \frac{(\overline{p}^2\projL+\widehat{p}^2\projR)m+\widehat{\slashed{p}}\overline{\slashed{p}}(\projL-\projR)m}{p^4-m^2\overline{p}^2}\\
            \frac{(\widehat{p}^2\projL+\overline{p}^2\projR)m+\widehat{\slashed{p}}\overline{\slashed{p}}(\projR-\projL)m}{p^4-m^2\overline{p}^2} & \frac{p^2\slashed{p}-m^2\overline{\slashed{p}}\projR}{p^4-m^2\overline{p}^2}\\
        \end{pmatrix}.
\end{equation}
In the massless limit, the denominators reduce to the familiar $D$-dimensional ones and the off-diagonal terms vanish, such that the propagators coincide with those of Option~\ref{Opt:Option1} (no further complications).
For nonzero masses, however, the denominators pick up additional $4$-dimensional terms of the form $m^2\overline{p}^2$ alongside the $D$-dimensional momenta.
Such denominators lie outside the standard toolkit of integration techniques in DReg and considerably complicate loop calculations.
Combined with the general proliferation of terms, this is the main drawback of Option~\ref{Opt:Opt2-GroupRef} for massive theories such as the EWSM in the broken phase.

\paragraph{Fermion-Gauge Boson Interaction:} 
As emphasised in Sec.~\ref{Sec:Origin_of_Symmetry_Breaking_in_BMHV_Scheme}, a significant subtlety in DReg is the ambiguous nature of extending gauge interaction terms from 4 to $D$ dimensions; this continuation is not unique.
To explore this, we continue to use the electron and a neutrino as illustrative examples, noting that the discussion carries over to arbitrary fermions.

For a vector-like theory such as standard QED, using Option~\ref{Opt:Option1} for the electron spinor, a natural choice --- in fact the standard treatment --- for the electron-photon interaction in $D$ dimensions is $\overline{e}\gamma^\mu e A_\mu$, which preserves the vector-like structure.
Decomposing this interaction into its chiral components reveals both 4- and $(D-4)$-dimensional parts:
\begin{align}
  \label{Eq:LeegammaQED}
\mathcal{L}_{ee\gamma} =
  \overline{e}\projR\overline{\gamma}^\mu \projL e\overline{A}_\mu
  +
  \overline{e}\projL\overline{\gamma}^\mu \projR e\overline{A}_\mu
  +
  \overline{e}\projR\widehat{\gamma}^\mu \projR e\widehat{A}_\mu
  +
  \overline{e}\projL\widehat{\gamma}^\mu \projL e\widehat{A}_\mu.
\end{align}
This expression shows that the conventional treatment introduces not only the expected chirality-preserving interactions with the 4-dimensional gauge field $\overline{A}_\mu$, but also chirality-violating evanescent interactions involving the $(D-4)$-dimensional photon $\widehat{A}_\mu$.\footnote{As mentioned before, the evanescent terms would vanish in a scheme with naively anticommuting $\gamma_5$.}
While the presence of such evanescent interactions is not problematic in vector-like gauge theories, where left- and right-handed components carry the same charge, they lead to the violation of some global symmetries, such as global hypercharge conservation, in chiral gauge theories by coupling fields that carry different gauge quantum numbers.
This observation provides the motivation --- depending on the context --- to chose a regularised Lagrangian in which the evanescent interaction terms are deliberately omitted.

\begin{optionblockroman}
As implied, the situation is less straightforward for chiral interactions, such as the flavour-changing charged current $\overline{e}\,\overline{\gamma}^\mu\projL\nu W^-_\mu$ (in 4 dimensions), which is the interaction of the electron and the neutrino with $W^-$ in the EWSM.
Subject to hermiticity and formal Lorentz invariance, the structure $\overline{\gamma}^\mu\projL$ may in principle be modified in $D$ dimensions by adding any evanescent linear combination of the form $a\widehat{\gamma}^\mu\projL+b\widehat{\gamma}^\mu\projR$.
Two concrete, motivated $D$-dimensional extensions (again using Option~\ref{Opt:Option1} for the electron) are:
\begin{option}[$\overline{\gamma}^\mu\projL \longrightarrow \gamma^\mu\projL=\projR\overline{\gamma}^\mu\projL+\projL\widehat{\gamma}^\mu\projL$]\label{Opt:Option-(i)}\
    \begin{equation}\label{Eq:W-Treatment-Option-(i)}
        \begin{aligned}
            \mathcal{L}_{e\nu W} =
            \overline{e}\projR\overline{\gamma}^\mu\projL\nu \overline{W}^-_\mu+
            \overline{e}\projL\widehat{\gamma}^\mu\projL\nu  \widehat{W}^-_\mu + \mathrm{h.c.},
        \end{aligned}
    \end{equation}
\end{option}
\begin{option}[$\overline{\gamma}^\mu\projL \longrightarrow \overline{\gamma}^\mu\projL =\projR\overline{\gamma}^\mu\projL$]\label{Opt:Option-(ii)}\
    \begin{equation}\label{Eq:W-Treatment-Option-(ii)}
        \begin{aligned}
            \mathcal{L}_{e\nu W} =
            \overline{e}\projR\overline{\gamma}^\mu\projL\nu \overline{W}^-_\mu + \mathrm{h.c.}.
        \end{aligned}
    \end{equation}
\end{option}
The key difference between these two options is the presence of an evanescent interaction term in Option~\ref{Opt:Option-(i)} (the ``naive'' $D$-dimensional continuation), which couples the right-handed electron $e_R$ to the left-handed neutrino $\nu_L$ via the $(D-4)$-dimensional gauge boson $\widehat{W}^-_\mu$.
The exclusion of such a term in Option~\ref{Opt:Option-(ii)}, which involves only the 4-dimensional $\overline{\gamma}^\mu$, appears to be the more advantageous choice here, as it avoids introducing interactions between fields of mismatched chirality.
\end{optionblockroman}

Crucially, the treatment of fermion-gauge interactions is independent of the specific construction chosen for the $D$-dimensional Dirac spinors.
For instance, the standard DReg prescription for QED can be recast in Option~\ref{Opt:Option2}, where $e_1=e_L+e_R^{\mathrm{st}}$ and $e_2=e_R+e_L^{\mathrm{st}}$, as
\begin{equation*}
    \begin{aligned}
        \mathcal{L}_{ee\gamma} =
        \overline{e}_1 \projR \overline{\slashed{A}} \projL e_1 
        +
        \overline{e}_2 \projL \overline{\slashed{A}} \projR e_2
        +
        \overline{e}_1 \projR \widehat{\slashed{A}} \projR e_2 
        +
        \overline{e}_2 \projL \widehat{\slashed{A}} \projL e_1,
    \end{aligned}
\end{equation*}
which is equivalent to Eq.~\eqref{Eq:LeegammaQED} but expressed solely in terms of $e_{1,2}$.

This analysis illustrates that there is no universal prescription for treating gauge interactions in $D$ dimensions.
The decision of whether to include evanescent interaction terms, and in which form, depends on the phenomenological context.
For some interactions, such as the neutral current interactions with the $Z$ boson in the EWSM, the optimal choice is often not immediately apparent.
Specific physics applications may favour one convention over another;
for instance, for the construction of effective Hamiltonians for $B$-meson decays (see Refs.~\cite{Ciuchini:1993fk,Buchalla:1995vs}), strictly 4-dimensional weak vertices, as in Eq.~\eqref{Eq:W-Treatment-Option-(ii)}, have often been employed.

\paragraph{General Extension of the Fermionic Lagrangian:}
The concepts discussed so far using the electron and a neutrino as specific examples extend straightforwardly to a generic Abelian chiral gauge theory --- with non-Abelian complications set aside as mentioned above --- encompassing both left- and right-handed interactions and arbitrary fermion multiplets.
In particular, the fermion multiplets $\psi_L$ and $\psi_R$ can be populated with both physical and sterile fields, so that both Option~\ref{Opt:Option1} and Option~\ref{Opt:Opt2-GroupRef} are admissible.

In 4 dimensions, the gauge interaction is encoded in the covariant derivative
\begin{align}\label{Eq:4-Dim-Covariant-Derivative}
    \overline{D}_{\mu} \psi = \big(\overline{\partial}_{\mu} + i g \hypL
      \overline{B}_{\mu} \projL + i g \hypR \overline{B}_{\mu} \projR\big) \psi,
\end{align}
where $\hypL$ and $\hypR$ are the hypercharge matrices, coupling the fermions to the gauge boson $B_\mu$.
The 4-dimensional fermionic Lagrangian reads
\begin{equation}
  \begin{aligned}
  \label{Eq:L_fermion_4D_General_Abelian_Model}
        \mathcal{L}^{4D}_{\mathrm{fermion}} = \overline{\psi}_i i \overline{\slashed{D}}_{ij} \psi_j = \overline{\psi}_i i \overline{\slashed{\partial}} \psi_i 
        &- g {\hypR}_{ij} \overline{\psi}_i \projL \overline{\slashed{B}} \projR \psi_j
        - g {\hypL}_{ij} \overline{\psi}_i \projR \overline{\slashed{B}} \projL \psi_j.
\end{aligned}
\end{equation}

Accommodating the ambiguities discussed above, a general $D$-dimensional Ansatz for the regularised fermionic Lagrangian can be formulated as
\begin{equation}
  \begin{aligned}
  \label{Eq:L_fermion_general_D-dim_Ansatz}
        \mathcal{L}_{\mathrm{fermion}} = \overline{\psi}_i i \slashed{\partial} \psi_i 
        &- g {\hypR}_{ij} \overline{\psi}_i \projL \overline{\slashed{B}} \projR \psi_j
        - g {\hypL}_{ij} \overline{\psi}_i \projR \overline{\slashed{B}} \projL \psi_j\\
        &- g {\hypRL}_{ij} \overline{\psi}_i \projL \widehat{\slashed{B}} \projL \psi_j
        - g {\hypLR}_{ij} \overline{\psi}_i \projR \widehat{\slashed{B}} \projR \psi_j.
\end{aligned}
\end{equation}
The first line contains the standard terms: a fully $D$-dimensional fermion kinetic term, which is necessary for regularisation, and the 4-dimensional gauge interactions naturally arising from Eq.~\ref{Eq:4-Dim-Covariant-Derivative}. 
The hypercharge matrices $\hypL$ and $\hypR$ are the same as in 4 dimensions, such that the 4-dimensional part of the regularised action manifestly preserves BRST-invariance.
The second line of Eq.~\eqref{Eq:L_fermion_general_D-dim_Ansatz} introduces a general set of evanescent interactions.
These couple the fermions to the $(D-4)$-dimensional component of the gauge field $\widehat{B}_\mu$ via evanescent hypercharges $\hypLR$ and $\hypRL$.
Keeping the $D$-dimensional extension as general as possible, the evanescent hypercharge matrices need not be tied to the $4$-dimensional gauge interaction matrices $\hypL$ and $\hypR$.
They can generally be independent and may in particular also be off-diagonal.
Hermiticity, however, imposes the restriction
\begin{align}\label{Eq:Requirement-YLR-YRL}
  \hypRL = \hypLR^{\dagger}.
\end{align}
Unlike the 4-dimensional gauge interactions, the evanescent interactions couple left-handed to right-handed fermions and thus flip chirality.  
As a consequence, they may link fields with different hypercharges and thereby break local and even global hypercharge conservation, in direct analogy to the evanescent parts of the kinetic term.
In order to preserve other global symmetries, such as electromagnetic charge $Q$ or colour charge, one may constrain the evanescent hypercharges to commute with $Q$ and respect colour charge conservation.
Further information on the evanescent hypercharges and specific applications are provided in chapter~\ref{Chap:General_Abelian_Chiral_Gauge_Theory}, especially sections~\ref{Sec:Definition_of_the_Theory_General_Abelian_Case} and \ref{Sec:Special_Cases_of_general_Abelian_Model}.

\subsection{General Structure of the Tree-Level Breaking}\label{Sec:General_Structure_of_the_Symmetry_Breaking}

Having discussed the dimensional ambiguities and outlined the possible extensions of the fermionic Lagrangian to $D$ dimensions, we now examine their impact on the violation of BRST symmetry.
As in Sec.~\ref{Sec:Origin_of_Symmetry_Breaking_in_BMHV_Scheme}, the fermionic part of the regularised Lagrangian remains the only source of symmetry breaking in the BMHV framework, owing to the modified algebra --- the BMHV algebra.
In the following, we consolidate the effects of the various extensions by evaluating the general structure of the tree-level breaking in an Abelian chiral gauge theory with the general $D$-dimensional Ansatz for the regularised fermionic Lagrangian provided in Eq.~\eqref{Eq:L_fermion_general_D-dim_Ansatz}:
\begin{equation}\label{Eq:GeneralTreeLevelBreaking_Abelian}
    \begin{aligned}
        \DeltaHat &= \STIopD\Big(\Dintx\,\mathcal{L}\Big) = \STIopD\Big(\Dintx\,\mathcal{L}_\mathrm{fermion}\Big) = \STIopD\Big(\Dintx\,\mathcal{L}_\mathrm{fermion,evan}\Big)\\
        &= - g \Dintx \, c \,
        \Bigg\{
        \overline{\psi}_i 
        \bigg[
        \projR 
        \bigg(
        {\hypR}_{ij} \overset{\leftarrow}{\widehat{\slashed{\partial}}}
        + {\hypL}_{ij} \overset{\rightarrow}{\widehat{\slashed{\partial}}}
        - {\hypLR}_{ij} \Big( \overset{\leftarrow}{\widehat{\slashed{\partial}}} 
                            + \overset{\rightarrow}{\widehat{\slashed{\partial}}} \Big)
        \bigg)
        \projR\\
        &\hspace{3cm} +
        \projL 
        \bigg(
        {\hypR}_{ij} \overset{\rightarrow}{\widehat{\slashed{\partial}}}
        + {\hypL}_{ij} \overset{\leftarrow}{\widehat{\slashed{\partial}}}
        - {\hypRL}_{ij} \Big( \overset{\leftarrow}{\widehat{\slashed{\partial}}} 
                            + \overset{\rightarrow}{\widehat{\slashed{\partial}}} \Big)
        \bigg)
        \projL
        \bigg]
        \psi_j
        \Bigg\}\\
        &\phantom{= } + i g^2 \Dintx \, c \,
        \Bigg\{
        \overline{\psi}_i 
        \bigg[
        \Big(
        {\hypRL}_{ik}{\hypL}_{kj} - {\hypR}_{ik}{\hypRL}_{kj}
        \Big)
        \projL \widehat{\slashed{B}} \projL\\
        &\hspace{3.2cm} +
        \Big(
        {\hypLR}_{ik}{\hypR}_{kj} - {\hypL}_{ik}{\hypLR}_{kj}
        \Big)
        \projR \widehat{\slashed{B}} \projR
        \bigg]
        \psi_j
        \Bigg\}\\
        &\eqqcolon \widehat{\Delta}_1\big[c,\overline{\psi},\psi\big] + \widehat{\Delta}_2\big[c,B,\overline{\psi},\psi\big],
    \end{aligned}
\end{equation}
where $\STIopD$ denotes the $D$-dimensional Slavnov-Taylor operator.
Importantly, Eq.~\eqref{Eq:GeneralTreeLevelBreaking_Abelian} indeed captures the general structure of the tree-level breaking, as an analogous form is also obtained in the non-Abelian case (see Sec.~\ref{Sec:SymmetryBreaking_in_SM}).

The $\hypL$- and $\hypR$-part of the first contribution, $\widehat{\Delta}_1$, originates from the evanescent part of the fermion kinetic term, as usual in the BMHV framework.
Its $\hypR$-part reproduces exactly the breaking in Eq.~\eqref{Eq:Delta_Hat_Introductory_Example} (see Sec.~\ref{Sec:Origin_of_Symmetry_Breaking_in_BMHV_Scheme}).
The $\hypL$-part is completely analogous and appears because the multiplet $\psi_L$ is here not generally assumed to be sterile but may also be populated with physical fields and thus couples with $\hypL$ to the gauge boson.

Beyond the familiar symmetry breaking, the evanescent gauge interactions in the second line of Eq.~\eqref{Eq:L_fermion_general_D-dim_Ansatz} generate two additional contributions.
First, $\widehat{\Delta}_1$ acquires extra terms involving the chirality-violating hypercharge matrices $\hypLR$ and $\hypRL$. 
Second, a gauge boson dependent breaking $\widehat{\Delta}_2$ appears at tree-level, which is new compared to the familiar case of Eq.~\eqref{Eq:Delta_Hat_Introductory_Example}.

Evidently, trying to set the breaking~\eqref{Eq:GeneralTreeLevelBreaking_Abelian} to zero by suitably adjusting the newly introduced evanescent hypercharges $\hypLR$ and $\hypRL$ fails, as it would require both of them to be equal to $\hypL$ and $\hypR$ at the same time depending on the particular term in Eq.~\eqref{Eq:GeneralTreeLevelBreaking_Abelian}.
This is clearly impossible, unless $\hypL=\hypR$, which however amounts to the trivial case of a vector-like gauge theory and thus excluded in this consideration.
However, we can analyse how different choices influence the symmetry breaking in Eq.~\eqref{Eq:GeneralTreeLevelBreaking_Abelian} and try to find a specific choice that leads to a minimal breaking.
An explicit analysis of this is to be found in chapter~\ref{Chap:General_Abelian_Chiral_Gauge_Theory}.
However, here we already set the stage for the explicit analysis.

Recall that, although BRST invariance is broken in $D$ dimensions due to the regularisation, the Lagrangian must remain hermitian, i.e.\ $\mathcal{L}^{\dagger}=\mathcal{L}$, even in $D$ dimensions.
This requirement implies the hermiticity condition $\hypRL = \hypLR^{\dagger}$ shown in Eq.~\eqref{Eq:Requirement-YLR-YRL}.
Consequently, every choice for $\hypLR$ and $\hypRL$ in the general Ansatz for the fermion-gauge interactions displayed in Eq.~\eqref{Eq:L_fermion_general_D-dim_Ansatz} must satisfy this constraint.
Furthermore, there are four independent covariants\footnote{
In contrast, only two such covariants appear in $4$ dimensions or in naive schemes with anticommuting $\gamma_5$.} in the general $D$-dimensional Ansatz in Eq.~\eqref{Eq:L_fermion_general_D-dim_Ansatz}.
We may rewrite the interaction terms as
\begin{equation}\label{Eq:GaugeInt-Vector-Axial-Current}
    \begin{aligned}
        &- g B_{\mu} \overline{\psi}_i \bigg[
        {\hypR}_{ij} \projL \overline{\gamma}^{\mu} \projR
        + {\hypL}_{ij} \projR \overline{\gamma}^{\mu} \projL
        + {\hypRL}_{ij} \projL \widehat{\gamma}^{\mu} \projL
        + {\hypLR}_{ij} \projR \widehat{\gamma}^{\mu} \projR 
        \bigg] \psi_j\\
        =&
        -g B_{\mu} \overline{\psi}_i \bigg[
        \frac{{\hypR}_{ij}+{\hypL}_{ij}}{2} \overline{\gamma}^{\mu}
        + \frac{{\hypLR}_{ij}+{\hypLR^{*}}_{ji}}{2} \widehat{\gamma}^{\mu}\\
        &\qquad\qquad\,\,\,\,
        + \frac{{\hypR}_{ij}-{\hypL}_{ij}}{2} \overline{\gamma}^{\mu}\gamma_5
        + \frac{{\hypLR}_{ij}-{\hypLR^{*}}_{ji}}{2} \widehat{\gamma}^{\mu}\gamma_5
        \bigg] \psi_j,
    \end{aligned}
\end{equation}
where in the second line we used the hermiticity condition~\eqref{Eq:Requirement-YLR-YRL} to write ${\hypRL}_{ij}={\hypLR^{*}}_{ji}$.
The first expression is written in terms of left- and right-handed currents, while the second version exhibits the 4-dimensional vector and axial vector currents and their evanescent counterparts.

The hermiticity properties of the fermion currents in Eq.~\eqref{Eq:GaugeInt-Vector-Axial-Current} are summarised in Tab.~\ref{Tab:Currents-Hermiticity} and constrain how the theory can be extended to $D$ dimensions.
According to these properties, the prefactors of $\overline{\gamma}^{\mu}$, $\widehat{\gamma}^{\mu}$ and $\overline{\gamma}^{\mu}\gamma_5$ are hermitian, whereas the prefactor of $\widehat{\gamma}^{\mu}\gamma_5$ is anti-hermitian.
\begin{table}[t!]
    \centering
    \begin{tabular}{|c|c|c|} \hline 
         fermion current & $\mathrm{h.c.}$ & hermiticity \\ \hline \hline
         \rule{0pt}{1.1em}$\overline{\psi}\overline{\gamma}^{\mu}\psi$ & $\overline{\psi}\overline{\gamma}^{\mu}\psi$ & hermitian \\ \hline
         \rule{0pt}{1.1em}$\overline{\psi}\widehat{\gamma}^{\mu}\psi$ & $\overline{\psi}\widehat{\gamma}^{\mu}\psi$ & hermitian \\ \hline
         \rule{0pt}{1.1em}$\overline{\psi}\overline{\gamma}^{\mu}\gamma_5\psi$ & $\overline{\psi}\overline{\gamma}^{\mu}\gamma_5\psi$ & hermitian \\ \hline
         \rule{0pt}{1.1em}$\overline{\psi}\widehat{\gamma}^{\mu}\gamma_5\psi$ & $-\,\overline{\psi}\widehat{\gamma}^{\mu}\gamma_5\psi$ & anti-hermitian \\ \hline
         \rule{0pt}{1.1em}$\overline{\psi}\projR\overline{\gamma}^{\mu}\projL\psi$ & $\overline{\psi}\projR\overline{\gamma}^{\mu}\projL\psi$ & hermitian \\ \hline
         \rule{0pt}{1.1em}$\overline{\psi}\projL\overline{\gamma}^{\mu}\projR\psi$ & $\overline{\psi}\projL\overline{\gamma}^{\mu}\projR\psi$ & hermitian \\ \hline
         \rule{0pt}{1.1em}$\overline{\psi}\projL\widehat{\gamma}^{\mu}\projL\psi$ & $\overline{\psi}\projR\widehat{\gamma}^{\mu}\projR\psi$ & --- \\ \hline
         \rule{0pt}{1.1em}$\overline{\psi}\projR\widehat{\gamma}^{\mu}\projR\psi$ & $\overline{\psi}\projL\widehat{\gamma}^{\mu}\projL\psi$ & --- \\ \hline
    \end{tabular}
    \caption{Hermiticity properties of fermion currents appearing in $D$ dimensions.}
    \label{Tab:Currents-Hermiticity}
\end{table}
As a consequence, assigning identical coefficients to $\widehat{\gamma}^\mu$ and $\overline{\gamma}^\mu$ allows the vector component of the gauge interaction to be extended uniformly to $D$ dimensions.
In contrast, the axial part cannot be extended uniformly to $D$ dimensions, since $\overline{\gamma}^{\mu}\gamma_5$ and $\widehat{\gamma}^{\mu}\gamma_5$ require coefficients of opposite hermiticity --- hermitian for the former and anti-hermitian for the latter.
This explains why it is impossible to adjust the evanescent hypercharges such that the breaking in Eq.~\eqref{Eq:GeneralTreeLevelBreaking_Abelian} vanishes, except in the trivial case of a vector-like gauge theory where all hypercharges are equal.
The primary obstacle to a uniform $D$-dimensional extension of the gauge interaction thus originates from the hermiticity properties of its axial component.

The structure of the fermion-gauge interaction in Eq.~\eqref{Eq:GaugeInt-Vector-Axial-Current} suggests several well-motivated choices for the evanescent hypercharges $\hypLR$:
\begin{itemize}
    \item $\hypLR=\hypRL=0$: This is the simplest choice.
    It eliminates the evanescent gauge interactions entirely, removing them as a potential source of global hypercharge violation and reducing the interaction to its purely 4-dimensional form: 
    \begin{align}
        -g B_{\mu}
            \overline{\psi}_i \bigg[
            \frac{{\hypR}_{ij}+{\hypL}_{ij}}{2} \overline{\gamma}^{\mu}
            + \frac{{\hypR}_{ij}-{\hypL}_{ij}}{2} \overline{\gamma}^{\mu}\gamma_5
            \bigg] \psi_j.  
    \end{align}
    \item $\hypLR=\hypRL=\frac{\hypR+\hypL}{2}$: This choice extends the vector component of the interaction to a fully $D$-dimensional form,
    \begin{align}
        -g B_{\mu}
            \overline{\psi}_i \bigg[
            \frac{{\hypR}_{ij}+{\hypL}_{ij}}{2} \gamma^{\mu}
            + \frac{{\hypR}_{ij}-{\hypL}_{ij}}{2} \overline{\gamma}^{\mu}\gamma_5
            \bigg] \psi_j.
    \end{align}
    This uniform extension of the vector interaction to $D$ dimensions corresponds to the standard treatment of vector-like gauge theories such as QED (cf.\ Eq.~\eqref{Eq:LeegammaQED}). 
    \item Other scenarios: As explained above, the axial component cannot be extended uniformly to $D$-dimensions, making other choices less simple.
    A noteworthy case is obtained by setting $\hypLR=Q$ (with $Q$ denoting the electric charge), which is relevant for the Standard Model where one might wish to treat the photon in a fully $D$-dimensional manner (see chapter~\ref{Chap:The_Standard_Model}). 
    In particular, with the SM electric and right-handed hypercharges, $Q=\hypR$, this choice leads to the fermion-gauge interaction
    \begin{align}
        -g B_{\mu}
            \overline{\psi}_i \bigg[
            \frac{-{\hypR}_{ij}+{\hypL}_{ij}}{2} \overline{\gamma}^{\mu} + Q_{ij}\gamma^\mu
            + \frac{{\hypR}_{ij}-{\hypL}_{ij}}{2} \overline{\gamma}^{\mu}\gamma_5
            \bigg] \psi_j.
    \end{align}
    Further possible choices include $\hypLR=\hypL$ or $\hypLR=(\hypR-\hypL)/2$.
\end{itemize}
In chapter~\ref{Chap:General_Abelian_Chiral_Gauge_Theory}, we explicitly determined all counterterm coefficients for SM values of the physical hypercharges $\hypL$ and $\hypR$, as well as for all such motivated choices of the evanescent hypercharges. 
The results for the most representative cases --- corresponding to $\hypLR$ is set to $0$, $\frac{\hypR+\hypL}{2}$, $Q$, $\hypL$, respectively --- are summarised in Sec.~\ref{Sec:TheRoleOf-YLR-YRL} in two tables:
in Tab.~\ref{Tab:YLR-Specialisations-Option-1} for Option~\ref{Opt:Option1} and in Tab.~\ref{Tab:YLR-Specialisations-Option-2} for Option~\ref{Opt:Option2}.
Across the columns, the tables list the different choices of evanescent hypercharges, while the rows display the corresponding counterterm coefficients for each choice.

\paragraph{Concluding Comments on Symmetry Breaking:}
The evanescent gauge interactions introduced in Sec.~\ref{Sec:Dimensional_Ambiguities_and_Evanescent_Shadows} do not preserve chirality, and therefore constitute an additional source of both local and global symmetry breaking, alongside the evanescent contributions from the fermion kinetic term (see Sec.~\ref{Sec:Origin_of_Symmetry_Breaking_in_BMHV_Scheme}).
The general structure of the tree-level breaking in Eq.~\eqref{Eq:GeneralTreeLevelBreaking_Abelian}, makes these two sources for the spurious breaking of BRST invariance and global symmetry --- induced by a non-anticommuting $\gamma_5$ within the BMHV scheme --- particularly transparent.
In summary, we identify two sources of symmetry breaking:
\begin{itemize}
    \item 
    \emph{evanescent parts of fermion kinetic terms}, whose structure depends on the specific dimensional continuation of the fermion fields to $D$ dimensions (see Sec.~\ref{Sec:Dimensional_Ambiguities_and_Evanescent_Shadows});
    \item 
    \emph{evanescent gauge interactions}, governed by the evanescent hypercharges, which may be set to zero or assigned different non-vanishing values (see  Sec.~\ref{Sec:Dimensional_Ambiguities_and_Evanescent_Shadows} and Eq.~(\ref{Eq:L_fermion_general_D-dim_Ansatz})).
\end{itemize}
Both of them mix fields of different chiralities and gauge quantum numbers, thereby acting as potential sources of symmetry breaking.

The evanescent gauge interactions are constrained by hermiticity and, as a result, the axial component of the gauge interaction cannot be extended uniformly to $D$ dimensions.
Thus, the evanescent hypercharges cannot be adjusted to completely cancel the tree-level breaking.
Nevertheless, several theoretically well-motivated choices for the evanescent hypercharges have been identified.
Their respective implications are analysed in detail in chapters~\ref{Chap:General_Abelian_Chiral_Gauge_Theory} (see particularly Sec.~\ref{Sec:Results-Shedding_Light_on_Evanescent_Shadows}) and \ref{Chap:The_Standard_Model}.

\section{Symmetry Restoration Procedure}\label{Sec:Symmetry_Restoration_Procedure}

In the previous section, we established explicitly that the consistent treatment of $\gamma_5$ within the BMHV scheme of DReg inevitably breaks BRST invariance.
This breaking is reflected by a violation of the corresponding Slavnov-Taylor identity, which must subsequently be compensated by suitable symmetry-restoring counterterms in order to obtain a properly renormalised quantum theory, as discussed in Sec.~\ref{Sec:Renormalisation-of-GaugeTheories}.
Accordingly, the ultimate symmetry requirement imposed on the fully renormalised theory is expressed by the Slavnov–Taylor identity in the form
\begin{align}\label{Eq:UltimateSymmetryRequirement}
    \mathop{\mathrm{LIM}}_{D \, \to \, 4} \, (\mathcal{S}_D(\Gamma_\mathrm{DRen})) = 0,
\end{align}
where $\Gamma_\mathrm{DRen}$ denotes the fully renormalised effective quantum action in $D$ dimensions, as introduced in Eq.~\eqref{Eq:Dimensionally-Renormalised-Effective-Action-in-D-dim}.
This condition ensures that the fully renormalised 4-dimensional theory --- the physically relevant case --- satisfies BRST invariance (cf.\ Sec.~\ref{Sec:Regularised-QAP}).
The corresponding symmetry-restoring counterterms required to fulfil Eq.~\eqref{Eq:UltimateSymmetryRequirement} are guaranteed to exist by the methods of algebraic renormalisation (see Sec.~\ref{Sec:Algebraic_Renormalisation}), provided that the theory is free of gauge anomalies.
Moreover, these counterterms are compatible with the gauge-fixing condition and the ghost equation, as demonstrated in Ref.~\cite{Piguet:1995er}.

The central problem that remains is the explicit determination of the breaking and, subsequently, of the required symmetry-restoring counterterms.
In the following, we first outline two possible strategies for constructing these counterterms and discuss a central structural feature of the breaking: its lowest-order contribution --- the tree-level breaking $\widehat{\Delta}$ --- is evanescent, which restricts the determination of higher-order symmetry breakings to power-counting divergent Green functions.
We then focus on the approach based on the quantum action principle, which is employed in all computations throughout this thesis, and discuss its practical implementation, including its specialisation to the Abelian case.
For this discussion, we mainly follow our previous presentations in Refs.~\cite{Stockinger:2023ndm,vonManteuffel:2025swv,Belusca-Maito:2023wah}.

\paragraph{Strategies and Preliminaries:}
The symmetry-restoration procedure is inherently iterative and proceeds order by order in perturbation theory.
The central object in this context is the dimensionally renormalised effective quantum action $\Gamma_\mathrm{DRen}$ introduced in Eq.~\eqref{Eq:Dimensionally-Renormalised-Effective-Action-in-D-dim}.
At the $n$-loop level, suppose that the theory has already been successfully renormalised up to order $(n-1)$ and that all necessary counterterms --- including both divergent and finite symmetry-restoring ones --- have been constructed to that order.
The corresponding subrenormalised effective action is then denoted by $\Gamma_\mathrm{subren}^{(\leq n)}$, in line with the discussions in Sec.~\ref{Sec:Notation_and_Organisation_of_Dimensional_Renormalisation}.
To complete the renormalisation at order $n$, the divergent and the finite symmetry-restoring counterterms at this loop level must be determined, leading to
\begin{align}\label{Eq:Gamma_DRen_Def_for_Sym_Restoration}
    \Gamma_\mathrm{DRen}^{(\leq n)} = \Gamma_\mathrm{subren}^{(\leq n)} + S_\mathrm{sct}^{(n)} + S_\mathrm{fct}^{(n)},
\end{align}
which defines the dimensionally renormalised effective quantum action up to order $n$.
As discussed in Sec.~\ref{Sec:Notation_and_Organisation_of_Dimensional_Renormalisation}, the divergent counterterms $S_\mathrm{sct}^{(n)}$ generally decompose into invariant and non-invariant components, $S_\mathrm{sct}^{(n)}=S_\mathrm{sct,inv}^{(n)}+S_\mathrm{sct,break}^{(n)}$, which is particularly relevant in the renormalisation of chiral gauge theories within the BMHV scheme, where symmetry breaking occurs.
These divergent counterterms are unambiguously determined by Eq.~\eqref{Eq:Determination-of-Singular-Counterterms}.
After subtracting the divergences, the theory becomes finite at the $n$-loop level, and the remaining task is to determine $S_\mathrm{fct}^{(n)}$ to restore the broken symmetry.
Two conceptually distinct strategies can be employed for this purpose:
\begin{enumerate}[label={(\arabic*)}]
    \item \emph{Direct evaluation:} Compute all Green functions in $\mathcal{S}_D(\Gamma_\mathrm{subren}^{(\leq n)}+S_\mathrm{sct}^{(n)})$, including their finite parts, insert them into the Slavnov-Taylor identity, and determine the potential non-vanishing breaking at order $n$.
    \item \emph{Quantum action principle approach:} Employ the quantum action principle introduced in Sec.~\ref{Sec:Regularised-QAP} (see Eq.~\eqref{Eq:Regularised-QAP-Formulation-wrt-Gamma}) to express the breaking as an operator insertion and evaluate the corresponding operator-inserted 1PI Green functions.
\end{enumerate}

The first strategy is conceptually straightforward and operates only on ordinary Green functions.
However, this approach suffers from several significant drawbacks.
First, the number of possible identities among Green functions is, in principle, infinite, which complicates any systematic treatment.
Moreover, the procedure involves evaluating finite and non-local contributions to the Green functions, which are typically far more challenging to compute than the divergent parts.
In particular, these finite terms cannot be obtained using the tadpole decomposition method introduced in Sec.~\ref{Sec:Tadpole_Decomposition}.
Since most finite contributions --- especially those non-polynomial in the external momenta --- are consistent with the symmetry and thus cancel in $\mathcal{S}_D(\Gamma_\mathrm{subren}^{(\leq n)}+S_\mathrm{sct}^{(n)})$, this strategy becomes unnecessarily complicated for practical calculations.
As this approach was not employed in the computations presented in this thesis, it will not be discussed further here.
For an explicit example of this method and a direct comparison with the second strategy based on the quantum action principle, we refer the reader to Sec.~7 of Ref.~\cite{Belusca-Maito:2023wah}.

The second strategy employs the quantum action principle to reformulate the breaking as the insertion of the corresponding composite operator $\Delta$ into the effective quantum action, denoted by $\Delta\cdot\Gamma_\mathrm{DRen}$.
Consequently, the breaking can be computed in terms of the operator-inserted 1PI Green functions derived from $\Delta\cdot\Gamma_\mathrm{DRen}$, as introduced in Sec.~\ref{Sec:GreenFunctions_and_GeneratingFunctionals} (see in particular Eq.~\eqref{Eq:1PI-Green-Functions-MomentumSpace-with-DeltaInsertion}).
The advantage of this method is that it substantially restricts possible non-vanishing contributions, leading to a more efficient computation.
The central property underlying this simplification is that the lowest-order contribution of the breaking operator $\Delta=\widehat{\Delta}+\Delta_\mathrm{ct}$, i.e.\ the tree-level braking $\widehat{\Delta}$ (cf.\ Eqs.~\eqref{Eq:Delta_Hat_Introductory_Example} and \eqref{Eq:GeneralTreeLevelBreaking_Abelian}), is purely evanescent.
As a result, only power-counting divergent Green functions need to be considered, since only their UV-divergent part can produce non-vanishing contributions to the breaking in the physical limit.
This follows from the fact that, in the limit $\mathop{\mathrm{LIM}}_{D \, \to \, 4}$, evanescent quantities can contribute only when multiplied by a $1/\epsilon$ pole, whereas finite evanescent terms vanish.
Finite and non-evanescent contributions thus emerge exclusively from contractions or other manipulations of evanescent numerators that generate factors of $\epsilon$, which then combine with $1/\epsilon$ divergences --- schematically, terms of the form $\epsilon/\epsilon$.
This is a crucial feature: it ensures that, even at high loop orders, symmetry restoration requires evaluating only the divergent parts of a limited set of operator-inserted Green functions.
We summarise this in the following statement:
\begin{theorem}[Insertion Behaviour of Evanescent Operators]\label{Thm:Insertion_Behaviour_of_Evanescent_Operators}\ \\
    Let $\Delta=\widehat{\Delta}+\Delta_\mathrm{ct}$ be a local composite operator in formally $D$-dimensional spacetime $(\mathbb{M}_D,\eta)$, whose lowest-order contribution $\widehat{\Delta}$ is evanescent.
    Then, any finite and non-evanescent contribution arising from the insertion of this operator into dimensionally regularised Green functions can only originate from the divergent part of those Green functions.
\end{theorem}

At the 1-loop level, the validity of this theorem is straightforward.
Only the insertion of the purely evanescent operator $\widehat{\Delta}$, which can be rewritten as $\widehat{\Delta}=\widehat{\eta}^{\mu\nu}\Delta^{(0)}_{\mu\nu}$, contributes.
Using the Bonneau identities established in Refs.~\cite{Bonneau:1979jx,Bonneau:1980zp}, one finds that the required finite and non-evanescent contributions are obtained from the residue of simple pole in $(4-D)$ of the Green functions with an insertion of $\widehat{\Delta}$, as discussed in detail in Refs.~\cite{Belusca-Maito:2020ala,Belusca-Maito:2020jiz}.
This realises precisely the mechanism described above: the insertion of $\widehat{\Delta}$ into 1-loop diagrams generates finite terms through combinations of evanescent numerators with $1/\epsilon$ poles.

At higher orders, the situation becomes less transparent because the counterterm contributions $\Delta_\mathrm{ct}^{(\geq1)}$ generally contain also 4-dimensional components (see chapters~\ref{Chap:General_Abelian_Chiral_Gauge_Theory}--\ref{Chap:The_Standard_Model}).
Nevertheless, the validity of the theorem remains plausible.
The leading contribution of the breaking operator $\widehat{\Delta}$ is always evanescent and determines the structure of the genuine $n$-loop diagrams arising from the operator-inserted Green functions, whereas the higher-order corrections $\Delta_\mathrm{ct}^{(n-k)}$ (for some $1\leq k<n$) appear only in the $k$-loop counterterm diagrams required for subrenormalisation.
Moreover, since the breaking originates from regularisation and the latter is necessitated solely by divergences, it is natural to expect that the corresponding symmetry-restoring counterterms $S_\mathrm{fct}^{(n)}$ emerge entirely from the divergent parts of the theory.
These considerations, while compelling, remain plausibility arguments rather than a formal proof.
A justification of the theorem’s validity at the 2-loop level is provided in Ref.~\cite{Kuehler:2021}, and a general all-order proof is currently under development, primarily pursued by the author of that reference.

\paragraph{Symmetry Restoration utilising the Quantum Action Principle:}
In this thesis, we employ the strategy based on the quantum action principle, formulated within the framework of DReg in Sec.~\ref{Sec:Regularised-QAP}.
Starting from Eq.~\eqref{Eq:Regularised-QAP-Formulation-wrt-Gamma} for the $D$-dimensional bare action $S=S_0+S_\mathrm{ct}$ --- which includes the counterterms and the BRST transformations $\delta\phi_i(x)$ coupled to the external sources $K_i(x)$ (cf.\ Eq.~\eqref{Eq:Full-Action-For-Reg-QAP}) --- we obtain the quantum action principle
\begin{align}\label{Eq:QAP-for-Breaking-in-BMHV}
    \int d^Dx \frac{\delta\Gamma_\mathrm{DRen}}{\delta\phi_i(x)}\frac{\delta\Gamma_\mathrm{DRen}}{\delta K_i(x)} = \bigg[\int d^Dx \frac{\delta S}{\delta\phi_i(x)}\delta\phi_i(x)\bigg]\cdot\Gamma_\mathrm{DRen},
\end{align}
for the dimensionally renormalised effective quantum action $\Gamma_\mathrm{DRen}$.
This identity provides an elegant and powerful tool for studying the consequences of field transformations that do not leave the action invariant --- here, the regularisation-induced breaking of BRST symmetry --- at the quantum level.
The operator inserted into $\Gamma_\mathrm{DRen}$ on the RHS of Eq.~\eqref{Eq:QAP-for-Breaking-in-BMHV} corresponds to the integrated variation of the action under the considered field transformations.
For compactness, we define the local composite operator
\begin{align}
    \Delta(x)=\frac{\delta S}{\delta\phi_i(x)}\delta\phi_i(x) = \frac{\delta (S_0+S_\mathrm{ct})}{\delta\phi_i(x)} \delta\phi_i(x).
\end{align}
The field variations $\delta\phi_i(x)$ are coupled to the external sources $K_i(x)$ and can thus be expressed as
\begin{align}
    \delta \phi_i(x) = \frac{\delta (S_0+S_\mathrm{ct})}{\delta K_i(x)} = s_D \phi_i(x) + \frac{\delta S_\mathrm{ct}}{\delta K_i(x)},
\end{align}
where, in the second equality, we have used that these variations correspond to the BRST transformations.
At tree-level, they are generated by the $D$-dimensional BRST operator $s_D$ as $s_D\phi_i(x)$, while the last term accounts for the fact that BRST transformations are local composite operators and therefore renormalise themselves. 
Integrating over spacetime yields
\begin{equation}
\begin{aligned}\label{Eq:Definition-of-Delta-Operator-as-Breaking}
    \Delta &= \int d^Dx \, \Delta(x) = \int d^Dx \, \frac{\delta (S_0+S_\mathrm{ct})}{\delta\phi_i(x)} \delta\phi_i(x) = \int d^Dx \, \frac{\delta (S_0+S_\mathrm{ct})}{\delta\phi_i(x)} \frac{\delta (S_0+S_\mathrm{ct})}{\delta K_i(x)}
    \\ 
    &= \mathcal{S}_D(S_0+S_\mathrm{ct}),
\end{aligned}
\end{equation}
where, in the last step, we used Eq.~\eqref{Eq:D-dim-ST-Operator} to express the expression terms of the $D$-dimensional Slavnov-Taylor operator.
This quantity represents the integrated local composite operator introduced above (cf.\ theorem~\ref{Thm:Insertion_Behaviour_of_Evanescent_Operators}) and is precisely the operator inserted into $\Gamma_\mathrm{DRen}$ on the RHS of Eq.~\eqref{Eq:QAP-for-Breaking-in-BMHV}.

Using Eq.~\eqref{Eq:Definition-of-Delta-Operator-as-Breaking}, the quantum action principle in Eq.~\eqref{Eq:QAP-for-Breaking-in-BMHV} can be rewritten in compact form
\begin{align}\label{Eq:QAPinDReg_Symmetry_Restoration}
    \mathcal{S}_D(\Gamma_\mathrm{DRen})=\Delta\cdot\Gamma_\mathrm{DRen},
\end{align}
which constitutes the central identity for the symmetry restoration procedure.
In this way, the possible breaking of the Slavnov–Taylor identity is represented as an operator insertion of the local composite operator $\Delta=\widehat{\Delta}+\Delta_{\mathrm{ct}}$.
At tree-level, the breaking corresponds to the evanescent operator $\widehat{\Delta}=\mathcal{S}_D(S_0)$ (cf.\ Eq.~\eqref{Eq:Tree-Level-BRST-Breaking-Introductory_Example}), while the inclusion of counterterms yields the general expression for the breaking
\begin{align}\label{Eq:DefDeltaBreaking}
    \Delta = \widehat{\Delta} + \Delta_{\mathrm{ct}} = \mathcal{S}_D(S_{0} + S_{\mathrm{ct}}).
\end{align}
The single operator insertion of $\Delta$ into $\Gamma_\mathrm{DRen}$ in Eq.~\eqref{Eq:QAPinDReg_Symmetry_Restoration} is defined in Sec.~\ref{Sec:GreenFunctions_and_GeneratingFunctionals}.
The field monomials appearing in $\Delta\cdot\Gamma_\mathrm{DRen}$ have ghost number~$1$ and are bounded by mass dimension~$4$; their vector space is spanned by a finite basis formed by any quantum extension of the corresponding classical insertions (see proposition~\ref{Prop:Basis_of_Insertions} in Sec.~\ref{Sec:Algebraic_Renormalisation}).

As a composite operator, $\Delta$ is itself subject to renormalisation.
Analogous to the counterterm action (see Eq.~\eqref{Eq:Counterterm-Action-Perturbative-Expansion}), it admits a perturbative expansion of the form
\begin{align}\label{Eq:DeltaOperator-PerturbativeExpansion}
    \Delta = \widehat{\Delta} + \sum_{n=1}^\infty \Delta^{(n)}_\mathrm{ct} =
    \widehat{\Delta} + \Delta^{(1)}_\mathrm{ct}
    + \Delta^{(2)}_\mathrm{ct} + \Delta^{(3)}_\mathrm{ct}
    + \Delta^{(4)}_\mathrm{ct} + \mathcal{O}(\hbar^5),
\end{align}
where $\widehat{\Delta}$ denotes the tree-level breaking, and $\Delta^{(n)}_\mathrm{ct}$ represents the $n$-loop contribution to the breaking of BRST symmetry.
Expanding the RHS of Eq.~\eqref{Eq:QAPinDReg_Symmetry_Restoration} in loop orders accordingly yields
\begin{align}\label{Eq:PerturbativeExpansion-DeltaDotGamma}
    \Delta\cdot\Gamma_\mathrm{DRen}=\widehat{\Delta} + \sum_{n=1}^\infty \hbar^n \bigg(\widehat{\Delta}\cdot\Gamma_\mathrm{DRen}^{(n)}+\sum_{k=1}^{n-1}\Delta^{(k)}_\mathrm{ct}\cdot\Gamma^{(n-k)}_\mathrm{DRen}+\Delta^{(n)}_\mathrm{ct}\bigg).
\end{align}
Inserting Eq.~\eqref{Eq:QAPinDReg_Symmetry_Restoration} into the ultimate symmetry requirement in Eq.~\eqref{Eq:UltimateSymmetryRequirement} then gives the perturbative condition
\begin{align}\label{Eq:PerturbativeRequirementAndStartingPoint}
    \mathop{\mathrm{LIM}}_{D \, \to \, 4} \, \bigg(\widehat{\Delta}\cdot\Gamma_\mathrm{DRen}^{(n)}+\sum_{k=1}^{n-1}\Delta^{(k)}_\mathrm{ct}\cdot\Gamma^{(n-k)}_\mathrm{DRen}+\Delta^{(n)}_\mathrm{ct}\bigg)=0, \hspace{0.75cm} \forall \, n \geq 1,
\end{align}
where $n$ denotes the loop order.
This equation serves as the starting point of the iterative symmetry restoration procedure based on the quantum action principle.

At a given loop order $n$, supposing the theory has been subrenormalised up to order $(n-1)$, with $\Gamma_\mathrm{subren}^{(\leq n)}$ constructed as outlined above, the counterterm action $S_\mathrm{ct}^{(\leq n-1)}$ and the breaking operator $\Delta_\mathrm{ct}^{(\leq n-1)}$ are known.
The first two terms in Eq.~\eqref{Eq:PerturbativeRequirementAndStartingPoint}, being fully subrenormalised, can then be computed unambiguously, which in turn allows for the determination of $\Delta_\mathrm{ct}^{(n)}$.
Subsequently, the counterterm action $S_\mathrm{ct}^{(n)}$ must be adjusted such that $[\mathcal{S}_D(S_0+S_\mathrm{ct}^{(\leq n)})]^{(n)}=\Delta_\mathrm{ct}^{(n)}$, thereby cancelling the contribution of the first two terms and ensuring that Eq.~\eqref{Eq:PerturbativeRequirementAndStartingPoint} is satisfied.
This iterative procedure determines the finite-symmetry restoring counterterms $S_\mathrm{fct}^{(n)}$ and the non-invariant singular counterterms $S_\mathrm{sct,break}^{(n)}$.

The former are not uniquely fixed, as Eq.~\eqref{Eq:PerturbativeRequirementAndStartingPoint} determines only their BRST variation and is therefore insensitive to finite symmetric differences.
This freedom allows for the addition of finite symmetric counterterms --- either to adjust the finite symmetry-restoring counterterms or to impose renormalisation conditions.
Likewise, Eq.~\eqref{Eq:PerturbativeRequirementAndStartingPoint} is blind to finite and evanescent counterterms, as these vanish upon taking the limit $\mathop{\mathrm{LIM}}_{D \, \to \, 4}$, providing additional flexibility to optimise the counterterm action.
The implications of these residual freedoms are discussed in more detail in Sec.~6.3.3 of Ref.~\cite{Belusca-Maito:2023wah}.

Since the full singular counterterm action $S_\mathrm{sct}^{(n)}=S_\mathrm{sct,inv}^{(n)}+S_\mathrm{sct,break}^{(n)}$ is obtained from the evaluation of ordinary subrenormalised 1PI Green functions, the non-invariant singular counterterms $S_\mathrm{sct,break}^{(n)}$ can be verified using Eq.~\eqref{Eq:PerturbativeRequirementAndStartingPoint} as a consistency check.
Although each term in this equation may contain both divergent and finite contributions, the total expression must be finite by construction.
Hence, the cancellation of divergences serves as a stringent internal consistency check of the calculation.

\paragraph{Practical Symmetry Restoration:}
Having established the theoretical foundations of the symmetry-restoration procedure, we now turn to its concrete realisation.
The following step-by-step algorithm summarises the practical implementation, which we subsequently illustrate and discuss at the 1-loop and 4-loop level.
\begin{algorithm}[Symmetry Restoration utilising the Quantum Action Principle]\ \\
At loop order $n$, given a fully subrenormalised theory described by $\Gamma_\mathrm{subren}^{(\leq n)}$, then proceed as follows:
    \begin{enumerate}[label={(\arabic*)}]
        \item \emph{UV renormalisation:} Compute the divergent part of all relevant subrenormalised ordinary 1PI Green functions $\langle \phi_{i_1}(p_1)\ldots\phi_{i_n}(p_n) \rangle^{\mathrm{1PI}}$ (cf.\ Eq.~\eqref{Eq:1PI-Green-Functions-MomentumSpace}) at order~$n$. 
        These divergences unambiguously determine the required singular counterterms $S_\mathrm{sct}^{(n)}$ by means of Eq.~\eqref{Eq:Determination-of-Singular-Counterterms}.
        \item \emph{Computation of the BRST breaking:} Using the quantum action principle, evaluate all relevant subrenormalised operator-inserted 1PI Green functions $\langle \Delta \phi_{i_1}(p_1)\ldots\phi_{i_n}(p_n) \rangle^{\mathrm{1PI}}$ (cf.\ Eq.~\eqref{Eq:1PI-Green-Functions-MomentumSpace-with-DeltaInsertion}) at order~$n$, where the local composite operator $\Delta=\widehat{\Delta}+\Delta_\mathrm{ct}$ represents the breaking of BRST invariance.
        This requires:
        \begin{enumerate}[label={(\alph*)}]
            \item genuine $n$-loop diagrams with one insertion of the tree-level breaking operator $\widehat{\Delta}$,
            \item $k$-loop counterterm diagrams with one insertion of the order-$(n-k)$ breaking operator $\Delta_\mathrm{ct}^{(n-k)}$, for all $1\leq k < n$.
        \end{enumerate}
        In this way, the first two terms of Eq.~\eqref{Eq:PerturbativeRequirementAndStartingPoint} are computed.
        \item \emph{Consistency Check:} Verify that all UV divergences in Eq.~\eqref{Eq:PerturbativeRequirementAndStartingPoint} cancel.
        Specifically, the BRST variation of the singular counterterms obtained in step~(1) must exactly cancel the divergent part of the total breaking computed in step~(2).
        In other words, step~(1) and step~(2) must yield identical results for $S_\mathrm{sct,break}^{(n)}$.
        \item \emph{Determination of finite symmetry-restoring counterterms:} Collect all finite contributions to Eq.~\eqref{Eq:PerturbativeRequirementAndStartingPoint} obtained in step~(2) and identify appropriate field monomials $X$ such that $b_DX$ cancels them.
        This determines the finite symmetry-restoring counterterms $S_\mathrm{fct}^{(n)}$, so that $[\mathcal{S}_D(S_0+S_\mathrm{fct}^{(\leq n)})]^{(n)}=\Delta_\mathrm{fct}^{(n)}$, modulo finite symmetric and finite evanescent contributions.
        Here, we decomposed $\Delta_\mathrm{ct}^{(n)}=\Delta_\mathrm{sct}^{(n)}+\Delta_\mathrm{fct}^{(n)}$.
        Together with the singular counterterms $S_\mathrm{sct}^{(n)}$, this ensures that Eq.~\eqref{Eq:PerturbativeRequirementAndStartingPoint} is satisfied at this order.
        \item \emph{Fixing the renormalisation scheme:} Impose renormalisation conditions to determine finite symmetric counterterms, and --- if convenient and desired --- introduce finite evanescent counterterms to optimise the renormalisation procedure, thereby uniquely fixing any remaining freedom.
        The complete order-$n$ counterterm action $S_\mathrm{ct}^{(n)}=S_\mathrm{sct}^{(n)}+S_\mathrm{fct}^{(n)}$ is then unambiguously specified.
    \end{enumerate}
Completing this procedure yields the fully renormalised effective quantum action at order $n$:
\begin{align}\label{Eq:FullyRenormalisedEffectiveQuantumActionAfterCompleteBMHVRenormalisation}
    \Gamma^{(\leq n)} \equiv \Gamma^{(\leq n)}_\mathrm{ren} \equiv \mathop{\mathrm{LIM}}_{D \, \to \, 4} \, \Big(\Gamma_\mathrm{subren}^{(\leq n)} + S_\mathrm{sct}^{(n)}+S_\mathrm{fct}^{(n)}\Big).
\end{align}
\end{algorithm}

Before turning to the explicit 1- and 4-loop cases, it is useful to examine the evaluation of the RHS of Eq.~\eqref{Eq:DefDeltaBreaking} in more detail.
As introduced in Sec.~\ref{Sec:The-Slavnov-Taylor-Identity}, the Slavnov-Taylor operator $\mathcal{S}_D$ is nonlinear (see Eqs.~\eqref{Eq:Slavnov-Taylor-Operator-Formal-Def} and \eqref{Eq:D-dim-ST-Operator}).
A perturbative evaluation of $\mathcal{S}_D(S_{0} + S_{\mathrm{ct}})$ thus requires expanding 
\begin{align}
    \mathcal{S}_D(S_0 + S_\mathrm{ct}) = \int d^D x \, \frac{\delta(S_0+S_\mathrm{ct})}{\delta \phi_i(x)} \frac{\delta(S_0+S_\mathrm{ct})}{\delta K_i(x)},
\end{align}
order by order in $\hbar$.
Carrying out this expansion explicitly gives
\begin{align}\label{Eq:STop-on-S_0-and-S_ct-up-to-order-hbar5}
    \mathcal{S}_D(S_0 + S_\mathrm{ct}) 
    = \, &s_D S_0 
    + b_D S_\mathrm{ct}^{(1)} 
    + \Big[ b_D S_\mathrm{ct}^{(2)} + \mathcal{S}_D(S_\mathrm{ct}^{(1)}) \Big] \nonumber\\
    + \bigg[ &b_D S_\mathrm{ct}^{(3)} + \int d^D x \bigg\{ \frac{\delta S_\mathrm{ct}^{(1)}}{\delta \phi_i(x)} \frac{\delta S_\mathrm{ct}^{(2)}}{\delta K_i(x)} + \frac{\delta S_\mathrm{ct}^{(2)}}{\delta \phi_i(x)} \frac{\delta S_\mathrm{ct}^{(1)}}{\delta K_i(x)} \bigg\} \bigg]\\
    + \bigg[ &b_D S_\mathrm{ct}^{(4)} + \mathcal{S}_D(S_\mathrm{ct}^{(2)}) + \int d^D x \bigg\{ \frac{\delta S_\mathrm{ct}^{(1)}}{\delta \phi_i(x)} \frac{\delta S_\mathrm{ct}^{(3)}}{\delta K_i(x)} + \frac{\delta S_\mathrm{ct}^{(3)}}{\delta \phi_i(x)} \frac{\delta S_\mathrm{ct}^{(1)}}{\delta K_i(x)} \bigg\} \bigg]
    + \mathcal{O}(\hbar^5). \nonumber
\end{align}
Here, $b_D$ denotes the linearised Slavnov-Taylor operator w.r.t.\ the classical action $S_0$ in $D$-dimensions (cf.\ Eq.~\eqref{Eq:b-The-Linearised-Slavnov-Taylor-Operator-wrt-Classical-Action}), given by
\begin{align}\label{Eq:bD-The-Linearised-Slavnov-Taylor-Operator-wrt-Classical-Action-in-D-dim}
    b_D = \int d^Dx \bigg( \frac{\delta S_0}{\delta K_i(x)} \frac{\delta}{\delta\phi_i(x)} + \frac{\delta S_0}{\delta\phi_i(x)}\frac{\delta}{\delta K_i(x)} \bigg) = s_D + \int d^Dx \frac{\delta S_0}{\delta\phi_i(x)}\frac{\delta}{\delta K_i(x)}.
\end{align}
As discussed in Ref.~\cite{Belusca-Maito:2020ala}, in contrast to its 4-dimensional analogue $b$, the $D$-dimensional operator $b_D$ is here not nilpotent, i.e.\ $b_D^2\neq0$, because the $D$-dimensional classical $S_0$ is not BRST invariant in the presence of regularisation-induced symmetry breaking.\footnote{A formally nilpotent operator $b_D^{\mathrm{nilpotent}}$ could be defined using the invariant part of the $D$-dimensional action $S_{0,\mathrm{inv}}$ (cf.\ Eq.~\eqref{Eq:Split-of-S_0-into-inv-and-evan-Part}), but this is not the operator relevant to the present analysis.}
This loss of nilpotency is an intrinsic subtlety of regularisation schemes that violate the symmetry, such as the BMHV scheme of DReg for chiral gauge theories.
In contrast, for symmetry-preserving regularisations --- in particular, for DReg applied to vector-like gauge theories --- the corresponding $D$-dimensional operator remains nilpotent.
Importantly, the 4-dimensional operator $b$ is nilpotent in either case, as the 4-dimensional classical action $S_0^{(4D)}$ is BRST invariant.

In what follows, we omit the subscript ``$\mathrm{subren}$'' on the effective quantum action for brevity, as no ambiguity arises. 
We begin at the 1-loop level and consider the subrenormalised (i.e.\ unrenormalised at the 1-loop level) effective action $\Gamma^{(\leq 1)}$. 
In order to obtain the renormalised effective action $\Gamma_\mathrm{DRen}^{(\leq 1)}$ (see Eq.~\eqref{Eq:Gamma_DRen_Def_for_Sym_Restoration}), the 1-loop counterterm action $S_\mathrm{ct}^{(1)}=S_\mathrm{sct}^{(1)}+S_\mathrm{fct}^{(1)}$ must be determined.
Following the algorithm outlined above, we first compute the singular counterterms by requiring the cancellation of all 1-loop divergences,
\begin{align}\label{Eq:1-loop-S_sct-Determination}
    \Gamma^{(1)}|_\mathrm{div} + S_\mathrm{sct}^{(1)} = 0,
\end{align}
where the divergences are extracted from ordinary subrenormalised 1PI Green functions.
Next, we compute the breaking at the 1-loop level via subrenormalised operator-inserted 1PI Green functions.
Using the perturbative symmetry-restoration requirement in Eq.~\eqref{Eq:PerturbativeRequirementAndStartingPoint}, the 1-loop breaking operator $\Delta_\mathrm{ct}^{(1)}$ is obtained as
\begin{equation}\label{Eq:1-loop-BRST-Breaking}
    \begin{aligned}
        {\big( \Delta \cdot \Gamma \big)}{}^{(1)} = 
        \widehat{\Delta} \cdot \Gamma^{(1)}
        = - \Delta^{(1)}_\mathrm{ct} + \mathcal{O}(\hat{.}),
    \end{aligned}
\end{equation}
where $\mathcal{O}(\hat{.})$ denotes finite evanescent terms that vanish in the limit $\mathop{\mathrm{LIM}}_{D\,\to\,4}$.
With the breaking operator decomposed as $\Delta_\mathrm{ct}^{(1)} = \Delta_\mathrm{sct}^{(1)} + \Delta_\mathrm{fct}^{(1)}$, its divergent part $\Delta_\mathrm{sct}^{(1)}$ and finite part $\Delta_\mathrm{fct}^{(1)}$ implicitly determine the corresponding counterterms $S_\mathrm{sct,break}^{(1)}$ and $S_\mathrm{fct}^{(1)}$ through Eq.~\eqref{Eq:STop-on-S_0-and-S_ct-up-to-order-hbar5}.
In particular, evaluating Eq.~\eqref{Eq:STop-on-S_0-and-S_ct-up-to-order-hbar5} strictly at the 1-loop level (in combination with Eq.~\eqref{Eq:DefDeltaBreaking}) yields
\begin{align}
    \Delta_\mathrm{ct}^{(1)} = [\mathcal{S}_D(S_0 + S_\mathrm{ct})]^{(1)} = b_D S_\mathrm{ct}^{(1)},
\end{align}
which serves as the determining equation for the non-invariant counterterms --- both divergent and finite.
In this way, the 1-loop symmetry-restoring counterterms are obtained from the operator-inserted 1PI Green functions generated by $\widehat{\Delta} \cdot \Gamma^{(1)}$.
The non-invariant divergent contributions $S_\mathrm{sct,break}^{(1)}$ can be compared with those derived from Eq.~\eqref{Eq:1-loop-S_sct-Determination}, providing the aforementioned consistency check.

The 2-loop case has been discussed in detail in Ref.~\cite{Belusca-Maito:2023wah}; here, we skip the 3-loop case and proceed directly to the 4-loop level.
We consider the subrenormalised effective action $\Gamma^{(\leq 4)}$ and aim to determine the 4-loop counterterm action $S_\mathrm{ct}^{(4)}=S_\mathrm{sct}^{(4)}+S_\mathrm{fct}^{(4)}$ required for the complete renormalisation at this order.
As before, the singular counterterms are obtained from the divergent part of ordinary subrenormalised 1PI Green functions:
\begin{align}
    \Gamma^{(4)}|_\mathrm{div} + S_\mathrm{sct}^{(4)} = 0.
\end{align}
Employing Eq.~\eqref{Eq:PerturbativeRequirementAndStartingPoint} as the starting point for symmetry restoration, the symmetry breaking at the 4-loop level reads
\begin{equation}\label{Eq:4-loop-BRST-Breaking}
    \begin{aligned}
        {\big( \Delta \cdot \Gamma \big)}{}^{(4)} = 
        \widehat{\Delta} \cdot \Gamma^{(4)}
        + \Delta^{(1)}_\mathrm{ct} \cdot \Gamma^{(3)}
        + \Delta^{(2)}_\mathrm{ct} \cdot \Gamma^{(2)}
        + \Delta^{(3)}_\mathrm{ct} \cdot \Gamma^{(1)} 
        = - \Delta^{(4)}_\mathrm{ct} + \mathcal{O}(\hat{.}),
    \end{aligned}
\end{equation}
which is obtained from the corresponding subrenormalised 1PI Green functions with one insertion of the respective lower-order breaking operator.
This yields the 4-loop breaking operator $\Delta_\mathrm{ct}^{(4)}$ modulo finite evanescent terms.
Evaluating Eq.~\eqref{Eq:STop-on-S_0-and-S_ct-up-to-order-hbar5} together with Eq.~\eqref{Eq:DefDeltaBreaking} strictly at the 4-loop level provides the determining equation for the non-invariant counterterms at this order:
\begin{equation}
\begin{aligned}\label{Eq:STI-Strictly-Evaluated-at-4-loop}
    \Delta_\mathrm{ct}^{(4)}&=[\mathcal{S}_D(S_0 + S_\mathrm{ct})]^{(4)}\\ 
    &= \bigg[ b_D S_\mathrm{ct}^{(4)} + \mathcal{S}_D(S_\mathrm{ct}^{(2)}) + \int d^D x \bigg\{ \frac{\delta S_\mathrm{ct}^{(1)}}{\delta \phi_i(x)} \frac{\delta S_\mathrm{ct}^{(3)}}{\delta K_i(x)} + \frac{\delta S_\mathrm{ct}^{(3)}}{\delta \phi_i(x)} \frac{\delta S_\mathrm{ct}^{(1)}}{\delta K_i(x)} \bigg\} \bigg].
\end{aligned}
\end{equation}
As in the 1-loop case, the non-invariant 4-loop counterterms --- in particular, the symmetry-restoring counterterms $S_\mathrm{fct}^{(4)}$ --- are implicitly determined by the 4-loop breaking operator, which itself is obtained as the negative of the 4-loop breaking ${\big( \Delta \cdot \Gamma \big)}{}^{(4)}$ (see Eq.~\eqref{Eq:4-loop-BRST-Breaking}).
Equation~\eqref{Eq:STI-Strictly-Evaluated-at-4-loop} also illustrates that the determination of counterterms becomes in general increasingly complicated at higher orders, as it involves lower-order counterterms due to the nonlinear nature of the Slavnov-Taylor operator.

In fact, this last observation highlights the main disadvantage of this strategy based on the quantum action principle.
Finite symmetry-restoring counterterms are determined only implicitly through the evaluation of the symmetry breaking at the respective order, expressed in terms of fields monomials with ghost number~$1$.
Constructing the corresponding field monomials of ghost number~$0$ whose BRST variation cancels this breaking can become highly nontrivial --- especially in non-Abelian gauge theories such as the Standard Model, which involve various types of fields and interaction structures.
Two main difficulties arise: first, the BRST transformations themselves are subject to renormalisation in non-Abelian theories; the variations of distinct ghost-number-$0$ monomials may contribute to identical ghost-number-$1$ monomials, leading to systems of equations that must be solved to identify the correct counterterm structure.

\paragraph{Symmetry Restoration of Abelian Gauge Invariance:}
In the context of symmetry restoration, the key property of Abelian gauge theories is that the associated BRST transformations do not renormalise, as discussed in sections~\ref{Sec:The-Slavnov-Taylor-Identity} and \ref{Sec:Peculiarities_of_Abelian_Gauge_Theories}.
Consequently, $\mathcal{L}_\mathrm{ext}$ does not receive any quantum corrections, implying that the variation of the counterterm action w.r.t.\ the external sources, $\delta S_\mathrm{ct}^{(n)}/\delta K_i(x)$, vanishes identically at all orders.
This feature significantly simplifies Eq.~\eqref{Eq:STop-on-S_0-and-S_ct-up-to-order-hbar5} and thus reduces the evaluation of the non-invariant counterterms described above to the application of the BRST operator $s_D$.

At any loop order~$n$, the non-invariant counterterms --- both the divergent part $S_\mathrm{sct,break}^{(n)}$ and the finite symmetry-restoring contributions $S_\mathrm{fct}^{(n)}$ --- are therefore implicitly determined by the $n$-loop breaking according to
\begin{align}\label{Eq:Determination-of-Symmetry-Restoring-CT-Abelian-Case}
    s_D S_\mathrm{ct}^{(n)} = \Delta_\mathrm{ct}^{(n)} = - {\big( \Delta \cdot \Gamma \big)}{}^{(n)} + \mathcal{O}(\hat{.}),
\end{align}
modulo finite evanescent terms.

The familiar Ward identities arise as a direct consequence of the Slavnov–Taylor identity, as discussed in Sec.~\ref{Sec:Peculiarities_of_Abelian_Gauge_Theories}.
Once the Slavnov–Taylor identity has been restored, the functional Ward identity introduced in Eq.~\eqref{Eq:FunctionalAbelianWardId} follows from it and must hold in the fully renormalised theory.
In particular, Eq.~\eqref{Eq:FunctionalAbelianWardId} implies the well-known Ward identities, including the transversality of the gauge boson self energy, the transversality of multi gauge boson interactions, and the relation between the fermion self energy to the fermion--gauge boson interaction, all of which must be satisfied order by order in the renormalised theory.
In chapter~\ref{Chap:BMHV_at_Multi-Loop_Level}, we carry out the renormalisation of an Abelian chiral gauge theory up to the 4-loop level and examine how these relations are affected by the regularisation-induced symmetry breaking.

In addition to the validity of the Slavnov-Taylor identity, the linear relations given in Eq.~\eqref{Eq:RquireAllOrdersRelations} must hold, as discussed in Sec.~\ref{Sec:Peculiarities_of_Abelian_Gauge_Theories}.
Because of their linear structure, these relations do not receive nontrivial quantum corrections and hence do not renormalise.
Any counterterm introduced by hand into these linear relations can be removed, as exemplified by the gauge-fixing condition and ghost equation in Sec.~\ref{Sec:Algebraic_Renormalisation}.

\paragraph{Concluding Remark on Renormalisability:}
Within the BMHV scheme of DReg, the renormalisation procedure is stable in the sense of Def.~\ref{Def:Stability_under_Renormalisation}: a power-counting renormalisable theory remains renormalisable, and the number of required counterterms --- including symmetry-restoring ones --- does not increase indefinitely with the loop order.
In particular, upon completion of the full renormalisation programme, including the restoration of symmetries as outlined above, and after taking the limit $\mathop{\mathrm{LIM}}_{D \, \to \, 4}$, the resulting renormalised theory described by the effective action $\Gamma^{(\leq n)}_\mathrm{ren}$ (see Eq.~\eqref{Eq:FullyRenormalisedEffectiveQuantumActionAfterCompleteBMHVRenormalisation}) is entirely expressed through reparametrisations of the fields and parameters already present in the classical action.
Here, reparametrisation is to be understood in a general sense, extending beyond simple multiplicative renormalisation, which alone is clearly not sufficient in the BMHV scheme.

A central role in ensuring this stability is played by proposition~\ref{Prop:Basis_of_Insertions}, which ensures that all possible breakings can be expanded in a finite basis of local ghost-number-$1$ field monomials bounded by power counting (see also Refs.~\cite{Piguet:1995er,Cornella:2022hkc}).
At higher loop orders, additional monomials may appear beyond those present at tree-level (see Eq.~\eqref{Eq:GeneralTreeLevelBreaking_Abelian}), such as, for instance, the ghost--triple gauge boson structure $c\overline{\partial}_\mu({\overline{B}}{}^\mu\overline{B}_\nu{\overline{B}}{}^\nu)$, which give rise to new $\Delta$-type interaction vertices appearing as insertions in counterterm diagrams (see Sec.~\ref{Sec:BMHV-Specific_Challenges}, Fig.~\ref{Fig:cAAA-3LCT1-Diagrams}).
Nevertheless, the number of such monomials and possible structures remains finite, constrained by ghost number~$1$ and mass dimension~$4$, thereby ensuring that the predictive power of the theory is maintained at any order in perturbation theory.

\section{BMHV-Specific Challenges}\label{Sec:BMHV-Specific_Challenges}

The application of the BMHV scheme within dimensional regularisation to chiral gauge theories introduces several technical challenges that substantially increase the computational complexity compared to vector-like gauge theories.
These complications stem from the regularisation-induced breaking of gauge and BRST symmetry and from the modified algebraic structure underlying this breaking, as explained in Sec.~\ref{Sec:Regularisation-Induced_Symmetry_Breaking}.
In what follows, we outline the most relevant of these difficulties, using the Abelian chiral gauge theory discussed in chapter~\ref{Chap:BMHV_at_Multi-Loop_Level} as an illustrative example, and contrast the situation with that in ordinary vector-like QED regularised in naive DReg.
We then discuss the implications for physical observables and clarify why the BMHV framework, despite its complexity, remains of practical relevance.
Following our presentation in Ref.~\cite{vonManteuffel:2025swv}, the two primary computational challenges in the BMHV scheme are:
\begin{enumerate}
    \item[$(i)$] \emph{Symmetry Breaking:} Ward and Slavnov-Taylor identities are violated and cannot be used in intermediates steps of the calculation.
    \item[$(ii)$] \emph{Proliferation of Lorentz Covariants:} More Lorentz covariants appear, involving both $D$- and $(D-4)$-dimensional components.
\end{enumerate}

First, the broken and subsequently to-be-restored Ward and Slavnov-Taylor identities imply the absence of symmetry relations among Green functions in the regularised theory, which therefore cannot be used to circumvent the explicit computation of multi-leg ($\geq3$) 1PI Green functions, as is typically done (see e.g.\ Refs.~\cite{Zoller:2015tha,Zoller:2016sgq,Chetyrkin:2017mwp,Herzog:2017ohr,Luthe:2017ttg,Davies:2019onf,Davies:2021mnc,Herren:2021vdk}).
Consequently, a complete renormalisation of a chiral gauge theory in the BMHV scheme at a given loop order requires the evaluation of all relevant 1PI Green functions, including those with up to five external legs (see chapters~\ref{Chap:General_Abelian_Chiral_Gauge_Theory} and \ref{Chap:The_Standard_Model}, as well as Refs.~\cite{Ebert:2024xpy,Kuhler:2025znv}).
In the Abelian chiral gauge theory considered in chapter~\ref{Chap:BMHV_at_Multi-Loop_Level}, Green functions with up to four external legs must be included.
This considerably increases the total number of amplitudes to be computed, posing a substantial challenge at higher orders and in theories with multiple types of fields.

The second complication concerns the algebraic tensor structure of the amplitudes --- specifically, the proliferation of Lorentz covariants.
The BMHV algebra necessarily introduces $4$- and $(D-4)$-dimensional components of Lorentz covariants in the numerators of the amplitudes, in addition to the manifestly $D$-dimensional ones.
Combined with the necessity of computing multi-leg ($\geq3$) 1PI Green functions, caused by the broken symmetry cf.~$(i)$, which generally possess more intricate Lorentz structures, this leads to significantly more involved tensors integrals and a larger number of independent Lorentz covariants.
This issue is apparent in all previous results published in Refs.~\cite{Belusca-Maito:2020ala,Belusca-Maito:2021lnk,Belusca-Maito:2023wah,Stockinger:2023ndm,Kuhler:2024fak,Ebert:2024xpy,Kuhler:2025znv,vonManteuffel:2025swv}, as well as in the results presented in this thesis (see chapters~\ref{Chap:General_Abelian_Chiral_Gauge_Theory}--\ref{Chap:The_Standard_Model}).
Our objective is to work exclusively with standard $D$-dimensional scalar Feynman integrals, which are subsequently processed via IBP reduction, as described in Sec.~\ref{Sec:IBP-Reduction}.
In ordinary vector-like QED, this can be accomplished straightforwardly by contracting all Lorentz indices using suitable projectors at the level of the Green functions.
In contrast, such contractions in the BMHV scheme can generate unwanted terms involving $4$- and $(D-4)$-dimensional loop momenta, yielding numerator polynomials of mixed dimensionality that obstruct the reduction to scalar integrals using standard IBP software tools.
This problem arises from contractions with non-$D$-dimensional indices.
To avoid this, we perform a fully $D$-dimensional tensor reduction of all tensor integrals, as explained in Sec.~\ref{Sec:Tensor_Reduction}, which requires a highly efficient implementation due to the occurrence of high tensor ranks.

An alternative strategy, not pursued in this thesis, is to treat the $(D-4)$-dimensional scalar products of the form $\widehat{k}_i\cdot\widehat{k}_j$ (commonly referred to as ``$\mu$-terms'') as additional numerator structures in the loop integrals.
However, such integrals lie outside the scope of standard IBP solvers and require specialised reduction techniques.
These methods have been formulated and applied at 1- and 2-loop order (see Refs.~\cite{Bern:1995db,Bern:2002tk,Heller:2020owb}), but no systematic generalisation to higher orders is currently available.
For this reason, we consistently employ a fully $D$-dimensional tensor reduction (Sec.~\ref{Sec:Tensor_Reduction}) to obtain strictly $D$-dimensional scalar integrals.

Overall, the renormalisation of chiral gauge theories within the BMHV framework requires substantially greater computational effort than that of vector-like gauge theories.
The complexity arises not only from the larger number of 1PI Green functions that must be computed but also from the increased algebraic intricacy of each individual amplitude.
In addition, the explicit breaking of gauge and BRST invariance necessitates the construction of finite symmetry-restoring counterterms, as described in Sec.~\ref{Sec:Symmetry_Restoration_Procedure}.
Although our method for deriving these counterterms is conceptually straightforward and algorithmically efficient, it nonetheless involves additional Green functions compared to vector-like cases.

\begin{figure}[t!]
    \centering
    \includegraphics[width=0.38\textwidth]{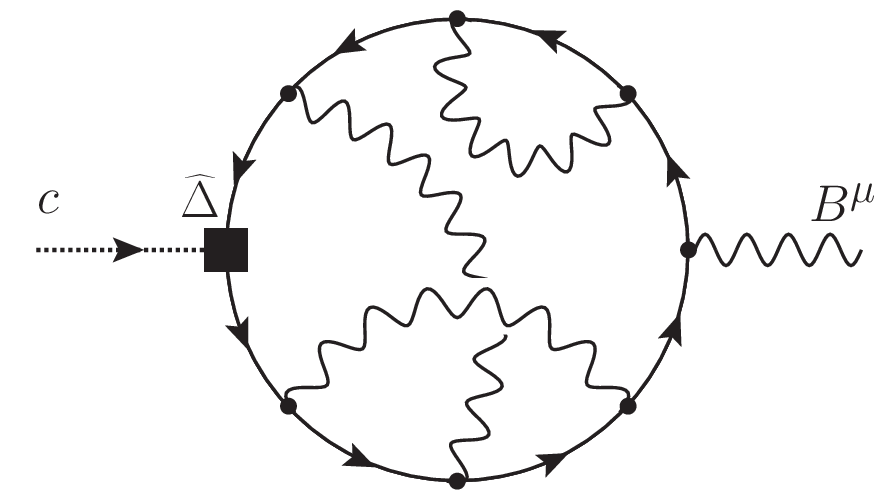}
    \hspace{3.5em}
    \includegraphics[width=0.38\textwidth]{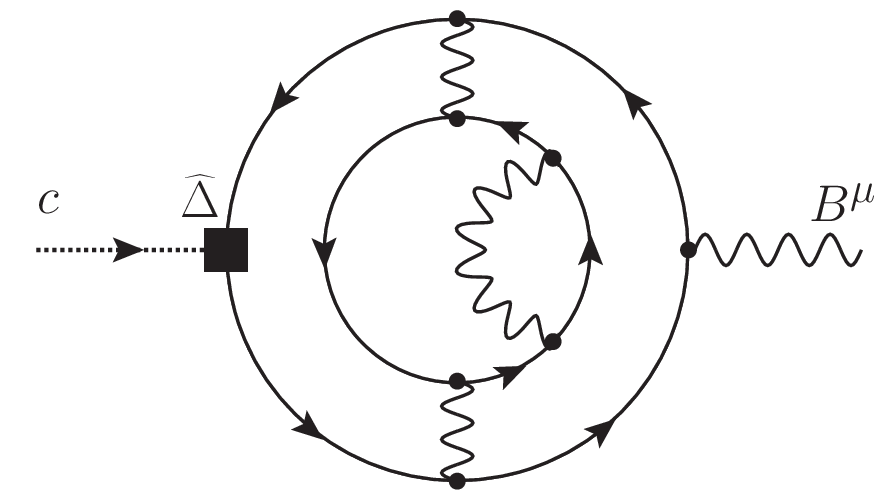}
    \caption{Representative 4-loop Feynman diagrams contributing to 
    the $\Delta$-operator-inserted 1PI Green function $i{\big(\Delta\cdot\Gamma\big)}{}_{B_{\mu}c}^{(4)}$.}
    \label{Fig:cA-4-Loop-Diagrams}
\end{figure}

\begin{figure}[t!]
    \centering
    \includegraphics[width=0.38\textwidth]{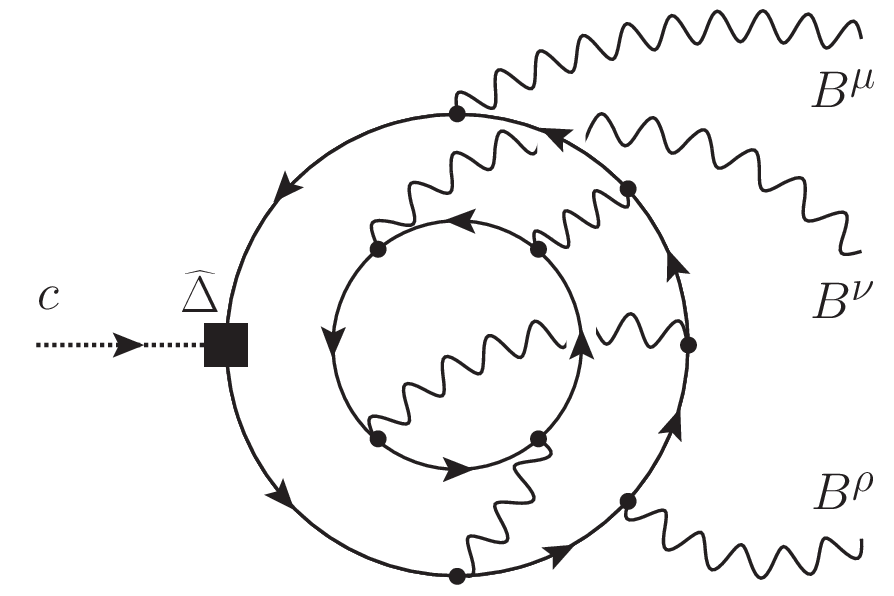}
    \hspace{3.5em}
    \includegraphics[width=0.44\textwidth]{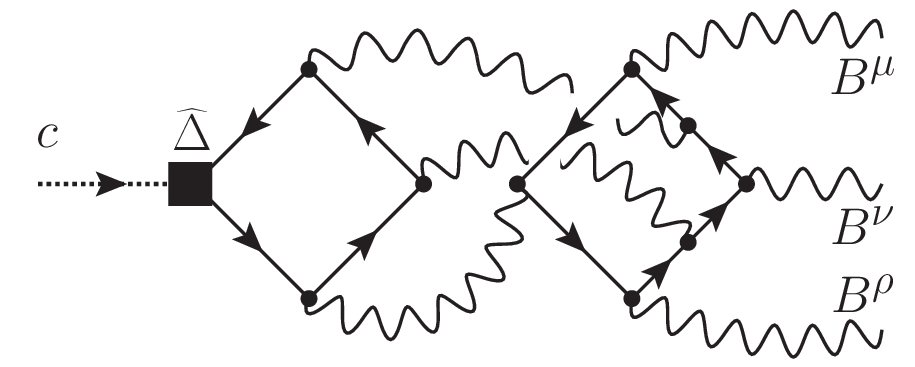}
    \caption{Representative 4-loop Feynman diagrams contributing to 
    the $\Delta$-operator-inserted 1PI Green function $i{\big(\Delta\cdot\Gamma\big)}{}_{B_{\rho}B_{\nu}B_{\mu}c}^{(4)}$.}
    \label{Fig:cAAA-4-Loop-Diagrams}
\end{figure}

To illustrate these challenges, we compare the renormalisation of ordinary vector-like QED with that of the BMHV-regularised Abelian chiral gauge theory introduced in Sec.~\ref{Sec:Right-Handed-Model-Definition}.
In vector-like QED, Ward identities relate all renormalisation constants to the gauge boson and fermion self energies, so that only these 2-point functions are required for a complete renormalisation, thereby drastically simplifying the calculation.
In contrast, the BMHV renormalisation of the Abelian chiral gauge theory requires the evaluation of nine different Green functions: five standard Green functions and four with an insertion of the local composite operator $\Delta$.
All relevant Green functions are listed in Sec.~\ref{Sec:Results-Right-Handed-Model}, including the 4-point functions $\langle B^{\sigma}B^{\rho}B^{\nu}B^{\mu} \rangle^{\mathrm{1PI}}$ and $\langle \Delta B^{\rho}B^{\nu}B^{\mu}c \rangle^{\mathrm{1PI}}$, as well as the unusual 2-point function $\langle \Delta B^{\mu}c \rangle^{\mathrm{1PI}}$.
The Green functions containing an insertion of the $\Delta$ operator are particularly computationally expensive.
Specifically, the Green function $\langle \Delta B^{\mu}c \rangle^{\mathrm{1PI}}$ has an overall degree of divergence of $\omega=3$, resulting in a large number of terms after tadpole decomposition (see Sec.~\ref{Sec:Tadpole_Decomposition}).
Two representative 4-loop diagrams contributing to this Green function are displayed in Fig.~\ref{Fig:cA-4-Loop-Diagrams}.
The Green function $\langle \Delta B^{\rho}B^{\nu}B^{\mu}c \rangle^{\mathrm{1PI}}$, while exhibiting a smaller degree of divergence of $\omega=1$, generates a significantly larger number of Feynman diagrams; two 4-loop examples are shown in Fig.~\ref{Fig:cAAA-4-Loop-Diagrams}.
At the 4-loop level, both of these Green functions give rise to tensor integrals of rank~$12$.

\begin{figure}[t!]
    \centering
    \includegraphics[width=0.38\textwidth]{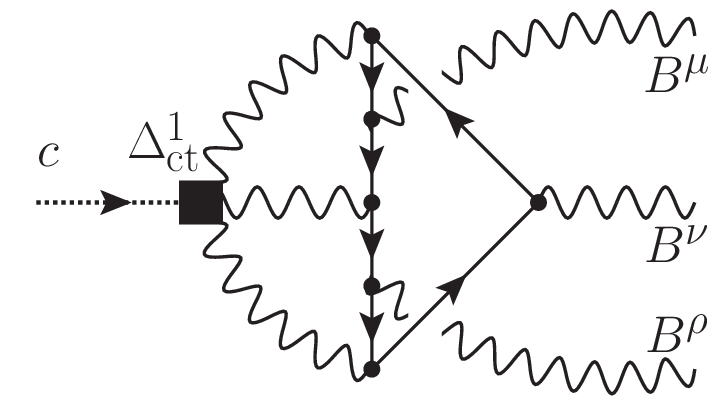}
    \hspace{3.5em}
    \includegraphics[width=0.38\textwidth]{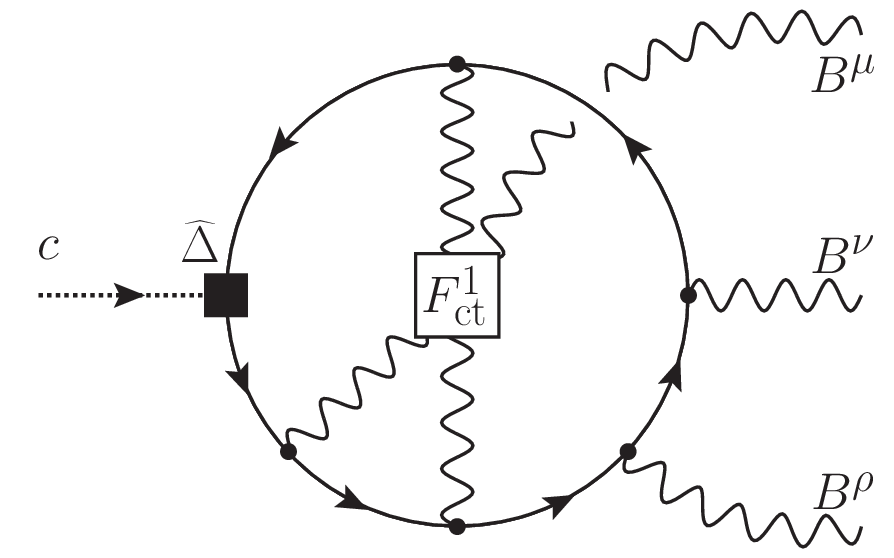}
    \caption{Representative 3-loop Feynman diagrams with counterterm insertions contributing to 
    the 4-loop, subrenormalised, $\Delta$-operator-inserted 1PI Green function $i{\big(\Delta\cdot\Gamma\big)}{}_{B_{\rho}B_{\nu}B_{\mu}c}^{(4)}$. The left diagram features the insertion of a new $\Delta$-vertex resulting from a 1-loop correction to the $\Delta$-operator, while the right diagram involves the insertion of a 1-loop, finite, symmetry-restoring counterterm.}
    \label{Fig:cAAA-3LCT1-Diagrams}
\end{figure}

It is worth emphasising that the number of genuine $L$-loop diagrams
in standard Green functions of order~$L$ is identical to that in other regularisation schemes.
This is because the BMHV scheme does not introduce additional tree-level vertices in the standard Lagrangian.
The only new tree-level interaction vertex is the $\widehat{\Delta}$-vertex (cf.\ Eq.~\eqref{Eq:Delta-Tree-Level-Interaction-Vertex}), which does not appear in the standard Lagrangian but only in the operator-inserted Green functions.
However, another source of computational complexity in the BMHV scheme arises from the increased number of $(<\!L)$-loop counterterm-inserted diagrams.
These additional diagrams originate from two sources: first, symmetry-restoring counterterms required at each order, and second, higher-order corrections to the breaking operator $\Delta$, which lead to counterterm diagrams containing insertions of these higher-order contributions.
For example, in the 4-loop calculation of the Green function $\langle \Delta B^{\rho}B^{\nu}B^{\mu}c \rangle^{\mathrm{1PI}}$, all $(<\!4)$-loop diagrams with such counterterm insertions must be taken into account.
Fig.~\ref{Fig:cAAA-3LCT1-Diagrams} illustrates two representative examples (both with 3 loops): the left diagram incorporates a 1-loop correction to the $\Delta$-operator, while the right diagram features the insertion of a 1-loop symmetry-restoring counterterm.
Neither of these counterterm vertices has a tree-level analogue.
Their existence highlights the increased structural complexity introduced by the symmetry-restoration required in the BMHV scheme --- affecting not only the operator-inserted Green functions but, through the symmetry-restoring counterterms, also the standard Green functions.

Despite these additional complications, the BMHV scheme remains of practical relevance not only due to its theoretical consistency but also because the complete renormalisation procedure, including symmetry restoration, must be performed only once.
Subsequently, physical observables can be computed in the usual way, as in any other regularisation scheme, provided that the BMHV algebra is implemented consistently.
Neither the number of Green functions required for the evaluation of an observable nor the number of their corresponding genuine $L$-loop diagrams increases.
Moreover, the $\Delta$-vertices play no role in the computation of observables, as they are solely required during the symmetry-restoration procedure.
Although the total number of counterterms is larger than in vector-like theories, these remain fully manageable: as shown in Sec.~\ref{Sec:Results-Right-Handed-Model}, all required counterterms can be expressed in a compact form suitable for computer implementation, even at the 4-loop level.
In summary, the BMHV scheme is not only fully self-consistent but also practically applicable and well suited for higher-order computations in chiral gauge theories.

\chapter{Abelian Chiral Gauge Theories in $D$ Dimensions}\label{Chap:General_Abelian_Chiral_Gauge_Theory}

In this chapter, we investigate the renormalisation of a massless Abelian chiral gauge theory with left- and right-handed fermions as well as complex scalar fields.
The central objective is to examine a broad class of possible realisations of the BMHV implementation and to analyse the dimensional ambiguities that arise from the non-uniqueness of the $D$-dimensional extension, as discussed in Sec.~\ref{Sec:Dimensional_Ambiguities_and_Evanescent_Shadows}.
We explore how these ambiguities affect the general structure of the symmetry breaking that appears in intermediate steps of the calculation and, consequently, how they influence the construction of the symmetry-restoring counterterms.
To this end, we omit non-Abelian complications and focus on general $D$-dimensional formulations of the fermionic sector, including evanescent gauge interactions and different continuations of the fermions, as outlined in Sec.~\ref{Sec:Dimensional_Ambiguities_and_Evanescent_Shadows}.
In this way, the analysis encompasses all realisations of the BMHV scheme discussed in the literature as special cases and further generalisations.
The following discussion largely follows the presentation we published in Ref.~\cite{Ebert:2024xpy}.

In Sec.~\ref{Sec:Definition_of_the_Theory_General_Abelian_Case}, we define the Abelian chiral gauge theory under consideration and provide a general $D$-dimensional extension of the underlying Lagrangian that accommodates several different realisations of the BMHV framework.
A comparison with approaches from the literature and a discussion of how they can be recovered by specialising our general setup are provided in Sec.~\ref{Sec:Special_Cases_of_general_Abelian_Model}.
Section~\ref{Sec:STIs-in-Abelian-Chiral-Gauge-Theory} provides a comprehensive list of the Slavnov-Taylor identities together with their corresponding breakings, derived from Eq.~\eqref{Eq:QAPinDReg_Symmetry_Restoration}, which govern the symmetry restoration procedure.
The complete 1-loop renormalisation of the theory, including both divergent and finite symmetry-restoring counterterms, is presented in Sec.~\ref{Sec:One-Loop-Renormalisation-Abelian-Chiral-Gauge-Theory}. 
Finally, in Sec.~\ref{Sec:Results-Shedding_Light_on_Evanescent_Shadows}, we analyse several representative special cases.
In particular, for left- and right-handed Standard Model hypercharges --- thus modelling the Abelian sector of the SM --- we investigate how different dimensional extensions of the fermionic action affect the counterterm structure, thereby explicitly revealing the dependence on evanescent aspects of the dimensional continuation.

\section{Definition of the Theory}\label{Sec:Definition_of_the_Theory_General_Abelian_Case}

We have already discussed the different realisations of the fermionic sector in Sec.~\ref{Sec:Dimensional_Ambiguities_and_Evanescent_Shadows}, where different options for the $D$-dimensional continuation of the fermions and their gauge interactions are presented.
The key feature of the model defined here is that it represents a general extension accommodating all these possibilities, thereby enabling the study of a broad class of $D$-dimensional realisations of the regularised theory. 
The corresponding $D$-dimensional classical Lagrangian may be written as
\begin{equation}\label{Eq:Model-Lagrangian-General-Abelian-Theory}
    \begin{aligned}
        \mathcal{L} = 
        \mathcal{L}_{\mathrm{fermion}} + \mathcal{L}_{\mathrm{gauge}}
        + \mathcal{L}_{\mathrm{scalar}} + \mathcal{L}_{\mathrm{Yukawa}}
        + \mathcal{L}_{\mathrm{ghost+fix}} + \mathcal{L}_{\mathrm{ext}},
    \end{aligned}
\end{equation}
which serves as the starting point of our analysis.
The $4$-dimensional theory is recovered as usual in the limit
\begin{equation}\label{Eq:4D-Limit-for-Model-Lagrangian}
    \begin{aligned}
        \mathop{\mathrm{LIM}}_{D \, \to \, 4} \mathcal{L} = \mathcal{L}^{(4D)}.
    \end{aligned}
\end{equation}
In the following, we describe in detail each contribution to the Lagrangian in Eq.~\eqref{Eq:Model-Lagrangian-General-Abelian-Theory}, thereby defining the full structure of the model used in this analysis.

\paragraph{Physical Field Content and Basic Symmetries:}
The theory under consideration is a massless Abelian chiral gauge theory based on a $U(1)$ (hypercharge) gauge group, associated with the gauge field $B^\mu$.
Its fermionic sector consists of a set of Dirac fermions $\psi_i$, each of which can be decomposed into chiral components according to $\psi_i={\psi_{L}}_i+{\psi_{R}}_i$.
Depending on the particular realisation, both chirality components of a given fermion may represent physical degrees of freedom, or one of them may act as fictitious, non-interacting sterile partner field (see Def.~\ref{Def:Sterile_Quantum_Field}).
The gauge interactions are governed by real and diagonal hypercharge matrices $\hypL{}_{ij}$ and $\hypR{}_{ij}$, which determine the couplings of the left- and right-handed fermions, respectively, to the $U(1)$ gauge boson.
In addition to the fermionic content, the model includes a set of complex scalar fields $\phi_a$ carrying a universal, real-valued hypercharge $\hypS$.
Beyond the local $U(1)$ gauge symmetry, we impose two additional global symmetries: a $U(1)_Q$ symmetry associated with the electromagnetic charge $Q$, and a $SU(3)_c$ colour symmetry.
Both of these are diagonal in the flavour space and implemented in a vector-like manner, meaning that for each fermion species $\psi_i$, the electromagnetic and colour charges of the left- and right-handed components coincide.
Furthermore, we restrict the Yukawa sector to interactions that conserve fermion number.

This setup can reproduce the structure of the Abelian sector of the Standard Model and of SM-like extensions such as the two-Higgs doublet model (2HDM), albeit with the electromagnetic symmetry $U(1)_Q$ and the colour symmetry $SU(3)_c$ treated as purely global rather than gauged.

\paragraph{Fermion Kinetic and Gauge Boson Interaction Terms:} 
To accommodate the most general class of $D$-dimensional realisations of the BMHV regularisation --- including all options discussed in Sec.~\ref{Sec:Dimensional_Ambiguities_and_Evanescent_Shadows} --- we adopt the generalised extension of the fermionic Lagrangian introduced there and employ the Ansatz provided in Eq.~\eqref{Eq:L_fermion_general_D-dim_Ansatz}.
Accordingly, the fermionic part of the $D$-dimensional Lagrangian is defined as
\begin{equation}
  \begin{aligned}
  \label{Eq:L_Fermion_General_Abelian_Chiral_Gauge_Theory}
        \mathcal{L}_{\mathrm{fermion}} = \overline{\psi}_j i \slashed{\partial} \psi_j 
        &- g {\hypR}_{ij} \overline{\psi}_i \projL \overline{\slashed{B}} \projR \psi_j
        - g {\hypL}_{ij} \overline{\psi}_i \projR \overline{\slashed{B}} \projL \psi_j\\
        &- g {\hypRL}_{ij} \overline{\psi}_i \projL \widehat{\slashed{B}} \projL \psi_j
        - g {\hypLR}_{ij} \overline{\psi}_i \projR \widehat{\slashed{B}} \projR \psi_j,
\end{aligned}
\end{equation}
where the 4-dimensional part can be expressed through the covariant derivative
\begin{equation}
    \begin{aligned}
        &\overline{D}_{\mu} \psi = \big(\overline{\partial}_{\mu} + i g \hypL
      \overline{B}_{\mu} \projL + i g \hypR \overline{B}_{\mu} \projR\big) \psi,
    \end{aligned}
\end{equation}
with the 4-dimensional gauge interactions governed by the physical hypercharges $\hypL$ and $\hypR$.

The details of this extension have already been discussed in the last paragraph of Sec.~\ref{Sec:Dimensional_Ambiguities_and_Evanescent_Shadows}; here, we briefly summarise the essential points.
The physical hypercharge matrices $\hypL$ and $\hypR$ are real and diagonal, and coincide with their counterparts in 4 dimensions, ensuring that the 4-dimensional part of the $D$-dimensional action manifestly preserves BRST invariance.
In contrast, the hypercharge matrices $\hypLR$ and $\hypRL$, which govern the evanescent gauge interactions appearing in the last line of Eq.~\eqref{Eq:L_Fermion_General_Abelian_Chiral_Gauge_Theory}, are not fixed by the physical hypercharges.
They may in general be independent and can potentially contain off-diagonal elements.
The evanescent hypercharges are only constrained by hermiticity: $\hypRL = \hypLR^{\dagger}$.
In order to preserve the global symmetries $U(1)_Q$ and $SU(3)_c$ introduced earlier, we further impose that $\hypLR$ and $\hypRL$ commute with the electric charge $Q$ and respect colour charge conservation.
Any additional assumptions or restrictions on these evanescent hypercharges $\hypLR$ and $\hypRL$ will be specified in the sections where explicit results are presented, particularly in Sec.~\ref{Sec:One-Loop-Renormalisation-Abelian-Chiral-Gauge-Theory} and \ref{Sec:Results-Shedding_Light_on_Evanescent_Shadows}.
These evanescent gauge interaction terms present a key novelty of our model, which we published in Ref.~\cite{Ebert:2024xpy}, and extend beyond previous studies in the literature (cf.\ Sec.~\ref{Sec:Special_Cases_of_general_Abelian_Model}).

The physical hypercharge matrices $\hypL$ and $\hypR$ are further constrained by the anomaly cancellation condition  
\begin{equation}\label{Eq:AnomalyCancellationCondition-GeneralAbelianTheory}
    \begin{aligned}
        \mathrm{Tr}\big(\hypR^3\big) - \mathrm{Tr}\big(\hypL^3\big) = 0,
    \end{aligned}
\end{equation}
which is assumed to hold as part of the definition of the theory (see Sec.~\ref{Sec:Anomalies}, Eq.~\eqref{Eq:Abelian-Anomaly-Cancellation-Condition}).
Finally, sterile fermionic components $\psi^{\mathrm{st}}_{X,k}$ ($X=L,R$), as defined in Def.~\ref{Def:Sterile_Quantum_Field}, are characterised by the complete absence of interactions and higher-order corrections.
Accordingly, all elements of the hypercharge matrices --- both physical and evanescent --- that correspond to sterile fields must vanish identically.

\paragraph{Gauge Boson, Scalar and Yukawa Terms:}
The remaining contributions to the Lagrangian are straightforward and do not introduce any conceptual difficulties.
The kinetic term for the Abelian gauge field follows its standard form,
\begin{equation}
    \begin{aligned}
        \mathcal{L}_{\mathrm{gauge}} = - \frac{1}{4} F^{\mu\nu}F_{\mu\nu},
    \end{aligned}
\end{equation}
where the field strength tensor is given by $F_{\mu\nu} = \partial_{\mu}B_{\nu} -
\partial_{\nu}B_{\mu}$, continued to $D$ dimensions in the natural way.

In the scalar sector, we introduce a set of complex scalar fields $\phi_a$, which are eigenstates of electric charge $Q$,\footnote{
Real scalars are not eigenstates of $Q$ and, in general, also not of the hypercharge.} 
and all carry the same real-valued hypercharge $\mathcal{Y}_S$.
The corresponding 4-dimensional covariant derivative reads
\begin{align}
  \overline{D}_{\mu}\phi_{a} = \big(\overline{\partial}_{\mu} 
  + i g \hypS \overline{B}_{\mu}\big)\phi_a, 
\end{align}
and the natural $D$-dimensional extension of the scalar Lagrangian, containing the kinetic, gauge interaction, and potential terms, takes the form
\begin{equation}
    \begin{aligned}
        \mathcal{L}_{\mathrm{scalar}} &=        
        (\partial_{\mu}\phi_a^{\dagger})(\partial^{\mu}\phi_a)
        - ig\hypS B^{\mu} \big( \phi_a^{\dagger} \partial_{\mu}\phi_a -
        \phi_a \partial_{\mu} \phi_a^{\dagger} \big)
        + g^2 \hypS^2 B^{\mu}B_{\mu}\phi_a^{\dagger}\phi_a\\
        &- \frac{\lambda_{klmn}}{6} \phi^{\dagger}_k \phi_l \phi^{\dagger}_m \phi_n .
    \end{aligned}
\end{equation}
Hermiticity requires the potential parameters to satisfy $\lambda_{klmn} = \lambda_{nmlk}^{*}$, and non-vanishing couplings are only allowed if the electric charge is conserved,
\begin{equation}
    \begin{aligned}
        \lambda_{klmn} \equiv 0, \quad \text{unless} \quad -Q_{\phi_k}+Q_{\phi_l}-Q_{\phi_m}+Q_{\phi_n}=0.
    \end{aligned}
\end{equation}
Furthermore, the coupling constants are symmetric under index permutations, $\lambda_{klmn}=\lambda_{mlkn}=\lambda_{knml}=\lambda_{mnkl}$.

Fermion number is assumed to be conserved perturbatively, ensuring that a well-defined fermion flow can be assigned to all Feynman diagrams.
As a consequence, Yukawa interactions involving charge-conjugated fermions $\psi^C_i$ --- which were permitted in the purely right-handed model of Ref.~\cite{Belusca-Maito:2020ala} --- are excluded.
Only Yukawa terms built from the fundamental fermion fields $\psi_i$ are allowed, leading to the interaction Lagrangian
\begin{equation}
    \begin{aligned}
        \mathcal{L}_{\mathrm{Yukawa}} = 
        - G^{a}_{ij} {\overline{\psi_L}}_i \phi_{a} {\psi_R}_j
        - K^{a}_{ij} {\overline{\psi_L}}_i \phi_{a}^{\dagger} {\psi_R}_j
        + \mathrm{h.c.}
    \end{aligned}
\end{equation}
The Yukawa matrices $G^{a}_{ij}$ and $K^{a}_{ij}$ are restricted by several conservation laws.
Global hypercharge conservation and BRST invariance impose
\begin{equation}\label{Eq:YukawaHyperchargeBRSTCondition}
    \begin{aligned}
        {\hypL}_{ik} G^{a}_{kj} - G^{a}_{ik} {\hypR}_{kj} - \hypS G^{a}_{ij} &= 0,\\
        {\hypL}_{ik} K^{a}_{kj} - K^{a}_{ik} {\hypR}_{kj} + \hypS K^{a}_{ij} &= 0,
    \end{aligned}
\end{equation}
while electric charge conservation requires
\begin{equation}
    \begin{aligned}
        G^{a}_{ij} &\equiv 0, \quad \text{unless} \quad -Q_{{\psi_L}_i}+Q_{\phi_a}+Q_{{\psi_R}_j} = 0,\\
        K^{a}_{ij} &\equiv 0, \quad \text{unless} \quad -Q_{{\psi_L}_i}-Q_{\phi_a}+Q_{{\psi_R}_j} = 0,
    \end{aligned}
\end{equation}
and the conservation of colour charge must hold in addition.

\paragraph{Ghost, Gauge Fixing and External Field Lagrangian:}
The gauge fixing and ghost contributions to the Lagrangian can be extended straightforwardly to $D$ dimensions without complications:
\begin{equation}\label{Eq:Ddim-LGaugeFixing+Ghost-GeneralAbelianTheory}
    \begin{aligned}
        \mathcal{L}_{\mathrm{ghost+fix}} = 
        - \frac{1}{2\xi} (\partial^{\mu}B_{\mu})^2
        - \overline{c} \Box c.
    \end{aligned}
\end{equation}
In our framework of renormalising chiral gauge theories, working with the Slavnov-Taylor identity and the quantum action principle, the BRST transformations of the fields are coupled to corresponding external sources (cf.\ Sec.~\ref{Sec:BRST-Symmetry}).
We therefore introduce the set of sources $\{\rho^{\mu},R_i,\Upsilon^a,\zeta\}$, which are incorporated through the Lagrangian term
\begin{equation}\label{Eq:Ddim-Lext-GeneralAbelianTheory}
    \begin{aligned}
        \mathcal{L}_{\mathrm{ext}} = 
        \rho^{\mu}s_DB_{\mu} 
        + {\overline{R}}{}^{i}s_D{\psi}_i + (s_D\overline{\psi}_i) R^{i}
        + {\Upsilon^{a}}^{\dagger}s_D\phi_{a} + \Upsilon^{a}s_D\phi_{a}^{\dagger}
        + \zeta s_Dc .
    \end{aligned}
\end{equation}

The corresponding BRST transformations, extended in a straightforward way to $D$ dimensions, are provided by 
\begin{equation}\label{Eq:Ddim-BRSTTrafos-GeneralAbelianTheory}
    \begin{aligned}
        s_DB_{\mu} &= \partial_{\mu}c,\\
        s_D\psi_i &= s_D{\psi_R}_i + s_D{\psi_L}_i = -igc\Big({\hypR}_{ij}{\psi_{R}}_j+{\hypL}_{ij}{\psi_{L}}_j\Big),\\
        s_D\overline{\psi}_i &= s_D{\overline{\psi_R}}_i + s_D{\overline{\psi_L}}_i = igc\Big({\overline{\psi_{R}}}_j{\hypR}_{ji}+{\overline{\psi_{L}}}_j{\hypL}_{ji}\Big),\\
        s_D\phi_a &= -igc \hypS \phi_a,\\
        s_Dc&=0,\\
        s_D\overline{c}&=\mathcal{B}= - \frac{1}{\xi} \partial^{\mu}B_{\mu},\\
        s_D\mathcal{B}&=0,
    \end{aligned}
\end{equation}
where $\mathcal{B}$ denotes the Nakanishi-Lautrup auxiliary field, which should not be confused with the gauge field $B_{\mu}$.

The tree-level breaking of gauge and BRST invariance in the present model, induced by the BMHV regularisation, coincides precisely with that given in Eq.~\eqref{Eq:GeneralTreeLevelBreaking_Abelian}, as shown and discussed in Sec.~\ref{Sec:Dimensional_Ambiguities_and_Evanescent_Shadows}.

\section{Special Cases}\label{Sec:Special_Cases_of_general_Abelian_Model}

Our general model, introduced in the previous section and defined by the tree-level Lagrangian in Eq.~\eqref{Eq:Model-Lagrangian-General-Abelian-Theory}, provides a versatile framework that unifies and extends several other approaches.
Before proceeding to its renormalisation, we first illustrate its flexibility by demonstrating how it can be specialised to concrete models of interest.

The realisation of a specific model depends on the particular field content, the detailed arrangement, and the chosen symmetries and conservation laws that constrain the allowed interactions.
In particular, the fermion multiplets $\psi_L$ and $\psi_R$ of the general model may be populated with physical and sterile fields, depending on the situation and the chosen realisation.
Additionally, an arbitrary number $N_S$ of complex scalar fields $\phi_a$ can be incorporated, although their inclusion is optional.
Table~\ref{Tab:SpecialCasesOverview} summarises the corresponding parameter choices required to recover the most relevant cases.
\begin{table}[h!]
    \centering
    \begin{tabular}{|c||c|c|c|c|c|c|c|c|} \hline 
        Theory & $\hypR$ & $\hypL$ & $\hypLR$ & $\hypRL$ & $\hypS$ & $\lambda$ & $G$ & $K$\\ \hline \hline
        QED & $Q$ & $Q$ & $Q$ & $Q$ & $0$ & $0$ & $0$ & $0$\\ \hline
        $\chi$QED \cite{Belusca-Maito:2021lnk,Belusca-Maito:2023wah,Stockinger:2023ndm,vonManteuffel:2025swv} & $\hypR$ & $0$ & $0$ & $0$ & $0$ & $0$ & $0$ & $0$\\ \hline
        Martin et al. \cite{Martin:1999cc} & (\ref{Eq:CPMartin-YLYR}) & (\ref{Eq:CPMartin-YLYR}) & $0$ & $0$ & $0$ & $0$ & $0$ & $0$\\ \hline
        Cornella et al. \cite{Cornella:2022hkc} & $T_R^a$ & $T_L^a$ & $0$ & $0$ & $0$ & $0$ & $0$ & $0$\\ \hline
        ASM & (\ref{Eq:ASSM-YLYR}) & (\ref{Eq:ASSM-YLYR}) & (\ref{Eq:ASSM-YLR}) & (\ref{Eq:ASSM-YLR}) & (\ref{Eq:ASSM-YS}) & (\ref{Eq:ASSM-lambda}) & (\ref{Eq:ASSM-Yuk-G}) & (\ref{Eq:ASSM-Yuk-K})\\ \hline
        A2HDM & (\ref{Eq:ASTHDM-YLYR}) & (\ref{Eq:ASTHDM-YLYR}) & (\ref{Eq:ASTHDM-YLR}) & (\ref{Eq:ASTHDM-YLR}) & (\ref{Eq:ASTHDM-YS}) & (\ref{Eq:ASTHDM-lambda}) & (\ref{Eq:ASTHDM-Yuk-G}) & (\ref{Eq:ASTHDM-Yuk-K})\\ \hline
    \end{tabular}
    \caption{Special cases of the Abelian chiral gauge theory defined in Sec.~\ref{Sec:Definition_of_the_Theory_General_Abelian_Case}.}
    \label{Tab:SpecialCasesOverview}
\end{table}

A brief overview of models discussed in the literature is provided in Sec.~\ref{Sec:ModelsFromLiterature}, followed by a detailed analysis of two specific realisations of particular phenomenological relevance.
The first corresponds to the Abelian sector of the Standard Model (ASM), discussed in Sec.~\ref{Sec:ASSM} (see the fifth row of Tab.~\ref{Tab:SpecialCasesOverview}), while the second concerns the Abelian sector of the two-Higgs doublet model (A2HDM), analysed in Sec.~\ref{Sec:ASTHDM} (see the sixth row of Tab.~\ref{Tab:SpecialCasesOverview}).
Although only the Abelian subsectors of these theories are considered here, they already provide valuable case studies of the renormalisation behaviour of chiral gauge theories in the BMHV framework, focusing on different dimensional realisations while avoiding additional non-Abelian complications.

\subsection{Selected Models from the Literature}\label{Sec:ModelsFromLiterature}

We begin with the first row of Tab.~\ref{Tab:SpecialCasesOverview}: a vector-like Abelian gauge theory such as QED can be recovered from our general setup by identifying all hypercharge matrices as equal, i.e.\ $\hypL=\hypR=\hypLR=\hypRL=Q$.
The inclusion of the evanescent gauge interactions governed by $\hypLR$ and $\hypRL$ thus makes it possible to embed the standard DReg treatment of QED (see Eq.~\eqref{Eq:LeegammaQED}) as a special case of our general model.
To obtain the pure QED limit, all Yukawa and scalar couplings are set to zero, as indicated in Tab.~\ref{Tab:SpecialCasesOverview}.

The purely right-handed Abelian models analysed in Refs.~\cite{Belusca-Maito:2021lnk,Belusca-Maito:2023wah,Stockinger:2023ndm,vonManteuffel:2025swv} --- here collectively referred to as ``chiral QED'' --- are recovered by populating only the right-handed multiplet $\psi_R$ with physical fermions, while filling the left-handed multiplet $\psi_L$ exclusively with sterile fields.
In the terminology of Sec.~\ref{Sec:Dimensional_Ambiguities_and_Evanescent_Shadows}, this corresponds to treating all fermions according to Option~\ref{Opt:Option2b}.
Since these models do not admit evanescent gauge interactions, we impose $\hypLR = \hypRL = 0$.
Moreover, all scalar couplings are omitted in this configuration.
The resulting setup corresponds to the second row of Tab.~\ref{Tab:SpecialCasesOverview}.

Moreover, our general framework also reproduces the Abelian special case of the chiral gauge theories analysed in Refs.~\cite{Martin:1999cc} and \cite{Cornella:2022hkc}.
Both of these publications carried out a 1-loop renormalisation of chiral theories with a compact, simple gauge group using the BMHV scheme and without scalar fields.
Our setup (see Sec.~\ref{Sec:Definition_of_the_Theory_General_Abelian_Case}) replicates their models when the considered gauge group is simply $U(1)$.

In the case of Ref.~\cite{Martin:1999cc}, the model can be realised within our framework by arranging the fermion multiplet $\psi$ in block structure form, so that only one half of the left-handed component $\psi_L$ and the opposite half of the right-handed component $\psi_R$ are physical, while their respective chiral counterparts are sterile.
This configuration corresponds to Option~\ref{Opt:Option2} introduced in Sec.~\ref{Sec:Dimensional_Ambiguities_and_Evanescent_Shadows}.
The associated coupling matrices share the same block structure and take the form
\begin{equation}\label{Eq:CPMartin-YLYR}
    \begin{aligned}
            \hypR =
            \begin{pmatrix}
                0 & 0\\
                0 & T_R^a
            \end{pmatrix},
            \quad
            \hypL =
            \begin{pmatrix}
                T_L^a & 0\\
                0 & 0
            \end{pmatrix}.
    \end{aligned}
\end{equation}
No evanescent gauge couplings are included in this model, and thus $\hypRL=\hypLR=0$.
Furthermore, the theory does not contain any scalar fields, so that all Yukawa and scalar couplings are set to zero (see row three of Tab.~\ref{Tab:SpecialCasesOverview}).

In contrast, the formulation of Ref.~\cite{Cornella:2022hkc} follows the spirit of Option~\ref{Opt:Option1}, where physical left- and right-handed fermions are combined into Dirac spinors whenever possible.
In this scenario, our general model reproduces their Abelian limit without requiring a block structure for the fermion multiplets.
As before, the evanescent gauge interactions are absent, implying $\hypRL=\hypLR=0$, and all couplings involving scalars vanish.
The corresponding configuration is summarised in row four of Tab.~\ref{Tab:SpecialCasesOverview}.

More recently, Ref.~\cite{OlgosoRuiz:2024dzq} extended the setup of Ref.~\cite{Cornella:2022hkc} by including a scalar sector and by considering the presence of evanescent gauge interactions.
However, the authors considered only a restricted form of these interactions, which in our notation corresponds to
\begin{equation}
    \begin{aligned}
        \hypLR = \frac{c \hypR + c^{*} \hypL}{2}, \qquad \hypRL = \frac{c \hypL + c^{*} \hypR}{2},
    \end{aligned}
\end{equation}
where $c$ is a complex constant.
Moreover, in their explicit 1-loop computations, they set $c=0$, thereby eliminating the evanescent contributions from the analysis altogether.

To the best of our knowledge, a systematic analysis of the impact of evanescent gauge interactions on the counterterm structure within the BMHV scheme has first been carried out in our publication~\cite{Ebert:2024xpy}.
In the following subsections, we present two phenomenologically motivated examples --- the Abelian sectors of the Standard Model and the two-Higgs doublet model --- and investigate the effects of evanescent gauge interactions, focusing in particular on the Standard Model case, in Sec.~\ref{Sec:Results-Shedding_Light_on_Evanescent_Shadows}.

\subsection{Abelian Sector of the Standard Model}\label{Sec:ASSM}

The Abelian sector of the Standard Model arises as a particular realisation of the general framework introduced in Sec.~\ref{Sec:Definition_of_the_Theory_General_Abelian_Case}. 
For simplicity, we restrict the discussion to a single fermion generation.
In the following, we analyse this model within both Option~\ref{Opt:Option1} and Option~\ref{Opt:Option2}, as defined in Sec.~\ref{Sec:Dimensional_Ambiguities_and_Evanescent_Shadows}, regarding the treatment of fermions in $D$ dimensions.

\paragraph{ASM with Fermion Multiplets according to Option~\ref{Opt:Option1}:}
In this case, the fermion multiplets are explicitly given by 
\begin{equation}\label{Eq:ASSM-FermionMultiplets}
    \begin{aligned}
        \psi_L &= 
        \begin{pmatrix}
            \nu_L, & e_L, & u^r_L, & u^g_L, & u^b_L, & d^r_L, & d^g_L, & d^b_L
        \end{pmatrix}^\mathsf{T},\\
        \psi_R &= 
        \begin{pmatrix}
            \nu_R^{\mathrm{st}}, & e_R, & u^r_R, & u^g_R, & u^b_R, & d^r_R, & d^g_R, & d^b_R
        \end{pmatrix}^\mathsf{T},
    \end{aligned}
\end{equation}
where only a sterile right-handed neutrino $\nu_R^{\mathrm{st}}$ is introduced.
Each pair $({\psi_L}_i,{\psi_R}_i)$ forms a single Dirac spinor $\psi_i$, combining physical fermions whenever possible, as proposed in Option~\ref{Opt:Option1}.
For example, the electron field is represented by a single Dirac spinor $e$ that contains both physical components $e_L$ and $e_R$.
The hypercharge matrices corresponding to this fermionic matter content read
\begin{equation}\label{Eq:ASSM-YLYR}
    \begin{aligned}
        \hypR =
        \begin{pmatrix}
            0 & 0 & 0 & 0\\
            0 & -1 & 0 & 0\\
            0 & 0 & \frac{2}{3} \mathbb{1}_{3\times3} & 0\\
            0 & 0 & 0 & -\frac{1}{3} \mathbb{1}_{3\times3}
        \end{pmatrix},
        \quad
        \hypL =
        \begin{pmatrix}
            -\frac{1}{2} & 0 & 0 & 0\\
            0 & -\frac{1}{2} & 0 & 0\\
            0 & 0 & \frac{1}{6} \mathbb{1}_{3\times3} & 0\\
            0 & 0 & 0 & \frac{1}{6} \mathbb{1}_{3\times3}
        \end{pmatrix},
    \end{aligned}
\end{equation}
and reproduce the familiar hypercharge assignments of the Standard Model fermions.
These matrices satisfy the anomaly cancellation condition of Eq.~\eqref{Eq:AnomalyCancellationCondition-GeneralAbelianTheory}. 

For a single fermion generation with multiplets as given in Eq.~\eqref{Eq:ASSM-FermionMultiplets}, the evanescent hypercharges $\hypLR$ and $\hypRL$ must be diagonal to ensure conservation of electric and colour charge.
Choosing them to be real and imposing hermiticity (Eq.~\eqref{Eq:Requirement-YLR-YRL}) implies $\hypLR=\hypRL$, which can be expressed as
\begin{equation}\label{Eq:ASSM-YLR}
    \begin{aligned}
        \hypLR = \hypRL =
        \begin{pmatrix}
            0 & 0 & 0 & 0\\
            0 & \mathcal{Y}_{LR}^{e} & 0 & 0\\
            0 & 0 & \mathcal{Y}_{LR}^{u} \mathbb{1}_{3\times3} & 0\\
            0 & 0 & 0 & \mathcal{Y}_{LR}^{d} \mathbb{1}_{3\times3}
        \end{pmatrix}.
    \end{aligned}
\end{equation}
The individual diagonal entries $\mathcal{Y}_{LR}^{e,u,d}$ remain free parameters, corresponding to different $D$-dimensional extensions of the theory.
The first entry must vanish because the right-handed neutrino $\nu_{R}^{\mathrm{st}}$ is sterile and therefore non-interacting.
The implications of different values for these parameters are discussed in Sec.~\ref{Sec:TheRoleOf-YLR-YRL}.

As outlined in Sec.~\ref{Sec:Definition_of_the_Theory_General_Abelian_Case}, the theory is required to respect global, vector-like $U(1)_{Q}$ and $SU(3)_c$ symmetries.
With the field content defined above, the corresponding generators are represented by
\begin{equation}\label{Eq:ASSM-Q}
    \begin{aligned}
        Q =
        \begin{pmatrix}
            0 & 0 & 0 & 0\\
            0 & -1 & 0 & 0\\
            0 & 0 & \frac{2}{3} \mathbb{1}_{3\times3} & 0\\
            0 & 0 & 0 & -\frac{1}{3} \mathbb{1}_{3\times3}
        \end{pmatrix},
    \end{aligned}
\end{equation}
for electric charge, and 
\begin{equation}\label{Eq:ASSM-Colour}
    \begin{aligned}
        T^{a}_c =
        \begin{pmatrix}
            0 & 0 & 0 & 0\\
            0 & 0 & 0 & 0\\
            0 & 0 & \frac{\lambda^{a}}{2} & 0\\
            0 & 0 & 0 & \frac{\lambda^{a}}{2}
        \end{pmatrix},
    \end{aligned}
\end{equation}
for colour charge, where $\lambda^a$ denote the Gell-Mann matrices.
These assignments correctly identify the fields $\nu$, $e$, $u$ and $d$
as the physical neutrino, electron, up-quark and down-quark, respectively.

We further introduce $N_S=2$ complex scalar fields $\{\phi_a\}_{a=1}^{2}$, each carrying the same hypercharge
\begin{equation}\label{Eq:ASSM-YS}
    \begin{aligned}
        \hypS = \frac{1}{2}.
    \end{aligned}
\end{equation}
The fields are distinguished by their electric charges, $Q_{\phi_1}=1$ and $Q_{\phi_2}=0$.
To reproduce the scalar potential of the Standard Model, we impose
\begin{equation}\label{Eq:ASSM-lambda}
    \begin{aligned}
        \lambda_{1111} &= \lambda_{2222} = 6 \lambda_{SM},\\
        \lambda_{1122} &= \lambda_{1221} = \lambda_{2112} = \lambda_{2211} = 3 \lambda_{SM},\\
        \lambda_{klmn} &\equiv 0, \quad \text{else},
    \end{aligned}
\end{equation}
which yields the Higgs potential
\begin{equation}\label{Eq:ASSM-HiggsPotential}
    \begin{aligned}
        V(\Phi) = \lambda_{SM} \big(\Phi^{\dagger}\Phi\big)^2,
    \end{aligned}
\end{equation}
with the scalar doublet
\begin{equation}
    \begin{aligned}
        \Phi =
        \begin{pmatrix}
            \phi_1\\
            \phi_2
        \end{pmatrix}.
    \end{aligned}
\end{equation}
The two components are identified with the physical fields $\phi_1=G^{+}$ and 
$\phi_2=(\varphi+iG^{0})/\sqrt{2}$.
Since we consider a massless version of the theory, the scalar mass term $\mu^2|\Phi|^2$ is omitted from Eq.~\eqref{Eq:ASSM-HiggsPotential}.
This omission has no impact on the renormalisation of dimension-$4$ operators, the regularisation-induced breaking of BRST invariance, or the running of the couplings.

The Yukawa sector for one fermion generation is specified by the coupling matrices
\begin{equation}\label{Eq:ASSM-Yuk-G}
    \begin{aligned}
        G^1 =
        \begin{pmatrix}
            0 & y_e & 0 & 0\\
            0 & 0 & 0 & 0\\
            0 & 0 & 0 & y_d \mathbb{1}_{3\times3}\\
            0 & 0 & 0 & 0
        \end{pmatrix},
        \quad
        G^2 =
        \begin{pmatrix}
            0 & 0 & 0 & 0\\
            0 & y_e & 0 & 0\\
            0 & 0 & 0 & 0\\
            0 & 0 & 0 & y_d \mathbb{1}_{3\times3}
        \end{pmatrix},
    \end{aligned}
\end{equation}
and
\begin{equation}\label{Eq:ASSM-Yuk-K}
    \begin{aligned}
        K^1 =
        \begin{pmatrix}
            0 & 0 & 0 & 0\\
            0 & 0 & 0 & 0\\
            0 & 0 & 0 & 0\\
            0 & 0 & -y_u \mathbb{1}_{3\times3} & 0
        \end{pmatrix},
        \quad
        K^2 =
        \begin{pmatrix}
            0 & 0 & 0 & 0\\
            0 & 0 & 0 & 0\\
            0 & 0 & y_u \mathbb{1}_{3\times3} & 0\\
            0 & 0 & 0 & 0
        \end{pmatrix},
    \end{aligned}
\end{equation}
with real Yukawa couplings $y_e, \, y_u, \, y_d \in \mathbb{R}$.

With this choice of matter fields and couplings, the model exhibits an additional global $SU(2)_L$ symmetry, which is, in general, not present in the theory defined in Sec.~\ref{Sec:Definition_of_the_Theory_General_Abelian_Case}.
This global symmetry is essential to reproduce the Abelian sector of the SM.
However, unlike in the full Standard Model, $SU(2)_L$ and $SU(3)_c$ appear here only as global symmetries, since non-Abelian gauge fields are excluded in the present analysis.

\paragraph{ASM with Fermion Multiplets according to Option~\ref{Opt:Option2}:}
In this realisation, every physical chiral fermion is paired with a sterile partner field.
In particular, the left- and right-handed multiplets are chosen as
\begin{equation}\label{Eq:ASSM-FermionMultiplets-Option2}
    \begin{aligned}
        \psi_L &= 
        \begin{pmatrix}
            \nu_L, & e_L, & u_L, & d_L,
          & \nu_L^{\mathrm{st}}, & e_L^{\mathrm{st}}, & u^{\mathrm{st}}_L, & d^{\mathrm{st}}_L
        \end{pmatrix}^\mathsf{T},\\
        \psi_R &= 
        \begin{pmatrix}
            \nu_R^{\mathrm{st}}, & e_R^{\mathrm{st}}, & u^{\mathrm{st}}_R, & d^{\mathrm{st}}_R,
          & \nu_R^{\mathrm{st}}, & e_R, & u_R, & d_R
        \end{pmatrix}^\mathsf{T},
    \end{aligned}
\end{equation}
with the up- and down-type quarks understood to appear in all three colours $\{r,g,b\}$, as in Eq.~\eqref{Eq:ASSM-FermionMultiplets}.
According to Eq.~\eqref{Eq:Electron-2-Spinors-Option2}), we define the two Dirac spinor multiplets
\begin{equation}\label{Eq:Fermions-psi1-psi2-Option2}
    \begin{aligned}
        \psi_1 &=
        \begin{pmatrix}
            \nu_L + \nu_R^{\mathrm{st}}, & 
            e_L + e_R^{\mathrm{st}}, & 
            u_L + u^{\mathrm{st}}_R, & 
            d_L + d^{\mathrm{st}}_R
        \end{pmatrix}^\mathsf{T},\\
        \psi_2 &=
        \begin{pmatrix}
            \nu_L^{\mathrm{st}} + \nu_R^{\mathrm{st}}, & 
            e_L^{\mathrm{st}} + e_R, & 
            u^{\mathrm{st}}_L + u_R, & 
            d^{\mathrm{st}}_L + d_R
        \end{pmatrix}^\mathsf{T},
    \end{aligned}
\end{equation}
so that $\psi_1$ collects the left-handed physical fermion fields of the SM (each combined with an appropriate sterile partner), while $\psi_2$ analogously contains the right-handed physical fermions.\footnote{Note that $\psi_2$ contains a fully sterile Dirac neutrino; this choice yields a particularly systematic block structure in the coupling matrices.}
The full Dirac multiplet is then
\begin{equation}\label{Eq:Dirac-Multiplet-Decomposition}
    \begin{aligned}
        \psi=\begin{pmatrix}\psi_1, \,\psi_2\end{pmatrix}^\mathsf{T},
    \end{aligned}
\end{equation}
which reproduces Eq.~\eqref{Eq:ASSM-FermionMultiplets-Option2} after projection with $\mathbb{P}_{\mathrm{L/R}}$.

With this enlarged fermionic field content, the SM coupling matrices acquire a block structure aligned with the decomposition in Eq.~\eqref{Eq:Dirac-Multiplet-Decomposition}.
Because sterile fields do not interact (see Def.~\ref{Def:Sterile_Quantum_Field}), the blocks associated with purely sterile fermions vanish, whereas the blocks acting on physical fields coincide with those used for Option~\ref{Opt:Option1}.

For the $4$-dimensional $16 \times 16$ hypercharge matrices we obtain
\begin{equation}\label{Eq:ASSM-YLYR-Option2}
    \begin{aligned}
        \hypR =
        \begin{pmatrix}
            0 & 0\\
            0 & \hypR^{\text{ASM,opt\ref{Opt:Option1}}}
        \end{pmatrix},
        \quad
        \hypL =
        \begin{pmatrix}
            \hypL^{\text{ASM,opt\ref{Opt:Option1}}} & 0\\
            0 & 0 
        \end{pmatrix},
    \end{aligned}
\end{equation}
with $\hypL^{\text{ASM,opt\ref{Opt:Option1}}}$ and $\hypR^{\text{ASM,opt\ref{Opt:Option1}}}$ given in Eq.~\eqref{Eq:ASSM-YLYR}.
As before, we assign identical real hypercharges to the evanescent $LR$- and the $RL$-interaction currents for each physical fermion.
However, in this case, the full evanescent hypercharge matrices are block off-diagonal, i.e.\ 
\begin{equation}\label{Eq:ASSM-YLR-YRL-Option2}
    \begin{aligned}
        \hypLR =
        \begin{pmatrix}
            0 & \hypLR^{\text{ASM,opt\ref{Opt:Option1}}}\\
            0 & 0
        \end{pmatrix},
        \quad
        \hypRL =
        \begin{pmatrix}
            0 & 0\\
            \hypRL^{\text{ASM,opt\ref{Opt:Option1}}} & 0 
        \end{pmatrix},
    \end{aligned}
\end{equation}
with $\hypLR^{\text{ASM,opt\ref{Opt:Option1}}}=\hypRL^{\text{ASM,opt\ref{Opt:Option1}}}$, as specified in Eq.~\eqref{Eq:ASSM-YLR}.
Consequently, in Option~\ref{Opt:Option2}, we have $\hypRL\neq\hypLR$ for the full $16\times16$ matrices, while the hermiticity condition in Eq.~\eqref{Eq:Requirement-YLR-YRL} remains satisfied.
Allowing nonzero evanescent hypercharges $\hypLR$ and $\hypRL$ is, however, in conflict with the original motivation of Option~\ref{Opt:Option2}, which aims to preserve at least global hypercharge conservation.
The reason is that the evanescent gauge interactions flip chirality and can therefore violate also global hypercharge conservation, as discussed in Sec.~\ref{Sec:Dimensional_Ambiguities_and_Evanescent_Shadows}.
Considering the electron as an illustrative example, the corresponding evanescent gauge interactions read
\begin{equation}\label{Eq:ChiralityViolation-YLR-YRL-Electron}
    \begin{aligned}
        - g \hypRL^e \overline{e_2} \projL \widehat{\slashed{B}} \projL e_1
        - g \hypLR^e \overline{e_1} \projR \widehat{\slashed{B}} \projR e_2
        =
        - g \hypRL^e \overline{e_R} \widehat{\slashed{B}} e_L
        - g \hypLR^e \overline{e_L} \widehat{\slashed{B}} e_R,
    \end{aligned}
\end{equation}
which explicitly mix $e_L$ (hypercharge $-1/2$) and $e_R$ (hypercharge $-1$).\footnote{Note that we work with real evanescent hypercharges, i.e.\ $\hypRL^e=(\hypLR^e)^{*}=\hypLR^e\in\mathbb{R}$.}
In what follows, we nevertheless keep $\hypLR,\hypRL\neq 0$, as in Eq.~\eqref{Eq:ASSM-YLR-YRL-Option2}, in order to retain generality, and we examine the consequences in Sec.~\ref{Sec:AnalysisOfASSMResults}.

The Yukawa sector follows an analogous block structure and reads
\begin{equation}\label{Eq:ASSM-Yuk-G-Option2}
    \begin{aligned}
        G^1 =
        \begin{pmatrix}
            0 & G^1_{\text{ASM,opt\ref{Opt:Option1}}}\\
            0 & 0
        \end{pmatrix},
        \quad
        G^2 =
        \begin{pmatrix}
            0 & G^2_{\text{ASM,opt\ref{Opt:Option1}}}\\
            0 & 0 
        \end{pmatrix},
    \end{aligned}
\end{equation}
and
\begin{equation}\label{Eq:ASSM-Yuk-K-Option2}
    \begin{aligned}
        K^1 =
        \begin{pmatrix}
            0 & K^1_{\text{ASM,opt\ref{Opt:Option1}}}\\
            0 & 0
        \end{pmatrix},
        \quad
        K^2 =
        \begin{pmatrix}
            0 & K^2_{\text{ASM,opt\ref{Opt:Option1}}}\\
            0 & 0 
        \end{pmatrix},
    \end{aligned}
\end{equation}
with $G^a_{\text{ASM,opt\ref{Opt:Option1}}}$ and $K^a_{\text{ASM,opt\ref{Opt:Option1}}}$ as defined in Eqs.~\eqref{Eq:ASSM-Yuk-G} and \eqref{Eq:ASSM-Yuk-K}, respectively.

All remaining ingredients of the construction --- most notably the scalar sector --- are unaffected by the choice of the $D$-dimensional continuation of the fermions and therefore coincide with those in Option~\ref{Opt:Option1}.

A comparative discussion of the 1-loop renormalisation for both options is presented in Sec.~\ref{Sec:AnalysisOfASSMResults}.

\subsection{Abelian Sector of the Two-Higgs Doublet Model}\label{Sec:ASTHDM}

The Abelian sector of the 2HDM can likewise be obtained as a specific realisation of the general framework introduced in Sec.~\ref{Sec:Definition_of_the_Theory_General_Abelian_Case}.
The construction proceeds analogously to that of the ASM discussed in Sec.~\ref{Sec:ASSM}, but now includes all three fermion generations. 
In the following, we restrict ourselves to the $D$-dimensional fermion treatment according to Option~\ref{Opt:Option1}; the extension to Option~\ref{Opt:Option2} is conceptually straightforward.
With an appropriate choice of fermion multiplets within this setup, the physical hypercharges of the model are given by
\begin{equation}\label{Eq:ASTHDM-YLYR}
    \begin{aligned}
        \hypR =
        \begin{pmatrix}
            0 & 0 & 0 & 0\\
            0 & - \mathbb{1}_{3\times3} & 0 & 0\\
            0 & 0 & \frac{2}{3} \mathbb{1}_{9\times9} & 0\\
            0 & 0 & 0 & -\frac{1}{3} \mathbb{1}_{9\times9}
        \end{pmatrix},
        \quad
        \hypL =
        \begin{pmatrix}
            -\frac{1}{2} \mathbb{1}_{3\times3} & 0 & 0 & 0\\
            0 & -\frac{1}{2} \mathbb{1}_{3\times3} & 0 & 0\\
            0 & 0 & \frac{1}{6} \mathbb{1}_{9\times9} & 0\\
            0 & 0 & 0 & \frac{1}{6} \mathbb{1}_{9\times9}
        \end{pmatrix},
    \end{aligned}
\end{equation}
which reproduce the standard hypercharge assignments for all three generations of leptons and quarks.
The evanescent hypercharges are chosen in analogy with Eq.~\eqref{Eq:ASSM-YLR} as
\begin{equation}\label{Eq:ASTHDM-YLR}
    \begin{aligned}
        \hypLR = \hypRL =
        \begin{pmatrix}
            0 & 0 & 0 & 0\\
            0 & \mathcal{Y}_{LR}^{e} \mathbb{1}_{3\times3} & 0 & 0\\
            0 & 0 & \mathcal{Y}_{LR}^{u} \mathbb{1}_{9\times9} & 0\\
            0 & 0 & 0 & \mathcal{Y}_{LR}^{d} \mathbb{1}_{9\times9}
        \end{pmatrix}.
    \end{aligned}
\end{equation}
In principle, when several generations are included, the evanescent hypercharge matrices could also contain off-diagonal elements that mix generations of leptons or quarks, while electric and colour charge conservation remain preserved.
The definitions of the electric charge $Q$ and the colour generators $T_c^a$ remain identical to the ASM case, consistent with global $U(1)_{Q}$ and $SU(3)_c$ symmetry and the fermionic field content of the 2HDM.

The scalar sector of the A2HDM involves $N_S=4$ complex scalar fields, $\{\phi_a\}_{a=1}^4$, each carrying the same hypercharge,
\begin{equation}\label{Eq:ASTHDM-YS}
    \begin{aligned}
        \hypS = \frac{1}{2}.
    \end{aligned}
\end{equation}
Among these, $\phi_1$ and $\phi_3$ are electrically charged, while $\phi_2$ and $\phi_4$ are neutral.
To reproduce the scalar potential of the 2HDM, the quartic couplings $\lambda_{klmn}$ are restricted such that the scalar self-interaction term yields
\begin{equation}\label{Eq:ASTHDM-lambda}
    \begin{aligned}
        V(\Phi_1,\Phi_2) 
        &= \frac{\lambda_1}{2} \big(\Phi_1^{\dagger}\Phi_1\big)^2
        + \frac{\lambda_2}{2} \big(\Phi_2^{\dagger}\Phi_2\big)^2
        + \lambda_3 \big(\Phi_1^{\dagger}\Phi_1\big)\big(\Phi_2^{\dagger}\Phi_2\big)
        + \lambda_4 \big(\Phi_1^{\dagger}\Phi_2\big)\big(\Phi_2^{\dagger}\Phi_1\big)\\
        &+ \bigg[
        \frac{\lambda_5}{2} \big(\Phi_1^{\dagger}\Phi_2\big)^2
        + \lambda_6 \big(\Phi_1^{\dagger}\Phi_1\big)\big(\Phi_1^{\dagger}\Phi_2\big)
        + \lambda_7 \big(\Phi_2^{\dagger}\Phi_2\big)\big(\Phi_1^{\dagger}\Phi_2\big)
        + \mathrm{h.c.}
        \bigg],
    \end{aligned}
\end{equation}
where the four scalar fields are grouped into two Higgs doublets,
\begin{equation}
    \begin{aligned}
        \Phi_1 =
        \begin{pmatrix}
            \phi_1\\
            \phi_2
        \end{pmatrix},
        \quad
        \Phi_2 =
        \begin{pmatrix}
            \phi_3\\
            \phi_4
        \end{pmatrix}.
    \end{aligned}
\end{equation}
As in the ASM case, the scalar mass terms are omitted since the analysis is restricted to massless theories, as discussed in Sec.~\ref{Sec:ASSM}.

The 2HDM, unlike the ASM, possesses two independent sets of Yukawa interactions, each corresponding to one of the two scalar doublets.
The Yukawa matrices corresponding to $\Phi_1$ are denoted by $G^{1,2}$ and $K^{1,2}$, while those associated with $\Phi_2$ are represented by $G^{3,4}$ and $K^{3,4}$.
Their explicit forms are
\begin{equation}\label{Eq:ASTHDM-Yuk-G}
    \begin{aligned}
        G^1 &=
        \begin{pmatrix}
            0 & Y_e & 0 & 0\\
            0 & 0 & 0 & 0\\
            0 & 0 & 0 & Y_d\\
            0 & 0 & 0 & 0
        \end{pmatrix},
        \quad
        G^2 =
        \begin{pmatrix}
            0 & 0 & 0 & 0\\
            0 & Y_e & 0 & 0\\
            0 & 0 & 0 & 0\\
            0 & 0 & 0 & Y_d
        \end{pmatrix},\\
        G^3 &=
        \begin{pmatrix}
            0 & Z_e & 0 & 0\\
            0 & 0 & 0 & 0\\
            0 & 0 & 0 & Z_d\\
            0 & 0 & 0 & 0
        \end{pmatrix},
        \quad
        G^4 =
        \begin{pmatrix}
            0 & 0 & 0 & 0\\
            0 & Z_e & 0 & 0\\
            0 & 0 & 0 & 0\\
            0 & 0 & 0 & Z_d
        \end{pmatrix},
    \end{aligned}
\end{equation}
\begin{equation}\label{Eq:ASTHDM-Yuk-K}
    \begin{aligned}
        K^1 &=
        \begin{pmatrix}
            0 & 0 & 0 & 0\\
            0 & 0 & 0 & 0\\
            0 & 0 & 0 & 0\\
            0 & 0 & -Y_u & 0
        \end{pmatrix},
        \quad
        K^2 =
        \begin{pmatrix}
            0 & 0 & 0 & 0\\
            0 & 0 & 0 & 0\\
            0 & 0 & Y_u & 0\\
            0 & 0 & 0 & 0
        \end{pmatrix},\\
        K^3 &=
        \begin{pmatrix}
            0 & 0 & 0 & 0\\
            0 & 0 & 0 & 0\\
            0 & 0 & 0 & 0\\
            0 & 0 & -Z_u & 0
        \end{pmatrix},
        \quad
        K^4 =
        \begin{pmatrix}
            0 & 0 & 0 & 0\\
            0 & 0 & 0 & 0\\
            0 & 0 & Z_u & 0\\
            0 & 0 & 0 & 0
        \end{pmatrix},
    \end{aligned}
\end{equation}
where the coupling matrices $Y_e$, $Y_u$, $Y_d$, $Z_e$, $Z_u$, and $Z_d$ are, in general, complex $3\times3$ matrices.

The resulting model, with the specified field content and coupling structure, exhibits a global $SU(2)_L$ symmetry, mirroring the symmetry properties of the ASM discussed in Sec.~\ref{Sec:ASSM}.
As in the ASM case, the $SU(2)_L$ and $SU(3)_c$ symmetries are treated as purely global, since non-Abelian gauge dynamics are not included in the analysis presented in this chapter.

\section{Slavnov-Taylor Identities}\label{Sec:STIs-in-Abelian-Chiral-Gauge-Theory}

Our methodology for determining the complete set of counterterms --- including the finite symmetry-restoring ones --- necessary and sufficient for the full renormalisation of a chiral gauge theory in the BMHV framework is described in detail in Sec.~\ref{Sec:Symmetry_Restoration_Procedure}.
In essence, these counterterms are fixed by imposing both finiteness of the renormalised theory and the validity of the Slavnov–Taylor identity, as formulated in Eq.~\eqref{Eq:UltimateSymmetryRequirement}, thereby ensuring a consistent quantum theory.

Using the quantum action principle, the breaking of BRST invariance can be expressed as an insertion of a local composite operator $\Delta$ into the effective quantum action, as shown in Eq.~\eqref{Eq:QAPinDReg_Symmetry_Restoration}. 
The operator $\Delta$ encodes the symmetry breaking, whose lowest-order contribution --- the tree-level breaking $\widehat{\Delta}=\mathcal{S}_D(S_0)$ --- is purely evanescent and given explicitly in Eq.~\eqref{Eq:GeneralTreeLevelBreaking_Abelian} for the considered model defined in Sec.~\ref{Sec:Definition_of_the_Theory_General_Abelian_Case}.

In the following, we present the complete set of identities derived from Eq.~\eqref{Eq:QAPinDReg_Symmetry_Restoration}, which are relevant for the renormalisation procedure.
These relations connect power-counting divergent 1PI Green functions: on the LHS, products of ordinary Green functions generated by the Slavnov-Taylor operator appear, while on the RHS, they are matched by Green functions with an insertion of the breaking operator $\Delta$.
The resulting identities serve as nontrivial consistency conditions for both divergent and finite parts of the amplitudes.

Since, in Abelian gauge theories, the BRST transformations --- and consequently the associated source terms in $\mathcal{L}_\mathrm{ext}$ (see Eq.~\eqref{Eq:Ddim-Lext-GeneralAbelianTheory}) --- do not renormalise, we can directly replace the source terms by their tree-level expressions.
With this substitution, the quantum action principle yields the following explicit set of relations among the Green functions relevant for renormalisation:\footnote{
For readability, subscripts, superscripts, and tildes indicating that the expressions are evaluated in momentum space have been omitted, as no ambiguity arises from this simplification.
All identities are valid at each order in perturbation theory, since the BRST transformations do not renormalise in the Abelian case.}

\begin{align}\label{Eq:STI-QAP-Bc}
    p^{\nu} \, \Gamma_{B^{\mu}(-p)B^{\nu}(p)}
    &= i \big( \Delta \cdot \Gamma \big)_{B^{\mu}(-p) c(p)} 
\end{align}
\begin{align}
\begin{split}\label{Eq:STI-QAP-BBc}
    q^{\rho} \, \Gamma_{B^{\rho}(q)B^{\nu}(p_2)B^{\mu}(p_1)}
    &= i \big( \Delta \cdot \Gamma \big)_{B^{\nu}(p_2)B^{\mu}(p_1) c(q)}
\end{split}\\[1.5ex]
\begin{split}\label{Eq:STI-QAP-FFc}
    q^{\mu} \, \Gamma_{\psi_{j,\beta}(p_2)\overline{\psi}_{i,\alpha}(p_1)B^{\mu}(q)}
    &+ g \big({\hypL}{}_{,lj}{\projL}{}_{,\delta\beta}+{\hypR}{}_{,lj}{\projR}{}_{,\delta\beta}\big)
    \Gamma_{\psi_{l,\delta}(-p_1)\overline{\psi}_{i,\alpha}(p_1)}\\
    &- g \big({\hypL}{}_{,il}{\projR}{}_{,\alpha\delta}+{\hypR}{}_{,il}{\projL}{}_{,\alpha\delta}\big)
    \Gamma_{\psi_{j,\beta}(p_2)\overline{\psi}_{l,\delta}(-p_2)}\\
    &= i \big( \Delta \cdot \Gamma \big)_{\psi_{j,\beta}(p_2)\overline{\psi}_{i,\alpha}(p_1)c(q)}
\end{split}\\[1.5ex]
\begin{split}\label{Eq:STI-QAP-SSbarc}
    q^{\mu} \, \Gamma_{\phi_{b}(p_2)\phi^{\dagger}_{a}(p_1)B^{\mu}(q)}
    &+ g \hypS
    \Gamma_{\phi_{b}(-p_1)\phi^{\dagger}_{a}(p_1)}
    - g \hypS
    \Gamma_{\phi_b(p_2)\phi^{\dagger}_{a}(-p_2)}\\
    &= i \big( \Delta \cdot \Gamma \big)_{\phi_{b}(p_2)\phi^{\dagger}_{a}(p_1)c(q)}
\end{split}\\[1.5ex]
\begin{split}\label{Eq:STI-QAP-SSc}
    q^{\mu} \, \Gamma_{\phi_{b}(p_2)\phi_{a}(p_1)B^{\mu}(q)}
    &+ g \hypS
    \Gamma_{\phi_{a}(p_1)\phi_{b}(-p_1)}
    + g \hypS
    \Gamma_{\phi_b(p_2)\phi_{a}(-p_2)}\\
    &= i \big( \Delta \cdot \Gamma \big)_{\phi_{b}(p_2)\phi_{a}(p_1)c(q)}
\end{split}\\[1.5ex]
\begin{split}\label{Eq:STI-QAP-SbarSbarc}
    q^{\mu} \, \Gamma_{\phi^{\dagger}_{b}(p_2)\phi^{\dagger}_{a}(p_1)B^{\mu}(q)}
    &- g \hypS
    \Gamma_{\phi^{\dagger}_{a}(p_1)\phi^{\dagger}_{b}(-p_1)}
    - g \hypS
    \Gamma_{\phi^{\dagger}_b(p_2)\phi^{\dagger}_{a}(-p_2)}\\
    &= i \big( \Delta \cdot \Gamma \big)_{\phi^{\dagger}_{b}(p_2)\phi^{\dagger}_{a}(p_1)c(q)}
\end{split}
\end{align}
\begin{align}
\begin{split}\label{Eq:STI-QAP-BBBc}
    q^{\sigma} \, \Gamma_{B^{\sigma}(q)B^{\rho}(p_3)B^{\nu}(p_2)B^{\mu}(p_1)}
    &= i \big( \Delta \cdot \Gamma \big)_{B^{\rho}(p_3)B^{\nu}(p_2)B^{\mu}(p_1) c(q)}
\end{split}\\[1.5ex]
\begin{split}\label{Eq:STI-QAP-FFbarBc}
 g \big({\hypL}{}_{,lj}{\projL}{}_{,\delta\beta}+{\hypR}{}_{,lj}{\projR}{}_{,\delta\beta}\big)
    &\Gamma_{\psi_{l,\delta}(-p_1-p_2)\overline{\psi}_{i,\alpha}(p_2)B^{\mu}(p_1)}\\
    - g \big({\hypL}{}_{,il}{\projR}{}_{,\alpha\delta}+{\hypR}{}_{,il}{\projL}{}_{,\alpha\delta}\big)
    &\Gamma_{\psi_{j,\beta}(p_3)\overline{\psi}_{l,\delta}(-p_1-p_3)B^{\mu}(p_1)}\\
    &= i \big( \Delta \cdot \Gamma \big)_{\psi_{j,\beta}(p_3)\overline{\psi}_{i,\alpha}(p_2)B^{\mu}(p_1)c(q)}
\end{split}\\[1.5ex]
\begin{split}\label{Eq:STI-QAP-FFbarSc}
g\hypS &\Gamma_{\psi_{j,\beta}(p_3)\overline{\psi}_{i,\alpha}(p_2)\phi_a(-p_2-p_3)}\\
    + g \big({\hypL}{}_{,lj}{\projL}{}_{,\delta\beta}+{\hypR}{}_{,lj}{\projR}{}_{,\delta\beta}\big)
    &\Gamma_{\psi_{l,\delta}(-p_1-p_2)\overline{\psi}_{i,\alpha}(p_2)\phi_a(p_1)}\\
    -  g \big({\hypL}{}_{,il}{\projR}{}_{,\alpha\delta}+{\hypR}{}_{,il}{\projL}{}_{,\alpha\delta}\big)
    &\Gamma_{\psi_{j,\beta}(p_3)\overline{\psi}_{l,\delta}(-p_1-p_3)\phi_a(p_1)}\\
    &= i \big( \Delta \cdot \Gamma \big)_{\psi_{j,\beta}(p_3)\overline{\psi}_{i,\alpha}(p_2)\phi_a(p_1)c(q)}
\end{split}\\[1.5ex]
\begin{split}\label{Eq:STI-QAP-FFbarSbarc}
    -g\hypS &\Gamma_{\psi_{j,\beta}(p_3)\overline{\psi}_{i,\alpha}(p_2)\phi^{\dagger}_a(-p_2-p_3)}\\
    + g \big({\hypL}{}_{,lj}{\projL}{}_{,\delta\beta}+{\hypR}{}_{,lj}{\projR}{}_{,\delta\beta}\big)
    &\Gamma_{\psi_{l,\delta}(-p_1-p_2)\overline{\psi}_{i,\alpha}(p_2)\phi^{\dagger}_a(p_1)}\\
    - g \big({\hypL}{}_{,il}{\projR}{}_{,\alpha\delta}+{\hypR}{}_{,il}{\projL}{}_{,\alpha\delta}\big)
    &\Gamma_{\psi_{j,\beta}(p_3)\overline{\psi}_{l,\delta}(-p_1-p_3)\phi^{\dagger}_a(p_1)}\\
    &= i \big( \Delta \cdot \Gamma \big)_{\psi_{j,\beta}(p_3)\overline{\psi}_{i,\alpha}(p_2)\phi^{\dagger}_a(p_1)c(q)}
\end{split}\\[1.5ex]
\begin{split}\label{Eq:STI-QAP-SSbarBc}
    q^{\nu}\,\Gamma_{\phi_b(p_3)\phi_{a}^{\dagger}(p_2)B^{\mu}(p_1)B^{\nu}(q)}
    &+ g \hypS 
    \Gamma_{\phi_b(-p_1-p_2)\phi^{\dagger}_{a}(p_2)B^{\mu}(p_1)}\\
    &- g \hypS
    \Gamma_{\phi_b(p_3)\phi^{\dagger}_a(-p_1-p_3)B^{\mu}(p_1)}\\
    &= i \big( \Delta \cdot \Gamma \big)_{\phi_b(p_3)\phi^{\dagger}_a(p_2)B^{\mu}(p_1)c(q)} 
\end{split}\\[1.5ex]
\begin{split}\label{Eq:STI-QAP-SSBc}
    q^{\nu}\,\Gamma_{\phi_b(p_3)\phi_{a}(p_2)B^{\mu}(p_1)B^{\nu}(q)}
    &+ g \hypS 
    \Gamma_{\phi_b(-p_1-p_2)\phi_{a}(p_2)B^{\mu}(p_1)}\\
    &+ g \hypS
    \Gamma_{\phi_b(p_3)\phi_a(-p_1-p_3)B^{\mu}(p_1)}\\
    &= i \big( \Delta \cdot \Gamma \big)_{\phi_b(p_3)\phi_a(p_2)B^{\mu}(p_1)c(q)}
\end{split}\\[1.5ex]
\begin{split}\label{Eq:STI-QAP-SbarSbarBc}
    q^{\nu}\,\Gamma_{\phi_b^{\dagger}(p_3)\phi_{a}^{\dagger}(p_2)B^{\mu}(p_1)B^{\nu}(q)}
    &- g \hypS 
    \Gamma_{\phi_b^{\dagger}(-p_1-p_2)\phi^{\dagger}_{a}(p_2)B^{\mu}(p_1)}\\
    &- g \hypS
    \Gamma_{\phi_b^{\dagger}(p_3)\phi^{\dagger}_a(-p_1-p_3)B^{\mu}(p_1)}\\
    &= i \big( \Delta \cdot \Gamma \big)_{\phi_b^{\dagger}(p_3)\phi^{\dagger}_a(p_2)B^{\mu}(p_1)c(q)}
\end{split}
\end{align}
\begin{align}
\begin{split}\label{Eq:STI-QAP-SSbarBBc}
    g \hypS 
    \Gamma_{\phi_b(-p_1-p_2-p_3)\phi^{\dagger}_{a}(p_3)B^{\nu}(p_2)B^{\mu}(p_1)}
    &- g \hypS
    \Gamma_{\phi_b(p_4)\phi^{\dagger}_a(-p_1-p_2-p_4)B^{\nu}(p_2)B^{\mu}(p_1)}\\
    &= i \big( \Delta \cdot \Gamma \big)_{\phi_b(p_4)\phi^{\dagger}_a(p_3)B^{\nu}(p_2)B^{\mu}(p_1)c(q)} 
\end{split}\\[1.5ex]
\begin{split}\label{Eq:STI-QAP-SSBBc}
    g \hypS 
    \Gamma_{\phi_b(-p_1-p_2-p_3)\phi_{a}(p_3)B^{\nu}(p_2)B^{\mu}(p_1)}
    &+ g \hypS
    \Gamma_{\phi_b(p_4)\phi_a(-p_1-p_2-p_4)B^{\nu}(p_2)B^{\mu}(p_1)}\\
    &= i \big( \Delta \cdot \Gamma \big)_{\phi_b(p_4)\phi_a(p_3)B^{\nu}(p_2)B^{\mu}(p_1)c(q)}
\end{split}\\[1.5ex]
\begin{split}\label{Eq:STI-QAP-SbarSbarBBc}
    - g \hypS 
    \Gamma_{\phi_b^{\dagger}(-p_1-p_2-p_3)\phi^{\dagger}_{a}(p_3)B^{\nu}(p_2)B^{\mu}(p_1)}
    &- g \hypS
    \Gamma_{\phi^{\dagger}_b(p_4)\phi^{\dagger}_a(-p_1-p_2-p_4)B^{\nu}(p_2)B^{\mu}(p_1)}\\
    &= i \big( \Delta \cdot \Gamma \big)_{\phi^{\dagger}_b(p_4)\phi^{\dagger}_a(p_3)B^{\nu}(p_2)B^{\mu}(p_1)c(q)}
\end{split}\\[1.5ex]
\begin{split}\label{Eq:STI-QAP-SSSbarSbarc}
    g \hypS 
    \Gamma_{\phi_d(-p_3-p_2-p_1)\phi_c(p_3)\phi_b^{\dagger}(p_2)\phi_a^{\dagger}(p_1)}
    &+ g \hypS 
    \Gamma_{\phi_d(p_4)\phi_c(-p_4-p_2-p_1)\phi_b^{\dagger}(p_2)\phi_a^{\dagger}(p_1)}\\
    - g \hypS 
    \Gamma_{\phi_d(p_4)\phi_c(p_3)\phi_b^{\dagger}(-p_4-p_3-p_1)\phi_a^{\dagger}(p_1)}
    &- g \hypS 
    \Gamma_{\phi_d(p_4)\phi_c(p_3)\phi_b^{\dagger}(p_2)\phi_a^{\dagger}(-p_4-p_3-p_2)}\\
    &= i \big( \Delta \cdot \Gamma \big)_{\phi_d(p_4)\phi_c(p_3)\phi_b^{\dagger}(p_2)\phi_a^{\dagger}(p_1)c(q)}
\end{split}\\[1.5ex]
\begin{split}\label{Eq:STI-QAP-SSSSc}
    + g \hypS 
    \Gamma_{\phi_d(-p_3-p_2-p_1)\phi_c(p_3)\phi_b(p_2)\phi_a(p_1)}
    &+ g \hypS 
    \Gamma_{\phi_d(p_4)\phi_c(-p_4-p_2-p_1)\phi_b(p_2)\phi_a(p_1)}\\
    + g \hypS 
    \Gamma_{\phi_d(p_4)\phi_c(p_3)\phi_b(-p_4-p_3-p_1)\phi_a(p_1)}
    &+ g \hypS 
    \Gamma_{\phi_d(p_4)\phi_c(p_3)\phi_b(p_2)\phi_a(-p_4-p_3-p_2)}\\
    &= i \big( \Delta \cdot \Gamma \big)_{\phi_d(p_4)\phi_c(p_3)\phi_b(p_2)\phi_a(p_1)c(q)}
\end{split}\\[1.5ex]
\begin{split}\label{Eq:STI-QAP-SSSSbarc}
    g \hypS 
    \Gamma_{\phi_d(-p_3-p_2-p_1)\phi_c(p_3)\phi_b(p_2)\phi_a^{\dagger}(p_1)}
    &+ g \hypS 
    \Gamma_{\phi_d(p_4)\phi_c(-p_4-p_2-p_1)\phi_b(p_2)\phi_a^{\dagger}(p_1)}\\
    + g \hypS 
    \Gamma_{\phi_d(p_4)\phi_c(p_3)\phi_b(-p_4-p_3-p_1)\phi_a^{\dagger}(p_1)}
    &- g \hypS 
    \Gamma_{\phi_d(p_4)\phi_c(p_3)\phi_b(p_2)\phi_a^{\dagger}(-p_4-p_3-p_2)}\\
    &= i \big( \Delta \cdot \Gamma \big)_{\phi_d(p_4)\phi_c(p_3)\phi_b(p_2)\phi_a^{\dagger}(p_1)c(q)}
\end{split}\\[1.5ex]
\begin{split}\label{Eq:STI-QAP-SSbarSbarSbarc}
    g \hypS 
    \Gamma_{\phi_d(-p_3-p_2-p_1)\phi_c^{\dagger}(p_3)\phi_b^{\dagger}(p_2)\phi_a^{\dagger}(p_1)}
    &- g \hypS 
    \Gamma_{\phi_d(p_4)\phi_c^{\dagger}(-p_4-p_2-p_1)\phi_b^{\dagger}(p_2)\phi_a^{\dagger}(p_1)}\\
    - g \hypS 
    \Gamma_{\phi_d(p_4)\phi_c^{\dagger}(p_3)\phi_b^{\dagger}(-p_4-p_3-p_1)\phi_a^{\dagger}(p_1)}
    &- g \hypS 
    \Gamma_{\phi_d(p_4)\phi_c^{\dagger}(p_3)\phi_b^{\dagger}(p_2)\phi_a^{\dagger}(-p_4-p_3-p_2)}\\
    &= i \big( \Delta \cdot \Gamma \big)_{\phi_d(p_4)\phi_c^{\dagger}(p_3)\phi_b^{\dagger}(p_2)\phi_a^{\dagger}(p_1)c(q)}
\end{split}\\[1.5ex]
\begin{split}\label{Eq:STI-QAP-SbarSbarSbarSbarc}
    - g \hypS 
    \Gamma_{\phi_d^{\dagger}(-p_3-p_2-p_1)\phi_c^{\dagger}(p_3)\phi_b^{\dagger}(p_2)\phi_a^{\dagger}(p_1)}
    &- g \hypS 
    \Gamma_{\phi_d^{\dagger}(p_4)\phi_c^{\dagger}(-p_4-p_2-p_1)\phi_b^{\dagger}(p_2)\phi_a^{\dagger}(p_1)}\\
    - g \hypS 
    \Gamma_{\phi_d^{\dagger}(p_4)\phi_c^{\dagger}(p_3)\phi_b^{\dagger}(-p_4-p_3-p_1)\phi_a^{\dagger}(p_1)}
    &- g \hypS 
    \Gamma_{\phi_d^{\dagger}(p_4)\phi_c^{\dagger}(p_3)\phi_b^{\dagger}(p_2)\phi_a^{\dagger}(-p_4-p_3-p_2)}\\
    &= i \big( \Delta \cdot \Gamma \big)_{\phi_d^{\dagger}(p_4)\phi_c^{\dagger}(p_3)\phi_b^{\dagger}(p_2)\phi_a^{\dagger}(p_1)c(q)}
\end{split}
\end{align}

\section{Renormalisation at the 1-Loop Level}\label{Sec:One-Loop-Renormalisation-Abelian-Chiral-Gauge-Theory}

In this section, we present the complete 1-loop renormalisation of the general Abelian chiral gauge theory defined by the $D$-dimensional Lagrangian in Eq.~\eqref{Eq:Model-Lagrangian-General-Abelian-Theory} (see Sec.~\ref{Sec:Definition_of_the_Theory_General_Abelian_Case}).
For this purpose, we closely follow our presentation in Ref.~\cite{Ebert:2024xpy}, where we reported these results first.
The discussion focuses on results at the generic level; model-specific results, obtained by specialising to the theories introduced in Sec.~\ref{Sec:Special_Cases_of_general_Abelian_Model} and summarised in Tab.~\ref{Tab:SpecialCasesOverview}, are provided in the subsequent section.
The analysis incorporates the evanescent gauge interactions associated with the hypercharges $\hypLR$ and $\hypRL$ (see Eq.~\eqref{Eq:L_Fermion_General_Abelian_Chiral_Gauge_Theory}).
All computations are performed in general $R_{\xi}$-gauge, cf.\ Eq.~\eqref{Eq:Ddim-LGaugeFixing+Ghost-GeneralAbelianTheory}, without restricting the gauge parameter $\xi$, and for $D=4-2\epsilon$.

To obtain compact expressions, we have employed the constraints on the Yukawa couplings that arise from 4-dimensional gauge invariance (see Eq.~\eqref{Eq:YukawaHyperchargeBRSTCondition}), as well as relations derived from them, together with the commutation properties of the hypercharge matrices. 
For the generic analysis presented here, we further restrict to the case $\hypRL=\hypLR$, taking these matrices to be hermitian and mutually commuting with $\hypL$ and $\hypR$, while respecting electric and colour charge conservation.
This choice includes, for instance, the evanescent hypercharge assignment in Eq.~\eqref{Eq:ASSM-YLR}, but not the one defined in Eq.~\eqref{Eq:ASSM-YLR-YRL-Option2}.
The following section (see Sec.~\ref{Sec:Results-Shedding_Light_on_Evanescent_Shadows}) will discuss the corresponding generalisation beyond this restriction.

The renormalisation procedure follows the methodology described in Sec.~\ref{Sec:Symmetry_Restoration_Procedure}.
A notable and conceptually important aspect of the renormalisation in the BMHV scheme for this theory is that it cannot be restricted to hypercharge-conserving Green functions alone.
As discussed in Sec.~\ref{Sec:Dimensional_Ambiguities_and_Evanescent_Shadows}, the presence of evanescent terms that mix chiralities may lead to a violation of global hypercharge conservation depending on the assigned gauge quantum numbers.
Consequently, also Green functions such as $\langle \phi\phi \rangle^\mathrm{1PI}$ and related ones must be included in the analysis.

All results have been obtained using the \texttt{Mathematica}-based computational framework introduced in Sec.~\ref{Sec:Computational_Setup} and employing the methods summarised in Chapter~\ref{Chap:Multi-Loop_Calculations}.
In particular, the UV divergent parts of all Feynman integrals were extracted through a tadpole decomposition, as described in Sec.~\ref{Sec:Tadpole_Decomposition}.
A subset of the results has been cross-checked using the \texttt{FORM}-based setup discussed in Sec.~\ref{Sec:Computational_Setup}, yielding perfect agreement.
Furthermore, the UV divergent BRST-breaking terms were computed both from ordinary 1PI Green functions and from 1PI Green functions containing a single insertion of the breaking operator $\Delta$ (see Sec.~\ref{Sec:Symmetry_Restoration_Procedure}).
The agreement between these two calculations provides a strong internal consistency check and the resulting 1PI Green functions satisfy all relations listed in Sec.~\ref{Sec:STIs-in-Abelian-Chiral-Gauge-Theory}, offering additional verification of correctness.

For clarity and compactness, all results in this section are expressed in terms of the coefficients defined in App.~\ref{App:1LoopCTCoeffs-GeneralAbelianChiralGaugeTheory}.
In Sec.~\ref{Sec:RelevantGreenFunctions-GeneralAbelianChiralGaugeTheory}, we comment on the 1PI Green functions relevant for the renormalisation procedure, whose explicit expressions are collected in App.~\ref{App:1LoopGreenFunctions-GeneralAbelianChiralGaugeTheory}.
The complete 1-loop counterterm action is then presented in the subsequent subsections: 
the singular part in Sec.~\ref{Sec:S_sct-GeneralAbelianChiralGaugeTheory}, and the finite symmetry-restoring contribution in Sec.~\ref{Sec:S_fct-GeneralAbelianChiralGaugeTheory}.

\subsection{Relevant Green Functions}\label{Sec:RelevantGreenFunctions-GeneralAbelianChiralGaugeTheory}

All relevant 1PI Green functions required for the complete renormalisation are listed in App.~\ref{App:1LoopGreenFunctions-GeneralAbelianChiralGaugeTheory}. 
The standard Green functions, which provide all divergences, are collected in App.~\ref{App:1LoopStandardGreenFunctions-GeneralAbelianChiralGaugeTheory}, while the $\Delta$-operator-inserted Green functions giving rise to the BRST breaking contributions --- both divergent and finite --- are presented in App.~\ref{App:1LoopBreakingGreenFunctions-GeneralAbelianChiralGaugeTheory}.
In this subsection, we identify the subset of power-counting divergent Green functions that vanish identically and therefore do not contribute, and we comment on the reasons for their vanishing.

We begin with the ordinary 1PI Green functions.
Except for those listed in Tab.~\ref{Tab:NonContributingOrdinaryGreenFunctions}, all power-counting divergent Green functions must be evaluated (see App.~\ref{App:1LoopStandardGreenFunctions-GeneralAbelianChiralGaugeTheory}).
The Green functions shown in Tab.~\ref{Tab:NonContributingOrdinaryGreenFunctions}, although power-counting divergent, yield no non-vanishing UV divergent contributions and can therefore be omitted.
The underlying reasons are summarised in the remarks provided in the third column of the table.
\begin{table}[t!]
    \centering
    \begin{tabular}{|c|c|c|} \hline
         Degree of div. & Green function & Remark \\
         \hline\hline
         \rule{0pt}{2.3em} 2 & \makecell{$\langle B^{\mu} \phi \rangle^\mathrm{1PI}$,\\ $\langle B^{\mu} \phi^{\dagger} \rangle^\mathrm{1PI}$} & \makecell{Green func.\ $\sim \mathcal{O}(p^2)$ due to power counting.\\Field monomial has an open Lorentz index\\that needs to be saturated.} \\
         \hline
         \rule{0pt}{1.7em} 1 & \makecell{$\langle B^{\nu} B^{\mu} \phi \rangle^\mathrm{1PI}$,\\ $\langle B^{\nu} B^{\mu} \phi^{\dagger} \rangle^\mathrm{1PI}$} & \makecell{Green func.\ $\sim \mathcal{O}(p)$ due to power counting.\\Field monomial has no open Lorentz index.} \\
         \hline
         \rule{0pt}{3em} 1 & \makecell{$\langle \phi \phi \phi \rangle^\mathrm{1PI}$,\\ $\langle \phi \phi \phi^{\dagger} \rangle^\mathrm{1PI}$,\\ $\langle \phi \phi^{\dagger} \phi^{\dagger} \rangle^\mathrm{1PI}$,\\ $\langle \phi^{\dagger} \phi^{\dagger} \phi^{\dagger} \rangle^\mathrm{1PI}$} & \makecell{Green func.\ $\sim \mathcal{O}(p)$ due to power counting.\\Field monomial has no open Lorentz index.} \\
         \hline
         \rule{0pt}{2.3em} 0 & \makecell{$\langle B^{\rho} B^{\nu} B^{\mu} \phi \rangle^\mathrm{1PI}$,\\ $\langle B^{\rho} B^{\nu} B^{\mu} \phi^{\dagger} \rangle^\mathrm{1PI}$} & \makecell{Green func.\ $\sim \mathcal{O}(p^0)$ due to power counting.\\Field monomial has an open Lorentz index\\that needs to be saturated.} \\
         \hline
         \rule{0pt}{3em} 0 & \makecell{$\langle B^{\mu} \phi \phi \phi \rangle^\mathrm{1PI}$,\\ $\langle B^{\mu} \phi \phi \phi^{\dagger} \rangle^\mathrm{1PI}$,\\ $\langle B^{\mu} \phi \phi^{\dagger} \phi^{\dagger} \rangle^\mathrm{1PI}$,\\ $\langle B^{\mu} \phi^{\dagger} \phi^{\dagger} \phi^{\dagger} \rangle^\mathrm{1PI}$} & \makecell{Green func.\ $\sim \mathcal{O}(p^0)$ due to power counting.\\Field monomial has an open Lorentz index\\that needs to be saturated.} \\
         \hline
    \end{tabular}
    \caption{Power-counting divergent 1PI Green functions that vanish a priori, and thus do not contribute.
    The first column lists the overall degree of divergence, the second lists the specific Green functions, and the final column provides a remark implying the reason why each contribution vanishes.
    In a massless theory, momenta provide the only scales, and divergences are restricted by the power-counting degree of divergence and must correspond to field monomials with saturated Lorentz indices.}
    \label{Tab:NonContributingOrdinaryGreenFunctions}
\end{table}

Next, we turn to the 1PI Green functions with one insertion of the breaking operator $\DeltaHat$.
As before, all power-counting divergent Green functions must, in principle, be considered (see App.~\ref{App:1LoopBreakingGreenFunctions-GeneralAbelianChiralGaugeTheory}), except for those that vanish in the first place.
These non-contributing Green functions are listed in Tab.~\ref{Tab:NonContributingDeltaGreenFunctions}, with explanations given in the third column.
The underlying reasons are largely analogous to those discussed for the ordinary Green functions (cf.\ Tab.~\ref{Tab:NonContributingOrdinaryGreenFunctions}), except for the final entry.
The Green function in the last row of Tab.~\ref{Tab:NonContributingDeltaGreenFunctions} appears initially to be admissible.
However, in an Abelian gauge theory, such a contribution to $\Delta\cdot\Gamma$ could only arise from the BRST transformation of a non-renormalisable operator $\propto BBBBB$, which is excluded in a renormalisable theory.
This situation differs in non-Abelian gauge theories, where the more involved BRST structure, particularly the transformations involving the gauge bosons, can generate such terms.
\begin{table}[t!]
    \centering
    \begin{tabular}{|c|c|c|} \hline
         Degree of div. & Green function & Remark \\
         \hline\hline
         \rule{0pt}{1.7em} 3 & \makecell{$\langle \Delta \phi c \rangle^\mathrm{1PI}$,\\$\langle \Delta \phi^{\dagger} c \rangle^\mathrm{1PI}$} & \makecell{Green func.\ $\sim \mathcal{O}(p^3)$ due to power counting.\\Field monomial has no open Lorentz index.} \\
         \hline
         \rule{0pt}{2.25em} 2 & \makecell{$\langle \Delta B^{\mu} \phi c \rangle^\mathrm{1PI}$,\\$\langle \Delta B^{\mu} \phi^{\dagger} c \rangle^\mathrm{1PI}$} & \makecell{Green func.\ $\sim \mathcal{O}(p^2)$ due to power counting.\\Field monomial has an open Lorentz index\\that needs to be saturated.} \\
         \hline
         \rule{0pt}{2.9em} 1 & \makecell{$\langle \Delta \phi \phi \phi c \rangle^\mathrm{1PI}$,\\$\langle \Delta \phi \phi \phi^{\dagger} c \rangle^\mathrm{1PI}$,\\$\langle \Delta \phi \phi^{\dagger} \phi^{\dagger} c \rangle^\mathrm{1PI}$,\\$\langle \Delta \phi^{\dagger} \phi^{\dagger} \phi^{\dagger} c \rangle^\mathrm{1PI}$} & \makecell{Green func.\ $\sim \mathcal{O}(p)$ due to power counting.\\Field monomial has no open Lorentz index.} \\
         \hline
         \rule{0pt}{1.7em} 1 & \makecell{$\langle \Delta B^{\nu} B^{\mu} \phi c \rangle^\mathrm{1PI}$,\\$\langle \Delta B^{\nu} B^{\mu} \phi^{\dagger} c \rangle^\mathrm{1PI}$} & \makecell{Green func.\ $\sim \mathcal{O}(p)$ due to power counting.\\Field monomial has no open Lorentz index.} \\
         \hline
         \rule{0pt}{2.9em} 0 & \makecell{$\langle \Delta B^{\mu} \phi \phi \phi c \rangle^\mathrm{1PI}$,\\$\langle \Delta B^{\mu} \phi \phi \phi^{\dagger} c \rangle^\mathrm{1PI}$,\\$\langle \Delta B^{\mu} \phi \phi^{\dagger} \phi^{\dagger} c \rangle^\mathrm{1PI}$,\\$\langle \Delta B^{\mu} \phi^{\dagger} \phi^{\dagger} \phi^{\dagger} c \rangle^\mathrm{1PI}$} & \makecell{Green func.\ $\sim \mathcal{O}(p^0)$ due to power counting.\\Field monomial has an open Lorentz index\\that needs to be saturated.} \\
         \hline
         \rule{0pt}{2.25em} 0 & \makecell{$\langle \Delta B^{\rho} B^{\nu} B^{\mu} \phi c \rangle^\mathrm{1PI}$,\\$\langle \Delta B^{\rho} B^{\nu} B^{\mu} \phi^{\dagger} c \rangle^\mathrm{1PI}$} & \makecell{Green func.\ $\sim \mathcal{O}(p^0)$ due to power counting.\\Field monomial has an open Lorentz index\\that needs to be saturated.} \\
         \hline
         \rule{0pt}{1.6em} 0 & \makecell{$\langle \Delta B^{\sigma} B^{\rho} B^{\nu} B^{\mu} c \rangle^\mathrm{1PI}$} &
         \makecell{Could only emerge from $s(BBBBB)$,\\which is non-renormalisable.} \\
         \hline
    \end{tabular}
    \caption{Power-counting divergent, single $\Delta$-operator inserted 1PI Green functions that vanish a priori.
    The first column indicates the overall degree of divergence, the second lists the specific operator-inserted Green functions, and the final column specifies the reason for their vanishing.
    In a massless theory, momenta constitute the only available scales, and allowed field monomials are restricted by both Lorentz invariance and renormalisability.}
    \label{Tab:NonContributingDeltaGreenFunctions}
\end{table}

\subsection{Singular Counterterm Action at the 1-Loop Level}\label{Sec:S_sct-GeneralAbelianChiralGaugeTheory}

Here we present the 1-loop singular counterterm action, adjusted so that it cancels all UV divergences arising in the Green functions shown in App.~\ref{App:1LoopStandardGreenFunctions-GeneralAbelianChiralGaugeTheory}.
We decompose it into a BRST-invariant and a non-invariant, symmetry-restoring part as
\begin{equation}\label{Eq:Ssct_1-Loop-GeneralAbelianTheory}
    \begin{aligned}
        S^{(1)}_{\mathrm{sct}} = S^{(1)}_{\mathrm{sct,inv}} + S^{(1)}_{\mathrm{sct,break}}.
    \end{aligned}
\end{equation}
The BRST-invariant part of the singular counterterm action is given by
\begin{equation}\label{Eq:Ssct_1-Loop_inv-GeneralAbelianTheory}
    \begin{aligned}
        S^{(1)}_{\mathrm{sct,inv}} = 
        - \frac{1}{16\pi^2} \, \frac{1}{\epsilon} &\Dintx 
        \bigg[  
        g^2 \! \prescript{}{F}{\overline{\mathcal{A}}}_{BB}^{1,\mathrm{inv}}
        \Big(-\frac{1}{4}\overline{F}^{\mu\nu}\overline{F}_{\mu\nu}\Big)
        + g^2
        \prescript{}{S}{\mathcal{A}}_{BB}^{1,\mathrm{inv}}
        \Big(-\frac{1}{4}F^{\mu\nu}F_{\mu\nu}\Big)\\
        &+ \overline{\mathcal{A}}_{\psi\overline{\psi},\mathrm{R},kj}^{1,\mathrm{inv}} \overline{\psi}_i i \overline{\slashed{D}}_{\mathrm{R},ik} \psi_j
        + \overline{\mathcal{A}}_{\psi\overline{\psi},\mathrm{L},kj}^{1,\mathrm{inv}} \overline{\psi}_i i \overline{\slashed{D}}_{\mathrm{L},ik} \psi_j\\
        &+ \Big( \overline{\psi}_i 
        \Big[
        \overline{\mathcal{A}}_{\psi\overline{\psi}\phi,\mathrm{R},ij}^{1,\mathrm{inv},a} \,
        \phi_a
        + 
        {\overline{\mathcal{A}}_{\psi\overline{\psi}\phi,\mathrm{L},ji}^{1,\mathrm{inv},a}}^{\hspace{-0.29cm}\dagger} \hspace{0.2cm} 
        \phi_a^{\dagger}
        \Big]
        \mathbb{P}_{\mathrm{R}}
        \psi_j 
        + \mathrm{h.c.}\Big)\\
        &+ g^2 \! \prescript{}{S}{\mathcal{A}}_{\phi\phi^{\dagger},ab}^{1,\mathrm{inv}} 
        \big(D^{\mu}\phi_a\big)^{\dagger}\big(D_{\mu}\phi_b\big)
        + \prescript{}{Y}{\overline{\mathcal{A}}}_{\phi\phi^{\dagger},ab}^{1,\mathrm{inv}}
        \big(\overline{D}^{\mu}\phi_a\big)^{\dagger}\big(\overline{D}_{\mu}\phi_b\big)\\
        &+ \frac{1}{4} \, \mathcal{A}_{\phi\phi^{\dagger}\phi\phi^{\dagger},abcd}^{1,\mathrm{inv}}
        \, \phi_a^{\dagger} \phi_b \phi_c^{\dagger} \phi_d
        \bigg],
    \end{aligned}
\end{equation}
with covariant derivatives
\begin{equation}
    \begin{aligned}
        \overline{D}^{\mu}_{\mathrm{L/R},ij} &=
        \Big( \overline{\partial}^{\mu} \delta_{ij} + i g \mathcal{Y}_{L/R,ij} \overline{B}^{\mu} \Big) \mathbb{P}_{\mathrm{L/R}},\\
        D^{\mu} \phi_a &=
        \Big( \partial^{\mu} + i g \hypS B^{\mu} \Big) \phi_a.
    \end{aligned}
\end{equation}
This part of the counterterm action is manifestly BRST invariant, it mirrors the symmetric tree-level structure (cf.\ Sec.~\ref{Sec:Definition_of_the_Theory_General_Abelian_Case}) and corresponds to multiplicative renormalisation.

The non-invariant, symmetry-restoring part of the singular counterterm action reads
\begin{align}\label{Eq:Ssct_1-Loop_break-GeneralAbelianTheory}
        S&^{(1)}_{\mathrm{sct,break}} = \frac{1}{16\pi^2} \frac{1}{\epsilon} \Dintx
        \bigg[
        \frac{g^2}{2}
        \bigg\{ \!
        \Big(
        \prescript{}{F}{\widehat{\mathcal{A}}}_{BB,1}^{1,\mathrm{break}}
        + \prescript{}{F}{\widehat{\mathcal{A}}}_{BB,3}^{1,\mathrm{break}}
        \Big)
        \widehat{B}_{\mu}\overline{\Box}\widehat{B}^{\mu}
        + 
        \prescript{}{F}{\widehat{\mathcal{A}}}_{BB,4}^{1,\mathrm{break}}
        \overline{B}_{\mu}\widehat{\Box}\overline{B}^{\mu}
        \nonumber\\
        &+
        \frac{2}{3} \prescript{}{F}{\widehat{\mathcal{A}}}_{BB,3}^{1,\mathrm{break}}
        \widehat{B}_{\mu}\widehat{\Box}\widehat{B}^{\mu}
        - 2 \prescript{}{F}{\widehat{\mathcal{A}}}_{BB,2}^{1,\mathrm{break}}
        \big(\overline{\partial}\cdot\overline{B}\big) \big(\widehat{\partial}\cdot\widehat{B}\big)
        - 2 \prescript{}{F}{\widehat{\mathcal{A}}}_{BB,1}^{1,\mathrm{break}}
        \big(\widehat{\partial}\cdot\widehat{B}\big)^2
        \bigg\}
        \nonumber\\
        &- \Big[ \overline{\psi}_i i
        \Big(
        \widehat{\slashed{\partial}} \widehat{\mathcal{A}}_{\psi\overline{\psi},ij}^{1,\mathrm{break}}
        + i g \widehat{\slashed{B}} \widehat{\mathcal{A}}_{\psi\overline{\psi}B,ij}^{1,\mathrm{break}}
        \Big) \projR
        \psi_j + \mathrm{h.c.} \Big]
        \\
        &+
        {\widehat{\mathcal{A}}}_{\phi\phi^{\dagger},ab}^{1,\mathrm{break}} \phi_{a}^{\dagger} \widehat{\Box} \phi_b
        - i g {\widehat{\mathcal{A}}}_{\phi\phi^{\dagger}B,ab}^{1,\mathrm{break}}
        \Big[ \big(\widehat{\partial}^{\mu}\phi_{a}^{\dagger}\big)\phi_b 
        - \phi_{a}^{\dagger}\big(\widehat{\partial}^{\mu}\phi_b\big) \Big] \widehat{B}_{\mu}
        - \frac{g^2}{2} {\widehat{\mathcal{A}}}_{\phi\phi^{\dagger}BB,ab}^{1,\mathrm{break}}
        \phi_{a}^{\dagger}\phi_{b} \widehat{B}^{\mu} \widehat{B}_{\mu}
        \nonumber\\
        &+
        \frac{1}{2} \bigg\{ \!
        {\widehat{\mathcal{A}}}_{\phi\phi,ab}^{1,\mathrm{break}} \phi_{a} \widehat{\Box} \phi_b
        - i g {\widehat{\mathcal{A}}}_{\phi\phi B,ab}^{1,\mathrm{break}}
        \Big[ \big(\widehat{\partial}^{\mu}\phi_{a}\big)\phi_b
        - \phi_{a}\big(\widehat{\partial}^{\mu}\phi_b\big) \Big] \widehat{B}_{\mu}
        \nonumber\\
        &
        - \frac{g^2}{2} {\widehat{\mathcal{A}}}_{\phi\phi BB,ab}^{1,\mathrm{break}}
        \phi_{a} \phi_{b} \widehat{B}^{\mu} \widehat{B}_{\mu} + \mathrm{h.c.}
        \bigg\}
        \bigg].\nonumber
\end{align}
All singular non-symmetric 1-loop counterterms are, as expected, evanescent, since they originate from the purely evanescent breaking $\widehat{\Delta}$.
The first two lines contain bilinear gauge boson terms and feature several structures that vanish for $\hypLR\equiv0$, leaving only the contribution $\propto\overline{B}_{\mu}\widehat{\Box}\overline{B}^{\mu}$.

The fermionic terms in the third line of Eq.~\eqref{Eq:Ssct_1-Loop_break-GeneralAbelianTheory} cannot be expressed in terms of a covariant derivative because the kinetic and the gauge interaction parts carry different coefficients.
These terms persist for $\hypLR\equiv0$, but they simplify considerably (see Eqs.~\eqref{App-Eq:DivCoeffRelations} and \eqref{App-Eq:FermionCoeffFermionGauge}--\eqref{App-Eq:FermionGaugeBosonCoeff}).

The scalar and scalar--gauge boson terms displayed in the last three lines of Eq.~\eqref{Eq:Ssct_1-Loop_break-GeneralAbelianTheory} also cannot be recast into a covariant derivative and break BRST invariance.
Furthermore, the last two lines violate global hypercharge conservation, in agreement with the discussion of Sec.~\ref{Sec:Dimensional_Ambiguities_and_Evanescent_Shadows}.
The regularisation-induced violation of global hypercharge conservation in the BMHV scheme has two sources:
the evanescent part of a fermion kinetic term that mixes physical left- and right-handed fermions with different gauge quantum numbers (realised according to Option~\ref{Opt:Option1}), and the evanescent gauge interactions.

The global hypercharge-violating scalar kinetic terms with coefficient ${\widehat{\mathcal{A}}}_{\phi\phi,ab}^{1,\mathrm{break}}$ are independent of $\hypLR$ and thus arise solely from the evanescent piece of the fermion kinetic term.
Therefore, they vanish only in the sterile-partner construction of Option~\ref{Opt:Opt2-GroupRef} (see Sec.~\ref{Sec:Dimensional_Ambiguities_and_Evanescent_Shadows}), and assigning suitable quantum numbers to the sterile partner fields.
In contrast, the scalar--gauge boson interaction terms, with coefficients ${\widehat{\mathcal{A}}}_{\phi\phi B,ab}^{1,\mathrm{break}}$ and ${\widehat{\mathcal{A}}}_{\phi\phi BB,ab}^{1,\mathrm{break}}$, emerge entirely from the evanescent gauge interactions and thus vanish for $\hypLR\equiv0$ (cf.\ Eqs.~\eqref{App-Eq:Div2ScalarCoeffs}, \eqref{App-Eq:SSBCoeff} and \eqref{App-Eq:SSBBCoeffs}).

Overall, the divergent symmetry-restoring part of the counterterm action $S^{(1)}_{\mathrm{sct,break}}$, shown in Eq.~\eqref{Eq:Ssct_1-Loop_break-GeneralAbelianTheory}, simplifies considerably for vanishing evanescent gauge interactions, i.e.\ $\hypLR\equiv0$.

\subsection{Finite Symmetry-Restoring Counterterm Action at the 1-Loop Level}\label{Sec:S_fct-GeneralAbelianChiralGaugeTheory}

The finite symmetry-restoring counterterms $S^{(1)}_{\mathrm{fct}}$ are determined by the condition that the Slavnov-Taylor identity is fulfilled after renormalisation, as required by Eq.~\ref{Eq:UltimateSymmetryRequirement} (see Sec.~\ref{Sec:Symmetry_Restoration_Procedure} for a detailed discussion).
Since BRST transformations do not renormalise in the Abelian case, this requirement together with the quantum action principle imply the implicit relation $\DeltaHat\cdot\Gamma\big|_{\mathrm{fin}}^{(1)}=-s_DS^{(1)}_{\mathrm{fct}}$ for the finite counterterms (cf.\ Eq.~\eqref{Eq:Determination-of-Symmetry-Restoring-CT-Abelian-Case}).
The corresponding $\DeltaHat$-inserted Green functions are listed in App.~\ref{App:1LoopBreakingGreenFunctions-GeneralAbelianChiralGaugeTheory}.
Solving the implicit relation yields the 1-loop finite symmetry-restoring counterterm action
\begin{equation}\label{Eq:Sfct_1-Loop-GeneralAbelianTheory}
    \begin{aligned}
        S^{(1)}_{\mathrm{fct}} &= \frac{1}{16\pi^2} \intx \bigg[
        \frac{g^2}{2} \mathcal{F}_{BB}^{1,\mathrm{break}} \, \overline{B}_{\mu} \overline{\Box} \, \overline{B}^{\mu}
        + \frac{g^4}{8} \mathcal{F}_{BBBB}^{1,\mathrm{break}} \, \overline{B}_{\mu} \overline{B}^{\mu} \overline{B}_{\nu} \overline{B}^{\nu}\\
        &+ g \, \overline{\psi}_i \overline{\slashed{B}} \Big[
        \mathcal{F}_{\psi\overline{\psi}B,\mathrm{R},ij}^{1,\mathrm{break}}
        \mathbb{P}_{\mathrm{R}}
        +
        \mathcal{F}_{\psi\overline{\psi}B,\mathrm{L},ij}^{1,\mathrm{break}}
        \mathbb{P}_{\mathrm{L}}
        \Big] \psi_j\\
        &-
        \Big(
        \overline{\psi}_i 
        \Big[ 
        \mathcal{F}_{\psi\overline{\psi}\phi,ij}^{1,\mathrm{break},a} \phi_{a}
        + \mathcal{F}_{\psi\overline{\psi}\phi^{\dagger},ij}^{1,\mathrm{break},a} \phi_{a}^{\dagger}
        \Big]
        \projR
        {\psi}_j
        + \mathrm{h.c.} \Big)
        \\ 
        &+ ig\, \mathcal{F}_{\phi\phi^{\dagger}B,ab}^{1,\mathrm{break}}
        \Big[
        \big(\overline{\partial}^{\mu}\phi_{a}^{\dagger}\big)\phi_b
        - \phi_{a}^{\dagger}\big(\overline{\partial}^{\mu}\phi_b\big) 
        \Big] 
        \overline{B}_{\mu}
        + \frac{g^2}{2} \mathcal{F}_{\phi\phi^{\dagger}BB,ab}^{1,\mathrm{break}}
        \phi_{a}^{\dagger}\phi_{b}\overline{B}^{\mu}\overline{B}_{\mu}\\
        &- \frac{1}{2} 
        \Big(
        \mathcal{F}_{\phi\phi,ab}^{1,\mathrm{break}}
        \Big[
        \overline{\partial}_{\mu}\phi_{a}\overline{\partial}^{\mu}\phi_{b}
        - 3 i g \mathcal{Y}_{S} \overline{\partial}_{\mu}\big(\phi_a\phi_b\big)\overline{B}^{\mu}
        + 3 g^2 \mathcal{Y}_{S}^2 \phi_a \phi_b \overline{B}_{\mu} \overline{B}^{\mu}
        \Big]
        + \mathrm{h.c.} \Big)\\
        &-
        \Big(
        \frac{1}{96} \, \mathcal{F}_{\phi\phi\phi\phi,abcd}^{1,\mathrm{break}} \,
        \phi_{a}\phi_{b}\phi_{c}\phi_{d}
        + \frac{1}{12} \, \mathcal{F}_{\phi^{\dagger}\phi\phi\phi,abcd}^{1,\mathrm{break}} \,
        \phi_{a}^{\dagger}\phi_{b}\phi_{c}\phi_{d}
        + \mathrm{h.c.} \Big)
        \bigg].
    \end{aligned}
\end{equation}
It contains eleven independent coefficients, given in Eqs.~\eqref{App-Eq:FiniteFermionGaugeBosonCoeffsFull}, \eqref{App-Eq:FiniteGaugeBosonCoeffs}, and \eqref{App-Eq:FiniteYukawaCoeffs}--\eqref{App-Eq:FiniteScalarGaugeBosonCoeffs}.
Unlike the divergent symmetry-breaking terms in Eq.~\eqref{Eq:Ssct_1-Loop_break-GeneralAbelianTheory}, none of these vanish entirely for $\hypLR\equiv0$.

Following the standard procedure (see Sec.~\ref{Sec:Symmetry_Restoration_Procedure}, and Sec.~6.3.3 of Ref.~\cite{Belusca-Maito:2023wah} regarding the residual freedom in finite counterterms), one may freely add finite but symmetric counterterms to $S^{(1)}_{\mathrm{fct}}$.
Here, the breaking related to the fermion self energy and the fermion--gauge boson interaction is fully assigned to the latter --- rather than distributed or attributed solely to the self energy as in Refs.~\cite{Belusca-Maito:2021lnk,Stockinger:2023ndm}.
This choice is motivated because the additional breaking from the Green function $\langle\Delta \psi \overline{\psi} B c\rangle^\mathrm{1PI}$ (see Eq.~\eqref{Eq:FFbarBc-GreenFunc-GeneralAbelianTheory}) is associated exclusively to the gauge interaction vertex and not to the self energy, as seen from Eq.~\eqref{Eq:STI-QAP-FFbarBc}, implying that it cannot be attributed to the self energy.
This breaking contribution, Eq.~\eqref{Eq:FFbarBc-GreenFunc-GeneralAbelianTheory}, arises from evanescent gauge interactions $\propto\hypLR$ and from the $\widehat{\Delta}_1\big[c,\overline{\psi},\psi\big]$-part of the breaking (see Eq.~\ref{Eq:GeneralTreeLevelBreaking_Abelian}) involving scalar couplings.
In contrast, the breaking from $\langle\Delta \psi \overline{\psi} c\rangle^\mathrm{1PI}$ (see Eq.~\eqref{Eq:FbarFc-GreenFunc-GeneralAbelianTheory}) involves both, as expressed in Eq.~\eqref{Eq:STI-QAP-FFc}.

Analogously, the scalar--gauge boson breaking is entirely attributed to the corresponding interaction vertices rather than to the scalar self energy.
As shown by Eqs.~\eqref{Eq:GhostDoubleScalar_1-Loop-GeneralAbelianTheory} and \eqref{Eq:GreenFunc-c-B-phiDagger-phi-GeneralAbelianTheory}, non-vanishing breakings originate from $\langle\Delta \phi \phi^\dagger c\rangle^\mathrm{1PI}$ and $\langle\Delta \phi \phi^\dagger B c\rangle^\mathrm{1PI}$, while the contribution from $\langle\Delta \phi \phi^\dagger B B c\rangle^\mathrm{1PI}$ vanishes (see Eq.~\eqref{Eq:SSbarBBc-DeltaGreenFunc-GeneralAbelianTheory}).
Since $\langle\Delta \phi \phi^\dagger B c\rangle^\mathrm{1PI}$ is linked solely to the scalar--gauge boson interaction (see Eq.~\eqref{Eq:STI-QAP-SSbarBc}), the breaking cannot be shifted to the scalar self energy.

As in the divergent case, the finite counterterms include contributions that break global hypercharge conservation.
This is most evident in the last two lines of Eq.~\eqref{Eq:Sfct_1-Loop-GeneralAbelianTheory}.
The relevant coefficients, given in Eqs.~\eqref{App-Eq:Finite2ScalarCoeff}, \eqref{App-Eq:Finite4ScalarCoeff-SSSS}, and \eqref{App-Eq:Finite4ScalarCoeff-SdaggerSSS}, are independent of the evanescent hypercharges and vanish only when adopting Option~\ref{Opt:Opt2-GroupRef}, where sterile partner fields are introduced for each fermion.
Since all these terms violate global hypercharge conservation, no symmetric counterterm can be added to redistribute the breaking without also breaking BRST invariance.
Among the scalar terms, the quartic interaction $\phi^{\dagger}\phi\phi^{\dagger}\phi$ is BRST invariant by itself, implying that any BRST-breaking contribution involving four scalars must also violate global hypercharge.

Similarly, the finite Yukawa breakings in the third line of Eq.~\eqref{Eq:Sfct_1-Loop-GeneralAbelianTheory} break global hypercharge conservation because their coefficients (see Eq.~\eqref{App-Eq:FiniteYukawaCoeffs}) combine fields whose hypercharges do not sum to zero unless the term is BRST invariant and hence removable by a symmetric counterterm.
Unlike the scalar case, however, these Yukawa breakings receive contributions from both evanescent sources --- the fermion kinetic term and the gauge interactions --- and can only be eliminated when both are suppressed, i.e.\ by applying Option~\ref{Opt:Opt2-GroupRef} and setting $\hypLR\equiv0$.

Together with the non-invariant divergent counterterms in Eq.~\eqref{Eq:Ssct_1-Loop_break-GeneralAbelianTheory}, the finite symmetry-restoring shown in Eq.~\eqref{Eq:Sfct_1-Loop-GeneralAbelianTheory} reestablish the validity of the Slavnov-Taylor identity at the 1-loop level, thereby restoring local gauge and BRST invariance as well as global hypercharge conservation.

\section{Shedding Light on Evanescent Shadows}\label{Sec:Results-Shedding_Light_on_Evanescent_Shadows}

In Sec.~\ref{Sec:Dimensional_Ambiguities_and_Evanescent_Shadows}, we discussed different admissible options for extending the fermion sector to $D$ dimensions and the accompanying dimensional ambiguities caused by the non-uniqueness of the $D$-dimensional continuation.
In Sec.~\ref{Sec:General_Structure_of_the_Symmetry_Breaking}, we then derived the general structure of the tree-level breaking and examined the implications of these ambiguities in general terms.
We showed that the evanescent gauge interactions cannot be adjusted to remove the tree-level breaking entirely, because the axial part of the interaction current cannot be extended uniformly to $D$ dimensions without violating hermiticity.
Nonetheless, we motivated several consistent choices for the evanescent hypercharges.

In the present section, we extend this analysis to the 1-loop level and study how different $D$-dimensional continuations of the fermions and evanescent hypercharge assignments affect the symmetry breaking.
For this purpose, we focus on specific specialisations of the general model.
We first consider the fermionic sector in the absence of scalar fields in Sec.~\ref{Sec:FermionicSectorNoScalars}, and focus particularly on the breaking arising from the fermion kinetic term.
Moreover, this serves as a consistency check, allowing comparison with established results (see Sec.~\ref{Sec:ModelsFromLiterature}).
In Sec.~\ref{Sec:AnalysisOfASSMResults}, we then turn to the Abelian sector of the SM (see Sec.~\ref{Sec:ASSM}) owing to its phenomenological relevance.
Here, both sources of symmetry breaking are included, and we allow for general, non-universal evanescent hypercharges, using Eq.~\eqref{Eq:ASSM-YLR} for Option~\ref{Opt:Option1} and Eq.~\eqref{Eq:ASSM-YLR-YRL-Option2} for Option~\ref{Opt:Opt2-GroupRef}.
Finally, Sec.~\ref{Sec:TheRoleOf-YLR-YRL} focuses on the chiral gauge interactions in $D$ dimensions and explores several concrete assignments of the evanescent hypercharges, using the Abelian sector of the SM as an illustrative case study.

\subsection{Exploring Options for the Fermionic Sector}\label{Sec:FermionicSectorNoScalars}

We first compare our results with Refs.~\cite{Martin:1999cc} and \cite{Cornella:2022hkc} in the corresponding special cases as discussed in Sec.~\ref{Sec:ModelsFromLiterature}.
For this purpose, we discard all scalar contributions and the evanescent gauge interactions, i.e.\ $\hypLR=\hypRL=0$, which leaves only the evanescent part of the fermion kinetic term as a source of symmetry breaking.
Under these assumptions, the singular and finite counterterm actions of Eqs.~\eqref{Eq:Ssct_1-Loop_inv-GeneralAbelianTheory}, \eqref{Eq:Ssct_1-Loop_break-GeneralAbelianTheory} and \eqref{Eq:Sfct_1-Loop-GeneralAbelianTheory} reduce to
\begin{align}
    \begin{split}\label{Eq:Ssct_YLYR_1-Loop-GeneralAbelianTheory}
        S^{(1)}_{\mathrm{sct,\mathrm{R+L}}} &= - \frac{g^2}{16\pi^2} \, \frac{1}{\epsilon} \Dintx 
        \bigg[
        \frac{2}{3} \mathrm{Tr}\big(\hypR^2+\hypL^2\big)
        \Big(-\frac{1}{4}\overline{F}^{\mu\nu}\overline{F}_{\mu\nu}\Big)\\
        &\qquad\qquad\qquad\;\;\;
        + \xi (\hypR^2)_{kj}
        \overline{\psi}_i i \overline{\slashed{D}}_{\mathrm{R},ik} \psi_j
        + \xi (\hypL^2)_{kj}
        \overline{\psi}_i i \overline{\slashed{D}}_{\mathrm{L},ik} \psi_j\\
        &\qquad\qquad\qquad\;\;\;
        + \frac{1}{3} \mathrm{Tr}\big((\hypR+\hypL)^2\big) \frac{1}{2} \overline{B}_{\mu}\widehat{\Box}\overline{B}^{\mu}
        + 2 \frac{2+\xi}{3} (\hypR\hypL)_{ij} \overline{\psi}_i i \widehat{\slashed{\partial}} \psi_j
        \bigg],
    \end{split}\\[1.5ex]
    \begin{split}\label{Eq:Sfct_YLYR_1-Loop-GeneralAbelianTheory}
        S^{(1)}_{\mathrm{fct,\mathrm{R+L}}} &=
        \frac{1}{16\pi^2} \intx 
        \bigg[-\frac{g^2}{3} \mathrm{Tr}\big((\hypR-\hypL)^2\big) \frac{1}{2} \overline{B}_{\mu} \overline{\Box} \, \overline{B}^{\mu}\\
        &\qquad\qquad\qquad\;\;\;\;
        + \frac{2 g^4}{3} \mathrm{Tr}\big((\hypR-\hypL)^4\big) \frac{1}{8} \overline{B}_{\mu} \overline{B}^{\mu} \overline{B}_{\nu} \overline{B}^{\nu}\\
        &\qquad\qquad\qquad\;\;\;\;
        + \frac{5+\xi}{6} g^3 (\hypR-\hypL)_{ik} \overline{\psi}_i \overline{\slashed{B}} 
        \Big( (\hypR^2)_{kj} \mathbb{P}_{\mathrm{R}}
        - (\hypL^2)_{kj} \mathbb{P}_{\mathrm{L}} \Big) \psi_j
        \bigg].
    \end{split}
\end{align}
Only the physical hypercharges $\hypL$ and $\hypR$ are left.
Equation~\eqref{Eq:Sfct_YLYR_1-Loop-GeneralAbelianTheory} exactly reproduces the finite symmetry-restoring counterterms of Ref.~\cite{Cornella:2022hkc} in the Abelian limit (where the considered gauge group is $U(1)$), and, in general, agrees with Refs.~\cite{Martin:1999cc,Cornella:2022hkc} for block-diagonal or non-block-diagonal hypercharge and fermion multiplet assignments (see rows three and four of Tab.~\ref{Tab:SpecialCasesOverview}).

These results also expose the impact of the specific $D$-dimensional realisation of the fermions (see Sec.~\ref{Sec:Dimensional_Ambiguities_and_Evanescent_Shadows}).
In Option~\ref{Opt:Opt2-GroupRef} the block structure (cf.\ Eq.~\eqref{Eq:ASSM-YLYR-Option2}) implies vanishing 
``left- and right-mixing'' contributions, i.e.\ $\hypL\hypR=0$.
Consequently, the evanescent and symmetry-breaking fermion term in the last line of Eq.~\eqref{Eq:Ssct_YLYR_1-Loop-GeneralAbelianTheory} drops out.
In contrast, in Option~\ref{Opt:Option1} (cf.\ Eq.~\eqref{Eq:ASSM-YLYR}), we have $\hypL\hypR\neq0$, and thus the corresponding fermion term remains, which gives Option~\ref{Opt:Opt2-GroupRef} a slight advantage.
For the gauge boson term in the last line of Eq.~\eqref{Eq:Ssct_YLYR_1-Loop-GeneralAbelianTheory} and for the entire finite symmetry-restoring counterterm action in Eq.~\eqref{Eq:Sfct_YLYR_1-Loop-GeneralAbelianTheory}, the choice of the fermion realisation affects only their prefactors, without causing any term to vanish.
The symmetric part of the singular counterterms (first two lines of Eq.~\eqref{Eq:Ssct_YLYR_1-Loop-GeneralAbelianTheory}) is identical for both options, as it does not originate from chirality-mixing sources.

As a further specialisation, we take $\hypL=0$ (row two of Tab.~\ref{Tab:SpecialCasesOverview}), or equivalently recast all left-handed fermions as right-handed fields with opposite charge.
Following Option~\ref{Opt:Option2b}, we add sterile left-handed partner fields and obtain
\begin{align}
    \begin{split}\label{Eq:Ssct_chiralQED_1-Loop-GeneralAbelianTheory}
        S^{(1)}_{\mathrm{sct,\chi QED}} &= - \frac{g^2}{16\pi^2} \, \frac{1}{\epsilon} \Dintx 
        \bigg[
        \frac{2}{3} \mathrm{Tr}\big(\hypR^2\big)
        \Big(-\frac{1}{4}\overline{F}^{\mu\nu}\overline{F}_{\mu\nu}\Big)
        + \xi (\hypR^2)_{kj}
        \overline{\psi}_i i \overline{\slashed{D}}_{\mathrm{R},ik} \psi_j\\
        &\qquad\qquad\qquad\quad\;\;\;
        + \frac{1}{3} \mathrm{Tr}\big(\hypR^2\big) \frac{1}{2} \overline{B}_{\mu}\widehat{\Box}\overline{B}^{\mu}
        \bigg],
    \end{split}\\[1.5ex]
    \begin{split}\label{Eq:Sfct_chiralQED_1-Loop-GeneralAbelianTheory}
        S^{(1)}_{\mathrm{fct,\chi QED}} &=
        \frac{1}{16\pi^2} \intx \bigg[
        -\frac{g^2}{3} \mathrm{Tr}\big(\hypR^2\big) \frac{1}{2} \overline{B}_{\mu} \overline{\Box} \, \overline{B}^{\mu}
        + \frac{2 g^4}{3} \mathrm{Tr}\big(\hypR^4\big) \frac{1}{8} \overline{B}_{\mu} \overline{B}^{\mu} \overline{B}_{\nu} \overline{B}^{\nu}\\
        &\qquad\qquad\qquad\;
        + \frac{5+\xi}{6} g^3 (\hypR^3)_{ij} \overline{\psi}_i \overline{\slashed{B}} \mathbb{P}_{\mathrm{R}} \psi_j
        \bigg],
    \end{split}
\end{align}
for the counterterm action, in agreement with right-handed chiral QED (see Refs.~\cite{Belusca-Maito:2021lnk,Belusca-Maito:2023wah,Stockinger:2023ndm,vonManteuffel:2025swv} and chapter~\ref{Chap:BMHV_at_Multi-Loop_Level}).
Here we have assigned the finite fermionic breaking entirely to the fermion--gauge boson vertex (last line of Eq.~\eqref{Eq:Sfct_chiralQED_1-Loop-GeneralAbelianTheory}); adding the symmetric finite counterterm $\frac{5+\xi}{6}g^2(\hypR^2)_{ik}\overline{\psi}_{i}i\overline{\slashed{D}}_{\mathrm{R},kj}\psi_j$ moves it to the self energy, reproducing exactly the results of Refs.~\cite{Belusca-Maito:2021lnk,Belusca-Maito:2023wah,Stockinger:2023ndm,vonManteuffel:2025swv}.
Consistently, the evanescent divergent fermion term in Eq.~\eqref{Eq:Ssct_YLYR_1-Loop-GeneralAbelianTheory} vanishes because $\hypL\hypR=0$ in this limit.

We further specialise to ordinary QED, serving as another consistency check of our results.
To this end, we restore the evanescent gauge couplings and take all fermion hypercharges equal to the electric charge $Q$ (first row of Tab.~\ref{Tab:SpecialCasesOverview}), which yields
\begin{equation}\label{Eq:Sct_QED_1-Loop-GeneralAbelianTheory}
    \begin{aligned}
        S^{(1)}_{\mathrm{ct,QED}} = 
        - \frac{g^2}{16\pi^2} \, \frac{1}{\epsilon} \Dintx 
        \bigg[  
        \frac{4}{3} \mathrm{Tr}\big(Q^2\big)
        \Big(-\frac{1}{4}F^{\mu\nu}F_{\mu\nu}\Big)
        + \xi (Q^2)_{kj}
        \overline{\psi}_i i \slashed{D}_{ik} \psi_j
        \bigg],
    \end{aligned}
\end{equation}
with the usual covariant derivative $D_{\mu}=\partial_{\mu}+igQB_{\mu}$.
This is the familiar 1-loop counterterm action of ordinary QED, fully BRST invariant and independent of the fermion realisation (as above, cf.\ symmetric part of Eq.~\eqref{Eq:Ssct_YLYR_1-Loop-GeneralAbelianTheory}).
Nonzero $\hypLR$ and $\hypRL$ are required to reproduce this standard form.

Finally, we consider the scenario $\hypR=\hypL=\mathcal{Y}$ with $\hypLR=\hypRL=0$, which corresponds to QED but keeping the fermion--gauge boson vertex strictly 4-dimensional.
Gauge invariance is then broken at the regularised level, so $\widehat{\Delta}\neq0$ and some divergent, evanescent breakings from the 1-loop Green functions $\langle \Delta B c \rangle^\mathrm{1PI}$, $\langle \Delta \psi \overline{\psi} c \rangle^\mathrm{1PI}$ and $\langle \Delta \phi \phi^{\dagger} c \rangle^\mathrm{1PI}$ (see Eqs.~\eqref{Eq:GhostGaugeBoson_1-Loop-GeneralAbelianTheory}, \eqref{Eq:FbarFc-GreenFunc-GeneralAbelianTheory} and \eqref{Eq:GhostDoubleScalar_1-Loop-GeneralAbelianTheory}) survive; all other 1-loop breakings vanish.
Crucially, all finite symmetry breakings vanish entirely, as noted in Ref.~\cite{Cornella:2022hkc}, which focuses exclusively on finite breakings.\footnote{%
Some coefficients in App.~\ref{App:1LoopCTCoeffs-GeneralAbelianChiralGaugeTheory} vanish trivially as they depend on $(\hypR-\hypL)$, while others --- such as the Yukawa and scalar coefficients in Eqs.~\eqref{App-Eq:FiniteYukawaCoeffs} and \eqref{App-Eq:Finite2ScalarCoeff}--\eqref{App-Eq:Finite4ScalarCoeff-SdaggerSSS} --- do not.
The latter vanish or reduce to symmetric finite counterterms for $\hypL=\hypR$ once the BRST constraint of Eq.~\eqref{Eq:YukawaHyperchargeBRSTCondition}, relating Yukawa and hypercharge matrices, is applied.
}
This follows from the tree-level breaking (obtained from Eq.~\eqref{Eq:GeneralTreeLevelBreaking_Abelian} for $\hypL=\hypR$, $\hypLR=\hypRL=0$),
\begin{equation}\label{Eq:BRST-Breaking-YL=YR}
        \begin{aligned}
            \widehat{\Delta}_{\mathrm{L}=\mathrm{R}}
            &= - g \mathcal{Y}_{ij} \Dintx \, c \widehat{\partial}_{\mu} \Big( \overline{\psi}_i \widehat{\gamma}^{\mu} \psi_j \Big)
            \,\,\,\,\Longrightarrow
        \end{aligned}
    \begin{tabular}{rl}
        \raisebox{-31.5pt}{\includegraphics[scale=0.5]{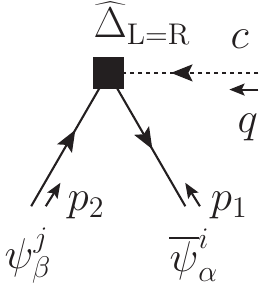}}&
        $\begin{aligned}
            &= g \mathcal{Y}_{ij} \widehat{\slashed{q}}.
        \end{aligned}$
    \end{tabular}
\end{equation}
Every $\widehat{\Delta}_{\mathrm{L}=\mathrm{R}}$-inserted diagram vanishes for purely 4-dimensional external ghost momentum $q$.
At the 1-loop level this removes all 4-dimensional finite symmetry-restoring counterterms; subrenormalisation at higher orders (especially in non-Abelian theories, where the ghost can be internal and hence the momentum $D$-dimensional) may change this conclusion (further investigations are required).
Here, in the Abelian case at the 1-loop level, divergent evanescent breakings remain, such as breaking contributions $\propto\widehat{q}^2\overline{q}^{\mu}$ from
$\langle \Delta B c \rangle^\mathrm{1PI}$ and $\propto\widehat{\slashed{q}}$ from $\langle \Delta \psi \overline{\psi} c \rangle^\mathrm{1PI}$.

In summary, changes in the $D$-dimensional fermion realisation only affect counterterms tied to chirality-mixing sources of breaking; the purely symmetric (multiplicative) renormalisation pieces are insensitive to the chosen option.

\subsection{The Abelian Sector of the Standard Model}\label{Sec:AnalysisOfASSMResults}

In this section, the general model of Sec.~\ref{Sec:Definition_of_the_Theory_General_Abelian_Case} is specialised to the Abelian sector of the SM (see Sec.~\ref{Sec:ASSM}).
We treat fermions in $D$ dimensions using both approaches (Option~\ref{Opt:Option1} and \ref{Opt:Opt2-GroupRef}) and allow for a general realisation of evanescent gauge interactions (see Sec.~\ref{Sec:Dimensional_Ambiguities_and_Evanescent_Shadows}).
In each case, we provide the complete counterterm action --- including divergent and finite symmetry-restoring counterterms --- for general evanescent hypercharge assignments.

We begin with the ``natural'' treatment of fermions, Option~\ref{Opt:Option1}, where the two chiral components of each fermion are combined into a single Dirac spinor.
The fermion content and evanescent hypercharges are those of Eqs.~\eqref{Eq:ASSM-FermionMultiplets} and \eqref{Eq:ASSM-YLR}.
We then turn to Option~\ref{Opt:Option2}, in which each fermion is coupled to a sterile partner field to avoid the breaking of global hypercharge conservation by the fermion kinetic term.
The fermion content is specified by the multiplets in Eq.~\eqref{Eq:ASSM-FermionMultiplets-Option2}, while the evanescent gauge interactions are governed by the hypercharges in Eq.~\eqref{Eq:ASSM-YLR-YRL-Option2}.

\paragraph{ASM with Fermion Multiplets according to Option~\ref{Opt:Option1}:}
As announced, we begin with Option~\ref{Opt:Option1} for the fermions, which allows diagonal evanescent hypercharges $\hypLR=\hypRL$ (see Eq.~\eqref{Eq:ASSM-YLR}). 
We first present the complete counterterm action, decomposed into three pieces as in Sec.~\ref{Sec:One-Loop-Renormalisation-Abelian-Chiral-Gauge-Theory}, and provide explicit definitions of the coefficients. 
Subsequently, we discuss the role of the evanescent couplings $\hypLR$.

The invariant divergent counterterm action for the Abelian sector of the SM reads
\begin{equation}\label{Eq:Ssct_1-Loop_inv_ASSM_Option1}
    \begin{aligned}
        S^{(1)}_{\mathrm{sct,inv}} &= 
        - \frac{1}{16\pi^2} \, \frac{1}{\epsilon} \Dintx 
        \bigg[  
        \frac{20 g^2}{9}
        \Big(-\frac{1}{4}\overline{F}^{\mu\nu}\overline{F}_{\mu\nu}\Big)
        + \frac{g^2}{6}
        \Big(-\frac{1}{4}F^{\mu\nu}F_{\mu\nu}\Big)\\
        &+ \overline{\psi}_f i \overline{\slashed{\partial}} 
        \Big(\delta Z^{f}_{\overline{\psi}\psi,\mathrm{R}} \projR
        + \delta Z^{f}_{\overline{\psi}\psi,\mathrm{L}} \projL \Big) \psi_f
        - g \overline{\psi}_f \overline{\slashed{B}} 
        \Big( \delta Z^{f}_{\overline{\psi}B\psi,\mathrm{R}} \projR
        + \delta Z^{f}_{\overline{\psi}B\psi,\mathrm{L}} \projL \Big) \psi_f\\
        &- \Big\{
        \delta Z_{Y}^{e} y_e \big(\overline{\nu_{L}} \phi_{1} e_{R} 
        + \overline{e_{L}} \phi_{2} e_{R} \big)
        + \delta Z_{Y}^{d} y_d \big( \overline{u_{L}} \phi_{1} d_{R}
        + \overline{d_{L}} \phi_{2} d_{R} \big)\\
        &- \delta Z_{Y}^{u} y_u \big( \overline{d_{L}} \phi_{1}^{\dagger} u_{R}
        - \overline{u_{L}} \phi_{2}^{\dagger} u_{R} \big)
        +\mathrm{h.c.} \Big\}\\
        &- \frac{3-\xi}{4} g^2 
        \big(D^{\mu}\phi_a\big)^{\dagger}\big(D_{\mu}\phi_a\big)
        + \delta y_2
        \big(\overline{D}^{\mu}\phi_a\big)^{\dagger}\big(\overline{D}_{\mu}\phi_a\big)\\
        &+ \frac{\delta \lambda}{4} \Big( \phi_1^{\dagger} \phi_1 \phi_1^{\dagger} \phi_1
        + 2 \, \phi_1^{\dagger} \phi_1 \phi_2^{\dagger} \phi_2 + \phi_2^{\dagger} \phi_2 \phi_2^{\dagger} \phi_2 \Big)
        \bigg],
    \end{aligned}
\end{equation}
with fermionic counterterm coefficients
\begin{equation}\label{Eq:ASSM-Ssctinv-Coeff-FF}
    \begin{aligned}
        \delta Z^{f}_{\overline{\psi}\psi,\mathrm{R}} =
        \begin{cases}
            0, &f=\nu\\
            \xi g^2 + y_e^2, &f=e\\
            \frac{4\xi}{9} g^2 + y_u^2, &f=u\\
            \frac{\xi}{9} g^2 + y_d^2, &f=d
        \end{cases},
        \quad
        \delta Z^{f}_{\overline{\psi}\psi,\mathrm{L}} =
        \begin{cases}
            \frac{\xi}{4} g^2 + \frac{y_e^2}{2}, &f\in\{\nu,e\}\\
            \frac{\xi}{36} g^2 + \frac{y_u^2+y_d^2}{2}, &f\in\{u,d\},
        \end{cases}
    \end{aligned}
\end{equation}
the fermion--gauge interaction coefficients
\begin{equation}\label{Eq:ASSM-Ssctinv-Coeff-FBF}
    \begin{aligned}
        \delta Z^{f}_{\overline{\psi}B\psi,\mathrm{R}} =
        \begin{cases}
            0, &f=\nu\\
            - \xi g^2 - y_e^2, &f=e\\
            \frac{8\xi}{27} g^2 + \frac{2 y_u^2}{3}, &f=u\\
            -\frac{\xi}{27} g^2 - \frac{y_d^2}{3}, &f=d
        \end{cases},
        \quad
        \delta Z^{f}_{\overline{\psi}B\psi,\mathrm{L}} =
        \begin{cases}
            -\frac{\xi}{8} g^2 - \frac{y_e^2}{4}, &f\in\{\nu,e\}\\
            \frac{\xi}{216} g^2 + \frac{y_u^2+y_d^2}{12}, &f\in\{u,d\},
        \end{cases}
    \end{aligned}
\end{equation}
the Yukawa counterterm coefficients
\begin{equation}\label{Eq:ASSM-Ssctinv-Coeff-Yukawa}
    \begin{aligned}
        \delta Z_{Y}^{e} =\frac{6+3\xi}{4} g^2, \qquad
        \delta Z_{Y}^{u} =\frac{12+13\xi}{36} g^2 + y_d^2, \qquad
        \delta Z_{Y}^{d} =-\frac{6-7\xi}{36} g^2 + y_u^2,
    \end{aligned}
\end{equation}
and the Yukawa coupling combinations
\begin{equation}\label{Eq:ASSM-Ssctinv-Coeff-YukawaCombinations}
    \begin{aligned}
        \delta y_2 = y_e^2 + 3 \big(y_u^2+y_d^2\big), \qquad
        \delta y_4 = y_e^4 + 3 \big(y_u^4+y_d^4\big).
    \end{aligned}
\end{equation}
The scalar self-interaction counterterm is given by
\begin{equation}\label{Eq:ASSM-Ssctinv-Coeff-Scalar}
    \begin{aligned}
        \delta \lambda &= 48\lambda_{SM}^2 - 2\xi g^2 \lambda_{SM} + \frac{3}{4} g^4 - 4 \delta y_4.
    \end{aligned}
\end{equation}
This result matches the structure expected from renormalisation transformations in the ASM.

The divergent symmetry-restoring counterterm action for Option~\ref{Opt:Option1} in the ASM is
\begin{equation}\label{Eq:Ssct_1-Loop_break_ASSM_Option1}
    \begin{aligned}
        S^{(1)}_{\mathrm{sct,break}} &= \frac{1}{16\pi^2} \, \frac{1}{\epsilon} \Dintx
        \bigg[
        - \frac{7g^2}{9}
        \overline{B}_{\mu}\widehat{\Box}\overline{B}^{\mu}\\
        &+ \Big( \frac{2}{3} \delta y_2 - 2 y_u y_d \Big) \phi_{1}^{\dagger} \widehat{\Box} \phi_1
        + \frac{2}{3} \delta y_2 \phi_{2}^{\dagger} \widehat{\Box} \phi_2
        + \frac{1}{6} \Big( \delta y_2 \phi_{2} \widehat{\Box} \phi_2 + \mathrm{h.c.} \Big)\\
        &+ \delta\widehat{X}^{\text{opt1}}_{\psi,f} \overline{\psi}_f i \widehat{\slashed{\partial}} \psi_f
        - g \, \delta\widehat{X}^{\text{opt1}}_{G,f} \overline{\psi}_f \widehat{\slashed{B}} \psi_f\\
        &+ \frac{g^2}{3} \delta \widehat{H}_1
        \big(\overline{\partial}\cdot\overline{B}\big) \big(\widehat{\partial}\cdot\widehat{B}\big)
        -\frac{2g^2}{3} \delta \widehat{H}_{2} \Big(
        \widehat{B}_{\mu}\overline{\Box}\widehat{B}^{\mu}
        +\widehat{B}_{\mu}\widehat{\Box}\widehat{B}^{\mu}
        +\big(\widehat{\partial}\cdot\widehat{B}\big)^2
        \Big)\\
        &+ i \frac{2g}{3} \delta \widehat{S}^{\text{opt1}}_3 
        \Big[ \big(\widehat{\partial}^{\mu}\phi_{1}^{\dagger}\big)\phi_1 
        - \phi_{1}^{\dagger}\big(\widehat{\partial}^{\mu}\phi_1\big) \Big] \widehat{B}_{\mu}
        + \frac{2g^2}{3} \delta \widehat{S}^{\text{opt1}}_4
        \phi_{1}^{\dagger}\phi_{1} \widehat{B}^{\mu} \widehat{B}_{\mu}
        \bigg],
    \end{aligned}
\end{equation}
with $\delta y_2$ as defined in Eq.~\eqref{Eq:ASSM-Ssctinv-Coeff-YukawaCombinations}, as well as fermion kinetic and fermion--gauge interaction coefficients given by
\begin{equation}\label{Eq:ASSM-Ssctbreak-Coeff-FF}
    \begin{aligned}
        \delta\widehat{X}^{\text{opt1}}_{\psi,f} &=
        \begin{cases}
            0, &f=\nu\\
            - g^2 \Big( \frac{2+\xi}{3} + \frac{1-\xi}{2} \hypLR^e - \frac{2+\xi}{3} (\hypLR^e)^2 \Big), &f=e\\
            - g^2 \Big( \frac{4+2\xi}{27} - 5\frac{1-\xi}{18} \hypLR^u - \frac{2+\xi}{3} (\hypLR^u)^2 \Big) + \frac{y_u y_d}{2}, &f=u\\
            g^2 \Big( \frac{2+\xi}{27} - \frac{1-\xi}{18} \hypLR^d + \frac{2+\xi}{3} (\hypLR^d)^2 \Big) + \frac{y_u y_d}{2}, &f=d,
        \end{cases}
    \end{aligned}
\end{equation}
and
\begin{equation}\label{Eq:ASSM-Ssctbreak-Coeff-FBF}
    \begin{aligned}
        \delta\widehat{X}^{\text{opt1}}_{G,f} &=
        \begin{cases}
            0, &f=\nu\\
            - g^2 \hypLR^e \Big( \frac{2+\xi}{3} + \frac{1-\xi}{2} \hypLR^e - \frac{2+\xi}{3} (\hypLR^e)^2 \Big), &f=e\\
            - g^2 \hypLR^u \Big( \frac{4+2\xi}{27} - 5\frac{1-\xi}{18} \hypLR^u - \frac{2+\xi}{3} (\hypLR^u)^2 \Big) + \frac{y_u y_d}{2} \Big( \frac{1}{2} + \hypLR^d \Big), &f=u\\
            g^2 \hypLR^d \Big( \frac{2+\xi}{27} - \frac{1-\xi}{18} \hypLR^d + \frac{2+\xi}{3} (\hypLR^d)^2 \Big) - \frac{y_u y_d}{2} \Big( \frac{1}{2} - \hypLR^u \Big), &f=d,
        \end{cases}
    \end{aligned}
\end{equation}
respectively.
The purely evanescent gauge boson contributions are governed by
\begin{equation}\label{Eq:ASSM-Ssctbreak-Coeff-BB}
    \begin{aligned}
        \delta \widehat{H}_1 &= 3 \hypLR^e - 5 \hypLR^u + \hypLR^d,\\
        \delta \widehat{H}_2 &= \lVert \hypLR \rVert^2 = (\hypLR^e)^2 + 3 \big[(\hypLR^u)^2+(\hypLR^d)^2\big],
    \end{aligned}
\end{equation}
consisting of only evanescent hypercharges, and the evanescent scalar--gauge boson interactions come with
\begin{equation}\label{Eq:ASSM-Ssctbreak-Coeff-SSB-SSBB}
    \begin{aligned}
        \delta \widehat{S}^{\text{opt1}}_3 &= y_e^2 \hypLR^e - 3 \big( y_u^2+y_d^2-y_uy_d \big) \big( \hypLR^u-\hypLR^d \big),\\
        \delta \widehat{S}^{\text{opt1}}_4 &= - y_e^2 (\hypLR^e)^2 - 3 \big( y_u^2+y_d^2-y_uy_d \big) \big( \hypLR^u-\hypLR^d \big)^2.
    \end{aligned}
\end{equation}

Evidently, evanescent hypercharges $\hypLR$ significantly influence the divergent symmetry-restoring counterterm action.
The first two lines of Eq.~\eqref{Eq:Ssct_1-Loop_break_ASSM_Option1} are independent of $\hypLR$, the third line contains both $\hypLR$-dependent and -independent terms, and the last two lines vanish entirely for $\hypLR=0$.
Hence, the divergent symmetry-restoring counterterms simplify considerably when evanescent gauge interactions are switched off.

Finally, the finite symmetry-restoring counterterm action is given by
\begin{equation}\label{Eq:Sfct_1-Loop_ASSM_Option1}
    \begin{aligned}
        S^{(1)}_{\mathrm{fct}} &= \frac{1}{16\pi^2} \intx \bigg[
        -\frac{g^2}{3} \overline{B}_{\mu} \overline{\Box} \, \overline{B}^{\mu}
        + \frac{g^4}{24} \overline{B}_{\mu} \overline{B}^{\mu} \overline{B}_{\nu} \overline{B}^{\nu}\\
        &- i \frac{g}{3} \delta y_2
        \Big[
        \big(\overline{\partial}^{\mu}\phi_{2}^{\dagger}\big)\phi_2
        - \phi_{2}^{\dagger}\big(\overline{\partial}^{\mu}\phi_2\big) 
        \Big] 
        \overline{B}_{\mu}
        - \frac{g^2}{4} \delta y_2 \Big[
        \phi_{1}^{\dagger}\phi_{1}\overline{B}^{\mu}\overline{B}_{\mu}
        + \frac{5}{3} \phi_{2}^{\dagger}\phi_{2}\overline{B}^{\mu}\overline{B}_{\mu} \Big]\\
        &+ g \, \overline{\psi}_f \overline{\slashed{B}} \Big[
        \delta F_{\mathrm{R},f}^{\text{opt1}}
        \mathbb{P}_{\mathrm{R}}
        +
        \delta F_{\mathrm{L},f}^{\text{opt1}}
        \mathbb{P}_{\mathrm{L}}
        \Big] \psi_f\\
        &- \Big\{
        \delta Y_{e,2}^{\text{opt1}} y_e \overline{e_{L}} \phi_{2}^{\dagger} e_{R}
        + \delta Y_{u,2}^{\text{opt1}} y_u \overline{u_{L}} \phi_{2} u_{R}
        + \delta Y_{d,2}^{\text{opt1}} y_d \overline{d_{L}} \phi_{2}^{\dagger} d_{R}
        +\mathrm{h.c.} \Big\}\\
        &+ \frac{1}{6} \delta y_2
        \Big[
        \overline{\partial}_{\mu}\phi_{2}\overline{\partial}^{\mu}\phi_{2}
        - \frac{3}{2} i g \overline{\partial}_{\mu}\big(\phi_2\phi_2\big)\overline{B}^{\mu}
        + \frac{3}{4} g^2 \phi_2 \phi_2 \overline{B}_{\mu} \overline{B}^{\mu} + \mathrm{h.c.}
        \Big]\\
        &- \Big(
        \frac{1}{12} \delta y_4 \,
        \phi_{2}\phi_{2}\phi_{2}\phi_{2}
        + \frac{2}{3} \big( \delta y_4 - \delta y_{ud} \big)
        \phi_{1}^{\dagger}\phi_{1}\phi_{2}\phi_{2}
        + \frac{2}{3} \delta y_4 \, 
        \phi_{2}^{\dagger}\phi_{2}\phi_{2}\phi_{2}
        + \mathrm{h.c.} \Big)
        \bigg],
    \end{aligned}
\end{equation}
with coefficients $\delta y_2$ and $\delta y_4$ from Eq.~\eqref{Eq:ASSM-Ssctinv-Coeff-YukawaCombinations}, and
\begin{equation}\label{Eq:ASSM-Sfct-Coeff-YukawaCombinations}
    \begin{aligned}
        \delta y_{ud} &= \frac{3}{2} y_u y_d \big( y_u^2 + y_d^2 \big).
    \end{aligned}
\end{equation}
Finite fermion--gauge boson interaction counterterms are governed by
\begin{equation}\label{Eq:ASSM-Sfct-Coeff-FBF-R-Option1}
    \begin{aligned}
        \delta F_{\mathrm{R},f}^{\text{opt1}} &=
        \begin{cases}
            0, &f=\nu\\
            - g^2 \Big( \frac{5+\xi}{12} - \frac{1-\xi}{6} \hypLR^e + \frac{5+\xi}{12} (\hypLR^e)^2 \Big), &f=e\\
            g^2 \Big( \frac{5+\xi}{27} + \frac{1-\xi}{9} \hypLR^u + \frac{5+\xi}{12} (\hypLR^u)^2 \Big), &f=u\\
            - g^2 \Big( \frac{5+\xi}{108} - \frac{1-\xi}{18} \hypLR^d + \frac{5+\xi}{12} (\hypLR^d)^2 \Big), &f=d,
        \end{cases}
    \end{aligned}
\end{equation}
and
\begin{equation}\label{Eq:ASSM-Sfct-Coeff-FBF-L-Option1}
    \begin{aligned}
        \delta F_{\mathrm{L},f}^{\text{opt1}} &=
        \begin{cases}
            - g^2 \frac{5+\xi}{48} - \frac{y_e^2}{4}, &f=\nu\\
            g^2 \Big( \frac{5+\xi}{48} - \frac{1-\xi}{12} \hypLR^e + \frac{5+\xi}{12} (\hypLR^e)^2 \Big) - \frac{y_e^2}{4}, &f=e\\
            - g^2 \Big( \frac{5+\xi}{432} + \frac{1-\xi}{36} \hypLR^u + \frac{5+\xi}{12} (\hypLR^u)^2 \Big) + \frac{y_u^2 - y_d^2}{4}, &f=u\\
            g^2 \Big( \frac{5+\xi}{432} + \frac{1-\xi}{36} \hypLR^d + \frac{5+\xi}{12} (\hypLR^d)^2 \Big) + \frac{y_u^2 - y_d^2}{4}, &f=d,
        \end{cases}
    \end{aligned}
\end{equation}
while the finite Yukawa counterterm coefficients are
\begin{equation}\label{Eq:ASSM-Sfct-Coeff-Yukawas-Option1}
    \begin{aligned}
        \delta Y_{e,2}^{\text{opt1}} &=g^2 \Big( \frac{5+\xi}{6} - \frac{1-\xi}{2} \hypLR^e + \frac{5+\xi}{3} (\hypLR^e)^2 \Big),\\
        \delta Y_{u,2}^{\text{opt1}} &=g^2 \Big( \frac{5+\xi}{27} + 5 \frac{1-\xi}{18} \hypLR^u + \frac{5+\xi}{3} (\hypLR^u)^2 \Big) + \frac{y_d^2}{2},\\
        \delta Y_{d,2}^{\text{opt1}} &=-g^2 \Big( \frac{5+\xi}{54} + \frac{1-\xi}{18} \hypLR^d - \frac{5+\xi}{3} (\hypLR^d)^2 \Big) + \frac{y_u^2}{2}.
    \end{aligned}
\end{equation}

Although $\hypLR$ affects these counterterms noticeable, none of the terms in Eq.~\eqref{Eq:Sfct_1-Loop_ASSM_Option1} vanish entirely for $\hypLR=\hypRL=0$; nonetheless, all fermionic counterterms simplify considerably for zero evanescent hypercharges (see Eqs.~\eqref{Eq:ASSM-Sfct-Coeff-FBF-R-Option1}--\eqref{Eq:ASSM-Sfct-Coeff-Yukawas-Option1}).

As expected for Option~\ref{Opt:Option1}, global hypercharge conservation is violated at the regularised level. 
Accordingly, Eqs.~\eqref{Eq:Ssct_1-Loop_break_ASSM_Option1} and \eqref{Eq:Sfct_1-Loop_ASSM_Option1} contain terms that preserve electric and colour charge but break hypercharge.
The terms responsible for the violation of global hypercharge are the evanescent term $\phi_2\widehat{\Box}\phi_2+\mathrm{h.c.}$ in Eq.~\eqref{Eq:Ssct_1-Loop_break_ASSM_Option1} and the finite 4-dimensional terms in the last three lines of Eq.~\eqref{Eq:Sfct_1-Loop_ASSM_Option1}.
Among these are Yukawa counterterms with the ``wrong'' scalar field, some depend on evanescent hypercharges and others are independent (see Eq.~\eqref{Eq:ASSM-Sfct-Coeff-Yukawas-Option1}).
The remaining global hypercharge-violating terms in the last two lines of Eq.~\eqref{Eq:Sfct_1-Loop_ASSM_Option1} are purely bosonic, containing products of scalar fields such as $\phi_2\phi_2$, and arise solely from the chirality-violating evanescent piece of the fermion kinetic term (see Eq.~\eqref{Eq:L_fermion-evan_2}), independent of the evanescent hypercharges $\hypLR$.
Consequently, none of the global hypercharge-violating contributions disappears for $\hypLR=\hypRL=0$.

In summary, although Option~\ref{Opt:Option1} is conceptually natural --- combining physical chiral fermion components into a single Dirac spinor whenever possible --- it entails global hypercharge violation arising from the evanescent part of the fermion kinetic term, which constitutes the main drawback of this approach (see Sec.~\ref{Sec:Dimensional_Ambiguities_and_Evanescent_Shadows}).

\paragraph{ASM with Fermion Multiplets according to Option~\ref{Opt:Option2}:}
We now turn to the Abelian SM formulated according to Option~\ref{Opt:Option2}, where each fermion is accompanied by a sterile partner field. 
This construction removes the fermion kinetic term as a source of global hypercharge violation. 
As in the previous case, we present the full counterterm structure and analyse the dependence on the evanescent hypercharges $\hypLR$ and $\hypRL$.
As shown in Sec.~\ref{Sec:FermionicSectorNoScalars}, the invariant singular counterterm action coincides with that of Option~\ref{Opt:Option1}, shown in Eq.~\eqref{Eq:Ssct_1-Loop_inv_ASSM_Option1}, while both the divergent and finite symmetry-restoring counterterms differ.

For the divergent symmetry-restoring counterterms, we obtain
\begin{equation}\label{Eq:Ssct_1-Loop_break_ASSM_Option2}
    \begin{aligned}
        &S^{(1)}_{\mathrm{sct,break}} = \frac{1}{16\pi^2} \, \frac{1}{\epsilon} \Dintx
        \bigg[
        - \frac{5g^2}{9}
        \overline{B}_{\mu}\widehat{\Box}\overline{B}^{\mu}
        + \frac{2}{3} \delta y_2 \phi_{a}^{\dagger} \widehat{\Box} \phi_a\\
        &+ \delta\widehat{X}^{\text{opt2}}_{\psi,f} \Big(
        \overline{\psi_{2}}_{f} i \widehat{\slashed{\partial}} \projL {\psi_{1}}_{f}
        + \overline{\psi_{1}}_{f} i \widehat{\slashed{\partial}} \projR {\psi_{2}}_{f}
        \Big)
        - g \delta\widehat{X}^{\text{opt2}}_{G,f} \Big(
        \overline{\psi_{2}}_{f} \widehat{\slashed{B}} \projL {\psi_{1}}_{f}
        + \overline{\psi_{1}}_{f} \widehat{\slashed{B}} \projR {\psi_{2}}_{f}
        \Big)\\
        &- g^2 \delta \widehat{H}_2 \Big( \widehat{B}^{\mu}\overline{\Box}\widehat{B}_{\mu}
        + \frac{2}{3} \widehat{B}^{\mu}\widehat{\Box}\widehat{B}_{\mu} \Big)
        + g^4 \delta \widehat{H}_2 \Big( 
        \frac{1}{4} \overline{B}^{\mu}\overline{B}_{\mu}\widehat{B}^{\nu}\widehat{B}_{\nu}
        + \widehat{B}^{\mu}\widehat{B}_{\mu}\widehat{B}^{\nu}\widehat{B}_{\nu}
        \Big)\\
        &+ \frac{g^2}{2} \Big(\delta \widehat{S}_{4a}^{\text{opt2}} \phi_1^{\dagger}\phi_1 \widehat{B}^{\mu}\widehat{B}_{\mu}
        + \delta \widehat{S}_{4b}^{\text{opt2}} \phi_2^{\dagger}\phi_2 \widehat{B}^{\mu}\widehat{B}_{\mu}\Big)
        - \frac{g^2}{4} \Big( \frac{\delta \widehat{S}_{4b}^{\text{opt2}}}{2} \phi_2\phi_2 \widehat{B}^{\mu}\widehat{B}_{\mu} +\mathrm{h.c.}\Big)
        \bigg],
    \end{aligned}
\end{equation}
where the fermion fields $\psi_1$ and $\psi_2$ follow Eq.~\eqref{Eq:Fermions-psi1-psi2-Option2}, while $\delta y_2$ and $\delta \widehat{H}_2$ are defined in Eqs.~\eqref{Eq:ASSM-Ssctinv-Coeff-YukawaCombinations} and \eqref{Eq:ASSM-Ssctbreak-Coeff-BB}, respectively.
The fermionic counterterm coefficients are given by
\begin{equation}\label{Eq:ASSM-Ssctbreak-Coeff-FF-Option2}
    \begin{aligned}
        \delta\widehat{X}^{\text{opt2}}_{\psi,f} &=
        \begin{cases}
            0, &f=\nu\\
            - g^2 \frac{1-\xi}{2} \hypLR^e, &f=e\\
            g^2 \frac{5-5\xi}{18} \hypLR^u, &f=u\\
            - g^2 \frac{1-\xi}{18} \hypLR^d, &f=d,
        \end{cases}
    \end{aligned}
\end{equation}
and
\begin{equation}\label{Eq:ASSM-Ssctbreak-Coeff-FBF-Option2}
    \begin{aligned}
        \delta\widehat{X}^{\text{opt2}}_{G,f} &=
        \begin{cases}
            0, &f=\nu\\
            - g^2 \hypLR^e \Big( \frac{3+\xi}{2} - \frac{5+\xi}{3} (\hypLR^e)^2 \Big), &f=e\\
            - g^2 \hypLR^u \Big( \frac{3+\xi}{9} - \frac{5+\xi}{3} (\hypLR^u)^2 \Big) + \hypLR^d y_u y_d, &f=u\\
            g^2 \hypLR^d \Big( \frac{3+\xi}{18} + \frac{5+\xi}{3} (\hypLR^d)^2 \Big) + \hypLR^u y_u y_d, &f=d,
        \end{cases}
    \end{aligned}
\end{equation}
and the evanescent scalar--gauge interaction coefficients by
\begin{equation}\label{Eq:ASSM-Ssctbreak-Coeff-SSBB-Option2}
    \begin{aligned}
        \delta \widehat{S}_{4a}^{\text{opt2}} &= 4 y_e^2 (\hypLR^e)^2 + 12 \Big[
        \big( y_u^2 + y_d^2 \big) \big( (\hypLR^u)^2 + (\hypLR^d)^2 \big) + 2 y_u y_d \hypLR^u \hypLR^d \big) \Big],\\
        \delta \widehat{S}_{4b}^{\text{opt2}} &= 8 y_e^2 (\hypLR^e)^2 + 24 \Big[ y_u^2 (\hypLR^u)^2 + y_d^2 (\hypLR^d)^2\Big].
    \end{aligned}
\end{equation}
With $\hypLR=\hypRL=0$, only the first line of Eq.~\eqref{Eq:Ssct_1-Loop_break_ASSM_Option2} remains nonzero; in particular, all fermionic terms vanish.
The divergent symmetry-restoring counterterm action therefore simplifies considerably compared to Option~\ref{Opt:Option1} (cf.\ Eq.~\eqref{Eq:Ssct_1-Loop_break_ASSM_Option1}).
For nonzero evanescent hypercharges, the differences between both options reduce mainly to modified coefficients and alternative field monomials.

Despite this simplification, Eq.~\eqref{Eq:Ssct_1-Loop_break_ASSM_Option2} still exhibits potential global hypercharge violation through chirality-mixing induced by the evanescent hypercharges. 
The term $\phi_2\phi_2\widehat{B}^{\mu}\widehat{B}_{\mu}+\mathrm{h.c.}$ in the last line explicitly breaks global hypercharge but vanishes identically for $\hypLR=\hypRL=0$, unlike the $\phi_2\widehat{\Box}\phi_2+\mathrm{h.c.}$ term found for Option~\ref{Opt:Option1}.

The gauge boson sector also differs between the two realisations. 
Option~\ref{Opt:Option2} (see Eq.~\eqref{Eq:Ssct_1-Loop_break_ASSM_Option2}) contains fewer bilinear contributions but includes an additional quartic gauge boson counterterm (absent in Option~\ref{Opt:Option1}), as seen in the second term of line three. 
All these terms are entirely determined by the evanescent hypercharges, reflecting the different structure of the hypercharge matrices in Eqs.~\eqref{Eq:ASSM-YLR} (for Option~\ref{Opt:Option1}) and \eqref{Eq:ASSM-YLR-YRL-Option2} (for Option~\ref{Opt:Option2}); also noted in Eq.~\eqref{Eq:BBBB-GreenFunc-GeneralAbelianTheory}.

The finite symmetry-restoring counterterm action for Option~\ref{Opt:Option2} reads
\begin{equation}\label{Eq:Sfct_1-Loop_ASSM_Option2}
    \begin{aligned}
        S^{(1)}_{\mathrm{fct}} &= \frac{1}{16\pi^2} \intx \bigg[
        -\frac{5g^2}{9} \overline{B}_{\mu} \overline{\Box} \, \overline{B}^{\mu}
        + \frac{95g^4}{648} \overline{B}_{\mu} \overline{B}^{\mu} \overline{B}_{\nu} \overline{B}^{\nu}\\
        &- i \frac{g}{6} \delta y_2
        \Big[
        \big(\overline{\partial}^{\mu}\phi_{a}^{\dagger}\big)\phi_a
        - \phi_{a}^{\dagger}\big(\overline{\partial}^{\mu}\phi_a\big) 
        \Big] 
        \overline{B}_{\mu}
        - \frac{g^2}{4} \delta y_2^{\text{opt2}}
        \phi_{a}^{\dagger}\phi_{a}\overline{B}^{\mu}\overline{B}_{\mu}\\
        &+ g \Big( \delta F_{\mathrm{R},f}^{\text{opt2}} \, \overline{\psi_2}_f \overline{\slashed{B}} \projR {\psi_2}_f 
        + \delta F_{\mathrm{L},f}^{\text{opt2}} \, \overline{\psi_1}_f \overline{\slashed{B}} \projL {\psi_1}_f \Big)\\
        &- \Big\{
        \delta Y_{e,2}^{\text{opt2}} y_e \overline{e_{L}} \phi_{2}^{\dagger} e_{R}
        + \delta Y_{u,2}^{\text{opt2}} y_u \overline{u_{L}} \phi_{2} u_{R} 
        + \delta Y_{d,2}^{\text{opt2}} y_d \overline{d_{L}} \phi_{2}^{\dagger} d_{R}
        +\mathrm{h.c.} \Big\}
        \bigg],
    \end{aligned}
\end{equation}
with $\delta y_2$ as in Eq.~\eqref{Eq:ASSM-Ssctinv-Coeff-YukawaCombinations} and
\begin{equation}\label{Eq:ASSM-Sfct-Coeff-YukawaCombinations-Option2}
    \begin{aligned}
        \delta y_2^{\text{opt2}} = 3 y_e^2 + \frac{13}{3} y_u^2 + \frac{7}{3} y_d^2.
    \end{aligned}
\end{equation}
The finite fermion--gauge boson interactions under Option~\ref{Opt:Option2} are governed by
\begin{equation}\label{Eq:ASSM-Sfct-Coeff-FBF-R-Option2}
    \begin{aligned}
        \delta F_{\mathrm{R},f}^{\text{opt2}} &=
        \begin{cases}
            0, &f=\nu\\
            - g^2 \Big( \frac{5+\xi}{6} + \frac{1+\xi}{4} (\hypLR^e)^2 \Big) - \frac{y_e^2}{2}, &f=e\\
            g^2 \Big( \frac{20+4\xi}{81} + \frac{11+7\xi}{36} (\hypLR^u)^2 \Big) + \frac{y_u^2}{6}, &f=u\\
            - g^2 \Big( \frac{5+\xi}{162} + \frac{13+5\xi}{36} (\hypLR^d)^2 \Big) + \frac{y_d^2}{6}, &f=d,
        \end{cases}
    \end{aligned}
\end{equation}
\begin{equation}\label{Eq:ASSM-Sfct-Coeff-FBF-L-Option2}
    \begin{aligned}
        \delta F_{\mathrm{L},f}^{\text{opt2}} &=
        \begin{cases}
            - g^2 \frac{5+\xi}{48} - \frac{y_e^2}{2}, &f=\nu\\
            - g^2 \Big( \frac{5+\xi}{48} - \frac{1}{2} (\hypLR^e)^2 \Big) - \frac{y_e^2}{2}, &f=e\\
            g^2 \Big( \frac{5+\xi}{1296} - \frac{8+\xi}{18} (\hypLR^u)^2 \Big) + \frac{2 y_u^2 - y_d^2}{6}, &f=u\\
            g^2 \Big( \frac{5+\xi}{1296} + \frac{7+2\xi}{18} (\hypLR^d)^2 \Big) + \frac{2 y_u^2 - y_d^2}{6}, &f=d,
        \end{cases}
    \end{aligned}
\end{equation}
while the coefficients of the finite Yukawa counterterms are given by
\begin{equation}\label{Eq:ASSM-Sfct-Coeff-Yukawas-Option2}
    \begin{aligned}
        \delta Y_{f,2}^{\text{opt2}} &=g^2 \frac{5+\xi}{3} (\hypLR^f)^2, \quad &&\text{with} \quad f\in\{e,u,d\}.
    \end{aligned}
\end{equation}
In contrast to Eq.~\eqref{Eq:Sfct_1-Loop_ASSM_Option1} for Option~\ref{Opt:Option1}, all finite global hypercharge-violating Yukawa terms (last line of Eq.~\eqref{Eq:Sfct_1-Loop_ASSM_Option2}) vanish for $\hypLR=\hypRL=0$, since they are entirely induced by evanescent gauge interactions --- the only possible source of hypercharge violation in Option~\ref{Opt:Option2}.
The scalar field contributions responsible for global hypercharge breaking in Option~\ref{Opt:Option1} (last two lines of Eq.~\eqref{Eq:Sfct_1-Loop_ASSM_Option1}) are fully absent here.
In summary, Option~\ref{Opt:Option2} yields a substantially more compact finite symmetry-restoring counterterm structure, free from global hypercharge violation in the limit $\hypLR=\hypRL=0$.

\paragraph{Concluding Comments:}
As discussed in Sec.~\ref{Sec:General_Structure_of_the_Symmetry_Breaking}, there are two potential sources of regularisation-induced symmetry breaking arising from the non-anticommuting $\gamma_5$ in the BMHV scheme.
With this in mind, the main findings of this section can be summarised as follows:
\begin{itemize}
    \item The choice between Option~\ref{Opt:Option1} and Option~\ref{Opt:Option2} does not affect the symmetric singular counterterm action $S_\mathrm{sct,inv}$.
    \item The evanescent part of the fermion kinetic term breaks global hypercharge conservation in Option~\ref{Opt:Option1}, but not in Option~\ref{Opt:Option2} with sterile partner fields.
    Consequently, Option~\ref{Opt:Option2} yields a more compact finite symmetry-restoring counterterm action.
    \item For nonzero evanescent hypercharges ($\hypLR\neq0$), the structural differences in the divergent symmetry-restoring part are minor and essentially limited to distinct evanescent field monomials. 
    For instance, Option~\ref{Opt:Option1} produces a counterterm $\phi_2\widehat{\Box}\phi_2+\mathrm{h.c.}$, independent of $\hypLR$, whereas Option~\ref{Opt:Option2} gives $\phi_2\phi_2\widehat{B}^{\mu}\widehat{B}_{\mu}+\mathrm{h.c.}$, entirely governed by $\hypLR$. 
    Both terms violate global hypercharge conservation.
    \item When evanescent gauge interactions vanish ($\hypLR=0$), both options simplify considerably; however, Option~\ref{Opt:Option2} has the additional benefit of eliminating both sources for global hypercharge violation, thereby removing the need for corresponding restoration counterterms and resulting in an even more compact symmetry-restoring counterterm structure.
\end{itemize}
While these aspects make Option~\ref{Opt:Option2} appealing, it suffers from complications in the treatment of massive fermion propagators (see Sec.~\ref{Sec:General_Structure_of_the_Symmetry_Breaking}, Eq.~\eqref{Eq:MassivePropagatorMatrix}), which makes the choice between the two approaches nontrivial.
However, in view of phenomenologically relevant theories with massive particles --- such as the EWSM after spontaneous symmetry breaking --- Option~\ref{Opt:Option1}, which generally introduces global symmetry violation, appears most promising for two reasons:
it avoids the complications associated with massive fermion propagators in Option~\ref{Opt:Option2}, and restoring local BRST invariance automatically generates also the corresponding restoration counterterms required for global hypercharge conservation.

In summary, exploring specific values of the evanescent hypercharges remains important, with the case $\hypLR=\hypRL=0$ being of particular relevance. 
This analysis is carried out in the next subsection for both options.

\subsection{The Impact of the Evanescent Hypercharges in the Abelian SM}\label{Sec:TheRoleOf-YLR-YRL}

In the previous section, we discussed the 1-loop renormalisation of the Abelian sector of the SM for general evanescent hypercharges.
Here, we consider specific assignments of the evanescent hypercharges and present the corresponding counterterm coefficients introduced in Sec.~\ref{Sec:AnalysisOfASSMResults} for the representative choices identified in Sec.~\ref{Sec:General_Structure_of_the_Symmetry_Breaking}, namely setting $\hypLR$ equal to $0$, $\frac{\hypR+\hypL}{2}$, $Q$, $\hypL$, respectively.
All counterterm coefficients have been computed explicitly, and the results for these cases are summarised in two tables: in Tab.~\ref{Tab:YLR-Specialisations-Option-1} for Option~\ref{Opt:Option1} and in Tab.~\ref{Tab:YLR-Specialisations-Option-2} for Option~\ref{Opt:Option2}.
Each table lists the various coefficients in the rows and the corresponding values of the evanescent hypercharges in the columns.

\begin{table}[t!]
    \tiny
    \centering
    \begin{tabular}{|c||c|c|c|c|c|} \hline
         & $0$ & $\frac{\hypR+\hypL}{2}$ & $Q=\hypR$ & $\hypL$  \\ \hline \hline
        $\delta \widehat{H}_1$ & $0$ & $-\frac{53}{12}$ & $-\frac{20}{3}$ & $-\frac{13}{6}$  \\ \hline
        $\delta \widehat{H}_2$ & $0$ & $+\frac{53}{48}$ & $+\frac{8}{3}$ & $+\frac{5}{12}$  \\ \hline\hline
        $\delta\widehat{X}^{\text{opt1}}_{\psi,\nu}$ & $0$ & $0$ & $0$ & $0$ \\ \hline
        $\delta\widehat{X}^{\text{opt1}}_{\psi,e}$ & $-g^2$ & $-\frac{7}{16}g^2$ & $0^*$ & $-\frac{3}{4}g^2$  \\ \hline
        $\delta\widehat{X}^{\text{opt1}}_{\psi,u}$ & $-\frac{2}{9}g^2 + \frac{y_u y_d}{2}$ & $-\frac{7}{144}g^2 + \frac{y_u y_d}{2}$ & $+\frac{2}{9}g^2 + \frac{y_u y_d}{2}$ & $-\frac{7}{36}g^2 + \frac{y_u y_d}{2}$ \\ \hline
        $\delta\widehat{X}^{\text{opt1}}_{\psi,d}$ & $+\frac{1}{9}g^2+\frac{y_u y_d}{2}$ & $+\frac{17}{144}g^2+\frac{y_u y_d}{2}$ & $+\frac29g^2+\frac{y_u y_d}{2}$ & $+\frac{5}{36}g^2+\frac{y_u y_d}{2}$  \\ \hline\hline
        $\delta\widehat{X}^{\text{opt1}}_{G,\nu}$ & $0$ & $0$ & $0$ & $0$  \\ \hline
        $\delta\widehat{X}^{\text{opt1}}_{G,e}$ & $0$ & $+\frac{21}{64}g^2$ & $0^*$ & $+\frac38 g^2$  \\ \hline
        $\delta\widehat{X}^{\text{opt1}}_{G,u}$ & $+\frac14{y_uy_d}$ & $-\frac{35}{1728}g^2 + \frac{5}{24}y_u y_d$ & $+ \frac{4}{27}g^2 + \frac{1}{12}y_u y_d$ & $-\frac{7}{216}g^2 + \frac{1}{3}y_u y_d$  \\ \hline
        $\delta\widehat{X}^{\text{opt1}}_{G,d}$ & $-\frac14{y_uy_d}$ & $- \frac{17}{1728}g^2 - \frac{1}{24}y_u y_d$ & $- \frac{2}{27}g^2 + \frac{1}{12}y_u y_d$ & $+ \frac{5}{216}g^2 - \frac{1}{6}y_u y_d$  \\ \hline\hline
        $\delta\widehat{S}^{\text{opt1}}_{3}$ & $0$ & $-\frac{3}{2}(\frac12 y_e^2 + y_u^2 + y_d^2 - y_u y_d )$ & $-y_e^2 -3(y_u^2 + y_d^2 - y_u y_d)$ & $- \frac12 y_e^2$  \\ \hline
        $\delta\widehat{S}^{\text{opt1}}_{4}$ & $0$ & $-\frac{3}{4}(\frac34 y_e^2 + y_u^2 + y_d^2 - y_u y_d)$ & $-y_e^2 -3(y_u^2 + y_d^2 - y_u y_d)$ & $- \frac14 y_e^2$  \\ \hline\hline
        $\delta F_{\mathrm{R},\nu}^{\text{opt1}}$ & $0$ & $0$ & $0$ & $0$  \\ \hline
        $\delta F_{\mathrm{R},e}^{\text{opt1}}$ & $-\frac12 g^2$ & $-\frac{25}{32}g^2$ & $-g^2$ & $-\frac{5}{8}g^2$  \\ \hline
        $\delta F_{\mathrm{R},u}^{\text{opt1}}$ & $+\frac29 g^2$ & $+\frac{89}{288}g^2$ & $+\frac49g^2$ & $+\frac{17}{72}g^2$ \\ \hline
        $\delta F_{\mathrm{R},d}^{\text{opt1}}$ & $-\frac{1}{18} g^2$ & $-\frac{17}{288}g^2$ & $-\frac{1}{9}g^2$ & $-\frac{5}{72}g^2$  \\ \hline\hline
        $\delta F_{\mathrm{L},\nu}^{\text{opt1}}$ & $- \frac18 g^2 - \frac14 y_e^2$ & $- \frac18 g^2 - \frac14 y_e^2$ & $-\frac18g^2 - \frac14 y_e^2$ & $-\frac18g^2 - \frac14 y_e^2$  \\ \hline
        $\delta F_{\mathrm{L},e}^{\text{opt1}}$ & $+ \frac18 g^2 - \frac14 y_e^2$ & $+ \frac{13}{32} g^2 - \frac14 y_e^2$  & $+\frac58g^2 - \frac14 y_e^2$ & $+\frac14g^2 - \frac14 y_e^2$ \\ \hline
        $\delta F_{\mathrm{L},u}^{\text{opt1}}$ & $-\frac{1}{72}g^2 + \frac14(y_u^2 - y_d^2)$ & $-\frac{29}{288}g^2 + \frac14(y_u^2 - y_d^2)$ & $-\frac{17}{72}g^2 + \frac14(y_u^2 - y_d^2)$ & $-\frac{1}{36}g^2 + \frac14(y_u^2 - y_d^2)$  \\ \hline
        $\delta F_{\mathrm{L},d}^{\text{opt1}}$ & $+\frac{1}{72}g^2 + \frac14(y_u^2 - y_d^2)$ & $+\frac{5}{288}g^2 + \frac14(y_u^2 - y_d^2)$ & $+\frac{5}{72}g^2 + \frac14(y_u^2 - y_d^2)$ & $+\frac{1}{36}g^2 + \frac14(y_u^2 - y_d^2)$  \\ \hline\hline
        $\delta Y_{e,2}^{\text{opt1}}$ & $+g^2$ & $+\frac{17}{8}g^2$ & $+3g^2$ & $+\frac32 g^2$  \\ \hline
        $\delta Y_{u,2}^{\text{opt1}}$ & $+\frac{2}{9}g^2 + \frac{y_d^2}{2}$ & $+\frac{41}{72}g^2 + \frac{y_d^2}{2}$ & $+\frac{10}{9}g^2 + \frac{y_d^2}{2}$ & $+\frac{5}{18}g^2 + \frac{y_d^2}{2}$  \\ \hline
        $\delta Y_{d,2}^{\text{opt1}}$ & $-\frac{1}{9}g^2 + \frac{y_u^2}{2}$ & $-\frac{7}{72}g^2 + \frac{y_u^2}{2}$ & $+\frac{1}{9}g^2 + \frac{y_u^2}{2}$ & $-\frac{1}{18}g^2 + \frac{y_u^2}{2}$ \\ \hline
    \end{tabular}
    \caption{Explicit results for the coefficients of the divergent and finite symmetry-restoring counterterms for Option~\ref{Opt:Option1}, as shown in Eqs.~\eqref{Eq:Ssct_1-Loop_break_ASSM_Option1} and \eqref{Eq:Sfct_1-Loop_ASSM_Option1}, respectively.
    The results are evaluated in Feynman gauge, i.e.\ for $\xi=1$, and the asterisk indicates cases where the coefficient vanishes only for $\xi=1$ but not for other gauge choices.
    }
    \label{Tab:YLR-Specialisations-Option-1}
\end{table}
\begin{table}[t!]
    \tiny
    \centering
    \begin{tabular}{|c||c|c|c|c|c|c|} \hline 
         & $0$ & $\frac{\hypR+\hypL}{2}$ & $Q=\hypR$ & $\hypL$ \\ \hline \hline
        $\delta \widehat{H}_1$ & $0$ & $-\frac{53}{12}$ & $-\frac{20}{3}$ & $-\frac{13}{6}$ \\ \hline
        $\delta \widehat{H}_2$ & $0$ & $\frac{53}{48}$ & $\frac{8}{3}$ & $\frac{5}{12}$ \\ \hline\hline
        $\delta\widehat{X}^{\text{opt2}}_{\psi,\nu}$ & $0$ & $0$ & $0$ & $0$ \\ \hline
        $\delta\widehat{X}^{\text{opt2}}_{\psi,e}$ & $0$ & $0^*$ & $0^*$ & $0^*$ \\ \hline
        $\delta\widehat{X}^{\text{opt2}}_{\psi,u}$ & $0$ & $0^*$ & $0^*$ & $0^*$ \\ \hline
        $\delta\widehat{X}^{\text{opt2}}_{\psi,d}$ & $0$ & $0^*$ & $0^*$ & $0^*$ \\ \hline\hline
        $\delta\widehat{X}^{\text{opt2}}_{G,\nu}$ & $0$ & $0$ & $0$ & $0$ \\ \hline
        $\delta\widehat{X}^{\text{opt2}}_{G,e}$ & $0$ & $+ \frac{21}{32}g^2$ & $0^*$ & $+\frac34 g^2$ \\ \hline
        $\delta\widehat{X}^{\text{opt2}}_{G,u}$ & $0$ & $- \frac{35}{864}g^2 - \frac{1}{12}y_u y_d$ & $+\frac{8}{27}g^2 - \frac{1}{3}y_uy_d$ & $-\frac{7}{108}g^2 + \frac{1}{6}y_uy_d$ \\ \hline
        $\delta\widehat{X}^{\text{opt2}}_{G,d}$ & $0$ & $- \frac{17}{864}g^2 + \frac{5}{12}y_u y_d$ & $-\frac{4}{27}g^2 + \frac{2}{3}y_uy_d$ & $+\frac{5}{108}g^2 + \frac{1}{6}y_uy_d$ \\ \hline\hline
        $\delta\widehat{S}^{\text{opt2}}_{4a}$ & $0$ & $\frac94 y_e^2 + \frac{13}{6}(y_u^2 + y_d^2) - \frac{5}{6} y_u y_d$ & $4 y_e^2 + \frac{20}{3}(y_u^2 +y_d^2) - \frac{16}{3} y_u y_d$ & $y_e^2 + \frac23(y_u^2 + y_d^2 + y_u y_d)$ \\ \hline
        $\delta\widehat{S}^{\text{opt2}}_{4b}$ & $0$ & $\frac92 y_e^2 + \frac16(25 y_u^2 + y_d^2)$ & $8y_e^2 + \frac13(32y_u^2 + 8y_d^2)$ & $2y_e^2 + \frac23(y_u^2 + y_d^2)$ \\ \hline\hline
        $\delta F_{\mathrm{R},\nu}^{\text{opt2}}$ & $0$ & $0$ & $0$ & $0$ \\ \hline
        $\delta F_{\mathrm{R},e}^{\text{opt2}}$ & $-g^2 -\frac12 y_e^2$ & $-\frac{41}{32}g^2 -\frac12 y_e^2$ & $-\frac32g^2 - \frac12 y_e^2$ & $-\frac98g^2 - \frac12 y_e^2$ \\ \hline
        $\delta F_{\mathrm{R},u}^{\text{opt2}}$ & $+\frac{8}{27} g^2 + \frac16 y_u^2$ & $+\frac{331}{864}g^2 +\frac16 y_u^2$ & $+\frac{14}{27}g^2 + \frac16 y_u^2$ & $+\frac{67}{216}g^2 + \frac16 y_u^2$ \\ \hline
        $\delta F_{\mathrm{R},d}^{\text{opt2}}$ & $-\frac{1}{27}g^2 + \frac16 y_d^2$ & $-\frac{35}{864}g^2 +\frac16 y_d^2$ & $-\frac{5}{54}g^2 + \frac16 y_d^2$ & $-\frac{11}{216}g^2 + \frac16 y_d^2$ \\ \hline\hline
        $\delta F_{\mathrm{L},\nu}^{\text{opt2}}$ & $-\frac18g^2 - \frac12y_e^2$ & $-\frac18g^2 - \frac12y_e^2$ & $-\frac18g^2 - \frac12y_e^2$ & $-\frac18g^2 - \frac12y_e^2$ \\ \hline
        $\delta F_{\mathrm{L},e}^{\text{opt2}}$ & $-\frac18g^2 - \frac12y_e^2$ & $+\frac{5}{32}g^2 - \frac12 y_e^2$ & $+\frac38g^2 - \frac12 y_e^2$ & $-\frac12 y_e^2$ \\ \hline
        $\delta F_{\mathrm{L},u}^{\text{opt2}}$ & $\frac{1}{216}g^2 +\frac13 y_u^2 - \frac16 y_d^2$ & $-\frac{71}{864}g^2 + \frac13y_u^2 - \frac16y_d^2$ & $-\frac{47}{216}g^2 +\frac13 y_u^2 - \frac16 y_d^2$ & $-\frac{1}{108}g^2 +\frac13 y_u^2 - \frac16 y_d^2$ \\ \hline
        $\delta F_{\mathrm{L},d}^{\text{opt2}}$ & $\frac{1}{216}g^2 +\frac13 y_u^2 - \frac16 y_d^2$ & $+\frac{7}{864}g^2 +\frac13 y_u^2 - \frac16 y_d^2$ & $+\frac{13}{216}g^2+\frac13 y_u^2 - \frac16 y_d^2$ & $+\frac{1}{54}g^2 +\frac13 y_u^2 - \frac16 y_d^2$ \\ \hline\hline
        $\delta Y_{e,2}^{\text{opt2}}$ & $0$ & $+\frac98 g^2$ & $+2g^2 $ & $+\frac12 g^2$ \\ \hline
        $\delta Y_{u,2}^{\text{opt2}}$ & $0$ & $+\frac{25}{72}g^2$ & $+\frac89g^2$ & $+\frac{1}{18}g^2$ \\ \hline
        $\delta Y_{d,2}^{\text{opt2}}$ & $0$ & $+\frac{1}{72}g^2$ & $+\frac29g^2$ & $+\frac{1}{18}g^2$ \\ \hline
    \end{tabular}
    \caption{Explicit results for the coefficients of the divergent and finite symmetry-restoring counterterms for Option~\ref{Opt:Option2}, as shown in Eqs.~\eqref{Eq:Ssct_1-Loop_break_ASSM_Option2} and \eqref{Eq:Sfct_1-Loop_ASSM_Option2}, respectively.
    The results are evaluated in Feynman gauge, i.e.\ for $\xi=1$, and the asterisk indicates cases where the coefficient vanishes only for $\xi=1$ but not for other gauge choices.
    }
    \label{Tab:YLR-Specialisations-Option-2}
\end{table}

The tables confirm that the choice $\hypLR=\hypRL=0$ yields the simplest results. 
In this case, the bosonic contributions governed by $\delta\widehat{H}_1$, $\delta\widehat{H}_2$, $\delta\widehat{S}^{\text{opt1}}_{3}$, $\delta\widehat{S}^{\text{opt1}}_{4}$, $\delta\widehat{S}^{\text{opt2}}_{4a}$ and $\delta\widehat{S}^{\text{opt2}}_{4b}$ vanish entirely.
For Option~\ref{Opt:Option1}, where physical left- and right-handed fermions are combined, this simplification primarily concerns the divergent counterterms.
The benefit becomes even more pronounced for Option~\ref{Opt:Option2} with sterile partner fields, as global hypercharge conservation is then manifestly maintained.
None of the alternative assignments for the evanescent hypercharges leads to comparable simplifications.

These results further demonstrate that the symmetry-restoring counterterms depend both on the chosen $D$-dimensional treatment of the fermions and their kinetic terms, and on the specific assignments of the couplings governing evanescent gauge interactions.
The general analysis already indicated that the most compact counterterm structure arises for vanishing evanescent hypercharges, as the inclusion of evanescent gauge interactions leads to a significant proliferation of terms.
The explicit evaluation for representative hypercharge assignments confirms this, identifying $\hypLR=\hypRL=0$ as the simplest and most economical choice.
For most applications, it is therefore advantageous to omit evanescent gauge interactions altogether.
Even in QED, which emerges as a vector-like gauge theory from spontaneous symmetry breaking in the EWSM, a purely 4-dimensional treatment of the photon appears to be the most pragmatic approach, particularly in light of the discussion at the end of Sec.~\ref{Sec:FermionicSectorNoScalars}.
If, however, one aims to formulate QED in $D$ dimensions within the EWSM framework, appropriately defined evanescent gauge interactions must be introduced.
Potential benefits of such an approach cannot be fully assessed within the present study, which is restricted to an Abelian gauge theory with a single gauge boson.
The corresponding situation in the full Standard Model is addressed in Chapter~\ref{Chap:The_Standard_Model}.

\chapter{Non-Anticommuting $\gamma_5$ at the Multi-Loop Level}\label{Chap:BMHV_at_Multi-Loop_Level}

In this chapter, we investigate the multi-loop behaviour of a non-anticommuting $\gamma_5$ within the framework of the BMHV scheme.
To this end, we specialise the general Abelian chiral gauge theory introduced in the previous chapter to a simplified model containing only physical right-handed fermions, while excluding evanescent gauge interactions and scalar fields.
This right-handed model represents the simplest nontrivial realisation of the general Abelian framework and provides a minimal setting in which the essential multi-loop features of the BMHV scheme --- the regularisation-induced symmetry breaking and its restoration through symmetry-restoring counterterms --- can be analysed in detail.

The following discussion largely follows the analysis we published in Refs.~\cite{Stockinger:2023ndm,vonManteuffel:2025swv}.
It constitutes the first complete 4-loop application of the BMHV scheme and marks the highest-order computation achieved so far within this framework (first reported in Ref.~\cite{vonManteuffel:2025swv}).
Carrying out this computation posed a major computational challenge and required a highly efficient and fully automated computational framework.
In particular, the calculation involved various 4-loop 1PI Green functions, some of which produced billions of intermediate algebraic terms per Feynman diagram.
To meet these demands, we developed a completely new and fully automated \texttt{FORM}-based setup dedicated to this project.
A detailed description of this framework and the computational techniques employed is provided in chapter~\ref{Chap:Multi-Loop_Calculations}.
Its successful application in the present context serves as a comprehensive validation of both its reliability and efficiency, establishing a foundation for future high-precision multi-loop computations.

The structure of this chapter is as follows.
In Sec.~\ref{Sec:Right-Handed-Model-Definition}, we summarise the definition of the right-handed Abelian chiral gauge theory as a special case of the general model introduced in the previous chapter.
Section~\ref{Sec:Results-Right-Handed-Model} then presents the complete multi-loop renormalisation of this theory up to the 4-loop level, including the explicit determination and analysis of both divergent and finite counterterms.

\section{Definition of the Theory}\label{Sec:Right-Handed-Model-Definition}

The right-handed model considered in this chapter is obtained as a special case of the general Abelian theory defined in Sec.~\ref{Sec:Definition_of_the_Theory_General_Abelian_Case}, by discarding all left-handed and evanescent gauge interactions as well as all scalar field contributions (see the second row of Tab.~\ref{Tab:SpecialCasesOverview} in Sec.~\ref{Sec:Special_Cases_of_general_Abelian_Model}).
Theories containing physical left-handed fermionic degrees of freedom can be embedded by expressing them in terms of equivalent right-handed spinor fields obtained via charge conjugation of the original left-handed ones (see Sec.~\ref{Sec:S_fct_4-loop-SM-RHTheory}), corresponding to Option~\ref{Opt:Option2b} for the treatment of fermions.
The $D$-dimensional regularisation and all structural aspects discussed in sections~\ref{Sec:Regularisation-Induced_Symmetry_Breaking} and \ref{Sec:Definition_of_the_Theory_General_Abelian_Case} remain fully applicable.
In particular, Sec.~\ref{Sec:Origin_of_Symmetry_Breaking_in_BMHV_Scheme} already introduced the fermionic sector of this right-handed model as an explicit example illustrating the regularisation-induced symmetry breaking in the BMHV scheme.

We briefly recall the essential definitions, closely following the discussions in the aforementioned sections.
The complete $D$-dimensional classical Lagrangian is given by 
\begin{equation}\label{Eq:TheLagrangian-RHTheory}
    \begin{aligned}
        \mathcal{L} = 
        \mathcal{L}_{\mathrm{fermion}} + \mathcal{L}_{\mathrm{gauge}}
        + \mathcal{L}_{\mathrm{ghost+fix}} + \mathcal{L}_{\mathrm{ext}},
    \end{aligned}
\end{equation}
whose spacetime integral defines the classical action $S_0$.
In the present model, all physical fermionic degrees of freedom are contained in the right-handed fermion multiplet $\psi_R$.
For each of these, a sterile left-handed partner field is introduced, forming the left-handed multiplet $\psi_L$, so that we can construct Dirac spinors of the form $\psi_i={\psi^{\mathrm{st}}_{L}}_{\!i}+{\psi_{R}}_i$.
The fermion kinetic term then takes the form
\begin{align}\label{Eq:LfermionKin-RHTheory}
    \mathcal{L}_{\mathrm{fermion,kin}} = 
        \overline{\psi}_i i \slashed{\partial} \psi_i =
                  {\overline{\psi_{R}}}_i i
                  \overline{\slashed{\partial}} {\psi_{R}}_i
                  +
                  {\overline{\psi^{\mathrm{st}}_{L}}}_{\!i} i
                  \overline{\slashed{\partial}} {\psi^{\mathrm{st}}_{L}}_{\!i}
                  +
                  {\overline{\psi_{R}}}_i i
                  \widehat{\slashed{\partial}} {\psi^{\mathrm{st}}_{L}}_{\!i}
                  +
                  {\overline{\psi^{\mathrm{st}}_{L}}}_{\!i} i
                  \widehat{\slashed{\partial}} {\psi_{R}}_i,
\end{align}
as discussed in Sec.~\ref{Sec:Origin_of_Symmetry_Breaking_in_BMHV_Scheme}.
The interaction of the right-handed fermions ${\psi_{R}}_i$ with the Abelian gauge boson $B^{\mu}$ is governed by the hypercharge matrix $\hypR$. 
Together with the kinetic term this yields the full fermionic Lagrangian
\begin{equation}\label{Eq:Lfermion-RHTheory}
    \begin{aligned}
        \mathcal{L}_{\mathrm{fermion}} = \overline{\psi}_i i \slashed{\partial} \psi_i 
        - g {\hypR}_{ij} \overline{\psi}_i \projL \overline{\slashed{B}} \projR \psi_j.
    \end{aligned}
\end{equation}
Its 4-dimensional right-handed part can be expressed using the right-handed covariant derivative
\begin{align}\label{Eq:Right-Handed-Covariant-Derivative-Definition}
    {{\overline{D}}{}^\mu_{R}}_{ij} = \big( {\overline{\partial}}{}^\mu \delta_{ij} + i g {\mathcal{Y}_R}_{ij} {\overline{B}}{}^\mu \big)\projR,
\end{align}
as shown in Eq.~\eqref{Eq:L_fermion_inv_with_cov-div}.
With only right-handed fermions present, the anomaly cancellation condition in Eq.~\eqref{Eq:Abelian-Anomaly-Cancellation-Condition} reduces to
\begin{equation}\label{Eq:AnomalyCancellationCondition-RHTheory}
    \begin{aligned}
        \mathrm{Tr}(\hypR^3)\equiv0,
    \end{aligned}
\end{equation}
which is imposed on the right-handed hypercharge matrix $\hypR$.

The gauge boson Lagrangian retains its standard form,
\begin{equation}
    \begin{aligned}
        \mathcal{L}_{\mathrm{gauge}} = - \frac{1}{4} F^{\mu\nu}F_{\mu\nu},
    \end{aligned}
\end{equation}
with field strength tensor $F_{\mu\nu} = \partial_{\mu}B_{\nu} - \partial_{\nu}B_{\mu}$.
Gauge fixing and ghost terms are included as in the general Abelian theory,
\begin{equation}\label{Eq:Ddim-LGaugeFixing+Ghost-RHTheory}
    \begin{aligned}
        \mathcal{L}_{\mathrm{ghost+fix}} = 
        - \frac{1}{2\xi} (\partial^{\mu}B_{\mu})^2
        - \overline{c} \Box c.
    \end{aligned}
\end{equation}

The BRST transformations follow as an appropriate specialisation of Eq.~\eqref{Eq:Ddim-BRSTTrafos-GeneralAbelianTheory}:
\begin{equation}\label{Eq:BRST-Trafos-RHTheory}
    \begin{aligned}
        s_D B_{\mu}(x) &= \partial_{\mu} c(x),\\
        s_D\psi_{i}(x) &= 
        s_D{\psi_R}_i(x) = - i g c(x) {\hypR}_{ij} {\psi_R}_j(x),\\
        s_D\overline{\psi}_i(x) &= s_D\overline{\psi_R}_i(x) = i g c(x) \overline{\psi_R}_j(x) {\hypR}_{ji},\\
        s_D c(x) &= 0, \\
        s_D\overline{c}(x) &= \mathcal{B}(x) = - \frac{1}{\xi} \partial^{\mu}B_{\mu}(x),\\
        s_D \mathcal{B}(x) &= 0.
    \end{aligned}
\end{equation}
Coupling to the external sources $\{\rho^{\mu},R^i,\zeta\}$ gives
\begin{equation}\label{Eq:Ddim-Lext-RHTheory}
    \begin{aligned}
        \mathcal{L}_{\mathrm{ext}} = 
        \rho^{\mu} s_D B_{\mu} 
        + {\overline{R}}{}^{i} s_D {\psi_R}_i + (s_D \overline{\psi_R}_i) R^{i}
        + \zeta s_D c.
    \end{aligned}
\end{equation}

The dimensional regularisation within the framework of the BMHV scheme introduces a spurious breaking of gauge and BRST symmetry, exactly as described in Sec.~\ref{Sec:Origin_of_Symmetry_Breaking_in_BMHV_Scheme}.
For the present right-handed model, the breaking originates solely from the evanescent part of the fermion kinetic term in Eq.~\eqref{Eq:LfermionKin-RHTheory} and is given by
\begin{equation} \label{Eq:Delta_Hat-Tree_Level_Breaking_RHTheory}
    \begin{aligned}
        \widehat{\Delta} = - g \,  {\hypR}_{ij} \Dintx \, c \, 
                    \Big\{
                    \overline{\psi}_i \Big(
                    \overset{\leftarrow}{\widehat{\slashed{\partial}}} \projR 
                    + \overset{\rightarrow}{\widehat{\slashed{\partial}}} \projL
                    \Big) \psi_j 
                    \Big\},
    \end{aligned}
\end{equation}
identical to the tree-level breaking introduced in Eq.~\eqref{Eq:Delta_Hat_Introductory_Example}.

\section{Renormalisation up to the 4-Loop Level}\label{Sec:Results-Right-Handed-Model}

In this section, we present the complete renormalisation of the right-handed Abelian chiral gauge theory defined in the previous section, and regularised in DReg within the BMHV scheme, up to the 4-loop level.
We provide explicit results for all counterterms and construct the complete counterterm action, which establishes both UV finiteness and the validity of the Slavnov-Taylor identity (see Eq.~\eqref{Eq:UltimateSymmetryRequirement}), order by order up to 4 loops.
For this purpose, we closely follow our presentations in Refs.~\cite{Stockinger:2023ndm,vonManteuffel:2025swv}, where we reported these results first.

As before, the renormalisation procedure follows the methodology described in Sec.~\ref{Sec:Symmetry_Restoration_Procedure}.
The results up to the 3-loop level were obtained independently using both computational setups introduced in Sec.~\ref{Sec:Computational_Setup}, i.e.\ the \texttt{Mathematica}- and the \texttt{FORM}-based setup.
In contrast, the 4-loop calculation required the more performant \texttt{FORM} implementation, and thus relied exclusively on the \texttt{FORM}-based framework.
All computations employ the techniques discussed in chapter~\ref{Chap:Multi-Loop_Calculations}, including the tadpole decomposition method (see Sec.~\ref{Sec:Tadpole_Decomposition}) to extract UV divergent contributions of Feynman diagrams.
The 1- and 2-loop computations were performed in general $R_\xi$-gauge (see Eq.~\eqref{Eq:Ddim-LGaugeFixing+Ghost-RHTheory}), without restricting the gauge parameter $\xi$, whereas the 3- and 4-loop results were computed in Feynman gauge, i.e.\ for $\xi=1$.
The anomaly cancellation condition in Eq.~\eqref{Eq:AnomalyCancellationCondition-RHTheory} was imposed throughout, dimensional regularisation was parametrised as usual with $D=4-2\epsilon$, and all external momenta are defined as incoming.

We first present the results for all relevant ordinary 1PI Green functions in Sec.~\ref{Sec:Ordinary1PIGF-RHTheory}, and for the operator-inserted 1PI Green functions in Sec.~\ref{Sec:OperatorInserted1PIGF-RHTheory}, both at each considered loop order.
From these we derive the complete counterterm action up to the 4-loop level, which admits the perturbative expansion
\begin{align}
    S_\mathrm{ct}^{(\leq4)} = S_\mathrm{ct}^{(1)} + S_\mathrm{ct}^{(2)} + S_\mathrm{ct}^{(3)} + S_\mathrm{ct}^{(4)}.
\end{align}
The singular counterterms $S^{(n)}_\mathrm{sct}=S^{(n)}_\mathrm{sct,inv}+S^{(n)}_\mathrm{sct,break}$, which render the theory finite, are presented in Sec.~\ref{Sec:SingularCountertermAction-RHTheory}, while the finite symmetry-restoring counterterms $S^{(n)}_\mathrm{fct}$ are discussed in Sec.~\ref{Sec:FiniteCountertermAction-RHTheory}.
The latter, combined with the BRST-breaking part of the singular counterterm action $S^{(n)}_\mathrm{sct,break}$, restore the broken BRST symmetry and thus reestablish the validity of the Slavnov-Taylor identity at the respective order.
Subsequently, in Sec.~\ref{Sec:S_fct_4-loop-SM-RHTheory}, we assign explicit Standard Model hypercharge values to $\hypR$ and examine the 4-loop finite symmetry-restoring counterterm action in the Abelian fermionic sector of the SM (see Sec.~\ref{Sec:ASSM}).
Afterwards, Sec.~\ref{Sec:AuxiliaryCTs-RHTheory} lists all auxiliary counterterms required within the tadpole decomposition framework (see Sec.~\ref{Sec:Tadpole_Decomposition}), determined up to the 4-loop level.
Finally, in Sec.~\ref{Sec:AbelianWIs-RHTheory}, we discuss the Abelian Ward identities, which must hold for the fully renormalised theory at each order.

For compactness and clarity, all counterterms are expressed in terms of coefficients defined explicitly in App.~\ref{App:Results-RightHandedAbelianTheory}.
This representation highlights the structure of the counterterms and allows a concise and transparent discussion of the results.
For readability, we omit subscripts, superscripts, and tildes that indicate momentum-space evaluation on the effective quantum action whenever no ambiguity arises.

As outlined in Sec.~\ref{Sec:Computational_Setup}, the computational framework was thoroughly validated.
The UV-divergent BRST-breaking contributions provide an additional internal consistency check of our results, as they were extracted independently from both the ordinary and the $\Delta$-inserted 1PI Green functions, yielding perfect agreement.
Furthermore, all counterterms are local polynomials in external momenta at each order, as required by general principles, and subrenormalisation consistently works up to the 4-loop order.
This constitutes a strong confirmation of, and provides further confidence in, the correctness of our results.

\subsection{Ordinary 1PI Green Functions}\label{Sec:Ordinary1PIGF-RHTheory}

We begin with the ordinary 1PI Green functions and present all relevant results to determine the singular counterterms at each loop order $n\in\{1,2,3,4\}$.
In the right-handed model under consideration, the power-counting divergent, subrenormalised ordinary 1PI Green functions at the $n$-loop level are given by
\begin{align}
    \begin{split}\label{Eq:GF-BB-RHTheory}
        i \Gamma_{B_{\nu}(-p)B_{\mu}(p)} \big|_{\mathrm{div}}^{(n)} &= 
        - \frac{i g^{2n}}{(16 \pi^2)^n} \,
        \delta Z^{(n)}_{B}
        \Big(\overline{p}^{\mu} \overline{p}^{\nu} - \overline{p}^2 \overline{\eta}^{\mu\nu}\Big)\\
        &\phantom{-\,\,} + \frac{i g^{2n}}{(16 \pi^2)^n} \Big(
        \delta \widehat{X}^{(n)}_{BB} \,
        \widehat{p}^2 \, \overline{\eta}^{\mu\nu}
        + \delta \overline{X}^{(n)}_{BB} \,
        \overline{p}^2 \, \overline{\eta}^{\mu\nu} \Big),
    \end{split}\\
    \begin{split}
        i \Gamma_{\psi_{j}(p)\overline{\psi}_{i}(-p)} \big|_{\mathrm{div}}^{(n)} &= 
        - \frac{i g^{2n}}{(16 \pi^2)^n} 
        \Big( 
        \delta Z^{(n)}_{\psi,ij}
        + \delta \overline{X}^{(n)}_{\overline{\psi}\psi,ij}
        \Big) \overline{\slashed{p}} \, \mathbb{P}_{\mathrm{R}},
    \end{split}\\
    \begin{split}
        i \Gamma_{\psi_j(p_2)\overline{\psi}_i(p_1)B_{\mu}(q)}\big|_{\mathrm{div}}^{(n)} &= 
        \frac{i g^{2n+1}}{(16 \pi^2)^n} \,
        \big(\mathcal{Y}_R\big)_{ik} \delta Z^{(n)}_{\psi,kj} \, 
        \overline{\gamma}^{\mu} \mathbb{P}_{\mathrm{R}},
    \end{split}\\
    \begin{split}
        i \Gamma_{B_{\mu}(q)B_{\nu}(p_1)B_{\rho}(p_2)} \big|_{\mathrm{div}}^{(n)} &= 0,
    \end{split}\\
    \begin{split}\label{Eq:GF-BBBB-RHTheory}
        i \Gamma_{B_{\mu}(p_2)B_{\nu}(p_1)B_{\rho}(p_4)B_{\sigma}(p_3)} \big|_{\mathrm{div}}^{(n)} &= 
        - \frac{i g^{2n+2}}{(16 \pi^2)^n} \,
        \delta \overline{X}^{(n)}_{BBBB}
        \Big( \overline{\eta}^{\mu\nu} \, \overline{\eta}^{\rho\sigma} + \overline{\eta}^{\mu\rho} \, \overline{\eta}^{\nu\sigma} + \overline{\eta}^{\mu\sigma} \, \overline{\eta}^{\nu\rho} \Big).
    \end{split}
\end{align}
All results are expressed in terms of suitably defined renormalisation constants, where divergent BRST-invariant counterterms are denoted by $\delta Z$, and divergent BRST-breaking counterterms by $\delta X$.
In the following, we provide their Laurent expansions in $\epsilon$.

\paragraph{1-Loop Results:}
For $n=1$, the BRST-invariant singular counterterm coefficients read
\begin{align}
    \begin{split}\label{Eq:CountertermCoeff-GaugeBoson-Inv-1Loop-RHTheory}
        \delta Z^{(1)}_{B} &= 
          \mathcal{A}_{BB}^{1,\mathrm{inv}} \, \frac{1}{\epsilon},
    \end{split}\\
    \begin{split}\label{Eq:CountertermCoeff-Fermion-Inv-1Loop-RHTheory}
        \delta Z^{(1)}_{\psi,ij} &= 
          \mathcal{A}_{\overline{\psi}\psi, ij}^{1,\mathrm{inv}} \, \frac{1}{\epsilon},
    \end{split}
\end{align}
while the singular BRST-breaking contributions are given by
\begin{align}
    \begin{split}\label{Eq:CountertermCoeff-GaugeBoson-Break-Evanescent-1Loop-RHTheory}
        \delta \widehat{X}^{(1)}_{BB} &=  
          \widehat{\mathcal{A}}_{BB}^{1,\mathrm{break}} \, \frac{1}{\epsilon},
    \end{split}\\
    \begin{split}\label{Eq:CountertermCoeff-GaugeBoson-Break-4dim-1Loop-RHTheory}
        \delta \overline{X}^{(1)}_{BB} &= 0,
    \end{split}\\
    \begin{split}\label{Eq:CountertermCoeff-BBBB-Break-1Loop-RHTheory}
        \delta \overline{X}^{(1)}_{BBBB} &= 0,
    \end{split}\\
    \begin{split}\label{Eq:CountertermCoeff-Fermion-Break-1Loop-RHTheory}
        \delta \overline{X}^{(1)}_{\overline{\psi}\psi,ij} &= 0.
    \end{split}
\end{align}
The explicit coefficients are listed in App.~\ref{App:Divergent1LoopCTCoeffs-RightHandedAbelianTheory}.
The only non-vanishing divergent contribution to the 1-loop breaking represents the breaking of transversality of the gauge boson self energy, $\delta \widehat{X}^{(1)}_{BB}$ (see Eq.~\eqref{Eq:CountertermCoeff-GaugeBoson-Break-Evanescent-1Loop-RHTheory}).
This breaking is purely evanescent, as expected, since the tree-level BRST breaking $\widehat{\Delta}$, from which this 1-loop breaking originates, is itself evanescent and can therefore induce only divergent evanescent breaking terms.
Any 4-dimensional breaking at the 1-loop level must be finite (see Sec.~\ref{Sec:OperatorInserted1PIGF-RHTheory} with coefficients in App.~\ref{App:Finite1LoopCTCoeffs-RightHandedAbelianTheory}), where the evanescent breaking structure is not determining the structure of the field monomial but instead contracted and introduces a factor of $\epsilon$ in the numerator, cancelling the UV pole (see Sec.~\ref{Sec:Symmetry_Restoration_Procedure}).

\paragraph{2-Loop Results:}
For $n=2$, the BRST-invariant singular counterterm coefficients are
\begin{align}
    \begin{split}\label{Eq:CountertermCoeff-GaugeBoson-Inv-2Loop-RHTheory}
        \delta Z^{(2)}_{B} &= 
          \mathcal{A}_{BB}^{2,\mathrm{inv}} \, \frac{1}{\epsilon},
    \end{split}\\
    \begin{split}\label{Eq:CountertermCoeff-Fermion-Inv-2Loop-RHTheory}
        \delta Z^{(2)}_{\psi,ij} &= 
          \mathcal{A}_{\overline{\psi}\psi, ij}^{2,\mathrm{inv}} \, \frac{1}{\epsilon}
          + \mathcal{B}_{\overline{\psi}\psi, ij}^{2,\mathrm{inv}} \, \frac{1}{\epsilon^2},
    \end{split}
\end{align}
and the singular BRST-breaking contributions are provided by
\begin{align}
    \begin{split}\label{Eq:CountertermCoeff-GaugeBoson-Break-Evanescent-2Loop-RHTheory}
        \delta \widehat{X}^{(2)}_{BB} &=  
          \widehat{\mathcal{A}}_{BB}^{2,\mathrm{break}} \, \frac{1}{\epsilon}
          + \widehat{\mathcal{B}}_{BB}^{2,\mathrm{break}} \, \frac{1}{\epsilon^2},
    \end{split}\\
    \begin{split}\label{Eq:CountertermCoeff-GaugeBoson-Break-4dim-2Loop-RHTheory}
        \delta \overline{X}^{(2)}_{BB} &= 0,
    \end{split}\\
    \begin{split}\label{Eq:CountertermCoeff-BBBB-Break-2Loop-RHTheory}
        \delta \overline{X}^{(2)}_{BBBB} &= 0,
    \end{split}\\
    \begin{split}\label{Eq:CountertermCoeff-Fermion-Break-2Loop-RHTheory}
        \delta \overline{X}^{(2)}_{\overline{\psi}\psi,ij} &= 
        \overline{\mathcal{A}}_{\overline{\psi}\psi, ij}^{2,\mathrm{break}} \, \frac{1}{\epsilon}.
    \end{split}
\end{align}
Explicit coefficients are given in App.~\ref{App:Divergent2LoopCTCoeffs-RightHandedAbelianTheory}.
At this loop-order, a non-vanishing 4-dimensional, i.e.\ non-evanescent, divergent breaking, given by $\delta \overline{X}^{(2)}_{\overline{\psi}\psi,ij}$ in Eq.~\eqref{Eq:CountertermCoeff-Fermion-Break-2Loop-RHTheory}, emerges. 
It originates from higher-order corrections, with contributions from the finite, 4-dimensional 1-loop corrections of $\Delta_\mathrm{ct}^{(1)}$ to the BRST breaking (see Sec.~\ref{Sec:OperatorInserted1PIGF-RHTheory}).
It is a divergent breaking of the Ward identity relating the fermion self energy and the fermion--gauge boson interaction.

\paragraph{3-Loop Results:}
For $n=3$, the singular BRST-invariant contributions read
\begin{align}
    \begin{split}\label{Eq:CountertermCoeff-GaugeBoson-Inv-3Loop-RHTheory}
        \delta Z^{(3)}_{B} &= 
          \mathcal{A}_{BB}^{3,\mathrm{inv}} \, \frac{1}{\epsilon}
          + \mathcal{B}_{BB}^{3,\mathrm{inv}} \, \frac{1}{\epsilon^2},
    \end{split}\\
    \begin{split}\label{Eq:CountertermCoeff-Fermion-Inv-3Loop-RHTheory}
        \delta Z^{(3)}_{\psi,ij} &= 
          \mathcal{A}_{\overline{\psi}\psi, ij}^{3,\mathrm{inv}} \, \frac{1}{\epsilon}
          + \mathcal{B}_{\overline{\psi}\psi, ij}^{3,\mathrm{inv}} \, \frac{1}{\epsilon^2}
          + \mathcal{C}_{\overline{\psi}\psi, ij}^{3,\mathrm{inv}} \, \frac{1}{\epsilon^3},
    \end{split}
\end{align}
while the BRST-breaking singular counterterm coefficients are given by
\begin{align}
    \begin{split}\label{Eq:CountertermCoeff-GaugeBoson-Break-Evanescent-3Loop-RHTheory}
        \delta \widehat{X}^{(3)}_{BB} &=  
          \widehat{\mathcal{A}}_{BB}^{3,\mathrm{break}} \, \frac{1}{\epsilon}
          + \widehat{\mathcal{B}}_{BB}^{3,\mathrm{break}} \, \frac{1}{\epsilon^2}
          + \widehat{\mathcal{C}}_{BB}^{3,\mathrm{break}} \, \frac{1}{\epsilon^3},
    \end{split}\\
    \begin{split}\label{Eq:CountertermCoeff-GaugeBoson-Break-4dim-3Loop-RHTheory}
        \delta \overline{X}^{(3)}_{BB} &= \overline{\mathcal{A}}_{BB}^{3,\mathrm{break}} \, \frac{1}{\epsilon},
    \end{split}\\
    \begin{split}\label{Eq:CountertermCoeff-BBBB-Break-3Loop-RHTheory}
        \delta \overline{X}^{(3)}_{BBBB} &= \overline{\mathcal{A}}_{BBBB}^{3,\mathrm{break}} \, \frac{1}{\epsilon},
    \end{split}\\
    \begin{split}\label{Eq:CountertermCoeff-Fermion-Break-3Loop-RHTheory}
        \delta \overline{X}^{(3)}_{\overline{\psi}\psi,ij} &= 
        \overline{\mathcal{A}}_{\overline{\psi}\psi, ij}^{3,\mathrm{break}} \, \frac{1}{\epsilon}
        + \overline{\mathcal{B}}_{\overline{\psi}\psi, ij}^{3,\mathrm{break}} \, \frac{1}{\epsilon^2}.
    \end{split}
\end{align}
Explicit coefficients are provided in App.~\ref{App:Divergent3LoopCTCoeffs-RightHandedAbelianTheory}.
At the 3-loop level, a 4-dimensional divergent breaking of the transversality of the gauge boson self energy, $\delta \overline{X}^{(3)}_{BB}$ (see Eq.~\eqref{Eq:CountertermCoeff-GaugeBoson-Break-4dim-3Loop-RHTheory}), appears in addition to the evanescent divergent contribution $\delta \widehat{X}^{(3)}_{BB}$.
Moreover, the transversality of the quartic gauge boson interaction receives a 4-dimensional divergent breaking $\delta \overline{X}^{(3)}_{BBBB}$ (see Eq.~\eqref{Eq:CountertermCoeff-BBBB-Break-3Loop-RHTheory}) for the first time at this order.

\paragraph{4-Loop Results:}
For $n=4$, the BRST-invariant singular counterterms are governed by
\begin{align}
    \begin{split}\label{Eq:CountertermCoeff-GaugeBoson-Inv-4Loop-RHTheory}
        \delta Z^{(4)}_{B} &= 
          \mathcal{A}_{BB}^{4,\mathrm{inv}} \, \frac{1}{\epsilon}
        + \mathcal{B}_{BB}^{4,\mathrm{inv}} \, \frac{1}{\epsilon^2} 
        + \mathcal{C}_{BB}^{4,\mathrm{inv}} \, \frac{1}{\epsilon^3},
    \end{split}\\
    \begin{split}\label{Eq:CountertermCoeff-Fermion-Inv-4Loop-RHTheory}
        \delta Z^{(4)}_{\psi,ij} &= 
          \mathcal{A}_{\overline{\psi}\psi, ij}^{4,\mathrm{inv}} \, \frac{1}{\epsilon}
        + \mathcal{B}_{\overline{\psi}\psi, ij}^{4,\mathrm{inv}} \, \frac{1}{\epsilon^2}
        + \mathcal{C}_{\overline{\psi}\psi, ij}^{4,\mathrm{inv}} \, \frac{1}{\epsilon^3}
        + \mathcal{D}_{\overline{\psi}\psi, ij}^{4,\mathrm{inv}} \, \frac{1}{\epsilon^4},
    \end{split}
\end{align}
and for the singular BRST-breaking counterterms we obtain
\begin{align}
    \begin{split}\label{Eq:CountertermCoeff-GaugeBoson-Break-Evanescent-4Loop-RHTheory}
        \delta \widehat{X}^{(4)}_{BB} &=  
          \widehat{\mathcal{A}}_{BB}^{4,\mathrm{break}} \, \frac{1}{\epsilon}
        + \widehat{\mathcal{B}}_{BB}^{4,\mathrm{break}} \, \frac{1}{\epsilon^2} 
        + \widehat{\mathcal{C}}_{BB}^{4,\mathrm{break}} \, \frac{1}{\epsilon^3}
        + \widehat{\mathcal{D}}_{BB}^{4,\mathrm{break}} \, \frac{1}{\epsilon^4},
    \end{split}\\
    \begin{split}\label{Eq:CountertermCoeff-GaugeBoson-Break-4dim-4Loop-RHTheory}
        \delta \overline{X}^{(4)}_{BB} &= 
          \overline{\mathcal{A}}_{BB}^{4,\mathrm{break}} \, \frac{1}{\epsilon}
        + \overline{\mathcal{B}}_{BB}^{4,\mathrm{break}} \, \frac{1}{\epsilon^2},
    \end{split}\\
    \begin{split}\label{Eq:CountertermCoeff-BBBB-Break-4Loop-RHTheory}
        \delta \overline{X}^{(4)}_{BBBB} &=
          \overline{\mathcal{A}}_{BBBB}^{4,\mathrm{break}} \, \frac{1}{\epsilon}
        + \overline{\mathcal{B}}_{BBBB}^{4,\mathrm{break}} \, \frac{1}{\epsilon^2},
    \end{split}\\
    \begin{split}\label{Eq:CountertermCoeff-Fermion-Break-4Loop-RHTheory}
        \delta \overline{X}^{(4)}_{\overline{\psi}\psi,ij} &= 
          \overline{\mathcal{A}}_{\overline{\psi}\psi, ij}^{4,\mathrm{break}} \, \frac{1}{\epsilon}
        + \overline{\mathcal{B}}_{\overline{\psi}\psi, ij}^{4,\mathrm{break}} \, \frac{1}{\epsilon^2}
        + \overline{\mathcal{C}}_{\overline{\psi}\psi, ij}^{4,\mathrm{break}} \, \frac{1}{\epsilon^3}.
    \end{split}
\end{align}
Explicit coefficients are given in App.~\ref{App:Divergent4LoopCTCoeffs-RightHandedAbelianTheory}.
At the 4-loop level, no new singular counterterm structures appear for the first time.
As in the previous orders and as usual, the existing singular counterterms merely receive higher-order $\epsilon$-poles in the Laurent expansion of their coefficients.

\subsection{Operator-Inserted 1PI Green Functions and Symmetry Breaking}\label{Sec:OperatorInserted1PIGF-RHTheory}

Next, we consider the power-counting divergent 1PI Green functions with a single insertion of the local composite operator $\Delta=\widehat{\Delta}+\Delta_\mathrm{ct}$, from which the breaking of gauge and BRST invariance at each loop order $n\in\{1,2,3,4\}$ is obtained, as described in Sec.~\ref{Sec:Symmetry_Restoration_Procedure}.
In the right-handed Abelian chiral gauge theory under consideration, the relevant subrenormalised $\Delta$-operator-inserted 1PI Green functions at the $n$-loop level are given by
\begin{align}
    \begin{split}\label{Eq:GF-cB}
        i \big(\Delta \cdot \Gamma \big)_{B_{\mu}(-p) c(p)}^{(n)} &= \frac{g^{2n}}{(16 \pi^2)^n} 
        \bigg[ 
        \delta \widehat{X}^{(n)}_{BB} \, 
        \widehat{p}^{2} \, \overline{p}^{\mu}
        + \Big( 
        \delta \overline{X}^{(n)}_{BB}  
        + \delta F^{(n)}_{BB}
        \Big) 
        \overline{p}^{2} \, \overline{p}^{\mu}
        \bigg],
    \end{split}\\
    \begin{split}
        i \big( \Delta \cdot \Gamma \big)_{\psi_j(p_2) \overline{\psi}_i(p_1) c(q)}^{(n)}
        &= \frac{g^{2n+1}}{(16 \pi^2)^n} \big(\mathcal{Y}_R\big)_{ik} 
        \Big(
        \delta \overline{X}^{(n)}_{\overline{\psi}\psi,kj}
        + \delta F^{(n)}_{\overline{\psi}\psi,kj} \Big) 
        \big( \overline{\slashed{p}}_1 + \overline{\slashed{p}}_2 \big) \mathbb{P}_{\mathrm{R}},
    \end{split}\\
    \begin{split}
        i \big( \Delta \cdot \Gamma \big)_{B_{\nu}(p_2) B_{\mu}(p_1) c(q)}^{(n)} &= 0,
    \end{split}\\
    \begin{split}\label{Eq:GF-cBBB}
        i \big( \Delta \cdot \Gamma \big)_{B_{\rho}(p_3) B_{\nu}(p_2) B_{\mu}(p_1) c(q)}^{(n)} &= 
        \frac{g^{2n+2}}{(16 \pi^2)^n} 
        \Big( 
        \delta \overline{X}^{(n)}_{BBBB} 
        + \delta F^{(n)}_{BBBB}
        \Big) \\
        &\times \big(\overline{p}_1 + \overline{p}_2 + \overline{p}_3\big)_{\sigma} \, \big( \overline{\eta}^{\mu\nu} \, \overline{\eta}^{\rho\sigma} + \overline{\eta}^{\mu\rho} \, \overline{\eta}^{\nu\sigma} + \overline{\eta}^{\mu\sigma} \, \overline{\eta}^{\nu\rho} \big),
    \end{split}
\end{align}
where the divergent counterterm coefficients are defined in Sec.~\ref{Sec:Ordinary1PIGF-RHTheory} and finite evanescent terms are omitted.
Explicit results for the finite counterterm coefficients are provided in Apps.~\ref{App:Finite1LoopCTCoeffs-RightHandedAbelianTheory}, \ref{App:Finite2LoopCTCoeffs-RightHandedAbelianTheory}, \ref{App:Finite3LoopCTCoeffs-RightHandedAbelianTheory} and \ref{App:Finite4LoopCTCoeffs-RightHandedAbelianTheory} for the 1-, 2-, 3- and 4-loop level, respectively.

The symmetry breakings in Eqs.~\eqref{Eq:GF-cB}--\eqref{Eq:GF-cBBB} correspond to violations of the well-known Abelian Ward identities, which express the transversality of the gauge boson self energy, the relation between the fermion self energy to the fermion--gauge boson interaction, and the transversality of multi gauge boson interactions, as derived from Eq.~\eqref{Eq:FunctionalAbelianWardId}.

Additionally, there are two further power-counting divergent operator-inserted 1PI Green functions,
\begin{align}
    \begin{split}\label{Eq:GF-cFBF}
        i \big( \Delta \cdot \Gamma \big)_{\psi_{j}(p_3) \overline{\psi}_i(p_2) B_{\mu}(p_1) c(q)}^{(n)} &\equiv 0,
    \end{split}\\
    \begin{split}\label{Eq:GF-cBBBB}
        i \big( \Delta \cdot \Gamma \big)_{B_{\sigma}(p_4) B_{\rho}(p_3) B_{\nu}(p_2) B_{\mu}(p_1) c(q)}^{(n)} &\equiv 0,
    \end{split}
\end{align}
which, however, vanish identically and therefore require no further consideration.
In the Abelian model considered here, which contains only a right-handed interaction current, this results from a cancellation of the leading power-counting term in the respective integrands.
For the latter Green function, see Eq.~\eqref{Eq:GF-cBBBB}, this also follows from renormalisability arguments, as discussed in Sec.~\ref{Sec:RelevantGreenFunctions-GeneralAbelianChiralGaugeTheory}. 

As described in Sec.~\ref{Sec:Symmetry_Restoration_Procedure}, these Green functions completely capture the breaking of BRST symmetry at each order $n\in\{1,2,3,4\}$, including both divergent and finite contributions, as reflected in the violation of the Slavnov–Taylor identity according to Eq.~\eqref{Eq:QAPinDReg_Symmetry_Restoration}.
Consequently, the full $n$-loop breaking of BRST symmetry can be written as
\begin{equation}\label{Eq:Full-BRST-Breaking-n-Loop}
    \begin{aligned}
        {\big( \Delta \cdot \Gamma \big)}{}^{(n)}
        = \frac{g^{2n}}{(16 \pi^2)^n} &\Dintx \, \bigg\{
        \delta \widehat{X}^{(n)}_{BB} \, c {\widehat{\partial}}{}^2 {\overline{\partial}}{}^{\mu} \overline{B}_{\mu}
        + \Big[ \delta {\overline{X}}{}^{(n)}_{BB} + \delta F^{(n)}_{BB} \Big]
        c {\overline{\partial}}{}^2 {\overline{\partial}}{}^{\mu} \overline{B}_{\mu}\\
        &\qquad\,\,
        + g^2 \Big[ \delta \overline{X}^{(n)}_{BBBB} + \delta F^{(n)}_{BBBB} \Big] 
        \frac{1}{2} c \, \overline{\partial}_{\mu} \big({\overline{B}}{}^{\mu}\overline{B}_{\nu}{\overline{B}}{}^{\nu}\big)\\
        &\qquad\,\,
        + g {\hypR}_{ik}
        \Big[ \delta {\overline{X}}{}^{(n)}_{\overline{\psi}\psi,kj} + \delta F^{(n)}_{\overline{\psi}\psi,kj} \Big]
        c \, \overline{\partial}_{\mu} \big( \overline{\psi}_i {\overline{\gamma}}{}^{\mu} \projR \psi_j \big)
        \bigg\}
         + \mathcal{O}(\hat{.}),
    \end{aligned}
\end{equation}
where finite evanescent terms, indicated by $\mathcal{O}(\hat{.})$, are omitted.
From this result, we obtain the $n$-loop correction to the breaking, $\Delta^{(n)}_\mathrm{ct}$, via Eq.~\eqref{Eq:PerturbativeRequirementAndStartingPoint} (see Eqs.~\eqref{Eq:1-loop-BRST-Breaking} and \eqref{Eq:4-loop-BRST-Breaking} for the explicit 1- and 4-loop cases, respectively).

The divergent contributions of ${\big( \Delta \cdot \Gamma \big)}{}^{(n)}$ correctly reproduce the BRST-breaking part of the singular contributions to the ordinary Green functions discussed in Sec.~\ref{Sec:Ordinary1PIGF-RHTheory}.
Equivalently, the divergent part of $\Delta^{(n)}_\mathrm{ct}$ arises from the BRST variation of the non-invariant singular counterterm action $S^{(n)}_\mathrm{sct,break}$ (see Sec.~\ref{Sec:SingularCountertermAction-RHTheory}) via Eq.~\eqref{Eq:DefDeltaBreaking}.
This provides the aforementioned internal consistency check of our results.

Evidently, all finite BRST-breaking structures in Eq.~\eqref{Eq:Full-BRST-Breaking-n-Loop} already appear at the 1-loop level, as can be seen from the explicit results for the finite counterterm coefficients in App.~\ref{App:Results-RightHandedAbelianTheory}. 
Hence, no new finite BRST-breaking structures emerge at higher orders.
The number of field monomials required to capture the complete finite breaking of gauge and BRST invariance is finite.
In particular, in the right-handed model under consideration, all finite BRST breakings in ${\big( \Delta \cdot \Gamma \big)}{}^{(n)}$  can be expressed in terms of the finite basis
\begin{align}\label{Eq:Finite-Basis-Breaking-RHTheory}
    \Big\{ 
    c {\overline{\partial}}{}^2 {\overline{\partial}}{}^{\mu} \overline{B}_{\mu}, \,
    c \, \overline{\partial}_{\mu} \big({\overline{B}}{}^{\mu}\overline{B}_{\nu}{\overline{B}}{}^{\nu}\big), \,
    c \, \overline{\partial}_{\mu} \big( \overline{\psi}_i {\overline{\gamma}}{}^{\mu} \projR \psi_j \big) 
    \Big\},
\end{align}
constrained by power-counting and ghost number~$1$.
Here, we also refer to the 1-loop study of Ref.~\cite{Cornella:2022hkc}, where they exclusively focused on finite contributions and derived a general basis of field monomials for theories with compact, simple (not necessarily Abelian) gauge groups.

The situation for divergent symmetry-breaking contributions is analogous, although additional field monomials are required due to the presence of evanescent structures (first term of Eq.~\eqref{Eq:Full-BRST-Breaking-n-Loop}).
Nonetheless, the total number of field monomials needed to parametrise all possible divergent symmetry breakings does not grow indefinitely but instead remains finite.
Importantly, while new divergent symmetry-breaking structures were observed at the 2- and 3-loop levels in Sec.~\ref{Sec:Ordinary1PIGF-RHTheory}, governed by the 4-dimensional $\delta \overline{X}$'s, the inspection of the complete symmetry breaking ${\big( \Delta \cdot \Gamma \big)}{}^{(n)}$ in Eq.~\eqref{Eq:Full-BRST-Breaking-n-Loop} shows that these do not represent genuinely new field monomials but rather divergent corrections to the finite symmetry breakings already present at the 1-loop level.

In general, the number of power-counting divergent $\Delta$-operator-inserted 1PI Green functions in a given power-counting renormalisable theory is finite, as is the number of Lorentz covariants required to parametrise their results.
Consequently, the number of field monomials contributing to the breaking ${\big( \Delta \cdot \Gamma \big)}{}^{(n)}$ is finite.
In agreement with proposition~\ref{Prop:Basis_of_Insertions} and the discussion at the end of Sec.~\ref{Sec:Symmetry_Restoration_Procedure}, this ensures the renormalisability of the theory in the BMHV scheme, since the symmetry breaking can always be compensated by a finite number of symmetry-restoring counterterms.

\subsection{Singular Counterterm Action}\label{Sec:SingularCountertermAction-RHTheory}

From the divergent parts of the ordinary 1PI Green functions presented in Sec.~\ref{Sec:Ordinary1PIGF-RHTheory}, we now derive the complete singular counterterm action.
As before, this action is decomposed into a BRST-invariant and a BRST-breaking component.
At the $n$-loop level, the BRST-invariant part takes the form
\begin{equation}\label{Eq:Ssct_inv_n-Loop}
    \begin{aligned}
        S^{(n)}_{\mathrm{sct,inv}} = 
        \frac{g^{2n}}{(16 \pi^2)^n} \Dintx 
        \bigg\{ 
        \delta Z^{(n)}_{B} \Big( -\frac{1}{4} \, \overline{F}^{\mu\nu} \, \overline{F}_{\mu\nu} \Big)
        + \delta Z^{(n)}_{\psi,kj} \overline{\psi}_i i {\overline{\slashed{D}}_{R}}_{ik} \psi_j
        \bigg\},
    \end{aligned}
\end{equation}
where the right-handed covariant derivative is defined in Eq.~\eqref{Eq:Right-Handed-Covariant-Derivative-Definition}.
The BRST-breaking part of the singular counterterm action is given by
\begin{equation}\label{Eq:Ssct_break_n-Loop}
    \begin{aligned}
        S^{(n)}_{\mathrm{sct,break}} = 
        \frac{g^{2n}}{(16 \pi^2)^n} \Dintx 
        \bigg\{ 
        &\delta \widehat{X}^{(n)}_{BB} \, \frac{1}{2} \overline{B}_{\mu} \widehat{\partial}^2 \overline{B}^{\mu}
        + \delta \overline{X}^{(n)}_{BB} \, \frac{1}{2} \overline{B}_{\mu} \overline{\partial}^2 \overline{B}^{\mu}\\
        + \, &g^2 \delta \overline{X}^{(n)}_{BBBB} \, \frac{1}{8} \overline{B}_{\mu} \overline{B}^{\mu} \overline{B}_{\nu} \overline{B}^{\nu}
        + \delta \overline{X}^{(n)}_{\overline{\psi}\psi,ij} \Big( \overline{\psi}_i i \overline{\slashed{\partial}} \projR \psi_j \Big)
        \bigg\}.
    \end{aligned}
\end{equation}

It is evident from the field structure in Eq.~\eqref{Eq:Ssct_inv_n-Loop} that $S_\mathrm{sct,inv}^{(n)}$ is manifestly BRST-invariant.
Furthermore, its form remains unchanged at all loop orders and mirrors the structure of the counterterm action in ordinary vector-like QED (cf.\ App.~\ref{App:Renormalisation_of_QED}), differing only by the use of a right-handed covariant derivative and modified coefficients.
This is because $S_\mathrm{sct,inv}^{(n)}$ involves only the symmetric field monomials already present in the classical Lagrangian.
Consequently, it corresponds to a multiplicative field and parameter renormalisation of the gauge and fermion fields, as well as of the gauge coupling, such that the product $g B^{\mu}$ does not renormalise --- precisely as in ordinary QED.

The BRST-breaking part of the singular counterterm action, Eq.~\eqref{Eq:Ssct_break_n-Loop}, displays a richer structure.
New field monomials emerge successively with increasing loop order up to $n=3$ (see Sec.~\ref{Sec:Ordinary1PIGF-RHTheory}).
At the 1-loop level, only the evanescent bosonic breaking (the first term) is present.
At 2 loops, a 4-dimensional fermionic breaking (the last term) appears, while at the 3-loop level two additional 4-dimensional bosonic breakings (the second and third terms) arise for the first time.
At the 4-loop level, no further field monomials occur --- the structure of $S^{(4)}_{\mathrm{sct,break}}$ is identical to that at 3 loops, apart from different counterterm coefficients and the appearance of higher-order $\epsilon$-poles, as expected at each successive order.
This saturation of structures follows from the finiteness of the basis of field monomials for the operator insertions underlying the symmetry breakings (see proposition~\ref{Prop:Basis_of_Insertions} in Sec.~\ref{Sec:Algebraic_Renormalisation}, as well as sections~\ref{Sec:Symmetry_Restoration_Procedure} and \ref{Sec:OperatorInserted1PIGF-RHTheory}).
Moreover, as noted in Sec.~\ref{Sec:OperatorInserted1PIGF-RHTheory}, the 4-dimensional non-invariant singular counterterm contributions, though new within the singular counterterm action, already appear in the finite counterterm action at the 1-loop level and therefore represent only divergent corrections to existing finite breakings.

Power counting and renormalisability constrain the number of possible local operators that can appear in divergent terms.
Moreover, in the right-handed model considered here, the absence of evanescent gauge or Yukawa interactions (in contrast to the general theory discussed in chapter~\ref{Chap:General_Abelian_Chiral_Gauge_Theory}) further limits the number of admissible structures.
Consequently, beyond a certain loop order, all possible field monomials contributing to BRST-breaking counterterms are exhausted --- a point that will be discussed further in the next section on finite breakings.

Following the reasoning of Refs.~\cite{Stockinger:2023ndm,vonManteuffel:2025swv}, we attribute the BRST breaking entirely to the bilinear fermionic terms (see Eq.~\eqref{Eq:Ssct_break_n-Loop}).
In the present context this convention, however, is not unique: the breaking could equally well be assigned to the fermion--gauge boson vertex correction, or distributed between both, through a suitable rearrangement.
Explicitly, we may rewrite the corresponding terms as
\begin{equation}
    \begin{aligned}
        \frac{g^{2n}}{(16 \pi^2)^n} \Dintx 
        \bigg\{
        \Big( 
        \delta Z^{(n)}_{\psi,kj} 
        + \delta \overline{X}^{(n)}_{\overline{\psi}\psi,kj} \Big) 
        \overline{\psi}_i i {\overline{\slashed{D}}_{R}}_{ik} \psi_j
        + g {\hypR}_{ik}
        \delta \overline{X}^{(n)}_{\overline{\psi}\psi,kj} 
        \overline{\psi}_i \overline{\slashed{B}} \projR \psi_j
        \bigg\},
    \end{aligned}
\end{equation}
which yields the same total singular counterterm action $S^{(n)}_{\mathrm{sct}}$, but with the BRST breaking fully attributed to the gauge interaction.
This redistribution freedom is model dependent and will be revisited in Sec.~\ref{Sec:FiniteCountertermAction-RHTheory}.

\subsection{Finite Symmetry-Restoring Counterterm Action}\label{Sec:FiniteCountertermAction-RHTheory}

From the finite contributions to the BRST breaking in Eq.~\eqref{Eq:Full-BRST-Breaking-n-Loop}, we now construct the finite symmetry-restoring counterterms at each loop order $n\in\{1,2,3,4\}$.
Following the procedure described in Sec.~\ref{Sec:Symmetry_Restoration_Procedure} for the Abelian theory under consideration, we determine the finite symmetry-restoring counterterms such that they satisfy $s_D S_\mathrm{fct}^{(n)}=-{\big( \Delta \cdot \Gamma \big)}{}^{(n)}_\mathrm{fin}$ (see Eq.~\eqref{Eq:Determination-of-Symmetry-Restoring-CT-Abelian-Case}).
At the $n$-loop level, the corresponding finite symmetry-restoring counterterm action is then given by
\begin{equation}\label{Eq:Sfct-n-Loop}
    \begin{aligned}
        S^{(n)}_{\mathrm{fct}} = 
        \frac{g^{2n}}{(16 \pi^2)^n} \intx 
        \bigg\{ 
        &\delta F_{BB}^{(n)} \, \frac{1}{2} \overline{B}_{\mu} \overline{\partial}^2 \overline{B}^{\mu} + g^2 \delta F_{BBBB}^{(n)} \, \frac{1}{8} \overline{B}_{\mu} \overline{B}^{\mu} \overline{B}_{\nu} \overline{B}^{\nu}\\
        + \, &
        \delta F_{\overline{\psi}\psi,ij}^{(n)} \, \overline{\psi}_i i \overline{\slashed{\partial}} \projR \psi_j
        \bigg\},
    \end{aligned}
\end{equation}
with explicit coefficients listed in App.~\ref{App:Results-RightHandedAbelianTheory}.
This expression constitutes the central result of this chapter.

Together with $S^{(n)}_{\mathrm{sct,break}}$ in Eq.~\eqref{Eq:Ssct_break_n-Loop}, these counterterms restore the broken BRST symmetry and thereby ensure the validity of the Slavnov-Taylor identity (cf.\ Eq.~\eqref{Eq:UltimateSymmetryRequirement}) at the $n$-loop level.
In contrast to the non-invariant singular counterterms $S^{(n)}_{\mathrm{sct,break}}$, no new field monomials appear at higher orders: all relevant structures already occur at the 1-loop level.
Consequently, the basic structure of the finite symmetry-restoring counterterm action remains unchanged at higher orders, differing only in the values of the coefficients (see also Refs.~\cite{Stockinger:2023ndm,vonManteuffel:2025swv}).
The admissible local operators for the finite symmetry-restoring counterterms constitute a finite set and are constrained by power-counting and renormalisability, as discussed in Sec.~\ref{Sec:OperatorInserted1PIGF-RHTheory}.

Specifically, the finite BRST-breaking contributions to ${\big( \Delta \cdot \Gamma \big)}{}^{(n)}$, which determine the corresponding symmetry-restoring counterterms, can be expressed in terms of the finite basis of ghost number~$1$ field monomials in Eq.~\eqref{Eq:Finite-Basis-Breaking-RHTheory}.
As a consequence, we can identify the finite basis of ghost number~$0$ field monomials
\begin{align}\label{Eq:Finite-Basis-S_fct-RHTheory}
    \Big\{ 
    \overline{B}_{\mu} {\overline{\partial}}{}^2 {\overline{B}}{}^{\mu}, \,
    \overline{B}_\mu{\overline{B}}{}^{\mu}\overline{B}_{\nu}{\overline{B}}{}^{\nu}, \,
    \overline{\psi}_i \overline{\slashed{\partial}} \projR \psi_j
    \Big\},
\end{align}
which is sufficient to parametrise all finite symmetry-restoring counterterms.
This agrees with proposition~\ref{Prop:Basis_of_Insertions}, according to which the basis for the local operator insertions encoding the symmetry breaking is finite.
Consequently, the finite breaking of gauge and BRST invariance at any loop order can always be compensated by a finite set of local counterterms.
Together with the preceding discussion of the non-invariant singular counterterms in Sec.~\ref{Sec:SingularCountertermAction-RHTheory}, this establishes renormalisability of the theory in the BMHV scheme, as already noted in Sec.~\ref{Sec:OperatorInserted1PIGF-RHTheory}.
For related discussions on renormalisability and the general structure of symmetry-restoring counterterms, we refer to Ref.~\cite{Piguet:1995er} for a general analysis of the finiteness of the insertion basis, to Ref.~\cite{Cornella:2022hkc} for an explicit derivation of the finite breaking and corresponding counterterm bases at the 1-loop level, and to Ref.~\cite{Kuhler:2025znv} for a 2-loop study of the symmetry-restoring counterterms in a non-Abelian chiral gauge theory.

The symmetry-restoring counterterm action is not unique, since only its BRST variation in $\mathop{\mathrm{LIM}}_{D \, \to \, 4}$ is fixed (see Eq.~\eqref{Eq:UltimateSymmetryRequirement}).
Here, we adopt the same convention as in Refs.~\cite{Stockinger:2023ndm,vonManteuffel:2025swv} to parametrise $S^{(n)}_{\mathrm{fct}}$:
first, we restrict ourselves to purely 4-dimensional operators; second, the entire fermionic breaking is attributed to the fermion bilinear term, omitting field monomials of the form $\overline{\psi}B\psi$.
This choice makes use of the freedom to add finite, symmetric counterterms without affecting BRST restoration (see Sec.~\ref{Sec:Symmetry_Restoration_Procedure}).
For example, adding the finite and symmetric counterterm
\begin{equation}
    \begin{aligned}
        - \frac{g^{2n}}{(16 \pi^2)^n} \intx \, 
        \delta F_{\overline{\psi}\psi,ij}^{(n)} \,
        \overline{\psi}_i i {\overline{\slashed{D}}_{R}}_{ik} \psi_j
    \end{aligned}
\end{equation}
shifts the breaking entirely to the fermion--gauge boson interaction.
In this way, we effectively perform a basis change, replacing the field monomial $\overline{\psi}_i \overline{\slashed{\partial}} \projR \psi_j$ in Eq.~\eqref{Eq:Finite-Basis-S_fct-RHTheory} with $\overline{\psi}_i \overline{\slashed{B}} \projR \psi_j$.
Alternatively, the breaking may be distributed between both structures by adding a different finite symmetric counterterm.

Importantly, assigning the complete fermionic breaking to the fermion bilinear terms alone is not always possible and depends on the specific model under consideration.
In more general chiral gauge theories, such as the one discussed in Sec.~\ref{Sec:One-Loop-Renormalisation-Abelian-Chiral-Gauge-Theory}, this assignment is not possible due to non-vanishing contributions from the 1PI Green function $\langle \Delta \psi \overline{\psi} B c \rangle^\mathrm{1PI}$ (cf.\ Eq.~\eqref{Eq:GF-cFBF}), which arise from evanescent gauge and Yukawa interactions absent in the present model.
Consequently, in Sec.~\ref{Sec:One-Loop-Renormalisation-Abelian-Chiral-Gauge-Theory}, the fermionic breaking was attributed to the fermion--gauge boson interaction, omitting fermion bilinear terms to obtain a minimal expression for $S_{\mathrm{fct}}$ (cf.\ Eq.~\eqref{Eq:Sfct_1-Loop-GeneralAbelianTheory}).

\subsection{Application to the Abelian Fermionic Sector of the SM}\label{Sec:S_fct_4-loop-SM-RHTheory}

The fermionic content of the Standard Model can be embedded into the framework of a purely right-handed theory that contains only right-handed physical fermions.
The key idea is to express the left-handed SM fields ${\psi_{L}}_i$ in terms of equivalent right-handed spinors obtained through charge conjugation, $({\psi_L}_i)^C$.
This corresponds to Option~\ref{Opt:Option2b} for the treatment of fermions in $D$ dimensions, as discussed in Sec.~\ref{Sec:Dimensional_Ambiguities_and_Evanescent_Shadows}.
In this formulation, the right-handed multiplet $\psi_R$ collects all physical SM fermions: the neutrinos $\nu_I\in\{\nu_e,\nu_{\mu},\nu_{\tau}\}$,\footnote{%
Here, we include sterile right-handed neutrinos ${\nu^\mathrm{st}_{R,I}}$ and physical left-handed neutrinos $({\nu_{L,I}})^C$, implemented through charge conjugation.} 
the charged leptons $e_I\in\{e,\mu,\tau\}$, and the quarks $u^i_I\in\{u^i,c^i,t^i\}$ and $d^i_I\in\{d^i,s^i,b^i\}$, where $I$ denotes the generation index and $i$ the colour index.
This setup allows to analyse the Abelian fermionic sector of the SM (cf.\ Sec.~\ref{Sec:AnalysisOfASSMResults}) within the right-handed model studied in this chapter.

In this framework, the hypercharge assignments of the SM fermions are encoded in the matrix
\begin{equation}\label{Eq:ASM-Hypercharges-RHTheory}
    \begin{aligned}
        \hypR^{\mathrm{SM}} =
        \begin{pmatrix}
            0 & 0 & 0 & 0 &  &  &  & \\
            0 & - \mathbb{1}_{3\times3} & 0 & 0 &  &  &  & \\
            0 & 0 & \frac{2}{3} \mathbb{1}_{9\times9} & 0 &  &  &  & \\
            0 & 0 & 0 & -\frac{1}{3} \mathbb{1}_{9\times9} &  &  &  & \\
             &  &  &  & \frac{1}{2} \mathbb{1}_{3\times3} & 0 & 0 & 0\\
             &  &  &  & 0 & \frac{1}{2} \mathbb{1}_{3\times3} & 0 & 0\\
             &  &  &  & 0 & 0 & -\frac{1}{6} \mathbb{1}_{9\times9} & 0\\
             &  &  &  & 0 & 0 & 0 & -\frac{1}{6} \mathbb{1}_{9\times9}
        \end{pmatrix},
    \end{aligned}
\end{equation}
where the hypercharges of the right-handed fermions appear in the upper-left block, and those of the charge-conjugated left-handed fermions in the lower-right block.
This hypercharge matrix satisfies the anomaly cancellation condition in Eq.~\eqref{Eq:AnomalyCancellationCondition-RHTheory}, as well as the gauge--gravitational anomaly condition $\mathrm{Tr}(\hypR)=0$.
Moreover, $\mathrm{Tr}(\hypR^2)=10$, which reflects the inclusion of all three generations in a single multiplet and is associated to the familiar normalisation factor $\sqrt{5/3}$ for the hypercharge coupling in Grand Unified Theories.

For the finite symmetry-restoring counterterm action at the 4-loop level, inserting the explicit SM hypercharges $\hypR^{\mathrm{SM}}$ from Eq.~\eqref{Eq:ASM-Hypercharges-RHTheory} into the finite counterterm coefficients of Eqs.~\eqref{Eq:Finite-BB-Break_4-Loop}, \eqref{Eq:Finite-BBBB-Break_4-Loop} and \eqref{Eq:Finite-FF-Break_4-Loop}, we obtain
\begin{equation}\label{Eq:Sfct-4-Loop-SM}
    \begin{aligned}
        S^{(4)}_{\mathrm{fct}}\Big|_{\hypR=\hypR^{\mathrm{SM}}} &= 
        \frac{g^8}{(16 \pi^2)^4} \intx 
        \bigg\{ 
        \delta F_{BB}^{(4),\mathrm{SM}} \, \frac{1}{2} \overline{B}_{\mu} \overline{\partial}^2 \overline{B}^{\mu} 
        + g^2 \delta F_{BBBB}^{(4),\mathrm{SM}} \, \frac{1}{8} \overline{B}_{\mu} \overline{B}^{\mu} \overline{B}_{\nu} \overline{B}^{\nu}\\
        &\qquad\qquad + \delta F_{\overline{e}e}^{(4),\mathrm{SM}} \, \overline{e_{R_I}} i \overline{\slashed{\partial}} e_{R_I} 
        + \delta F_{\overline{u}u}^{(4),\mathrm{SM}} \, \overline{u^i_{R_I}} i \overline{\slashed{\partial}} u^i_{R_I}
        + \delta F_{\overline{d}d}^{(4),\mathrm{SM}} \, \overline{d^i_{R_I}} i \overline{\slashed{\partial}} d^i_{R_I}\\
        &\qquad\qquad + \delta F_{\overline{L}L}^{(4),\mathrm{SM}} \, \overline{\big(L_{L_I}\big)^C} i \overline{\slashed{\partial}} \big(L_{L_I}\big)^C
        + \delta F_{\overline{Q}Q}^{(4),\mathrm{SM}} \, \overline{\big(Q^i_{L_I}\big)^C} i \overline{\slashed{\partial}} \big(Q^i_{L_I}\big)^C
        \bigg\},
    \end{aligned}
\end{equation}
where the explicit 4-loop coefficients are
{\small
\begin{alignat*}{5}
        \delta F_{BB}^{(4),\mathrm{SM}} &= \frac{105574465087}{604661760} &&+ \frac{2665684621}{6998400} \zeta(3) &&- \frac{499349}{174960} \zeta(4) &&- \frac{99009133}{209952} \zeta(5)
        &&\approx 140.38,\\
        \delta F_{BBBB}^{(4),\mathrm{SM}} &= \frac{1740074889071}{27209779200} &&- \frac{3393600941}{7873200} \zeta(3) &&+ \frac{151740709}{3149280} \zeta(4) &&- \frac{50011855}{944784} \zeta(5)
        &&\approx -456.91,\\
        \delta F_{\overline{e}e}^{(4),\mathrm{SM}} &= \frac{27665962754029}{75582720000} &&+ \frac{3480124631}{17496000} \zeta(3) &&+ \frac{211030883}{583200} \zeta(4) &&- \frac{20725669}{19440} \zeta(5)
        &&\approx -108.73,\\
        \delta F_{\overline{u}u}^{(4),\mathrm{SM}} &= \frac{39860981881231}{680244480000} &&+ \frac{22394315941}{88573500} \zeta(3) &&+ \frac{66753553}{1312200} \zeta(4) &&- \frac{231672547}{787320} \zeta(5)
        &&\approx 112.46,\\
        \delta F_{\overline{d}d}^{(4),\mathrm{SM}} &= \frac{2388867184943}{340122240000} &&- \frac{67097069}{1417176000}\zeta(3) &&+ \frac{291946387}{5248800} \zeta(4) &&- \frac{93959699}{1574640} \zeta(5)
        &&\approx 5.29,\\
        \delta F_{\overline{L}L}^{(4),\mathrm{SM}} &= \frac{62763071106811}{4837294080000} &&+ \frac{144119506409}{1119744000} \zeta(3) &&+ \frac{619731137}{37324800} \zeta(4) &&- \frac{34588759}{311040} \zeta(5)
        &&\approx 70.35,\\
        \delta F_{\overline{Q}Q}^{(4),\mathrm{SM}} &= \frac{72769737196819}{43535646720000} &&- \frac{437815601831}{90699264000} \zeta(3) &&+ \frac{7110059113}{335923200} \zeta(4) &&- \frac{476640839}{25194240} \zeta(5)
        &&\approx -0.84.
\end{alignat*}
}

\subsection{Auxiliary Counterterms of the Tadpole Decomposition Method}\label{Sec:AuxiliaryCTs-RHTheory}

Throughout this chapter, UV divergences of the relevant Green functions (see sections~\ref{Sec:Ordinary1PIGF-RHTheory} and \ref{Sec:OperatorInserted1PIGF-RHTheory}) were extracted using the tadpole decomposition method introduced in Sec.~\ref{Sec:Tadpole_Decomposition}.
Following the algorithm outlined there, the procedure requires auxiliary counterterms proportional to the artificial mass scale $M$, which compensate for the omission of numerator terms $\propto M^2$ in intermediate steps of the calculation.

In the right-handed Abelian model considered here, only two 1PI Green functions generate such auxiliary counterterms, both possessing a power-counting degree of divergence $\omega\geq2$: the gauge boson self energy $\langle B^\nu B^\mu\rangle^{\mathrm{1PI}}$ and the $\Delta$-operator-inserted ghost--gauge boson 2-point function $\langle \Delta B^\mu c \rangle^{\mathrm{1PI}}$.
The inclusion of these counterterms is essential for subrenormalisation, ensuring that the discarded $M^2$ numerator terms are properly accounted for.
However, they have no physical significance; they solely serve as a mathematical device inherent to the tadpole decomposition framework and do not belong to the renormalised theory itself.
Consequently, they do not contribute to the BRST breaking ${\big( \Delta \cdot \Gamma \big)}{}^{(n)}$ or to the counterterm action $S_\mathrm{ct}^{(n)}$.
A detailed discussion of their role and interpretation is provided in Sec.~\ref{Sec:Tadpole_Decomposition}.

The explicit $M^2$-contributions of these two Green functions have been computed up to the 4-loop level, with the corresponding coefficients listed in App.~\ref{App:AuxiliaryCTs-RightHandedAbelianTheory}.
At the $n$-loop level, the auxiliary contributions read\footnote{The auxiliary counterterms are given by the negative of these contributions, ensuring their cancellation.}
\begin{align}
    \begin{split}\label{Eq:Aux-CT-GaugeBoson-SelfEnergy}
        i \Gamma_{B_{\nu}(-p)B_{\mu}(p)} \big|_{\mathrm{mass}}^{(n)} &= \frac{ig^{2n}}{(16\pi^2)^n} \delta {\overline{M}}{}^{(n)}_{BB} M^2 \overline{\eta}^{\mu\nu},
    \end{split}\\
    \begin{split}\label{Eq:Aux-CT-Ghost-GaugeBoson}
        i {\big( \Delta \cdot \Gamma \big)_{B_{\mu}(-p) c(p)}^{(n)}}\Big|^{(n)}_\mathrm{mass} &= \frac{g^{2n}}{(16\pi^2)^n} \delta {\overline{M}}{}^{(n)}_{cB} M^2 \overline{p}^{\mu}.
    \end{split}
\end{align}
Here, $\delta {\overline{M}}{}^{(n)}_{BB}$ and $\delta {\overline{M}}{}^{(n)}_{cB}$ are the auxiliary counterterm coefficients at loop order $n\in\{1,2,3,4\}$.
As in other $\Delta$-operator-inserted Green functions (cf.\ Sec.~\ref{Sec:OperatorInserted1PIGF-RHTheory}), the coefficient $\delta {\overline{M}}{}^{(n)}_{cB}$ also contains finite parts originating from the insertion of the local composite operator $\Delta$, whose lowest-order contribution is purely evanescent.

The corresponding auxiliary counterterm coefficients are:

\paragraph{1-Loop Counterterm Coefficients:}

\begin{align}
    \begin{split}
        \delta {\overline{M}}{}^{(1)}_{BB} &= {\overline{\mathcal{A}}}{}^{1,\mathrm{mass}}_{BB} \, \frac{1}{\epsilon},
    \end{split}\\
    \begin{split}
        \delta {\overline{M}}{}^{(1)}_{cB} &= {\overline{\mathcal{F}}}{}^{1,\mathrm{mass}}_{cB}.
    \end{split}
\end{align}

\paragraph{2-Loop Counterterm Coefficients:}

\begin{align}
    \begin{split}
        \delta {\overline{M}}{}^{(2)}_{BB} &= {\overline{\mathcal{A}}}{}^{2,\mathrm{mass}}_{BB} \, \frac{1}{\epsilon} + {\overline{\mathcal{B}}}{}^{2,\mathrm{mass}}_{BB} \, \frac{1}{\epsilon^2},
    \end{split}\\
    \begin{split}
        \delta {\overline{M}}{}^{(2)}_{cB} &= {\overline{\mathcal{F}}}{}^{2,\mathrm{mass}}_{cB} + {\overline{\mathcal{A}}}{}^{2,\mathrm{mass}}_{cB} \, \frac{1}{\epsilon}.
    \end{split}
\end{align}

\paragraph{3-Loop Counterterm Coefficients:}

\begin{align}
    \begin{split}
        \delta {\overline{M}}{}^{(3)}_{BB} &= {\overline{\mathcal{A}}}{}^{3,\mathrm{mass}}_{BB} \, \frac{1}{\epsilon} + {\overline{\mathcal{B}}}{}^{3,\mathrm{mass}}_{BB} \, \frac{1}{\epsilon^2} + {\overline{\mathcal{C}}}{}^{3,\mathrm{mass}}_{BB} \, \frac{1}{\epsilon^3},
    \end{split}\\
    \begin{split}
        \delta {\overline{M}}{}^{(3)}_{cB} &= {\overline{\mathcal{F}}}{}^{3,\mathrm{mass}}_{cB} + {\overline{\mathcal{A}}}{}^{3,\mathrm{mass}}_{cB} \, \frac{1}{\epsilon} + {\overline{\mathcal{B}}}{}^{3,\mathrm{mass}}_{cB} \, \frac{1}{\epsilon^2}.
    \end{split}
\end{align}

\paragraph{4-Loop Counterterm Coefficients:}

\begin{align}
    \begin{split}
        \delta {\overline{M}}{}^{(4)}_{BB} &= {\overline{\mathcal{A}}}{}^{4,\mathrm{mass}}_{BB} \, \frac{1}{\epsilon} + {\overline{\mathcal{B}}}{}^{4,\mathrm{mass}}_{BB} \, \frac{1}{\epsilon^2} + {\overline{\mathcal{C}}}{}^{4,\mathrm{mass}}_{BB} \, \frac{1}{\epsilon^3} + {\overline{\mathcal{D}}}{}^{4,\mathrm{mass}}_{BB} \, \frac{1}{\epsilon^4},
    \end{split}\\
    \begin{split}
        \delta {\overline{M}}{}^{(4)}_{cB} &= {\overline{\mathcal{F}}}{}^{4,\mathrm{mass}}_{cB} +  {\overline{\mathcal{A}}}{}^{4,\mathrm{mass}}_{cB} \, \frac{1}{\epsilon} + {\overline{\mathcal{B}}}{}^{4,\mathrm{mass}}_{cB} \, \frac{1}{\epsilon^2} + {\overline{\mathcal{C}}}{}^{4,\mathrm{mass}}_{cB} \, \frac{1}{\epsilon^3}.
    \end{split}
\end{align}

\subsection{Abelian Ward Identities}\label{Sec:AbelianWIs-RHTheory}

As discussed in Sec.~\ref{Sec:Symmetry_Restoration_Procedure} and demonstrated explicitly for the right-handed model in Sec.~\ref{Sec:OperatorInserted1PIGF-RHTheory}, the BMHV scheme breaks the well-known Abelian Ward identities that follow from Eq.~\eqref{Eq:FunctionalAbelianWardId}.
This regularisation-induced breaking affects, in particular, the transversality of the gauge boson self energy and of multi gauge boson interactions, as well as the relation between the fermion self energy and the fermion--gauge boson interaction.

At the $n$-loop level, the corresponding violations of these identities at the subrenormalised level can be expressed as
\begin{subequations}
    \begin{align}
        p_\nu \Big(i \Gamma^{(n)}_{B_{\nu}(-p)B_{\mu}(p)}\Big) &= i \Big(i\Delta \cdot \Gamma \big|_{B_{\mu}(-p) c(p)}^{(n)}\Big),
        \\
        -q_\sigma \Big( i \Gamma^{(n)}_{B_{\mu}(p_1)B_{\nu}(p_2)B_{\rho}(p_3)B_{\sigma}(q)}\Big) &=-i\Big(i \Delta \cdot \Gamma \big|_{B_{\rho}(p_3) B_{\nu}(p_2) B_{\mu}(p_1) c(q)}^{(n)}\Big),
        \\
        \Big( i \Gamma^{(n)}_{\psi_j(p)\overline{\psi}_i(-p)B_{\mu}(0)}\Big) + g \mathcal{Y}_R \frac{\partial}{\partial p_\mu} \Big( i \Gamma^{(n)}_{\psi_{j}(p)\overline{\psi}_{i}(-p)} \Big) &= i \frac{\partial}{\partial q_\mu} \Big( i \Delta \cdot \Gamma \big|_{\psi_j(p_2) \overline{\psi}_i(p_1) c(q)}^{(n)} \Big)\Big|_{q=0},
    \end{align}
\end{subequations}
with $q=-p_1-p_2-p_3$.
These relations coincide with those presented in Sec.~\ref{Sec:STIs-in-Abelian-Chiral-Gauge-Theory}, specialised to the present right-handed model and, in the case of the fermionic identity, expressed in differential form.
They provide a direct interpretation of the symmetry breaking (as noted in Sec.~\ref{Sec:Peculiarities_of_Abelian_Gauge_Theories}) and serve as an internal consistency check of the computations, since both sides can be evaluated independently following the procedure described in Sec.~\ref{Sec:Symmetry_Restoration_Procedure}.

Once the symmetry restoration procedure is completed, and the full counterterm action $S_\mathrm{ct}^{(n)}=S_\mathrm{sct}^{(n)}+S_\mathrm{fct}^{(n)}$ is added to the subrenormalised effective quantum action, these breakings must cancel.
The Ward identities are then restored, meaning that for the fully renormalised effective quantum action $\Gamma_\mathrm{ren}$ at a given loop order, the RHS of the above equations vanish.
Explicitly, we recover the familiar Ward identities:
the transversality of the gauge boson self energy,
\begin{align}
    p_\nu \frac{i\delta^2 \Gamma_\mathrm{ren}}{\delta B_\nu(-p)\delta B_\mu(p)} = 0,
\end{align}
the transversality of multi gauge boson interactions (shown here for the 4-point function),
\begin{align}
    -q_\sigma\, \frac{i\delta^4 \Gamma_\mathrm{ren}}{\delta B_\mu(p_1)\delta B_\nu(p_2)\delta B_\rho(p_3) \delta B_\sigma(q)} = 0,
\end{align}
and the relation of the fermion self energy and the fermion--gauge boson interaction,
\begin{align}
    \frac{i\delta^3 \Gamma_\mathrm{ren}}{\delta \psi_j(p)\delta \overline{\psi}_i(-p)\delta B_\mu(0)} + g \mathcal{Y}_R \frac{\partial}{\partial p_\mu} \frac{i\delta^2 \Gamma_\mathrm{ren}}{\delta \psi_j(p)\delta \overline{\psi}_i(-p)} = 0.
\end{align}
Hence, the inclusion of the complete counterterm action restores the Slavnov-Taylor identity at the renormalised level and, consequently, the Abelian Ward identities order by order in perturbation theory.

\chapter{The Standard Model in the BMHV Scheme}\label{Chap:The_Standard_Model}

In this chapter, we apply our methodology to the Standard Model of particle physics and present the first step towards a fully self-consistent multi-loop renormalisation of the SM with a non-anticommuting $\gamma_5$ in the BMHV scheme.
So far, multi-loop computations within the SM have typically relied on additional external arguments to resolve ambiguities arising in regularisation schemes that employ a naively anticommuting $\gamma_5$ and/or were limited to specific applications (see Sec.~\ref{Sec:Alternative-g5-Schemes} and the references therein).
A comprehensive and fully consistent treatment within the BMHV framework at the multi-loop level has not yet been achieved.

The overarching goal of our research programme is to perform such a multi-loop renormalisation of the SM, employing a no-compromise approach to a fully consistent treatment of $\gamma_5$ within the BMHV scheme and providing all required counterterms, thereby enabling high-precision electroweak phenomenology without ambiguities and conceptual limitations.
While this endeavour is still ongoing, we have reached an important milestone by completing the full 1-loop renormalisation of the SM in the BMHV scheme, which we discuss in this chapter.
This achievement represents a crucial first step and provides the foundation for subsequent higher-order computations.

In contrast to chapters~\ref{Chap:General_Abelian_Chiral_Gauge_Theory} and \ref{Chap:BMHV_at_Multi-Loop_Level}, the present chapter does not present a completed project but rather reports intermediate results from an ongoing research effort.
Accordingly, we keep the discussion concise and defer a more detailed analysis to forthcoming publications, which are currently in preparation.
The results shown here are thus new and not yet published.
We note that a first discussion of the finite symmetry-restoring 1-loop counterterms in this context was provided in Ref.~\cite{OlgosoRuiz:2024dzq}, which, however, was restricted exclusively to the finite counterterms, employed a rather compact presentation not directly suited for phenomenological applications, and did not include evanescent gauge interactions.

In what follows, we present the first complete 1-loop renormalisation of the full Standard Model within the BMHV scheme, including all required counterterms --- both divergent and finite --- which render the SM finite and reestablish the validity of the Slavnov-Taylor identity at the 1-loop level.
We begin in Sec.~\ref{Sec:SM-Definition-and-Setup} with a brief formulation of the SM, introducing our notation and discussing its $D$-dimensional extension within the BMHV scheme of DReg.
Section~\ref{Sec:SymmetryBreaking_in_SM} then analyses the tree-level symmetry breaking induced by the BMHV regularisation.
Finally, in Sec.~\ref{Sec:SM_1-Loop_Renormalisation}, we present the complete 1-loop counterterm action, encompassing both the singular and the finite symmetry-restoring counterterms.
Moreover, we provide an initial exploration of the implications of evanescent gauge interactions within the framework of the SM, continuing the analysis of chapter~\ref{Chap:General_Abelian_Chiral_Gauge_Theory}, particularly Sec.~\ref{Sec:Results-Shedding_Light_on_Evanescent_Shadows}.

\section{The Standard Model of Particle Physics}\label{Sec:SM-Definition-and-Setup}

The Standard Model is a chiral gauge theory in which left- and right-handed fermions interact differently with the electroweak gauge bosons, as discussed in Sec.~\ref{Sec:Chiral_Gauge_Theories}.
Its gauge group (already introduced in Eq.~\eqref{Eq:The-SM-Gauge-Group}) is given by
\begin{align}\label{Eq:The-SM-Gauge-Group-YetAgain}
    G_{\mathrm{SM}}=U(1)_Y \times SU(2)_L \times SU(3)_c \, .
\end{align}
Only left-handed fermions transform nontrivially under $SU(2)_L$, whereas right-handed fermions appear as $SU(2)_L$ singlets.
In addition, left- and right-handed fermions carry different $U(1)_Y$ hypercharges.
As a consequence, the chiral nature of the SM requires a consistent treatment of $\gamma_5$ within the regularisation and renormalisation procedure.

In the following, we introduce our notation, as well as the SM fields and generators in Sec.~\ref{Sec:Fields_Generators_Notation}.
In Sec.~\ref{Sec:SM-Lagrangian}, we then discuss the complete SM Lagrangian, regularised in the BMHV of DReg.
Finally, Sec.~\ref{Eq:SM-GroupInvariants-CouplingStructures} introduces several group invariants and coupling structures constructed from the SM generators and matrices, which will be used to express the counterterm results.

\subsection{Fields, Generators and Notation}\label{Sec:Fields_Generators_Notation}

We denote fermion generation (or flavour) indices by $I,J,K,\ldots\in\{1,2,3\}$, accounting for the three generations of the SM, $N_{f}=3=\delta_{II}$.
Fundamental colour indices associated with $SU(3)_c$ are written as $i,j,k,\ldots\in\{1,2,3\}$, with $N_c=3=\delta_{ii}$, while doublet indices of $SU(2)_L$ are denoted by $a,b,c,\ldots\in\{1,2\}$, where $N_L=2=\delta_{aa}$.
In this way, leptons and quarks can be expressed in terms of doublets as 
\begin{align}\label{Eq:SM-Fermion-Doublets}
    l_I^{a} =
    {\begin{pmatrix}
        \nu_I\\
        e_I
    \end{pmatrix}}^a ,
    \qquad \quad
    q_I^{i,a} =
    {\begin{pmatrix}
        u^i_I\\
        d^i_I
    \end{pmatrix}}^a .
\end{align}
The scalar sector of the SM contains two complex fields $\phi_a$, forming the Higgs doublet
\begin{equation}\label{Eq:SM-Higgs-Doublet}
    \begin{aligned}
        \Phi =
        \begin{pmatrix}
            \phi_1\\
            \phi_2
        \end{pmatrix},
    \end{aligned}
\end{equation}
where $\phi_1=G^{+}$ and $\phi_2=(\varphi+iG^{0})/\sqrt{2}$, with $G^0$ and $G^\pm$ the Goldstone bosons and $\varphi=H+v$ representing the Higgs boson $H$ and its vacuum expectation value $v$.
The SM contains twelve gauge bosons in total: one gauge boson $B_\mu$ associated with $U(1)_Y$, three $W_\mu^A$ for $SU(2)_L$, and eight gluons for $SU(3)_c$.
For compactness, we collectively denote all gauge bosons by $\vgb_\mu^A$ with adjoint colour indices $A,B,C,\ldots\in\{1,\ldots,12\}$, such that $A=1$ corresponds to $U(1)_Y$, $A\in\{2,3,4\}$ to $SU(2)_L$, and $A\in\{5,\ldots,12\}$ to $SU(3)_c$.
After spontaneous symmetry breaking, the gauge fields $B_\mu$ and $W_\mu^A$ give rise to the physical mass eigenstates --- the photon $A_\mu$ and the weak gauge bosons $Z_{\mu}$ and $W_\mu^\pm$.
The relations between the flavour and mass eigenstates are given by
\begin{subequations}\label{Eq:EWSM-Gauge-Boson-Flavour-and-Mass-Eigenstates}
    \begin{align}
        W^1_\mu &= \frac{W_\mu^+ + W_\mu^-}{\sqrt{2}},\\
        W^2_\mu &= i \frac{W_\mu^+ - W_\mu^-}{\sqrt{2}},\\
        W^3_\mu &= Z_\mu \cos(\theta_W) + A_\mu \sin(\theta_W),\\
        B_\mu &= - Z_\mu \sin(\theta_W) + A_\mu \cos(\theta_W),
    \end{align}
\end{subequations}
where $\theta_W$ denotes the weak mixing angle.
Note that an adjoint index marked with a prime, e.g.\ $A'$, is not implicitly summed over but serves solely as a label.

In the following, we discuss the generators of the SM gauge group (see Eq.~\eqref{Eq:The-SM-Gauge-Group-YetAgain}).
For $U(1)_Y$, the hypercharges of left- and right-handed fermions are given by
\begin{equation}\label{Eq:SM-Hypercharges}
    \begin{alignedat}{2}
        &\mathcal{Y}^l_{L} = 
            \begin{pmatrix}
                \mathcal{Y}_l & 0\\
                0 & \mathcal{Y}_l
            \end{pmatrix} =
            \begin{pmatrix}
                -\frac{1}{2} & 0\\
                0 & -\frac{1}{2}
            \end{pmatrix},
            \qquad
        &&\mathcal{Y}^l_{R} = 
            \begin{pmatrix}
                \mathcal{Y}_{\nu} & 0\\
                0 & \mathcal{Y}_e
            \end{pmatrix} =
            \begin{pmatrix}
                0 & 0\\
                0 & -1
            \end{pmatrix},\\
        &\mathcal{Y}^q_{L} = 
            \begin{pmatrix}
                \mathcal{Y}_q & 0\\
                0 & \mathcal{Y}_q
            \end{pmatrix} =
            \begin{pmatrix}
                \frac{1}{6} & 0\\
                0 & \frac{1}{6}
            \end{pmatrix},
            \qquad
        &&\mathcal{Y}^q_{R} = 
            \begin{pmatrix}
                \mathcal{Y}_u & 0\\
                0 & \mathcal{Y}_d
            \end{pmatrix} =
            \begin{pmatrix}
                \frac{2}{3} & 0\\
                0 & -\frac{1}{3}
            \end{pmatrix},
    \end{alignedat}
\end{equation}
whereas the universal hypercharge of the scalars $\phi_a$ reads
\begin{equation}\label{Eq:SM-Scalar-Hypercharge}
    \begin{aligned}
        \mathcal{Y}_{S} = \frac{1}{2}.
    \end{aligned}
\end{equation}
We denote the weak generators of $SU(2)_L$ by $t_L^A$ and the strong generators of $SU(3)_c$ by $t^A_s$, so that
\begin{equation}\label{Eq:SM-Generators}
    \begin{aligned}
        t^{A}_L = \frac{\sigma^{(A-1)}}{2}, 
        \qquad &A\in\{2,3,4\},\\
        t^A_s = \frac{\lambda^{(A-4)}}{2},
        \qquad &A\in\{5,\ldots,12\},
    \end{aligned}
\end{equation}
where $\sigma^A$ denote the Pauli matrices, 
\begin{align}
    \sigma^1=
    \begin{pmatrix}
        0 & 1\\
        1 & 0
    \end{pmatrix},
    \qquad
    \sigma^2=
    \begin{pmatrix}
        0 & -i\\
        i & 0
    \end{pmatrix},
    \qquad
    \sigma^3=
    \begin{pmatrix}
        1 & 0\\
        0 & -1
    \end{pmatrix},
\end{align}
and $\lambda^A$ the Gell-Mann matrices.
The weak and strong generators obey the familiar Lie algebra relations
\begin{equation}\label{Eq:SM-CommutatorRelations}
    \begin{aligned}
        \big[t_L^A,t_L^B\big] = i \varepsilon^{ABC} t_L^C,
        \qquad\quad
        \big[t_s^A,t_s^B\big] = i f^{ABC} t_s^C,
    \end{aligned}
\end{equation}
where $\varepsilon^{ABC}$ denotes the totally antisymmetric Levi-Civita symbol and $f^{ABC}$ the totally antisymmetric structure constants of $SU(3)_c$.
For later use, we define the combinations
\begin{align}
    \sigma_{12}^\pm \coloneqq \frac{\sigma^1 \pm i \sigma^2}{2},
    \qquad\quad
    \sigma_{03}^\pm \coloneqq \frac{\mathbb{1}_{2\times2} \pm  \sigma^3}{2}.
\end{align}
With the specific matter content of the Standard Model, the left- and right-handed hypercharge assignments given in Eq.~\eqref{Eq:SM-Hypercharges}, and the generators defined in Eq.~\eqref{Eq:SM-Generators}, the anomaly cancellation condition of Eq.~\eqref{Eq:General-Anomaly-Cancellation-Condition} is fulfilled.
Consequently, the Standard Model is free of gauge anomalies (see Sec.~\ref{Sec:Anomalies} and references therein).

To streamline notation, we collect all generators into unified objects, analogous to the treatment of the gauge bosons.
To this end, we define the lepton generators as
\begin{equation}\label{Eq:Lepton-Super-Generator}
    \begin{aligned}
        \myGenLe^A_{\overline{\alpha}\beta,ab} =
        \begin{cases}
            g_Y \, \mathcal{Y}^l_{\overline{\alpha}\beta,ab}, \quad &A = 1\\
            g_W \, t^A_{W,\overline{\alpha}\beta,ab}, \quad &A \in \{2,3,4\}\\
            0, \quad &A \in \{5,\ldots,12\}
        \end{cases}
    \end{aligned},
\end{equation}
the quark generators as
\begin{equation}\label{Eq:Quark-Super-Generator}
    \begin{aligned}
        \myGenQu^A_{\overline{\alpha}\beta,ab,ij} =
        \begin{cases}
            g_Y \, \mathcal{Y}^q_{\overline{\alpha}\beta,ab} \, \delta_{ij}, \quad &A = 1\\
            g_W \, t^{A}_{W,\overline{\alpha}\beta,ab} \, \delta_{ij}, \quad &A \in \{2,3,4\}\\
            g_s \, t^{A}_{S,\overline{\alpha}\beta,ij} \, \delta_{ab}, \quad &A \in \{5,\ldots,12\}
        \end{cases}
    \end{aligned},
\end{equation}
and the scalar generators as
\begin{equation}\label{Eq:Scalar-Super-Generator}
    \begin{aligned}
        \myGenScl^A_{ab} =
        \begin{cases}
            g_Y \, \mathcal{Y}_{S} \, \delta_{ab}, \quad &A = 1\\
            g_W \, t^A_{L,ab}, \quad &A \in \{2,3,4\}\\
            0, \quad &A \in \{5,\ldots,12\}
        \end{cases}
    \end{aligned}.
\end{equation}
Note that we employ the compact chirality notation introduced in Sec.~\ref{Sec:Practical_Implementation_of_Counterterms}, where the indices $\alpha$ and $\beta$ denote chirality, and an overbar indicates opposite handedness, i.e.\ $\overline{\alpha}=R$ for $\alpha=L$ and vice versa.
In this ``chirality space'', the generators take the explicit form
\begin{equation}\label{Eq:Generators-in-Projectorspace}
    \begin{aligned}
        \big( \mathcal{Y}^{f}_{\alpha\beta,ab} \big) 
        &=   \begin{pmatrix}
                \mathcal{Y}^f_{L,ab} & \widehat{\mathcal{Y}}^{f}_{ab}\\
                \big({\widehat{\mathcal{Y}}^{f}}{}^{\dagger}\big)_{ab} & \mathcal{Y}^f_{R,ab}
            \end{pmatrix},
        \qquad f\in\{l,q\},
        \\
        \big( t^A_{W,\alpha\beta,ab} \big) 
        &=   \begin{pmatrix}
                t^A_{L,ab} & \widehat{t}^A_{ab}\\
                \big({\widehat{t}^A}{}^{\dagger}\big)_{ab} & 0
            \end{pmatrix},
        \\
        \big( t^A_{S,\alpha\beta,ij} \big) 
        &=   \begin{pmatrix}
                t^A_{s,ij} & \widehat{t}^A_{s,ij}\\
                \big({\widehat{t}^A_{s}}{}^{\dagger}\big)_{ij} & t^A_{s,ij}
            \end{pmatrix},
    \end{aligned}
\end{equation}
where $\widehat{\mathcal{Y}}^{f}_{ab}$, $\widehat{t}^A_{ab}$, and $\widehat{t}^A_{s,ij}$ denote the generators of the evanescent gauge interactions (see Sec.~\ref{Sec:Dimensional_Ambiguities_and_Evanescent_Shadows} for details), whose specific form is determined in Sec.~\ref{Sec:SM-Lagrangian}.
The electric charges of the SM fermions are given by (cf.\ Sec.~\ref{Sec:ASSM}),
\begin{align}
    Q^l = \begin{pmatrix}
            0 & 0\\
            0 & -1
        \end{pmatrix},
    \qquad\quad
    Q^q = \begin{pmatrix}
            \frac{2}{3} & 0\\
            0 & -\frac{1}{3}
        \end{pmatrix}.
\end{align}
Using the generators introduced above, the electric charge generator is defined as
\begin{align}\label{Eq:Electric-Charge-Generator-SM}
    \mathcal{Q}_{\overline{\alpha}\beta,ab}^f = t^{3}_{W,\overline{\alpha}\beta,ab} + \mathcal{Y}^f_{\overline{\alpha}\beta,ab},
\end{align}
and the weak generators associated with the charged gauge bosons $W_\mu^\pm$ as
\begin{align}
    \tau^\pm_{\overline{\alpha}\beta,ab} = t^{1}_{W,\overline{\alpha}\beta,ab} \pm i t^{2}_{W,\overline{\alpha}\beta,ab}.
\end{align}
The electromagnetic coupling constant is given by
\begin{align}
    e = g_Y \cos(\theta_W) = g_W \sin(\theta_W) = \frac{g_Y g_W}{\sqrt{g_Y^2 + g_W^2}}.
\end{align}
Analogous to the generators, we also introduce the generalised structure constants as
\begin{equation}\label{Eq:Super-Structure-Constants}
    \begin{aligned}
        \mathscr{C}^{ABC} = 
        \begin{cases}
            g_W \, \varepsilon^{ABC}, \quad &A,B,C \in \{2,3,4\},\\
            g_s \, f^{ABC}, \quad &A,B,C \in \{5,\ldots,12\},\\
            0, \quad &\mathrm{else}.
        \end{cases}
    \end{aligned}
\end{equation}

\subsection{The Standard Model Lagrangian}\label{Sec:SM-Lagrangian}

The complete Standard Model Lagrangian in $D$ dimensions is given by 
\begin{align}\label{Eq:SM-Lagrangian-Complete}
    \mathcal{L} = \mathcal{L}_\mathrm{fermion} + \mathcal{L}_\mathrm{gauge} + \mathcal{L}_\mathrm{Higgs} + \mathcal{L}_\mathrm{Yukawa} + \mathcal{L}_\mathrm{ghost+fix} + \mathcal{L}_\mathrm{ext},
\end{align}
and serves as the starting point for the renormalisation procedure.
Each term corresponds to one of the sectors of the Standard Model, which will be briefly introduced in the following paragraphs.

\paragraph{Fermionic Sector:}

Following the discussion in Sec.~\ref{Sec:Dimensional_Ambiguities_and_Evanescent_Shadows}, the general $D$-dimensional Ansatz for the fermionic Lagrangian of the SM may be written as
\begin{align}\label{Eq:L_fermion-in-the-SM}
    \mathcal{L}_\mathrm{fermion} = 
    {\overline{l}}{}^a_I i \slashed{\partial} l_I^a - \myGenLe_{\overline{\alpha}\beta,ab}^A {\overline{l}}{}^a_I \,\mathbb{P}_{\alpha} {\slashed{\vgb}}{}^A \mathbb{P}_{\beta} \, l_I^b
    + {\overline{q}}{}^{i,a}_I i \slashed{\partial} q_I^{i,a} - \myGenQu_{\overline{\alpha}\beta,ab,ij}^A {\overline{q}}{}^{i,a}_I \,\mathbb{P}_{\alpha} {\slashed{\vgb}}{}^A \mathbb{P}_{\beta} \, q_I^{j,b},
\end{align}
with generators as defined in Eqs.~\eqref{Eq:Lepton-Super-Generator} and \eqref{Eq:Quark-Super-Generator}.

This Ansatz includes the physical 4-dimensional gauge interaction currents and general evanescent gauge interactions, which remain unspecified at this stage except for the requirement of hermiticity.
As discussed in Sec.~\ref{Sec:Results-Shedding_Light_on_Evanescent_Shadows}, omitting evanescent interactions typically yields the most compact results and is thus often the preferred choice.
In the Standard Model, however, the situation is more subtle due to the mixing of the electroweak gauge fields $B_\mu$ and $W_\mu^A$ into mass eigenstates $A_\mu$, $Z_\mu$ and $W_\mu^\pm$ (see Eq.~\eqref{Eq:EWSM-Gauge-Boson-Flavour-and-Mass-Eigenstates}).
Given that electromagnetism and the strong interaction are vector-like, it is natural to ask whether the photon and gluon interactions can be treated in a fully $D$-dimensional way.
To achieve this, all four chiral components of the fermionic currents must couple to the corresponding gauge boson through identical generators.

For the strong interaction, this condition is readily satisfied by identifying the corresponding evanescent generator with the strong generator, $\widehat{t}_s^A=t_{s}^A$.
The electromagnetic case is less straightforward: here, the electric charge generator (see Eq.~\eqref{Eq:Electric-Charge-Generator-SM}) must be identical for all chiral covariants, i.e.\ equal to the electric charge itself, leading to the condition
\begin{align}\label{Eq:SM-Condition-on-Electric-Charge-Generator}
    \big( \mathcal{Q}_{\overline{\alpha}\beta,ab}^f \big) =
    \begin{pmatrix}
        \mathcal{Y}^f_{L,ab} + t^3_{L,ab} & \widehat{\mathcal{Y}}^{f}_{ab} + \widehat{t}^3_{ab}\\
        \big({\widehat{\mathcal{Y}}^{f}}{}^{\dagger}\big)_{ab} + \big({\widehat{t}^3}{}^{\dagger}\big)_{ab} & \mathcal{Y}^f_{R,ab}
    \end{pmatrix}
    \overset{!}{=}
    \begin{pmatrix}
        Q^f_{ab} & Q^f_{ab}\\
        Q^f_{ab} & Q^f_{ab}
    \end{pmatrix},
    \qquad f\in\{l,q\}.
\end{align}
Hence, only the combination of the hypercharge and the third weak generator is constrained.

Further insight arises from examining the weak interactions of neutrinos.
Using Eq.~\eqref{Eq:Generators-in-Projectorspace} in Eq.~\eqref{Eq:L_fermion-in-the-SM}, the corresponding contribution reads
\begin{align}
        &g_W\, t^{A}_{L,11} \overline{\nu_{L}}_{,I} \slashed{W}^A {\nu_{L,I}}  
        + g_W t^{A}_{L,12} \overline{\nu_{L}}_{,I} \slashed{W}^A {e_{L,I}} 
        + g_W t^{A}_{L,21} \overline{e_{L}}_{,I} \slashed{W}^A {\nu_{L,I}} 
        + g_W t^{A}_{L,22} \overline{e_{L}}_{,I} \slashed{W}^A {e_{L,I}} \nonumber\\
        +\, &g_W\, \widehat{t}^{A}_{11} \overline{\nu_{L}}_{,I} \slashed{W}^A {\nu^\mathrm{st}_{R,I}}  
        + g_W \widehat{t}^{A}_{12} \overline{\nu_{L}}_{,I} \slashed{W}^A {e_{R,I}} 
        + g_W \widehat{t}^{A}_{21} \overline{e_{L}}_{,I} \slashed{W}^A {\nu^\mathrm{st}_{R,I}} 
        + g_W \widehat{t}^{A}_{22} \overline{e_{L}}_{,I} \slashed{W}^A {e_{R,I}}\\
        +\, &g_W\, {\widehat{t}^{A}}{}^*_{\!\!\!\!11} \overline{\nu^\mathrm{st}_{R}}_{\!, I} \slashed{W}^A {\nu_{L,I}}  
        + g_W {\widehat{t}^{A}}{}^*_{\!\!\!\!21} \overline{\nu^\mathrm{st}_{R}}_{\!, I} \slashed{W}^A {e_{L,I}} 
        + g_W {\widehat{t}^{A}}{}^*_{\!\!\!\!12} \overline{e_{R}}_{,I} \slashed{W}^A {\nu_{L,I}} 
        + g_W {\widehat{t}^{A}}{}^*_{\!\!\!\!22} \overline{e_{R}}_{,I} \slashed{W}^A {e_{L,I}} \nonumber
\end{align}
Some of these terms involve the sterile right-handed neutrinos ${\nu^\mathrm{st}_{R,I}}$, which by definition must not participate in any interaction.
Consequently, such couplings must vanish identically, implying that the evanescent generator in the lepton sector must take the form
\begin{align}
    \widehat{t}^A =
    \begin{pmatrix}
        0 & \widehat{t}^A_{12}\\
        0 & \widehat{t}^A_{22}
    \end{pmatrix}.
\end{align}
This generator does not satisfy the $\mathfrak{su}(2)$ algebra, which, while not necessary for evanescent generators, complicates computations and is thus undesirable.
We therefore adopt the simplifying choice $\widehat{t}^A\equiv0$.
Under this condition, Eq.~\eqref{Eq:SM-Condition-on-Electric-Charge-Generator} requires the evanescent hypercharges to coincide with the right-handed hypercharges, i.e.\ $\widehat{\mathcal{Y}}^{f}_{ab}=\mathcal{Y}^f_{R,ab}$.

In this project, we do not pursue the most general form of evanescent gauge interactions.
Such general considerations were already analysed in the Abelian case in chapter~\ref{Chap:General_Abelian_Chiral_Gauge_Theory}, where it was found that nonzero evanescent couplings tend to produce significantly more involved expressions and thereby complicate the computation.
To nevertheless explore their potential impact in the SM --- in particular, the effect of treating QED and/or QCD as fully 4- or $D$-dimensional --- we adopt the following parametrisation:
\begin{equation}\label{Eq:Evanescent-Generators-Choice}
    \begin{aligned}
        \widehat{\mathcal{Y}}^{f}_{ab} &= c_{\mathrm{QED}} \, \mathcal{Y}^f_{R,ab}, \qquad f\in\{l,q\},\\
        \widehat{t}^A_{ab} &\equiv 0,\\
        \widehat{t}^A_{s,ij} &= c_{\mathrm{QCD}} \, t^A_{s,ij},
    \end{aligned}
\end{equation}
with $c_{\mathrm{QED}}, c_{\mathrm{QCD}} \in \mathbb{R}$.
This choice ensures $\big({\widehat{\mathcal{Y}}^{f}}{}^{\dagger}\big) = \widehat{\mathcal{Y}}^{f}$ and $\big({\widehat{t}^A_{s}}{}^{\dagger}\big)=\widehat{t}^A_{s}$, leading to $\myGenLe^A_{LR}=\myGenLe^A_{RL}=\widehat{\myGenLe}^A$ and $\myGenQu^A_{LR}=\myGenQu^A_{RL}=\widehat{\myGenQu}^A$.

Inserting Eq.~\eqref{Eq:Evanescent-Generators-Choice} into Eq.~\eqref{Eq:L_fermion-in-the-SM} and transforming the gauge bosons to mass eigenstates according to Eq.~\eqref{Eq:EWSM-Gauge-Boson-Flavour-and-Mass-Eigenstates}, we obtain the following expressions for the gauge interaction terms:
\begin{itemize}
    \item photon interaction current
    \begin{equation}\label{Eq:Photon-Interaction-Current-SM}
        \begin{aligned}
            &- e Q^l_{ab} {\overline{l}}{}^{a}_I \slashed{A} l_I^{b} + (1-c_\mathrm{QED}) e Q^l_{ab} {\overline{l}}{}^{a}_I \widehat{\slashed{A}} l_I^{b}\\
            &- \, e Q^q_{ab} {\overline{q}}{}^{i,a}_I \slashed{A} q_I^{i,b} + (1-c_\mathrm{QED}) e Q^q_{ab} {\overline{q}}{}^{i,a}_I \widehat{\slashed{A}} q_I^{i,b},
        \end{aligned}
    \end{equation}
    \item $Z_\mu$ interaction current
    \begin{equation}\label{Eq:Z-Interaction-Current-SM}
        \begin{aligned}
            &\frac{s_W}{c_W} e Q^l_{ab} {\overline{l}}{}^{a}_I \slashed{Z} l_I^{b}
            -\frac{g_W}{c_W} t^3_{L,ab} {\overline{l}}{}^{a}_I \overline{\slashed{Z}} \mathbb{P}_{\mathrm{L}} l_I^{b}
            -(1-c_\mathrm{QED})\frac{s_W}{c_W} e Q^l_{ab} {\overline{l}}{}^{a}_I \widehat{\slashed{Z}} l_I^{b}\\
            +\,&\frac{s_W}{c_W} e Q^q_{ab} {\overline{q}}{}^{i,a}_I \slashed{Z} q_I^{i,b}
            -\frac{g_W}{c_W} t^3_{L,ab} {\overline{q}}{}^{i,a}_I \overline{\slashed{Z}} \mathbb{P}_{\mathrm{L}} q_I^{i,b}
            -(1-c_\mathrm{QED})\frac{s_W}{c_W} e Q^q_{ab} {\overline{q}}^{i,a}_I \widehat{\slashed{Z}} q_I^{i,b},
        \end{aligned}
    \end{equation}
    \item $W_\mu^\pm$ interaction current
    \begin{equation}\label{Eq:W-Interaction-Current-SM}
        \begin{aligned}
            &- \frac{g_W}{\sqrt{2}} \tau^+_{ab}\, {\overline{l}}{}^{a}_I {\overline{\slashed{W}}}{}^+ \mathbb{P}_{\mathrm{L}} l_I^{b}
            - \frac{g_W}{\sqrt{2}} \tau^-_{ab} \, {\overline{l}}{}^{a}_I {\overline{\slashed{W}}}{}^- \mathbb{P}_{\mathrm{L}} l_I^{b}\\
            &- \frac{g_W}{\sqrt{2}} \tau^+_{ab}\, {\overline{q}}{}^{i,a}_I {\overline{\slashed{W}}}{}^+ \mathbb{P}_{\mathrm{L}} q_I^{i,b}
            - \frac{g_W}{\sqrt{2}} \tau^-_{ab} \, {\overline{q}}{}^{i,a}_I {\overline{\slashed{W}}}{}^- \mathbb{P}_{\mathrm{L}} q_I^{i,b},
        \end{aligned}
    \end{equation}
    \item gluon interaction current
    \begin{equation}\label{Eq:Gluon-Interaction-Current-SM}
        \begin{aligned}
            - g_s t^A_{s,ij} {\overline{q}}{}^{i,a}_I {\slashed{G}}{}^A q_I^{j,a} + (1-c_\mathrm{QCD}) g_s t^A_{s,ij} {\overline{q}}{}^{i,a}_I {\widehat{\slashed{G}}}{}^A q_I^{j,a}.
        \end{aligned}
    \end{equation}
\end{itemize}
Two special cases of the parametrisation in Eq.~\eqref{Eq:Evanescent-Generators-Choice} are of particular interest.
For $c_{\mathrm{QED}}=c_{\mathrm{QCD}}=0$, all gauge interactions are purely 4-dimensional.
For $c_{\mathrm{QED}}=1$ and/or $c_{\mathrm{QCD}}=1$, the photon and/or gluon interaction current, respectively, become fully $D$-dimensional, corresponding to the standard treatment of QED/QCD in naive DReg.
In contrast, the $Z_\mu$ interaction current cannot be extended to a fully $D$-dimensional form due to its purely 4-dimensional component $\propto t_{L,ab}^3 \overline{\gamma}^\mu \projL$, while the $W_\mu^\pm$ interaction currents remain intrinsically 4-dimensional in the chosen parametrisation. 
Indeed, no consistent choice of evanescent generators can render the $Z_\mu$ or $W_\mu^\pm$ currents fully $D$-dimensional, as their axial components cannot be continued uniformly to $D$-dimensions, in agreement with the discussion in Sec.~\ref{Sec:General_Structure_of_the_Symmetry_Breaking}.

\paragraph{Gauge Sector:}
The Lagrangian of the gauge fields takes the standard form
\begin{equation}
\begin{aligned}
    \mathcal{L}_\mathrm{gauge} &= - \frac{1}{4} \vgb_{\mu\nu}^A \vgb^{A,\mu\nu}\\
    &= - \frac{1}{4} \big( \partial_\mu \vgb_\nu^A - \partial_\nu \vgb_\mu^A \big)^2
    + \mathscr{C}^{ABC} \partial^\mu \vgb^{A,\nu} \vgb^{B}_{\mu} \vgb^{C}_{\nu}
    - \frac{1}{4} \mathscr{C}^{ABC} \mathscr{C}^{ADE} \vgb_\mu^B \vgb_\nu^C \vgb^{D,\mu} \vgb^{E,\nu},
\end{aligned}
\end{equation}
with the field strength tensor $\vgb_{\mu\nu}^A=\partial_\mu \vgb_\nu^A - \partial_\nu \vgb_\mu^A - \mathscr{C}^{ABC} \vgb_\mu^B \vgb_\nu^C$, and the structure constants $\mathscr{C}^{ABC}$ as defined in Eq.~\eqref{Eq:Super-Structure-Constants}.

\paragraph{Higgs Sector:}
The Higgs Lagrangian reads
\begin{align}
    \mathcal{L}_\mathrm{Higgs} = (D^\mu \Phi)^\dagger(D_\mu\Phi) - V(\Phi),
\end{align}
with the covariant derivative
\begin{align}
    D_\mu\Phi = \big( \partial_\mu + i \myGenScl^A \vgb_\mu^A \big)\Phi,
\end{align}
acting on the Higgs doublet $\Phi$, and scalar generator $\myGenScl^A$ defined in Eq.~\eqref{Eq:Scalar-Super-Generator}.
The Higgs potential takes the usual form
\begin{align}
    V(\Phi) = \mu^2 \Phi^\dagger \Phi + \lambda (\Phi^\dagger\Phi)^2,
\end{align}
where $\mu^2 <0$ represents the only mass scale of the SM.
The Higgs boson acquires a non-vanishing vacuum expectation value $v$, and in the broken phase the doublet can be parametrised as
\begin{align}\label{Eq:SM-Higgs-Doublet-broken-phasen}
    \Phi =
        \begin{pmatrix}
            G^{+}\\
            \frac{v+H+iG^{0}}{\sqrt{2}}
        \end{pmatrix},
\end{align}
such that 
\begin{align}
    \Phi_\mathrm{min} \equiv \langle 0 | \Phi | 0 \rangle = 
        \begin{pmatrix}
            0\\
            \frac{v}{\sqrt{2}}
        \end{pmatrix},
\end{align}
with $v^2=-\mu^2/\lambda$.
Spontaneous symmetry breaking then generates the Higgs mass $M^2_H=2\lambda v^2$, along with the masses of other SM particles.
In the present work, however, we perform the analysis in the unbroken phase and therefore use the parametrisation of Eq.~\eqref{Eq:SM-Higgs-Doublet}.

\paragraph{Yukawa Sector:}
The Yukawa Lagrangian is given by
\begin{align}
    \mathcal{L}_\mathrm{Yukawa} = - \Big( 
    Y_{l,IJ} {\overline{l}}{}^{a}_I \phi_a \projR l_J^2 + 
    Y_{d,IJ} {\overline{q}}{}^{i,a}_I \phi_a \projR q_J^{i,2} + 
    Y_{u,IJ} {\overline{q}}{}^{i,a}_I \varepsilon_{ab} \phi_b^\dagger \projR q_J^{i,1} + 
    \mathrm{h.c.} \Big),
\end{align}
where $\varepsilon_{ab}$ denotes the totally antisymmetric Levi-Civita symbol in 2 dimensions ($\varepsilon_{12}=-\varepsilon_{21}=1$).
Here we exclusively use the fermion doublets introduced in Eq.~\eqref{Eq:SM-Fermion-Doublets}; the right-handed singlets appearing in the Yukawa sector are accessed through the corresponding components of these doublets.

In general, the Yukawa matrices are complex $3\times3$ matrices $Y'_{l}$, $Y'_{d}$, and $Y'_{u}$.
Using the freedom of field redefinitions, two of them can be diagonalised.
In the unbroken phase, both components of a left-handed fermion doublet must be transformed identically due to $SU(2)_L$ symmetry, whereas the right-handed components may be transformed independently.
As a result, $Y'_{d}$ and $Y'_{u}$ cannot be diagonalised simultaneously.
Without loss of generality, we choose to diagonalise $Y'_{u}$, leading to
\begin{subequations}
    \begin{align}
        Y_l &= U^l_L Y'_l (U^l_R)^\dagger = \mathrm{diag}(y_e,y_\mu,y_\tau),\\
        Y_u &= U^u_L Y'_u (U^u_R)^\dagger = \mathrm{diag}(y_u,y_c,y_t),\\
        Y_d &= U^u_L Y'_d (U^d_R)^\dagger = U^u_L (U^d_L)^\dagger U^d_L Y'_d (U^d_R)^\dagger = V_\mathrm{CKM} Y_d^\mathrm{diag},
    \end{align}
\end{subequations}
where we introduced the CKM-matrix $V_\mathrm{CKM}=U^u_L (U^d_L)^\dagger$ and $Y_d^\mathrm{diag}=\mathrm{diag}(y_d,y_s,y_b)$.

\paragraph{Ghosts and Gauge Fixing:}
As before, we work in the $R_\xi$-gauge, for which the combined ghost and gauge-fixing Lagrangian reads
\begin{equation}\label{Eq:SM_Gauge-Fixing_and_Ghost_Lagrangian}
    \begin{aligned}
        \mathcal{L}_\mathrm{ghost+fix} &= \frac{\xi}{2} \mathcal{B}^A\mathcal{B}^A + \mathcal{B}^A\partial^\mu \vgb_\mu^A - {\overline{c}}{}^A \partial^\mu D_\mu^{AC} c^C\\
        &= - \frac{1}{2\xi} (\partial^\mu\vgb^A_\mu)^2 - {\overline{c}}{}^A \partial^\mu D_\mu^{AC} c^C\\
        &= - \frac{1}{2\xi} (\partial^\mu\vgb^A_\mu)^2 - {\overline{c}}{}^A \Box c^A - \mathscr{C}^{ABC} (\partial^\mu {\overline{c}}{}^A) \vgb^B_\mu c^C,
    \end{aligned}
\end{equation}
where, in the second line, the Nakanishi–Lautrup auxiliary field has been integrated out according to
\begin{align}
    \mathcal{B}^A= - \frac{1}{\xi} \partial^\mu \vgb_\mu^A.
\end{align}
We introduce the modified gauge parameter $\xi_F \coloneqq 1 - \xi$, such that Feynman gauge corresponds to $\xi_F=0$.
This choice is made for computational convenience, as it reduces the number of terms generated by the gauge boson propagator.

\paragraph{BRST Transformations and External Sources:}
The BRST transformations in the SM are given by
\begin{subequations}\label{Eq:SM-BRST-Transformations}
    \begin{align}
        s_D \vgb_\mu^A &= (D_\mu c)^A = \partial_\mu c^A - \mathscr{C}^{ABC} \vgb_\mu^B c^C,\\
        s_D l^a_{R,I} &= - i c^A \myGenLe_{R,ab}^A l^b_{R,I},\\
        s_D l^a_{L,I} &= - i c^A \myGenLe_{L,ab}^A l^b_{L,I},\\
        s_D q^{i,a}_{R,I} &= - i c^A \myGenQu_{R,ab,ij}^A q^{j,b}_{R,I},\\
        s_D q^{i,a}_{L,I} &= - i c^A \myGenQu_{L,ab,ij}^A q^{j,b}_{L,I},\\
        s_D \phi_a &= - i c^A \myGenScl^A_{ab} \phi_b,\\
        s_D c^A &= \frac{1}{2} \mathscr{C}^{ABC} c^B c^C,\\
        s_D \overline{c}^A &= \mathcal{B}^A = - \frac{1}{\xi} \partial^\mu \vgb_\mu^A,\\
        s_D \mathcal{B}^A &= 0,
    \end{align}
\end{subequations}
where the generators are defined in Eqs.~\eqref{Eq:Lepton-Super-Generator}--\eqref{Eq:Scalar-Super-Generator}, and the structure constants are given in Eq.~\eqref{Eq:Super-Structure-Constants}.

In contrast to the Abelian theories discussed in chapters~\ref{Chap:General_Abelian_Chiral_Gauge_Theory} and \ref{Chap:BMHV_at_Multi-Loop_Level}, the BRST transformations in the SM receive nontrivial quantum corrections, i.e.\ they renormalise, as expected in a non-Abelian gauge theory.
As before and as introduced in Sec.~\ref{Sec:BRST-Symmetry}, we couple the BRST transformations in Eq.~\eqref{Eq:SM-BRST-Transformations} to external sources, yielding
\begin{equation}\label{Eq:SM-L_ext-BRST-Trafos-coupled-to-sources}
    \begin{aligned}
        \mathcal{L}_{\mathrm{ext}} &= 
        \rho^{\mu}_A s_D \vgb^A_{\mu} 
        + {\overline{R}}{}^{a}_{l,I} s_D {l}^a_I + (s_D{\overline{l}}{}^a_I) R^{a}_{l,I}
        + {\overline{R}}{}^{i,a}_{q,I} s_D {q}^{i,a}_I + (s_D{\overline{q}}{}^{i,a}_I) R^{i,a}_{q,I}\\
        &+ {\Upsilon^{a}}^{\dagger} s_D \phi_{a} + \Upsilon^{a} s_D \phi_{a}^{\dagger}
        + \zeta_A s_D c^A .        
    \end{aligned}
\end{equation}
Since the BRST transformations renormalise in the non-Abelian case, higher-order corrections to $\mathcal{L}_{\mathrm{ext}}$ are expected.

Note that we do not couple the BRST variation of the antighost field ${\overline{c}}{}^A$ to a source, i.e.\ we omit a term of the form $\chi_A s_D {\overline{c}}{}^A$.
The reason is that this term does not renormalise: the equation of motion of $\chi_A$ is linear in the quantum fields and can be imposed to hold at all orders (see Sec.~\ref{Sec:Peculiarities_of_Abelian_Gauge_Theories} for more details).
Moreover, any related counterterm that might be added by hand can be removed, as discussed in Sec.~\ref{Sec:Algebraic_Renormalisation} and Ref.~\cite{Piguet:1995er}.
For further discussions of the renormalisation of external source terms $\mathcal{L}_{\mathrm{ext}}$ in non-Abelian gauge theories at higher orders, we refer to Ref.~\cite{Kuhler:2025znv}.

\subsection{Group Invariants and Coupling Structures}\label{Eq:SM-GroupInvariants-CouplingStructures}

Here, we collect a number of structures constructed from the generators and coupling matrices of the Standard Model, such as group invariants, which will be used in the presentation of the counterterm results.

We begin with the group invariants associated with the fundamental generators of the SM gauge groups.
In particular, the Dynkin indices are defined as
\begin{equation}\label{Eq:Dynkin-Indices}
    \begin{alignedat}{2}
        S_{2}\big(F_L\big) \delta^{AB} &= \myTrbig{ t_L^A t_L^B } &&= \frac{1}{2} \, \delta^{AB},\\
        S_{2}\big(F_s\big) \delta^{AB} &= \myTrbig{ t_s^A t_s^B } &&= \frac{1}{2} \, \delta^{AB},
    \end{alignedat}
\end{equation}
and the quadratic Casimirs by
\begin{equation}\label{Eq:Quadratic-Casimirs}
    \begin{alignedat}{3}
        C_{2}\big(F_L\big) \delta_{ab} &= t^A_{L,ac} t^A_{L,cb} &&= \frac{3}{4} \, \delta_{ab}, &&\\
        C_{2}\big(F_s\big) \delta_{ij} &= t^A_{s,ik} t^A_{s,kj} &&= \frac{4}{3} \, \delta_{ij}, &&\\
        C_{2}\big(G_L\big) \delta^{AB} &= \varepsilon^{ACD}\varepsilon^{BCD} &&= N_L \, \delta^{AB} &&= 2 \, \delta^{AB},\\
        C_{2}\big(G_s\big) \delta^{AB} &= f^{ACD}f^{BCD} &&= N_c \, \delta^{AB} &&= 3 \, \delta^{AB}.
    \end{alignedat}
\end{equation}

Next, we extend these definitions to the larger generator structures introduced in Eqs.~\eqref{Eq:Lepton-Super-Generator}--\eqref{Eq:Scalar-Super-Generator} and the structure constants in Eq.~\eqref{Eq:Super-Structure-Constants}.
Analogously to the above, we define generalised Dynkin indices.
For the scalar generators, we have
\begin{equation}\label{Eq:Scalar-Super-Dynkin-Indices}
    \begin{alignedat}{3}
        \mathcal{S}_2\big(S^{A'}\big) \delta^{AB} 
        &= \myTrbig{\Theta^A\Theta^B} &&= 
        \begin{cases}
            g_Y^2 \, \mathcal{Y}_S^2 \, N_L 
            , \quad &A,B = 1\\
            g_W^2 \, S_{2}\big(F_L\big) \, \delta^{AB}, \quad &A,B\in\{2,3,4\}\\
            0, \quad &\mathrm{else}.
        \end{cases}
    \end{alignedat}
\end{equation}
For the lepton generators, we define
\begin{equation}\label{Eq:Lepton-Super-Dynkin-Indices}
    \begin{alignedat}{2}
        \mathcal{S}_2\big(F^{l,A'}_R\big) \delta^{AB} 
        &= \myTrbig{\myGenLe_R^A\myGenLe_R^B} &&= 
        \begin{cases}
            g_Y^2 \, \myTrbig{(\mathcal{Y}^l_R)^2} 
            , \quad &A,B = 1\\
            0, \quad &\mathrm{else},
        \end{cases}\\
        \mathcal{S}_2\big(F^{l,A'}_L\big) \delta^{AB} 
        &= \myTrbig{\myGenLe_L^A\myGenLe_L^B} &&= 
        \begin{cases}
            g_Y^2 \, \myTrbig{(\mathcal{Y}^l_L)^2} 
            , \quad &A,B = 1\\
            g_W^2 \, S_{2}\big(F_L\big) \delta^{AB}, \quad &A,B\in\{2,3,4\}\\
            0, \quad &\mathrm{else}.
        \end{cases}
    \end{alignedat}
\end{equation}
Similarly, for the quark generators, we find
\begin{equation}\label{Eq:Quark-Super-Dynkin-Indices}
    \begin{alignedat}{2}
        \mathcal{S}_2\big(F^{q,A'}_R\big) \delta^{AB} 
        &= \myTrbig{\myGenQu_R^A\myGenQu_R^B} &&= 
        \begin{cases}
            g_Y^2 \, N_c \, \myTrbig{(\mathcal{Y}^q_R)^2} 
            , \quad &A,B = 1\\
            g_s^2 \, N_L \, S_{2}\big(F_s\big) \delta^{AB}, \quad &A,B\in\{5,\ldots,12\}\\
            0, \quad &\mathrm{else},
        \end{cases}\\
        \mathcal{S}_2\big(F^{q,A'}_L\big) \delta^{AB} 
        &= \myTrbig{\myGenQu_L^A\myGenQu_L^B} &&= 
        \begin{cases}
            g_Y^2 \, N_c \, \myTrbig{(\mathcal{Y}^q_L)^2} 
            , \quad &A,B = 1\\
            g_W^2 \, N_c \, S_{2}\big(F_L\big) \delta^{AB}, \quad &A,B\in\{2,3,4\}\\
            g_s^2 \, N_L \, S_{2}\big(F_s\big) \delta^{AB}, \quad &A,B\in\{5,\ldots,12\}\\
            0, \quad &\mathrm{else}.
        \end{cases}
    \end{alignedat}
\end{equation}

At the 1-loop level and beyond, one also encounters mixed traces involving both left- and right-handed generators.
For the leptons, we define
\begin{equation}\label{Eq:Lepton-Super-Dynkin-Indices-Mixed}
    \begin{alignedat}{2}
        \mathcal{S}^{AB}_2\big(F^{l}_R,F^{l}_L\big) 
        &= \myTrbig{\myGenLe_R^A\myGenLe_L^B} &&= 
        \begin{cases}
            g_Y^2 \, \myTrbig{\mathcal{Y}^l_R\mathcal{Y}^l_L} 
            , \quad &A,B = 1\\
            g_Y g_W \, \myTrbig{ \mathcal{Y}^l_R t_L^B }, \quad &A=1,B\in\{2,3,4\}\\
            0, \quad &\mathrm{else},
        \end{cases}\\
        \mathcal{S}^{AB}_2\big(F^{l}_L,F^{l}_R\big) 
        &= \myTrbig{\myGenLe_L^A\myGenLe_R^B} &&= \mathcal{S}^{BA}_2\big(F^{l}_R,F^{l}_L\big).
    \end{alignedat}
\end{equation}
Analogously, for the quarks we obtain
\begin{equation}\label{Eq:Quark-Super-Dynkin-Indices-Mixed}
    \begin{alignedat}{2}
        \mathcal{S}^{AB}_2\big(F^{q}_R,F^{q}_L\big) 
        &= \myTrbig{\myGenQu_R^A\myGenQu_L^B} &&= 
        \begin{cases}
            g_Y^2 \, N_c \, \myTrbig{\mathcal{Y}^q_R\mathcal{Y}^q_L} 
            , \quad &A,B = 1\\
            g_Y g_W \, N_c \, \myTrbig{ \mathcal{Y}^q_R t_L^B }, \quad &A=1,B\in\{2,3,4\}\\
            g_s^2 \, N_L \, S_{2}\big(F_s\big) \delta^{AB}, \quad &A,B\in\{5,\ldots,12\}\\
            0, \quad &\mathrm{else},
        \end{cases}\\
        \mathcal{S}^{AB}_2\big(F^{q}_L,F^{q}_R\big) 
        &= \myTrbig{\myGenQu_L^A\myGenQu_R^B} &&= \mathcal{S}^{BA}_2\big(F^{q}_R,F^{q}_L\big).
    \end{alignedat}
\end{equation}

Since the analysis also includes evanescent generators (see Eq.~\eqref{Eq:Evanescent-Generators-Choice}), traces involving these must likewise be considered.
For the leptons, we find
\begin{equation}\label{Eq:Lepton-Super-Dynkin-Indices-Evanescent}
    \begin{alignedat}{2}
        \mathcal{S}_2\big(F^{l,A'}_{R},F^{l,B'}_{LR}\big) \delta^{AB} 
        &= \myTrbig{\myGenLe^A_R\widehat{\myGenLe}^B} &&= 
        \begin{cases}
            g_Y^2 \, c_{\mathrm{QED}} \myTrbig{(\mathcal{Y}^l_R)^2}, \quad &A,B = 1\\
            0, \quad &\mathrm{else},
        \end{cases}\\
        \mathcal{S}_2\big(F^{l,A'}_{LR},F^{l,B'}_{R}\big) \delta^{AB}
        &= \myTrbig{\widehat{\myGenLe}^A\myGenLe^B_R} &&= \mathcal{S}_2\big(F^{l,A'}_{R},F^{l,B'}_{LR}\big) \delta^{AB},\\
        \mathcal{S}^{AB}_2\big(F^{l}_{L},F^{l}_{LR}\big) 
        &= \myTrbig{\myGenLe^A_L\widehat{\myGenLe}^B} &&= 
        \begin{cases}
            g_Y^2 \, c_{\mathrm{QED}} \myTrbig{\mathcal{Y}^l_L\mathcal{Y}^l_R}, \quad &A,B = 1\\
            g_Y g_W \, c_{\mathrm{QED}} \myTrbig{t^A_L\mathcal{Y}^l_R}, \quad &A \in \{2,3,4\}, B = 1\\
            0, \quad &\mathrm{else},
        \end{cases}\\
        \mathcal{S}^{AB}_2\big(F^{l}_{LR},F^{l}_{L}\big)
        &= \myTrbig{\widehat{\myGenLe}^A\myGenLe^B_L} &&= \mathcal{S}^{BA}_2\big(F^{l}_{L},F^{l}_{LR}\big),\\
        \mathcal{S}_2\big(F^{l,A'}_{LR}\big) \delta^{AB}
        &= \myTrbig{\widehat{\myGenLe}^A\widehat{\myGenLe}^B} &&= 
        \begin{cases}
            g_Y^2 \, c_{\mathrm{QED}}^2 \myTrbig{(\mathcal{Y}^l_R)^2}, \quad &A,B = 1\\
            0, \quad &\mathrm{else}.
        \end{cases}
    \end{alignedat}
\end{equation}
For the quarks, we analogously obtain
\begin{equation}\label{Eq:Quarks-Super-Dynkin-Indices-Evanescent}
    \begin{alignedat}{2}
        \mathcal{S}_2\big(F^{q,A'}_{R},F^{q,B'}_{LR}\big) \delta^{AB} 
        &= \myTrbig{\myGenQu^A_R\widehat{\myGenQu}^B} &&= 
        \begin{cases}
            g_Y^2 \, c_{\mathrm{QED}} N_c \, \myTrbig{(\mathcal{Y}^q_R)^2}, \quad &A,B = 1\\
            g_s^2 \, c_{\mathrm{QCD}} N_L \, S_{2}\big(F_s\big) \delta^{AB}, \quad &A,B \in \{5,\ldots,12\}\\
            0, \quad &\mathrm{else},
        \end{cases}\\
        \mathcal{S}_2\big(F^{q,A'}_{LR},F^{q,B'}_{R}\big) \delta^{AB}
        &= \myTrbig{\widehat{\myGenQu}^A\myGenQu^B_R} &&= \mathcal{S}_2\big(F^{q,A'}_{R},F^{q,B'}_{LR}\big) \delta^{AB},\\
        \mathcal{S}^{AB}_2\big(F^{q}_{L},F^{q}_{LR}\big) 
        &= \myTrbig{\myGenQu^A_L\widehat{\myGenQu}^B} &&= 
        \begin{cases}
            g_Y^2 \, c_{\mathrm{QED}} N_c \, \myTrbig{\mathcal{Y}^q_L\mathcal{Y}^q_R}, \quad &A,B = 1\\
            g_Y g_W \, c_{\mathrm{QED}} N_c \, \myTrbig{t^A_L\mathcal{Y}^q_R}, \quad &A \in \{2,3,4\}, B = 1\\
            g_s^2 \, c_{\mathrm{QCD}} N_L \, S_{2}\big(F_s\big) \delta^{AB}, \quad &A,B \in \{5,\ldots,12\}\\
            0, \quad &\mathrm{else},
        \end{cases}\\
        \mathcal{S}^{AB}_2\big(F^{q}_{LR},F^{q}_{L}\big)
        &= \myTrbig{\widehat{\myGenQu}^A\myGenQu^B_L} &&= \mathcal{S}^{BA}_2\big(F^{q}_{L},F^{q}_{LR}\big),\\
        \mathcal{S}_2\big(F^{q,A'}_{LR}\big) \delta^{AB}
        &= \myTrbig{\widehat{\myGenQu}^A\widehat{\myGenQu}^B} &&= 
        \begin{cases}
            g_Y^2 \, c_{\mathrm{QED}}^2 N_c \, \myTrbig{(\mathcal{Y}^q_R)^2}, 
            \quad &A,B = 1\\
            g_s^2 \, c_{\mathrm{QCD}}^2 N_L \, S_{2}\big(F_s\big) \delta^{AB}, 
            \quad &A,B\in\{5,\ldots,12\}\\
            0, \quad &\mathrm{else}.
        \end{cases}
    \end{alignedat}
\end{equation}

Furthermore, for traces involving four generators, we define
\begin{align}
    \mathcal{S}_{4}^{ABCD}\big(X_1,X_2,X_3,X_4\big) &= \myTrbig{X_1^A X_2^B X_3^C X_4^D} + \myTrbig{X_2^B X_1^A X_3^C X_4^D} + \myTrbig{X_1^A X_2^B X_4^D X_3^C}\nonumber\\
    &+ \myTrbig{X_2^B X_1^A X_4^D X_3^C} - \myTrbig{X_1^A X_3^C X_2^B X_4^D} - \myTrbig{X_2^B X_3^C X_1^A X_4^D}\nonumber\\
    &- \myTrbig{X_1^A X_4^D X_2^B X_3^C} - \myTrbig{X_2^B X_4^D X_1^A X_3^C}.
\end{align}
Here, we use $F^{f}_Y$, with $Y\in\{L,R\}$, to indicate the corresponding generator.
For instance,
\begin{equation}
    \begin{aligned}
        \mathcal{S}_{4}^{ABCD}&\big(F^{q}_R+F_L^q,F^{q}_R+F_L^q,F_{LR}^q,F_{LR}^q\big)
        = \myTrbig{\big(\myGenQu_R^A + \myGenQu_L^A\big) \big(\myGenQu_R^B + \myGenQu_L^B\big) \widehat{\myGenQu}^C \widehat{\myGenQu}^D} + \ldots 
    \end{aligned}
\end{equation}

Likewise, we also define generalised quadratic Casimirs.
For the scalar generators,
\begin{equation}\label{Eq:Quadratic-Super-Casimir-Scalar}
    \begin{aligned}
        \mathcal{C}_2\big(S\big) \delta_{ab} &= \myGenScl^A_{ac}\myGenScl^A_{cb} 
        = \Big[ g_Y^2 \mathcal{Y}_S^2 + g_W^2 C_{2}\big(F_L\big) \Big] \delta_{ab}
        = \frac{g_Y^2 + 3 g_W^2}{4} \, \delta_{ab}.
    \end{aligned}
\end{equation}
For purely left- or right-handed generators, the quadratic lepton Casimirs are
\begin{equation}\label{Eq:Quadratic-Super-Casimir-Leptons}
    \begin{alignedat}{2}
        \mathcal{C}^{ab}_2\big(F^l_R\big) &= \myGenLe^A_{R,ac}\myGenLe^A_{R,cb} &&= g_Y^2 \big((\mathcal{Y}^l_R)^2\big)_{ab},\\
        \mathcal{C}_2\big(F^l_L\big) \delta_{ab} &= \myGenLe^A_{L,ac}\myGenLe^A_{L,cb} &&= \Big[ g_Y^2 \mathcal{Y}_l^2 + g_W^2 C_{2}(F_L) \Big] \delta_{ab},
    \end{alignedat}
\end{equation}
and for quarks
\begin{equation}\label{Eq:Quadratic-Super-Casimir-Quarks}
    \begin{alignedat}{2}
        \mathcal{C}^{ab}_2\big(F^q_R\big) \delta_{ij} &= \myGenQu^A_{R,ac,ik}\myGenQu^A_{R,cb,kj} &&= \Big[ g_Y^2 \big((\mathcal{Y}^q_R)^2\big)_{ab} + g_s^2 C_{2}(F_s) \delta_{ab} \Big] \delta_{ij},\\
        \mathcal{C}_2\big(F^q_L\big) \delta_{ab} \, \delta_{ij} &= \myGenQu^A_{L,ac,ik}\myGenQu^A_{L,cb,kj} &&= \Big[ g_Y^2 \mathcal{Y}_q^2 + g_W^2 C_{2}(F_L) + g_s^2 C_{2}(F_s) \Big] \delta_{ab} \, \delta_{ij}.
    \end{alignedat}
\end{equation}
For mixed chiralities, the quadratic lepton and quark Casimirs read
\begin{equation}\label{Eq:Super-C_2-Leptons-Quarks-mixed}
    \begin{alignedat}{3}
        \mathcal{C}^{ab}_2\big(F^l_R,F^l_L\big) &= \myGenLe^A_{R,ac}\myGenLe^A_{L,cb} &&= g_Y^2 \big(\mathcal{Y}^l_R \mathcal{Y}^l_L\big)_{ab} &&= \mathcal{C}^{ab}_2\big(F^l_L,F^l_R\big),\\
        \mathcal{C}^{ab}_2\big(F^q_R,F^q_L\big) \delta_{ij} &= \myGenQu^A_{R,ac,ik}\myGenQu^A_{L,cb,kj} &&= \Big[ g_Y^2 \big(\mathcal{Y}^q_R \mathcal{Y}^q_L\big)_{ab} + g_s^2 C_{2}(F_s) \delta_{ab} \Big] \delta_{ij} &&= \mathcal{C}^{ab}_2\big(F^q_L,F^q_R\big) \delta_{ij}.
    \end{alignedat}
\end{equation}
Including evanescent generators (see Eq.~\eqref{Eq:Evanescent-Generators-Choice}), we obtain for leptons
\begin{equation}\label{Eq:Super-C_2-Leptons-Evanescent}
    \begin{alignedat}{3}
        \mathcal{C}^{ab}_2\big(F^l_R,F^l_{LR}\big) &=
        \myGenLe^A_{R,ac}\widehat{\myGenLe}^A_{cb} &&= 
        c_{\mathrm{QED}} g_Y^2 \big((\mathcal{Y}^l_R)^2\big)_{ab} &&= 
        \mathcal{C}^{ab}_2\big(F^l_{LR},F^l_{R}\big),\\
        \mathcal{C}^{ab}_2\big(F^l_L,F^l_{LR}\big) &=
        \myGenLe^A_{L,ac}\widehat{\myGenLe}^A_{cb} &&= 
        c_{\mathrm{QED}} g_Y^2 \big(\mathcal{Y}^l_L \mathcal{Y}^l_R\big)_{ab} &&= 
        \mathcal{C}^{ab}_2\big(F^l_{LR},F^l_{L}\big),\\
        \mathcal{C}^{ab}_2\big(F^l_{LR}\big) &=
        \widehat{\myGenLe}^A_{ac}\widehat{\myGenLe}^A_{cb} &&= 
        c_{\mathrm{QED}}^2 g_Y^2 \big((\mathcal{Y}^l_R)^2\big)_{ab} &&,
    \end{alignedat}
\end{equation}
and for quarks
\begin{equation}\label{Eq:Super-C_2-Quarks-Evanescent}
    \begin{alignedat}{2}
        \mathcal{C}^{ab}_2\big(F^q_R,F^q_{LR}\big) \delta_{ij} &=
        \mathcal{C}^{ab}_2\big(F^q_{LR},F^q_{R}\big) \delta_{ij} &&=
        \myGenQu^A_{R,ac,ik}\widehat{\myGenQu}^A_{cb,kj}\\
        & &&= \Big[
        c_{\mathrm{QED}} g_Y^2 \big((\mathcal{Y}^q_R)^2\big)_{ab} +
        c_{\mathrm{QCD}} g_s^2 C_{2}(F_s) \delta_{ab} \Big] \delta_{ij},\\
        \mathcal{C}^{ab}_2\big(F^q_L,F^q_{LR}\big) \delta_{ij} &= 
        \mathcal{C}^{ab}_2\big(F^q_{LR},F^q_{L}\big) \delta_{ij} &&=
        \myGenQu^A_{L,ac,ik}\widehat{\myGenQu}^A_{cb,kj}\\
        & &&= \Big[
        c_{\mathrm{QED}} g_Y^2 \big(\mathcal{Y}^q_L \mathcal{Y}^q_R\big)_{ab} +
        c_{\mathrm{QCD}} g_s^2 C_{2}(F_s) \delta_{ab} \Big] \delta_{ij},\\
        \mathcal{C}^{ab}_2\big(F^q_{LR}\big) \delta_{ij} &=
        \widehat{\myGenQu}^A_{ac,ik}\widehat{\myGenQu}^A_{cb,kj} &&= \Big[
        c_{\mathrm{QED}}^2 g_Y^2 \big((\mathcal{Y}^q_R)^2\big)_{ab} +
        c_{\mathrm{QCD}}^2 g_s^2 C_{2}(F_s) \delta_{ab} \Big] \delta_{ij}.
    \end{alignedat}
\end{equation}
The generalised quadratic Casimir in the adjoint representation is
\begin{equation}\label{Eq:Super-C_A}
    \begin{aligned}
        \mathcal{C}_2\big(G^{A'}\big) \delta^{AB} = \mathscr{C}^{ACD}\mathscr{C}^{BCD} =
        \begin{cases}
            g_W^2 \, C_{2}\big(G_L\big) \delta^{AB}, \quad &A,B,\ldots\in\{2,3,4\}\\
            g_s^2 \, C_{2}\big(G_s\big) \delta^{AB}, \quad &A,B,\ldots\in\{5,\ldots,12\}\\
            0, \quad &\mathrm{else}.
        \end{cases}
    \end{aligned}
\end{equation}

For traces of Yukawa matrices, we define
\begin{equation}\label{Eq:Yukawa-Traces}
    \begin{aligned}
        H_2(L) &= \myTrbig{Y_l^{\dagger}Y_l},\\
        H_2(D) &= \myTrbig{Y_d^{\dagger}Y_d},\\
        H_2(U) &= \myTrbig{Y_u^{\dagger}Y_u},
    \end{aligned}
\end{equation}
\begin{equation}\label{Eq:Yukawa-4-Traces}
    \begin{aligned}
        H_4(L) &= \myTrbig{Y_l^{\dagger}Y_l Y_l^{\dagger}Y_l},\\
        H_4(D) &= \myTrbig{Y_d^{\dagger}Y_d Y_d^{\dagger}Y_d},\\
        H_4(U) &= \myTrbig{Y_u^{\dagger}Y_u Y_u^{\dagger}Y_u},
    \end{aligned}
\end{equation}
as well as the useful combinations
\begin{equation}\label{Eq:Yukawa-Trace-Combi}
    \begin{aligned}
        \mathcal{H}_2(Y) &= H_2(L) + N_c \big[ H_2(D) + H_2(U) \big],\\
        \mathcal{H}_4(Y) &= H_4(L) + N_c \big[ H_4(D) + H_4(U) \big],
    \end{aligned}
\end{equation}
and
\begin{equation}\label{Eq:Yukawa-Trace-Combi-2}
    \begin{aligned}
        \mathcal{H}_2(D,U) &= N_c \Big[ \myTrbig{Y_d Y_u} + \myTrbig{Y_d^\dagger Y_u^\dagger} \Big],\\
        \mathcal{H}_2(L,D,\overline{U}) &= \myTrbig{Y_lY_l} + N_c \Big[ \myTrbig{Y_d Y_d} + \myTrbig{Y_u^\dagger Y_u^\dagger} \Big],\\
        \mathcal{H}_4(L,D,\overline{U}) &= \myTrbig{Y_lY_lY_lY_l} + N_c \Big[ \myTrbig{Y_d Y_dY_d Y_d} + \myTrbig{Y_u^\dagger Y_u^\dagger Y_u^\dagger Y_u^\dagger} \Big].
    \end{aligned}
\end{equation}
For matrix products, we introduce
\begin{equation}\label{Eq:Yukawa-Products}
    \begin{alignedat}{2}
        \mathbf{Y}_2(L) &= Y_l^{\dagger}Y_l, 
        \qquad 
        \mathbf{Y}_2(\overline{L}) &&= Y_l Y_l^{\dagger},\\
        \mathbf{Y}_2(D) &= Y_d^{\dagger}Y_d, 
        \qquad 
        \mathbf{Y}_2(\overline{D}) &&= Y_d Y_d^{\dagger},\\
        \mathbf{Y}_2(U) &= Y_u^{\dagger}Y_u, 
        \qquad 
        \mathbf{Y}_2(\overline{U}) &&= Y_u Y_u^{\dagger}.
    \end{alignedat}
\end{equation}


\section{Regularisation-Induced Symmetry Breaking in the SM}\label{Sec:SymmetryBreaking_in_SM}

As in chapters~\ref{Chap:General_Abelian_Chiral_Gauge_Theory} and \ref{Chap:BMHV_at_Multi-Loop_Level}, the consistent treatment of $\gamma_5$ in the BMHV scheme leads to a spurious breaking of gauge and BRST symmetry.
This breaking originates solely from the evanescent part of the fermion kinetic term and the evanescent gauge interactions, as explained in detail in Sec.~\ref{Sec:Regularisation-Induced_Symmetry_Breaking}.
Accordingly, only the fermionic part $\mathcal{L}_\mathrm{fermion}$ (see Eq.~\eqref{Eq:L_fermion-in-the-SM}) contributes to the regularisation-induced symmetry breaking, while the rest of the Lagrangian in Eq.~\eqref{Eq:SM-Lagrangian-Complete} remains manifestly BRST invariant in $D$ dimensions.
At tree-level, the general BRST breaking in the Standard Model takes the form
\begin{align}\label{Eq:SM-General-BRST-Breaking-Tree-Level}
        \widehat{\Delta} &= \mathcal{S}_D(S_0) = \mathcal{S}_D \big( \Dintx \, \mathcal{L}_\mathrm{fermion} \big)
        \nonumber\\
        &= - \Dintx \, c^A \,
        \Bigg\{
        {\overline{l}}{}^a_I 
        \bigg[
        \projR 
        \bigg(
        \myGenLe^A_{R,ab} \overset{\leftarrow}{\widehat{\slashed{\partial}}}
        + \myGenLe^A_{L,ab} \overset{\rightarrow}{\widehat{\slashed{\partial}}}
        - \myGenLe^A_{LR,ab} \Big( \overset{\leftarrow}{\widehat{\slashed{\partial}}} 
                            + \overset{\rightarrow}{\widehat{\slashed{\partial}}} \Big)
        \bigg)
        \projR
        \nonumber\\
        &\phantom{= \Dintx \, c^A \;
        \Bigg\{
        {\overline{l}}{}^a_I 
        \bigg[} +
        \projL 
        \bigg(
        \myGenLe^A_{R,ab} \overset{\rightarrow}{\widehat{\slashed{\partial}}}
        + \myGenLe^A_{L,ab} \overset{\leftarrow}{\widehat{\slashed{\partial}}}
        - \myGenLe^A_{RL,ab} \Big( \overset{\leftarrow}{\widehat{\slashed{\partial}}} 
                            + \overset{\rightarrow}{\widehat{\slashed{\partial}}} \Big)
        \bigg)
        \projL
        \bigg]
        l_I^b
        \nonumber\\
        &\phantom{= \Dintx \, c^A \,\,} + {\overline{q}}{}^{i,a}_I 
        \bigg[
        \projR 
        \bigg(
        \myGenQu^A_{R,ab,ij} \overset{\leftarrow}{\widehat{\slashed{\partial}}}
        + \myGenQu^A_{L,ab,ij} \overset{\rightarrow}{\widehat{\slashed{\partial}}}
        - \myGenQu^A_{LR,ab,ij} \Big( \overset{\leftarrow}{\widehat{\slashed{\partial}}} 
                            + \overset{\rightarrow}{\widehat{\slashed{\partial}}} \Big)
        \bigg)
        \projR
        \nonumber\\
        &\phantom{= \Dintx \, c^A \;
        \Bigg\{
        {\overline{l}}{}^a_I 
        \bigg[} +
        \projL 
        \bigg(
        \myGenQu^A_{R,ab,ij} \overset{\rightarrow}{\widehat{\slashed{\partial}}}
        + \myGenQu^A_{L,ab,ij} \overset{\leftarrow}{\widehat{\slashed{\partial}}}
        - \myGenQu^A_{RL,ab,ij} \Big( \overset{\leftarrow}{\widehat{\slashed{\partial}}} 
                            + \overset{\rightarrow}{\widehat{\slashed{\partial}}} \Big)
        \bigg)
        \projL
        \bigg]
        q_I^{j,b}
        \Bigg\}
        \nonumber\\
        &\phantom{= } - i \Dintx \, c^A \,
        \Bigg\{
        {\overline{l}}{}^a_I
        \bigg[
        \Big(
        \myGenLe^A_{R,ac}\myGenLe^B_{RL,cb} - \myGenLe^B_{RL,ac}\myGenLe^A_{L,cb} - i \mathscr{C}^{ABC} \myGenLe^C_{RL,ab}
        \Big)
        \projL {\widehat{\slashed{\vgb}}}{}^B \projL
        \nonumber\\
        &\phantom{= i \Dintx \, c^A
        \Bigg\{
        {\overline{l}}{}^a_I
        \bigg[} +
        \Big(
        \myGenLe^A_{L,ac}\myGenLe^B_{LR,cb} - \myGenLe^B_{LR,ac}\myGenLe^A_{R,cb} - i \mathscr{C}^{ABC} \myGenLe^C_{LR,ab}
        \Big)
        \projR {\widehat{\slashed{\vgb}}}{}^B \projR
        \bigg]
        l_I^b
        \nonumber\\
        &\phantom{i \Dintx \, c^A \,\,
        \Bigg\{} + {\overline{q}}{}^{i,a}_I 
        \bigg[
        \Big(
        \myGenQu^A_{R,ac,ik}\myGenQu^B_{RL,cb,kj} - \myGenQu^B_{RL,ac,ik}\myGenQu^A_{L,cb,kj} - i \mathscr{C}^{ABC} \myGenQu^C_{RL,ab,ij}
        \Big)
        \projL {\widehat{\slashed{\vgb}}}{}^B \projL
        \nonumber\\
        &\phantom{= i \Dintx \, c^A
        \Bigg\{
        {\overline{l}}{}^a_I
        \bigg[} +
        \Big(
        \myGenQu^A_{L,ac,ik}\myGenQu^B_{LR,cb,kj} - \myGenQu^B_{LR,ac,ik}\myGenQu^A_{R,cb,kj} - i \mathscr{C}^{ABC} \myGenQu^C_{LR,ab,ij}
        \Big)
        \projR {\widehat{\slashed{\vgb}}}{}^B \projR
        \bigg]
        q_I^{j,b}
        \Bigg\}
        \nonumber\\
        &\eqqcolon \widehat{\Delta}_1\big[c,\overline{l},l\big] + \widehat{\Delta}_1\big[c,\overline{q},q\big] + \widehat{\Delta}_2\big[c,\vgb,\overline{l},l\big] + \widehat{\Delta}_2\big[c,\vgb,\overline{q},q\big],
\end{align}
where the evanescent generators have not yet been specified according to Eq.~\eqref{Eq:Evanescent-Generators-Choice}.
Its structure matches that of Eq.~\eqref{Eq:GeneralTreeLevelBreaking_Abelian}, as discussed in Sec.~\ref{Sec:General_Structure_of_the_Symmetry_Breaking}, with the only modification being the additional non-Abelian contributions $\propto i\mathscr{C}^{ABC}$ to the ghost--fermion--gauge boson part of the breaking.
These terms, however, do not introduce new breaking structures but rather add to the existing contribution $\widehat{\Delta}_2$, and vanish identically in the Abelian limit.

Specifying the evanescent generators according to Eq.~\eqref{Eq:Evanescent-Generators-Choice} and expanding the breaking into its $U(1)_Y$, $SU(2)_L$ and $SU(3)_c$ components yields
\begin{align}\label{Eq:SM-Specified-BRST-Breaking-Tree-Level}
    \widehat{\Delta} = \Dintx \Big[ 
    \widehat{\Delta}_Y(x) + \widehat{\Delta}_W(x) + \widehat{\Delta}_S(x)
    + \widehat{\Delta}_{YW}(x) + \widehat{\Delta}_{YS}(x) + \widehat{\Delta}_{WS}(x) \Big],
\end{align}
where the individual contributions are presented below.

From the $U(1)_Y$ sector, with $U(1)_Y$ gauge boson $B_\mu$ and ghost $c^1\equiv c_B$, the contribution reads
\begin{align}\label{Eq:SM-Breaking-Y}
    \widehat{\Delta}_Y(x) &= -g_Y c_B \Bigg\{ 
        {\overline{l}}{}^a_I 
        \bigg[
        \projR 
        \bigg( (1-c_\mathrm{QED}) \mathcal{Y}^l_{R,ab} \overset{\leftarrow}{\widehat{\slashed{\partial}}}
        + \big(\mathcal{Y}^l_{L,ab}-c_\mathrm{QED}\mathcal{Y}^l_{R,ab}\big) \overset{\rightarrow}{\widehat{\slashed{\partial}}} \bigg)
        \projR
        \nonumber\\
        &\phantom{= g_Y c_B \Bigg\{ 
        {\overline{l}}{}^a_I 
        \bigg[} + 
        \projL 
        \bigg( (1-c_\mathrm{QED}) \mathcal{Y}^l_{R,ab} \overset{\rightarrow}{\widehat{\slashed{\partial}}}
        + \big(\mathcal{Y}^l_{L,ab}-c_\mathrm{QED}\mathcal{Y}^l_{R,ab}\big) \overset{\leftarrow}{\widehat{\slashed{\partial}}} \bigg)
        \projL
        \bigg]
        l^b_I
        \Bigg\}
        \nonumber\\
        &\phantom{= \;}
        + i c_\mathrm{QED} g_Y^2 \mathcal{Y}^l_{R,ac} \big(\mathcal{Y}^l_{R}-\mathcal{Y}^l_{L}\big)_{cb} c_B {\overline{l}}{}^a_I \Big[ \projR \widehat{\slashed{B}} \projR - \projL \widehat{\slashed{B}} \projL \Big] l_I^b
        \nonumber\\
        &\phantom{= \;} 
        -g_Y c_B \Bigg\{ 
        {\overline{q}}{}^{i,a}_I 
        \bigg[
        \projR 
        \bigg( (1-c_\mathrm{QED}) \mathcal{Y}^q_{R,ab} \overset{\leftarrow}{\widehat{\slashed{\partial}}}
        + \big(\mathcal{Y}^q_{L,ab}-c_\mathrm{QED}\mathcal{Y}^q_{R,ab}\big) \overset{\rightarrow}{\widehat{\slashed{\partial}}} \bigg)
        \projR
        \nonumber\\
        &\phantom{= g_Y c_B \Bigg\{ 
        {\overline{l}}{}^a_I 
        \bigg[} + 
        \projL 
        \bigg( (1-c_\mathrm{QED}) \mathcal{Y}^q_{R,ab} \overset{\rightarrow}{\widehat{\slashed{\partial}}}
        + \big(\mathcal{Y}^q_{L,ab}-c_\mathrm{QED}\mathcal{Y}^q_{R,ab}\big) \overset{\leftarrow}{\widehat{\slashed{\partial}}} \bigg)
        \projL
        \bigg]
        q^{i,b}_I
        \Bigg\}
        \nonumber\\
        &\phantom{= \;}
        + i c_\mathrm{QED} g_Y^2 \mathcal{Y}^q_{R,ac} \big(\mathcal{Y}^q_{R}-\mathcal{Y}^q_{L}\big)_{cb} c_B {\overline{q}}{}^{i,a}_I \Big[ \projR \widehat{\slashed{B}} \projR - \projL \widehat{\slashed{B}} \projL \Big] q_I^{i,b},
\end{align}
where we note that $\projR \widehat{\slashed{B}} \projR - \projL \widehat{\slashed{B}} \projL=\widehat{\slashed{B}}\gamma_5$.
From the $SU(2)_L$ sector, with weak ghosts $c^A$ ($A\in\{2,3,4\}$), we obtain 
\begin{align}\label{Eq:SM-Breaking-W}
    \widehat{\Delta}_W(x) &= 
    - g_W t_{L,ab}^A c^A \bigg[ 
    {\overline{l}}{}^a_I \Big( \projR \overset{\rightarrow}{\widehat{\slashed{\partial}}} \projR + \projL \overset{\leftarrow}{\widehat{\slashed{\partial}}} \projL \Big) l_I^b
    + {\overline{q}}{}^{i,a}_I \Big( \projR \overset{\rightarrow}{\widehat{\slashed{\partial}}} \projR + \projL \overset{\leftarrow}{\widehat{\slashed{\partial}}} \projL \Big) q_I^{i,b}
    \bigg].
\end{align}
From the $SU(3)_c$ sector, with strong ghosts $c^A$ ($A\in\{5,\ldots,12\}$), one finds
\begin{align}\label{Eq:SM-Breaking-S}
    \widehat{\Delta}_S(x) &= - (1-c_\mathrm{QCD}) g_s t_{s,ij}^A c^A \widehat{\partial}_\mu \big( 
    {\overline{q}}{}^{i,a}_I \widehat{\gamma}^\mu q_I^{j,a} \big).
\end{align}
In addition, we also have mixed contributions.
In particular, we obtain a breaking with $SU(2)_L$ ghosts $c^A$ ($A\in\{2,3,4\}$) and $U(1)_Y$ gauge boson $B_\mu$ of the form
\begin{align}\label{Eq:SM-Breaking-YW}
    \widehat{\Delta}_{YW}(x) = &- i c_\mathrm{QED} g_Y g_W \Big[
    \big(t_L^A \mathcal{Y}_R^l\big)_{ab} c^A {\overline{l}}{}^a_I \projR \widehat{\slashed{B}} \projR l_I^b
    - \big(\mathcal{Y}_R^l t_L^A\big)_{ab} c^A {\overline{l}}{}^a_I \projL \widehat{\slashed{B}} \projL l_I^b
    \Big]\nonumber\\
    &- i c_\mathrm{QED} g_Y g_W \Big[
    \big(t_L^A \mathcal{Y}_R^q\big)_{ab} c^A {\overline{q}}{}^{i,a}_I \projR \widehat{\slashed{B}} \projR q_I^{i,b}
    - \big(\mathcal{Y}_R^q t_L^A\big)_{ab} c^A {\overline{q}}{}^{i,a}_I \projL \widehat{\slashed{B}} \projL q_I^{i,b}
    \Big].
\end{align}
Similarly, for $U(1)_Y$ ghost $c^1\equiv c_B$ and $SU(3)_c$ gauge bosons $G_\mu^A$ ($A\in\{5,\ldots,12\}$), we obtain
\begin{align}\label{Eq:SM-Breaking-YS}
    \widehat{\Delta}_{YS}(x) &= i c_\mathrm{QCD} g_Y g_s \big( \mathcal{Y}_R^q - \mathcal{Y}_L^q \big)_{ab} t_{s,ij}^A c_B {\overline{q}}{}^{i,a}_I \Big[ \projR {\widehat{\slashed{G}}}{}^A \projR - \projL {\widehat{\slashed{G}}}{}^A \projL \Big] q_I^{j,b}.
\end{align}
Finally, for $SU(2)_L$ ghosts $c^A$ ($A\in\{2,3,4\}$) and $SU(3)_c$ gauge bosons $G_\mu^B$ ($B\in\{5,\ldots,12\}$), the contribution reads
\begin{align}\label{Eq:SM-Breaking-WS}
    \widehat{\Delta}_{WS}(x) &= - i c_\mathrm{QCD} g_W g_s t_{L,ab}^A t^B_{s,ij} c^A {\overline{q}}{}^{i,a}_I \Big[ \projR {\widehat{\slashed{G}}}{}^A \projR - \projL {\widehat{\slashed{G}}}{}^A \projL \Big] q_I^{j,b}.
\end{align}
Note again $\projR {\widehat{\slashed{G}}}{}^A \projR - \projL {\widehat{\slashed{G}}}{}^A \projL = {\widehat{\slashed{G}}}{}^A \gamma_5$.

Setting the evanescent gauge interactions to zero ($c_\mathrm{QED}=c_\mathrm{QCD}=0$) eliminates all mixed contributions (see Eqs.~\eqref{Eq:SM-Breaking-YW}--\eqref{Eq:SM-Breaking-WS}).
Particularly, in this limit, all tree-level breakings involving gauge bosons vanish, leaving only the contributions from $\widehat{\Delta}_1$ (cf.\ Eq.~\eqref{Eq:SM-General-BRST-Breaking-Tree-Level}).
Thus, omitting evanescent gauge interactions considerably simplifies the tree-level breaking.
However, in this case, a breaking term $\propto g_s$ from the $SU(3)_c$ sector remains (see Eq.~\eqref{Eq:SM-Breaking-S}); it vanishes only when the strong interaction current is treated fully $D$-dimensional, corresponding to $c_\mathrm{QCD}=1$.

Importantly, there is no choice of $c_\mathrm{QED}$ and $c_\mathrm{QCD}$ that eliminates all strong contributions to the breaking $\propto g_s$.
For $c_\mathrm{QCD}=1$ (fully $D$-dimensional strong interaction current), Eq.~\eqref{Eq:SM-Breaking-S} vanishes, while the mixed breakings in Eqs.~\eqref{Eq:SM-Breaking-YS}--\eqref{Eq:SM-Breaking-WS} persist.
Conversely, for $c_\mathrm{QCD}=0$ (purely $4$-dimensional strong interaction current), the mixed contributions vanish, whereas Eq.~\eqref{Eq:SM-Breaking-S} remains.
Hence, a symmetry breaking $\propto g_s$ always occurs already at tree level, even though QCD is a vector-like gauge theory.
Consequently, the QCD sector cannot be omitted in the symmetry-restoration procedure: the full Standard Model must be considered.

\section{Renormalisation of the SM at the 1-Loop Level}\label{Sec:SM_1-Loop_Renormalisation}

In this section, we present the complete 1-loop renormalisation of the full Standard Model within the BMHV scheme.
As emphasised at the beginning of this chapter, this work represents an ongoing research effort rather than a finalised project.
Accordingly, the focus here is on the explicit results for the counterterms, while a detailed analysis and discussion are deferred to forthcoming publications, which are currently in preparation.

As before, the renormalisation procedure follows the general strategy described in Sec.~\ref{Sec:Symmetry_Restoration_Procedure}.
We employ Option~\ref{Opt:Option1} for the treatment of fermions in $D$ dimensions and include evanescent gauge interactions with generators parametrised according to Eq.~\eqref{Eq:Evanescent-Generators-Choice}.
As a consequence, evanescent terms mix left- and right-handed fermions carrying different gauge quantum numbers, which leads to a violation of global hypercharge conservation.
This issue, discussed in Sec.~\ref{Sec:Dimensional_Ambiguities_and_Evanescent_Shadows} and further analysed in chapter~\ref{Chap:General_Abelian_Chiral_Gauge_Theory}, requires the computation of additional Green functions such as $\langle \phi_b \phi_a \rangle^{\mathrm{1PI}}$, $\langle \Delta \phi_b \phi_a c^A \rangle^{\mathrm{1PI}}$ and related ones to obtain the full set of symmetry-restoring counterterms.

All computations were carried out using the \texttt{FORM}-based computational framework introduced in Sec.~\ref{Sec:Computational_Setup} and the methods described in chapter~\ref{Chap:Multi-Loop_Calculations}.
As before, the UV-divergent parts of the Feynman diagrams were extracted using the tadpole decomposition method (see Sec.~\ref{Sec:Tadpole_Decomposition}).
Calculations were performed in general $R_\xi$-gauge, cf.\ Eq.~\eqref{Eq:SM_Gauge-Fixing_and_Ghost_Lagrangian}.
Throughout, we used the modified gauge parameter $\xi_F=1-\xi$ (see Sec.~\ref{Sec:SM-Lagrangian}) and expressed all results in terms of $\xi_F$ rather than $\xi$.
Dimensional regularisation was parametrised with $D=4-2\epsilon$, and all external momenta were defined as incoming.
Moreover, the anomaly cancellation condition is explicitly satisfied in the Standard Model, as discussed in Sec.~\ref{Sec:Fields_Generators_Notation}.

Unlike in chapters~\ref{Chap:General_Abelian_Chiral_Gauge_Theory} and \ref{Chap:BMHV_at_Multi-Loop_Level}, we did not compute the UV divergent BRST-breaking terms from both ordinary and operator-inserted Green functions.
Here, we determined them solely from ordinary Green functions, and rely on the internal consistency of our computational framework, which has been thoroughly tested and successfully validated throughout all previous calculations.
Explicit cross-checks using operator-inserted Green functions will be conducted in future work.

The complete 1-loop counterterm action of the Standard Model in the BMHV scheme reads
\begin{align}
    S_\mathrm{ct}^{(1)} = S^{(1)}_\mathrm{sct} + S^{(1)}_\mathrm{fct}.
\end{align}
The singular counterterms $S^{(1)}_\mathrm{sct}$ are presented in Sec.~\ref{Sec:SM_1-Loop_Renormalisation_Singular_CT}, while the finite symmetry-restoring counterterms $S^{(1)}_\mathrm{fct}$ are discussed in Sec.~\ref{Sec:SM_1-Loop_Renormalisation_Finite_CT}.
All counterterms are expressed in terms of coefficients defined in App.~\ref{App:Results-TheStandardModel}: 
divergent counterterm coefficients are collected in App.~\ref{App:SM-Divergent1LoopCTCoeffs}, and the coefficients of the finite symmetry-restoring counterterms in App.~\ref{App:SM-Finite1LoopCTCoeffs}.
This compact representation highlights the structure of the counterterms while keeping the presentation concise.

\subsection{Singular Counterterm Action at the 1-Loop Level}\label{Sec:SM_1-Loop_Renormalisation_Singular_CT}

Here we present the complete 1-loop singular counterterm action, which renders the Standard Model defined by the Lagrangian in Eq.~\eqref{Eq:SM-Lagrangian-Complete} UV finite at the 1-loop level.
These counterterms were obtained from the divergent parts of all relevant power-counting divergent 1PI Green functions.
In total, we identified and computed 36 such Green functions (counting Green functions involving leptons and quarks separately) that yield non-vanishing contributions.

In contrast to chapters~\ref{Chap:General_Abelian_Chiral_Gauge_Theory} and \ref{Chap:BMHV_at_Multi-Loop_Level}, we do not decompose the singular counterterm action into an invariant and a non-invariant part. 
Instead, we organise it according to the sectors of the SM Lagrangian, writing
\begin{align}
    S_\mathrm{sct}^{(1)} = S^{(1)}_{\mathrm{sct,fermion}} + S^{(1)}_{\mathrm{sct,gauge}} + S^{(1)}_{\mathrm{sct,Higgs}} + S^{(1)}_{\mathrm{sct,Yukawa}} + S^{(1)}_{\mathrm{sct,ghost}} + S^{(1)}_{\mathrm{sct,ext}}.
\end{align}
The explicit results for each sector are presented in the following paragraphs.

\paragraph{Fermionic Sector:}
We begin with the singular counterterms of the fermionic sector, which take the form
\begin{equation}\label{Eq:SM-S_sct-fermion}
    \begin{aligned}
        S^{(1)}_{\mathrm{sct,fermion}} &= \frac{1}{16\pi^2} \Dintx \bigg\{
        \delta {\overline{Z}}{}^{l,(1)}_{R,ac,IJ} \, {\overline{l}}{}_I^{a} i {\overline{\slashed{D}}}{}^{l}_{R,cb} \, l_J^b
        + \delta {\overline{Z}}{}^{l,(1)}_{L,IJ} \, {\overline{l}}{}_I^{a} i {\overline{\slashed{D}}}{}^{l}_{L,ab} \, l_J^b
        \\
        &
        + \delta {\overline{Z}}{}^{q,(1)}_{R,ac,IJ} \, {\overline{q}}{}_I^{i,a} i {\overline{\slashed{D}}}{}^{q}_{R,cb,ij} \, q_J^{j,b}
        + \delta {\overline{Z}}{}^{q,(1)}_{L,IJ} \, {\overline{q}}{}_I^{i,a} i {\overline{\slashed{D}}}{}^{q}_{L,ab,ij} \, q_J^{j,b}\\
        &
        - \delta {\overline{X}}{}^{(1)}_{\overline{f}\vgb f} 
        \Big[ 
        \myGenLe^{A}_{L,ab} {\overline{l}}{}_I^{a} {\overline{\slashed{\vgb}}}{}^{A} \projL l_I^b
        + \myGenQu^{A}_{R,ab,ij} {\overline{q}}{}_I^{i,a} {\overline{\slashed{\vgb}}}{}^{A} \projR q_I^{j,b}
        + \myGenQu^{A}_{L,ab,ij} {\overline{q}}{}_I^{i,a} {\overline{\slashed{\vgb}}}{}^{A} \projL q_I^{j,b}
        \Big]
        \\
        &
        + \delta {\widehat{X}}{}^{l,(1)}_{ac} \, {\overline{l}}{}_I^{a} i {\widehat{\slashed{D}}}{}^{l}_{cb} l_I^b
        + \Big[ \delta {\widehat{X}}{}^{q,(1)}_{ac,IJ} \, {\overline{q}}{}_I^{i,a} i {\widehat{\slashed{D}}}{}^{q}_{cb,ij} \projR q_J^{j,b} 
        - \delta {\widehat{X}}{}^{{\overline{q}}{}\vgb q,(1),A}_{ab,ij,IJ} \, {\overline{q}}{}_I^{i,a} {\widehat{\slashed{\vgb}}}{}^{A} \projR q_J^{j,b}
        \mathrlap{ + \mathrm{h.c.} \Big]
        \bigg\},} \;\;
    \end{aligned}
\end{equation}
with covariant derivatives
\begin{equation}
\begin{alignedat}{4}
        &{\overline{D}}{}^{l,\mu}_{R,ab} &&= 
        \big( \overline{\partial}^{\mu} \delta_{ab} + i \myGenLe_{R,ab}^A \overline{\vgb}^{A,\mu} \big) \projR,
        \qquad 
        &&{\overline{D}}{}^{q,\mu}_{R,ab,ij} &&= 
        \big( \overline{\partial}^{\mu} \delta_{ab} \delta_{ij} + i \myGenQu_{R,ab,ij}^A \overline{\vgb}^{A,\mu} \big) \projR,\\
        &{\overline{D}}{}^{l,\mu}_{L,ab} &&= 
        \big( \overline{\partial}^{\mu} \delta_{ab} + i \myGenLe_{L,ab}^A \overline{\vgb}^{A,\mu} \big) \projL,
        \qquad 
        &&{\overline{D}}{}^{q,\mu}_{L,ab,ij} &&= 
        \big( \overline{\partial}^{\mu} \delta_{ab} \delta_{ij} + i \myGenQu_{L,ab,ij}^A \overline{\vgb}^{A,\mu} \big) \projL,\\
        &{\widehat{D}}{}^{l,\mu}_{ab} &&= 
        \big( \widehat{\partial}^{\mu} \delta_{ab} + i \widehat{\myGenLe}_{ab}^A \widehat{\vgb}^{A,\mu} \big), 
        \qquad
        &&{\widehat{D}}{}^{q,\mu}_{ab,ij} &&= 
        \big( \widehat{\partial}^{\mu} \delta_{ab} \delta_{ij} + i \widehat{\myGenQu}_{ab,ij}^A \widehat{\vgb}^{A,\mu} \big),
\end{alignedat}
\end{equation}
and counterterm coefficients listed in App.~\ref{App:SM-Divergent1LoopCTCoeffs}.

The first two lines of Eq.~\eqref{Eq:SM-S_sct-fermion} correspond to the 4-dimensional, BRST-invariant divergent counterterms for leptons and quarks.
Their coefficients are independent of the evanescent interaction parameters $c_\mathrm{QED}$ and $c_\mathrm{QCD}$.
The third line contains additional 4-dimensional but non-invariant contributions, while the last line displays the evanescent BRST-breaking divergent counterterms of the fermionic sector.

As expected at the 1-loop level, only the evanescent non-invariant counterterms in the last line depend on the evanescent generators, as seen explicitly in Eqs.~\eqref{Eq:SM-Lepton-SE-Evanescent-CTs}, \eqref{Eq:SM-Quark-SE-Evanescent-CTs} and \eqref{Eq:SM-Evanescent-Quark-Gauge-Interaction-CTs}.
The counterterm $\delta {\widehat{X}}{}^{l,(1)}_{ab}$ depends only on $c_\mathrm{QED}$ and simplifies for both $c_\mathrm{QED}=0$ and $c_\mathrm{QED}=1$, with a slightly more compact expression in the case $c_\mathrm{QED}=0$.
The situation is similar for $\delta {\widehat{X}}{}^{q,(1)}_{ab,IJ}$, which depends on both $c_\mathrm{QED}$ and $c_\mathrm{QCD}$: it simplifies for any choice $c_\mathrm{QED},c_\mathrm{QCD}\in\{0,1\}$, with the simplest results obtained for ($c_\mathrm{QED}=0$, $c_\mathrm{QCD}=0$) and ($c_\mathrm{QED}=0$, $c_\mathrm{QCD}=1$).

For the counterterm $\delta {\widehat{X}}{}^{{\overline{q}}{}\vgb q,(1),A}_{ab,ij,IJ}$, a simplification occurs only for $c_\mathrm{QED}=1/2$ --- which is, however, not a motivated choice (see Sec.~\ref{Sec:SM-Lagrangian}).
For the physically motivated values $c_\mathrm{QED}\in\{0,1\}$, the $U(1)_Y$ part changes only by an overall sign, while all other parts remain unchanged.
Its dependence on $c_\mathrm{QCD}$ enters through $\kappa_\mathrm{QCD}$, defined in Eq.~\eqref{Eq:SM-Kappa_QCD}, which yields structurally the same expression --- differing only in the numerical prefactor of $\xi_F$ --- for both $c_\mathrm{QCD}=0$ and $c_\mathrm{QCD}=1$.

\paragraph{Gauge Sector:}
Next, we turn to the gauge sector.
The corresponding divergent counterterms are given by
\begin{align}\label{Eq:SM-S_sct_gauge}
        S^{(1)}_{\mathrm{sct,gauge}} &= \frac{1}{16\pi^2} \Dintx \bigg\{
        - \delta Z_{\vgb}^{(1)} \frac{1}{4} \big(\partial_{\mu}\vgb^A_{\nu}-\partial_{\nu}\vgb^A_{\mu}\big)^2
        + \delta Z_{3\vgb}^{(1)} \mathscr{C}^{ABC} \partial_{\mu} \vgb_{\nu}^A \vgb^{B,\mu} \vgb^{C,\nu}
        \nonumber\\
        &\qquad
        - \delta Z_{4\vgb}^{(1)} \frac{1}{4} \mathscr{C}^{ABC} \mathscr{C}^{ADE} \vgb_{\mu}^B \vgb_{\nu}^C \vgb^{D,\mu} \vgb^{E,\nu}
        - \delta \overline{Z}_{\vgb}^{(1)} \frac{1}{4} \overline{\vgb}^A_{\mu\nu} \overline{\vgb}^{A,\mu\nu}
        \nonumber\\
        &\qquad
        - \delta \widehat{Z}_{\vgb}^{(1)} \frac{1}{4} 
        \Big[ 
        \big(\widehat{\partial}_{\mu}\widehat{\vgb}^A_{\nu}-\widehat{\partial}_{\nu}\widehat{\vgb}^A_{\mu}\big)^2
        - 4 \, c_{\mathrm{QCD}} \mathscr{C}^{ABC} \widehat{\partial}_{\mu} \widehat{\vgb}_{\nu}^A \widehat{\vgb}^{B,\mu} \widehat{\vgb}^{C,\nu}
        \nonumber\\
        &\qquad\qquad\qquad + c_{\mathrm{QCD}}^2 \mathscr{C}^{ABC} \mathscr{C}^{ADE} \widehat{\vgb}_{\mu}^B \widehat{\vgb}_{\nu}^C \widehat{\vgb}^{D,\mu} \widehat{\vgb}^{E,\nu}
        \Big]
        \\
        &\qquad
        + \delta \widehat{Z}_{\vgb}^{(1)} \frac{1}{2} \widehat{\vgb}^A_{\mu} \overline{\partial}^2 \widehat{\vgb}^{A,\mu}
        + \delta \widehat{X}_{\vgb,12}^{(1),AB} \frac{1}{2} \overline{\vgb}^A_{\mu} \widehat{\partial}^2 \overline{\vgb}^{B,\mu}
        - \delta \widehat{X}_{\vgb,34}^{(1),AB} \overline{\vgb}^A_{\mu} \overline{\partial}^{\mu} \widehat{\partial}^{\nu} \widehat{\vgb}^{B}_{\nu}
        \nonumber\\
        &\qquad
        + \delta \widehat{Z}_{\vgb}^{(1)} \mathscr{C}^{ABC} \overline{\partial}_{\mu} \widehat{\vgb}_{\nu}^A \overline{\vgb}^{B,\mu} \widehat{\vgb}^{C,\nu}
        + \delta \widehat{X}^{(1),ABC}_{3\vgb} \widehat{\partial}_{\mu} \overline{\vgb}_{\nu}^A \widehat{\vgb}^{C,\mu} \overline{\vgb}^{B,\nu}
        \nonumber\\
        &\qquad
        - \delta \widehat{X}^{(1),ABCD}_{4\vgb} \frac{1}{4} \overline{\vgb}_{\mu}^A \overline{\vgb}^{B,\mu} \widehat{\vgb}^{C}_{\nu} \widehat{\vgb}^{D,\nu}
        \bigg\},
        \nonumber
\end{align}
with coefficients collected in App.~\ref{App:SM-Divergent1LoopCTCoeffs}.

The first two lines of Eq.~\eqref{Eq:SM-S_sct_gauge} contain the BRST-invariant counterterms, while the remaining lines comprise non-invariant evanescent contributions.
The second term in the second line, governed by $\delta \overline{Z}_{\vgb}^{(1)}$, is clearly BRST invariant, since it is given by a trace over two field strength tensors ${\overline{\vgb}}{}_{\mu\nu}^A$.
It is a purely 4-dimensional expression and contains only fermionic contributions.
This counterterm is fully analogous to the invariant gauge-sector counterterms encountered in chapters~\ref{Chap:General_Abelian_Chiral_Gauge_Theory} and \ref{Chap:BMHV_at_Multi-Loop_Level}.

For the first three counterterms proportional to $\delta Z_{\vgb}^{(1)}$, $\delta Z_{3\vgb}^{(1)}$ and $\delta Z_{4\vgb}^{(1)}$, the situation is less obvious.
These terms are fully $D$-dimensional and contain scalar contributions governed by the Dynkin index $\mathcal{S}_2\big(S^{A'}\big)$ (see Eq.~\eqref{Eq:Scalar-Super-Dynkin-Indices}), as well as gauge boson contributions governed by the adjoint quadratic Casimir $\mathcal{C}_2\big(G^{A'}\big)$ (see Eq.~\eqref{Eq:Super-C_A}).
While the scalar contributions of all three counterterms can again be rewritten as a trace of two field strength tensors, the gauge boson contributions differ in each case.
These gauge boson contributions are purely non-Abelian and originate from gauge boson self-interactions absent in the Abelian theories considered previously.
Nevertheless, the full combination of these counterterms is BRST invariant once one accounts for the renormalisation of the BRST transformations in non-Abelian theories, as reflected in the 1-loop contributions of the external source terms $S^{(1)}_{\mathrm{sct,ext}}$ (see Eq.~\eqref{Eq:SM-S_sct_ext}).

The counterterms in the third and fourth lines of Eq.~\eqref{Eq:SM-S_sct_gauge} are purely evanescent and proportional to $\delta \widehat{Z}_{\vgb}^{(1)}$.
For the choice $c_\mathrm{QED}=0$ and $c_\mathrm{QCD}=1$, they can be rewritten entirely in terms of two field strength tensors and thus become BRST invariant.
In this case, combining it with the $\delta \overline{Z}_{\vgb}^{(1)}$ counterterm, the corresponding fermionic QCD contribution becomes fully $D$-dimensional.\footnote{Choosing in addition $c_\mathrm{QED}=1$ renders one half of the right-handed $U(1)_Y$ contribution fully $D$-dimensional, while the other half stays evanescent; the left-handed contribution remains purely 4-dimensional.}
This may appear appealing at first sight.

However, for the alternative choice $c_\mathrm{QED}=0$ and $c_\mathrm{QCD}=0$, all evanescent gauge-sector counterterms, except for the one governed by $\delta \widehat{X}_{\vgb,12}^{(1),AB}$, vanish identically, since each of them involves at least one evanescent generator.
What remains are only the BRST-invariant counterterms of the first two lines and the single non-invariant evanescent term proportional to $\delta \widehat{X}_{\vgb,12}^{(1),AB}$.
This leads to a significant simplification and thus motivates the choice ($c_\mathrm{QED}=0$, $c_\mathrm{QCD}=0$) for practical applications.

Importantly, the counterterm $\propto \delta \widehat{X}_{\vgb,12}^{(1),AB}$ is completely independent of ($c_\mathrm{QED}$, $c_\mathrm{QCD}$).
It is a purely fermionic contribution, containing left-handed, right-handed and mixed components (see Eq.~\eqref{Eq:SM-GaugeBoson-Evanescent-Breaking-CT-indep-of-evan-generators}), and corresponds to the familiar BRST-breaking evanescent counterterm $\overline{\vgb}^A_{\mu} \widehat{\partial}^2 \overline{\vgb}^{B,\mu}$ already encountered in the results of chapters~\ref{Chap:General_Abelian_Chiral_Gauge_Theory} and \ref{Chap:BMHV_at_Multi-Loop_Level}.

\paragraph{Higgs Sector:}
The divergent counterterms of the Higgs sector take the form
\begin{align}\label{Eq:SM-S_sct_Higgs}
        S^{(1)}_{\mathrm{sct,Higgs}} &= \frac{1}{16\pi^2} \Dintx \bigg\{
        \delta Z_{\phi}^{(1)} \partial^{\mu} \phi_a^{\dagger} \partial_{\mu} \phi_a
        - \delta Z_{\mu}^{(1)} \mu^2 \phi_a^{\dagger} \phi_a
        + \delta \overline{Z}_{\phi}^{(1)} \overline{\partial}^{\mu} \phi_a^{\dagger} \overline{\partial}_{\mu} \phi_a
        \nonumber\\
        &+ \delta \widehat{X}_{\phi}^{(1)} \widehat{\partial}^{\mu} \phi_1^{\dagger} \widehat{\partial}_{\mu} \phi_1
        - \frac{1}{2} \Big[ \delta \widehat{X}_{\phi\phi}^{(1)} \, \phi_2 \widehat{\Box} \phi_2 + \mathrm{h.c.} \Big]
        \nonumber\\
        &- i \delta Z^{(1)}_{\phi^{\dagger}\phi\vgb} \myGenScl^{A}_{ab} \vgb_{\mu}^A \big[ \phi_a^{\dagger} (\partial^{\mu}\phi_b) - (\partial^{\mu}\phi^{\dagger}_a) \phi_b \big]
        - i \delta \overline{Z}^{(1)}_{\phi^{\dagger}\phi\vgb} \myGenScl^{A}_{ab} \overline{\vgb}_{\mu}^A \big[ \phi_a^{\dagger} (\overline{\partial}^{\mu}\phi_b) - (\overline{\partial}^{\mu}\phi^{\dagger}_a) \phi_b \big]
        \nonumber\\
        &- i \delta \widehat{X}^{(1),A}_{\phi^{\dagger}\phi\vgb} \widehat{\vgb}_{\mu}^A \big[ \phi_1^{\dagger} (\widehat{\partial}^{\mu}\phi_1) - (\widehat{\partial}^{\mu}\phi^{\dagger}_1) \phi_1 \big]
        \nonumber\\
        &+ \delta Z^{(1)}_{\phi^{\dagger}\phi\vgb\vgb} \myGenScl^{A}_{ac} \myGenScl^{B}_{cb} \vgb_{\mu}^A \vgb^{B,\mu} \phi_a^{\dagger} \phi_b
        + \delta \overline{Z}^{(1)}_{\phi^{\dagger}\phi\vgb\vgb} \myGenScl^{A}_{ac} \myGenScl^{B}_{cb} \overline{\vgb}_{\mu}^A \overline{\vgb}^{B,\mu} \phi_a^{\dagger} \phi_b
        \nonumber\\
        &+ \delta \widehat{X}^{(1),AB}_{\phi^{\dagger}\phi\vgb\vgb} \frac{1}{2} \widehat{\vgb}_{\mu}^A \widehat{\vgb}^{B,\mu} \phi_1^{\dagger} \phi_1
        - \delta Z^{(1)}_{\lambda} \phi_a^{\dagger} \phi_a \phi_b^{\dagger} \phi_b
        \bigg\},
\end{align}
with coefficients provided in App.~\ref{App:SM-Divergent1LoopCTCoeffs}.

Evidently, there is only one counterterm that violates global hypercharge conservation: the purely evanescent counterterm of the form $\phi_2 \widehat{\Box} \phi_2 + \mathrm{h.c.}$, appearing in the second line of Eq.~\eqref{Eq:SM-S_sct_Higgs}.
The corresponding coefficient $\delta \widehat{X}_{\phi\phi}^{(1)}$ is entirely independent of the evanescent interaction parameters $c_\mathrm{QED}$ and $c_\mathrm{QCD}$. 
Thus, this contribution originates solely from the chirality-violating part of the fermion kinetic term (cf.\ Eq.~\eqref{Eq:L_fermion-evan_2}).
Consequently, this term would vanish only in the sterile-partner construction of Option~\ref{Opt:Option2}.
Since here we work with Option~\ref{Opt:Option1}, this counterterm is unavoidable.
In fact, the situation is analogous to that in chapter~\ref{Chap:General_Abelian_Chiral_Gauge_Theory}; see particularly Sec.~\ref{Sec:AnalysisOfASSMResults}, where this same term was found to be the only divergent global hypercharge violating counterterm in the case of Option~\ref{Opt:Option1}. 
Importantly, all counterterms in Eq.~\eqref{Eq:SM-S_sct_Higgs} preserve electric and colour charge, as expected.

The only 1-loop counterterms in the Higgs sector that depend on the evanescent generators are those governed by $\delta \widehat{X}^{(1),A}_{\phi^{\dagger}\phi\vgb}$ (fourth line) and $\delta \widehat{X}^{(1),AB}_{\phi^{\dagger}\phi\vgb\vgb}$ (first term in the last line).
Both are purely evanescent, contribute only in the Abelian $U(1)_Y$ sector, and depend on $c_\mathrm{QED}$.
In particular, they vanish identically for $c_\mathrm{QED}=0$.
This further reinforces that choosing $c_\mathrm{QED}=0$ is technically advantageous and therefore the preferred option.

\paragraph{Yukawa Sector:}
We continue with the divergent 1-loop counterterms of the Yukawa sector, which are given by
\begin{equation}\label{Eq:SM-S_sct_Yukawa}
    \begin{aligned}
        S^{(1)}_{\mathrm{sct,Yukawa}} &= - \frac{1}{16\pi^2} \Dintx \Big\{ 
        \delta Y_{l,IJ}^{(1)} \overline{l}^{a}_{I} \phi_{a} \projR l_{J}^{2}
        + \delta Y_{d,IJ}^{(1)} \overline{q}^{i,a}_{I} \phi_{a} \projR q_{J}^{i,2}\\
        &\phantom{= \frac{1}{16\pi^2} \Dintx \Big\{}
        + \delta Y_{u,IJ}^{(1)} \varepsilon_{ab} \overline{q}^{i,a}_{I} \phi_{b}^{\dagger} \projR q_{J}^{i,1}
        + \mathrm{h.c.}
        \Big\},
    \end{aligned}
\end{equation}
with coefficients provided in App.~\ref{App:SM-Divergent1LoopCTCoeffs}.

As already observed in Sec.~\ref{Sec:AnalysisOfASSMResults}, the divergent 1-loop Yukawa counterterms are BRST invariant and completely independent of the evanescent generators.
In particular, they do not depend on $c_\mathrm{QED}$ or $c_\mathrm{QCD}$.

\paragraph{Ghost Sector:}
The divergent 1-loop counterterms of the ghost sector read
\begin{equation}\label{Eq:SM-S_sct_ghost}
    \begin{aligned}
        S^{(1)}_{\mathrm{sct,ghost}} &= \frac{1}{16\pi^2} \Dintx \Big\{
        \delta Z_{c}^{(1)} \partial^{\mu} \overline{c}^{A} \partial_{\mu} c^{A}
        - \delta Z^{(1)}_{\overline{c}\vgb c} \mathscr{C}^{ABC} \big(\partial^{\mu}\overline{c}^A\big) \vgb^{B}_{\mu} c^{C}
        \Big\},
    \end{aligned}
\end{equation}
with coefficients given in App.~\ref{App:SM-Divergent1LoopCTCoeffs}.
As established in Sec.~\ref{Sec:Algebraic_Renormalisation}, the gauge-fixing term does not renormalise; accordingly, only ghost counterterms appear in this sector.

The 1-loop ghost counterterms are fully $D$-dimensional and manifestly BRST invariant.
Furthermore, they are also entirely independent of the evanescent generators, i.e.\ they do not depend on $c_\mathrm{QED}$ or $c_\mathrm{QCD}$.

\paragraph{BRST Transformations and External Sources:}
Finally, we turn to the divergent 1-loop counterterms of the external source terms.
As discussed earlier, the BRST transformations receive nontrivial quantum corrections in non-Abelian gauge theories and therefore renormalise.
Since we coupled these transformations to external sources (see Eq.~\eqref{Eq:SM-L_ext-BRST-Trafos-coupled-to-sources}), their renormalisation is reflected by the following counterterms
\begin{equation}\label{Eq:SM-S_sct_ext}
    \begin{aligned}
        S^{(1)}_{\mathrm{sct,ext}} &= \frac{1}{16\pi^2} \Dintx \bigg\{
        \delta Z_{c}^{(1)} \rho_A^{\mu} \partial_{\mu} c^{A}
        - \delta Z^{(1)}_{\overline{c}\vgb c} \mathscr{C}^{ABC} \rho_A^{\mu} \vgb^{B}_{\mu} c^{C}\\
        &- \Big[ i \delta Z^{(1)}_{\overline{c}\vgb c} c^A 
        \Big( 
        \overline{R}_{l,I}^a \myGenLe_{L,ab}^A \projL l_I^b
        + \overline{R}_{q,I}^{i,a} \Big( \myGenQu_{R,ab,ij}^A \projR + \myGenQu_{L,ab,ij}^A \projL \Big) q_I^{j,b}
        \Big)
        + \mathrm{h.c.} \Big]\\
        &+ \Big[ i \delta Z^{(1)}_{\overline{c}\vgb c} \myGenScl^A_{ab} c^A \Upsilon_{a}^{\dagger} \phi_b + \mathrm{h.c.} \Big]
        + \frac{1}{2} \delta Z^{(1)}_{\overline{c}\vgb c} \mathscr{C}^{ABC} \zeta_A c^B c^C
        \bigg\}.
    \end{aligned}
\end{equation}
The corresponding coefficients are the same as those for the ghosts, see Eq.~\eqref{Eq:SM-S_sct_ghost}.

All of these counterterms are fully $D$-dimensional, independent of the evanescent generators, and coincide with the structure expected in a naive regularisation scheme.
Consequently, none of them contributes to the symmetry-restoration.

\subsection{Finite Symmetry-Restoring Counterterm Action at the 1-Loop Level}\label{Sec:SM_1-Loop_Renormalisation_Finite_CT}

In this section, we present the complete 1-loop finite symmetry-restoring counterterm action of the Standard Model.
Together with the non-invariant divergent counterterms from Sec.~\ref{Sec:SM_1-Loop_Renormalisation_Singular_CT}, these counterterms restore the BRST symmetry broken by the BMHV regularisation and thereby establish the validity of the Slavnov-Taylor identity at the 1-loop level according to Eq.~\eqref{Eq:UltimateSymmetryRequirement}.

The construction of the symmetry-restoring counterterms follows the procedure outlined in Sec.~\ref{Sec:Symmetry_Restoration_Procedure}.
In particular, we propose in each sector a general Ansatz of possible local field monomials and adjust their coefficients such that they satisfy $b_D S_\mathrm{fct}^{(1)} = - (\widehat{\Delta}\cdot\Gamma)_\mathrm{fin}^{(1)}$.
As before and as emphasised in Sec.~\ref{Sec:Symmetry_Restoration_Procedure}, the resulting finite counterterms are not unique: they may be modified by adding finite BRST-invariant counterterms without affecting the validity of the renormalisation or the restored symmetry.

To obtain the breaking contributions, we identified and evaluated 30 $\Delta$-operator-inserted 1PI Green functions (again counting leptonic and quark Green functions separately).
Together with the ordinary Green functions required for the results in Sec.~\ref{Sec:SM_1-Loop_Renormalisation_Singular_CT}, this amounts to a total of 66 computed Green functions, ranging from 2- to 5-point functions. 

Analogous to the singular counterterms, the finite symmetry-restoring counterterms are organised according to the sectors of the SM Lagrangian, yielding
\begin{align}
    S_\mathrm{fct}^{(1)} = S^{(1)}_{\mathrm{fct,fermion}} + S^{(1)}_{\mathrm{fct,gauge}} + S^{(1)}_{\mathrm{fct,Higgs}} + S^{(1)}_{\mathrm{fct,Yukawa}} + S^{(1)}_{\mathrm{fct,ext}}.
\end{align}
Here we have used that no finite symmetry-restoring counterterms arise in the ghost sector at 1-loop order.
The explicit expressions for each sector are presented in the following paragraphs.

\paragraph{Fermionic Sector:}
Again, we begin with the counterterms of the fermionic sector, given by
\begin{equation}\label{Eq:SM-S_fct_fermion}
\begin{aligned}
    S_\mathrm{fct,fermion}^{(1)} &= \frac{1}{16\pi^2} \int d^4x \Big\{ 
    \delta F^{(1)}_{l,ab} {\overline{l}}{}^a_{I} i {\overline{\slashed{\partial}}} l_I^b
    - \delta F^{(1),A}_{\overline{l}\vgb l,R,ab} {\overline{l}}{}^a_{I} {\overline{\slashed{\vgb}}}{}^A \projR l_I^b
    - \delta F^{(1),A}_{\overline{l}\vgb l,L,ab,IJ} {\overline{l}}{}^a_{I} {\overline{\slashed{\vgb}}}{}^A \projL l_J^b\\
    &
    + \delta F^{(1)}_{q,ab} {\overline{q}}{}^{i,a}_{I} i {\overline{\slashed{\partial}}} q_I^{i,b}
    - \delta F^{(1),A}_{\overline{q}\vgb q,R,ab,ij} {\overline{q}}{}^{i,a}_{I} {\overline{\slashed{\vgb}}}{}^A \projR q_I^{j,b}
    - \delta F^{(1),A}_{\overline{q}\vgb q,L,ab,ij,IJ} {\overline{q}}{}^{i,a}_{I} {\overline{\slashed{\vgb}}}{}^A \projL q_J^{j,b}
    \Big\},
\end{aligned}
\end{equation}
with coefficients collected in App.~\ref{App:SM-Finite1LoopCTCoeffs}.

In contrast to the Abelian theories discussed before (see sections~\ref{Sec:S_fct-GeneralAbelianChiralGaugeTheory} and \ref{Sec:FiniteCountertermAction-RHTheory}), the finite fermionic breaking cannot be attributed exclusively to either the fermion self energies or the gauge interaction vertices alone.
The reason is that the counterterms $\delta F^{(1)}_{f,ab}$ ($f\in\{l,q\}$) associated to the fermion kinetic terms do not commute with the left-handed generators $\myGenLe_L^A$ or $\myGenQu_L^A$, and thus violate global $SU(2)_L$ symmetry.
They do not preserve the doublet structure and can be regarded as a globally non-invariant fermion field renormalisation.
Therefore, this breaking cannot be shifted to the vertex corrections by adding finite symmetric counterterms.
Similarly, there exists no finite symmetric counterterm that would allow the vertex-breaking to be absorbed entirely into the self energies.

All finite fermionic counterterms depend on the evanescent generators, as seen in Eqs.~\eqref{Eq:SM-finite-CT-Leptons-KineticTerm}--\eqref{Eq:SM-finite-CT-Quarks-Left-handed-Interaction}.
The lepton counterterms depend on $c_\mathrm{QED}$, while the quark counterterms depend on both $c_\mathrm{QED}$ and $c_\mathrm{QCD}$.
For all these terms, the expressions simplify considerably for any of the physically motivated choices $c_\mathrm{QED},c_\mathrm{QCD}\in\{0,1\}$.

For the right-handed lepton vertex correction $\delta F^{(1),A}_{\overline{l}\vgb l,R,ab}$, the results for $c_\mathrm{QED}=0$ and $c_\mathrm{QED}=1$ are structurally identical, differing only by the presence ($c_\mathrm{QED}=0$) or absence ($c_\mathrm{QED}=1$) of a $\myxi$-dependent prefactor of the $\myGenLe_R^B\myGenLe_L^A\myGenLe_R^B$ term.
For the left-handed vertex correction $\delta F^{(1),A}_{\overline{l}\vgb l,L,ab,IJ}$, the simplification is less significant, but $c_\mathrm{QED}=0$ yields the most compact form.

The behaviour of the quark vertex counterterms is similar to that of the leptons.
Both $\delta F^{(1),A}_{\overline{q}\vgb q,R,ab,ij}$ and $\delta F^{(1),A}_{\overline{q}\vgb q,L,ab,ij,IJ}$ simplify noticeably for any choice with $c_\mathrm{QED},c_\mathrm{QCD}\in\{0,1\}$.
The configurations ($c_\mathrm{QED}=0$, $c_\mathrm{QCD}=0$) and ($c_\mathrm{QED}=0$, $c_\mathrm{QCD}=1$) yield particularly compact expressions, with the latter especially simplifying the left-handed quark--gauge interaction counterterm.

For the fermion kinetic counterterms $\delta F^{(1)}_{l,ab}$ and $\delta F^{(1)}_{q,ab}$ the situation is unambiguous:
both vanish identically for the choice $c_\mathrm{QED}=0$ and $c_\mathrm{QCD}=0$.
As a consequence, this makes ($c_\mathrm{QED}=0$, $c_\mathrm{QCD}=0$) the technically most advantageous choice for practical applications, yielding the simplest form for the entire fermionic finite counterterm sector.

\paragraph{Gauge Sector:}
The finite symmetry-restoring 1-loop counterterms of the gauge sector take the form
\begin{equation}\label{Eq:SM-S_fct_gauge}
\begin{aligned}
    S_\mathrm{fct,gauge}^{(1)} &= \frac{1}{16\pi^2} \int d^4x \bigg\{ 
    \delta F^{(1),AB}_{2\vgb} {\overline{\vgb}}{}_\mu^A \overline{\Box} {\overline{\vgb}}{}^{B,\mu} 
    + \delta F^{(1),ABC}_{3\vgb\text{-1}} \big({\overline{\partial}}{}^\mu {\overline{\vgb}}{}^{A,\nu}\big) {\overline{\vgb}}{}^B_\mu {\overline{\vgb}}{}^C_\nu \\
    &\phantom{\frac{1}{16\pi^2} \int d^4x \,\,\,\,}
    + \delta F^{(1),ABC}_{3\vgb\text{-2}} {\overline{\vgb}}{}^{A,\nu} \big({\overline{\partial}}{}^\mu {\overline{\vgb}}{}^{B}_{\mu}\big) {\overline{\vgb}}{}^C_\nu 
    + \delta F^{(1),ABCD}_{4\vgb} {\overline{\vgb}}{}^{A}_{\mu} {\overline{\vgb}}{}^{B,\mu} {\overline{\vgb}}{}^C_\nu {\overline{\vgb}}{}^{D,\nu} \bigg\},
\end{aligned}
\end{equation}
with coefficients provided in App.~\ref{App:SM-Finite1LoopCTCoeffs}.

None of these finite gauge-sector counterterms depend on the evanescent generators, and are thus completely independent of $c_\mathrm{QED}$ and $c_\mathrm{QCD}$, as seen in Eqs.~\eqref{Eq:SM-finite-CT-gauge-boson-SelfEnergy}--\eqref{Eq:SM-finite-CT-gauge-boson-4Point}.

\paragraph{Higgs Sector:}
Next, we consider the Higgs sector. 
The corresponding finite symmetry-restoring 1-loop counterterms are given by
\begin{equation}\label{Eq:SM-S_fct_Higgs}
\begin{aligned}
    S_\mathrm{fct,Higgs}^{(1)} &= \frac{1}{16\pi^2} \int d^4x \bigg\{ 
    \delta F^{(1)}_{\phi^\dagger\phi,ab} \, \phi_a^\dagger \overline{\Box} \phi_b
    + i \delta F^{(1),A}_{-,ab} \big[ \big({\overline{\partial}}{}^\mu\phi_a^\dagger\big) \phi_b - \phi_a^\dagger \big({\overline{\partial}}{}^\mu\phi_b\big) \big] {\overline{\vgb}}{}^A_\mu\\
    &\phantom{\frac{1}{16\pi^2} \int d^4x \,\,\,\,}
    + i \delta F^{(1),A}_{+,ab} \big[ \big({\overline{\partial}}{}^\mu\phi_a^\dagger\big) \phi_b + \phi_a^\dagger \big({\overline{\partial}}{}^\mu\phi_b\big) \big] {\overline{\vgb}}{}^A_\mu
    + \delta F^{(1),AB}_{\phi^\dagger\phi\vgb\vgb,ab} \phi_a^\dagger \phi_b {\overline{\vgb}}{}^A_\mu {\overline{\vgb}}{}^{B,\mu}\\
    &\phantom{\frac{1}{16\pi^2} \int d^4x \,\,\,\,}
    + \Big[ \delta F^{(1)}_{\phi\phi,ab} \, {\overline{\partial}}{}^\mu \phi_a {\overline{\partial}}{}_\mu\phi_b 
    + i \delta F_{\phi\phi\vgb,ab}^A \big({\overline{\partial}}{}^\mu\phi_a\big) \phi_b {\overline{\vgb}}{}^A_\mu\\
    &\phantom{\frac{1}{16\pi^2} \int d^4x \,\,\,\, }
    + \delta F^{(1),AB}_{\phi\phi\vgb\vgb,ab} \phi_a \phi_b {\overline{\vgb}}{}^A_\mu {\overline{\vgb}}{}^{B,\mu} + \mathrm{h.c.} \Big]
    - \delta F^{(1)}_{\phi^\dagger\phi\phi^\dagger\phi,abcd} \phi_a^\dagger\phi_b\phi_c^\dagger\phi_d\\
    &\phantom{\frac{1}{16\pi^2} \int d^4x \,\,\,\, }
    - \Big[ \delta F^{(1)}_{\phi^\dagger\phi\phi\phi,abcd} \phi_a^\dagger\phi_b\phi_c\phi_d
    + \delta F^{(1)}_{\phi\phi\phi\phi,abcd} \phi_a\phi_b\phi_c\phi_d + \mathrm{h.c.} \Big]
    \bigg\},
\end{aligned}
\end{equation}
with coefficients listed in App.~\ref{App:SM-Finite1LoopCTCoeffs}.

As in the gauge sector, none of the counterterms in Eq.~\eqref{Eq:SM-S_fct_Higgs} depend on the evanescent gauge interactions, as seen in Eqs.~\eqref{Eq:SM-finite-CT-Scalar-SelfEnergy}--\eqref{Eq:SM-finite-CT-Scalar-DoubleScalar-DoubleGaugeBoson}, Eqs.~\eqref{Eq:SM-finite-CT-Scalar-phi-phi}--\eqref{Eq:SM-finite-CT-Scalar-phi-phi-V-V}, and Eqs.~\eqref{Eq:SM-finite-CT-Scalar-phidagger-phi-phidagger-phi}--\eqref{Eq:SM-finite-CT-Scalar-phi-phi-phi-phi}.

In contrast to the divergent counterterms in Eq.~\eqref{Eq:SM-S_sct_Higgs}, the finite Higgs-sector contains several counterterms that violate global hypercharge conservation.
These are the terms in the square bracket in the third and fourth lines, as well as the terms in the last line of Eq.~\eqref{Eq:SM-S_fct_Higgs}.
They are structurally identical to those encountered in Sec.~\eqref{Sec:AnalysisOfASSMResults}, with analogous properties.
Importantly, all counterterms in Eq.~\eqref{Eq:SM-S_fct_Higgs} preserve electric and colour charge.

Additionally, two new counterterms appear in the non-Abelian case, governed by the coefficients $\delta F^{(1),A}_{+,ab}$ and $\delta F^{(1)}_{\phi^\dagger\phi\phi^\dagger\phi,abcd}$.
These terms were absent in the Abelian theory discussed in Sec.~\eqref{Sec:AnalysisOfASSMResults}, as they vanish identically for the Abelian case with adjoint colour index $A=1$; they contribute only for $A\in\{2,3\}$.

\paragraph{Yukawa Sector:}
We continue with the finite symmetry-restoring 1-loop Yukawa counterterms, which are provided by
\begin{equation}\label{Eq:SM-S_fct_Yukawa}
\begin{aligned}
    S_\mathrm{fct,Yukawa}^{(1)} &= -\frac{1}{16\pi^2} \int d^4x \bigg\{ 
    Y_{l,IJ} \delta F^{l,(1)}_{Y,abc} {\overline{l}}{}^{a}_I \phi_b \projR l_J^c
    + \delta G^{l,(1)}_{Y,abc,IJ} {\overline{l}}{}^{a}_I \phi_b^\dagger \projR l_I^c\\
    &\phantom{- \frac{1}{16\pi^2} \int d^4x \bigg\{ \,}
    + Y_{d,IJ} \delta F^{d,(1)}_{Y,abc} {\overline{q}}{}^{i,a}_I \phi_b \projR q_J^{i,c}
    + \delta G^{d,(1)}_{Y,abc,IJ} {\overline{q}}{}^{i,a}_I \phi_b^\dagger \projR q_J^{i,c}\\
    &\phantom{- \frac{1}{16\pi^2} \int d^4x \bigg\{ \,}
    + Y_{u,IJ} \delta F^{u,(1)}_{Y,abc} {\overline{q}}{}^{i,a}_I \phi_b^\dagger \projR q_J^{i,c}
    + \delta G^{u,(1)}_{Y,abc,IJ} {\overline{q}}{}^{i,a}_I \phi_b \projR q_J^{i,c}
    + \mathrm{h.c.}
    \bigg\},
\end{aligned}
\end{equation}
with coefficients displayed in App.~\ref{App:SM-Finite1LoopCTCoeffs}.

There are two types of finite Yukawa counterterms.
The first consists of terms that break local gauge and BRST invariance but preserve global hypercharge conservation; these are governed by the coefficients $\delta F^{f,(1)}_{Y,abc}$ with $f\in\{l,d,u\}$.
The second type contains the counterterms that additionally violate global hypercharge conservation, governed by the coefficients $\delta G^{f,(1)}_{Y,abc,IJ}$ for $f\in\{l,d,u\}$.

The global hypercharge preserving counterterms $\propto \delta F^{f,(1)}_{Y,abc}$ contribute only to the non-Abelian part of the BRST breaking, i.e.\ for adjoint colour index $A\in\{2,3,4\}$, and vanish in the Abelian limit $A=1$.
Consequently, these terms were absent in Abelian model analysed in Sec.~\eqref{Sec:AnalysisOfASSMResults}, where only the global hypercharge violating contributions $\propto \delta G^{f,(1)}_{Y,abc,IJ}$ appeared.

All finite Yukawa counterterms in Eq.~\eqref{Eq:SM-S_fct_Yukawa} depend on the evanescent generators, as shown in Eqs.~\eqref{Eq:SM-finite-CT-Yukawa-Lepton-F}--\eqref{Eq:SM-finite-CT-Yukawa-Up-Quark-G}.
As before, these terms simplify for $c_\mathrm{QED},c_\mathrm{QCD}\in\{0,1\}$. 
The global hypercharge violating counterterms governed by $\delta G^{f,(1)}_{Y,abc,IJ}$ are structurally identical for all such choices, differing only by their prefactors.
In contrast, the global hypercharge conserving counterterms governed by $\delta F^{f,(1)}_{Y,abc}$ vanish entirely for the choice ($c_\mathrm{QED}=0$, $c_\mathrm{QCD}=0$).
Thus, also within the Yukawa sector, omitting evanescent gauge interactions leads to the most economical set of counterterms and therefore appears to be the preferable choice.

\paragraph{BRST Transformations and External Sources:}
Finally, we present the finite symmetry-restoring 1-loop counterterms of the external source terms, given by
\begin{equation}\label{Eq:SM-S_fct_ext}
\begin{aligned}
    S_\mathrm{fct,ext}^{(1)} &= - \frac{i \delta F^{(1)}_{R_f}}{16\pi^2} \int d^4x \Big\{ 
    c^A {\overline{R}}{}^a_{l,I} \big[ \myGenLe^A_{R,ab} \projR + \myGenLe^A_{L,ab} \projL \big] l_I^b\\
    &\phantom{- \frac{i \delta \mathcal{F}^{(1)}_{R_f}}{16\pi^2} \int d^4x \,\,\,\,}
    + c^A {\overline{R}}{}^{i,a}_{q,I} \big[ \myGenQu^A_{R,ab,ij} \projR + \myGenQu^A_{L,ab,ij} \projL \big]  q_I^{j,b}
    - c^A {\overline{R}}{}^{i,a}_{q,I} {\widehat{\myGenQu}}{}^A_{ab,ij} q_I^{j,b}
    + \mathrm{h.c.}
    \Big\},
\end{aligned}
\end{equation}
with coefficients provided in App.~\ref{App:SM-Finite1LoopCTCoeffs}.
As in the divergent case, these counterterms encode the renormalisation of the BRST transformations inherent to non-Abelian gauge theories.

Only the fermionic source terms require finite symmetry-restoring contributions.
Both the lepton and the quark terms are governed by the coefficient $\delta \mathcal{F}^{(1)}_{R_f}$, which is proportional to the adjoint quadratic Casimir and independent of the evanescent generators (see Eq.~\eqref{Eq:SM-finite-CT-external-Sources}).
For leptons, the simple Ansatz ${\overline{R}}{}^a_{l,I} (s_D l_I^b)$ is sufficient.
For quarks, however, an additional term involving the evanescent generator ${\widehat{\myGenQu}}{}^A$ is required.

Regarding the QCD sector, both choices $c_\mathrm{QCD}=0$ and $c_\mathrm{QCD}=1$ are motivated.
For $c_\mathrm{QCD}=0$, the additional term $\propto {\widehat{\myGenQu}}{}^A$ vanishes.
For $c_\mathrm{QCD}=1$, the entire QCD contribution cancels, because in this case the $SU(3)_c$ components of $\myGenQu_R^A$, $\myGenQu^A_L$ and ${\widehat{\myGenQu}}{}^A$ all coincide with the strong generator $t_s^A$ (cf.\ Eq.~\eqref{Eq:Quark-Super-Generator}).
Setting $c_\mathrm{QED}=1$, only the right-handed Abelian contribution cancels, while the left-handed term remains.

\paragraph{Comments on the Implications of Evanescent Gauge Interactions:}
A preliminary examination of the counterterms presented in this section and in Sec.~\ref{Sec:SM_1-Loop_Renormalisation_Singular_CT}, with coefficients collected in App.~\ref{App:Results-TheStandardModel}, indicates that the simplest and most economical expressions are obtained for $(c_\mathrm{QED}=0,c_\mathrm{QCD}=0)$, corresponding to vanishing evanescent gauge interactions.
The choice $(c_\mathrm{QED}=0,c_\mathrm{QCD}=1)$ also has appealing features, as it treats the QCD interaction current fully $D$-dimensional and leads to certain simplifications in the QCD sector.
In contrast, introducing evanescent $U(1)_Y$ gauge interactions, i.e.\ $c_\mathrm{QED}\neq0$, appears overall to be disadvantageous.
This conclusion aligns with the observations made in Sec.~\ref{Sec:Results-Shedding_Light_on_Evanescent_Shadows}.

Nevertheless, further investigation is required before a definitive choice of evanescent parameters can be recommended.
First, the electroweak and the QCD sectors should be analysed separately by rewriting the counterterm action in terms of gauge bosons $B_\mu$, $W^A_\mu$ and $G_\mu^A$, rather than in terms of the unified gauge bosons $\vgb_\mu^A$ used throughout this chapter.
While the unified notation yields a particularly compact and computationally efficient representation, it obscures detailed structural features that must be disentangled to fully assess the role of evanescent interactions.

Second, the impact of evanescent gauge interactions should be analysed in the broken phase of the SM, expressed in terms of mass eigenstates $A_\mu$, $Z_\mu$, $W_\mu^{\pm}$ and $G^A_\mu$.
Up to this point, our analysis has been restricted to the unbroken phase.
However, the evanescent $U(1)_Y$ gauge interaction governed by $c_\mathrm{QED}$ was introduced solely to achieve a fully $D$-dimensional photon interaction current (see Sec.~\ref{Sec:SM-Lagrangian}).
To meaningfully compare a purely 4-dimensional versus a fully $D$-dimensional photon interaction current, one must perform the analysis after electroweak symmetry breaking, where QED and QCD arise as the unbroken gauge subgroups of the SM.

\chapter{Conclusions}\label{Chap:Conclusions}

In this thesis, we have presented a comprehensive study of the renormalisation of chiral gauge theories in DReg with non-anticommuting $\gamma_5$, employing the mathematically consistent BMHV scheme.
Our work encompasses both conceptual foundations and high-precision multi-loop applications, thereby significantly advancing the state of the art in the consistent treatment of chiral gauge theories.

To this end, we provided a detailed discussion of the underlying theoretical concepts, including the general theory of renormalisation, its application to gauge theories, and DReg as the regularisation scheme of choice.
A central theme of this work is the systematic analysis and restoration of the regularisation-induced breaking of gauge and BRST invariance, originating from the modified algebraic relations of the BMHV scheme, as discussed in chapter~\ref{Chap:Practical_Symmetry_Restoration}.
We established a transparent and fully algorithmic procedure for determining the singular and finite symmetry-restoring counterterms required at each order of perturbation theory to obtain a fully renormalised theory (see particularly Sec.~\ref{Sec:Symmetry_Restoration_Procedure}).

Since the $D$-dimensional extension --- and thus the explicit implementation of the regularisation --- is not unique, we investigated the dimensional ambiguities and evanescent details inherent to DReg.
These ambiguities arise primarily in the fermionic sector, where both the construction of $D$-dimensional Dirac spinors and the presence of evanescent gauge interactions play an important role.
In particular, we analysed the freedom in the $D$-dimensional realisation of the fermion kinetic terms and gauge interactions in Sec.~\ref{Sec:Dimensional_Ambiguities_and_Evanescent_Shadows}, which essentially correspond to different implementations of the BMHV regularisation, and examined in chapter~\ref{Chap:General_Abelian_Chiral_Gauge_Theory} how different choices affect the structure of the induced symmetry breaking and the resulting counterterms.

Regarding the kinetic terms, we compared two conceptually distinct approaches of treating fermions in $D$-dimensions: combining physical chiral fermions into a single Dirac spinor whenever possible (see Option~\ref{Opt:Option1}) and the sterile-partner construction (see Option~\ref{Opt:Opt2-GroupRef}).
Each option has advantages and drawbacks.
While the sterile-partner approach avoids the violation of global symmetries caused by the evanescent part of the kinetic terms that mix left- and right-handed fermions with different gauge quantum numbers, it introduces substantial complications in the presence of massive fermions.
In particular, the propagator of massive fermions acquires a nonstandard denominator involving mixed $D$- and 4-dimensional momenta, which greatly complicates the evaluation of loop integrals.
These difficulties outweigh the benefits of preserving global symmetries, rendering this option impractical for applications where fermion masses cannot be neglected, such as low-energy electroweak precision phenomenology.

Concerning the fermion--gauge interactions, we allowed for general evanescent interaction terms that are only restricted by hermiticity and couple the gauge bosons to fermion currents of the form $\overline{\psi}\widehat{\gamma}^\mu\mathbb{P}_{\mathrm{R/L}}\psi$, thereby covering the most general $D$-dimensional interaction structure permitted in the BMHV scheme.
Our analysis shows that including such evanescent gauge interactions leads to a substantial proliferation of terms and significantly enlarges the counterterm structure. 
By examining several motivated prescriptions, we identified that omitting evanescent gauge interactions results in the most economical and transparent counterterm structure, and thus yields the practically most efficient realisation.

Overall, our study demonstrates explicitly how the choice of $D$-dimensional fermion kinetic terms and evanescent gauge interactions influences the regularisation-induced symmetry breaking and the complexity of the resulting counterterms.
These findings, which we first published in Ref.~\cite{Ebert:2024xpy}, provide valuable guidance for future multi-loop applications.

A major achievement of this thesis is the complete renormalisation of an Abelian chiral gauge theory within the BMHV scheme of DReg up to the 4-loop level (see chapter~\ref{Chap:BMHV_at_Multi-Loop_Level}), together with the development of a fully automated, high-performance computational framework capable of performing this calculation (see Sec.~\ref{Sec:Computational_Setup}).
We first reported the corresponding results in Refs.~\cite{Stockinger:2023ndm,vonManteuffel:2025swv}.
In contrast to non-self-consistent $\gamma_5$ prescriptions, the BMHV scheme with non-anticommuting $\gamma_5$ provides a mathematically consistent and fully self-contained framework, so that no external arguments were required at any stage of our computations.
The 4-loop computation represents the highest-order application of the BMHV scheme to date and demonstrates conclusively that a fully self-consistent treatment of $\gamma_5$ is achievable even in very high-order calculations. 

The feasibility of this calculation relied crucially on the highly optimised \texttt{FORM}-based computational setup developed in this thesis.
We implemented highly efficient symbolic procedures and optimised algorithms that incorporate the tadpole decomposition for extracting UV divergences, a powerful tensor-reduction algorithm based on the orbit partition approach, and streamlined routines for evaluating the Dirac algebra adapted to the BMHV scheme with non-anticommuting $\gamma_5$.
In chapter~\ref{Chap:Multi-Loop_Calculations}, we provided a detailed description of this framework and the methods employed therein.
The resulting computational performance enabled the computation of multi-loop Green functions, some of which involved 4-loop Feynman diagrams that generated billions of intermediate terms.

Beyond the explicit multi-loop results, two conceptual conclusions emerge.
First, we have explicitly demonstrated the renormalisability of a chiral gauge theory within the framework of the BMHV scheme up to the 4-loop level.
Crucially, the regularisation-induced symmetry breaking at each loop order can be parametrised in terms of a finite basis of local field monomials for the operator insertion $\Delta\cdot\Gamma$, constrained by power-counting and ghost number.
Consequently, all spurious symmetry breakings can be eliminated by a finite set of local counterterms, ensuring the restoration of the Slavnov–Taylor identity at every order.
The resulting counterterm action remains compact even at the 4-loop level and is well-suited for computer implementation in automated multi-loop calculations.
These findings are fully consistent with the general discussion of renormalisability and the finiteness of the insertion basis in Ref.~\cite{Piguet:1995er}, as well as with the explicit analyses of Refs.~\cite{Cornella:2022hkc,Kuhler:2025znv}.
In Ref.~\cite{Cornella:2022hkc}, they derived the bases for the finite breaking and the corresponding symmetry-restoring counterterms at the 1-loop level, while the 2-loop study of Ref.~\cite{Kuhler:2025znv} demonstrates that no new counterterm structures emerge at higher orders, thereby confirming the finiteness of the counterterm basis.

Second, the successful completion of the 4-loop computation validates the automated computational framework developed in this work.
Extensive tests during its development, together with its successful application to the projects presented here, demonstrate its robustness and efficiency.
The setup is sufficiently powerful to handle numerous 4-loop Green functions and thus establishes a solid computational foundation for future high-precision studies.

Building upon these developments, the thesis concludes with the complete 1-loop renormalisation of the full SM in the BMHV scheme, presented in chapter~\ref{Chap:The_Standard_Model}. 
In view of an application to electroweak physics in the spontaneously broken phase of the SM with massive fermions, we employed Option~\ref{Opt:Option1} for the $D$-dimensional treatment of fermions.
We included evanescent gauge interactions in the $U(1)_Y$ and $SU(3)_c$ sectors, parametrised by $c_\mathrm{QED}$ and $c_\mathrm{QCD}$, respectively.
This allows the photon and gluon interaction currents to be treated either purely 4-dimensionally or fully $D$-dimensionally.
Although the fully $D$-dimensional treatment has certain appealing features, particularly for the QCD sector, the most transparent and economical results are again obtained when evanescent gauge interactions are omitted, corresponding to $(c_\mathrm{QED}=0,c_\mathrm{QCD}=0)$.
This 1-loop renormalisation constitutes a crucial first step towards a fully consistent multi-loop renormalisation of the SM with non-anticommuting $\gamma_5$, and establishes the foundation for a long-term research programme in high-precision electroweak phenomenology.

Looking ahead, the methods and results developed here open several promising directions. 
The natural next step is to extend the BMHV renormalisation of the SM to the 2-loop level and beyond, and to incorporate an analysis of spontaneous symmetry breaking in this context.
These results will be essential for the computation of electroweak precision-observables, such as the mass of the $W^\pm$ boson, $Z$-decays into quarks ($Z\to q\overline{q}$), relevant $2\to2$ scattering processes, and Higgs production.
Beyond phenomenological applications, the techniques developed in this thesis provide the tools needed to address several conceptual questions.
These include the renormalisation group behaviour and explicit determination of $\beta$-functions in the framework of the BMHV scheme at higher orders, as exemplified by the Abelian 2-loop study in Ref.~\cite{Belusca-Maito:2022wem}.
Further promising directions include an analysis of the Weyl-consistency conditions --- used in the literature to resolve ambiguities arising in alternative $\gamma_5$ prescriptions --- within the BMHV scheme and the clarification to what extend such alternative $\gamma_5$ prescriptions may be related to, or constrained by, the fully consistent BMHV framework so as to eliminate any ambiguities arising in schemes with naively anticommuting $\gamma_5$.
Moreover, it is worthwhile to investigate the complete symmetry restoration procedure (see Sec.~\ref{Sec:Symmetry_Restoration_Procedure}) in the context of ``deformed'' Slavnov-Taylor identities, which may lead to modified or generalised symmetry relations that could supersede the usual BRST invariance and persist under regularisation, thereby simplifying the renormalisation procedure.
First considerations in this direction can be found in Ref.~\cite{OlgosoRuiz:2024dzq}.
Since any realistic extension of the SM involves chiral gauge interactions, the methods developed here are directly applicable to a wide range of BSM scenarios, including EFTs such as the SMEFT.

By resolving the $\gamma_5$-problem with mathematical rigour and algorithmic efficiency, this work establishes a consistent foundation for theoretical predictions that match the precision requirements of current and upcoming experiments.


\appendix

\chapter{Results --- General Abelian Chiral Gauge Theory}\label{App:Results-GeneralAbelianChiralGaugeTheory}

Here, we provide the explicit 1-loop results for the counterterm coefficients used in Sec.~\ref{Sec:One-Loop-Renormalisation-Abelian-Chiral-Gauge-Theory} (see App.~\ref{App:1LoopCTCoeffs-GeneralAbelianChiralGaugeTheory}) and for all 1PI Green functions (see App.~\ref{App:1LoopGreenFunctions-GeneralAbelianChiralGaugeTheory}) relevant for the renormalisation of the Abelian chiral gauge theory discussed in chapter~\ref{Chap:General_Abelian_Chiral_Gauge_Theory} (see Sec.~\ref{Sec:Definition_of_the_Theory_General_Abelian_Case} for the model definition).

\section{Explicit Results for the 1-Loop Counterterm Coefficients}\label{App:1LoopCTCoeffs-GeneralAbelianChiralGaugeTheory}

We present here the counterterm coefficients for the general results presented
in Sec.~\ref{Sec:One-Loop-Renormalisation-Abelian-Chiral-Gauge-Theory}, assuming $\hypRL=\hypLR$ as hermitian matrices commuting with $\hypL$ and $\hypR$ and conserving electric and colour charge.
The expressions are simplified using the BRST relations in Eq.~\eqref{Eq:YukawaHyperchargeBRSTCondition}, the above commutation properties, and trace cyclicity, yielding identities such as
\begin{equation}
    \begin{aligned}
        \mathrm{Tr}\big(G^a{K^b}^{\dagger}\big) &= 0,
        &
        \quad
        \mathrm{Tr}\big(MG^a{K^b}^{\dagger}\big) &= 0, 
        &
        \quad 
        \mathrm{Tr}\big(M{G^a}^{\dagger}K^b\big) &= 0,\\
        \mathrm{Tr}\big(M_1{G^a}^{\dagger}M_2K^b\big) &= 0,
        &
        \quad
        \mathrm{Tr}\big({G^a}^{\dagger}K^b{G^c}^{\dagger}K^d\big) &= 0,
        &
        \quad 
        \mathrm{Tr}\big({G^a}^{\dagger}G^b{G^c}^{\dagger}K^d\big) &= 0,
    \end{aligned}
\end{equation}
with $M\in\{\hypL,\hypR,\hypL^2,\hypL\hypR,\hypR^2\}$ and $M_1,\,M_2\in\{\hypL,\hypR\}$, 
among others, illustrating the types of relations employed to simplify the final expressions.

\subsection{Coefficients of Divergent Contributions}\label{App:Divergent1LoopCTCoeffs-GeneralAbelianChiralGaugeTheory}

We begin with the coefficients of the divergent contributions, and define the relations
\begin{equation}\label{App-Eq:DivCoeffRelations}
    \begin{aligned}
        \overline{\mathcal{A}}_{\psi\overline{\psi},\mathrm{R},ji}^{1,\mathrm{inv}} &= g^2 \prescript{}{F}{\overline{\mathcal{A}}}_{\psi\overline{\psi},\mathrm{R},ji}^{1,\mathrm{inv}}
        + \prescript{}{Y}{\overline{\mathcal{A}}}_{\psi\overline{\psi},\mathrm{R},ji}^{1,\mathrm{inv}},\\
        \overline{\mathcal{A}}_{\psi\overline{\psi},\mathrm{L},ji}^{1,\mathrm{inv}} &= g^2 \prescript{}{F}{\overline{\mathcal{A}}}_{\psi\overline{\psi},\mathrm{L},ji}^{1,\mathrm{inv}}
        + \prescript{}{Y}{\overline{\mathcal{A}}}_{\psi\overline{\psi},\mathrm{L},ji}^{1,\mathrm{inv}},\\
        \widehat{\mathcal{A}}_{\psi\overline{\psi},ji}^{1,\mathrm{break}} &= 
        g^2 \prescript{}{F}{\widehat{\mathcal{A}}}_{\psi\overline{\psi},\mathrm{LR},ji}^{1,\mathrm{break}}
        + \prescript{}{Y}{\widehat{\mathcal{A}}}_{\psi\overline{\psi},ji}^{1,\mathrm{break}},\\
        \widehat{\mathcal{A}}_{\psi\overline{\psi}B,ji}^{1,\mathrm{break}} &= 
        g^2 \Big(\mathcal{Y}_{LR} \prescript{}{F}{\widehat{\mathcal{A}}}_{\psi\overline{\psi},\mathrm{LR}}^{1,\mathrm{break}} \Big)_{ji}
        - \Big( \prescript{}{YS}{\widehat{\mathcal{A}}}_{\psi\overline{\psi}B,ji}^{1,\mathrm{break}}
        + \prescript{}{YF}{\widehat{\mathcal{A}}}_{\psi\overline{\psi}B,ji}^{1,\mathrm{break}}\Big),\\
        \overline{\mathcal{A}}_{\psi\overline{\psi}\phi,\mathrm{R},ji}^{1,\mathrm{inv},a} &= \prescript{}{Y}{\overline{\mathcal{A}}}_{\psi\overline{\psi}\phi,\mathrm{R},ji}^{1,\mathrm{inv},a} + g^2 \, \Big(\prescript{}{YF}{\overline{\mathcal{A}}}_{\psi\overline{\psi}\phi,\mathrm{R},ji}^{1,\mathrm{inv},a} + \prescript{}{YSF}{\overline{\mathcal{A}}}_{\psi\overline{\psi}\phi,\mathrm{R},ji}^{1,\mathrm{inv},a}\Big),\\
        \overline{\mathcal{A}}_{\psi\overline{\psi}\phi,\mathrm{L},ji}^{1,\mathrm{inv},a} &= \prescript{}{Y}{\overline{\mathcal{A}}}_{\psi\overline{\psi}\phi,\mathrm{L},ji}^{1,\mathrm{inv},a} + g^2 \, \Big(\prescript{}{YF}{\overline{\mathcal{A}}}_{\psi\overline{\psi}\phi,\mathrm{L},ji}^{1,\mathrm{inv},a} + \prescript{}{YSF}{\overline{\mathcal{A}}}_{\psi\overline{\psi}\phi,\mathrm{L},ji}^{1,\mathrm{inv},a}\Big),\\
        \mathcal{A}_{\phi\phi^{\dagger}\phi\phi^{\dagger},abcd}^{1,\mathrm{inv}} &= \Big( g^4 \prescript{}{S}{\mathcal{A}}_{\phi\phi^{\dagger}\phi\phi^{\dagger}}^{1,\mathrm{inv}} + g^2 \prescript{}{S\lambda}{\mathcal{A}}_{\phi\phi^{\dagger}\phi\phi^{\dagger}}^{1,\mathrm{inv}} + \prescript{}{Y}{\mathcal{A}}_{\phi\phi^{\dagger}\phi\phi^{\dagger}}^{1,\mathrm{inv}} + \prescript{}{\lambda}{\mathcal{A}}_{\phi\phi^{\dagger}\phi\phi^{\dagger}}^{1,\mathrm{inv}} \Big)_{abcd},
    \end{aligned}
\end{equation}
to combine all contributions associated with the same field monomials.
The coefficients are then presented below, grouped by field monomial type.

\paragraph{Gauge Boson Coefficients:} 
The gauge boson coefficients receive fermionic,
\begin{equation}\label{App-Eq:GaugeBosonCoeffsFermionic}
    \begin{aligned}
        \prescript{}{F}{\overline{\mathcal{A}}}_{BB}^{1,\mathrm{inv}} &= 
        \frac{2}{3} \Big[ \mathrm{Tr}\big(\mathcal{Y}_R^2\big) + \mathrm{Tr}\big(\mathcal{Y}_L^2\big) \Big],\\
        \prescript{}{F}{\widehat{\mathcal{A}}}_{BB,1}^{1,\mathrm{break}} &= 
        \frac{2}{3} \, \mathrm{Tr}\big(\mathcal{Y}_{LR}^2\big),\\
        \prescript{}{F}{\widehat{\mathcal{A}}}_{BB,2}^{1,\mathrm{break}} &= 
        \frac{2}{3} \, \mathrm{Tr}\Big(\big(\mathcal{Y}_R+\mathcal{Y}_L\big)\mathcal{Y}_{LR}\Big),\\
        \prescript{}{F}{\widehat{\mathcal{A}}}_{BB,3}^{1,\mathrm{break}} &= 
        - 2 \, \mathrm{Tr}\big(\mathcal{Y}_{LR}^2\big),\\
        \prescript{}{F}{\widehat{\mathcal{A}}}_{BB,4}^{1,\mathrm{break}} &= 
        - \frac{1}{3} \, \mathrm{Tr}\Big(\big(\mathcal{Y}_R+\mathcal{Y}_L\big)^2\Big),
    \end{aligned}
\end{equation}
and scalar contributions,
\begin{equation}\label{App-Eq:GaugeBosonCoeffsScalar}
    \begin{aligned}
        \prescript{}{S}{\mathcal{A}}_{BB}^{1,\mathrm{inv}} &= 
        \frac{N_S}{3} \, \mathcal{Y}_{S}^2.
    \end{aligned}
\end{equation}

\paragraph{Fermion Coefficients:} 
Fermion contributions arise from both gauge interactions,
\begin{equation}\label{App-Eq:FermionCoeffFermionGauge}
    \begin{aligned}
        \prescript{}{F}{\overline{\mathcal{A}}}_{\psi\overline{\psi},\mathrm{R},ji}^{1,\mathrm{inv}} &=
        \xi \big(\mathcal{Y}_{R}^2\big)_{ji},\\
        \prescript{}{F}{\overline{\mathcal{A}}}_{\psi\overline{\psi},\mathrm{L},ji}^{1,\mathrm{inv}} &=
        \xi \big(\mathcal{Y}_{L}^2\big)_{ji},\\
        \prescript{}{F}{\widehat{\mathcal{A}}}_{\psi\overline{\psi},\mathrm{LR},ji}^{1,\mathrm{break}} &=
        \frac{1}{3} 
        \Big[
        2(2+\xi) \big(\mathcal{Y}_{R}\mathcal{Y}_{L}\big)_{ji}
        -(1-\xi) \big[\big(\mathcal{Y}_{R}+\mathcal{Y}_{L}\big)\mathcal{Y}_{LR}\big]_{ji}
        -(2+\xi) \big(\mathcal{Y}_{LR}^2\big)_{ji}
        \Big],
    \end{aligned}
\end{equation}
and Yukawa interactions,
\begin{equation}\label{App-Eq:FermionCoeffYukawa}
    \begin{aligned}
        \prescript{}{Y}{\overline{\mathcal{A}}}_{\psi\overline{\psi},\mathrm{R},ji}^{1,\mathrm{inv}} &=
        \frac{1}{2} \Big[\big({G^a}^{\dagger}G^a\big)_{ji} + \big({K^a}^{\dagger}K^a\big)_{ji}\Big],\\
        \prescript{}{Y}{\overline{\mathcal{A}}}_{\psi\overline{\psi},\mathrm{L},ji}^{1,\mathrm{inv}} &=
        \frac{1}{2} \Big[\big(G^a{G^a}^{\dagger}\big)_{ji} + \big(K^a{K^a}^{\dagger}\big)_{ji}\Big],\\
        \prescript{}{Y}{\widehat{\mathcal{A}}}_{\psi\overline{\psi},ji}^{1,\mathrm{break}} &=
        \frac{1}{2} \Big[\big(G^aK^a\big)_{ji} + \big(K^aG^a\big)_{ji}\Big].
    \end{aligned}
\end{equation}

\paragraph{Fermion-Gauge Boson Coefficients:} 
Besides the usual vertex correction with gauge boson exchange, additional counterterm contributions arise from scalar--gauge boson and fermion--gauge boson diagrams with Yukawa interactions:
\begin{equation}\label{App-Eq:FermionGaugeBosonCoeff}
    \begin{aligned}
        \prescript{}{YS}{\widehat{\mathcal{A}}}_{\psi\overline{\psi}B,ji}^{1,\mathrm{break}} &=
        \frac{1}{2} \, \mathcal{Y}_{S} \Big[\big(K^aG^a\big)_{ji} - \big(G^aK^a\big)_{ji}\Big],\\
        \prescript{}{YF}{\widehat{\mathcal{A}}}_{\psi\overline{\psi}B,ji}^{1,\mathrm{break}} &=
        - \frac{1}{2} \Big[\big(G^a\mathcal{Y}_{LR}K^a\big)_{ji} + \big(K^a\mathcal{Y}_{LR}G^a\big)_{ji}\Big].
    \end{aligned}
\end{equation}

\paragraph{Yukawa Coefficients:}
The divergent Yukawa contributions are governed by
\begin{equation}\label{App-Eq:DivYukawaCoeffs}
    \begin{aligned}
        \prescript{}{Y}{\mathcal{A}}_{\psi\overline{\psi}\phi,\mathrm{R},ji}^{1,\mathrm{inv},a} &=
        \big(G^b{K^a}^{\dagger}K^b\big)_{ji}+\big(K^b{K^a}^{\dagger}G^b\big)_{ji},\\
        \prescript{}{YF}{\mathcal{A}}_{\psi\overline{\psi}\phi,\mathrm{R},ji}^{1,\mathrm{inv},a} &=
        -(3+\xi)\big(\mathcal{Y}_{L}G^a\mathcal{Y}_{R}\big)_{ji},\\
        \prescript{}{YSF}{\mathcal{A}}_{\psi\overline{\psi}\phi,\mathrm{R},ji}^{1,\mathrm{inv},a} &=
        \xi\,\mathcal{Y}_{S}\Big[\big(G^a\mathcal{Y}_{R}\big)_{ji}-\big(\mathcal{Y}_{L}G^a\big)_{ji}\Big],\\
        \prescript{}{Y}{\mathcal{A}}_{\psi\overline{\psi}\phi,\mathrm{L},ji}^{1,\mathrm{inv},a} &=
        \big({G^b}^{\dagger}G^a{K^b}^{\dagger}\big)_{ji}+\big({K^b}^{\dagger}G^a{G^b}^{\dagger}\big)_{ji},\\
        \prescript{}{YF}{\mathcal{A}}_{\psi\overline{\psi}\phi,\mathrm{L},ji}^{1,\mathrm{inv},a} &=
        -(3+\xi)\big(\mathcal{Y}_{R}{K^a}^{\dagger}\mathcal{Y}_{L}\big)_{ji},\\
        \prescript{}{YSF}{\mathcal{A}}_{\psi\overline{\psi}\phi,\mathrm{L},ji}^{1,\mathrm{inv},a} &=
        \xi\,\mathcal{Y}_{S}\Big[\big({K^a}^{\dagger}\mathcal{Y}_{L}\big)_{ji}-\big(\mathcal{Y}_{R}{K^a}^{\dagger}\big)_{ji}\Big].
    \end{aligned}
\end{equation}

\paragraph{Scalar Coefficients:}
For the bilinear scalar terms we find
\begin{equation}\label{App-Eq:Div2ScalarCoeffs}
    \begin{aligned}
        \prescript{}{S}{\mathcal{A}}_{\phi\phi^{\dagger},ab}^{1,\mathrm{inv}} &=
        -(3-\xi)\mathcal{Y}_{S}^2 \, \delta_{ab},\\
        \prescript{}{Y}{\overline{\mathcal{A}}}_{\phi\phi^{\dagger},ab}^{1,\mathrm{inv}} &=
        \mathrm{Tr}\big({G^a}^{\dagger}G^b\big)+\mathrm{Tr}\big({K^b}^{\dagger}K^a\big),\\
        {\widehat{\mathcal{A}}}_{\phi\phi^{\dagger},ab}^{1,\mathrm{break}} &=
        \frac{2}{3}\Big(\mathrm{Tr}\big({G^a}^{\dagger}G^b\big)+\mathrm{Tr}\big({K^b}^{\dagger}K^a\big)\Big)
        + \frac{1}{3}\Big(\mathrm{Tr}\big({G^a}^{\dagger}{K^b}^{\dagger}\big)+\mathrm{Tr}\big(K^aG^b\big)\Big),\\
        {\widehat{\mathcal{A}}}_{\phi\phi,ab}^{1,\mathrm{break}} &=
        \frac{1}{3}\Big(\mathrm{Tr}\big(G^aG^b\big)+\mathrm{Tr}\big({K^a}^{\dagger}{K^b}^{\dagger}\big)\Big),
    \end{aligned}
\end{equation}
while for quartic scalar contributions we obtain
\begin{equation}\label{App-Eq:DivSSSScoeffs}
    \begin{aligned}
        \prescript{}{S}{\mathcal{A}}_{\phi\phi^{\dagger}\phi\phi^{\dagger},abcd}^{1,\mathrm{inv}} &=
        6 \, \mathcal{Y}_{S}^4 \, \big(\delta_{ad}\,\delta_{bc}+\delta_{ac}\,\delta_{bd}\big),\\
        \prescript{}{S\lambda}{\mathcal{A}}_{\phi\phi^{\dagger}\phi\phi^{\dagger},abcd}^{1,\mathrm{inv}} &=
        \frac{2\xi}{3} \, \mathcal{Y}_{S}^2 \, \big( \lambda_{abcd} - \lambda_{adcb} - \lambda_{cbad} - \lambda_{cdab} \big),\\
        \prescript{}{Y}{\mathcal{A}}_{\phi\phi^{\dagger}\phi\phi^{\dagger},cadb}^{1,\mathrm{inv}} &=
        -2 \Big[ 
        \mathrm{Tr}\big(G^a{G^c}^{\dagger}G^b{G^d}^{\dagger}\big)
        + \mathrm{Tr}\big(G^a{G^d}^{\dagger}G^b{G^c}^{\dagger}\big)
        + \mathrm{Tr}\big(G^a{G^c}^{\dagger}K^d{K^b}^{\dagger}\big)\\
        &+ \mathrm{Tr}\big(G^a{G^d}^{\dagger}K^c{K^b}^{\dagger}\big)
        + \mathrm{Tr}\big(G^b{G^c}^{\dagger}K^d{K^a}^{\dagger}\big)
        + \mathrm{Tr}\big(G^b{G^d}^{\dagger}K^c{K^a}^{\dagger}\big)\\
        &+ \mathrm{Tr}\big(G^a{K^b}^{\dagger}K^c{G^d}^{\dagger}\big)
        + \mathrm{Tr}\big(G^a{K^b}^{\dagger}K^d{G^c}^{\dagger}\big)
        + \mathrm{Tr}\big(G^b{K^a}^{\dagger}K^c{G^d}^{\dagger}\big)\\
        &+ \mathrm{Tr}\big(G^b{K^a}^{\dagger}K^d{G^c}^{\dagger}\big)
        + \mathrm{Tr}\big(K^c{K^a}^{\dagger}K^d{K^b}^{\dagger}\big)
        + \mathrm{Tr}\big(K^d{K^a}^{\dagger}K^c{K^b}^{\dagger}\big)
        \Big],\\
        \prescript{}{\lambda}{\mathcal{A}}_{\phi\phi^{\dagger}\phi\phi^{\dagger},cadb}^{1,\mathrm{inv}} &=
        \frac{2}{9} \big( 2\,\lambda_{calk}\,\lambda_{dbkl} + 2\,\lambda_{dalk}\,\lambda_{cbkl} + \lambda_{kalb}\,\lambda_{ckdl} \big).
    \end{aligned}
\end{equation}

\paragraph{Scalar-Gauge Boson Coefficients:}
Coefficients for single gauge boson interactions read
\begin{equation}\label{App-Eq:SSBCoeff}
    \begin{aligned}
        {\widehat{\mathcal{A}}}&_{\phi\phi^{\dagger}B,ab}^{1,\mathrm{break}}\\
        = \, &\frac{1}{3}
        \Big[
        \mathrm{Tr}\big(\mathcal{Y}_{LR}{K^b}^{\dagger}{G^a}^{\dagger}\big)
        - \mathrm{Tr}\big(\mathcal{Y}_{LR}{G^a}^{\dagger}{K^b}^{\dagger}\big)
        + \mathrm{Tr}\big(\mathcal{Y}_{LR}G^bK^a\big)
        -\mathrm{Tr}\big(\mathcal{Y}_{LR}K^aG^b\big)
        \Big]\\
        + \, &\frac{2}{3}
        \Big[
        \mathrm{Tr}\big(\mathcal{Y}_{LR}G^b{G^a}^{\dagger}\big)
        - \mathrm{Tr}\big(\mathcal{Y}_{LR}{G^a}^{\dagger}G^b\big)
        + \mathrm{Tr}\big(\mathcal{Y}_{LR}{K^b}^{\dagger}K^a\big)
        -\mathrm{Tr}\big(\mathcal{Y}_{LR}K^a{K^b}^{\dagger}\big)
        \Big],\\
        {\widehat{\mathcal{A}}}&_{\phi\phi B,ab}^{1,\mathrm{break}}\\
        = \, &\frac{1}{3}
        \Big[
        \mathrm{Tr}\big(\mathcal{Y}_{LR}{K^b}^{\dagger}{K^a}^{\dagger}\big)
        - \mathrm{Tr}\big(\mathcal{Y}_{LR}{K^a}^{\dagger}{K^b}^{\dagger}\big)
        + \mathrm{Tr}\big(\mathcal{Y}_{LR}G^bG^a\big)
        -\mathrm{Tr}\big(\mathcal{Y}_{LR}G^aG^b\big)
        \Big],
    \end{aligned}
\end{equation}
while such with two gauge bosons are given by
\begin{equation}\label{App-Eq:SSBBCoeffs}
    \begin{aligned}
        {\widehat{\mathcal{A}}}&_{\phi\phi^{\dagger}BB,ab}^{1,\mathrm{break}}\\ 
        &= 
        \frac{2}{3} \Big[
        \mathrm{Tr}\Big(\mathcal{Y}_{LR}^2\big\{{G^{a}}^{\dagger},{K^{b}}^{\dagger}\big\}\Big)
        + \mathrm{Tr}\Big(\mathcal{Y}_{LR}^2\big\{G^{b},K^{a}\big\}\Big)
        + 2 \, \mathrm{Tr}\Big(\mathcal{Y}_{LR}^2\big\{{G^{a}}^{\dagger},G^{b}\big\}\Big)\\
        &+ 2 \, \mathrm{Tr}\Big(\mathcal{Y}_{LR}^2\big\{{K^{b}}^{\dagger},K^{a}\big\}\Big)
        - 2 \, \mathrm{Tr}\Big(\mathcal{Y}_{LR}{G^{a}}^{\dagger}\mathcal{Y}_{LR}{K^{b}}^{\dagger}\Big)
        - 2 \, \mathrm{Tr}\Big(\mathcal{Y}_{LR}G^{b}\mathcal{Y}_{LR}K^{a}\Big)\\
        &- 4 \, \mathrm{Tr}\Big(\mathcal{Y}_{LR}{G^{a}}^{\dagger}\mathcal{Y}_{LR}G^{b}\Big)
        - 4 \, \mathrm{Tr}\Big(\mathcal{Y}_{LR}{K^{b}}^{\dagger}\mathcal{Y}_{LR}K^{a}\Big)
        \Big],\\
       {\widehat{\mathcal{A}}}&_{\phi\phi BB,ab}^{1,\mathrm{break}} = 
        \frac{2}{3} \Big[
        \mathrm{Tr}\Big(\mathcal{Y}_{LR}^2\big\{G^{a},G^{b}\big\}\Big)
        + \mathrm{Tr}\Big(\mathcal{Y}_{LR}^2\big\{{K^{a}}^{\dagger},{K^{b}}^{\dagger}\big\}\Big)\\
        &\qquad\quad\, - 2\,\mathrm{Tr}\Big(\mathcal{Y}_{LR}G^{a}\mathcal{Y}_{LR}G^{b}\Big)
        - 2\,\mathrm{Tr}\Big(\mathcal{Y}_{LR}{K^{a}}^{\dagger}\mathcal{Y}_{LR}{K^{b}}^{\dagger}\Big)
        \Big],
    \end{aligned}
\end{equation}
where anticommutators are used for compactness.

\subsection{Coefficients of Finite Contributions}\label{App:Finite1LoopCTCoeffs-GeneralAbelianChiralGaugeTheory}

We next provide the coefficients for the finite contributions.
To organise the results for finite fermion--gauge boson contributions, we introduce
\begin{equation}\label{App-Eq:FiniteFermionGaugeBosonCoeffsFull}
    \begin{aligned}
        \mathcal{F}_{\psi\overline{\psi}B,\mathrm{R},ji}^{1,\mathrm{break}} &= 
        g^2 \Big(\mathcal{Y}_{R} \prescript{}{F}{\mathcal{F}}_{\psi\overline{\psi},\mathrm{R},ji}^{1,\mathrm{break}} 
        + \prescript{}{F}{\mathcal{F}}_{\psi\overline{\psi}B,ji}^{1,\mathrm{break}}\Big)
        + \prescript{}{YF}{\mathcal{F}}_{\psi\overline{\psi}B,\mathrm{R},ji}^{1,\mathrm{break}},\\
        \mathcal{F}_{\psi\overline{\psi}B,\mathrm{L},ji}^{1,\mathrm{break}} &= 
        g^2 \Big(\mathcal{Y}_{L} \prescript{}{F}{\mathcal{F}}_{\psi\overline{\psi},\mathrm{L},ji}^{1,\mathrm{break}} 
        - \prescript{}{F}{\mathcal{F}}_{\psi\overline{\psi}B,ji}^{1,\mathrm{break}}\Big)
        + \prescript{}{YF}{\mathcal{F}}_{\psi\overline{\psi}B,\mathrm{L},ji}^{1,\mathrm{break}}.
    \end{aligned}
\end{equation}

\paragraph{Gauge Boson Coefficients:}
The coefficients for bilinear and quartic finite gauge boson contributions are
\begin{equation}\label{App-Eq:FiniteGaugeBosonCoeffs}
    \begin{aligned}
        \mathcal{F}_{BB}^{1,\mathrm{break}} &= 
        - \frac{1}{3} \, \mathrm{Tr}\Big(\big(\mathcal{Y}_R-\mathcal{Y}_L\big)^2\Big),\\
        \mathcal{F}_{BBBB}^{1,\mathrm{break}} &= 
        \frac{2}{3} \, \mathrm{Tr}\Big(\big(\mathcal{Y}_R-\mathcal{Y}_L\big)^4\Big).
    \end{aligned}
\end{equation}

\paragraph{Fermion Coefficients:}
Finite fermion contributions are governed by
\begin{equation}
    \begin{aligned}
        \prescript{}{F}{\mathcal{F}}_{\psi\overline{\psi},\mathrm{R},ji}^{1,\mathrm{break}} &=
        \big(\mathcal{Y}_R-\mathcal{Y}_L\big)_{jk}\Big(\frac{5+\xi}{6}\mathcal{Y}_R+\frac{1-\xi}{3}\mathcal{Y}_{LR}\Big)_{ki},\\
        \prescript{}{F}{\mathcal{F}}_{\psi\overline{\psi},\mathrm{L},ji}^{1,\mathrm{break}} &=
        -\big(\mathcal{Y}_R-\mathcal{Y}_L\big)_{jk}\Big(\frac{5+\xi}{6}\mathcal{Y}_L+\frac{1-\xi}{3}\mathcal{Y}_{LR}\Big)_{ki}.
    \end{aligned}
\end{equation}

\paragraph{Fermion-Gauge Boson Coefficients:}
Besides the standard gauge boson exchange vertex correction with couplings $\hypL$ and $\hypR$, the finite fermion--gauge boson counterterms are determined by the coefficients
\begin{equation}
    \begin{aligned}
        \prescript{}{F}{\mathcal{F}}_{\psi\overline{\psi}B,ji}^{1,\mathrm{break}} &=
        \frac{5+\xi}{6}\big(\mathcal{Y}_R-\mathcal{Y}_L\big)_{jk}\big(\mathcal{Y}_{LR}^2\big)_{ki},\\
        \prescript{}{YF}{\mathcal{F}}_{\psi\overline{\psi}B,\mathrm{R},ji}^{1,\mathrm{break}} &=
        -\frac{1}{2}\Big[\big({G^a}^{\dagger}\big(\mathcal{Y}_{R}-\mathcal{Y}_{L}\big)G^a\big)_{ji}
        +\big({K^a}^{\dagger}\big(\mathcal{Y}_{R}-\mathcal{Y}_{L}\big)K^a\big)_{ji}\Big],\\
        \prescript{}{YF}{\mathcal{F}}_{\psi\overline{\psi}B,\mathrm{L},ji}^{1,\mathrm{break}} &=
        \frac{1}{2}\Big[\big(G^a\big(\mathcal{Y}_{R}-\mathcal{Y}_{L}\big){G^a}^{\dagger}\big)_{ji}
        +\big(K^a\big(\mathcal{Y}_{R}-\mathcal{Y}_{L}\big){K^a}^{\dagger}\big)_{ji}\Big].
    \end{aligned}
\end{equation}

\paragraph{Yukawa Coefficients:}
Finite Yukawa counterterms are governed by
\begin{equation}\label{App-Eq:FiniteYukawaCoeffs}
\begin{alignedat}{2}
        \mathcal{F}_{\psi\overline{\psi}\phi,ji}^{1,\mathrm{break},a} 
        &=
        &&- \frac{1}{2}\big(G^bG^aK^b+K^bG^aG^b\big)_{ji} 
        + \frac{1-\xi}{3} g^2 \big(\mathcal{Y}_{L}{K^{a}}^{\dagger}\mathcal{Y}_{LR}+\mathcal{Y}_{LR}{K^{a}}^{\dagger}\mathcal{Y}_{R}\big)_{ji}\\
        &
        &&+ \frac{5+\xi}{3} g^2 \big(\mathcal{Y}_{L}{K^{a}}^{\dagger}\mathcal{Y}_{R}+\mathcal{Y}_{LR}{K^{a}}^{\dagger}\mathcal{Y}_{LR}\big)_{ji},\\
        \mathcal{F}_{\psi\overline{\psi}\phi^{\dagger},ji}^{1,\mathrm{break},a}
        &= 
        &&- \frac{1}{2}\big(K^bK^aG^b+G^bK^aK^b\big)_{ji} 
        + \frac{1-\xi}{3} g^2 \big(\mathcal{Y}_{L}{G^{a}}^{\dagger}\mathcal{Y}_{LR}+\mathcal{Y}_{LR}{G^{a}}^{\dagger}\mathcal{Y}_{R}\big)_{ji}\\
        &
        &&+ \frac{5+\xi}{3} g^2 \big(\mathcal{Y}_{L}{G^{a}}^{\dagger}\mathcal{Y}_{R}+\mathcal{Y}_{LR}{G^{a}}^{\dagger}\mathcal{Y}_{LR}\big)_{ji}.
\end{alignedat}
\end{equation}

\paragraph{Scalar Coefficients:}
For the coefficients of the finite scalar counterterms, contributing to global hypercharge violation, we find
\begin{equation}\label{App-Eq:Finite2ScalarCoeff}
    \begin{aligned}
        \mathcal{F}_{\phi\phi,ab}^{1,\mathrm{break}} &= 
        -\frac{1}{3}\Big(\mathrm{Tr}\big(G^aG^b\big)+\mathrm{Tr}\big({K^a}^{\dagger}{K^b}^{\dagger}\big)\Big),
    \end{aligned}
\end{equation}
and
\begin{align}
    \begin{split}\label{App-Eq:Finite4ScalarCoeff-SSSS}
        \mathcal{F}_{\phi\phi\phi\phi,abcd}^{1,\mathrm{break}} &=
        \bigg\{\frac{1}{3}\mathrm{Tr}\big(G^{a}G^{b}G^{c}G^{d}\big)
        +\frac{2}{3}\Big[\mathrm{Tr}\big(G^{a}G^{b}G^{c}{K^{d}}^{\dagger}\big)+\mathrm{Tr}\big(G^{a}G^{b}{K^{c}}^{\dagger}G^{d}\big)\\
        &+\mathrm{Tr}\big(G^{a}{K^{b}}^{\dagger}G^{c}G^{d}\big)+\mathrm{Tr}\big({K^{a}}^{\dagger}G^{b}G^{c}G^{d}\big)\Big]\\
        &-\frac{2}{3}\Big[
        \mathrm{Tr}\big(G^{a}G^{b}{K^{c}}^{\dagger}{K^{d}}^{\dagger}\big)+\mathrm{Tr}\big(G^{a}{K^{b}}^{\dagger}G^{c}{K^{d}}^{\dagger}\big)
        +\mathrm{Tr}\big(G^{a}{K^{b}}^{\dagger}{K^{c}}^{\dagger}G^{d}\big)\\
        &+\mathrm{Tr}\big({K^{a}}^{\dagger}G^{b}G^{c}{K^{d}}^{\dagger}\big)+\mathrm{Tr}\big({K^{a}}^{\dagger}G^{b}{K^{c}}^{\dagger}G^{d}\big)
        +\mathrm{Tr}\big({K^{a}}^{\dagger}{K^{b}}^{\dagger}G^{c}G^{d}\big)\Big]\\
        &+\frac{2}{3}\Big[\mathrm{Tr}\big(G^{a}{K^{b}}^{\dagger}{K^{c}}^{\dagger}{K^{d}}^{\dagger}\big)
        +\mathrm{Tr}\big({K^{a}}^{\dagger}G^{b}{K^{c}}^{\dagger}{K^{d}}^{\dagger}\big)
        +\mathrm{Tr}\big({K^{a}}^{\dagger}{K^{b}}^{\dagger}G^{c}{K^{d}}^{\dagger}\big)\\
        &+\mathrm{Tr}\big({K^{a}}^{\dagger}{K^{b}}^{\dagger}{K^{c}}^{\dagger}G^{d}\big)\Big]
        + \frac{1}{3}\mathrm{Tr}\big({K^{a}}^{\dagger}{K^{b}}^{\dagger}{K^{c}}^{\dagger}{K^{d}}^{\dagger}\big)\bigg\}\\
        &+ \big(\text{$23$ permutations of the indices $(a,b,c,d)$}\big),
    \end{split}\\[1.5ex]
    \begin{split}\label{App-Eq:Finite4ScalarCoeff-SdaggerSSS}
        \mathcal{F}_{\phi^{\dagger}\phi\phi\phi,abcd}^{1,\mathrm{break}} &=
        \bigg\{\frac{4}{3}\Big[
        \mathrm{Tr}\big({G^{a}}^{\dagger}G^{b}G^{c}G^{d}\big)
        +\mathrm{Tr}\big({G^{a}}^{\dagger}G^{b}{K^{c}}^{\dagger}{K^{d}}^{\dagger}\big)
        +\mathrm{Tr}\big({G^{a}}^{\dagger}{K^{b}}^{\dagger}G^{c}{K^{d}}^{\dagger}\big)\\
        &+\mathrm{Tr}\big({G^{a}}^{\dagger}{K^{b}}^{\dagger}{K^{c}}^{\dagger}G^{d}\big)
        +\mathrm{Tr}\big(K^{a}G^{b}G^{c}{K^{d}}^{\dagger}\big)
        +\mathrm{Tr}\big(K^{a}G^{b}{K^{c}}^{\dagger}G^{d}\big)\\
        &+\mathrm{Tr}\big(K^{a}{K^{b}}^{\dagger}G^{c}G^{d}\big)
        +\mathrm{Tr}\big(K^{a}{K^{b}}^{\dagger}{K^{c}}^{\dagger}{K^{d}}^{\dagger}\big)
        -\mathrm{Tr}\big({G^{a}}^{\dagger}G^{b}G^{c}{K^{d}}^{\dagger}\big)\\
        &-\mathrm{Tr}\big({G^{a}}^{\dagger}G^{b}{K^{c}}^{\dagger}G^{d}\big)
        -\mathrm{Tr}\big({G^{a}}^{\dagger}{K^{b}}^{\dagger}G^{c}G^{d}\big)
        -\mathrm{Tr}\big(K^{a}G^{b}{K^{c}}^{\dagger}{K^{d}}^{\dagger}\big)\\
        &-\mathrm{Tr}\big(K^{a}{K^{b}}^{\dagger}G^{c}{K^{d}}^{\dagger}\big)
        -\mathrm{Tr}\big(K^{a}{K^{b}}^{\dagger}{K^{c}}^{\dagger}G^{d}\big)
        \Big]\\
        &+\frac{2}{3}\Big[
        \mathrm{Tr}\big({G^{a}}^{\dagger}{K^{b}}^{\dagger}{K^{c}}^{\dagger}{K^{d}}^{\dagger}\big)
        +\mathrm{Tr}\big(K^{a}G^{b}G^{c}G^{d}\big)\Big]\bigg\}\\
        &+ \big(\text{$5$ permutations of the indices $(b,c,d)$}\big).
    \end{split}
\end{align}

\paragraph{Scalar-Gauge Boson Coefficients:}
Finally, the finite scalar--gauge boson counterterms come with the coefficients
\begin{equation}\label{App-Eq:FiniteScalarGaugeBosonCoeffs}
    \begin{alignedat}{2}
        \mathcal{F}_{\phi\phi^{\dagger}B,ab}^{1,\mathrm{break}} &= 
        &&-\frac{1}{3}\Big[ 
        \mathrm{Tr}\Big(\mathcal{Y}_{L}\big[{G^a}^{\dagger}G^b-{K^b}^{\dagger}K^a\big]\Big)
        + \mathrm{Tr}\Big(\mathcal{Y}_{R}\big[K^a{K^b}^{\dagger}-G^b{G^a}^{\dagger}\big]\Big)\\
        &
        &&+\mathcal{Y}_{S} \Big( \mathrm{Tr}\big({G^a}^{\dagger}G^{b}\big) + \mathrm{Tr}\big({K^{b}}^{\dagger}K^{a}\big) \Big)
        \Big],\\
        \mathcal{F}_{\phi\phi^{\dagger}BB,ab}^{1,\mathrm{break}} &= 
        &&-\frac{2}{3}\Big[
        \mathrm{Tr}\Big(\mathcal{Y}_{L}^2\big[{G^a}^{\dagger}G^b+{K^b}^{\dagger}K^a\big]\Big)
        + 3 \, \mathrm{Tr}\Big(\mathcal{Y}_{L}^2\big[G^b{G^a}^{\dagger}+K^a{K^b}^{\dagger}\big]\Big)\\
        &
        &&- 4 \, \mathrm{Tr}\Big(\mathcal{Y}_{R}\mathcal{Y}_{L}\big[{G^a}^{\dagger}G^b+G^b{G^a}^{\dagger}+{K^b}^{\dagger}K^a+K^a{K^b}^{\dagger}\big]\Big)\\
        &
        &&+ 3 \, \mathrm{Tr}\Big(\mathcal{Y}_{R}^2\big[{G^a}^{\dagger}G^b+{K^b}^{\dagger}K^a\big]\Big)
        + \mathrm{Tr}\Big(\mathcal{Y}_{R}^2\big[G^b{G^a}^{\dagger}+K^a{K^b}^{\dagger}\big]\Big)\\
        &
        &&+ \mathrm{Tr}\Big(\big[\mathcal{Y}_{R}+\mathcal{Y}_{L}\big]{G^a}^{\dagger}\big[\mathcal{Y}_{R}+\mathcal{Y}_{L}\big]G^{b}\Big)
        - 4 \, \mathrm{Tr}\Big(\mathcal{Y}_{R}{G^a}^{\dagger}\mathcal{Y}_{L}G^b\Big)\\
        &
        &&+ \mathrm{Tr}\Big(\big[\mathcal{Y}_{R}+\mathcal{Y}_{L}\big]{K^b}^{\dagger}\big[\mathcal{Y}_{R}+\mathcal{Y}_{L}\big]K^a\Big)
        - 4 \, \mathrm{Tr}\Big(\mathcal{Y}_{R}{K^b}^{\dagger}\mathcal{Y}_{L}K^a\Big)
        \Big].
    \end{alignedat}
\end{equation}

\section{Green Functions at the 1-Loop Level}\label{App:1LoopGreenFunctions-GeneralAbelianChiralGaugeTheory}

In this section, we present the 1-loop results for all 1PI Green functions relevant for the renormalisation of the Abelian chiral gauge theory defined in Sec.~\ref{Sec:Definition_of_the_Theory_General_Abelian_Case} and discussed throughout chapter~\ref{Chap:General_Abelian_Chiral_Gauge_Theory}.
The UV divergent part of the ordinary, power-counting divergent 1PI Green functions are listed in App.~\ref{App:1LoopStandardGreenFunctions-GeneralAbelianChiralGaugeTheory}, while the BRST-breaking contributions --- obtained from the operator-inserted 1PI Green functions --- are collected in App.~\ref{App:1LoopBreakingGreenFunctions-GeneralAbelianChiralGaugeTheory}.
All computations have been carried out as described in Sec.~\ref{Sec:One-Loop-Renormalisation-Abelian-Chiral-Gauge-Theory}, with all external momenta defined as incoming.

\subsection{Ordinary 1PI Green Functions --- 1-Loop Divergences}\label{App:1LoopStandardGreenFunctions-GeneralAbelianChiralGaugeTheory}

We begin with all relevant ordinary, power-counting divergent 1PI Green functions, excluding the non-contributing ones summarised in Tab.~\ref{Tab:NonContributingOrdinaryGreenFunctions}.
From these, we derive the singular counterterms discussed in Sec.~\ref{Sec:S_sct-GeneralAbelianChiralGaugeTheory}.
For clarity, the results are organised according to the number of external legs, i.e.\ into 2-, 3- and 4-point Green functions.

\subsubsection{2-Point Functions}

Five non-vanishing $2$-point functions remain --- excluding the two in Tab.~\ref{Tab:NonContributingOrdinaryGreenFunctions} --- grouped into gauge boson, fermion, and scalar contributions.

\paragraph{(i) Gauge Boson Self Energy:}
\begin{equation}\label{Eq:GaugeBosonSelfEnergy_1-Loop-GeneralAbelianTheory}
    \begin{aligned}
        i \Gamma&_{B_{\mu}(p)B_{\nu}(-p)}\big|_{\mathrm{div}}^{(1)}\\
        &= 
        \frac{ig^2}{16 \pi^2} 
        \prescript{}{S}{\mathcal{A}}_{BB}^{1, \mathrm{inv}} \, \frac{1}{\epsilon} \,
        \Big(p^{\mu} p^{\nu} - p^2 \eta^{\mu\nu}\Big) \, 
        + \frac{ig^2}{16 \pi^2} 
        \prescript{}{F}{\overline{\mathcal{A}}}_{BB}^{1, \mathrm{inv}} \, \frac{1}{\epsilon} \,
        \Big(\overline{p}^{\mu} \overline{p}^{\nu} - \overline{p}^2 \overline{\eta}^{\mu\nu}\Big)\\
        &+ \frac{ig^2}{16 \pi^2} \, \frac{1}{\epsilon} \bigg\{
        \Big( 
        \prescript{}{F}{\widehat{\mathcal{A}}}_{BB,1}^{1, \mathrm{break}}
        + \prescript{}{F}{\widehat{\mathcal{A}}}_{BB,3}^{1, \mathrm{break}}
        \Big) \,
        \overline{p}^2 \widehat{\eta}^{\mu\nu}
        + \prescript{}{F}{\widehat{\mathcal{A}}}_{BB,4}^{1, \mathrm{break}} \,
        \widehat{p}^2 \overline{\eta}^{\mu\nu}\\
        &\qquad\qquad + \frac{2}{3}
        \prescript{}{F}{\widehat{\mathcal{A}}}_{BB,3}^{1, \mathrm{break}} \,
        \widehat{p}^2 \widehat{\eta}^{\mu\nu}
        +
        \prescript{}{F}{\widehat{\mathcal{A}}}_{BB,2}^{1, \mathrm{break}} \,
        \big(\overline{p}^{\mu} \widehat{p}^{\nu} + \widehat{p}^{\mu} \overline{p}^{\nu} \big) 
        + 2 \prescript{}{F}{\widehat{\mathcal{A}}}_{BB,1}^{1, \mathrm{break}} \,
        \widehat{p}^{\mu} \widehat{p}^{\nu}
        \bigg\}.
    \end{aligned}
\end{equation}
The first line of Eq.~\eqref{Eq:GaugeBosonSelfEnergy_1-Loop-GeneralAbelianTheory} represents the transversal part of the gauge boson self energy, while the remaining terms encode the divergent and BRST-breaking parts.
As shown in Refs.~\cite{Belusca-Maito:2021lnk,Stockinger:2023ndm,vonManteuffel:2025swv} for chiral QED, the fermionic contribution to the transversal part (second term in the first line) is purely 4-dimensional due to the chiral projectors, whereas the scalar contribution (first term) is fully $D$-dimensional.
Among the divergent BRST-breaking terms, only $\prescript{}{F}{\widehat{\mathcal{A}}}_{BB,4}^{1, \mathrm{break}} \, \widehat{p}^2 \overline{\eta}^{\mu\nu}$ (see Eqs.~\eqref{App-Eq:GaugeBosonCoeffsFermionic} and \eqref{App-Eq:GaugeBosonCoeffsScalar}) is independent of the evanescent hypercharges $\hypLR$; all others stem from the evanescent interactions.
These additional terms generate Lorentz structures absent in the 1-loop results of chiral QED and vanish for $\hypLR \equiv 0$.

\paragraph{(ii) Fermion Self Energy:}
\begin{equation}\label{Eq:FermionSelfEnergy_1-Loop-GeneralAbelianTheory}
    \begin{aligned}
        i \Gamma_{\psi_i(p)\overline{\psi}_j(-p)} \big|_{\mathrm{div}}^{(1)}
        &= \frac{i}{16 \pi^2} \, \frac{1}{\epsilon} \, \overline{\slashed{p}} \, \bigg[
        \overline{\mathcal{A}}_{\psi\overline{\psi},\mathrm{R},ji}^{1,\mathrm{inv}}
        \mathbb{P}_{\mathrm{R}} 
        +
        \overline{\mathcal{A}}_{\psi\overline{\psi},\mathrm{L},ji}^{1,\mathrm{inv}}
        \mathbb{P}_{\mathrm{L}}
        \bigg]\\
        &+ \frac{i}{16 \pi^2} \, \frac{1}{\epsilon} \, \widehat{\slashed{p}} \, \bigg[
        \widehat{\mathcal{A}}_{\psi\overline{\psi},ji}^{1,\mathrm{break}}
        \mathbb{P}_{\mathrm{R}} 
        + 
        {\widehat{\mathcal{A}}{}_{\psi\overline{\psi},ij}^{1,\mathrm{break}}}^{\dagger}
        \mathbb{P}_{\mathrm{L}}
        \Big].
    \end{aligned}
\end{equation}
Both the 4-dimensional and evanescent terms receive left- and right-handed contributions.
Each coefficient contains fermionic and Yukawa parts (see Eq.~\eqref{App-Eq:DivCoeffRelations}, as well as Eqs.~\eqref{App-Eq:FermionCoeffFermionGauge} and \eqref{App-Eq:FermionCoeffYukawa}).
Unlike in previous cases, none of the coefficients vanish for $\hypLR\equiv0$; the evanescent interactions therefore leave the Lorentz structure unchanged.

\paragraph{(iii) Scalar Self Energy:}
\begin{align}
    \begin{split}\label{Eq:ScalarSelfEnergy_1-Loop-GeneralAbelianTheory}
        i \Gamma_{\phi_a(p)\phi^{\dagger}_b(-p)} \big|_{\mathrm{div}}^{(1)} &= \frac{i}{16 \pi^2} \, \frac{1}{\epsilon} \, \Big(
        g^2
        \prescript{}{S}{\mathcal{A}}_{\phi\phi^{\dagger},ba}^{1,\mathrm{inv}}
        + \prescript{}{Y}{\overline{\mathcal{A}}}_{\phi\phi^{\dagger},ba}^{1,\mathrm{inv}}
        \Big) \, 
        \overline{p}^2\\ 
        &+ \frac{i}{16 \pi^2} \, \frac{1}{\epsilon} \, \Big(
        g^2
        \prescript{}{S}{\mathcal{A}}_{\phi\phi^{\dagger},ba}^{1,\mathrm{inv}}
        + {\widehat{\mathcal{A}}}_{\phi\phi^{\dagger},ba}^{1, \mathrm{break}}
        \Big) \, \widehat{p}^2,
    \end{split}\\[1.5ex]
    \begin{split}\label{Eq:phiphiSelfEnergy_1-Loop-GeneralAbelianTheory}
        i \Gamma_{\phi_a(p)\phi_b(-p)} \big|_{\mathrm{div}}^{(1)} &= 
        \frac{i}{16 \pi^2} \, \frac{1}{\epsilon} \,
        {\widehat{\mathcal{A}}}_{\phi\phi,ba}^{1,\mathrm{break}}\,
        \widehat{p}^2,
    \end{split}\\[1.5ex]
    \begin{split}\label{Eq:phidaggerphidaggerSelfEnergy_1-Loop-GeneralAbelianTheory}
        i \Gamma_{\phi^{\dagger}_a(p)\phi^{\dagger}_b(-p)} \big|_{\mathrm{div}}^{(1)} &= 
        \frac{i}{16 \pi^2} \, \frac{1}{\epsilon} \,
        {\widehat{\mathcal{A}}{}_{\phi\phi,ba}^{1,\mathrm{break}}}^{\dagger}\,
        \widehat{p}^2.
    \end{split}
\end{align}
The scalar sector receives contributions from the self energy in Eq.~\eqref{Eq:ScalarSelfEnergy_1-Loop-GeneralAbelianTheory} and from the hypercharge-violating Green functions in Eqs.~\eqref{Eq:phiphiSelfEnergy_1-Loop-GeneralAbelianTheory} and \eqref{Eq:phidaggerphidaggerSelfEnergy_1-Loop-GeneralAbelianTheory}.
These violations originate from the evanescent fermion kinetic term and the evanescent gauge interactions (see Sec.~\ref{Sec:Dimensional_Ambiguities_and_Evanescent_Shadows}).
The two hypercharge-violating Green functions depend only on the former (independent of $\hypLR$) and are purely evanescent.
In contrast, the standard scalar self energy includes both $4$- and $(D-4)$-dimensional parts: the purely scalar terms $\propto\prescript{}{S}{\mathcal{A}}_{\phi\phi^{\dagger},ba}^{1,\mathrm{inv}}$ are BRST-invariant and together form a fully $D$-dimensional contribution, whereas the Yukawa-induced terms differ between the $4$- and $(D-4)$-dimensional parts, thereby generating a BRST-breaking contribution.

\subsubsection{3-Point Functions}

For the 3-point sector, seven power-counting divergent Green functions remain after excluding the scalar--gauge and triple-scalar interactions listed in Tab.~\ref{Tab:NonContributingOrdinaryGreenFunctions}.

\paragraph{(iv) Fermion-Gauge Boson Interaction:}
\begin{equation}\label{Eq:FFB_1-Loop-GeneralAbelianTheory}
    \begin{aligned}
        i \Gamma_{\psi_i(p_2)\overline{\psi}_j(p_1)B_{\mu}(q)} \big|_{\mathrm{div}}^{(1)} =
        &-\frac{ig}{16\pi^2} \, \frac{1}{\epsilon} \, \overline{\gamma}^{\mu} 
        \Big[
        \Big( \mathcal{Y}_{R}
        \overline{\mathcal{A}}_{\psi\overline{\psi},\mathrm{R}}^{1, \mathrm{inv}}
        \Big)_{ji}
        \mathbb{P}_{\mathrm{R}}
        +
        \Big( \mathcal{Y}_{L}
        \overline{\mathcal{A}}_{\psi\overline{\psi},\mathrm{L}}^{1, \mathrm{inv}}
        \Big)_{ji} 
        \mathbb{P}_{\mathrm{L}}
        \Big]\\
        &-\frac{ig}{16 \pi^2} \, \frac{1}{\epsilon} \, \widehat{\gamma}^{\mu} 
        \Big[
        \widehat{\mathcal{A}}_{\psi\overline{\psi}B,ji}^{1,\mathrm{break}}\,
        \mathbb{P}_{\mathrm{R}}
        +
        {\widehat{\mathcal{A}}{}_{\psi\overline{\psi}B,ij}^{1,\mathrm{break}}}^{\dagger}\,
        \mathbb{P}_{\mathrm{L}}
        \Big].
    \end{aligned}
\end{equation}
Both the $4$- and $(D-4)$-dimensional parts receive fermionic and Yukawa contributions (see App.~\ref{App:1LoopCTCoeffs-GeneralAbelianChiralGaugeTheory}, Eq.~\eqref{App-Eq:DivCoeffRelations} and Eqs.~\eqref{App-Eq:FermionCoeffFermionGauge}--\eqref{App-Eq:FermionGaugeBosonCoeff}).
When $\hypLR \equiv 0$, only the Yukawa-induced evanescent terms persist.

\paragraph{(v) Triple Gauge Boson Interaction:}
As in chiral QED (see Refs.~\cite{Belusca-Maito:2021lnk,Stockinger:2023ndm,vonManteuffel:2025swv}), the triple gauge boson interaction yields no UV divergence, and thus
\begin{equation}
    \begin{aligned}
        i \Gamma_{B_{\mu}(q)B_{\nu}(p_1)B_{\rho}(p_2)} \big|_{\mathrm{div}}^{(1)} = 0.
    \end{aligned}
\end{equation}

\paragraph{(vi) Yukawa Interaction:}
The divergent 1-loop corrections to the Yukawa interactions read
\begin{align}
    \begin{split}\label{Eq:YukawaGreenFunc1-FFS-GeneralAbelianTheory}
        i \Gamma_{\psi_i(p_2)\overline{\psi}_j(p_1)\phi_a(q)} \big|_{\mathrm{div}}^{(1)} &=
        \frac{i}{16 \pi^2} \, \frac{1}{\epsilon} 
        \Big[
        \overline{\mathcal{A}}_{\psi\overline{\psi}\phi,\mathrm{R},ji}^{1,\mathrm{inv},a}\,
        \mathbb{P}_{\mathrm{R}}
        + 
        \overline{\mathcal{A}}_{\psi\overline{\psi}\phi,\mathrm{L},ji}^{1,\mathrm{inv},a}\,
        \mathbb{P}_{\mathrm{L}}
        \Big],
    \end{split}\\[1.5ex]
    \begin{split}\label{Eq:YukawaGreenFunc2-FFSbar-GeneralAbelianTheory}
        i \Gamma_{\psi_i(p_2)\overline{\psi}_j(p_1)\phi^{\dagger}_a(q)} \big|_{\mathrm{div}}^{(1)} &=
        \frac{i}{16 \pi^2} \, \frac{1}{\epsilon} 
        \Big[
        {\overline{\mathcal{A}}{}_{\psi\overline{\psi}\phi,\mathrm{L},ij}^{1,\mathrm{inv},a}}^{\!\!\!\!\!\!\dagger}\,\,\,\,
        \mathbb{P}_{\mathrm{R}}
        + 
        {\overline{\mathcal{A}}{}_{\psi\overline{\psi}\phi,\mathrm{R},ij}^{1,\mathrm{inv},a}}^{\!\!\!\!\!\!\!\dagger}\,\,\,\,
        \mathbb{P}_{\mathrm{L}}
        \Big],
    \end{split}
\end{align}
with coefficients given in App.~\ref{App:1LoopCTCoeffs-GeneralAbelianChiralGaugeTheory} in Eqs.~\eqref{App-Eq:DivCoeffRelations} and \eqref{App-Eq:DivYukawaCoeffs}.

\paragraph{(vii) Scalar-Gauge Boson Triple Interaction:}
\begin{align}
    \begin{split}\label{Eq:phi-phiDagger-B-GreenFunc-GeneralAbelianTheory}
        i \Gamma_{\phi_a(p_2)\phi^{\dagger}_b(p_1)B_{\mu}(q)} \big|_{\mathrm{div}}^{(1)} &=
        \frac{ig}{16\pi^2} \, \frac{\mathcal{Y}_{S}}{\epsilon} \Big(
        g^2
        \prescript{}{S}{\mathcal{A}}_{\phi\phi^{\dagger},ba}^{1,\mathrm{inv}}
        + \prescript{}{Y}{\overline{\mathcal{A}}}_{\phi\phi^{\dagger},ba}^{1,\mathrm{inv}} 
        \Big) \Big( \overline{p}_1^{\mu} - \overline{p}_2^{\mu} \Big)\\
        &+ \frac{ig}{16 \pi^2} \, \frac{1}{\epsilon} \Big(
        g^2 \, \mathcal{Y}_{S}
        \prescript{}{S}{\mathcal{A}}_{\phi\phi^{\dagger},ba}^{1,\mathrm{inv}}
        + {\widehat{\mathcal{A}}}_{\phi\phi^{\dagger}B,ba}^{1,\mathrm{break}} 
        \Big) \Big( \widehat{p}_1^{\mu} - \widehat{p}_2^{\mu} \Big),
    \end{split}\\[1.5ex]
    \begin{split}\label{Eq:phi-phi-B-GreenFunc-GeneralAbelianTheory}
        i \Gamma_{\phi_a(p_2)\phi_b(p_1) B_{\mu}(q)} \big|_{\mathrm{div}}^{(1)} &=
        \frac{ig}{16 \pi^2} \, \frac{1}{\epsilon}
        {\widehat{\mathcal{A}}}_{\phi\phi B, ba}^{1,\mathrm{break}} \, \Big( \widehat{p}_1^{\mu} - \widehat{p}_2^{\mu} \Big),
    \end{split}\\[1.5ex]
    \begin{split}\label{Eq:phiDagger-phiDagger-B-GreenFunc-GeneralAbelianTheory}
        i \Gamma_{\phi^{\dagger}_a(p_2)\phi^{\dagger}_b(p_1) B_{\mu}(q)} \big|_{\mathrm{div}}^{(1)} &=
        - \frac{ig}{16 \pi^2} \, \frac{1}{\epsilon} \,
        {\widehat{\mathcal{A}}{}_{\phi\phi B,ba}^{1,\mathrm{break}}}^{\dagger} \Big( \widehat{p}_1^{\mu} - \widehat{p}_2^{\mu} \Big).
    \end{split}
\end{align}
As in the scalar 2-point case, two additional global hypercharge-violating and purely evanescent Green functions appear (see Eqs.~\eqref{Eq:phi-phi-B-GreenFunc-GeneralAbelianTheory} and \eqref{Eq:phiDagger-phiDagger-B-GreenFunc-GeneralAbelianTheory}), alongside the standard $\phi^\dagger_a\phi_b B^\mu$ vertex correction in Eq.~\eqref{Eq:phi-phiDagger-B-GreenFunc-GeneralAbelianTheory}.
Unlike before, these vanish for $\hypLR\equiv0$ since they originate solely from evanescent gauge interactions, not from the fermion kinetic term.

\subsubsection{4-Point Functions}

Finally, nine power-counting divergent 4-point Green functions remain, excluding the scalar--triple gauge boson and triple scalar--single gauge boson cases listed in the last two rows of Tab.~\ref{Tab:NonContributingOrdinaryGreenFunctions}.

\paragraph{(viii) Quartic Gauge Boson Interaction:}
The divergent 1-loop correction to the quartic gauge boson self-interaction vanishes,
\begin{equation}\label{Eq:BBBB-GreenFunc-GeneralAbelianTheory}
    \begin{aligned}
        i \Gamma_{B_{\mu}(p_2)B_{\nu}(p_1)B_{\rho}(p_4)B_{\sigma}(p_3)} \big|_{\mathrm{div}}^{(1)} = 0.
    \end{aligned}
\end{equation}
Importantly, this result holds for $\hypRL=\hypLR$, but a nonzero divergence appears when $\hypRL\neq\hypLR$ (cf.\ Eq.~\eqref{Eq:ASSM-YLR-YRL-Option2}); see Sec.~\ref{Sec:AnalysisOfASSMResults} for details.

\paragraph{(ix) Scalar-Gauge Boson Quartic Interaction:}
\begin{align}
    \begin{split}\label{Eq:phi-phiDagger-B-B-GeneralAbelianTheory}
        i \Gamma_{\phi_a(p_2)\phi^{\dagger}_b(p_1) B_{\mu}(p_4)B_{\nu}(p_3)}\big|_{\mathrm{div}}^{(1)} &=
        \frac{ig^2}{16 \pi^2} \, \frac{2\,\mathcal{Y}_{S}^2}{\epsilon} 
        \Big( 
        g^2 \prescript{}{S}{\mathcal{A}}_{\phi\phi^{\dagger},ba}^{1,\mathrm{inv}}
        + \prescript{}{Y}{\overline{\mathcal{A}}}_{\phi\phi^{\dagger},ba}^{1,\mathrm{inv}}
        \Big) \,
        \overline{\eta}^{\mu\nu}\\
        &+ \frac{ig^2}{16 \pi^2} \, \frac{2}{\epsilon} \, 
        \Big( 
        g^2 \, \mathcal{Y}_{S}^2 \prescript{}{S}{\mathcal{A}}_{\phi\phi^{\dagger},ba}^{1,\mathrm{inv}}
        + \frac{1}{2} {\widehat{\mathcal{A}}}_{\phi\phi^{\dagger} BB,ba}^{1,\mathrm{break}}
        \Big) \,
        \widehat{\eta}^{\mu\nu},
    \end{split}\\[1.5ex]
    \begin{split}\label{Eq:phi-phi-B-B-GeneralAbelianTheory}
        i \Gamma_{\phi_a(p_2)\phi_b(p_1) B_{\mu}(p_4)B_{\nu}(p_3)} \big|_{\mathrm{div}}^{(1)} &=
        \frac{ig^2}{16 \pi^2} \, \frac{1}{\epsilon}
        {\widehat{\mathcal{A}}}_{\phi\phi BB, ba}^{1, \mathrm{break}} \, \widehat{\eta}^{\mu\nu},
    \end{split}\\[1.5ex]
    \begin{split}\label{Eq:phiDagger-phiDagger-B-B-GeneralAbelianTheory}
        i \Gamma_{\phi^{\dagger}_a(p_2)\phi^{\dagger}_b(p_1) B_{\mu}(p_4)B_{\nu}(p_3)} \big|_{\mathrm{div}}^{(1)} &=
        \frac{ig^2}{16 \pi^2} \, \frac{1}{\epsilon}
        {\widehat{\mathcal{A}}{}_{\phi\phi BB,ba}^{1, \mathrm{break}}}^{\!\!\!\!\!\dagger} \,\,\, \widehat{\eta}^{\mu\nu}.     
    \end{split}
\end{align}
Two purely evanescent, global hypercharge-violating scalar--gauge boson quartic Green functions appear (see Eqs.~\eqref{Eq:phi-phi-B-B-GeneralAbelianTheory} and \eqref{Eq:phiDagger-phiDagger-B-B-GeneralAbelianTheory}), alongside the standard contribution in Eq.~\eqref{Eq:phi-phiDagger-B-B-GeneralAbelianTheory}.
They originate solely from evanescent gauge interactions and vanish for $\hypLR\equiv0$ (see Eq.~\eqref{App-Eq:SSBBCoeffs}).

\paragraph{(x) Quartic Scalar Interaction:}
Only the global hypercharge-conserving Green function contributes, while all global hypercharge-violating ones vanish.
\begin{align}
    \begin{split}\label{Eq:SSSbarSbar-GreenFunc-GeneralAbelianTheory}
        i \Gamma_{\phi_a(p_2)\phi_b(p_1)\phi^{\dagger}_c(p_4)\phi^{\dagger}_d(p_3)}\big|_{\mathrm{div}}^{(1)}
        &=
        \frac{i}{16 \pi^2} \, \frac{1}{\epsilon} \, 
        \mathcal{A}_{\phi\phi^{\dagger}\phi\phi^{\dagger},cadb}^{1,\mathrm{inv}},
    \end{split}\\[1.5ex]
    \begin{split}\label{Eq:SSSSbar-GreenFunc-GeneralAbelianTheory}
        i \Gamma_{\phi_a(p_2)\phi_b(p_1)\phi_c(p_4)\phi^{\dagger}_d(p_3)} \big|_{\mathrm{div}}^{(1)} &= 0,
    \end{split}\\[1.5ex]
    \begin{split}\label{Eq:SSbarSbarSbar-GreenFunc-GeneralAbelianTheory}
        i \Gamma_{\phi_a(p_2)\phi^{\dagger}_b(p_1)\phi^{\dagger}_c(p_4)\phi^{\dagger}_d(p_3)} \big|_{\mathrm{div}}^{(1)} &= 0,
    \end{split}\\[1.5ex]
    \begin{split}\label{Eq:SSSS-GreenFunc-GeneralAbelianTheory}
        i \Gamma_{\phi_a(p_2)\phi_b(p_1)\phi_c(p_4)\phi_d(p_3)} \big|_{\mathrm{div}}^{(1)} &= 0,
    \end{split}\\[1.5ex]
    \begin{split}\label{Eq:SbarSbarSbarSbar-GreenFunc-GeneralAbelianTheory}
        i \Gamma_{\phi^{\dagger}_a(p_2)\phi^{\dagger}_b(p_1)\phi^{\dagger}_c(p_4)\phi^{\dagger}_d(p_3)} \big|_{\mathrm{div}}^{(1)} &= 0.
    \end{split}
\end{align}
Since the four-scalar interactions have zero power counting and contain no Lorentz structures, no evanescent BRST-breaking divergences induced by $\DeltaHat$ can arise.
However, $4$-dimensional hypercharge-breaking effects may appear at higher orders (cf.\ Refs.~\cite{Stockinger:2023ndm,vonManteuffel:2025swv}).

\subsection{Operator-Inserted 1PI Green Functions --- 1-Loop BRST Breaking}\label{App:1LoopBreakingGreenFunctions-GeneralAbelianChiralGaugeTheory}

We now turn to the operator-inserted, power-counting divergent 1PI Green functions relevant for the renormalisation.
As discussed in Sec.~\ref{Sec:Symmetry_Restoration_Procedure}, these generate both divergent and finite BRST breakings, from which we derive the symmetry-restoring counterterms.
Non-contributing Green functions listed in Tab.~\ref{Tab:NonContributingDeltaGreenFunctions} are omitted, and finite evanescent terms are suppressed since they vanish in the limit $\mathop{\mathrm{LIM}}_{D\,\to\,4}$.
As a consistency check, the divergent breaking terms perfectly match those obtained from the ordinary Green functions in App.~\ref{App:1LoopStandardGreenFunctions-GeneralAbelianChiralGaugeTheory}.
As before, the results are organised w.r.t.\ external legs, i.e.\ into 2-, 3-, 4- and 5-point Green functions.

\subsubsection{2-Point Functions}

For the 2-point case, only one non-vanishing Green function remains, excluding the two listed in the first row of Tab.~\ref{Tab:NonContributingDeltaGreenFunctions}.

\paragraph{(xi) Ghost-Gauge Boson Contribution:}
\begin{equation}\label{Eq:GhostGaugeBoson_1-Loop-GeneralAbelianTheory}
    \begin{aligned}
        i \big( \widehat{\Delta} \cdot \Gamma \big)_{B_{\mu}(-p) c(p)}^{(1)} 
        &= \frac{g^2}{16 \pi^2} \, \frac{1}{\epsilon} \, 
        \bigg[
        \Big( 
        \prescript{}{F}{\widehat{\mathcal{A}}}_{BB,1}^{1, \mathrm{break}} 
        + \prescript{}{F}{\widehat{\mathcal{A}}}_{BB,2}^{1, \mathrm{break}}
        + \prescript{}{F}{\widehat{\mathcal{A}}}_{BB,3}^{1, \mathrm{break}}
        \Big) \,
        \overline{p}^2 \widehat{p}^{\mu}\\
        &\qquad\qquad + \Big( 
        \prescript{}{F}{\widehat{\mathcal{A}}}_{BB,2}^{1, \mathrm{break}} 
        + \prescript{}{F}{\widehat{\mathcal{A}}}_{BB,4}^{1, \mathrm{break}}
        \Big) \,
        \widehat{p}^2 \overline{p}^{\mu} \bigg]\\
        &+ \frac{g^2}{16 \pi^2}
        {\mathcal{F}}_{BB}^{1, \mathrm{break}} \, 
        \overline{p}^2 \, \overline{p}^{\mu}.
    \end{aligned}
\end{equation}
Contracting Eq.~\eqref{Eq:GaugeBosonSelfEnergy_1-Loop-GeneralAbelianTheory} with $p_{\nu}$ reproduces the divergence in Eq.~\eqref{Eq:GhostGaugeBoson_1-Loop-GeneralAbelianTheory}, confirming the divergent part of the relation in Eq.~\eqref{Eq:STI-QAP-Bc}.

\subsubsection{3-Point Functions}

For the 3-point case, five contributing Green functions remain, excluding those in the second row of Tab.~\ref{Tab:NonContributingDeltaGreenFunctions}.

\paragraph{(xii) Ghost-Fermion-Fermion Contribution:}
The BRST breaking associated with the fermion self energy and the fermion--gauge boson interaction, as shown in Eq.~\eqref{Eq:STI-QAP-FFc}, is given by
\begin{equation}\label{Eq:FbarFc-GreenFunc-GeneralAbelianTheory}
    \begin{aligned}
        &i \big( \widehat{\Delta} \cdot \Gamma \big)_{\psi_i(p_2) \overline{\psi}_j(p_1) c(q)}^{(1)} =\\
        &- \frac{g}{16 \pi^2} \, \frac{1}{\epsilon} 
        \bigg\{
        \widehat{\slashed{p}}_1
        \bigg[
        \Big(
        \Big(\widehat{\mathcal{A}}_{\psi\overline{\psi}}^{1, \mathrm{break}}\mathcal{Y}_{R}\Big)_{ji}
        - \widehat{\mathcal{A}}_{\psi\overline{\psi}B,ji}^{1, \mathrm{break}}
        \Big) 
        \mathbb{P}_{\mathrm{R}}
        +
        \Big(
        \Big(\mathcal{Y}_{L}\widehat{\mathcal{A}}_{\psi\overline{\psi}}^{1, \mathrm{break}}\Big)^{\dagger}_{ij}
        - {\widehat{\mathcal{A}}{}_{\psi\overline{\psi}B,ij}^{1, \mathrm{break}}}^{\dagger}
        \Big) 
        \mathbb{P}_{\mathrm{L}}
        \bigg]\\
        &\qquad\qquad + 
        \widehat{\slashed{p}}_2
        \bigg[
        \Big(
        \Big(\mathcal{Y}_{L}\widehat{\mathcal{A}}_{\psi\overline{\psi}}^{1, \mathrm{break}}\Big)_{ji}
        - \widehat{\mathcal{A}}_{\psi\overline{\psi}B,ji}^{1, \mathrm{break}}
        \Big)
        \mathbb{P}_{\mathrm{R}}
        +
        \Big(
        \Big(\widehat{\mathcal{A}}_{\psi\overline{\psi}}^{1, \mathrm{break}}\mathcal{Y}_{R}\Big)^{\dagger}_{ij}
        - {\widehat{\mathcal{A}}{}_{\psi\overline{\psi}B,ij}^{1, \mathrm{break}}}^{\dagger}
        \Big) 
        \mathbb{P}_{\mathrm{L}}
        \bigg] \bigg\}\\
        &+ \frac{g}{16 \pi^2} \, \Big( \overline{\slashed{p}}_1 + \overline{\slashed{p}}_2 \Big) 
        \bigg\{
        \mathcal{F}_{\psi\overline{\psi}B,\mathrm{R},ji}^{1, \mathrm{break}}
        \mathbb{P}_{\mathrm{R}}
        + \mathcal{F}_{\psi\overline{\psi}B,\mathrm{L},ji}^{1, \mathrm{break}} 
        \mathbb{P}_{\mathrm{L}}
        \bigg\}.
    \end{aligned}
\end{equation}

\paragraph{(xiii) Ghost-double Gauge Boson Contribution:}
The ghost--double gauge boson 3-point Green function associated with the ABJ-anomaly (see Sec.~\ref{Sec:Anomalies}) via the Slavnov-Taylor identity in Eq.~\eqref{Eq:STI-QAP-BBc} vanishes once the anomaly cancellation condition in Eq.~\eqref{Eq:AnomalyCancellationCondition-GeneralAbelianTheory} is imposed, yielding
\begin{equation}\label{Eq:GhostDoubleGaugeBoson_1-Loop-GeneralAbelianTheory}
    \begin{aligned}
        i \big( \widehat{\Delta} \cdot \Gamma \big)_{B_{\nu}(p_2)B_{\mu}(p_1)c(q)}^{(1)} &=
        - \frac{ig^3}{16\pi^2} \, \frac{4}{3} \, \Big(\mathrm{Tr}\big(\mathcal{Y}_R^3\big)-\mathrm{Tr}\big(\mathcal{Y}_L^3\big)\Big) \, \overline{\varepsilon}^{\mu\nu\rho\sigma} \, \overline{p}_{1,\rho} \, \overline{p}_{2,\sigma} = 0.
    \end{aligned}
\end{equation}

\paragraph{(xiv) Ghost-double Scalar Contribution:}
The BRST breaking associated with the scalar 2-point and scalar--gauge boson 3-point Green functions, as presented in Eqs.~\eqref{Eq:STI-QAP-SSbarc}--\eqref{Eq:STI-QAP-SbarSbarc}, is given by
\begin{align}
    \begin{split}\label{Eq:GhostDoubleScalar_1-Loop-GeneralAbelianTheory}
        i \big( &\widehat{\Delta} \cdot \Gamma \big)_{\phi_b(p_2)\phi^{\dagger}_a(p_1) c(q)}^{(1)}\\
        = &- \frac{g}{16 \pi^2} \, \frac{1}{\epsilon} \, 
        \Big(
        {\widehat{\mathcal{A}}}_{\phi\phi^{\dagger}B,ab}^{1, \mathrm{break}} 
        - \mathcal{Y}_{S} 
        {\widehat{\mathcal{A}}}_{\phi\phi^{\dagger},ab}^{1, \mathrm{break}}
        \Big)
        \Big( \widehat{p}_1^2 - \widehat{p}_2^2 \Big)\\
        &+ \frac{g}{16 \pi^2} \, {\mathcal{F}}_{\phi\phi^{\dagger}B,ab}^{1,\mathrm{break}} \, \Big( \overline{p}_1^2 - \overline{p}_2^2 \Big),
    \end{split}\\[1.5ex]
    \begin{split}\label{Eq:Ghostphiphi_1-Loop-GeneralAbelianTheory}
        i \big( &\widehat{\Delta} \cdot \Gamma \big)_{\phi_b(p_2)\phi_a(p_1)c(q)}^{(1)}\\
        &= \frac{g}{16 \pi^2} \, \frac{1}{\epsilon} \,
        \Big[
        \Big(
        \mathcal{Y}_{S} 
        {\widehat{\mathcal{A}}}_{\phi\phi,ab}^{1, \mathrm{break}}
        - {\widehat{\mathcal{A}}}_{\phi\phi B,ab}^{1, \mathrm{break}}
        \Big) \, \widehat{p}_1^2
        +
        \Big(
        \mathcal{Y}_{S} 
        {\widehat{\mathcal{A}}}_{\phi\phi,ab}^{1, \mathrm{break}}
        + {\widehat{\mathcal{A}}}_{\phi\phi B,ab}^{1, \mathrm{break}}
        \Big) \, \widehat{p}_2^2
        \Big]\\
        &+ \,
        \frac{g}{16 \pi^2} \, 4 \, \mathcal{Y}_{S} \, {\mathcal{F}}_{\phi\phi,ab}^{1, \mathrm{break}}
        \Big(\overline{p}_1^2 + \overline{p}_2^2 + \frac{3}{2} \, \overline{p}_1 \cdot \overline{p}_2\Big),
    \end{split}\\[1.5ex]
    \begin{split}\label{Eq:Ghostphibarphibar_1-Loop-GeneralAbelianTheory}
        i \big( &\widehat{\Delta} \cdot \Gamma \big)_{\phi^{\dagger}_b(p_2)\phi^{\dagger}_a(p_1)c(q)}^{(1)}\\
        = &-\frac{g}{16 \pi^2} \, \frac{1}{\epsilon} \,
        \Big[
        \Big(
        \mathcal{Y}_{S} \,
        {\widehat{\mathcal{A}}{}_{\phi\phi,ab}^{1, \mathrm{break}}}^{\dagger}
        - {\widehat{\mathcal{A}}{}_{\phi\phi B,ab}^{1, \mathrm{break}}}^{\dagger}
        \Big) \, \widehat{p}_1^2
        +
        \Big(
        \mathcal{Y}_{S} \, 
        {\widehat{\mathcal{A}}{}_{\phi\phi,ab}^{1, \mathrm{break}}}^{\dagger}
        + {\widehat{\mathcal{A}}{}_{\phi\phi B,ab}^{1, \mathrm{break}}}^{\dagger}
        \Big) \, \widehat{p}_2^2
        \Big]\\
        &-\frac{g}{16 \pi^2} \, 4 \, \mathcal{Y}_{S} \, {{\mathcal{F}}_{\phi\phi,ab}^{1, \mathrm{break}}}^{\dagger}
        \Big(\overline{p}_1^2 + \overline{p}_2^2 + \frac{3}{2} \, \overline{p}_1 \cdot \overline{p}_2\Big),
    \end{split}
\end{align}
The last two breakings in Eqs.~\eqref{Eq:Ghostphiphi_1-Loop-GeneralAbelianTheory} and \eqref{Eq:Ghostphibarphibar_1-Loop-GeneralAbelianTheory} break global hypercharge conservation (cf.\ App.~\ref{App:1LoopStandardGreenFunctions-GeneralAbelianChiralGaugeTheory}) and introduce a finite hypercharge-violating contribution $\propto{\mathcal{F}}_{\phi\phi,ab}^{1, \mathrm{break}}$.

\subsubsection{4-Point Functions}

Seven 4-point Green functions yield non-vanishing contributions, excluding the six listed in the third and fourth rows of Tab.~\ref{Tab:NonContributingDeltaGreenFunctions}.

\paragraph{(xv) Ghost-triple Gauge Boson Contribution:}
For the BRST breaking of the quartic gauge boson Green function (see Eq.~\eqref{Eq:STI-QAP-BBBc}), we obtain
\begin{equation}\label{Eq:BBBc-GreenFunc-GeneralAbelianTheory}
    \begin{aligned}
        i \big( \widehat{\Delta} \cdot \Gamma& \big)_{B_{\rho}(p_3)B_{\nu}(p_2)B_{\mu}(p_1)c(q)}^{(1)}\\
        &= \frac{g^4}{16 \pi^2} 
        {\mathcal{F}}_{BBBB}^{1,\mathrm{break}}\,
        \big(\overline{p}_1 + \overline{p}_2 + \overline{p}_3\big)_{\sigma} \, 
        \Big( \overline{\eta}^{\rho\sigma} \, \overline{\eta}^{\mu\nu}
        + \overline{\eta}^{\nu\sigma} \, \overline{\eta}^{\mu\rho}
        + \overline{\eta}^{\mu\sigma} \, \overline{\eta}^{\nu\rho} \Big).
    \end{aligned}
\end{equation}
For $\hypRL=\hypLR$, this contribution is finite.
If $\hypRL\neq\hypLR$ (cf.\ Eq.~\eqref{Eq:ASSM-YLR-YRL-Option2}), an additional divergent BRST-breaking term arises (see Sec.~\ref{Sec:AnalysisOfASSMResults}).

\paragraph{(xvi) Ghost-Gauge Boson-Fermion-Fermion Contribution:}
Another BRST breaking associated to the fermion--gauge boson interaction (see Eq.~\eqref{Eq:STI-QAP-FFbarBc}) reads
\begin{equation}\label{Eq:FFbarBc-GreenFunc-GeneralAbelianTheory}
    \begin{aligned}
        i \big( \widehat{\Delta} \cdot \Gamma \big)_{\psi_i(p_2) \overline{\psi}_j(p_3) B_{\mu}(p_1) c(q)}^{(1)}
        = \frac{g^2}{16 \pi^2} \, \frac{1}{\epsilon} \, \widehat{\gamma}^{\mu} 
        \bigg[
        &\Big(
        \mathcal{Y}_{L}\,\widehat{\mathcal{A}}_{\psi\overline{\psi}B}^{1,\mathrm{break}}
        - \widehat{\mathcal{A}}_{\psi\overline{\psi}B}^{1,\mathrm{break}}\,\mathcal{Y}_{R}
        \Big)_{ji} \,
        \mathbb{P}_{\mathrm{R}}\\
        - &\Big(
        \mathcal{Y}_{L}\,\widehat{\mathcal{A}}_{\psi\overline{\psi}B}^{1,\mathrm{break}}
        - \widehat{\mathcal{A}}_{\psi\overline{\psi}B}^{1,\mathrm{break}}\,\mathcal{Y}_{R}
        \Big)_{ij}^{\dagger} \,
        \mathbb{P}_{\mathrm{L}}
        \bigg].
    \end{aligned}
\end{equation}
This contribution is purely evanescent.
From Eqs.~\eqref{App-Eq:DivCoeffRelations}, \eqref{App-Eq:FermionCoeffFermionGauge} and \eqref{App-Eq:FermionGaugeBosonCoeff} we see that it receives fermionic and Yukawa contributions from $\widehat{\Delta}_2\big[c,B,\overline{\psi},\psi\big]$-inserted diagrams, and only Yukawa contributions from $\widehat{\Delta}_1\big[c,\overline{\psi},\psi\big]$-inserted diagrams.
Hence, it vanishes in scalar-free models with $\hypLR=0$, as in chiral QED.

\paragraph{(xvii) Ghost-Yukawa Contribution:}
At the 1-loop level, Yukawa interactions induce only finite BRST-breaking contributions, given by
\begin{align}
    \begin{split}\label{Eq:YukawaDeltaGreenFunc1-cSFF-GeneralAbelianTheory}
        i \big( \widehat{\Delta} \cdot \Gamma \big)_{\psi_i(p_2) \overline{\psi}_j(p_3) \phi_{a}(p_1) c(q)}^{(1)}&\\ 
        = - \frac{g}{16 \pi^2} \, 
        \bigg[
        &\Big(
        \mathcal{Y}_{L}\mathcal{F}_{\psi\overline{\psi}\phi}^{1,\mathrm{break}}
        -\mathcal{F}_{\psi\overline{\psi}\phi}^{1,\mathrm{break}}\mathcal{Y}_{R}
        -\mathcal{Y}_{S}\mathcal{F}_{\psi\overline{\psi}\phi}^{1,\mathrm{break}}
        \Big)^{a}_{ji} \, \mathbb{P}_{\mathrm{R}}\\
        -
        &\Big(
        \mathcal{Y}_{L}\mathcal{F}_{\psi\overline{\psi}\phi^{\dagger}}^{1,\mathrm{break}}
        -\mathcal{F}_{\psi\overline{\psi}\phi^{\dagger}}^{1,\mathrm{break}}\mathcal{Y}_{R}
        +\mathcal{Y}_{S}\mathcal{F}_{\psi\overline{\psi}\phi^{\dagger}}^{1,\mathrm{break}}
        \Big)^{a \, \dagger}_{ij} \, \mathbb{P}_{\mathrm{L}}
        \bigg],
    \end{split}\\[1.5ex]
    \begin{split}\label{Eq:YukawaDeltaGreenFunc2-cSbarFF-GeneralAbelianTheory}
        i \big( \widehat{\Delta} \cdot \Gamma \big)_{\psi_i(p_2) \overline{\psi}_j(p_3) \phi^{\dagger}_{a}(p_1) c(q)}^{(1)}&\\
        = - \frac{g}{16 \pi^2} \, 
        \bigg[
        &\Big(
        \mathcal{Y}_{L}\mathcal{F}_{\psi\overline{\psi}\phi^{\dagger}}^{1,\mathrm{break}}
        -\mathcal{F}_{\psi\overline{\psi}\phi^{\dagger}}^{1,\mathrm{break}}\mathcal{Y}_{R}
        +\mathcal{Y}_{S}\mathcal{F}_{\psi\overline{\psi}\phi^{\dagger}}^{1,\mathrm{break}}
        \Big)^{a}_{ji} \, \mathbb{P}_{\mathrm{R}}\\
        -
        &\Big(
        \mathcal{Y}_{L}\mathcal{F}_{\psi\overline{\psi}\phi}^{1,\mathrm{break}}
        -\mathcal{F}_{\psi\overline{\psi}\phi}^{1,\mathrm{break}}\mathcal{Y}_{R}
        -\mathcal{Y}_{S}\mathcal{F}_{\psi\overline{\psi}\phi}^{1,\mathrm{break}}
        \Big)^{a\,\dagger}_{ij} \, \mathbb{P}_{\mathrm{L}}
        \bigg].
    \end{split}
\end{align}
The coefficients in Eqs.~\eqref{Eq:YukawaDeltaGreenFunc1-cSFF-GeneralAbelianTheory} and \eqref{Eq:YukawaDeltaGreenFunc2-cSbarFF-GeneralAbelianTheory} (see Eq.~\eqref{App-Eq:FiniteYukawaCoeffs}) are chosen such that the corresponding terms in the finite counterterm action, Eq.~\eqref{Eq:Sfct_1-Loop-GeneralAbelianTheory}, have a simple, directly readable from as BRST origin of these expressions.

\paragraph{(xviii) Ghost-Gauge boson-double Scalar Contribution:}
The BRST breakings corresponding to the triple and quartic scalar--gauge boson interactions, see Eqs.~\eqref{Eq:STI-QAP-SSbarBc}--\eqref{Eq:STI-QAP-SbarSbarBc}, are given by
\begin{equation}
    \begin{aligned}\label{Eq:GreenFunc-c-B-phiDagger-phi-GeneralAbelianTheory}
        i \big( \widehat{\Delta} &\cdot \Gamma \big)_{\phi_b(p_3) \phi_a^{\dagger}(p_2) B_{\mu}(p_1) c(q)}^{(1)}\\
        &= \frac{g^2}{16 \pi^2} \,
        \frac{1}{\epsilon}
        \Big(
        2 \,\mathcal{Y}_{S} {\widehat{\mathcal{A}}}_{\phi\phi^{\dagger}B,ab}^{1,\mathrm{break}} - {\widehat{\mathcal{A}}}_{\phi\phi^{\dagger}BB,ab}^{1,\mathrm{break}}\Big) 
        \Big(\widehat{p}_1^{\mu} + \widehat{p}_2^{\mu} + \widehat{p}_3^{\mu}\Big)\\
        &- \frac{g^2}{16 \pi^2}
        \Big(2\,\mathcal{Y}_{S}\, {\mathcal{F}}_{\phi\phi^{\dagger}B,ab}^{1, \mathrm{break}} 
        - {\mathcal{F}}_{\phi\phi^{\dagger}BB,ab}^{1, \mathrm{break}}\Big) 
        \Big(\overline{p}_1^{\mu} + \overline{p}_2^{\mu} + \overline{p}_3^{\mu}\Big),
    \end{aligned}
\end{equation}
\begin{equation}
    \begin{alignedat}{2}
        i \big( \widehat{\Delta} &\cdot \Gamma \big)_{\phi_b(p_3) \phi_a(p_2) B_{\mu}(p_1) c(q)}^{(1)}
        &&\\
        &= \frac{g^2}{16 \pi^2} \, 
        \frac{1}{\epsilon}
        \bigg[
        - {\widehat{\mathcal{A}}}_{\phi\phi BB,ab}^{1, \mathrm{break}} \, \widehat{p}_1^{\mu}
        &&+ \Big(
        2 \, \mathcal{Y}_{S} {\widehat{\mathcal{A}}}_{\phi\phi B,ab}^{1, \mathrm{break}}
        - {\widehat{\mathcal{A}}}_{\phi\phi BB,ab}^{1, \mathrm{break}}
        \Big) \, \widehat{p}_2^{\mu}\\
        &
        &&- \Big(
        2 \, \mathcal{Y}_{S} {\widehat{\mathcal{A}}}_{\phi\phi B,ab}^{1, \mathrm{break}}
        + {\widehat{\mathcal{A}}}_{\phi\phi BB,ab}^{1, \mathrm{break}}
        \Big) \, \widehat{p}_3^{\mu}
        \bigg]\\
        &-\mathrlap{\frac{g^2}{16 \pi^2} \, 6 \, \mathcal{Y}_{S}^2
        \, {\mathcal{F}}_{\phi\phi,ab}^{1,\mathrm{break}} \,
        \Big(\overline{p}_2^{\mu} + \overline{p}_3^{\mu}\Big),}
        &&
    \end{alignedat}
\end{equation}
\begin{equation}
    \begin{alignedat}{2}
        i \big( \widehat{\Delta} &\cdot \Gamma \big)_{\phi_b^{\dagger}(p_3) \phi_a^{\dagger}(p_2) B_{\mu}(p_1) c(q)}^{(1)}
        &&\\
        &= \frac{g^2}{16 \pi^2} \, 
        \frac{1}{\epsilon}
        \bigg[
        - {\widehat{\mathcal{A}}{}_{\phi\phi BB,ab}^{1, \mathrm{break}}}^{\!\!\!\!\!\dagger} \,\,\, \widehat{p}_1^{\mu}
        &&+ \Big(
        2 \, \mathcal{Y}_{S} \, {\widehat{\mathcal{A}}{}_{\phi\phi B,ab}^{1, \mathrm{break}}}^{\dagger}
        - {\widehat{\mathcal{A}}{}_{\phi\phi BB,ab}^{1, \mathrm{break}}}^{\!\!\!\!\!\dagger} \,\,
        \Big) \, \widehat{p}_2^{\mu}\\
        &
        &&- \Big(
        2 \, \mathcal{Y}_{S} \, {\widehat{\mathcal{A}}{}_{\phi\phi B,ab}^{1, \mathrm{break}}}^{\dagger}
        + {\widehat{\mathcal{A}}{}_{\phi\phi BB,ab}^{1, \mathrm{break}}}^{\!\!\!\!\!\dagger} \,\,
        \Big) \, \widehat{p}_3^{\mu}
        \bigg]\\
        &- \mathrlap{\frac{g^2}{16 \pi^2} \, 6 \, \mathcal{Y}_{S}^2
        \, {{\mathcal{F}}_{\phi\phi,ab}^{1,\mathrm{break}}}^{\dagger} \,
        \Big(\overline{p}_2^{\mu} + \overline{p}_3^{\mu}\Big).}
        &&
    \end{alignedat}
\end{equation}
These terms yield both divergent and finite breakings of BRST invariance and global hypercharge conservation.

\subsubsection{5-Point Functions}

Finally, due to the ghost's zero mass dimension, eight power-counting divergent divergent 5-point Green functions remain, excluding the seven listed in the last three rows of Tab.~\ref{Tab:NonContributingDeltaGreenFunctions}.

\paragraph{(ixx) Ghost-double Gauge boson-double Scalar Contribution:}
In addition to item (xviii), further breakings of the quartic scalar--gauge boson interaction according to Eqs.~\eqref{Eq:STI-QAP-SSbarBBc}--\eqref{Eq:STI-QAP-SbarSbarBBc} are given by
\begin{align}
    \begin{split}\label{Eq:SSbarBBc-DeltaGreenFunc-GeneralAbelianTheory}
        i \big( \widehat{\Delta} \cdot \Gamma &\big)_{\phi_b(p_4) \phi_a^{\dagger}(p_3) B_{\nu}(p_2) B_{\mu}(p_1) c(q)}^{(1)}
        = 0,
    \end{split}\\[1.5ex]
    \begin{split}\label{Eq:SSBBc-DeltaGreenFunc-GeneralAbelianTheory}
        i \big( \widehat{\Delta} \cdot \Gamma &\big)_{\phi_b(p_4) \phi_a(p_3) B_{\nu}(p_2) B_{\mu}(p_1) c(q)}^{(1)}\\
        &= \frac{g^3}{16 \pi^2}
        \bigg[
        \frac{2\,\mathcal{Y}_{S}}{\epsilon}
        {\widehat{\mathcal{A}}}_{\phi\phi BB,ab}^{1, \mathrm{break}} \, \widehat{\eta}^{\mu\nu}
        + 
        12 \, \mathcal{Y}_{S}^3 \,
        {\mathcal{F}}_{\phi\phi,ab}^{1,\mathrm{break}} \, \overline{\eta}^{\mu\nu}
        \bigg],
    \end{split}\\[1.5ex]
    \begin{split}\label{Eq:SbarSbarBBc-DeltaGreenFunc-GeneralAbelianTheory}
        i \big( \widehat{\Delta} \cdot \Gamma &\big)_{\phi_b^{\dagger}(p_4) \phi_a^{\dagger}(p_3) B_{\nu}(p_2) B_{\mu}(p_1) c(q)}^{(1)}\\
        &= - \frac{g^3}{16 \pi^2}
        \bigg[
        \frac{2\,\mathcal{Y}_{S}}{\epsilon}\,
       {\widehat{\mathcal{A}}{}_{\phi\phi BB,ab}^{1, \mathrm{break}}}^{\!\!\!\!\!\dagger} \,\,\, \widehat{\eta}^{\mu\nu}
        + 
        12 \, \mathcal{Y}_{S}^3 \,
        {{\mathcal{F}}_{\phi\phi,ab}^{1,\mathrm{break}}}^{\dagger} \, \overline{\eta}^{\mu\nu}
        \bigg].
    \end{split}
\end{align}
The global hypercharge-conserving term vanishes, Eq.~\eqref{Eq:SSbarBBc-DeltaGreenFunc-GeneralAbelianTheory}, as required by the momentum independence of the local part of the Green functions on the LHS of Eq.~\eqref{Eq:STI-QAP-SSbarBBc}.
By Contrast, the global hypercharge-violating Green functions in Eqs.~\eqref{Eq:SSBBc-DeltaGreenFunc-GeneralAbelianTheory} and \eqref{Eq:SbarSbarBBc-DeltaGreenFunc-GeneralAbelianTheory} produce finite and divergent contributions.

\paragraph{(xx) Ghost-quartic Scalar Contribution:}
Unlike the divergent part of four scalar self-interactions in Eqs.~\eqref{Eq:SSSbarSbar-GreenFunc-GeneralAbelianTheory}--\eqref{Eq:SbarSbarSbarSbar-GreenFunc-GeneralAbelianTheory}, the BRST breaking is zero for the global hypercharge-conserving contribution and nonzero for the global hypercharge-violating ones,
\begin{align}
    \begin{split}\label{Eq:SSSbarSbarc-DeltaGreenFunc-GeneralAbelianTheory}
        i \big( \widehat{\Delta} \cdot \Gamma \big)_{\phi_d(p_4) \phi_c(p_3) \phi_b^{\dagger}(p_2) \phi_a^{\dagger}(p_1) c(q)}^{(1)}
        &= 0,
    \end{split}\\[1.5ex]
    \begin{split}\label{Eq:SSSS-DeltaGreenFunc-GeneralAbelianTheory}
        i \big( \widehat{\Delta} \cdot \Gamma \big)_{\phi_d(p_4) \phi_c(p_3) \phi_b(p_2) \phi_a(p_1) c(q)}^{(1)}
        &= \frac{g}{16 \pi^2} \, \mathcal{Y}_{S} \, {\mathcal{F}}_{\phi\phi\phi\phi,abcd}^{1,\mathrm{break}},
    \end{split}\\[1.5ex]
    \begin{split}
        i \big( \widehat{\Delta} \cdot \Gamma \big)_{\phi_d(p_4) \phi_c(p_3) \phi_b(p_2) \phi_a^{\dagger}(p_1) c(q)}^{(1)}
        &= \frac{g}{16 \pi^2} \, \mathcal{Y}_{S} \, {\mathcal{F}}_{\phi^{\dagger}\phi\phi\phi,abcd}^{1,\mathrm{break}},
    \end{split}\\[1.5ex]
    \begin{split}
        i \big( \widehat{\Delta} \cdot \Gamma \big)_{\phi_d(p_4) \phi_c^{\dagger}(p_3) \phi_b^{\dagger}(p_2) \phi_a^{\dagger}(p_1) c(q)}^{(1)}
        &= - \frac{g}{16 \pi^2} \, \mathcal{Y}_{S} \, \Big({{\mathcal{F}}_{\phi^{\dagger}\phi\phi\phi}^{1,\mathrm{break}}}^{\dagger}\Big)_{dbca},
    \end{split}\\[1.5ex]
    \begin{split}\label{Eq:SbarSbarSbarSbar-DeltaGreenFunc-GeneralAbelianTheory}
        i \big( \widehat{\Delta} \cdot \Gamma \big)_{\phi_d^{\dagger}(p_4) \phi_c^{\dagger}(p_3) \phi_b^{\dagger}(p_2) \phi_a^{\dagger}(p_1) c(q)}^{(1)}
        &= - \frac{g}{16 \pi^2} \, \mathcal{Y}_{S} \, \Big({{\mathcal{F}}_{\phi\phi\phi\phi}^{1,\mathrm{break}}}^{\dagger}\Big)_{abcd}.
    \end{split}
\end{align}
The vanishing of Eq.~\eqref{Eq:SSSbarSbarc-DeltaGreenFunc-GeneralAbelianTheory} follows for the same reason as in Eq.~\eqref{Eq:SSbarBBc-DeltaGreenFunc-GeneralAbelianTheory}, cf.\ Eq.~\eqref{Eq:STI-QAP-SSSbarSbarc}.
The nonzero BRST breaking contributions that violate global hypercharge are purely finite.

\chapter{Results --- Right-Handed Abelian Gauge Theory}\label{App:Results-RightHandedAbelianTheory}

Here, we provide the explicit results for all counterterm coefficients used in Sec.~\ref{Sec:Results-Right-Handed-Model}, covering all contributions up to and including the four-loop level.
These are relevant for the renormalisation of the right-handed Abelian chiral gauge theory discussed in chapter~\ref{Chap:BMHV_at_Multi-Loop_Level} (see Sec.~\ref{Sec:Right-Handed-Model-Definition} for the model definition).
In particular, in Apps.~\ref{App:1LoopCTCoeffs-RightHandedAbelianTheory}--\ref{App:4LoopCTCoeffs-RightHandedAbelianTheory} present the coefficients of both the divergent and finite symmetry-restoring counterterms, while App.~\ref{App:AuxiliaryCTs-RightHandedAbelianTheory} lists all auxiliary counterterms required in the framework of the tadpole decomposition.

\section{Explicit Results for the 1-Loop Counterterm Coefficients}\label{App:1LoopCTCoeffs-RightHandedAbelianTheory}

In this section, we present the explicit coefficients of all counterterms contributing at the 1-loop level.

\subsection{Coefficients of Divergent 1-Loop Contributions}\label{App:Divergent1LoopCTCoeffs-RightHandedAbelianTheory}

The following coefficients govern the singular 1-loop counterterms, which remove UV divergences from the right-handed Abelian theory at this order.
We begin with the bosonic contributions and then turn to the fermionic sector.

\paragraph{Gauge Boson 1-Loop Coefficients:}
The coefficient of the BRST-invariant counterterm is
\begin{align}
    \mathcal{A}_{BB}^{1,\mathrm{inv}} = - \frac{2}{3} \mathrm{Tr}\big(\mathcal{Y}_R^2\big),
\end{align}
while the non-invariant, evanescent counterterm coefficient reads
\begin{align}
    \widehat{\mathcal{A}}_{BB}^{1,\mathrm{break}} = - \frac{1}{3} \mathrm{Tr}\big(\mathcal{Y}_R^2\big).
\end{align}

\paragraph{Fermion 1-Loop Coefficients:}
At the 1-loop level, only an invariant, divergent fermion counterterm arises, given by
\begin{align}
    \mathcal{A}_{\overline{\psi}\psi, ij}^{1,\mathrm{inv}} = - \xi  \big( \mathcal{Y}_R^2 \big)_{ij}.
\end{align}

\subsection{Coefficients of Finite 1-Loop Contributions}\label{App:Finite1LoopCTCoeffs-RightHandedAbelianTheory}

We now list the finite 1-loop counterterm coefficients, which restore BRST symmetry after subtraction of the divergent contributions.

\paragraph{Gauge Boson 1-Loop Coefficients:}
The finite bilinear gauge boson contributions are governed by
\begin{align}
    \delta F_{BB}^{(1)} = - \frac{1}{3} \mathrm{Tr}\big(\mathcal{Y}_R^2\big),
\end{align}
while the coefficient of the quartic 1-loop term reads
\begin{align}
    \delta F_{BBBB}^{(1)} = \frac{2}{3} \mathrm{Tr}\big(\mathcal{Y}_R^4\big).
\end{align}

\paragraph{Fermion 1-Loop Coefficients:}
The coefficient of the finite fermionic counterterm at the 1-loop level is
\begin{align}
    \delta F_{\overline{\psi}\psi, ij}^{(1)} = \frac{5+\xi}{6}  \big( \mathcal{Y}_R^2 \big)_{ij}.
\end{align}

\section{Explicit Results for the 2-Loop Counterterm Coefficients}\label{App:2LoopCTCoeffs-RightHandedAbelianTheory}

In this section, we collect the 2-loop counterterm coefficients required for the renormalisation of the right-handed Abelian model at this order.
Following the structure of the 1-loop case, we first present the results for the divergent contributions and then for the finite ones.
Within each subsection, the results of the bosonic sector are given first, followed by those of the fermionic sector.

\subsection{Coefficients of Divergent 2-Loop Contributions}\label{App:Divergent2LoopCTCoeffs-RightHandedAbelianTheory}

The coefficients below govern the singular 2-loop counterterms that cancel the remaining UV divergences at this order.

\paragraph{Gauge Boson 2-Loop Coefficients:}

\begin{align}
    \mathcal{A}_{BB}^{2,\mathrm{inv}} = - \frac{2}{3} \frac{2+\xi}{3} \mathrm{Tr}\big(\mathcal{Y}_R^4\big)
\end{align}

\begin{align}
    \begin{split}
        \widehat{\mathcal{A}}_{BB}^{2,\mathrm{break}} &= \frac{43-26\xi}{72} \mathrm{Tr}\big(\mathcal{Y}_R^4\big)
    \end{split}\\
    \begin{split}
        \widehat{\mathcal{B}}_{BB}^{2,\mathrm{break}} &= - \frac{\xi}{6} \mathrm{Tr}\big(\mathcal{Y}_R^4\big)
    \end{split}
\end{align}

\paragraph{Fermion 2-Loop Coefficients:}

\begin{align}
    \begin{split}
        \mathcal{A}_{\overline{\psi}\psi, ij}^{2,\mathrm{inv}} &= \frac{9(1+\xi)-\xi^2}{12} \big( \mathcal{Y}_R^4 \big)_{ij} - \frac{24\xi^2 - 3\xi - 1}{180} \mathrm{Tr}\big(\mathcal{Y}_R^2\big) \big( \mathcal{Y}_R^2 \big)_{ij}
    \end{split}\\
    \begin{split}
        \mathcal{B}_{\overline{\psi}\psi, ij}^{2,\mathrm{inv}} &= \frac{\xi^2}{2} \big( \mathcal{Y}_R^4 \big)_{ij}
    \end{split}
\end{align}

\begin{align}
    \overline{\mathcal{A}}_{\overline{\psi}\psi, ij}^{2,\mathrm{break}} = - \frac{(17+3\xi)\xi}{24} \big( \mathcal{Y}_R^4 \big)_{ij} + \frac{153+4\xi+3\xi^2}{720} \mathrm{Tr}\big(\mathcal{Y}_R^2\big) \big( \mathcal{Y}_R^2 \big)_{ij}
\end{align}

\subsection{Coefficients of Finite 2-Loop Contributions}\label{App:Finite2LoopCTCoeffs-RightHandedAbelianTheory}

The finite 2-loop counterterm coefficients presented here ensure the restoration of BRST symmetry after renormalisation.

\paragraph{Gauge Boson 2-Loop Coefficients:}

\begin{align}
    \delta F_{BB}^{(2)} = \frac{17+5\xi}{48} \mathrm{Tr}\big(\mathcal{Y}_R^4\big)
\end{align}

\begin{align}
    \delta F_{BBBB}^{(2)} = - \frac{3}{2} \frac{5+\xi}{6} \mathrm{Tr}\big(\mathcal{Y}_R^6\big)
\end{align}

\paragraph{Fermion 2-Loop Coefficients:}

\begin{align}
    \delta F_{\overline{\psi}\psi, ij}^{(2)} = - \frac{4558+519\xi+3\xi^2}{1440}  \big( \mathcal{Y}_R^4 \big)_{ij} + \frac{1221-92\xi+471\xi^2}{43200} \mathrm{Tr}\big(\mathcal{Y}_R^2\big) \big( \mathcal{Y}_R^2 \big)_{ij}
\end{align}

\section{Explicit Results for the 3-Loop Counterterm Coefficients}\label{App:3LoopCTCoeffs-RightHandedAbelianTheory}

This section summarises the explicit 3-loop counterterm coefficients contributing to the renormalisation of the right-handed Abelian theory at this order.
The structure and organisation of the results follow the same pattern as in the previous sections.
Starting from the 3-loop level, all results are presented in Feynman gauge ($\xi=1$).
Some coefficients differ by an overall sign compared to Ref.~\cite{Stockinger:2023ndm}, because the sign conventions for the coefficients have been adapted to those in Ref.~\cite{vonManteuffel:2025swv}.

\subsection{Coefficients of Divergent 3-Loop Contributions}\label{App:Divergent3LoopCTCoeffs-RightHandedAbelianTheory}

The following coefficients determine the divergent 3-loop counterterms needed to establish UV finiteness at this order.

\paragraph{Gauge Boson 3-Loop Coefficients:}

\begin{align}
    \begin{split}\label{Eq:A_BB_inv_3-loop}
        \mathcal{A}_{BB}^{3,\mathrm{inv}} &= - \frac{1}{1620} \Big[ 2552 \, \mathrm{Tr}\big(\mathcal{Y}_R^6\big) + 61 \, \mathrm{Tr}\big(\mathcal{Y}_R^4\big) \mathrm{Tr}\big(\mathcal{Y}_R^2\big) \Big]
    \end{split}\\
    \begin{split}\label{Eq:B_BB_inv_3-loop}
        \mathcal{B}_{BB}^{3,\mathrm{inv}} &= \frac{4}{162} \Big[ 3 \, \mathrm{Tr}\big(\mathcal{Y}_R^6\big) - 5 \, \mathrm{Tr}\big(\mathcal{Y}_R^4\big) \mathrm{Tr}\big(\mathcal{Y}_R^2\big) \Big]
    \end{split}
\end{align}

\begin{align}
    \begin{split}\label{Eq:A_BB_hat_break_3-Loop}
        \widehat{\mathcal{A}}_{BB}^{3,\mathrm{break}} &= - \frac{1}{64800} \Big[ \big( 156672 \, \zeta(3) - 49427 \big) \mathrm{Tr}\big(\mathcal{Y}_R^6\big) - 8374 \, \mathrm{Tr}\big(\mathcal{Y}_R^4\big) \mathrm{Tr}\big(\mathcal{Y}_R^2\big) \Big]
    \end{split}\\
    \begin{split}\label{Eq:B_BB_hat_break_3-Loop}
        \widehat{\mathcal{B}}_{BB}^{3,\mathrm{break}} &= \frac{1}{1080} \Big[ 529 \, \mathrm{Tr}\big(\mathcal{Y}_R^6\big) + 122 \, \mathrm{Tr}\big(\mathcal{Y}_R^4\big) \mathrm{Tr}\big(\mathcal{Y}_R^2 \Big]
    \end{split}\\
    \begin{split}\label{Eq:C_BB_hat_break_3-Loop}
        \widehat{\mathcal{C}}_{BB}^{3,\mathrm{break}} &= - \frac{1}{18} \, \mathrm{Tr}\big(\mathcal{Y}_R^6\big)
    \end{split}
\end{align}

\begin{align}\label{Eq:A_BB_bar_break_3-Loop}
    \overline{\mathcal{A}}_{BB}^{3,\mathrm{break}} &= \frac{1}{1080} \Big[ 18 \, \mathrm{Tr}\big(\mathcal{Y}_R^6\big) + 79 \, \mathrm{Tr}\big(\mathcal{Y}_R^4\big) \mathrm{Tr}\big(\mathcal{Y}_R^2\big) \Big]
\end{align}

\begin{align}\label{Eq:A_BBBB_bar_break_3-Loop}
    \overline{\mathcal{A}}_{BBBB}^{3,\mathrm{break}} &= - \frac{1}{54} \Big[ 6 \, \mathrm{Tr}\big(\mathcal{Y}_R^8\big) + 13 \, \mathrm{Tr}\big(\mathcal{Y}_R^6\big) \mathrm{Tr}\big(\mathcal{Y}_R^2\big) + 48 \, \big(\mathrm{Tr}\big(\mathcal{Y}_R^4\big)\big)^2 \Big]
\end{align}

\paragraph{Fermion 3-Loop Coefficients:}

\begin{align}
    \begin{split}\label{Eq:AfermionInv_3-Loop}
        \mathcal{A}_{\overline{\psi}\psi,ij}^{3,\mathrm{inv}} &= - \frac{1}{3888} \bigg[ 21843 \big(\mathcal{Y}_R^6\big)_{ij} - 4338 \big(\mathcal{Y}_R^4\big)_{ij} \mathrm{Tr}\big(\mathcal{Y}_R^2\big)\\
        &\hspace{1.3cm} - \Big( 2166 \mathrm{Tr}\big(\mathcal{Y}_R^4\big) - 91 \big(\mathrm{Tr}\big(\mathcal{Y}_R^2\big)\big)^2 \Big) \big(\mathcal{Y}_R^2\big)_{ij} + 2430 \mathrm{Tr}\big(\mathcal{Y}_R^5\big) \big(\mathcal{Y}_R\big)_{ij} \bigg]
    \end{split}\\
    \begin{split}\label{Eq:BfermionInv_3-Loop}
        \mathcal{B}_{\overline{\psi}\psi,ij}^{3,\mathrm{inv}} &= - \frac{1}{324} \Big[ 432 \big(\mathcal{Y}_R^6\big)_{ij} - 186 \big(\mathcal{Y}_R^4\big)_{ij} \mathrm{Tr}\big(\mathcal{Y}_R^2\big)\\
        &\hspace{1.3cm} - 6 \big(\mathcal{Y}_R^2\big)_{ij} \mathrm{Tr}\big(\mathcal{Y}_R^4\big) - \big(\mathcal{Y}_R^2\big)_{ij} \big(\mathrm{Tr}\big(\mathcal{Y}_R^2\big)\big)^2 \Big]
    \end{split}\\
    \begin{split}\label{Eq:CfermionInv_3-Loop}
        \mathcal{C}_{\overline{\psi}\psi,ij}^{3,\mathrm{inv}} &= - \frac{1}{6} \big(\mathcal{Y}_R^6\big)_{ij}
    \end{split}
\end{align}

\begin{align}
    \begin{split}\label{Eq:AfermionBreak_3-Loop}
        \overline{\mathcal{A}}_{\overline{\psi}\psi,ij}^{3,\mathrm{break}} &= \frac{1}{18} \bigg[ 79 \big(\mathcal{Y}_R^6\big)_{ij} - \frac{169}{6} \, \big(\mathcal{Y}_R^4\big)_{ij} \mathrm{Tr}\big(\mathcal{Y}_R^2\big)\\
        &\hspace{1.3cm} - \frac{\big(\mathcal{Y}_R^2\big)_{ij}}{108} \Big( 159 \mathrm{Tr}\big(\mathcal{Y}_R^4\big) - 113 \big(\mathrm{Tr}\big(\mathcal{Y}_R^2\big)\big)^2 \Big) + \frac{45}{4} \, \big(\mathcal{Y}_R\big)_{ij} \mathrm{Tr}\big(\mathcal{Y}_R^5\big) \bigg]
    \end{split}\\
    \begin{split}\label{Eq:BfermionBreak_3-Loop}
        \overline{\mathcal{B}}_{\overline{\psi}\psi,ij}^{3,\mathrm{break}} &= \frac{1}{3} \bigg[ \big(\mathcal{Y}_R^6\big)_{ij} - \frac{1}{2} \big(\mathcal{Y}_R^4\big)_{ij} \mathrm{Tr}\big(\mathcal{Y}_R^2\big) + \frac{\big(\mathcal{Y}_R^2\big)_{ij}}{54} \Big( 3 \mathrm{Tr}\big(\mathcal{Y}_R^4\big) + 13 \big(\mathrm{Tr}\big(\mathcal{Y}_R^2\big)\big)^2 \Big) \bigg]
    \end{split}
\end{align}

\subsection{Coefficients of Finite 3-Loop Contributions}\label{App:Finite3LoopCTCoeffs-RightHandedAbelianTheory}

The finite 3-loop counterterm coefficients listed here restore BRST symmetry after subtraction of the divergent terms.

\paragraph{Gauge Boson 3-Loop Coefficients:}

\begin{align}\label{Eq:F_BB_Break_3-Loop}
    \delta F^{(3)}_{BB} = - \frac{1}{21600} \Big[ \big( 35242 + 8448 \, \zeta(3) \big) \mathrm{Tr}\big(\mathcal{Y}_R^6\big) + 1639 \, \mathrm{Tr}\big(\mathcal{Y}_R^4\big) \mathrm{Tr}\big(\mathcal{Y}_R^2\big) \Big]
\end{align}

\begin{align}\label{Eq:F_BBBB_Break_3-Loop}
    \delta F^{(3)}_{BBBB} &= \frac{1}{54} \bigg[ \frac{1387+2592\,\zeta(3)}{10} \, \mathrm{Tr}\big(\mathcal{Y}_R^8\big) +  \frac{101}{20} \, \mathrm{Tr}\big(\mathcal{Y}_R^6\big) \mathrm{Tr}\big(\mathcal{Y}_R^2\big) + 51 \, \big(\mathrm{Tr}\big(\mathcal{Y}_R^4\big)\big)^2 \bigg]
\end{align}

\paragraph{Fermion 3-Loop Coefficients:}

\begin{align}\label{Eq:FfermionBreak_3-Loop}
    \begin{split}
    \delta F^{(3)}_{\overline{\psi}\psi,ij} &=  \bigg( \frac{775}{54} + \frac{58}{9} \, \zeta(3) \bigg) \big(\mathcal{Y}_R^6\big)_{ij} - \frac{10}{9} \, \big(\mathcal{Y}_R^4\big)_{ij} \mathrm{Tr}\big(\mathcal{Y}_R^2\big)\\
    &+ \big(\mathcal{Y}_R^2\big)_{ij} \Bigg[ \bigg( \frac{9725}{3888} + \frac{14}{3} \, \zeta(3) \bigg) \mathrm{Tr}\big(\mathcal{Y}_R^4\big) - \frac{1993}{23328} \, \big(\mathrm{Tr}\big(\mathcal{Y}_R^2\big)\big)^2 \Bigg]\\
    &- \big(\mathcal{Y}_R\big)_{ij} \bigg( \frac{215}{96} - 7 \, \zeta(3) \bigg) \mathrm{Tr}\big(\mathcal{Y}_R^5\big)
    \end{split}
\end{align}

\section{Explicit Results for the 4-Loop Counterterm Coefficients}\label{App:4LoopCTCoeffs-RightHandedAbelianTheory}

This section provides the explicit 4-loop counterterm coefficients required for the renormalisation of the right-handed Abelian theory at this order.
The structure and organisation of the results follow the same pattern as in the previous sections.
All 4-loop results are presented in Feynman gauge ($\xi=1$).

\subsection{Coefficients of Divergent 4-Loop Contributions}\label{App:Divergent4LoopCTCoeffs-RightHandedAbelianTheory}

The coefficients below describe the divergent 4-loop counterterms required for UV renormalisation at this order.

\paragraph{Gauge Boson Four-Loop Coefficients:}

We begin with the coefficients associated with the divergent, BRST-invariant bilinear gauge boson counterterm, see Eq.~\eqref{Eq:CountertermCoeff-GaugeBoson-Inv-4Loop-RHTheory},
\begin{align}
    \begin{split}
    \mathcal{A}_{BB}^{4,\mathrm{inv}} &= 
    \bigg( \frac{1322627}{97200} + \frac{4019 \, \zeta(3)}{675} \bigg) \mathrm{Tr}\big(\mathcal{Y}_R^8\big) + \bigg( \frac{104077}{116640} - \frac{1058 \, \zeta(3)}{2025} \bigg) \mathrm{Tr}\big(\mathcal{Y}_R^6\big) \mathrm{Tr}\big(\mathcal{Y}_R^2\big)\\
    &\quad + \bigg( \frac{537133}{583200} + \frac{34 \, \zeta(3)}{3} \bigg) \mathrm{Tr}\big(\mathcal{Y}_R^4\big)^2
    + \frac{4111}{145800} \, \mathrm{Tr}\big(\mathcal{Y}_R^4\big) \mathrm{Tr}\big(\mathcal{Y}_R^2\big)^2,
    \end{split}\\
    \begin{split}
    \mathcal{B}_{BB}^{4,\mathrm{inv}} &= 
    - \frac{13}{45} \, \mathrm{Tr}\big(\mathcal{Y}_R^8\big)
    - \frac{371}{1080} \, \mathrm{Tr}\big(\mathcal{Y}_R^6\big) \mathrm{Tr}\big(\mathcal{Y}_R^2\big) 
    - \frac{1607}{3240} \, \mathrm{Tr}\big(\mathcal{Y}_R^4\big)^2\\
    &\quad - \frac{707}{14580} \, \mathrm{Tr}\big(\mathcal{Y}_R^4\big) \mathrm{Tr}\big(\mathcal{Y}_R^2\big)^2,
    \end{split}\\
    \begin{split}
    \mathcal{C}_{BB}^{4,\mathrm{inv}} &= 
    \frac{1}{54} \, \mathrm{Tr}\big(\mathcal{Y}_R^6\big) \mathrm{Tr}\big(\mathcal{Y}_R^2\big)
    + \frac{1}{162} \, \mathrm{Tr}\big(\mathcal{Y}_R^4\big)^2
    - \frac{23}{486} \, \mathrm{Tr}\big(\mathcal{Y}_R^4\big) \mathrm{Tr}\big(\mathcal{Y}_R^2\big)^2.
    \end{split}
\end{align}
Next, we present the coefficients for the divergent, BRST-breaking and evanescent counterterm that is bilinear in gauge bosons, see Eq.~\eqref{Eq:CountertermCoeff-GaugeBoson-Break-Evanescent-4Loop-RHTheory},
\begin{align}
    \begin{split}
        \widehat{\mathcal{A}}_{BB}^{4,\mathrm{break}} &=  
        - \bigg( \frac{389961509}{15552000} + \frac{67 \, \pi^4}{9000} + \frac{1119401 \, \zeta(3)}{27000} - \frac{4079 \, \zeta(5)}{45} \bigg) \mathrm{Tr}\big(\mathcal{Y}_R^8\big)\\
        &\quad - \bigg( \frac{19047839}{11664000} + \frac{659 \, \pi^4}{40500} - \frac{8471 \, \zeta(3)}{6750} \bigg) \mathrm{Tr}\big(\mathcal{Y}_R^6\big) \mathrm{Tr}\big(\mathcal{Y}_R^2\big)\\
        &\quad - \bigg( \frac{131895011}{11664000} + \frac{27512 \, \zeta(3)}{675} - \frac{2954 \, \zeta(5)}{45} \bigg) \mathrm{Tr}\big(\mathcal{Y}_R^4\big)^2\\
        &\quad - \frac{102163}{1749600} \mathrm{Tr}\big(\mathcal{Y}_R^4\big) \mathrm{Tr}\big(\mathcal{Y}_R^2\big)^2,
    \end{split}\\
    \begin{split}
        \widehat{\mathcal{B}}_{BB}^{4,\mathrm{break}} &=  
        - \bigg( \frac{582931}{259200} + \frac{67 \, \zeta(3)}{150} \bigg) \mathrm{Tr}\big(\mathcal{Y}_R^8\big) 
        + \bigg( \frac{8263}{12960} - \frac{659 \, \zeta(3)}{675} \bigg) \mathrm{Tr}\big(\mathcal{Y}_R^6\big) \mathrm{Tr}\big(\mathcal{Y}_R^2\big)\\
        &\quad + \frac{361}{2700} \mathrm{Tr}\big(\mathcal{Y}_R^4\big)^2 
        + \frac{1399}{19440} \mathrm{Tr}\big(\mathcal{Y}_R^4\big) \mathrm{Tr}\big(\mathcal{Y}_R^2\big)^2,
    \end{split}\\
    \begin{split}
        \widehat{\mathcal{C}}_{BB}^{4,\mathrm{break}} &= \frac{1231}{4320} \mathrm{Tr}\big(\mathcal{Y}_R^8\big) 
        + \frac{83}{540} \mathrm{Tr}\big(\mathcal{Y}_R^6\big) \mathrm{Tr}\big(\mathcal{Y}_R^2\big) 
        + \frac{7}{1080} \mathrm{Tr}\big(\mathcal{Y}_R^4\big)^2 
        + \frac{\mathrm{Tr}\big(\mathcal{Y}_R^4\big) \mathrm{Tr}\big(\mathcal{Y}_R^2\big)^2}{27},
    \end{split}\\
    \begin{split}
        \widehat{\mathcal{D}}_{BB}^{4,\mathrm{break}} &= - \frac{\mathrm{Tr}\big(\mathcal{Y}_R^8\big)}{72},
    \end{split}
\end{align}
followed by the coefficients for the divergent, BRST-breaking and $4$-dimensional counterterm that is bilinear in gauge bosons, see Eq.~\eqref{Eq:CountertermCoeff-GaugeBoson-Break-4dim-4Loop-RHTheory},
\begin{align}
    \begin{split}
        \overline{\mathcal{A}}_{BB}^{4,\mathrm{break}} &= 
        - \bigg( \frac{2353}{16200} - \frac{4 \, \zeta(3)}{225} \bigg) \mathrm{Tr}\big(\mathcal{Y}_R^8\big)
        - \bigg( \frac{22999}{38880} + \frac{34 \, \zeta(3)}{675} \bigg) \mathrm{Tr}\big(\mathcal{Y}_R^6\big) \mathrm{Tr}\big(\mathcal{Y}_R^2\big)\\ 
        &\quad - \frac{2507}{64800} \mathrm{Tr}\big(\mathcal{Y}_R^4\big)^2 
        - \frac{10327}{1166400} \mathrm{Tr}\big(\mathcal{Y}_R^4\big) \mathrm{Tr}\big(\mathcal{Y}_R^2\big)^2,
    \end{split}\\
    \begin{split}
        \overline{\mathcal{B}}_{BB}^{4,\mathrm{break}} &= 
        \frac{\mathrm{Tr}\big(\mathcal{Y}_R^8\big)}{1080}
        + \frac{13}{3240} \mathrm{Tr}\big(\mathcal{Y}_R^6\big) \mathrm{Tr}\big(\mathcal{Y}_R^2\big)
        + \frac{\mathrm{Tr}\big(\mathcal{Y}_R^4\big)^2}{720}
        + \frac{449}{19440} \mathrm{Tr}\big(\mathcal{Y}_R^4\big) \mathrm{Tr}\big(\mathcal{Y}_R^2\big)^2.
    \end{split}
\end{align}

For the quartic gauge boson counterterm, the coefficients of the divergent, BRST-breaking, and $4$-dimensional contribution, see Eq.~\eqref{Eq:CountertermCoeff-BBBB-Break-4Loop-RHTheory}, are provided by 
\begin{align}
    \begin{split}
        \overline{\mathcal{A}}_{BBBB}^{4,\mathrm{break}} &= 
        \left( \frac{1742}{2025} + \frac{8\zeta(3)}{225} \right) \mathrm{Tr}\big(\mathcal{Y}_R^{10}\big)
        + \left( \frac{75677}{48600} + \frac{128\zeta(3)}{135} \right) \mathrm{Tr}\big(\mathcal{Y}_R^8\big) \mathrm{Tr}\big(\mathcal{Y}_R^2\big) \\
        &+ \frac{12359}{3240} \mathrm{Tr}\big(\mathcal{Y}_R^6\big) \mathrm{Tr}\big(\mathcal{Y}_R^4\big)
        - \frac{3539}{291600} \mathrm{Tr}\big(\mathcal{Y}_R^6\big) \mathrm{Tr}\big(\mathcal{Y}_R^2\big)^2
        - \frac{998}{1215} \mathrm{Tr}\big(\mathcal{Y}_R^5\big)^2\\
        &- \frac{407}{4860} \mathrm{Tr}\big(\mathcal{Y}_R^4\big)^2  \, \mathrm{Tr}\big(\mathcal{Y}_R^2\big),
    \end{split}\\
    \begin{split}
        \overline{\mathcal{B}}_{BBBB}^{4,\mathrm{break}} = 
        &- \frac{1}{270} \mathrm{Tr}\big(\mathcal{Y}_R^{10}\big)
        - \frac{19}{810} \mathrm{Tr}\big(\mathcal{Y}_R^8\big) \mathrm{Tr}\big(\mathcal{Y}_R^2\big)
        - \frac{1}{108} \mathrm{Tr}\big(\mathcal{Y}_R^6\big)\mathrm{Tr}\big(\mathcal{Y}_R^4\big) \\
        &- \frac{347}{4860} \mathrm{Tr}\big(\mathcal{Y}_R^6\big)\mathrm{Tr}\big(\mathcal{Y}_R^2\big)^2
        - \frac{2}{3} \mathrm{Tr}\big(\mathcal{Y}_R^4\big)^2 \, \mathrm{Tr}\big(\mathcal{Y}_R^2\big),
    \end{split}
\end{align}

\paragraph{Fermion Four-Loop Coefficients:}

Analogous to the case of the gauge boson contributions, we proceed to list all coefficients for the fermionic 4-loop counterterms.
We begin with those corresponding to the divergent, BRST-invariant fermionic contributions, see Eq.~\eqref{Eq:CountertermCoeff-Fermion-Inv-4Loop-RHTheory},
\begin{align}
    \begin{split}
        \mathcal{A}_{\overline{\psi}\psi, ij}^{4,\mathrm{inv}} &= 
        \bigg( \frac{156477439}{2592000} + \frac{32773 \zeta(3)}{300} - 160 \zeta(5) \bigg) \big( \mathcal{Y}_R^8 \big)_{ij}\\
        &\quad - \bigg( \frac{14415971}{2592000} + \frac{2357 \zeta(3)}{450} \bigg) \mathrm{Tr} \big( \mathcal{Y}_R^2 \big) \big( \mathcal{Y}_R^6 \big)_{ij}\\
        &\quad -  \bigg[  
        \bigg( \frac{28122617}{1555200} - \frac{3104 \zeta(3)}{225} \bigg) \mathrm{Tr} \big( \mathcal{Y}_R^4 \big)  
        - \frac{3213671}{11664000} \mathrm{Tr} \big( \mathcal{Y}_R^2 \big)^2 \bigg] \big( \mathcal{Y}_R^4 \big)_{ij}\\
        &\quad -
        \bigg( \frac{8177}{2880} - \frac{264 \zeta(3)}{25} \bigg) \mathrm{Tr} \big( \mathcal{Y}_R^5 \big) \big( \mathcal{Y}_R^3 \big)_{ij}\\
        &\quad - 
        \bigg[  
        \bigg( \frac{2252129}{777600} - \frac{547 \zeta(3)}{450} \bigg) \mathrm{Tr} \big( \mathcal{Y}_R^6 \big)
        + \frac{66841}{11664000} \mathrm{Tr} \big( \mathcal{Y}_R^2 \big)^3\\
        &\qquad + \bigg( \frac{161755}{186624} + \frac{1069 \zeta(3)}{450} \bigg) \mathrm{Tr} \big( \mathcal{Y}_R^4 \big) \mathrm{Tr} \big( \mathcal{Y}_R^2 \big) 
        \bigg] \big( \mathcal{Y}_R^2 \big)_{ij} \\
        &\quad + 
        \bigg[
        \bigg( \frac{49367}{9600} - \frac{2 \zeta(3)}{5} \bigg) \mathrm{Tr} \big( \mathcal{Y}_R^7 \big) 
        + \bigg( \frac{1271}{1440} - \frac{257 \zeta(3)}{75} \bigg) \mathrm{Tr} \big( \mathcal{Y}_R^5 \big)  \mathrm{Tr} \big( \mathcal{Y}_R^2 \big) \bigg] \big( \mathcal{Y}_R \big)_{ij} ,
    \end{split}\\
    \begin{split}
        \mathcal{B}_{\overline{\psi}\psi, ij}^{4,\mathrm{inv}} &= 
        \frac{29063}{4800} \big( \mathcal{Y}_R^8 \big)_{ij}
        - \frac{147353}{43200} \mathrm{Tr} \big( \mathcal{Y}_R^2 \big) \big( \mathcal{Y}_R^6 \big)_{ij}\\
        &\quad - \bigg( \frac{1}{90} \mathrm{Tr}\big( \mathcal{Y}_R^4 \big) - \frac{43811}{64800} \mathrm{Tr}\big( \mathcal{Y}_R^2 \big)^2 \bigg) \big( \mathcal{Y}_R^4 \big)_{ij} 
        + \frac{299}{240} \mathrm{Tr}\big( \mathcal{Y}_R^5 \big)  \big( \mathcal{Y}_R^3 \big)_{ij}\\
        &\quad - \bigg( \frac{611}{12960} \mathrm{Tr}\big( \mathcal{Y}_R^6 \big) 
        - \frac{13429}{38880} \mathrm{Tr} \big( \mathcal{Y}_R^4 \big)  \mathrm{Tr} \big( \mathcal{Y}_R^2 \big) 
        - \frac{157}{194400}  \mathrm{Tr}\big( \mathcal{Y}_R^2 \big)^3 \bigg) \big( \mathcal{Y}_R^2 \big)_{ij} \\
        &\quad  - \bigg( \frac{37}{480} \mathrm{Tr}\big( \mathcal{Y}_R^7 \big) 
        + \frac{191}{360} \mathrm{Tr} \big( \mathcal{Y}_R^5 \big) \mathrm{Tr} \big( \mathcal{Y}_R^2 \big) \bigg) \big( \mathcal{Y}_R \big)_{ij},
    \end{split}\\
    \begin{split}
        \mathcal{C}_{\overline{\psi}\psi,ij}^{4,\mathrm{inv}} &= 
        \frac{451}{720} \big( \mathcal{Y}_R^8 \big)_{ij}
        - \frac{1057}{2160} \mathrm{Tr}\big( \mathcal{Y}_R^2 \big)  \big( \mathcal{Y}_R^6 \big)_{ij} 
        - \bigg( \frac{7}{432} \mathrm{Tr}\big( \mathcal{Y}_R^4 \big)
        - \frac{109}{540} \mathrm{Tr}\big( \mathcal{Y}_R^2 \big)^2 \bigg) \big( \mathcal{Y}_R^4 \big)_{ij}\\
        &\quad + \bigg( \frac{\mathrm{Tr}\big( \mathcal{Y}_R^6 \big)}{216}
        + \frac{\mathrm{Tr}\big( \mathcal{Y}_R^4 \big) \mathrm{Tr}\big( \mathcal{Y}_R^2 \big)}{432}
        + \frac{\mathrm{Tr}\big( \mathcal{Y}_R^2 \big)^3}{3240} \bigg) \big( \mathcal{Y}_R^2 \big)_{ij},
    \end{split}\\
    \begin{split}
        \mathcal{D}_{\overline{\psi}\psi,ij}^{4,\mathrm{inv}} &= \frac{1}{24} \big( \mathcal{Y}_R^8 \big)_{ij},
    \end{split}
\end{align}
and continue with the divergent, non-invariant contributions given in Eq.~\eqref{Eq:CountertermCoeff-Fermion-Break-4Loop-RHTheory},
\begin{align}
    \begin{split}
        \overline{\mathcal{A}}_{\overline{\psi}\psi, ij}^{4,\mathrm{break}} &= 
        - \bigg( \frac{29481469}{1296000} + \frac{5461}{900} \zeta(3) \bigg)  
        \big( \mathcal{Y}_R^8 \big)_{ij}  
        +  
        \bigg( \frac{82599047}{7776000} + \frac{2411}{900} \zeta(3) \bigg) 
        \mathrm{Tr} \big( \mathcal{Y}_R^2 \big) \big( \mathcal{Y}_R^6 \big)_{ij}\\  
        &-  
        \bigg[ 
        \bigg( \frac{919571}{388800} + \frac{1198}{225} \zeta(3) \bigg) \mathrm{Tr} \big( \mathcal{Y}_R^4 \big) 
        + \frac{10432717}{11664000} \mathrm{Tr} \big( \mathcal{Y}_R^2 \big)^2 
        \bigg]
        \big( \mathcal{Y}_R^4 \big)_{ij}\\  
        &+  
        \bigg( \frac{8233}{14400} - \frac{707}{100} \zeta(3) \bigg)
        \mathrm{Tr} \big( \mathcal{Y}_R^5 \big) 
        \big( \mathcal{Y}_R^3 \big)_{ij}\\  
        &+ \bigg[  
        \bigg( \frac{1194907}{777600} - \frac{23}{90} \zeta(3) \bigg)
        \mathrm{Tr} \big( \mathcal{Y}_R^6 \big)
        - \frac{22139}{1296000} \mathrm{Tr} \big( \mathcal{Y}_R^2 \big)^3\\  
        &\quad +   
        \bigg( \frac{41257}{186624} + \frac{335}{108} \zeta(3) \bigg)
        \mathrm{Tr} \big( \mathcal{Y}_R^4 \big) \mathrm{Tr} \big( \mathcal{Y}_R^2 \big)
        \bigg] \big( \mathcal{Y}_R^2 \big)_{ij}\\  
        &- \bigg[  
        \bigg( \frac{49367}{9600} - \frac{2}{5} \zeta(3) \bigg)
        \mathrm{Tr} \big( \mathcal{Y}_R^7 \big)
        + \bigg( \frac{1271}{1440} - \frac{257}{75} \zeta(3) \bigg)
        \mathrm{Tr} \big( \mathcal{Y}_R^5 \big) \mathrm{Tr} \big( \mathcal{Y}_R^2 \big)
        \bigg] \big( \mathcal{Y}_R \big)_{ij},
    \end{split}\\
    \begin{split}
        \overline{\mathcal{B}}_{\overline{\psi}\psi, ij}^{4,\mathrm{break}} &= 
        - \frac{16609}{7200} \big( \mathcal{Y}_R^8 \big)_{ij}
        + \frac{254021}{129600} \mathrm{Tr} \big( \mathcal{Y}_R^2 \big) \big( \mathcal{Y}_R^6 \big)_{ij}
        - \frac{149}{240} \mathrm{Tr} \big( \mathcal{Y}_R^5 \big) \big( \mathcal{Y}_R^3 \big)_{ij}\\
        &\quad +  
        \bigg( \frac{121}{810} \mathrm{Tr} \big( \mathcal{Y}_R^4 \big) - \frac{145091}{194400} \mathrm{Tr} \big( \mathcal{Y}_R^2 \big)^2 \bigg) \big( \mathcal{Y}_R^4 \big)_{ij}\\
        &\quad -  
        \bigg( \frac{719}{12960} \mathrm{Tr} \big( \mathcal{Y}_R^6 \big)
        + \frac{107}{864} \mathrm{Tr} \big( \mathcal{Y}_R^4 \big) \mathrm{Tr} \big( \mathcal{Y}_R^2 \big)
        - \frac{2909}{64800} \mathrm{Tr} \big( \mathcal{Y}_R^2 \big)^3 \bigg) 
        \big( \mathcal{Y}_R^2 \big)_{ij}\\
        &\quad + 
        \bigg( \frac{37}{480} \mathrm{Tr} \big( \mathcal{Y}_R^7 \big)  
        + \frac{191}{360} \mathrm{Tr} \big( \mathcal{Y}_R^5 \big) \mathrm{Tr} \big( \mathcal{Y}_R^2 \big) \bigg) 
        \big( \mathcal{Y}_R \big)_{ij},   
    \end{split}\\
    \begin{split}
        \overline{\mathcal{C}}_{\overline{\psi}\psi, ij}^{4,\mathrm{break}} &= 
        - \frac{31}{360} \big( \mathcal{Y}_R^8 \big)_{ij} 
        + \frac{41}{720} \mathrm{Tr} \big( \mathcal{Y}_R^2 \big) \big( \mathcal{Y}_R^6 \big)_{ij}
        - \bigg( \frac{\mathrm{Tr} \big( \mathcal{Y}_R^4 \big)}{54}  
        + \frac{173}{3240} \mathrm{Tr} \big( \mathcal{Y}_R^2 \big)^2 \bigg) \big( \mathcal{Y}_R^4 \big)_{ij}\\
        &\quad + 
        \bigg( \frac{1}{216} \mathrm{Tr} \big( \mathcal{Y}_R^6 \big)  
        + \frac{5}{432} \mathrm{Tr} \big( \mathcal{Y}_R^4 \big) \mathrm{Tr} \big( \mathcal{Y}_R^2 \big)
        + \frac{37}{1080} \mathrm{Tr} \big( \mathcal{Y}_R^2 \big)^3 \bigg) \big( \mathcal{Y}_R^2 \big)_{ij}. 
    \end{split}
\end{align}

\subsection{Coefficients of Finite 4-Loop Contributions}\label{App:Finite4LoopCTCoeffs-RightHandedAbelianTheory}

The following coefficients govern the finite symmetry-restoring 4-loop counterterms that restore gauge and BRST invariance after renormalisation.

\paragraph{Gauge Boson Four-Loop Coefficients:}

The coefficients of the finite symmetry-restoring bilinear and quartic gauge-boson counterterms are
\begin{align}
    \begin{split}\label{Eq:Finite-BB-Break_4-Loop}
        \delta F^{(4)}_{BB} &= 
        \left(\frac{403759}{25920} + \frac{\pi^4}{3375} + \frac{30113\zeta(3)}{4500} - \frac{403\zeta(5)}{45} \right) \mathrm{Tr}\big(\mathcal{Y}_R^8\big)\\
        &+ \left(\frac{4525151}{11664000} - \frac{17\,\pi^4}{20250} - \frac{15569\zeta(3)}{40500} \right) \mathrm{Tr}\big(\mathcal{Y}_R^6\big)\, \mathrm{Tr}\big(\mathcal{Y}_R^2\big)\\
        &+ \left(\frac{38768057}{7776000} + \frac{12061\zeta(3)}{900} - \frac{713\zeta(5)}{45} \right) \mathrm{Tr}\big(\mathcal{Y}_R^4\big)^2\\
        &- \frac{4240349}{69984000} \mathrm{Tr}\big(\mathcal{Y}_R^4\big) \mathrm{Tr}\big(\mathcal{Y}_R^2\big)^2,
    \end{split}\\[1.5ex]
    \begin{split}\label{Eq:Finite-BBBB-Break_4-Loop}
        \delta F^{(4)}_{BBBB} &= 
        - \left( \frac{1123939}{18000} - \frac{2\,\pi^4}{3375} + \frac{43376\zeta(3)}{1125} - \frac{392\zeta(5)}{9} \right) \mathrm{Tr}\big(\mathcal{Y}_R^{10}\big) \\
        &+ \left( \frac{3544039}{2916000} + \frac{32\,\pi^4}{2025} - \frac{599\zeta(3)}{2025} \right) \mathrm{Tr}\big(\mathcal{Y}_R^8\big) \mathrm{Tr}\big(\mathcal{Y}_R^2\big) \\
        &- \left( \frac{666463}{388800} + \frac{10888\zeta(3)}{675} + \frac{40\zeta(5)}{9} \right) \mathrm{Tr}\big(\mathcal{Y}_R^6\big) \mathrm{Tr}\big(\mathcal{Y}_R^4\big) \\
        &+ \frac{6224207}{17496000} \mathrm{Tr}\big(\mathcal{Y}_R^6\big) \mathrm{Tr}\big(\mathcal{Y}_R^2\big)^2 
        + \frac{93341}{291600} \mathrm{Tr}\big(\mathcal{Y}_R^4\big)^2\, \mathrm{Tr}\big(\mathcal{Y}_R^2\big) \\
        &+ \left( \frac{730379}{72900} + \frac{7922\zeta(3)}{675} - \frac{536\zeta(5)}{15} \right) \mathrm{Tr}\big(\mathcal{Y}_R^5\big)^2.
    \end{split}
\end{align}

\paragraph{Fermion Four-Loop Coefficients:}

The finite fermionic counterterm coefficient is
\begin{align}\label{Eq:Finite-FF-Break_4-Loop}
    \begin{split}
        \delta F^{(4)}_{\overline{\psi}\psi,ij} &=
        - \bigg( 
        \frac{2296950361}{25920000} - \frac{113 \pi^4}{18000} + \frac{3803231}{54000} \zeta(3) - \frac{\zeta(5)}{15} 
        \bigg) \big( \mathcal{Y}_R^8 \big)_{ij} \\  
        &+ \bigg( 
        \frac{887862323}{155520000} + \frac{2411 \pi^4}{54000} + \frac{113087}{18000} \zeta(3) 
        \bigg) \mathrm{Tr} \big( \mathcal{Y}_R^2 \big) \big( \mathcal{Y}_R^6 \big)_{ij}\\  
        &+ \bigg[   
        \bigg( \frac{33905827}{583200} - \frac{37 \pi^4}{3375} + \frac{307901}{6750} \zeta(3) - \frac{7153}{45} \zeta(5) \bigg) 
        \mathrm{Tr} \big( \mathcal{Y}_R^4 \big)\\
        &\quad + \bigg( \frac{579755521}{699840000} + \frac{29}{540} \zeta(3) \bigg)
        \mathrm{Tr} \big( \mathcal{Y}_R^2 \big)^2 
        \bigg] \big( \mathcal{Y}_R^4 \big)_{ij} \\  
        &+ \bigg( 
        \frac{1593257}{172800} - \frac{7 \pi^4}{6000} - \frac{729797}{6000} \zeta(3) + \frac{479}{6} \zeta(5) 
        \bigg) 
        \mathrm{Tr} \big( \mathcal{Y}_R^5 \big) \big( \mathcal{Y}_R^3 \big)_{ij}\\  
        &+ \bigg[ 
        \bigg( \frac{975213329}{46656000} - \frac{1433 \pi^4}{81000} + \frac{811729}{27000} \zeta(3) - \frac{1141}{15} \zeta(5) \bigg) \mathrm{Tr} \big( \mathcal{Y}_R^6 \big) \\
        &\quad +
        \bigg( \frac{78192881}{139968000} + \frac{67 \pi^4}{1296} + \frac{299}{400} \zeta(3) \bigg) 
        \mathrm{Tr} \big( \mathcal{Y}_R^4 \big) \mathrm{Tr} \big( \mathcal{Y}_R^2 \big)\\  
        &\quad -  
        \bigg( \frac{3350431}{25920000} - \frac{37}{540} \zeta(3) \bigg)
        \mathrm{Tr} \big( \mathcal{Y}_R^2 \big)^3
        \bigg] \big( \mathcal{Y}_R^2 \big)_{ij}\\  
        &+ \bigg[  
        \bigg( \frac{42763633}{1728000} + \frac{\pi^4}{150} - \frac{5923}{300} \zeta(3) - \frac{263}{10} \zeta(5) \bigg)
        \mathrm{Tr} \big( \mathcal{Y}_R^7 \big)\\  
        &\quad - 
        \bigg( \frac{3833903}{1296000} - \frac{257 \pi^4}{4500} + \frac{1261}{1500} \zeta(3) \bigg) 
        \mathrm{Tr} \big( \mathcal{Y}_R^5 \big) \mathrm{Tr} \big( \mathcal{Y}_R^2 \big)
        \bigg] \big( \mathcal{Y}_R \big)_{ij}.
    \end{split}
\end{align}

\section{Explicit Results for the Auxiliary Counterterm Coefficients}\label{App:AuxiliaryCTs-RightHandedAbelianTheory}

Finally, we list the coefficients of all auxiliary counterterms introduced within the tadpole decomposition framework, as defined in Sec.~\ref{Sec:Tadpole_Decomposition}.
We begin with the 1-loop results and proceed successively up to the 4-loop level.
Within each paragraph, the results are presented first for the coefficient of the auxiliary counterterm arising from the gauge boson self energy (see Eq.~\eqref{Eq:Aux-CT-GaugeBoson-SelfEnergy}), followed by the coefficient associated with the auxiliary counterterm in the operator-inserted ghost--gauge boson 2-point function (see Eq.~\eqref{Eq:Aux-CT-GaugeBoson-SelfEnergy}).
All results are presented in Feynman gauge ($\xi=1$).

\paragraph{1-Loop Counterterm Coefficients:}
At the 1-loop level, the auxiliary counterterm of the gauge boson is governed by
\begin{align}
    {\overline{\mathcal{A}}}{}^{1,\mathrm{mass}}_{BB} &= 2 \,\mathrm{Tr}\big(\mathcal{Y}_R^2\big),
\end{align}
while the auxiliary counterterm arising from the operator-inserted ghost--gauge boson 2-point function reads
\begin{align}
    {\overline{\mathcal{F}}}{}^{1,\mathrm{mass}}_{cB} &= 2 \,\mathrm{Tr}\big(\mathcal{Y}_R^2\big).
\end{align}
We proceed in the same manner for the 2-, 3-, and 4-loop results.

\paragraph{2-Loop Counterterm Coefficients:}

\begin{align}
    \begin{split}
        {\overline{\mathcal{A}}}{}^{2,\mathrm{mass}}_{BB} &= - \frac{5}{2} \, \mathrm{Tr}\big(\mathcal{Y}_R^4\big)
    \end{split}\\
    \begin{split}
        {\overline{\mathcal{B}}}{}^{2,\mathrm{mass}}_{BB} &= \mathrm{Tr}\big(\mathcal{Y}_R^4\big)
    \end{split}
\end{align}

\begin{align}
    \begin{split}
        {\overline{\mathcal{F}}}{}^{2,\mathrm{mass}}_{cB} &= - \frac{25}{6} \, \mathrm{Tr}\big(\mathcal{Y}_R^4\big)
    \end{split}\\
    \begin{split}
        {\overline{\mathcal{A}}}{}^{2,\mathrm{mass}}_{cB} &= \mathrm{Tr}\big(\mathcal{Y}_R^4\big)
    \end{split}
\end{align}

\paragraph{3-Loop Counterterm Coefficients:}

\begin{align}
    \begin{split}
        {\overline{\mathcal{A}}}{}^{3,\mathrm{mass}}_{BB} &= \frac{1727}{108} \mathrm{Tr}\big(\mathcal{Y}_R^6\big) + \frac{91}{108} \mathrm{Tr}\big(\mathcal{Y}_R^4\big) \mathrm{Tr}\big(\mathcal{Y}_R^2\big)
    \end{split}\\
    \begin{split}
        {\overline{\mathcal{B}}}{}^{3,\mathrm{mass}}_{BB} &= - \frac{65}{18} \mathrm{Tr}\big(\mathcal{Y}_R^6\big) - \frac{37}{27} \mathrm{Tr}\big(\mathcal{Y}_R^4\big) \mathrm{Tr}\big(\mathcal{Y}_R^2\big)
    \end{split}\\
    \begin{split}
        {\overline{\mathcal{C}}}{}^{3,\mathrm{mass}}_{BB} &= \frac{1}{3} \mathrm{Tr}\big(\mathcal{Y}_R^6\big)
    \end{split}
\end{align}

\begin{align}
    \begin{split}
        {\overline{\mathcal{F}}}{}^{3,\mathrm{mass}}_{cB} &= \left(\frac{17531}{1296} + \frac{76 \zeta(3)}{9}\right) \mathrm{Tr}\big(\mathcal{Y}_R^6\big) + \frac{1535}{3888} \mathrm{Tr}\big(\mathcal{Y}_R^4\big) \mathrm{Tr}\big(\mathcal{Y}_R^2\big)
    \end{split}\\
    \begin{split}
        {\overline{\mathcal{A}}}{}^{3,\mathrm{mass}}_{cB} &= \frac{101}{108} \mathrm{Tr}\big(\mathcal{Y}_R^6\big) + \frac{37}{108} \mathrm{Tr}\big(\mathcal{Y}_R^4\big) \mathrm{Tr}\big(\mathcal{Y}_R^2\big)
    \end{split}\\
    \begin{split}
        {\overline{\mathcal{B}}}{}^{3,\mathrm{mass}}_{cB} &= - \frac{4}{9} \mathrm{Tr}\big(\mathcal{Y}_R^6\big) - \frac{22}{27} \mathrm{Tr}\big(\mathcal{Y}_R^4\big) \mathrm{Tr}\big(\mathcal{Y}_R^2\big)
    \end{split}
\end{align}

\paragraph{4-Loop Counterterm Coefficients:}

\begin{align}
    \begin{split}
        {\overline{\mathcal{A}}}{}^{4,\mathrm{mass}}_{BB} &= 
        - \bigg(\frac{8625613}{86400} + \frac{9283 \zeta(3)}{450}\bigg) \mathrm{Tr}\big(\mathcal{Y}_R^8\big) 
        + \bigg(\frac{6151}{64800} + \frac{297 \zeta(3)}{50}\bigg) \mathrm{Tr}\big(\mathcal{Y}_R^6\big) \mathrm{Tr}\big(\mathcal{Y}_R^2\big)\\
        &- \bigg(\frac{18455}{1728} + \frac{511 \zeta(3)}{18}\bigg) \mathrm{Tr}\big(\mathcal{Y}_R^4\big)^2 
        + \frac{552119}{466560} \mathrm{Tr}\big(\mathcal{Y}_R^4\big) \mathrm{Tr}\big(\mathcal{Y}_R^2\big)^2
    \end{split}\\
    \begin{split}
        {\overline{\mathcal{B}}}{}^{4,\mathrm{mass}}_{BB} &= 
        \frac{17153}{864} \mathrm{Tr}\big(\mathcal{Y}_R^8\big) 
        + \frac{3817}{648} \mathrm{Tr}\big(\mathcal{Y}_R^6\big) \mathrm{Tr}\big(\mathcal{Y}_R^2\big) 
        + \frac{37}{27} \mathrm{Tr}\big(\mathcal{Y}_R^4\big)^2 
        - \frac{773}{2592} \mathrm{Tr}\big(\mathcal{Y}_R^4\big) \mathrm{Tr}\big(\mathcal{Y}_R^2\big)^2
    \end{split}\\
    \begin{split}
        {\overline{\mathcal{C}}}{}^{4,\mathrm{mass}}_{BB} &= 
        - \frac{15}{8} \mathrm{Tr}\big(\mathcal{Y}_R^8\big) 
        - \frac{25}{27} \mathrm{Tr}\big(\mathcal{Y}_R^6\big) \mathrm{Tr}\big(\mathcal{Y}_R^2\big)
        - \frac{29}{54} \mathrm{Tr}\big(\mathcal{Y}_R^4\big)^2 
        - \frac{475}{648} \mathrm{Tr}\big(\mathcal{Y}_R^4\big) \mathrm{Tr}\big(\mathcal{Y}_R^2\big)^2
    \end{split}\\
    \begin{split}
        {\overline{\mathcal{D}}}{}^{4,\mathrm{mass}}_{BB} &= 
        \frac{1}{12} \mathrm{Tr}\big(\mathcal{Y}_R^8\big)
    \end{split}
\end{align}

\begin{align}
    \begin{split}
        {\overline{\mathcal{F}}}{}^{4,\mathrm{mass}}_{cB} &= 
        \bigg(\frac{82898299}{3888000} + \frac{127 \pi^4}{4500} - \frac{202753 \zeta(3)}{4500} - \frac{666 \zeta(5)}{5}\bigg) \mathrm{Tr}\big(\mathcal{Y}_R^8\big) \\
        &+ \frac{50075761 + 963072 \pi^4 - 101340288 \zeta(3)}{23328000} \, \mathrm{Tr}\big(\mathcal{Y}_R^6\big) \mathrm{Tr}\big(\mathcal{Y}_R^2\big)\\
        &+ \left(\frac{13295063}{583200} + \frac{7 \pi^4}{180} + \frac{19 \zeta(3)}{6} - \frac{381 \zeta(5)}{5}\right) \mathrm{Tr}\big(\mathcal{Y}_R^4\big)^2\\
        &+ \left(\frac{16372549}{13996800} - \frac{41 \zeta(3)}{54}\right) \mathrm{Tr}\big(\mathcal{Y}_R^4\big) \mathrm{Tr}\big(\mathcal{Y}_R^2\big)^2
    \end{split}\\
    \begin{split}
        {\overline{\mathcal{A}}}{}^{4,\mathrm{mass}}_{cB} &= 
        - \frac{533947 - 69216 \zeta(3)}{21600} \, \mathrm{Tr}\big(\mathcal{Y}_R^8\big) 
        + \frac{147677 + 321024 \zeta(3)}{129600} \, \mathrm{Tr}\big(\mathcal{Y}_R^6\big) \mathrm{Tr}\big(\mathcal{Y}_R^2\big) \\
        &- \bigg(\frac{92791}{15552} + 7 \zeta(3)\bigg) \mathrm{Tr}\big(\mathcal{Y}_R^4\big)^2 + \frac{161887}{233280} \mathrm{Tr}\big(\mathcal{Y}_R^4\big) \mathrm{Tr}\big(\mathcal{Y}_R^2\big)^2
    \end{split}\\
    \begin{split}
        {\overline{\mathcal{B}}}{}^{4,\mathrm{mass}}_{cB} &= 
        \frac{899}{216} \mathrm{Tr}\big(\mathcal{Y}_R^8\big) 
        + \frac{3173}{1296} \mathrm{Tr}\big(\mathcal{Y}_R^6\big) \mathrm{Tr}\big(\mathcal{Y}_R^2\big) 
        + \frac{41}{18} \mathrm{Tr}\big(\mathcal{Y}_R^4\big)^2 
        - \frac{49}{144} \mathrm{Tr}\big(\mathcal{Y}_R^4\big) \mathrm{Tr}\big(\mathcal{Y}_R^2\big)^2
    \end{split}\\
    \begin{split}
        {\overline{\mathcal{C}}}{}^{4,\mathrm{mass}}_{cB} &= 
        - \frac{1}{4} \mathrm{Tr}\big(\mathcal{Y}_R^8\big) 
        - \frac{5}{36} \mathrm{Tr}\big(\mathcal{Y}_R^6\big) \mathrm{Tr}\big(\mathcal{Y}_R^2\big) 
        - \frac{55}{108} \mathrm{Tr}\big(\mathcal{Y}_R^4\big)^2 
        - \frac{175}{324} \mathrm{Tr}\big(\mathcal{Y}_R^4\big) \mathrm{Tr}\big(\mathcal{Y}_R^2\big)^2
    \end{split}
\end{align}

\chapter{Results --- The Standard Model}\label{App:Results-TheStandardModel}

Here, we provide the explicit 1-loop results for the counterterm coefficients used in Sec.~\ref{Sec:SM_1-Loop_Renormalisation}.
The coefficients of the divergent counterterms discussed in Sec.~\eqref{Sec:SM_1-Loop_Renormalisation_Singular_CT} are collected in App.~\ref{App:SM-Divergent1LoopCTCoeffs}, while those of the finite-symmetry restoring counterterms in Sec.~\ref{Sec:SM_1-Loop_Renormalisation_Finite_CT} are listed in App.~\ref{App:SM-Finite1LoopCTCoeffs}.

All results were computed as described in Sec.~\ref{Sec:SM_1-Loop_Renormalisation}, are expressed in terms of the modified gauge parameter $\xi_F=1-\xi$ (see Sec.~\ref{Sec:SM-Lagrangian}), and make use of the group invariants and coupling structures defined in Sec.~\ref{Eq:SM-GroupInvariants-CouplingStructures}.
For compactness and computational efficiency, we adopt a notation in which the gauge bosons and generators of all gauge groups are collected into unified objects, as introduced in Sec.~\ref{Sec:Fields_Generators_Notation}.
The following examples illustrate this notation:
\begin{subequations}
\begin{align}
    \mathscr{C}^{ABC} \myGenQu_{L,ab,ij}^C &= 
    \begin{cases}
        g_W^2 \varepsilon^{ABC} t_{L,ab}^C \delta_{ij}, \quad &A,B,C\in\{2,3,4\},\\
        g_s^2 f^{ABC} t_{s,ij}^C \delta_{ab}, \quad &A,B,C\in\{5,\ldots,12\},\\
        0, \quad &\mathrm{else},
    \end{cases} \label{Eq:SM-Notation-Example-1} \\
    \mathcal{C}_2\big(G^{A'}\big) \myGenLe_{L,ab}^A &= 
    \begin{cases}
        g_W^3 C_2(G_L) t_{L,ab}^A, \quad &A\in\{2,3,4\},\\
        0, \quad &\mathrm{else},
    \end{cases} \label{Eq:SM-Notation-Example-2} \\
    \mathcal{C}_2\big(G^{A'}\big) \myGenQu_{L,ab,ij}^A &= 
    \begin{cases}
        g_W^3 C_2(G_L) t_{L,ab}^A \delta_{ij}, \quad &A\in\{2,3,4\},\\
        g_s^3 C_2(G_s) t_{s,ij}^A \delta_{ab}, \quad &A\in\{5,\ldots,12\},\\
        0, \quad &\mathrm{else}.
    \end{cases} \label{Eq:SM-Notation-Example-3}
\end{align}
\end{subequations}
Importantly, adjoint colour indices marked with a prime (such as $A'$ in Eqs.~\eqref{Eq:SM-Notation-Example-2} and \eqref{Eq:SM-Notation-Example-3}) are not implicitly summed over; they serve only as labels indicating which specific component of the adjoint quadratic Casimir (see Eq.~\eqref{Eq:Super-C_A}) --- or of any other object carrying such an index --- contributes.

\section{Counterterm Coefficients of Divergent 1-Loop Contributions}\label{App:SM-Divergent1LoopCTCoeffs}

We begin with the coefficients of the divergent 1-loop contributions.

\paragraph{Fermionic Sector:}
The divergent fermionic counterterm action is provided in Eq.~\eqref{Eq:SM-S_sct-fermion}.
For the leptons, we obtain
\begin{align}
    \begin{split}\label{Eq:SM-Lepton-SE-R-CTs}
        \delta\overline{Z}^{l,(1)}_{R,ab,IJ} = &- \frac{1}{\epsilon} \Big\{ (1-\myxi) \mathcal{C}_2^{ab}\big(F^l_R\big) \delta_{IJ} + \frac{\myNL}{2} \delta_{2a} \delta_{2b} \big(\mathbf{Y}_2(L)\big)_{IJ} \Big\},
    \end{split}\\[1.25ex]
    \begin{split}\label{Eq:SM-Lepton-SE-L-CTs}
        \delta\overline{Z}^{l,(1)}_{L,IJ} = &- \frac{1}{\epsilon} \Big\{ (1-\myxi) \mathcal{C}_2\big(F_L^l\big) \delta_{IJ} + \frac{1}{2} \big(\mathbf{Y}_2(\overline{L})\big)_{IJ} \Big\},
    \end{split}\\[1.25ex]
    \begin{split}\label{Eq:SM-Lepton-SE-Evanescent-CTs}
        \delta\widehat{X}^{l,(1)}_{ab} = &- \frac{1}{\epsilon} \bigg\{ \frac{6-2\myxi}{3} \mathcal{C}_2^{ab}\big(F_L^l,F_R^l\big)
        - \frac{\myxi}{3} \Big[ \mathcal{C}_2^{ab}\big(F_R^l,F_{LR}^l\big) + \mathcal{C}_2^{ab}\big(F_L^l,F_{LR}^l\big) \Big]\\
        &- \frac{3-\myxi}{3} \mathcal{C}_2^{ab}\big(F_{LR}^l\big) \bigg\},
    \end{split}
\end{align}
while for the quarks the coefficients are given by
\begin{align}
    \begin{split}\label{Eq:SM-Quark-SE-R-CTs}
        \delta\overline{Z}^{q,(1)}_{R,ab,IJ} = &- \frac{1}{\epsilon} \Big\{ (1-\myxi) \mathcal{C}_2^{ab}\big(F_R^q\big) \delta_{IJ} + \frac{\myNL}{2} \delta_{2a} \delta_{2b} \big(\mathbf{Y}_2(D)\big)_{IJ} + \delta_{1a} \delta_{1b} \big(\mathbf{Y}_2(U)\big)_{IJ} \Big\},
    \end{split}\\[1.25ex]
    \begin{split}\label{Eq:SM-Quark-SE-L-CTs}
        \delta\overline{Z}^{q,(1)}_{L,IJ}  = &- \frac{1}{\epsilon} \Big\{ (1-\myxi) \mathcal{C}_2\big(F_L^q\big) \delta_{IJ} + \frac{1}{2} \Big[ \big(\mathbf{Y}_2(\overline{D})\big)_{IJ} + \big(\mathbf{Y}_2(\overline{U})\big)_{IJ} \Big] \Big\} ,
    \end{split}\\[1.25ex]
    \begin{split}\label{Eq:SM-Quark-SE-Evanescent-CTs}
        \delta\widehat{X}^{q,(1)}_{ab,IJ} = &- \frac{1}{\epsilon} 
        \bigg\{ 
        \frac{6-2\myxi}{3} \mathcal{C}_2^{ab}\big(F_L^q,F_R^q\big) \delta_{IJ}
        - \frac{\myxi}{3} \Big[ 
        \mathcal{C}_2^{ab}\big(F_R^q,F_{LR}^q\big) + \mathcal{C}_2^{ab}\big(F_L^q,F_{LR}^q\big) 
        \Big] \delta_{IJ}\\
        &- \frac{3-\myxi}{3} \mathcal{C}_2^{ab}\big(F_{LR}^q\big) \delta_{IJ}
        - \frac{1}{2} \Big[ \delta_{2b} \varepsilon_{1a} \big( Y_u Y_d \big)_{IJ} - \delta_{1b} \varepsilon_{2a} \big( Y_d Y_u \big)_{IJ} \Big] \bigg\}.
    \end{split}
\end{align}
The 4-dimensional non-invariant fermion--gauge interactions are governed by
\begin{equation}\label{Eq:SM-Fermion-Gauge-Interaction-4d-CTs}
    \begin{aligned}
        \delta\overline{X}_{\overline{f}\vgb f}^{(1)} = 
        - \frac{4-\myxi}{4\epsilon} \mathcal{C}_2\big(G^{A'}\big).
    \end{aligned}
\end{equation}
The evanescent non-invariant quark--gauge interactions are
\begin{align}\label{Eq:SM-Evanescent-Quark-Gauge-Interaction-CTs}
        \delta&\widehat{X}^{\overline{q}\vgb q,(1),A}_{ab,ij,IJ} = - \kappa_{\mathrm{QCD}}\, \mathcal{C}_2\big(G^{A'}\big) \, \myGenQu^{A}_{R,ab,ij} \, \delta_{IJ}
        \nonumber\\
        &-\frac{1}{2} \myGenScl^{A}_{ce} \delta_{ij} \Big[ \varepsilon_{ea} \delta_{1c} \delta_{2b} \big(Y_uY_d\big)_{IJ} + \varepsilon_{2e} \delta_{ac} \delta_{1b} \big(Y_dY_u\big)_{IJ} \Big] 
        \times \begin{cases}
            (1-2 c_{\mathrm{QED}}) , \quad &A=1\\
            1, \quad &A\in\{2,3,4\}\\
            0, \quad &\mathrm{else}
        \end{cases}
        \nonumber\\
        &= -\frac{1}{\epsilon}
        \begin{cases}
            \frac{g_Y}{2} \mathcal{Y}_S (1-2 c_{\mathrm{QED}}) \delta_{ij} \Big[ \varepsilon_{1a} \delta_{2b} \big(Y_uY_d\big)_{IJ} + \varepsilon_{2a} \delta_{1b} \big(Y_dY_u\big)_{IJ} \Big], \quad &A=1\\
            \frac{g_W}{2} t^A_{L,ce} \delta_{ij} \Big[ \varepsilon_{ea} \delta_{1c} \delta_{2b} \big(Y_uY_d\big)_{IJ} + \varepsilon_{2e} \delta_{ac} \delta_{1b} \big(Y_dY_u\big)_{IJ} \Big], \quad &A\in\{2,3,4\}\\
            \kappa_{\mathrm{QCD}} \, g_s^3 \, C_{2}(G_s) \, t_{s,ij}^A \, \delta_{ab} \, \delta_{IJ}, \quad &A\in\{5,\ldots,12\},
        \end{cases}
\end{align}
with 
\begin{equation}\label{Eq:SM-Kappa_QCD}
    \begin{aligned}
        \kappa_{\mathrm{QCD}} &= \frac{1}{12} \big[ 12-2\myxi-4\myxi c_{\mathrm{QCD}} - (6-5\myxi) c_{\mathrm{QCD}}^2 + (6-2\myxi)c_{\mathrm{QCD}}^3 \big].
    \end{aligned}
\end{equation}

\paragraph{Gauge Sector:}
The divergent gauge-sector counterterm action is presented in Eq.~\eqref{Eq:SM-S_sct_gauge}, with explicit coefficients given by
\begin{align}
    \begin{split}\label{Eq:SM-GaugeBoson-SE-D-dim-CTs}
        \delta Z_{\vgb}^{(1)} &= \frac{1}{\epsilon} \bigg\{ 
        \frac{10+3\myxi}{6} \mathcal{C}_2\big(G^{A'}\big) - \frac{1}{3} \mathcal{S}_2\big(S^{A'}\big) 
        \bigg\},
    \end{split}\\[1.25ex]
    \begin{split}
        \delta\overline{Z}_{\vgb}^{(1)} &= -\frac{2}{3 \epsilon}  N_{f} \sum_{f\in\{l,q\}} \Big[ \mathcal{S}_2\big(F^{f,A'}_R\big) + \mathcal{S}_2\big(F^{f,A'}_L\big) \Big],
    \end{split}\\[1.25ex]
    \begin{split}
        \delta\widehat{Z}^{(1)}_{\vgb} &= -\frac{4}{3\epsilon} N_{f} \sum_{f\in\{l,q\}} \mathcal{S}_2\big(F^{f,A'}_{LR}\big),
    \end{split}\\[1.25ex]
    \begin{split}\label{Eq:SM-GaugeBoson-Evanescent-Breaking-CT-indep-of-evan-generators}
        \delta\widehat{X}^{(1),AB}_{\vgb,12} &= - \frac{1}{3\epsilon} N_{f}
        \bigg\{
        \sum_{f\in\{l,q\}} \Big[ \mathcal{S}_2\big(F^{f,A'}_R\big) + \mathcal{S}_2\big(F^{f,A'}_L\big) \Big] \delta^{AB}\\
        &\qquad\qquad + \sum_{f\in\{l,q\}} \Big[ \mathcal{S}^{AB}_2\big(F^f_R,F^f_L\big) + \mathcal{S}^{AB}_2\big(F^f_L,F^f_R\big) \Big]
        \bigg\}\\
        &= \frac{1}{2} \, \delta\overline{Z}_{\vgb}^{(1)} \delta^{AB} 
        - \frac{N_{f}}{3\epsilon} \sum_{f\in\{l,q\}} \Big[ \mathcal{S}^{AB}_2\big(F^f_R,F^f_L\big) + \mathcal{S}^{AB}_2\big(F^f_L,F^f_R\big) \Big]\\
        &= \delta\widehat{X}^{(1),BA}_{\vgb,12} \, ,
    \end{split}\\[1.25ex]
    \begin{split}
        \delta\widehat{X}^{(1),AB}_{\vgb,34} &= - \frac{2}{3\epsilon} N_{f} \sum_{f\in\{l,q\}} \Big[ \mathcal{S}_2\big(F^{f,A'}_R,F^{f,B'}_{LR}\big) \delta^{AB} + \mathcal{S}^{AB}_2\big(F^f_L,F^f_{LR}\big) \Big]
        = \delta\widehat{X}^{(1),BA}_{\vgb,43} \, ,
    \end{split}\\[1.25ex]
    \begin{split}
        \delta\widehat{X}^{(1),AB}_{\vgb,43} &= - \frac{2}{3\epsilon} N_{f} \sum_{f\in\{l,q\}} \Big[ \mathcal{S}_2\big(F^{f,A'}_{LR},F^{f,B'}_R\big) \delta^{AB} + \mathcal{S}^{AB}_2\big(F^{f}_{LR},F^f_L\big) \Big]
        = \delta\widehat{X}^{(1),BA}_{\vgb,34} \, ,
    \end{split}
\end{align}
and triple gauge boson interaction counterterms
\begin{align}\label{Eq:SM-Triple-Gauge-Interaction-CTs}
    \begin{split}
        \delta Z^{(1)}_{3\vgb} &=  \frac{1}{\epsilon} \bigg\{ 
        \frac{8+9\myxi}{12} \mathcal{C}_2\big(G^{A'}\big) - \frac{1}{3} \mathcal{S}_{2}\big(S^{A'}\big) \bigg\},
    \end{split}\\[1.25ex]
    \begin{split}
        \delta\widehat{X}_{3\vgb}^{(1),ABC} &= - \frac{\mathscr{C}^{ABD}}{2}\big( \delta\widehat{X}^{(1),DC}_{\vgb,34} + \delta\widehat{X}^{(1),CD}_{\vgb,34}\big).
    \end{split}
\end{align}
The coefficients for quartic gauge boson self-interactions read
\begin{align}\label{Eq:SM-4-Gauge-Self-Interaction-CTs}
    \begin{split}
        \delta Z^{(1)}_{4\vgb} 
        &= - \frac{1}{\epsilon} \bigg\{ \frac{1-3\myxi}{3} \, \mathcal{C}_{2}\big(G^{A'}\big) + \frac{1}{3} \, \mathcal{S}_{2}\big(S^{A'}\big) \bigg\},
    \end{split}\\[1.25ex]
    \begin{split}
        \delta \widehat{X}^{(1),ABCD}_{4\vgb} &= - \frac{N_f}{3\epsilon} \sum_{f\in\{l,q\}} \mathcal{S}_{4}^{ABCD}\big(F^{f}_L+F_R^f,F^{f}_L+F_R^f,F_{LR}^f,F_{LR}^f\big),
    \end{split}
\end{align}
and, for the purely evanescent term in the quartic gauge boson self-interaction, we find
\begin{align}
        \delta \widehat{X}^{(1),ABCD}_{4\vgb,\mathrm{ev}} &= - \frac{4 N_f}{3\epsilon} \sum_{f\in\{l,q\}} \mathcal{S}_{4}^{ABCD}\big(F_{LR}^f,F_{LR}^f,F_{LR}^f,F_{LR}^f\big)
        \nonumber\\
        &= -
        \begin{cases}
            \frac{4 N_f}{3\epsilon} \, c_{\mathrm{QCD}}^4 \, g_s^4 \, \big(f^{ACE}f^{BDE} + f^{ADE}f^{BCE}\big), \quad &A,B,C,D \in \{5,\ldots,12\}\\
            0, \quad &\mathrm{else}
        \end{cases}
        \nonumber\\
        &= c_{\mathrm{QCD}}^2 \, \delta\widehat{Z}^{(1)}_{\vgb} \, \Big(\mathscr{C}^{ACE}\mathscr{C}^{BDE} + \mathscr{C}^{ADE}\mathscr{C}^{BCE}\Big)
\end{align}

\paragraph{Higgs Sector:}
The divergent 1-loop counterterm action of the Higgs sector is shown in Eq.~\eqref{Eq:SM-S_sct_Higgs}.
The coefficients of the scalar kinetic terms read
\begin{align}
    \begin{split}\label{Eq:SM-Scalar-SE-CTs-phidagger-phi-D-dim}
        \delta Z^{(1)}_{\phi} &= \frac{1}{\epsilon} \Big[ (2+\myxi) \mathcal{C}_2\big(S\big) - \frac{2}{3} \, \mathcal{H}_2(Y) \Big],
    \end{split}\\[1.25ex]
    \begin{split}
        \delta\overline{Z}^{(1)}_{\phi} &= - \frac{1}{3\epsilon} \, \mathcal{H}_2(Y),
    \end{split}\\[1.25ex]
    \begin{split}
        \delta\widehat{X}^{(1)}_{\phi} \delta_{1a}\delta_{1b} &= - \frac{N_c}{3\epsilon} \Big[ \delta_{1b} \varepsilon_{2a} \trYY{u}{d} + \delta_{1a} \varepsilon_{2b} \trYbarYbar{u}{d} \Big] \\
        &= \frac{1}{3\epsilon} \mathcal{H}_2(D,U) \, \delta_{1a}\delta_{1b} \,,
    \end{split}\\[1.25ex]
    \begin{split}
        \delta Z^{(1)}_{\mu} &= - \frac{1}{\epsilon} \Big[ (1-\myxi) \mathcal{C}_2\big(S\big) - 2 (\myNL+1) \lambda \Big],
    \end{split}\\[1.25ex]
    \begin{split}
        \delta\widehat{X}^{(1)}_{\phi\phi} \delta_{2a}\delta_{2b} &= - \frac{1}{3\epsilon} \bigg\{ \Big[ \trYY{l}{l} + N_c \trYY{d}{d} \Big] \delta_{2a} \delta_{2b} + N_c \trYbarYbar{u}{u} \varepsilon_{1a} \varepsilon_{1b} \bigg\} \\
        &= - \frac{1}{3\epsilon} \mathcal{H}_2(L,D,\overline{U}) \, \delta_{2a}\delta_{2b} \, .
    \end{split}
\end{align}
For the scalar--gauge boson interactions, we find
\begin{align}\label{Eq:SM-Scalar-Gauge-Interaction-CTs}
    \begin{split}
        \delta Z^{(1)}_{\phi^{\dagger}\phi\vgb} 
        &= \frac{1}{\epsilon} \bigg\{ (2+\myxi) \mathcal{C}_2(S) - \frac{4-\myxi}{4} \mathcal{C}_{2}\big(G^{A'}\big) \bigg\},
    \end{split}\\[1.25ex]
    \begin{split}
        \delta \overline{Z}^{(1)}_{\phi^{\dagger}\phi\vgb} 
        &= - \frac{1}{\epsilon} \, \mathcal{H}_2(Y),
    \end{split}\\[1.25ex]
    \begin{split}
        \delta \widehat{X}^{(1),A}_{\phi^{\dagger}\phi\vgb}
        &=
        \begin{cases}
            \begin{aligned}
        - \frac{4}{3\epsilon} c_{\mathrm{QED}} g_Y \mathcal{Y}_S \Big[ \mathcal{H}_2(Y) - \frac{1}{2} \mathcal{H}_2(D,U) \Big], 
        \end{aligned} \quad &A = 1\\
        0, \quad &\mathrm{else}.
        \end{cases}
    \end{split}
\end{align}
The coefficients of the double scalar--double gauge boson interactions are given by
\begin{align}\label{Eq:SM-2-Scalar-2-GaugeBoson-Interaction-CTs}
    \begin{split}
        \delta Z^{(1)}_{\phi^{\dagger}\phi\vgb\vgb}
        &= - \frac{1}{\epsilon} 
        \bigg\{ 
        \frac{4-\myxi}{4} \Big[ \mathcal{C}_2\big(G^{A'}\big) + \mathcal{C}_2\big(G^{B'}\big) \Big] - (2+\myxi) \mathcal{C}_{2}\big(S\big)
        \bigg\},
    \end{split}\\[1.25ex]
    \begin{split}
        \delta \overline{Z}^{(1)}_{\phi^{\dagger}\phi\vgb\vgb} &= - \frac{1}{\epsilon} \, \mathcal{H}_2(Y),
    \end{split}\\[1.25ex]
    \begin{split}
        \delta\widehat{X}^{(1),AB}_{\phi^{\dagger}\phi\vgb\vgb} &= 
        \begin{cases}
            - \frac{16}{3\epsilon} \, c_{\mathrm{QED}}^2 \, g_Y^2 \, \mathcal{Y}_S^2 \Big[ \mathcal{H}_2(Y) + \frac{1}{2} \mathcal{H}_2(D,U) \Big], \quad &A,B=1\\
            0, \quad &\mathrm{else}.
        \end{cases}
    \end{split}
\end{align}
The divergent counterterm of the quartic scalar self-interaction is governed by
\begin{equation}
    \begin{aligned}
        \delta Z^{(1)}_{\lambda} = \frac{1}{\epsilon} \Big[ 2 g_Y^4 \mathcal{Y}_S^4 + \big(\mathcal{C}_2(S)\big)^2 - 2(1-\myxi)\lambda\,\mathcal{C}_2(S) + 12 \lambda^2 - \mathcal{H}_4(Y) \Big].
    \end{aligned}
\end{equation}

\paragraph{Yukawa Sector:}
The divergent Yukawa counterterm action is given in Eq.~\eqref{Eq:SM-S_sct_Yukawa}.
For the lepton Yukawa vertex corrections, we obtain
\begin{equation}\label{Eq:SM-Lepton-Yukawa-Interaction-CTs}
    \begin{aligned}
        \delta Y^{(1)}_{l,IJ} = - \frac{1}{\epsilon} \Big[ (1-\myxi) \mathcal{C}_2(S) + (4-\myxi) \mathcal{C}^{22}_2\big(F_R^l,F_L^l\big) \Big] Y_{l,IJ},
    \end{aligned}
\end{equation}
while the quark Yukawa corrections read
\begin{align}\label{Eq:SM-Quark-Yukawa-Interaction-CTs}
    \begin{split}
        \delta Y^{(1)}_{d,IJ} &= - \frac{1}{\epsilon} \Big[ (1-\myxi) \mathcal{C}_2(S) \, \delta_{IK} + (4-\myxi) \mathcal{C}^{22}_2\big(F_R^q,F_L^q\big) \, \delta_{IK}
        + \big(\mathbf{Y}_2(\overline{U})\big)_{IK} \Big] Y_{d,KJ},
    \end{split}\\[1.25ex]
    \begin{split}
        \delta Y^{(1)}_{u,IJ} &= - \frac{1}{\epsilon} \Big[ (1-\myxi) \mathcal{C}_2(S) \, \delta_{IK} + (4-\myxi) \mathcal{C}^{11}_2\big(F_R^q,F_L^q\big) \, \delta_{IK}
        + \big(\mathbf{Y}_2(\overline{D})\big)_{IK} \Big] Y_{u,KJ}.
    \end{split}
\end{align}

\paragraph{Ghost Sector:}
Finally, for the ghost sector, with counterterm action shown in Eq.~\eqref{Eq:SM-S_sct_ghost}, we find
\begin{align}
    \begin{split}\label{Eq:SM-Ghost-SE-CTs}
        \delta Z^{(1)}_{c} &= \frac{2+\myxi}{4\epsilon} \mathcal{C}_2\big(G^{A'}\big),
    \end{split}\\[1.25ex]
    \begin{split}
        \delta Z^{(1)}_{\overline{c}\vgb c} &= - \frac{1-\myxi}{2\epsilon} \mathcal{C}_2\big(G^{A'}\big).
    \end{split}
\end{align}

\section{Counterterm Coefficients of Finite 1-Loop Contributions}\label{App:SM-Finite1LoopCTCoeffs}

Here, we provide the coefficients of the finite symmetry-restoring 1-loop contributions.

\paragraph{Fermionic Sector:}
We begin with the fermionic sector.
The finite symmetry-restoring counterterm action is given in Eq.~\eqref{Eq:SM-S_fct_fermion}, and the corresponding lepton coefficients read
\begin{align}
    \begin{split}\label{Eq:SM-finite-CT-Leptons-KineticTerm}
        \delta F^{(1)}_{l,ab} &= \frac{3-\xi_F}{3} \mathcal{C}_2^{ab}(F^l_{LR}) - \frac{\xi_F}{3}\mathcal{C}_2^{ab}(F_{L}^l, F^l_{LR}),
    \end{split}\\[1.25ex]
    \begin{split}\label{Eq:SM-finite-CT-Leptons-Right-handed-Interaction}
        \delta F^{(1),A}_{\overline{l}\vgb l,R,ab} &=  
        - \frac{6-\xi_F}{6} \mathcal{C}_2^{ac}(F^l_R) \myGenLe^A_{R,cb} 
        - \frac{\xi_F}{3} \mathcal{C}_2^{ac}(F^l_R,F^l_{LR}) \myGenLe^A_{R,cb}
        - \frac{\xi_F}{6} \mathcal{C}_2^{ac}(F^l_{LR}) \myGenLe^A_{R,cb}\\ 
        &\quad\,\, 
        - \frac{\xi_F}{3} \mathcal{C}_2^{ac}(F^l_L,F^l_{LR}) \myGenLe^A_{R,cb}
        + \frac{6-\xi_F}{6}\big(\myGenLe^B_R \myGenLe^A_L \myGenLe^B_R\big)_{ab} 
        + \frac{6-\xi_F}{6}\big(\widehat{\myGenLe}^B \myGenLe^A_L \widehat{\myGenLe}^B\big)_{ab}\\
        &\quad\,\,  
        + \frac{\xi_F}{3}\big(\widehat{\myGenLe}^B \myGenLe^A_L \myGenLe^B_R\big)_{ab} ,
    \end{split}\\[1.25ex]
    \begin{split}\label{Eq:SM-finite-CT-Leptons-Left-handed-Interaction}
        \delta F^{(1),A}_{\overline{l}\vgb l,L,ab,IJ} &= \bigg\{
        \frac{6-\xi_F}{6} \mathcal{C}_2^{ac}(F^l_{LR}) \myGenLe^A_{R,cb}
        + \frac{\xi_F}{3} \mathcal{C}_2^{ac}(F^l_L,F^l_{LR}) \myGenLe^A_{R,cb}
        - \frac{6-\xi_F}{6} \mathcal{C}_2(F^l_L) \myGenLe^A_{L,ab} \\ 
        &\quad\,\,
        - \frac{\xi_F}{3} \Big[ \mathcal{C}_2^{ac}(F^l_L,F^l_{LR}) \myGenLe^A_{L,cb} + \myGenLe^A_{L,ac} \mathcal{C}_2^{cb}(F^l_L,F^l_{LR}) \Big]\\
        &\quad\,\,
        + \frac{6-\xi_F}{6}\big(\myGenLe^B_L \myGenLe^A_R \myGenLe^B_L\big)_{ab}
        - \frac{\xi_F}{6}\big(\widehat{\myGenLe}^B \myGenLe^A_L \widehat{\myGenLe}^B\big)_{ab} \bigg\} \delta_{IJ}
        + \frac{1}{2} \myGenScl^A_{22} \delta_{ab} \big(\mathbf{Y}_2(\overline{L})\big)_{IJ},
    \end{split}
\end{align}
while the quark coefficients are
\begin{align}
    \begin{split}\label{Eq:SM-finite-CT-Quarks-KineticTerm}
        \delta F^{(1)}_{q,ab} &= \frac{3-\xi_F}{3} \mathcal{C}_2^{ab}(F^q_{LR}) - \frac{\xi_F}{3}\mathcal{C}_2^{ab}(F_{L}^q, F^q_{LR}),
    \end{split}\\[1.25ex]
    \begin{split}\label{Eq:SM-finite-CT-Quarks-Right-handed-Interaction}
        \delta F^{(1),A}_{\overline{q}\vgb q,R,ab,ij} &=  
        - \frac{6-\xi_F}{6} \mathcal{C}_2^{ac}(F^q_R) \myGenQu^A_{R,cb,ij} 
        - \frac{\xi_F}{3} \mathcal{C}_2^{ac}(F^q_R,F^q_{LR}) \myGenQu^A_{R,cb,ij}
        - \frac{\xi_F}{6} \mathcal{C}_2^{ac}(F^q_{LR}) \mathrlap{ \myGenQu^A_{R,cb,ij} }\\ 
        &\quad\,\, 
        - \frac{\xi_F}{3} \mathcal{C}_2^{ac}(F^q_L,F^q_{LR}) \myGenQu^A_{R,cb,ij}
        + \frac{6-\xi_F}{6}\big(\myGenQu^B_R \myGenQu^A_L \myGenQu^B_R\big)_{ab,ij} \\
        &\quad\,\,  
        + \frac{6-\xi_F}{6}\big(\widehat{\myGenQu}^B \myGenQu^A_L \widehat{\myGenQu}^B\big)_{ab,ij}
        + \frac{\xi_F}{3}\big(\widehat{\myGenQu}^B \myGenQu^A_L \myGenQu^B_R\big)_{ab,ij} \\
        &\quad\,\,  
        + \frac{1+c_\mathrm{QCD}}{2} \mathcal{C}_2( G^{A'} ) {\widehat{\myGenQu}}{}^A_{ab,ij},
    \end{split}\\[1.25ex]
    \begin{split}\label{Eq:SM-finite-CT-Quarks-Left-handed-Interaction}
        \delta F^{(1),A}_{\overline{q}\vgb q,L,ab,ij,IJ} 
        &= \bigg\{
        \frac{6-\xi_F}{6} \mathcal{C}_2^{ac}(F^q_{LR}) \myGenQu^A_{R,cb,ij}
        + \frac{\xi_F}{3} \mathcal{C}_2^{ac}(F^q_L,F^q_{LR}) \myGenQu^A_{R,cb,ij}
        - \frac{6-\xi_F}{6} \mathcal{C}_2(F^q_L) \mathrlap{ \myGenQu^A_{L,ab,ij} } \\ 
        &\quad\,\,
        - \frac{\xi_F}{3} \Big[ \mathcal{C}_2^{ac}(F^q_L,F^q_{LR}) \myGenQu^A_{L,cb,ij} + \myGenQu^A_{L,ac,ij} \mathcal{C}_2^{cb}(F^q_L,F^q_{LR}) \Big]\\
        &\quad\,\,
        + \frac{6-\xi_F}{6}\big(\myGenQu^B_L \myGenQu^A_R \myGenQu^B_L\big)_{ab,ij}
        - \frac{\xi_F}{6}\big(\widehat{\myGenQu}^B \myGenQu^A_L \widehat{\myGenQu}^B\big)_{ab,ij} 
        + \frac{3-\xi_F}{6} \mathcal{C}_2( G^{A'} ) {\widehat{\myGenQu}}{}^A_{ab,ij} \bigg\} \delta_{IJ} \\
        &\quad\,\,
        + \frac{1}{2} \myGenScl^A_{22} \delta_{ab} \delta_{ij} \Big[ 
        \big(\mathbf{Y}_2(\overline{D})\big)_{IJ}
        - \big(\mathbf{Y}_2(\overline{U})\big)_{IJ}
        \Big].
    \end{split}
\end{align}

\paragraph{Gauge Sector:}
Next, we turn to the gauge sector, with counterterm action given in Eq.~\eqref{Eq:SM-S_fct_gauge}.
The corresponding coefficients take the form
\begin{align}
    \begin{split}\label{Eq:SM-finite-CT-gauge-boson-SelfEnergy}
        \delta F^{(1),AB}_{2\vgb} &= - \frac{N_f}{6} \!\! \sum_{f\in\{l,q\}} \!\!  \Big[ 
        \mathcal{S}_2\big(F_R^{f,A'}\big) \delta^{AB} + \mathcal{S}_2\big(F_L^{f,A'}\big) \delta^{AB}\\
        &\phantom{= \frac{N_f}{6} \!\! \sum_{f\in\{l,q\}} \!\!  \Big[} 
        - \mathcal{S}^{AB}_2\big(F^{f}_R,F^{f}_L\big) - \mathcal{S}^{BA}_2\big(F^{f}_R,F^{f}_L\big)
        \Big],
    \end{split}\\[1.25ex]
    \begin{split}\label{Eq:SM-finite-CT-gauge-boson-3Point-1}
        \delta F^{(1),ABC}_{3\vgb\text{-1}} &= \frac{N_f}{3} \!\! \sum_{f\in\{l,q\}} \!\! \Big[ 
        \mathcal{S}^{CD}_2\big(F^{f}_R,F^{f}_L\big) \mathscr{C}^{DAB} 
        - \mathcal{S}^{BD}_2\big(F^{f}_R,F^{f}_L\big) \mathscr{C}^{DAC}\\
        &\hspace{1.92cm}
        - 2 \,\mathcal{S}_2\big(F_L^{f,C'}\big) \mathscr{C}^{ABC}
        \Big],
    \end{split}\\[1.25ex]
    \begin{split}\label{Eq:SM-finite-CT-gauge-boson-3Point-2}
        \delta F^{(1),ABC}_{3\vgb\text{-2}} &= - \frac{N_f}{3} \!\! \sum_{f\in\{l,q\}} \!\!
        \mathcal{S}^{CD}_2\big(F^{f}_R,F^{f}_L\big) \mathscr{C}^{DAB},
    \end{split}\\[1.25ex]
    \begin{split}\label{Eq:SM-finite-CT-gauge-boson-4Point}
        \delta F^{(1),ABCD}_{4\vgb} &= - \frac{N_f}{12} \!\! \sum_{f\in\{l,q\}} \!\! \Big[
        5 \myTrbig{F_L^{f,A}, F_L^{f,D}, F_L^{f,B}, F_L^{f,C} }
        - 4 \myTrbig{ F_R^{f,A}, F_L^{f,D}, F_L^{f,B}, F_L^{f,C} }\\
        &\hspace{1.75cm}
        - 2 \myTrbig{ F_R^{f,A}, F_L^{f,D}, F_R^{f,B}, F_L^{f,C} }
        + 2 \myTrbig{ F_R^{f,A}, F_R^{f,D}, F_R^{f,B}, F_L^{f,C} }\\
        &\hspace{1.75cm}
        - \myTrbig{ F_R^{f,A}, F_R^{f,D}, F_R^{f,B}, F_R^{f,C} }
        - 6 \myTrbig{ F_L^{f,A}, F_L^{f,D}, F_L^{f,C}, F_L^{f,B} }\\
        &\hspace{1.75cm}
        + 8 \myTrbig{ F_R^{f,A}, F_L^{f,D}, F_L^{f,C}, F_L^{f,B} }
        + 4 \myTrbig{ F_R^{f,A}, F_L^{f,D}, F_R^{f,C}, F_L^{f,B} }\\
        &\hspace{1.75cm}
        + 2 \myTrbig{ F_R^{f,A}, F_R^{f,D}, F_R^{f,C}, F_L^{f,B} }
        - 8 \myTrbig{ F_R^{f,A}, F_R^{f,D}, F_L^{f,C}, F_L^{f,B} }
        \Big],
    \end{split}
\end{align}
where in the last equation $F_Y^{f,A}$ denotes the fermion generator of chirality $Y\in\{R,L\}$ with adjoint colour index $A$, such that for instance
\begin{align}
    \myTrbig{ F_R^{q,A}, F_L^{q,D}, F_R^{q,B}, F_L^{q,C} } \equiv \myTrbig{ \myGenQu_R^A, \myGenQu_L^D, \myGenQu_R^B, \myGenQu_L^C }.
\end{align}

\paragraph{Higgs Sector:}
The finite symmetry-restoring restoring counterterm action of the Higgs sector is provided by Eq.~\eqref{Eq:SM-S_fct_Higgs}.
The global hypercharge conserving scalar and scalar--gauge boson counterterm coefficients read
\begin{align}
    \begin{split}\label{Eq:SM-finite-CT-Scalar-SelfEnergy}
        \delta F^{(1)}_{\phi^\dagger\phi,ab} &= \frac{1}{6} \mathcal{H}_2(D,U) \sigma^3_{ab},
    \end{split}\\[1.25ex]
    \begin{split}
        \delta F^{(1),A}_{-,ab} &= \frac{1}{6}\bigg[
        \mathcal{H}_2(Y) - \frac{1}{2}\mathcal{H}_2(D,U) 
        \bigg] \{\myGenScl^A, \sigma^3\}_{ab} 
        - \frac{1}{3} \mathcal{H}_2(Y) \myGenScl^A_{ab},
    \end{split}\\[1.25ex]
    \begin{split}
        \delta F^{(1),A}_{+,ab} &= - \frac{1}{4} \mathcal{H}_2(D,U) [\myGenScl^A, \sigma^3]_{ab},
    \end{split}\\[1.25ex]
    \begin{split}\label{Eq:SM-finite-CT-Scalar-DoubleScalar-DoubleGaugeBoson}
        \delta F^{(1),AB}_{\phi^\dagger\phi\vgb\vgb,ab} &= 
        h_1 \{\myGenScl^A, \myGenScl^B\}_{ab} \\
        &
        + h_2 \Big[ 
        (\myGenScl^A \sigma^3 \myGenScl^B)_{ab} 
        + (\myGenScl^B \sigma^3 \myGenScl^A)_{ab} 
        - (\sigma^3 \myGenScl^A \myGenScl^B \sigma^3)_{ab} 
        - (\sigma^3 \myGenScl^B \myGenScl^A \sigma^3)_{ab} \Big]\\
        &
        + h_3 \Big[ 
        (\sigma^3 \myGenScl^A \myGenScl^B)_{ab} 
        + (\sigma^3 \myGenScl^B \myGenScl^A)_{ab} 
        + (\myGenScl^A \myGenScl^B \sigma^3)_{ab} 
        + (\myGenScl^B \myGenScl^A \sigma^3)_{ab} \Big] \\
        &
        + h_4 \Big[ 
        (\sigma^3 \myGenScl^A \sigma^3 \myGenScl^B)_{ab} 
        + (\sigma^3 \myGenScl^B \sigma^3 \myGenScl^A)_{ab} 
        + (\myGenScl^A \sigma^3 \myGenScl^B \sigma^3)_{ab} 
        + (\myGenScl^B \sigma^3 \myGenScl^A \sigma^3)_{ab} \\
        &\qquad
        + (\sigma^3 \myGenScl^A \sigma^3 \myGenScl^B \sigma^3)_{ab} 
        + (\sigma^3 \myGenScl^B \sigma^3 \myGenScl^A \sigma^3)_{ab}\\
        &\qquad
        - 8 (\sigma_{12}^{+} \myGenScl^A \sigma_{03}^{-} \myGenScl^B \sigma_{12}^{-} )_{ab}
        - 8 (\sigma_{12}^{+} \myGenScl^B \sigma_{03}^{-} \myGenScl^A \sigma_{12}^{-} )_{ab} \Big],
    \end{split}
\end{align}
where
\begin{subequations}
\begin{align}
    h_1 &= - \frac{19}{48} \mathcal{H}_2(Y) + \frac{1}{12} \mathcal{H}_2(D,U),\\
    h_2 &=   \frac{7}{48}  \mathcal{H}_2(Y) + \frac{1}{12} \mathcal{H}_2(D,U),\\
    h_3 &=   \frac{5}{48}  \mathcal{H}_2(Y) - \frac{1}{12} \mathcal{H}_2(D,U),\\
    h_4 &=   \frac{3}{48}  \mathcal{H}_2(Y).
\end{align}
\end{subequations}
The global hypercharge violating contributions are governed by 
\begin{align}
    \begin{split}\label{Eq:SM-finite-CT-Scalar-phi-phi}
        \delta F_{\phi \phi,ab}^{(1)} &= \frac{1}{6} \mathcal{H}_2(L,D,\overline{U}) (\sigma_{03}^{-})_{ab},
    \end{split}\\[1.25ex]
    \begin{split}
        \delta F_{\phi \phi \vgb,ab}^{(1),A} &=
        -\frac{1}{2} \mathcal{H}_2(L,D,\overline{U}) \bigg[ 
        \myGenScl^A_{ab}
        - \frac{1}{2}\{\myGenScl^A, \sigma^3\}_{ab} 
        + \big(\sigma_{12}^{+} \myGenScl^A \sigma_{12}^{+}\big)_{ab}
        - \big(\sigma_{03}^{+} \myGenScl^A \sigma_{03}^{-}\big)_{ab}\bigg]\\
        &\quad
        + \varrho_0 \bigg[ 
        \frac{1}{2}[\myGenScl^A, \sigma^3]_{ab} 
        + \big(\sigma_{12}^{+} \myGenScl^A \sigma_{12}^{+}\big)_{ab}
        - \big(\sigma_{03}^{+} \myGenScl^A \sigma_{03}^{-}\big)_{ab} \bigg],
    \end{split}\\[1.25ex]
    \begin{split}\label{Eq:SM-finite-CT-Scalar-phi-phi-V-V}
        \delta F_{\phi\phi \vgb \vgb,ab}^{(1),AB} &=
        \varrho_1 \Big[ 
        \big(\sigma_{12}^+ \myGenScl^A \myGenScl^B \sigma_{12}^+\big)_{ab} 
        + \sigma_{12}^+ \myGenScl^B \myGenScl^A \sigma_{12}^+ \big)_{ab} 
        \Big] \\
        & 
        + \varrho_2 \Big[ 
        \big( \sigma_{03}^- \myGenScl^A \myGenScl^B \sigma_{03}^- \big)_{ab} 
        + \big( \sigma_{03}^- \myGenScl^B \myGenScl^A \sigma_{03}^- \big)_{ab}\\
        &\qquad
        + \big( \sigma_{03}^- \myGenScl^A \myGenScl^B \sigma_{03}^+ \big)_{ab} 
        + \big( \sigma_{03}^- \myGenScl^B \myGenScl^A \sigma_{03}^+ \big)_{ab} 
        \Big]\\
        & 
        + \varrho_3 \Big[ 
        \big( \sigma_{12}^+ \myGenScl^A \sigma_{12}^+ \myGenScl^B \big)_{ab} 
        + \big( \sigma_{12}^+ \myGenScl^B \sigma_{12}^+ \myGenScl^A \big)_{ab}\\
        &\qquad
        - \big( \sigma_{03}^- \myGenScl^A \sigma_{03}^+ \myGenScl^B \big)_{ab} 
        - \big( \sigma_{03}^- \myGenScl^B \sigma_{03}^+ \myGenScl^A \big)_{ab}
        \Big],
    \end{split}
\end{align}
where
\begin{subequations}
\begin{align}
    \varrho_0 &= \frac{1}{6} \Big[\mathcal{H}_2(L,D,\overline{U})  - 4 N_c \mathrm{Tr}\big(Y_u^\dagger Y_d\big)\Big],\\
    \varrho_1 &= \frac{1}{6} \Big[\mathcal{H}_2(L,D,\overline{U})  -   N_c \mathrm{Tr}\big(Y_u^\dagger Y_d\big)\Big],\\
    \varrho_2 &= \frac{1}{4} \mathcal{H}_2(L,D,\overline{U}) ,\\
    \varrho_3 &= \frac{1}{12} \Big[\mathcal{H}_2(L,D,\overline{U}) + 2 N_c \mathrm{Tr}\big(Y_u^\dagger Y_d\big)\Big].
\end{align}
\end{subequations}
Finally, the coefficients of the finite quartic scalar self-interactions are given by 
\begin{align}
    \begin{split}\label{Eq:SM-finite-CT-Scalar-phidagger-phi-phidagger-phi}
        \delta F^{(1)}_{\phi^\dagger\phi\phi^\dagger\phi,abcd} &=
        \frac{1}{6} \Big[ 
        \mathcal{H}_{4}^{(1)} (\sigma_{03}^{+})_{ab} (\sigma_{03}^{+})_{cd} 
        + \mathcal{H}_{4}^{(2)} (\sigma_{03}^{-})_{ab} (\sigma_{03}^{-})_{cd} \Big],
    \end{split}\\[1.25ex]
    \begin{split}
        \delta F^{(1)}_{\phi^\dagger\phi\phi\phi,abcd} &= 
        \frac{1}{9}\big(\mathcal{H}_{4}^{(3)} - \mathcal{H}_{4}^{(4)}\big) 
        \Big[ 
        (\sigma_{12}^+)_{ab} (\sigma_{12}^+)_{cd} 
        + (\sigma_{12}^+)_{ab} (\sigma_{12}^-)_{cd} 
        + (\sigma_{03}^+)_{ab} (\sigma_{03}^-)_{cd}
        \Big]\\
        &
        + \frac{1}{3} \mathcal{H}_{4}^{(3)} (\sigma_{03}^-)_{ab} (\sigma_{03}^-)_{cd},
    \end{split}\\[1.25ex]
    \begin{split}\label{Eq:SM-finite-CT-Scalar-phi-phi-phi-phi}
        \delta F^{(1)}_{\phi\phi\phi\phi,abcd} &=
        \frac{1}{12} \mathcal{H}_4(L,D,\overline{U}) \, (\sigma_{03}^-)_{ab} (\sigma_{03}^-)_{cd}\,,
    \end{split}
\end{align}
where
\begin{align}
    \begin{split}
        \mathcal{H}_{4}^{(1)} &= 
        2\,\mathrm{Tr}\big( Y_l^\dagger Y_l^\dagger Y_l Y_l \big)
        + N_c \Big[
        \mathrm{Tr}\big( Y_d Y_d Y_u Y_u \big)
        + \mathrm{Tr}\big( Y_d Y_u Y_d Y_u \big)
        - 2\,\mathrm{Tr}\big( Y_d Y_u Y_u^\dagger Y_u \big)\\
        &
        - 2\,\mathrm{Tr}\big( Y_d^\dagger Y_d Y_d^\dagger Y_u^\dagger \big)
        - 2\,\mathrm{Tr}\big( Y_d^\dagger Y_d Y_u Y_d \big)
        - 2\,\mathrm{Tr}\big( Y_d^\dagger Y_d Y_u Y_u^\dagger \big)
        + 2\,\mathrm{Tr}\big( Y_d^\dagger Y_d^\dagger Y_d Y_d \big)\\
        &
        + \mathrm{Tr}\big( Y_d^\dagger Y_d^\dagger Y_u^\dagger Y_u^\dagger \big)
        + \mathrm{Tr}\big( Y_d^\dagger Y_u^\dagger Y_d^\dagger Y_u^\dagger \big)
        - 2\,\mathrm{Tr}\big( Y_d^\dagger Y_u^\dagger Y_u Y_d \big)
        - 2\,\mathrm{Tr}\big( Y_d^\dagger Y_u^\dagger Y_u Y_u^\dagger \big)\\
        &
        + 2\,\mathrm{Tr}\big( Y_u^\dagger Y_u^\dagger Y_u Y_u \big)
        \Big],
    \end{split}\\
    \begin{split}
        \mathcal{H}_{4}^{(2)} &= 
        - 2\,\mathrm{Tr}\big( Y_l^\dagger Y_l^\dagger Y_l Y_l \big)
        + N_c \Big[
        \mathrm{Tr}\big( Y_d Y_d Y_u Y_u \big)
        + 2\,\mathrm{Tr}\big( Y_d Y_u Y_u^\dagger Y_u \big)
        + 2\,\mathrm{Tr}\big( Y_d^\dagger Y_d Y_d^\dagger Y_u^\dagger \big)\\
        &
        + 2\,\mathrm{Tr}\big( Y_d^\dagger Y_d Y_u Y_d \big)
        + 2\,\mathrm{Tr}\big( Y_d^\dagger Y_d Y_u Y_u^\dagger \big)
        - 2\,\mathrm{Tr}\big( Y_d^\dagger Y_d^\dagger Y_d Y_d \big)
        + \mathrm{Tr}\big( Y_d^\dagger Y_d^\dagger Y_u^\dagger Y_u^\dagger \big)\\
        &
        + 2\,\mathrm{Tr}\big( Y_d^\dagger Y_u^\dagger Y_u Y_d \big)
        + 2\,\mathrm{Tr}\big( Y_d^\dagger Y_u^\dagger Y_u Y_u^\dagger \big)
        - 2\,\mathrm{Tr}\big( Y_u^\dagger Y_u^\dagger Y_u Y_u \big)
        \Big],
    \end{split}\\
    \begin{split}
        \mathcal{H}_{4}^{(3)} &= 
        2 \Big(
        \mathrm{Tr}\big( Y_l^\dagger Y_l Y_l Y_l \big)
        + N_c \Big[
        \mathrm{Tr}\big( Y_d^\dagger Y_d Y_d Y_d \big)
        + \mathrm{Tr}\big( Y_u^\dagger Y_u^\dagger Y_u^\dagger Y_u \big)
        \Big] \Big),
    \end{split}\\
    \begin{split}
        \mathcal{H}_{4}^{(4)} &= 
        N_c \Big[
        \mathrm{Tr}\big( Y_d Y_d Y_d Y_u \big)
        + \mathrm{Tr}\big( Y_d^\dagger Y_u^\dagger Y_u^\dagger Y_u^\dagger \big)
        \Big].
    \end{split}
\end{align}

\paragraph{Yukawa Sector:}
We continue with the Yukawa sector, with counterterm action given in Eq.~\eqref{Eq:SM-S_fct_Yukawa}.
The lepton contributions are given by
\begin{align}
    \begin{split}\label{Eq:SM-finite-CT-Yukawa-Lepton-F}
        \delta F^{l,(1)}_{Y,abc} &= - \bigg[ 
        \frac{3+\xi_F}{12}\mathcal{C}_2^{ab}(F^l_{R}, F^l_{LR}) + \frac{\xi_F}{6} \mathcal{C}_2^{ab}(F^l_{LR})
        \bigg] \delta_{c2} \, ,
    \end{split}\\
    \begin{split}
        \delta G^{l,(1)}_{Y,abc,IJ} &= Y_{l,JI}^* \bigg[
        \frac{6-\xi_F}{6} \mathcal{C}_2^{cb}(F^l_R)
        + \frac{\xi_F}{2}\mathcal{C}_2^{cb}(F^l_{R},F^l_{LR}) 
        + \frac{6-\xi_F}{3}\mathcal{C}_2^{cb}(F^l_{LR})
        \bigg] \delta_{a2} \, ,
    \end{split}
\end{align}
while the down-type quark contributions are governed by the coefficients
\begin{align}
    \begin{split}
        \delta F^{d,(1)}_{Y,abc} &= \bigg[ 
        \frac{3+\xi_F}{12}\mathcal{C}_2^{ab}(F^q_{R}, F^q_{LR}) 
        + \frac{\xi_F}{6} \mathcal{C}_2^{ab}(F^q_{LR})
        + (1-\xi_F)\mathcal{C}_2^{ab}(F^q_{L}, F^q_{LR}) 
        \bigg] \delta_{c2} \, ,
    \end{split}\\
    \begin{split}
        \delta G^{d,(1)}_{Y,abc,IJ} &=
        Y_{d,JI}^* \bigg[
        \frac{\xi_F}{3} \mathcal{C}_2^{cb}(F^q_{R},F^q_{LR})\delta_{a2} 
        - \frac{6-\xi_F}{3}\mathcal{C}_2^{cb}(F^q_{LR})\delta_{a2} 
        - \xi_F \mathcal{C}_2^{cb}(F^q_{L}, F^q_{LR})\delta_{a2}\\
        &\quad 
        + \frac{6-\xi_F}{3}\mathcal{C}_2^{cb}(F^q_{R}, F^q_L)\delta_{a2} 
        + \frac{4}{3}\xi_F\mathcal{C}_2^{2a}(F^q_L, F^q_{LR})\delta_{bc} 
        + 2 \, \mathcal{C}_2^{2a}(F^q_{R}, F^q_L)\delta_{bc}\\
        &\quad 
        + \frac{2}{3}(6-\xi_F)\mathcal{C}_2^{2b}(F^q_{LR})\delta_{ac} 
        - 2 \, \mathcal{C}_2^{2b}(F^q_{R}, F^q_L)\delta_{ac} \bigg]\\
        &\quad
        + \frac{1}{2} (\sigma_{12}^{-})_{bc} \delta_{a1} (Y_uY_uY_d)_{IJ}
        - \frac{1}{2} \Big[ (\sigma_{12}^{-})_{bc} \delta_{a1} - \delta_{bc} \delta_{a2} \Big] (Y_dY_uY_u)_{IJ} \, ,
    \end{split}
\end{align}
and the up-type quark coefficients take the form
\begin{align}
    \begin{split}
        \delta F^{u,(1)}_{Y,abc} &= 
        \frac{\xi_F}{3} \mathcal{C}_2^{ea}(F^q_{LR}) \varepsilon_{be} \delta_{c1} 
        + \frac{9-\xi_F}{6}\mathcal{C}_2^{ea}(F^q_{L}, F^q_{LR}) \varepsilon_{be}\delta_{c1}
        + 2\,\mathcal{C}_2^{ea}(F^q_{R}, F^q_L)\varepsilon_{be} \delta_{c1}\\
        &\quad
        + \frac{3}{2}\mathcal{C}_2^{ce}(F^q_R)\varepsilon_{eb}\delta_{a1} 
        - \mathcal{C}_2^{ce}(F^q_{R}, F^q_L)\varepsilon_{eb}\delta_{a1}
        + \frac{1}{2}\mathcal{C}_2^{1a}(F^q_R)\varepsilon_{bc} \, ,
    \end{split}\\
    \begin{split}\label{Eq:SM-finite-CT-Yukawa-Up-Quark-G}
        \delta G^{u,(1)}_{Y,abc,IJ} &=
        Y_{u,JI}^* \bigg[
        \frac{6-\xi_F}{3}\mathcal{C}_2^{ec}(F^q_R, F^q_L)\varepsilon_{eb}\delta_{a1}
        + \frac{6+5\xi_F}{6}\mathcal{C}_2^{ec}(F^q_R, F^q_{LR})\varepsilon_{eb}\delta_{a1} \\
        &\quad
        - \mathcal{C}_2^{ec}(F^q_{LR})\varepsilon_{eb}\delta_{a1} 
        - \frac{2+\xi_F}{2}\mathcal{C}_2^{a1}(F^q_R, F^q_{LR})\varepsilon_{cb} 
        - \frac{6-\xi_F}{3}\mathcal{C}_2^{a1}(F^q_{LR})\varepsilon_{cb}\\ 
        &\quad 
        + \frac{6+\xi_F}{3}\mathcal{C}_2^{a1}(F^q_L, F^q_{LR})\varepsilon_{cb}
        - \mathcal{C}_2^{ec}(F^q_{L}, F^q_{LR})\varepsilon_{eb}\delta_{a1} 
        + \frac{15-2\xi_F}{3}\mathcal{C}_2^{ac}(F^q_{LR})\varepsilon_{1b}\\ 
        &\quad 
        - \mathcal{C}_2^{ac}(F^q_{L}, F^q_{LR})\varepsilon_{1b} \bigg]
        - \frac{1}{2} (\sigma_{12}^{+})_{bc} \delta_{a1} (Y_uY_dY_d)_{IJ}
        + \frac{1}{2} (\sigma_{12}^{-})_{bc} \delta_{a1} (Y_dY_dY_u)_{IJ} \, .
    \end{split}
\end{align}

\paragraph{BRST Transformations and External Sources:}
Finally, the finite symmetry-restoring counterterm contribution to the BRST transformations is given by Eq.~\eqref{Eq:SM-S_fct_ext}, with coefficient
\begin{align}\label{Eq:SM-finite-CT-external-Sources}
    \delta F^{(1)}_{R_f} = - \frac{1-\xi_F}{4} \mathcal{C}_2(G^{A'}).
\end{align}

\chapter{Renormalisation of Quantum Electrodynamics}\label{App:Renormalisation_of_QED}

Here, we present the counterterms required for the complete renormalisation of ordinary quantum electrodynamics (QED) up to the 4-loop level.
QED is an Abelian vector-like gauge theory describing fermions $\psi_i$ interacting with the photon $A_\mu$, the gauge boson associated with the symmetry group $U(1)_Q$ and the mediator of the electromagnetic interaction.
The corresponding tree-level QED Lagrangian is given by
\begin{align}
    \mathcal{L} = \overline{\psi}_i i \slashed{D} \psi_i - \frac{1}{4} F^{\mu\nu} F_{\mu\nu} - \frac{1}{2\xi} (\partial^{\mu}A_{\mu})^2
        - \overline{c} \Box c,
\end{align}
which is analogous to that discussed in Sec.~\ref{Sec:Peculiarities_of_Abelian_Gauge_Theories}, except that here all fermions are assumed to carry the same electric charge $Q$.
The covariant derivative is $D_\mu = \partial_\mu + i e Q A_\mu$, the field strength tensor takes the familiar form $F_{\mu\nu}=\partial_\mu A_\nu-\partial_\nu A_\mu$, and $N_f$ denotes the number fermion flavours.

At the $n$-loop level, the counterterm action of QED reads
\begin{align}
    S^{(n)}_{\mathrm{ct,QED}} = 
        \frac{e^{2n}}{(16 \pi^2)^n} \Dintx 
        \bigg\{ 
        \delta Z^{(n)}_{A} \Big( -\frac{1}{4} \, \overline{F}^{\mu\nu} \, \overline{F}_{\mu\nu} \Big)
        + \delta Z^{(n)}_{\psi} \overline{\psi}_i i \slashed{D} \psi_i
        \bigg\}.
\end{align}
The renormalisation constants $Z^{(n)}_{A}$ and $Z^{(n)}_{\psi}$ are adjusted such that they cancel the UV-divergent part of the corresponding 1PI Green functions.
These divergent contributions are extracted using the tadpole decomposition method described in Sec.~\ref{Sec:Tadpole_Decomposition}.
In this method, auxiliary counterterms must be included at intermediate steps of the renormalisation.
At the $n$-loop level, the relevant auxiliary contribution in QED is
\begin{align}
    i \Gamma_{A_{\nu}(-p)A_{\mu}(p)} \big|_{\mathrm{mass}}^{(n)} &= \frac{ie^{2n}}{(16\pi^2)^n} \delta Z_{M}^{(n)} M^2 \eta^{\mu\nu}.
\end{align}
The corresponding auxiliary counterterm is given by the negative of this expression, ensuring its cancellation in the fully renormalised amplitudes.

The explicit results for all renormalisation constants at loop orders $n\in\{1,2,3,4\}$ are presented in the following.

\paragraph{1-Loop Renormalisation Constants:}
\begin{align}
     \begin{split}
          \delta Z_{A}^{(1)} &= -\frac{4}{3}N_{f}Q^{2}\frac{1}{\epsilon},
     \end{split}\\
     \begin{split}
          \delta Z_{\psi}^{(1)} &= -\xi Q^{2}\frac{1}{\epsilon},
     \end{split}\\
     \begin{split}
          \delta Z_{M}^{(1)} &= 4N_{f}Q^{2}\frac{1}{\epsilon}.
     \end{split}
\end{align}

\paragraph{2-Loop Renormalisation Constants:}
\begin{align}
     \begin{split}
          \delta Z_{A}^{(2)} &= -2N_{f}Q^{4}\frac{1}{\epsilon},
     \end{split}\\
     \begin{split}
          \delta Z_{\psi}^{(2)} &= Q^{4} \bigg[ \frac{\xi^{2}}{2}\frac{1}{\epsilon^{2}} + \frac{3+4N_{f}}{4}\frac{1}{\epsilon} \bigg],
     \end{split}\\
     \begin{split}
          \delta Z_{M}^{(2)} &= N_{f} \xi Q^{4} \bigg[ \frac{2}{\epsilon^{2}} - \frac{1}{\epsilon} \bigg].
     \end{split}
\end{align}

\paragraph{3-Loop Renormalisation Constants:}
\begin{align}
     \begin{split}
          \delta Z_{A}^{(3)} &= -\frac{8}{9}Q^{6} \Bigg[ 
          \frac{N_{f}^{2}}{\epsilon^{2}} 
          - \frac{(9+22N_{f})N_{f}}{12}\frac{1}{\epsilon} 
          \Bigg],
     \end{split}\\
     \begin{split}
          \delta Z_{\psi}^{(3)} &= -\frac{Q^{6}}{6} \Bigg[ 
          \frac{\xi^{3}}{\epsilon^{3}} 
          - \bigg( 4N_f\frac{3+4N_f}{3} - \frac{9+12N_f}{2}\xi \bigg) \frac{1}{\epsilon^{2}} 
          + \bigg(3+2N_f\frac{27+20N_f}{9}\bigg) \frac{1}{\epsilon} \Bigg],
     \end{split}\\
     \begin{split}
          \delta Z_{M}^{(3)} &= \frac{2}{3} N_{f} Q^{6} \Bigg[ 
          \frac{\xi^{2}}{\epsilon^{3}} 
          - \bigg( 3+4N_f + \frac{\xi^2}{2} \bigg) \frac{1}{\epsilon^{2}} 
          + \bigg( 5 N_f - \frac{1+\xi^2}{4} \bigg) \frac{1}{\epsilon} 
          \Bigg].
     \end{split}
\end{align}

\paragraph{4-Loop Renormalisation Constants:}
In Feynman gauge, i.e.\ for $\xi=1$, we find
\begin{align}
    \begin{split}
        \delta Z_{A}^{(4)} &= -\frac{16}{27} N_{f} Q^{8} \Bigg[
        \frac{N_{f}^{2}}{\epsilon^{3}} 
        + \frac{63-44 N_f}{24} N_f \frac{1}{\epsilon^{2}}
        - \bigg( \frac{621}{32} - \frac{95-312\zeta(3)}{8} N_f + \frac{77}{36} N_f^2 \bigg) \frac{1}{\epsilon} \Bigg],
    \end{split}\\
    \begin{split}
          \delta Z_{\psi}^{(4)} &= \frac{Q^{8}}{24} \Bigg[ 
          \frac{1}{\epsilon^{4}} 
          + \bigg(9 - 4 N_f - \frac{16}{3} N_f^2 + \frac{64}{3} N_f^3 \bigg) \frac{1}{\epsilon^3}\\
          &\qquad\quad
          + \bigg( \frac{75}{4} + 66 N_f + \frac{484}{9} N_f^2 - \frac{160}{9} N_f^3 \bigg) \frac{1}{\epsilon^2} \\
          &\qquad\quad
          + \bigg( \frac{3081}{4} + 2400 \zeta(3) - 3840 \zeta(5) - (920 - 384\zeta(3)) N_f\\
          &\qquad\quad
          - \frac{608-576\zeta(3)}{3} N_f^2 - \frac{560}{27} N_f^3 \bigg) \frac{1}{\epsilon} \Bigg],
    \end{split}\\
    \begin{split}
          \delta Z_{M}^{(4)} &= \frac{N_{f}Q^{8}}{6} \Bigg[ 
          \frac{1}{\epsilon^{4}} 
          - \bigg( 8 + 18N_f + \frac{32}{3}N_f^2 \bigg) \frac{1}{\epsilon^3} 
          + \bigg( \frac{43}{4} + \frac{76}{3} N_f + \frac{80}{3} N_f^2 \bigg) \frac{1}{\epsilon^2}\\
          &\qquad\quad 
          + \bigg( 30 + 12\zeta(3) + \frac{151-128\zeta(3)}{2}N_f - \frac{872}{27}N_f^2 \bigg) \frac{1}{\epsilon} \Bigg].
    \end{split}
\end{align}

\cleardoublepage

\printbibliography

\cleardoublepage
\ifodd\value{page}%
  \thispagestyle{empty}%
  \mbox{}%
  \newpage%
\fi

\thispagestyle{empty}

\begin{center}
  {\Large\bfseries Selbstständigkeitserklärung\par}
\end{center}

\vspace{1cm}

Hiermit erkläre ich, dass ich die vorliegende Dissertation mit dem Titel \emph{Renormalisation of Chiral Gauge Theories with Non-Anticommuting $\gamma_5$ at the Multi-Loop Level} selbstständig und ohne unzulässige Hilfe Dritter angefertigt habe.
Alle verwendeten Quellen, Hilfsmittel und Beiträge anderer sind im Text oder im Literaturverzeichnis vollständig und eindeutig gekennzeichnet.

\vspace{2cm}

\noindent\rule{6cm}{0.4pt}\\
Matthias Weißwange\\
Dresden, den 27.11.2025

\end{document}